%% file: Thesis.tex
\documentclass[a4paper,12pt,twoside]{book}

\usepackage{amsthm}  

\usepackage{amsmath,amssymb,amsfonts,color,graphicx,cite,feynarts}
\usepackage{axodraw,latexsym}
\usepackage{epsfig,psfrag,rotating,soul}
\usepackage{rotfloat}

\usepackage{hyperref}
\usepackage{fancyhdr}
\usepackage{enumitem}
\usepackage[utf8]{inputenc}
\usepackage{setspace}
\usepackage{textcomp}
\usepackage[retainorgcmds]{IEEEtrantools}
\usepackage{url}
\usepackage{minitoc}
\usepackage{tikz}
\usetikzlibrary{shapes,arrows}

\usepackage[a4paper,twoside, hmarginratio={1:1}]{geometry}
\include{paperdef}

\usepackage{eso-pic}
\begin{document}
\frontmatter
\dominitoc

\pagestyle{empty}
\pagenumbering{roman}

\begin{titlepage}
\thispagestyle{empty}

\begin{center}
{\Huge\bfseries Fenomenolog\'ia de Mezcla de Sabor en Modelos Supersim\'etricos\\[1ex]} 
\vspace*{10em}
\begin{figure}[ht!]
\begin{center}
\psfig{file=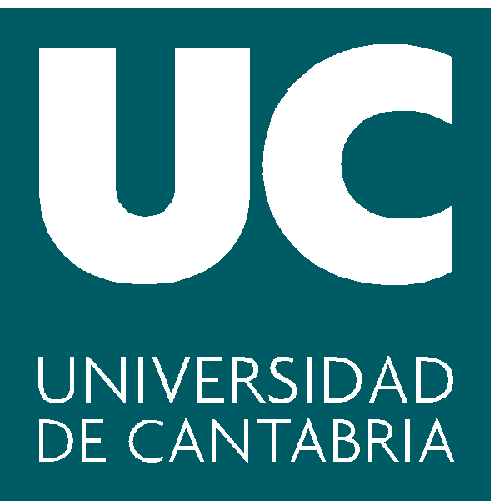  ,scale=0.70}
\end{center}
\end{figure}

\vspace*{15em}

%

%
{\LARGE\bfseries Muhammad Rehman}\\
\vspace*{7em}
{\Large \bf Instituto de F\'{\i}sica de Cantabria, }\\[.2cm]
{\Large \bf Universidad de Cantabria}

\end{center}
\end{titlepage}


\begin{titlepage}
\thispagestyle{empty}

\begin{center}
{\Huge\bfseries Flavor Mixing Phenomenology in Supersymmetric Models\\[1ex]} 
\vspace*{10em}
\begin{figure}[ht!]
\begin{center}
\psfig{file=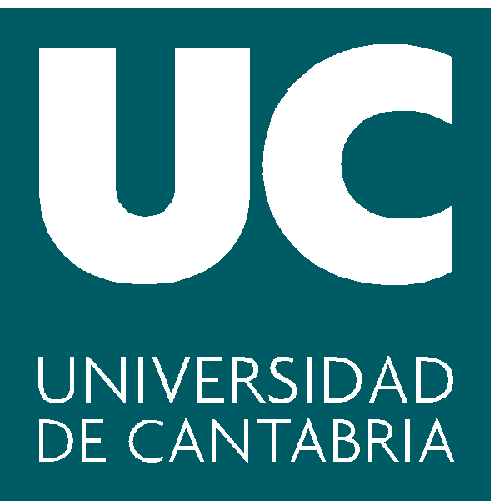,scale=0.70}
\end{center}
\end{figure}

\vspace*{15em}

%

%
{\LARGE\bfseries Muhammad Rehman}\\
\vspace*{7em}
{\Large \bf Instituto de F\'{\i}sica de Cantabria, }\\[.2cm]
{\Large \bf Universidad de Cantabria}

\end{center}
\end{titlepage}

\begin{titlepage}
\thispagestyle{empty}
\begin{center}

\vspace*{1em}
{\Large \bf Instituto de F\'{\i}sica de Cantabria}\\[.2cm]
{\Large \bf Universidad de Cantabr\'{\i}a}\\[7em]

{\Huge\bfseries Fenomenolog\'ia de Mezcla de Sabor en Modelos Supersim\'etricos\\[1ex]} 

\vspace*{3em}
Memoria de Tesis Doctoral realizada por\\
\vspace*{0.5em}
{\LARGE\bfseries Muhammad Rehman}\\
\vspace*{0.5em}
presentada en el Instituto de F\'{\i}sica de Cantabria,\\
Universidad de Cantabr\'{\i}a\\
%

\vfill
Trabajo dirigido por el\\
{\large\bfseries Dr. Sven Heinemeyer},\\
Investigador cient\'{\i}fico del Instituto de F\'{\i}sica de Cantabr\'{\i}a IFCA (CSIC-UC)\\[1em]
y por el\\
{\large\bfseries Dr. Mario E. G\'omez},\\
Investigador cient\'{\i}fico del Departamento de F\'{\i}sica Aplicada, Universidad de Huelva\\[1em]
{\large\bf Santander, Diciembre de 2015 }
%


\end{center}
\end{titlepage}
\begin{titlepage}
\thispagestyle{empty}
\begin{center}

\vspace*{1em}
{\Large \bf Instituto de F\'{\i}sica de Cantabria}\\[.2cm]
{\Large \bf Universidad de Cantabria}\\[7em]

{\Huge\bfseries Flavor Mixing Phenomenology in Supersymmetric Models\\[1ex]} 

\vspace*{3em}
A dissertation submitted in partial fulfillment of the requirements \\ 
for the degree of Doctor of Philosophy in Physics by \\
\vspace*{0.5em}
{\LARGE\bfseries Muhammad Rehman}\\
\vspace*{0.5em}
Presented to Instituto de F\'{\i}sica de Cantabria,\\
Universidad de Cantabria\\
%

\vfill
Work done under the supervision of\\
{\large\bfseries Dr. Sven Heinemeyer},\\
Scientific Investigator, Instituto de F\'{\i}sica de Cantabria IFCA (CSIC-UC)\\[1em]
and\\
{\large\bfseries Dr. Mario E. G\'omez},\\
Scientific Investigator, Departamento de F\'{\i}sica Aplicada, Universidad de Huelva\\[1em]
{\large\bf Santander, December 2015 }
%


\end{center}
\end{titlepage}







\frontmatter
\pagestyle{plain}
\null

%
\vspace*{15em}

\begin{center}
\begin{figure}[ht!]
\psfig{file=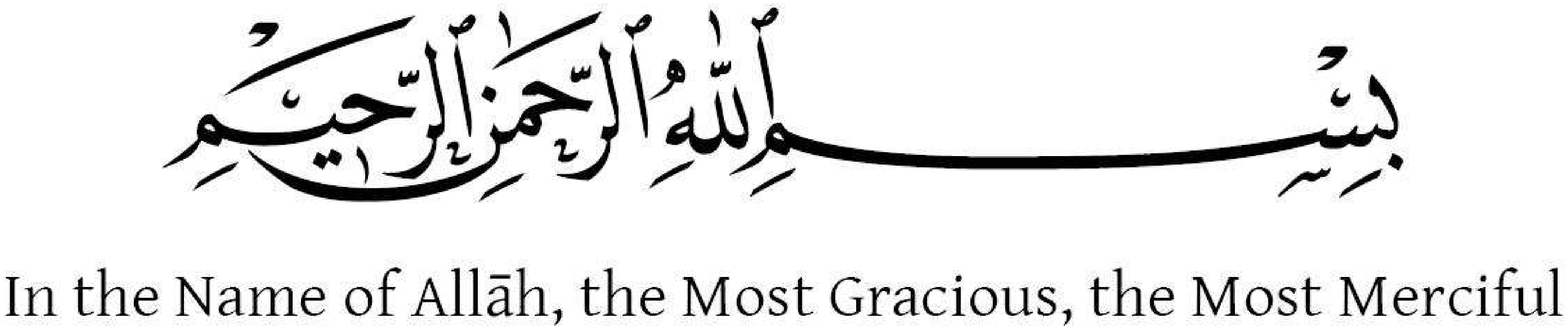  ,scale=0.40}
\end{figure} 
\end{center}

\newpage
\vspace*{20em}
\begin{center}
$\mathcal{MY\ SUCCESS\ CAN\ ONLY\ COME\ FROM\ ALLAH.}$
$\mathcal{IN\ HIM\ I\ TRUST\, AND\ UNTO\ HIM\ I\ LOOK.\ }$
\end{center}

\hspace{25em} (Al Quran 11:88)

\newpage
\vspace*{20em}
$\mathcal{TO\ } $

\begin{center}


$\mathcal{THE\ GUIDING\ LIGHT\ PROPHET\ MUHAMMAD\ (P.B.U.H)\ \&\ } $

$\mathcal{ MY\ FATHER\  MR.\ NAZIR\ HUSSAIN\ }$ 

\end{center}
\tableofcontents

\include{Acknow}

\listoftables
\addcontentsline{toc}{chapter}{List of Tables}

\listoffigures
\addcontentsline{toc}{chapter}{List of Figures}

\mainmatter
\pagenumbering{arabic}

\input intro_es

\input chapters

\input sum_es

\newpage
\renewcommand{\thesection}{\Alph{section}}

\newpage
\begin{spacing}{0.6}
\addcontentsline{toc}{chapter}{Bibliography}
\bibliography{biblio}

\end{spacing}

\end{document}

%% file: Acknow.tex
\chapter*{Acknowledgement}

\addcontentsline{toc}{chapter}{Acknowledgement}

I find no words to praise Allah, who is the sole owner of the treasures of knowledge. He, who inspired all the prophets and sent revelation to enable them to guide the humanity to the path of virtue, also inspired me, enlightened my mind, blessed me with sufficient wisdom and laid bare before me the hidden treasures of knowledge. He guided me through all the labyrinths of life and brought me to this stage that I have succeeded in doing something worthwhile. Whenever, I try to think that what I could do without His mercy, my faith in His benevolence, strengthens.

Infinite salutations be upon the Holy Prophet Muhammad (P.B.U.H) who brought Man out of the depth of ignorance and introduced him to the light of knowledge. In his sayings he laid great stress on acquiring knowledge and aroused deep love for it in his Ummah. His teachings encouraged me and proved to be a torch in the exploration of new realities.

Although on the cover page of this thesis only my name is mentioned but I acknowledge the services of so many people whose names may not all be enumerated. Their contributions and sincere motivations are gratefully acknowledged. I beg pardon to all those whose names I did not mention. But there are some names which I can never forget due to their relentless efforts.   

Bundles of thanks to my reverend supervisor Sven Heinemeyer who allowed me to share his vast knowledge and get maximum experience from his lifelong experience.  I could not have imagined having a better advisor and mentor for my Ph.D study. During the whole research he guided me beyond my expectations. Without his exemplary supervision, I could not do whatever I have done.  He showed an attitude and substance of a genius; he continually and convincingly conveyed a spirit of adventure in regard to research. I do not remember a single moment during the whole course of my study, when I found him non-cooperative or reluctant to help. I wish that one day I would become as good an advisor to my students as he has been to me.   
     
Special thanks to my co-supervisor Mario E Gomez, who provided me valuable information and sagacious suggestions during the entire course of my research. I am greatly indebted to him not only for his dynamic guidance and encouragement but also for his positive and constructive criticism. His lively remarks and encouragement refreshed my mind and gave me new energy to complete this cumbrous work. Without his persistent help this dissertation would not have been possible.

I am deeply thankful to Multidark mangement specially Susana Hernandez (Manager Multidark) and Carlos Mu\~{n}oz (Cooridinator Multidark) for their help and support during my Ph.D. I would like to pay my gratitude to Thomas Hahn, who provided me with every possible help whenever it was needed. This work would not have been possible without his help in the implimentation of new Feynman rules in \fa, \fc\ and \fh. I also would like to thank Barbara Chazin for her help in writing the Spanish part of the thesis. I also cannot forget the earnest cooperation of my friends Federico van der Pahlen and Leo Galeta, who proved very helpful to me during my entire PhD.

Sincere thanks to my brother Muhammad Qamar, sisters, nephews and nieces whose love and friendly gossips lessened my worries, refreshed my mind and encouraged me to complete my work.  Although I lived far from my relatives but communication with them provided emotional atmosphere and concentration that was needed for this cumbersome work. 

I would like to pledge my gratitude and deepest emotion to my wife, Ayesha Rehman for her moral support and cheerful inspiration which did not let me down whenever I found my task tiresome and unaccomplished. I will never forget the time when she presented a cup of hot tea and tasty snacks in the middle of the night, keeping a lively smile on her face. I also would like to mention the names of my cute little daughters Ayerah Rehman and Ayezah Rehman whose innocent smile and lively antics provided a soothing and balmy effect on my burdened mind. 

I bow my head in honor of my father Nazir Hussain (late) and my mother Hameeda Begum, who always loved  me, blessed me with their noble advice, prayed for my success and bestowed upon me sufficient material resources. They spent many sleepless nights while praying for my success and worrying about my health. They always prayed to see the bud of their wishes bloom into a flower. It is all due to them whatever I have achieved today. It is heart wrenching for me to think that when I will return to my country my father will not be there to hug me and appreciate my efforts. After his death, I feel as if I have lost a source of great motivation and strong support.

I also express my deepest and immeasurable appreciations to my dear homeland i.e. Pakistan whose holy soil gave birth to so many talented people. Its  serene environment and cheerful people prompted me to do something worthwhile that can add to its glory. 

Last but not least I would like to acknowledge Spanish Government for providing me financial support through the Spanish MICINN's Consolider-Ingenio 2010 Programme under grant MultiDark CSD2009-00064. I also express my sincere gratitude to the Instituto de Fisica de Cantabria and Universidad de Cantabria for letting my dreams of being a student there be fulfilled. They generously provided me all sufficient means necessary for my research requirements. 

Finally, I would like to say that, apart from me, this research will be immensely important for those who are interested to know about this subject. I hope they will find it valuable. I have tried heart and soul to gather all relevant documents regarding this subject. I do not know how far I am able to do that. Furthermore I do not claim that all the information included in this thesis is described perfectly. There may be shortcomings,  factual errors and mistakes which I confess entirely to be mine. I humbly invite positive criticism for future guidance.

\chapter*{List of Publications}

\addcontentsline{toc}{chapter}{List of Publications}
This thesis is based on the following puplications.
\begin{itemize}
\item M.E.~G{\'o}mez, S.~Heinemeyer and M.~Rehman, {\em The Quark Flavor Violating Higgs Decay \boldmath{\hbs} in the MSSM},
		   {\tt arXiv:1511.04342 [hep-ph]}.
\item M.~G{\'o}mez, S.~Heinemeyer and M.~Rehman, {\em Effects of Sfermion Mixing induced by RGE Running in the Minimal Flavor Violating CMSSM}, 
                    {\em Eur. Phys. J.} {\bf C} (2015) 9, 434
                    {\tt arXiv:1501.02258 [hep-ph]}.
\item M.~G\'omez, T.~Hahn, S.~Heinemeyer, M.~Rehman, {\em Higgs masses and Electroweak Precision Observables in Lepton Flavor Violating MSSM},
                  {\em Phys.\ Rev.} {\bf D 90} (2014) 074016 
                  {\tt arXiv:1408.0663 [hep-ph]}.
\end{itemize}

%% file: intro_es.tex
\chapter*{Introducci\'on}

\addcontentsline{toc}{chapter}{Introducci\'on}

El Modelo Est\'andar (ME) de la f\'{\i}sica de part\'{\i}culas \cite{Glashow, Weinberg, Salam}, fruto de un  inmenso esfuerzo tanto te\'orico como experimental, muestra  la naturaleza de los ingredientes que forman nuestro universo y c\'omo interact\'uan entre s\'{\i}. Seg\'un el  ME,  nuestro universo se compone de fermiones (part\'{\i}culas de spin 1/2), de los que seis son  leptones y otros seis  quarks,  contenidos  en tres familias. A cada fermi\'on le corresponde  una anti-part\'{\i}cula con n\'umeros cu\'anticos opuestos. Las part\'{\i}culas asociadas con los campos de interacci\'on son bosones (part\'{\i}culas de spin 1); los fotones $(\gamma )$ y los bosones l ($W^{\pm }$ y $Z$) se asocian a la interacci\'on electrod\'ebil,  los    gluones ($g$)  a la fuerte.  La gravedad no es parte del ME.  Las simetr\'{\i}as y los principios de invariancia determinan la forma de estas fuerzas, el  ME se basa en el grupo gauge $SU(3)_{\rm C}\ \times SU(2)_{\rm L}\times U(1)_{\rm Y}$.  La  renormalizabilidad e invariancia gauge exigen que la simetr\'{\i}a  $SU(2)_{\rm L}\times U(1)_{\rm Y}$  se rompa espont\'aneamente mediante el llamado mecanismo de Higgs. Todas las predicciones establecidas por ME se han confirmado experimentalmente. El descubrimiento de la \'ultima pieza que faltaba por conocer, el bos\'on de Higgs, se anunci\'o el 4 de julio de 2012 en el gran Colisionador de Hadrones (LHC) del Conseil Europ\'een pour la Recherche Nucleaire (CERN)\cite{ATLAS:2013mma,CMS:yva}. De este modo, el ME es la teor\'{\i}a m\'as precisa y elegante en la actualidad. Sin embargo, a pesar su \'exito,  hay buenas razones tanto  te\'oricas como experimentales que nos llevan m\'as all\'a del ME. De modo que puede pensarse que  ME  es un caso l\'{\i}mite de una teor\'{\i}a m\'as general. 

El primer problema no explicado por el ME est\'a relacionado con el sector de los neutrinos, considerados sin masa en el ME.  En varios experimentos se ha observado la  desaparici\'on del neutrinos electr\'onico o mu\'onico. Esto ha aportado  la evidencia suficiente para aceptar su oscilaci\'on de sabor \cite{Neutrino-Osc}. Esta observaci\'on ha confirmado que los neutrinos tienen masas diferentes  y que los tres sabores de neutrinos $\nu _{e},$ $\nu _{\mu }$ y $\nu _{\tau}$ se mezclan entre a s\'{\i} para formar tres estados propios de masa. Esto implica la no conservaci\'on del sabor lept\'onico, por lo tanto, la predicci\'on de procesos de violacion de sabor de leptones cagados (cLFV), como ocurre en el sector de los quarks.  

La extensi\'on más simple del  ME para acomodar las masas de los neutrinos consiste en introducir tres singletes fermi\'onicos $ SU(3)_{\rm C}\times SU(2)_{\rm L}\times U(1)_{\rm Y}$  y acoplarlos a neutrinos mediante  interacciones Yukawa, las masas de los neutrinos se generar\'{\i}an  a trav\'es de ruptura de simetr\'{\i}a electrod\'ebil (EWSB). Sin embargo esta extensi\'on del ME requiere accoplamientos de Yukawa extremadamente peque\~nos y violar el n\'umero lept\'onico a baja energ\'{\i}a. Como ni los neutrinos dextr\'ogiros ni la violaci\'on del n\'umero lept\'onico se han observado a esa escala, es preciso buscar un mecanismo que explique las masas de los neutrinos respetando ambas evidencias. Una de las soluciones al problema es incluir en la teor\'{\i}a un mecanismo ``see-saw'' (mecanismo de balanc\'{\i}n)  \cite{seesaw:I}  para generar las masas de los neutrinos, el cual no s\'olo permite  los acoplamientos de Yukawa, sino que adem\'as  explica porque los neutrinos lev\'ogiros son m\'as ligeros que los otros fermiones de ME. Estos mecanismos asumen  la existencia de neutrinos  my pesados del tipo de Majorana, los cuales se acoplan a los del ME mediante interacciones Yukawa. Las masas de los neutrinos son generadas por un operador efectivo de dimension 5. Esto da lugar a estados f\'{\i}sicos de neutrinos que mezclan el sabor  y en consecuencia predicen  violaci\'on del  sabor leptones (LFV).

Por otro lado ME tampoco explica suficientemente  el sector de Higgs. Aunque el ME es renormalizable,  se cree que es v\'alido s\'olo hasta cierta escala de energ\'{\i}a, la cual est\'a realcionada con la aparici\'on de f\'{\i}sica desconocida.  Si esta  escala  se asocia con la integraci\'on de la gravedad en teor\'{\i}a, deber\'{\i}a estar en torno a la masa de  Planck (10$^{19}$ GeV). De este modo, las correcciones a la masa del Higgs $M_H$ debidas a los fermiones ser\'{\i}an: 
\begin{equation}
\delta M_{H}^{2}(f)=- \frac{\left\vert \lambda_f \right\vert^2}{8 \pi^2}[\Lambda^2+.....],
\end{equation}%
donde $\lambda_f$ representa el acoplamiento del fermi\'on $f$ al campo de Higgs y $\Lambda$ es el corte ultravioleta utilizado para regular el integral.  Si \'este se reemplaza por  la masa de Planck se obtiene $\delta M_{H}^{2} \approx 10^{30} \gev^2$. Esta enorme correcci\'on se podr\'{\i}a cancelar con  una masa original  del mismo orden y signo opuesto. Sin embargo, estas dos contribuciones se deber\'{\i}an cancelar entre s\'{\i} con una precisi\'on de una parte en $10^{26}$ para explicar la masa del Higgs observada experimentalmete. Este es el llamado ``problema de la jerarqu\'{\i}a''. 

El tercer problema que el ME no explica es el de  la materia oscura (MO). Las primeras especulaciones sobre la existencia de  MO se debieron al astr\'onomo Zwicky. En 1933, observ\'o que la masa total de la materia luminosa procedente del c\'umulo de Coma es mucho menor que la masa total podemos suponer por movimiento de las galaxias que lo integran \cite{Zwicky}. En la actualidad hay diversas muestras de la presencia de MO en nuestro universo. El efecto de lente gravitatoria y las curvas de rotaci\'on  de las galaxias espirales son  observaciones que apuntan a la existencia de la llamada ``masa perdida'' en  el universo. Resultados recientes de los expeimentos WMAP \cite{WMAP} y Planck \cite{Planck} proporcionan un valor preciso de la masa del  universo y la forma en que se divide entre los diferentes tipos de materia y energ\'{\i}a. No hay ninguna part\'{\i}cula del ME que pueda servir como candidato MO.

Por las razones anteriores  hay que encontrar teor\'{\i}as renormalizables que puedan eliminar divergencias cuadr\'aticas en la masa del bos\'on de Higgs, proporcionar un candidato a MO  y explicar el la violaci\'on del sabor lept\'onico. Una  de \'estas,  la extensi\'on supersim\'etrica del ME,  cuya versi\'on m\'as simple es el  {\it Minimal Supersymmetric Standard Model (MSSM)} \cite{mssm}, puede hacer frente a los problemas mencionados. El  MSSM predice la existencia de una pareja  para cada una de las part\'{\i}culas fundamentales de la ME a las que se atribuye un esp\'{\i}n que se diferencia en media unidad de sus compa\~neros del ME. La presencia de las  super-part\'{\i}culas contribuye a  cancelar las divergencias cuadr\'aticas en el bos\'on masa de Higgs. Tambi\'en la part\'{\i}cula supersim\'etrica m\'as ligera (LSP) puede ser un candidato a MO.  Sin embargo, el MSSM, como el ME,  asumen neutrinos sin masa, por lo que el  MSSM tiene que ser ampliado para que sea consistente con las observaciones de las mezclas de sabor de \'estos \cite{Neutrino-Osc}.

Gran parte del esfuerzo del LHC  se ha dedicado a descubrir la supersimetr\'{\i}a (SUSY), pero hasta el momento ninguna part\'{\i}cula SUSY se ha observado\cite{Atlas:SUSY2015,CMS:SUSY2015}. Otro enfoque para descubrir SUSY procede del estudio de los efectos indirectos de las part\'{\i}culas SUSY en otros observables \cite{PomssmRep}. Las mezclas de  sabor ofrecen una perspectiva \'unica en este sentido, ya que la mayor parte de los efectos indirectos de las part\'{\i}culas SUSY proceden de observaciones en los que \'estas se producen.  La primera de ellas, es el proceso de cambio de sabor en el sector de los quarks en corrientes neutras  (FCNC). En ME, los procesos del tipo  FCNC est\'an ausentes a nivel de \'arbol y s\'olo pueden ocurrir en nivel de un bucle. La \'unica fuente de FCNC de en el ME es la matriz de Cabibbo, Kobayashi y Maskawa (CKM), sin embargo no es significativa debido a la cancelación entre las diversas  contribuciones (mecanismo  GIM ).  Por otro lado, en el MSSM, la posible desalineaci\'on entre las matrices de masa de los  quarks y sus parejas supersim\'eticas (squarks)  es otra fuente de violaci\'on de sabor, capaz de superar a  la contribuci\'on ME en varios \'ordenes de magnitud. Cualquier posible desviaci\'on experimental de la predicci\'on de ME para la FCNS  ser\'{\i}a una evidencia clara de nueva f\'{\i}sica y, posiblemente, un indicio del  MSSM. Del mismo modo, las  predicciones del MSSM  para cLFV son cero. Incluso las extensiones del tipo ``see-saw''  del ME no predicen
tasas considerables para estos procesos. Las tasas cLFV en esta extensi\'on del  ME son casi 40 \'ordenes de magnitud menor que las actuales l\'{\i}mites experimentales y, por consiguiente sin posibilidad de ser observadas. En cambio, la extensi\'on ``see-saw''  del MSSM, predice valores m\'as altos, cercanos a los l\'{\i}mites de observaci\'on actuales.  Por otro lado, tras el descubrimiento del bos\'on de Higgs con una masa en torno a los 125 GeV, es preciso incorporar correcciones radiativas grandes para su explicaci\'on.  Una masa superior a 1 TeV de la pareja supersim\'etrica del top, el s-top, podr\'{\i}a dar respuesta al problema, pero a costa de reintroducir un ajuste innatural de los par\'ametros. Sin embargo, esta inconveniencia puede evitarse con una mezcla fuerte entre las compontes quirales del stop o mediante una mezcla de sabor ente los s-quarks.

La forma m\'as general de introducir mezcla de sabor en el MSSM es a trav\'es de los par\'ametros que rompen la supersimetr\'{\i}a. Estos par\'ametros, dotan de masas moderamente grandes a las part\'{\i}culas supersim\'etricas. De este modo, no es posible con una \'unica rotaci\'on en el espacio del sabor diagonalizar simult\'aneamente  las masas de los fermiones y las de sus correspondientes parejas supersim\'etricas. 
Esta desalineaci\'on puede producirse por varias causas; un ejemplo son las ecuaciones de renormalizaci\'on (RGE): Aun partiendo de masas supersim\'etricas sin violaci\'on de sabor a una alta energ\'{\i}a, las RGE pueden inducirla debido a que contienen acoplamientos Yukawa no diagonales.  Este tipo de enfoque es conocido en la literatura como la violaci\'on de sabor m\'{\i}nima  (MFV) ~\cite{MFV1,MFV2}, donde se supone que el sabor y la violaci\'on de la simetr\'{\i}a  CP en el sector de quarks se describe en su totalidad por la matriz CKM.

Los escenerios del tipo  MFV est\'an bien motivados por el hecho de que no introducen  nuevas fuentes de violaci\'on de  sabor y de CP. Los cuales entrar\'{\i}an en  conflicto con los l\'{\i}mites experimentales en  los sectores de los  kaones y $B_d$,  descritos por el ME con una precisi\'on  del\% ~ 10 \cite{HFAgroup}. Para la primera y segunda generaci\'on squarks,  sensibles a los datos de $K^0 -\bar{K}^0$ y $D^0 -\bar{D}^0$, las restricciones son muy fuertes. Sin embargo, sistemas con la tercera generaci\'on est\'an menos limitados, ya que los datos de la mezcla   $B^0 -\bar{B}^0$ a\'un dejan lugar para nuevas fuentes de violaci\'on sabor. Esto abre la posibilidad para scenerios m\'as generales como los de  violaci\'on no m\'{\i}nima de sabor (NMFV), aparte de los de MFV.

En esta tesis, se presenta un estudio sistem\'atico y simult\'aneo de los efectos de la mezcla de sabor en diferentes observables utilizando el MFV y el NMFV. Como un primer paso estudiaremos las mezclas de  squarks y sleptones en el MSSM a baja energ\'{\i}a, sin utilizar un modelo espec\'{\i}fico (MI). Para el enfoque MI, introducimos arbitrariamente los par\'ametros de mezcla de sabor en las matrices de masa de los  sfermiones, sin tener en cuenta el origen de estos par\'ametros. Estudiamos los efectos de la mezcla de los  squark en los observables de la interacci\'on electrod\'ebiles medidos con gran precisi\'on (EWPO), la f\'{\i}sica del quark $b$  (BPO) y  las desintegraciones del bos\'on de Higgs que violan el sabor. Para la mezcla del sabor lept\'onico, estudiamos tambi\'en los efectos sobre EWPO, la masa del los bosones de Higgs y las desintegraciones de \'estos que violan el sabor lept\'onico (LFVHD). En segundo lugar,  extendemos  nuestro an\'alisis a la fuente de la mezcla de sabor.  Para ello analizamos  la mezcla de sabor inducida por las  RGE  en la evoulci\'on de los par\'ametros desde las escalas GUT y electrod\'ebil. En este estudio trabajamos con la hip\'otesis de MFV tanto para squarks como para sleptones. Por consiguiente, vamos a investigar dos modelos (en los siguientes cap\'{\i}tulos se introduciran m\'as definiciones y citas):

\begin{itemize}
\item[(i)] 
El modelo supersim\'etrico m\'{\i}nimo con rotura de la supersimetr\'{\i}a mediante
par\'ame- tros universales (CMSSM). En este caso solo hay violaci\'on
de sabor en los squarks.
\item[(ii)]
El modelo CMSSM ampliado mediante un mecanismo ``see-saw'' de tipo I \cite{seesaw:I},
llamado ``\CMSSMI'' 
\end{itemize}

En muchos an\'alisis del CMSSM o sus extensiones como el NUHM1 o NUHM2 (v\'ease la Ref. \cite{AbdusSalam:2011fc} y las referencias en \'el), se ha utilizado la  hip\'otesis de MFV , asumiendo que contribuciones procedentes de MFV son insignificantes tanto  para procesos FCNC como para otros  observables como EWPO y la masa del bos\'on de  Higgs masas ( ver por ejemplo \cite{CMSSM-NUHM}).  En este trabajo vamos a analizar si esta suposici\'on est\'a justificada, y si la inclusi\'on de los efectos MFV podr\'{\i}an conducir a restricciones adicionales del espacio de par\'ametros del CMSSM. En este sentido, vamos a evaluar en el CMSSM y en el CMSSMI  el  siguiente conjunto de observables: BPO, en particular, \bsg, \bmm\ y \dmbs;  EWPO, en particular, $\MW$ y el \'angulo de Weinberg efectivo, $\sweff$;  la masas de los bosones de Higgs neutros y cargados en el MSSM, as\'{\i} como cLFV y LFVHD.

La disposici\'on de la tesis es la siguiente. El cap\'{\i}tulo 1 contiene una  introducci\'on al  ME. En el cap\'{\i}tulo 2 presentamos MSSM y sus extensiones ``see-saw''.  El cap\'{\i}tulo 3 est\'a dedicado a la base de c\'alculo de los observables considerados en este trabajo. En el  cap\'{\i}tulo 4 presentaremos  los resultados en el caso de la mezcla de sabor squark en el enfoque MI y el estudio sus efectos para los observables BPO, EWPO y QFVHD. En el cap\'{\i}tulo 5, de estudian los efectos de la mezcla de los sleptones en  EWPO, las correciones a las masas de los Higgs y LFVHD en el contexto  MI. El cap\'{\i}tulo 6 se centrar\'a en el an\'alisis del CMSSM y CMSSMI, para los que presentamos los efectos de mezcla del  sabor en los observabels  EWPO, BPO, las predicciones de masas del bos\'on de Higgs, QFVHD, cLFV mezcla y LFVHD. El cap\'{\i}tulo 7  se reserva para las conclusiones.

%% file: chapters.tex
\chapter*{Introduction}

\addcontentsline{toc}{chapter}{Introduction}

The Standard Model (SM) of the fundamental interactions\cite{Glashow, Weinberg, Salam}, the results of immense experimental
and theoretical effort, elucidates the ingredients forming our universe and
how do they interact. SM asserts that our universe is made up of fermions
(spin 1/2 particles) interacting through fields of which they are the
sources. Among the fermions, there are six leptons and six quarks
categorized in three families and have their respective anti particles with
opposite quantum numbers. The particles associated with the interaction
fields are bosons (spin 1 particles) namely photon $(\gamma ),$ weak vector
boson ($W^{\pm }$, $Z$) and gluons ($g$) and a scalar particle Higgs ($H$). The gauge bosons act as
force carriers of electromagnetic, weak and strong interactions. Gravity is
not part of SM. Symmetries and invariance principles determine the form
of these forces. SM is based on the gauge group $SU(3)_{\rm C}\times SU(2)_{\rm L}\times U(1)_{\rm Y}$. The renormalizability and gauge invariance demands
the $SU(2)_{\rm L}\times U(1)_{\rm Y}$ symmetry to be spontaneously broken through
Higgs mechanism. All the predictions laid down by SM have been
experimentally confirmed. The discovery of the last missing piece namely the Higgs boson at large hadron collider (LHC) was
announced on 4th July 2012 at Conseil Europeen pour la Recherche Nucleaire (CERN) \cite{ATLAS:2013mma,CMS:yva}, proving SM\ the most accurate and elegant theory at present. In spite of all its successes SM is believed to be
a limiting case of a more general theory. There are well motivated theoretical
as well as experimental reasons which coerce us to go beyond the SM.

The first problem of the SM is related to the neutrino sector.
Neutrinos are strictly massless in the SM. Several key experiments with
solar, atmospheric, reactor and accelerator neutrinos observed the disappearance of electron or muon neutrinos, the
evidence enough for scientists to acquiesce neutrino oscillation\cite{Neutrino-Osc}. 
This observation has confirmed that neutrinos have distinct
masses and that 3 neutrino flavors $\nu _{e},$ $\nu _{\mu }$ and $\nu _{\tau
}$ mix among themselves to form 3 mass eigenstates. The fact that neutrinos
are massive and mix implies non-conservation of lepton flavor, hence charged
lepton flavor violating processes (cLFV) are expected in lepton sector just
as quark flavor violating processes arise in quark sector. 

The trivial extension to SM to accomodate neutrino masses is to introduce three 
fermionic $ SU(3)_{\rm C}\times SU(2)_{\rm L}\times U(1)_{\rm Y}$ singlets (missing right handed neutrinos) and write down
the neutrino Yukawa couplings which generates neutrino masses via electroweak symmetry breaking (EWSB).
However this extension of SM requires extremely small Yukawa couplings and violate lepton number at low energy scale.
As right handed neutrinos and lepton number violation has not been observed at low energy, one should look
for a mechanism that can generate masses for left handed neutrinos at low energy and also respect
the non observation of right handed neutrinos and lepton number violation.
One of the solutions to overcome this problem is the so called ``seesaw mechanism'' \cite{seesaw:I} which can be used
to generate neutrino masses which not only allow large neutrino Yukawa couplings but also explain why left handed
neutrinos are lighter compared to other SM fermions. These mechanisms assume the nature of neutrinos to
be Majorana and existance of very massive particle that couple to the neutrinos in Yukawa analogue.
The neutrino masses are then generated by an effective dimension 5 operator.  
This gives rise to neutrino physical states which are not flavor diagonal and consequently generate lepton flavor violation (LFV).

On the other hand SM also lack sufficient explaination in Higgs sector.
For example SM is renormalizable, yet it is believed that SM is valid only up to some
cut-off energy scale. This cut-off can be related to the scale where new physics appear,  for example the Planck scale (10$^{19}$ GeV) where quantum gravity becomes important. One-loop corrections to the Higgs mass $M_H$  due to a fermion $f$ in the loop are given by
\begin{equation}
\delta M_{H}^{2}(f)=- \frac{\left\vert \lambda_f \right\vert^2}{8 \pi^2}[\Lambda^2+.....],
\end{equation}%
where $\lambda_f$ represents the fermion coupling to the Higgs field and $\Lambda$ is the ultraviolet cutoff used to regulate the loop integral. If one replace the cutoff by the Planck mass one obtains $\delta M_{H}^{2} \approx 10^{30} \gev^2$. One could
cancel these large correction with a bare mass of the same order and opposite
sign. However, these two contributions should cancel with a precision of one
part in $10^{26}$  to provide the observed Higgs mass. This is the so-called ``hierarchy problem''.

Third and equally important issue is the Dark Matter (DM). First speculation about the DM was due to
astronomer Zwicky. In 1933, he observed that the mass from the luminous
matter coming from COMA cluster is much smaller than the total mass we can
derivate from the motion of the cluster member galaxies \cite{Zwicky}. There are now
several pieces of observational evidence for DM in our universe.
Gravitational lensing and the unexpected rotational curves of spiral
galaxies are among these observations that point to there being so-called
``missing mass'' throughout the universe. Recent results from WMAP\cite{WMAP} and PLANCK\cite{Planck}
give us our most accurate value for the total mass in the universe and how
it is divided among different types of matter and energy. There is no SM particle that can serve as a DM candidate.

Due to all these reasons one needs to find renormalizable theories that can
remove quadratic divergences in the Higgs boson mass, a theory that can provide us
with a DM candidate and can explain LFV. Supersymmetric extension of the SM\ namely
Minimal Supersymmetric Standard Model (MSSM)\cite{mssm}, is technically well equipped to
deal with above mentioned discrepancies. The MSSM predicts the existence of a
super-partner for each of the fundamental degree of freedom of the SM with spin
differing by half unit. The presence of super-partners called sparticles in
the loop cancels the quadratic divergences in the Higgs boson mass. Also
the lightest supersymmetric particle (LSP) can be a DM candidate. However, the MSSM, like SM, assume
neutrinos to be massless so simple version of MSSM has to be extended with a mechanism like the seesaw to make it consistent with experimental observation of neutrino masses and mixing\cite{Neutrino-Osc}.

Much of the effort has been devoted at the LHC to discover supersymmetry (SUSY). But as of yet no SUSY particle has been observed at the colliders\cite{Atlas:SUSY2015,CMS:SUSY2015}. Another approach to discover SUSY could be to study the indirect effects of the SUSY particles on other observables\cite{PomssmRep}. Flavor mixing offer a unique prospective in this regard since most of the indirect effects of the SUSY particles involve the flavor mixing observables. 
First among these are the Flavor Changing Neutral Current (FCNC) processes in the quark sector. In SM, FCNC processes are absent at tree level and can only occur at one-loop level. The only source of FCNC's in the SM is the Cabibbi Kobayashi Maskawa (CKM) matrix. However these processes are highly supressed due to GIM cancellations. On the other hand, in the MSSM, possible misalignment between the quark and squark mass matrices is another source of flavor violation that can dominate the SM contribution by several orders of magnitude. Any possible experimental deviation from the SM prediction for FCNS's would be a clear evidence of new physics and possibly a hint for MSSM.
Similarly, SM predictions for cLFV are zero. Even seesaw extensions of the SM do not predict
sizable rates for these processes, the cLFV rates in SM seesaw models are almost 40 orders of magnitude 
smaller than the present experimental bounds and consequently beyond the experimental reach. On the other hand seesaw extensions of MSSM are well capable of explaining the higher rates (touching the present experimental bounds) for these processes if observed. 
Also after the discovery of the Higgs boson with mass $M_{h}\approx 125 \gev$, one needs large radiative corrections. One obvious choice would be to choose scalar top mass heavier $\geq 1 \tev$. However this could go into the direction of (re-)introducing tuning. One can avoid this problem by choosing large left-right or flavor mixing (instead of assuming heavy scalar top mass). Consequently the issue of flavor mixing in SUSY needs to be explored in detail, which precisely is the aim of the thesis in hand.  

Within the MSSM, the possible presence of soft 
SUSY-breaking (SSB) parameters in the squark and slepton sector, which are
off-diagonal in flavor space (mass parameters as well as trilinear
couplings) are the most general way to introduce flavor mixing within the
MSSM. For example in MSSM, the off-diagonality in the sfermion mass matrix reflects the   
misalignment (in flavor space) between fermions and sfermions mass
matrices, that cannot be diagonalized simultaneously. 
This misalignment can be produced from various
origins. For instance, off-diagonal sfermion
mass matrix entries can be generated by Renormalization Group Equations
(RGE) running. Going from a high energy scale, where no flavor violation is
assumed, down to the electroweak (EW) scale can generate such entries due to
presence of non diagonal Yukawa matrices in RGE's. This kind of approach in the 
literature is known as the Minimal Flavor Violation (MFV)~\cite{MFV1,MFV2}, where flavor and $\cp$-violation in quark sector
is assumed to be entirely described by the CKM matrix, even in theories beyond the SM.

MFV scenerios are well motivated due to the fact that they do not introduce new sources of flavor and
$\cp$-violation, which can potentially lead to large non-standard effects in
flavor processes, in conflict with experimental bounds particularly from the kaon and $B_d$
sectors which are well described by the SM upto an accuracy of the $\sim 10\%$
level~\cite{HFAgroup}. For the first and second generation squarks which are
sensitive to the data on $K^0-\bar{K}^0$ and $D^0-\bar{D}^0$ the constraints are very tight. However the third generation system is,
in principle, less constrained, since present data on $B^0-\bar{B}^0$ mixing still leaves
some room for new sources of flavor violation. This opens the prospect for the more general scenerios, namely the Non Minimal Flavor Violation (NMFV) scenerios, other then the MFV ones. 

In this thesis we will present a systematic and simultanous study of the effects of flavor mixing on different observables in MFV as well as the NMFV scenerios. As a first step we will study squark and slepton mixing in the MSSM at low energy in Model-Independent (MI) way. For MI approach, we introduce flavor mixing parameters into the sfermion mass matrices by hand and do not consider the possible origin of these parameters. For the squark mixing we will be presenting the effects to electroweak precision observables (EWPO), B-Physics Observables (BPO) and quark flavor violating higgs decays (QFVHD). For slepton mixing we will study the effects to EWPO, higgs boson mass predictions and lepton flavor violating higgs decays (LFVHD). 

In the second step we will extend our analysis to the source of flavor mixing and will analyze the flavor mixing induced by RGE running from GUT to EW scale. In this study we will work within the MFV hypothesis for squarks as well as sleptons. Consequently, we will investigate two models (more detailed definitions and citations will follow in the next chapters):
\begin{itemize}
\item[(i)] 
the Constrained Minimal Supersymmetric Standard Model (CMSSM), where only flavor violation in the squark sector is present.
\item[(ii)]
the CMSSM augmented by the seesaw type~I mechanism\cite{seesaw:I}, 
called ``\CMSSMI'' below.
\end{itemize}

In many analyses of the CMSSM, or extensions such as the NUHM1 or NUHM2
(see \citere{AbdusSalam:2011fc} and references therein), 
the hypothesis of MFV has been used, and it has been assumed
that the contributions coming
from MFV are negligible not only for FCNC processes but for other observables like EWPO and Higgs masses as well, see, e.g.,
\citere{CMSSM-NUHM}. 
We will analyze whether this assumption is justified, and
whether including these MFV effects could lead to additional constraints
on the CMSSM parameter space. In this respect we will evaluate in the CMSSM
and in the \CMSSMI\ the following set of observables: BPO, in particular \bsg, \bmm\ and \dmbs; 
EWPO, in particular $\MW$ and the
effective weak leptonic mixing angle, $\sweff$; the masses of
the neutral and charged Higgs bosons in the MSSM, as well as cLFV and LFVHD.

The layout of the thesis is as follows. Chapter 1 contains the introduction to SM.
In chapter 2 we introduce MSSM and its seesaw extensions. Chapter 3 is devoted to 
the calculational basis of the observables considered in this work.
In chapter 4 we will be presenting our results in the case of squark flavor mixing in MI approach and study the effects to the BPO, EWPO and QFVHD. In chapter 5 slepton mixing effects to EWPO, Higgs mass predictions and LFVHD in MI approach will be presented.
Chapter 6 will be focusing on our analysis in CMSSM and CMSSM-seesaw~I where we present the flavor mixing effects to EWPO, BPO, Higgs boson mass predictions, QFVHD, cLFV and LFVHD. Chapter 7 is devoted to the summary and conclusions.  

\chapter{The Standard Model}

Symmetries play an important role in physics. Their presence in a particular
problem often simplifies the problem. Particle physicists, using the concept
of gauge symmetries, are able to build SM, which is a very successful model
to explain the fundamental particles and their interactions. The theory has
been formulated by writing the Lagrangian of the fundamental particles. The
Lagrangian has been written by using the concept of internal symmetries and
gauge invariance. All these aspects are discussed in detail hereafter and 
subsequent discussion follows closely~\citeres{GL-Kane,Hollik:2010id}.  

\section{Fundamental particles and forces}

Quarks and leptons (collectively called fermions, spin 1/2 particles) are (assumed to be)
elementary particles of nature. There are six types (flavors) of leptons and
quarks placed in three families. Fermions are chiral particles which
connotes that left and right handed fields transform differently. The left
handed components are placed in EW $SU(2)$ doublets and right
handed components are placed in EW singlets
\begin{eqnarray}
L &=&\binom{\nu _{e}}{e}_{L},\ \ \ \binom{\nu _{\mu }}{\mu }_{L},\ \ \ 
\binom{\nu _{\tau }}{\mu }_{L},\ \ \   \notag \\
&&e_{R},\ \ \ \mu _{R},\ \ \ \tau _{R},\   \label{lepton}
\end{eqnarray}

\begin{eqnarray}
Q &=&\binom{u}{d}_{L},\ \ \ \binom{c}{s}_{L},\ \ \ \binom{t}{b}_{L},\ \  
\notag \\
&&\ u_{R},\ \ \ d_{R},\ \ c_{R},\ \ \ s_{R},\ \ \ t_{R},\ \ \ b_{R}
\label{quarks}
\end{eqnarray}
L on the left represents lepton and in the subscript on the right it means
left-handed. Neutrinos being left handed are absent in the EW
singlets. For the quarks another index is required to describe how the
quarks transform under $SU(3)$ transformation.

\begin{equation}
Q_{\alpha }=\binom{u_{\alpha }}{d_{\alpha }}_{L}  \label{1.1}
\end{equation}

Quarks and leptons interact through unified EW and strong forces.
These forces are transmitted by the exchange of particles, called gauge
bosons $(\gamma $ , $W^{\pm }$ and $Z)$. \ These are the mediators of
the unified EW force and gluons are the mediators of strong force.
There is an additional particle called Higgs boson predicted by the SM,
which has implications with regard to the origin of mass. It was discovered
recently at LHC CERN \cite{ATLAS:2013mma,CMS:yva}.
\section{Gauge transformation and invariance}
All particles appear to have three kind of gauge invariances, $(U(1)$%
, $SU(2),$ $SU(3))$. The $U(1)$ is related to the electromagnetic charge,
the $SU(2)$ corresponds to the non-abelian weak isospin and $SU(3)$ is
associated with the non-abelian strong (color) charge. In 1961 Glashow \cite%
{Glashow} proposed $SU(2)_{\rm L}\times U(1)_{\rm Y}$ structure of the SM. Weinberg
and Salam \cite{Salam, Weinberg} extended his idea and employed the
hypothesis of spontaneous symmetry breaking in their gauge theory models to
generate masses of gauge boson and fermions. Later on the Glashow, Weinberg and
Salam model achieved the theoretical status when 't Hooft \cite{thooft}
demonstrated that the form of symmetry breaking would not spoil the
renormalizability possessed by the massless theory.
\section{The SM Lagrangian}
The complete Lagrangian for the SM can be written as
\begin{equation}
{\cal L} ={\cal L}_{\rm fermion}+{\cal L}_{\rm gauge}+{\cal L}_{\rm Higgs}
\label{fullLagr}
\end{equation}
${\cal L}_{\rm fermion}$ is given by the relation

\begin{equation}
{\cal L} _{\rm fermion}=\sum\limits_{f=L,Q}\bar{f}\iota \gamma ^{\mu }D_{\mu }f,
\label{Lfermi}
\end{equation}%
where L and Q are given in \refeq{lepton} and \refeq{quarks} and 
$D_{\mu } $ is a covariant derivative given by
\begin{equation}
D_{\mu }=\ \partial _{\mu }-\iota g_{1}\frac{Y}{2}B_{\mu }-\iota
g_{2}\frac{\sigma^{i}}{2}W_{\mu }^{i}-\iota g_{3}\frac{\lambda ^{\alpha }}{2}%
G_{\mu }^{\alpha }.  
\label{CovariantD}
\end{equation}

It is to be noted that whenever the terms in $ D_{\mu } $ act on a
fermionic state of different matrix form, they give zero, by definition. The
second term represents the $U(1)$ symmetry. $B_{\mu }$ is spin 1 field
needed to maintain gauge invariance and Y is the generator of $U(1)$
transformations, that is also called hypercharge. The $g_{1}$ is the $U(1)$
gauge coupling, the third and the fourth term represents $SU(2)$ and $SU(3)$
symmetries respectively, three W$_{\mu }^{i}$ for $SU(2)$ and eight $G_{\mu
}^{\alpha }$ for $SU(3)$, one for each generator ($\sigma ^{i},\lambda
^{\alpha })$ of transformation whereas $g_{2}$ and $g_{3}$ are the $SU(2)$
and $SU(3)$ gauge couplings respectively.

The Lagrangian for the $SU(2)_{\rm L}\times U(1)_{ \rm Y}$ gauge sector of the theory
is
\begin{equation}
{\cal L}_{\rm gauge}=-\frac{1}{4}W_{\mu \nu }^{i}W_{i}^{\mu \nu }-\frac{1}{4}%
F_{\mu \nu }F^{\mu \nu },
\label{Lgauge}
\end{equation}%
where $F_{\mu \nu }$ is the field strength tensor for $U(1)$ gauge boson $%
B_{\mu }$ and is given by%
\begin{equation}
F_{\mu \nu }=\partial _{\mu }B_{\nu }-\partial _{\nu }B_{\mu }  
\label{FStrengthT}
\end{equation}%
and 
\begin{equation}
W_{\mu \nu }^{i}=\partial _{\mu }W_{\nu }^{i}-\partial _{\nu }W_{\mu
}^{i}+g_{2}\varepsilon _{ijk}W_{\mu }^{j}W_{\nu }^{k}  
\label{WStrengthT}
\end{equation}%
is the field strength tensor for the $SU(2)$ gauge boson, $\varepsilon
_{ijk}$ in the third term of \refeq{WStrengthT} is structure constant and
this term appears due to non-abelian nature of the $SU(2)$ group.

${\cal L}$ does not contain any mass term. In order to generate masses for
fermions and bosons, Higgs mechanism is introduced which will be discussed in \refse{Higgs-Mech}. 
\section{Electroweak theory}
By using \refeq{Lfermi}, the $U(1)$ and $SU(2)$ terms for the
Lagrangian of the first generation of leptons can be written as
\begin{eqnarray}
{\cal L}_{\rm lepton} &=&\frac{g_{1}}{2}[Y_{L}(\bar{\nu}_{L}\gamma ^{\mu }\nu
_{L}+\bar{e}_{L}\gamma ^{\mu }e_{L})+Y_{R}\bar{e}_{R}\gamma ^{\mu
}e_{R})]B_{\mu }  \notag \\
&&-\frac{g_{2}}{2}[\bar{\nu}_{L}\gamma ^{\mu }\nu _{L}W_{\mu }^{o}-\bar{e}%
_{L}\gamma ^{\mu }e_{L}W_{\mu }^{o}-\sqrt{2}\bar{\nu}_{L}\gamma ^{\mu
}e_{L}W_{\mu }^{+}  \notag \\
&&-\sqrt{2}\bar{e}_{L}\gamma ^{\mu }\nu _{L}W_{\mu }^{-}],  
\label{Llepton}
\end{eqnarray}%
as neutrinos do not have electromagnetic interactions, the terms of the form 
$\frac{g_1}{2} Y_{L}\bar{\nu}_{L}\gamma^{\mu}\nu_{L}B_{\mu}$ must be
avoided. To do so the coefficient $Z_{\mu }\varpropto g_{1}Y_{L}B_{\mu
}-g_{2}W_{\mu }^{0}$ of the term $\bar{\nu}_{L}\gamma ^{\mu }\nu _{L}$ is
assumed to be orthogonal to the electromagnetic field $A_{\mu }$. After
diagonalization one gets
\begin{equation}
A_{\mu }=\frac{g_{2}B_{\mu }-g_{1}Y_{L}W_{\mu }^{0}}{\sqrt{%
g_{2}^{2}+g_{1}^{2}Y_{L}^{2}}}, 
\label{Amuterm}
\end{equation}

\begin{equation}
Z_{\mu }=\frac{g_{1}Y_{L}B_{\mu }-g_{2}W_{\mu }^{0}}{\sqrt{%
g_{2}^{2}+g_{1}^{2}Y_{L}^{2}}}.  
\label{Zmuterm}
\end{equation}%

Solving for $B_{\mu }$ and $W_{\mu }^{0}$, one gets
\begin{equation}
B_{\mu }=\frac{g_{2}A_{\mu }+g_{1}Y_{L}Z_{\mu }}{\sqrt{%
g_{2}^{2}+g_{1}^{2}Y_{L}^{2}}},  
\label{Bmuterm}
\end{equation}

\begin{equation}
W_{\mu }^{o}=\frac{g_{2}Z_{\mu }-g_{1}Y_{L}A_{\mu }}{\sqrt{%
g_{2}^{2}+g_{1}^{2}Y_{L}^{2}}}.  
\label{Wmuterm}
\end{equation}

With these definitions the neutral current interactions of the electrons in
\refeq{Llepton} are modified as
\begin{eqnarray}
&&-A_{\mu }[\bar{e}_{L}\gamma ^{\mu }e_{L}\frac{g_{1}g_{2}Y_{L}}{\sqrt{%
g_{2}^{2}+g_{1}^{2}Y_{L}^{2}}}+\bar{e}_{R}\gamma ^{\mu }e_{R}\frac{%
g_{1}g_{2}Y_{R}}{2\sqrt{g_{2}^{2}+g_{1}^{2}Y_{L}^{2}}}],  \notag \\
&&-Z_{\mu }[\bar{e}_{L}\gamma ^{\mu }e_{L}\frac{g_{1}^{2}Y_{L}^{2}-g_{2}^{2}%
}{\sqrt{g_{2}^{2}+g_{1}^{2}Y_{L}^{2}}}+\bar{e}_{R}\gamma ^{\mu }e_{R}\frac{%
g_{1}^{2}Y_{L}Y_{R}}{2\sqrt{g_{2}^{2}+g_{1}^{2}Y_{L}^{2}}}].  
\label{AmuZmu}
\end{eqnarray}%
This gives
\begin{equation}
e=\frac{-g_{1}g_{2}Y_{L}}{\sqrt{g_{2}^{2}+g_{1}^{2}Y_{L}^{2}}}  
\label{elecCharge1}
\end{equation}%
and 
\begin{equation}
e=\frac{-g_{1}g_{2}Y_{R}}{2\sqrt{g_{2}^{2}+g_{1}^{2}Y_{L}^{2}}}  
\label{elecCharge2}
\end{equation}%
From \refeq{elecCharge1} and \refeq{elecCharge2} it follows that
\begin{equation}
2Y_{L}=Y_{R}  
\label{Yldef}
\end{equation}%
As $g_{1}$ can be redefined to absorb any change in Y$_{L},$ Y$_{L}$ has
been set to $-1$ and \refeq{elecCharge1} is modified as
\begin{equation}
e=\frac{g_{1}g_{2}}{\sqrt{g_{2}^{2}+g_{1}^{2}}}  
\label{elecCharge3}
\end{equation}

\begin{equation}
e=g_{2}\sin \theta _{W}  
\label{elecCharge4}
\end{equation}%
where $\theta_{W}$ is EW mixing angle with $\sin ^{2}\theta _{W}=(%
\frac{g_{1}}{\sqrt{g_{2}^{2}+g_{1}^{2}}})^{2}$.
\section{Spontaneous symmetry breaking}
\label{SSB-SM}
The Lagrangian in \refeq{Lfermi} does not contain any mass term and
mass terms can not be added explicitly by hand as it would break gauge
invariance. Mass terms are included in SM\ Lagrangian by the Higgs
mechanism, using the idea of spontaneous symmetry breaking. Consider the
Lagrangian for a scalar field $\phi$
\begin{equation}
{\cal L} =\frac{1}{2}\partial _{\mu }\phi \partial ^{\mu }\phi -(\frac{1}{2}%
\mu ^{2}\phi ^{2}+\frac{1}{4}\lambda \phi ^{4});\text{ \ \ \ \ \ \ \ \ \ \ }%
\lambda >0 
\label{Lscalar}
\end{equation}%
Here
\begin{equation}
V=\frac{1}{2}\mu ^{2}\phi ^{2}+\frac{1}{4}\lambda \phi ^{4}.
\label{Potential}
\end{equation}%
If $\mu ^{2}>0$ then the vacuum corresponds to $\phi_{0} =0$ but if $\mu ^{2}<0$
then the minimum of the potential is
\begin{equation*}
\frac{\partial V}{\partial \phi }=0
\end{equation*}

\begin{equation}
\phi_0 (\mu ^{2}+\lambda \phi_0 ^{2})=0
\label{1.22}
\end{equation}

\begin{equation}
\phi_0 =\pm \sqrt{\frac{-\mu ^{2}}{\lambda }}=v
\label{1.23}
\end{equation}%
where $v$ is vacuum expectation value (VEV) of Higgs field $\phi $. To
determine the particle spectrum one must study the theory in the region of
the minimum by putting $\phi =v +\eta (x)$ and expanding around $\eta =0.$
Using $\phi =v +\eta (x)$ and \refeq{1.23} in \refeq{Lscalar} yields
\begin{equation}
{\cal L}=\frac{1}{2}\partial _{\mu } \eta \partial ^{\mu } \eta-(\lambda v \eta
^{2}+\lambda v \eta ^{3}+\frac{1}{4}\lambda \eta ^{4})+{\rm const.}
\label{1.24}
\end{equation}
The term in $\eta ^{2}$ has the correct sign so it can be interpreted  as mass square and the vacuum does not
have the reflection symmetry of the original Lagrangian. This is called
spontaneous symmetry breaking.
\section{Higgs mechanism}
\label{Higgs-Mech}
The renormalizability and gauge invariance of the theory demands that the
symmetry $SU(2)_{\rm L}\times U(1)_{\rm Y}$ be spontaneously broken through Higgs
mechanism. For this purpose a complex weak doublet of Higgs scalar with hypercharge $Y=1$, 
\begin{equation}
\Phi (x) =%
\begin{pmatrix}
\phi ^{+}(x) \\ 
\phi ^{0}(x) %
\end{pmatrix}
\label{Phi-Def1}
\end{equation}
is introduced which is coupled to the gauge fields through
\begin{equation}
{\cal L}_{\rm Higgs} = (D_{\mu } \Phi)^{\dagger}(D_{\mu } \Phi)-V (\Phi). 
\label{LHiggs}
\end{equation}
In this case the covariant derivative is given by
\begin{equation}
D_{\mu }=\ \partial _{\mu }-\iota \frac {g_{1}}{2}B_{\mu }-\iota
g_{2}\frac{\sigma ^{i}}{2}W_{\mu }^{i}.  
\label{CovariantD-Higgs}
\end{equation}
The Higgs field self-interaction enters through the Higgs potential with constants $\mu^2$ and $\lambda$,
\begin{equation}
V(\Phi)=-\mu^2 \Phi^{\dagger} \Phi+\frac{\lambda}{4}(\Phi^{\dagger} \Phi)^2.  
\label{Higgs-Pot}
\end{equation}
In the ground state, the vacuum, the potential has a minimum. For $\mu^2$, $\lambda > 0 $, the minimum does not
occur for $\Phi = 0$; instead, V is minimized by all non-vanishing field configurations with $\Phi^{\dagger} \Phi =2 \mu^2\ \lambda$.
Selecting the one which is real and electrically neutral, one gets the VEV
\begin{equation}
\langle\Phi\rangle=\frac{1}{\sqrt 2} 
\begin{pmatrix}
0 \\
v
\end{pmatrix}.
\label{VEV-SMHiggs}
\end{equation}
Although the Lagrangian is symmetric under gauge transformations of the full $SU(2) \times U(1)$ group,
the vacuum configuration $\langle\Phi\rangle$ does not have this symmetry: the symmetry has been spontaneously
broken. $\langle\Phi\rangle$ is still symmetric under transformations of the electromagnetic subgroup $U(1)_{\rm em}$, which
is generated by the charge Q, thus preserving the electromagnetic gauge symmetry.

The scalar feld in \refeq{Phi-Def1} can be written as
\begin{equation}
\Phi (x) =%
\begin{pmatrix}
\phi ^{+}(x) \\ 
\left( v+H(x)+\iota \chi(x)\right)/{\sqrt 2} %
\end{pmatrix},
\label{Phi-Def2}
\end{equation}
where the components $\phi^+$, $H$, $\chi$ have vacuum expectation values zero. Expanding the potential in \refeq{Higgs-Pot}
around the vacuum configuration in terms of the components yields a mass term for $H$, whereas $\phi^+$ and
$\chi$ are massless. Exploiting the invariance of the Lagrangian, the components $\phi^+$ and $\chi$ can be eliminated
by a suitable gauge transformation; this means that they are unphysical degrees of freedom (called Higgs
ghosts or would-be Goldstone bosons). Choosing this particular gauge where $\phi^+ = \chi = 0$, denoted as
the unitary gauge, the Higgs doublet field has the simple form
\begin{equation}
\Phi (x) =%
\frac{1}{{\sqrt 2}}\begin{pmatrix}
0 \\ 
v+H(x) %
\end{pmatrix}.
\label{Phi-Def3}
\end{equation}
The real field $H(x)$ thus describes physical neutral scalar particles, the Higgs bosons, with mass
\begin{equation}
M_{H}={\sqrt 2} \mu= {\sqrt \lambda v}.
\end{equation}

The gauge invariant Higgs–gauge field interaction in the kinetic part
of \refeq{LHiggs} gives rise to mass terms for the vector bosons in the non-diagonal form
\begin{equation}
\frac{1}{2}\left(\frac{g_2}{2}v\right)^2 (W^2_1+W^2_2)+\frac{1}{2}\left(\frac{v}{2}\right)^2
\begin{pmatrix}
W_\mu^3,B_\mu
\end{pmatrix}
\begin{pmatrix}
g_2^2   & g_1 g_2\\
g_1 g_2 & g_1^2
\end{pmatrix}
\begin{pmatrix}
W^{\mu,3}\\
B_\mu
\end{pmatrix}.
\label{LHiggsKinetic}
\end{equation}
The first term can be written as
\begin{equation}
\left(\frac{g_2}{2}v\right)^2 W^+_\mu W^{-\mu}.
\end{equation}
For the charged boson the expected mass term for the Lagrangian would be $m^2 W^+ W^-$, so we can conclude that the charged $W$ boson has indeed acquired a mass
\begin{equation}
M_{W}=\frac{1}{2}g_2 v.
\label{Wmass}
\end{equation}%
The second term in the \refeq{LHiggsKinetic} is not diagonal and we have to define new eigenvalues to find the particles with definite mass. In fact, we already have the answer in hand, because the combination of $B$ and $W^3$ appearing in \refeq{LHiggsKinetic} is just the combination we have called $Z_\mu$ (see \refeq{Zmuterm}).  
From \refeq{LHiggsKinetic} and normalization of $Z$ in \refeq{Zmuterm}, we can conclude that the neutral gauge boson $Z$ acquires mass
\begin{equation}
M_{Z}=\frac{v }{2}\sqrt{g_{2}^{2}+g_{1}^{2}}=\frac{M_{W}}{\cos
\theta _{W}},  
\label{1.27}
\end{equation}%
while the photon remains massless. 

In SM, all quarks and charged fermions get their
masses through the Yukawa couplings with the Higgs field $\Phi$:
\BE
{\cal L}_{\rm Yukawa} = (Y^{u})_{ij}\bar{Q}_{L_{i}}\Phi^{*} u_{R_{j}}  
 + (Y^{d})_{ij}\bar{Q}_{L_{i}}\Phi d_{R_{j}}  
 + (Y^{e})_{ij}\bar{L}_{L_{i}}\Phi e_{R_{j}}, 
\label{Lyukawa}
\EE
where $Y^{u}$, $Y^{d}$ and $Y^{e}$ are up-quark down-quark and charged leptons Yukawa coupling, $Q_{L}$ and $L_{L}$ are left handed quark and lepton doublets, $u_{R}$, $d_{R}$ and $e_{R}$ are $SU(2)_{L}$ -
singlet right-handed fields of up-type quarks, down-type quarks and charged leptons respectively and $i$, $j$ are the generation indices.
After the EW symmetry is broken by a nonzero VEV $v$
of the Higgs field, the Yukawa terms in \refeq{Lyukawa} yield the mass matrices of quarks and charged
leptons
\BEA
(m_{u})_{ij} = (Y^{u})_{ij} v ,\: (m_{d})_{ij} = (Y^{d})_{ij} v ,\: (m_{e})_{ij} = (Y^{e})_{ij} v
\label{yukawamassterms}
\EEA
Neutrinos are massless in the SM. They cannot have Dirac masses
because there are no $SU(2)_{\rm L}$ - singlet (“sterile”) right-handed neutrinos $\nu_{R}$ in the SM.
\section{The CKM matrix}
\label{Sec:CKM}
The quark doublets introduced in \refeq{quarks} can have up-down
transitions of the form $u_{i}\rightarrow d_{i}$ mediated by the $W^{\pm },$
where $u_{i}$ can be any up type quark and $d_{i}$ represents any down type
quark. These kind of interactions are absent among lepton doublets due to
conservation of lepton flavor. The mixing among different generations
indicated by rare kaon decay led Cabibbo \cite{cabibbo} to introduce the
mixing angle $\theta _{c}$ called cabibbo angle so that the quark doublet
given in \refeq{quarks} is modified to

\begin{equation}
\binom{u}{d^{\prime }}=\binom{u}{d\cos \theta _{c}+s\sin \theta _{c}}.
\label{1.48}
\end{equation}

This means that the weak eigenstate $d^{\prime }$ is a linear combination
of real mass eigenstates $d$ and $s$. This concept was modified by S.L.
Glashow, J. Iliopoulos and L. Maiani \cite{GIM}. They were able to predict
the existence of charm quark even before its discovery. This completed the
two quark doublets. They explained the mixing with the help of $2\times 2$
unitary matrix. As the concept of quark mixing was indicated through the
rare kaon decays, there was also indication of $\cp$ violation in these decays.
So it was believed that the $\cp$ violation has its origin in quark mixing.
This idea was adopted by Kobayashi and Maskawa \cite{KM} to introduce the
third quark doublet, as $\cp$ violation cannot be accommodated by two
doublets, consequently, they proposed the 3$\times 3$ unitary matrix called
CKM matrix given by
\begin{equation}
V_{\rm CKM}=%
\begin{pmatrix}
V_{ud} & V_{us} & V_{ub} \\ 
V_{cd} & V_{cs} & V_{cb} \\ 
V_{td} & V_{ts} & V_{tb}%
\end{pmatrix}.
\label{CKM-Vud}
\end{equation}

Thus the rotation from the $SU(2)$ interaction eigenstate basis, $q^{\rm int}_{L,R}$, to the physical mass  eigenstate basis, 
$q_{L,R}^{\rm phys}$, is performed by the unitary transformations, $V^{u,d}_{L,R}$:

\begin{equation}
\VL u^{\rm phys}_{L,R} \\ c^{\rm phys}_{L,R} \\ t^{\rm phys}_{L,R} \VR =
V^u_{L,R} \VL u^{\rm int}_{L,R} \\ c^{\rm int}_{L,R} \\ t^{\rm int}_{L,R} \VR~,~~~~
\VL d^{\rm phys}_{L,R} \\ s^{\rm phys}_{L,R} \\ b^{\rm phys}_{L,R} \VR =
V^d_{L,R} \VL d^{\rm int}_{L,R} \\ s^{\rm int}_{L,R} \\ b^{\rm int}_{L,R} \VR~,
\end{equation}
such that the quark mass matrices in the physical basis are:
\begin{eqnarray}
\frac{v}{\sqrt 2}V^u_L Y^{u*}V^{u\dagger}_R&=&{\rm diag}\left(m_u,m_c,m_t\right),
\\
\frac{v}{\sqrt 2}V^d_L Y^{d*}V^{d\dagger}_R&=&{\rm diag}\left(m_d,m_s,m_b\right).
\end{eqnarray} 
In short, the quark flavour mixing is encoded in the CKM matrix, 
\begin{equation}
\VCKM= V^u_L V^{d\dagger}_L.
\end{equation}

There are nine parameters in the CKM matrix as shown in \refeq{CKM-Vud} which can be reduced to four in the standard
parametrization \cite{ckmparameters}. The three Euler angles $\theta _{12},\
\theta _{13},\ \theta _{23}$ and one phase factor $\delta ,$ which accounts
for the $\cp$ violation. The CKM matrix in standard parameterization is given by
\begin{equation}
V_{\rm CKM}=%
\begin{pmatrix}
c_{12}c_{13} & s_{12}c_{13} & s_{13}e^{-\iota \delta } \\ 
-s_{12}c_{23}-c_{12}s_{13}e^{\iota \delta } & 
c_{12}c_{23}-s_{12}s_{23}s_{13}e^{\iota \delta } & s_{23}c_{13} \\ 
s_{12}s_{23}-c_{12}c_{23}s_{13}e^{\iota \delta } & 
-c_{12}s_{23}-s_{12}c_{23}s_{13}e^{\iota \delta } & c_{13}c_{23}%
\end{pmatrix}
\label{1.50}
\end{equation}
where $c_{ij}=\cos \theta _{ij}$ and $s_{ij}=\sin \theta _{ij}$ ($i,j=1,2,3)$. The elements of the CKM matrix exhibit a pronounced
hierarchy. While the diagonal elements are close to unity, the off-diagonal elements are small, such that e.g. $V_{ud}\gg V_{us}\gg V_{ub}$. 
In  terms  of  the  angles $\theta_{ij}$ we  have $s_{12}\gg s_{23}\gg s_{13}$. This fact is usually expressed in terms of the
Wolfenstein parameterization \cite{wolf-parameter}, which can be understood as an expansion in $\lambda=|V_{us}|$. This reads up to order $\lambda^3$
\begin{equation}
V_{\rm CKM}=%
\begin{pmatrix}
1-\frac{\lambda ^{2}}{2} & \lambda & A\lambda ^{3}\left( \rho -\iota \eta
\right) \\ 
-\lambda & 1-\frac{\lambda ^{2}}{2} & A\lambda ^{2} \\ 
A\lambda ^{3}\left( 1-\rho -\iota \eta \right) & A\lambda ^{2} & 1%
\end{pmatrix}
\label{1.51}
\end{equation}%
with parameters $A,\rho \ $\ and $\eta$ are assumed to be of order 1. The current values of the CKM elements, obtained from a global fit using all the available measurements and imposing the SM constraints, are collected in the following matrix \cite{Olive:CKM}:
\begin{equation}
V_{\rm CKM}=
\left(\begin{array}{ccc}
0.97427 \pm 0.00014 & 0.22536 \pm 0.00061 & 0.00355 \pm 0.00015 \\
0.22522 \pm 0.00061 & 0.97343 \pm 0.00015 & 0.0414 \pm 0.0012  \\
0.00886^{+0.00033}_{-0.00032} & 0.0405^{+0.0011}_{-0.0012} & 0.99914 \pm 0.00005
\end{array}\right).
\end{equation}

\chapter{Supersymmetry \& Its Seesaw Extention}

A SUSY transformation turns a bosonic state into a fermionic state, and vice versa. The
operator $Q$ that generates such transformations must be an anticommuting spinor, with
\begin{equation}
Q \left|{\rm Boson}\right\rangle=\left|{\rm Fermion}\right\rangle, \quad Q\left|{\rm Fermion}\right\rangle=\left|{\rm Boson}\right\rangle.
\label{SUSY-Trans1}
\end{equation}

Spinors are intrinsically complex objects, so $Q^{\dagger}$ (the hermitian conjugate of $Q$) is also a symmetry
generator. Because $Q$ and $Q^{\dagger}$ are fermionic operators, they carry spin angular momentum 1/2, so it is
clear that SUSY must be a spacetime symmetry. 
The No-go theorem \cite{mandula} asserts that it is impossible to mix internal
and Lorentz space time symmetries (when described by the commutators only) in a non-trivial way. If one wants to
extend the space-time structure, one will be left with the only choice of
SUSY with graded Lie algebra. The simplest realization is given by

\begin{equation}
\{Q_{\alpha },Q_{\overset{.}{\alpha }}^{\dagger }\}=2\sigma _{\alpha \overset%
{.}{\alpha }}^{\mu }P_{\mu },
\label{1.35}
\end{equation}

\begin{equation}
\{Q_{\alpha },Q_{\beta }\}=\{Q_{\overset{.}{\alpha }},Q_{\overset{.}{\beta }%
}\}=0 ,
\label{1.36}
\end{equation}

\begin{equation}
\lbrack Q_{\alpha },P_{\mu }]=[Q_{\overset{.}{\alpha }},P_{\mu }]=0,
\label{1.37}
\end{equation}%
where $P^{\mu }$ is the
momentum generator of space-time translations and $\sigma ^{\mu }$ $%
=(1,\sigma ^{1},$ $\sigma ^{2},$ $\sigma ^{3})$. In the following sections, we will give some motivation for SUSY and review main aspacts of the SUSY, in particular the MSSM and its seesaw extension. The subsequent discussion follows closely~\citeres{Martin:1997ns,PomssmRep}. 

\section{Motivation}
\label{motivation}
SUSY can successfully explain some of the major deficiencies of SM, as discussed in the introduction,
in a more natural way. 

\begin{itemize}
\item \textbf {Hierarchy problem:}
The simplest form of SUSY can solve the hierarchy problem mentioned
in the introduction. Quadratic divergences appearing at one loop level in Higgs mass
vanish due to cancellation between bosons and fermions. Consider for example coupling of the Higgs field $H$ to a Dirac fermion $f$ with
a term in the Lagrangian $-\lambda_f H{\bar f}f$. The one-loop radiative corrections to the Higgs mass $M_H$  will be of the form
\begin{equation}
\delta M_{H}^{2}(f)=- \frac{\left\vert \lambda_f \right\vert^2}{8 \pi^2}[\Lambda^2-2m^2_f\ {\rm ln} \frac{\Lambda}{m_f}+.....]
\label{delMH-fermi}
\end{equation}%
where $m_f$ is the mass of the fermion in the loop. As can be seen from above equation Higgs boson mass is  quadratically divergent. In the case of fermion (gauge boson), the chiral (gauge) symmetry constitutes the ``natural barrier'' preventing their masses to become arbitrarily
large. In the case of Higgs boson, there is no symmetry that protects the scalar mass and in the limit $M_H \rightarrow 0$, the symmetry of the model is not increased. SUSY constitutes
so far the most interesting answer to hierarchy problem. As we have mentioned in the introduction, SUSY associates a scalar particle
with every fermionic degree of freedom in the theory with, in principle, identical masses and gauge
quantum numbers. Therefore, in a supersymmetric theory we would have a new
contribution to the Higgs mass at one loop given by
\begin{equation}
\delta M_{H}^{2}({\tilde f})=- \frac{\lambda_{\tilde f}}{8 \pi^2}[\Lambda^2-2m^2_{\tilde f}\ {\rm ln} \frac{\Lambda}{m_{\tilde f}}+.....],
\label{delMH-Sfermi}
\end{equation}%
where $\lambda_{\tilde f}$ is the SUSY particle coupling to the Higgs field, $m_{\tilde f}$ is the mass of the SUSY particle in the loop. If we compare \refeq{delMH-fermi} and \refeq{delMH-Sfermi} we see that with $|\lambda_f|^2=-\lambda_{\tilde f}$ and $m_f = m_{\tilde f}$ we obtain a total correction $\delta M_H^2(f ) + \delta M_H^2({\tilde f} ) = 0 $, i.e. quadratic divergence cancels exactly. If SUSY was an exact symmetry of nature, particles and their
superpartners would have the same mass, and therefore the superpartners should have
been observed in collider experiments. However we have not found scalars exactly degenerate with the SM fermions. This means SUSY can not be an exact symmetry of nature, it must be a broken symmetry. By comparing \refeq{delMH-fermi} and \refeq{delMH-Sfermi}, we can see that we must
still require $|\lambda_f|^2=-\lambda_{\tilde f}$ if we want to ensure the cancellation of quadratic
divergences. SUSY can be broken only in couplings with positive mass dimension, as for instance the
masses. This is called ``soft SUSY-breaking'' \cite{Girardello-SSB}. Now if we take $m^2_{\tilde f}= m^2_f+ \delta^2 $ we
obtain a correction to the Higgs mass,
\begin{equation}
\delta M_H^2(f ) + \delta M_H^2({\tilde f} ) =\frac{\left\vert \lambda_f \right\vert^2}{8 \pi^2} \delta^2 {\rm ln} \frac{\Lambda}{m_{\tilde f}}+.....,
\label{delMH-SSB}
\end{equation}%
and this is only logarithmically divergent and proportional to mass difference between fermion and its scalar partner and is, therefor, under control.
\item \textbf {Gauge coupling unification:}
The idea of gauge unification gets simplified by the SUSY. The coupling
constants $\alpha _{1},$ $\alpha _{2}$ and $\alpha _{3}$ vary with energy
and the rate of the variation of these coupling constants depends on the particle
content of the theory. If these coupling constants are extrapolated to the
higher energies using the particle content of the SM, these do not meet at
the same point. However, when the same extrapolation is repeated using the
particle contents of the SUSY, the three coupling constants meet at the same
point \cite{Dimopoulos-GCU,Marciano-GCU,Amaldi-GCU}. The ``exact'' unification of the gauge couplings within the MSSM
may or may not be an accident. But it provides enough reasons to consider supersymmetric
models seriously as it links SUSY and grand
unification in an inseparable manner\cite{Mohapatra-GCU}.
\item \textbf {Dark Matter candidate:}
There is no particle in the SM that can serve as a DM candidate. However most of the SUSY models provide a particle which might explain missing mass in the universe. For example lightest neutralino could be a DM candidate in the MSSM (see details below).
\item \textbf{Supergravity:}
If SUSY is formulated as a local symmetry, a spin 2 particle corresponding to the graviton, the hypothetical particle
that mediates gravity, is introduced. Then the supersymmetric models of gravity called
supergravity have the elegant feature to link the SM fundamental interactions with gravity\cite{Nieuwenhuizen-SG}.
\end{itemize}

\section{Superpotential}

In this section we will describe the concept of superpotential. The aim is to arrive
at a recipe that will allow to write down the allowed interaction terms of a general
supersymmetric theory, so that later these results can be applied to the special case of the MSSM (see, e.g., the discussion in \citere{Martin:1997ns}). 

The single-particle states of a supersymmetric theory fall into irreducible representations of the
SUSY algebra, called supermultiplets. Each supermultiplet contains both fermion and boson
states, which are commonly known as superpartners of each other.
Each supermultiplet contains an equal number of fermionic and bosonic degrees of freedom.

The minimum fermion content of any theory in four dimensions consists
of a single left-handed two-component Weyl fermion $\psi$. Since this is an
intrinsically complex object, it seems sensible to choose as its superpartner
a complex scalar field $\phi$. This combination of a two-component Weyl fermion and a complex scalar field
is called a chiral or matter or scalar supermultiplet. 

The next-simplest possibility for a supermultiplet contains a spin-1 vector boson. If the theory is to
be renormalizable, this must be a gauge boson that is massless, at least before the gauge symmetry is
spontaneously broken. Its superpartner is therefore a massless spin-1/2 Weyl fermion, called gaugino. 
Such a combination of spin-1/2 gauginos and spin-1
gauge bosons is called a gauge or vector supermultiplet.

The simplest action one can write down for chiral supermultiplet
just consists of kinetic energy terms for scalar and fermionic fields.
\begin{equation}
S=\int d^{4}x({\cal L}_{\rm scalar}+{\cal L}_{\rm fermion})
\label{1.38}
\end{equation}%
with 
\begin{equation}
{\cal L}_{\rm scalar}=\partial ^{\mu }\phi ^{\ast }\partial _{\mu
}\phi ,\qquad {\cal L}_{\rm fermion}=\iota {\bar \psi} \bar{\sigma}^{\mu
}\partial _{\mu }\psi. 
\end{equation}
This is called the massless, non-interacting Wess-Zumino model.
The number of fermionic degrees of freedom must be equal to the number of
bosonic degrees of freedom. But scalar field contains one degree of freedom
and one can add one more if a complex scalar field is introduced. However, a fermionic
field carries at least four components. Two of these degrees of freedom can be fixed by Dirac equation. It means, the algebra of SUSY only closes on-shell in this formulation. This can be fixed by a trick. One can invent a new complex scalar field F, which does not
have a kinetic term. Such fields are called auxiliary, and they are really just book-keeping devices that
allow the SUSY algebra to close off-shell. Thus the free part of the
Lagrangian is
\begin{equation}
{\cal L}_{\rm free}=\partial ^{\mu }\phi ^{\ast i}\partial _{\mu
}\phi_{i}+\iota {\bar \psi} ^{i}\bar{\sigma}^{\mu
}\partial _{\mu }\psi_i+ F^{*i}F_{i}~,
\end{equation}
where it is summed over repeated indices $i$ (not to be confused with the
suppressed spinor indices). Now the most general set of renormalizable interactions for
these fields that is consistent with SUSY must have
dynamical field content with mass dimension $\leq$ 4. So, the candidate terms that are also SUSY invariant 
are:
\begin{equation}
{\cal L}_{\rm int}=-\frac{1}{2} W^{ij} \psi_{i}\psi_{j}+ W^{i}F_{i}+c.c.
\end{equation}

A very useful object $W$ called the superpotential is introduced.
\begin{equation}
W=\frac{1}{2} M^{ij} \phi_{i}\phi_{j}+ \frac{1}{6} y^{ijk} \phi_{i}\phi_{j}\phi_{k}
\end{equation}
where $M^{ij}$ is a symmetric mass matrix for the fermion fields, and $y^{ijk}$ is
the Yukawa coupling of the scalar $\phi_k$. One can write
\begin{equation}
W^{ij}=\frac{\partial^2}{\partial \phi_i \partial \phi_j} W, \quad  W^{i}=\frac{\partial W}{\partial \phi_i}. 
\end{equation}

The auxiliary fields $F_i$
and $F^{\ast i}$ can be eliminated using their classical equations of motion.
The part of ${\cal L}_{\rm free}$ + ${\cal L}_{\rm int}$ that contains the auxiliary fields is $F_iF^{\ast i}$ + $W^iF_i$ +
$W^{\ast}_{i} F^{\ast i}$, leading to the equations of motion

\begin{equation}
F_{i}=-W^*_i; \quad  F^{*i}=-W^{i}. 
\label{EQM-F}
\end{equation}
After making the replacement \refeq{EQM-F} in ${\cal L}_{\rm free}$ + ${\cal L}_{\rm int}$, one obtains the Lagrangian density
\begin{equation}
{\cal L}=\partial ^{\mu }\phi ^{\ast i}\partial _{\mu
}\phi_{i}+\iota {\bar \psi} ^{i}\bar{\sigma}^{\mu
}\partial _{\mu }\psi_i-\frac{1}{2} (W^{ij} \psi_{i}\psi_{j}+W^{*}_{ij} {\bar \psi}^{i}{\bar \psi}^{j})-W^{i} W^{*}_{i}.
\end{equation}
In short, the most general non-gauge interactions
for chiral supermultiplets are determined by a single analytic function of
the complex scalar fields, the superpotential $W$. 

\section{The MSSM}

The supersymmetric version of SM is called the MSSM with $N=1$ generators, where N refer to 
the number of distinct copies of $Q$ and $Q^{\dagger}$.
According to the MSSM each of the fundamental particle of SM has a superpartner
with spin differing by half unit. These particles are placed either in
chiral or gauge supermultiplet as shown in \refta{tab:chiral} and \refta{tab:gauge}. 

\renewcommand{\arraystretch}{1.4}
\begin{table}[tb]
\begin{center}
\begin{tabular}{|c|c|c|c|}
\hline
Superfields & spin 0 & spin 1/2 & $(SU(3)_C, \> SU(2)_L,\> U(1)_Y)$ 
\\  
\hline
\hline
$\hat Q$ & $({\stilde u}_L\>\>\>{\stilde d}_L )$&
 $(u_L\>\>\>d_L)$ & $(\>{\bf 3},\>{\bf 2}\>,\>{1\over 6})$
\\
$\hat U$
&${\stilde u}^*_R$ & $u^\dagger_R$ & 
$(\>{\bf \overline 3},\> {\bf 1},\> -{2\over 3})$
\\ $\hat D$ &${\stilde d}^*_R$ & $d^\dagger_R$ & 
$(\>{\bf \overline 3},\> {\bf 1},\> {1\over 3})$
\\  \hline
$\hat L$ &$({\stilde \nu}\>\>{\stilde e}_L )$&
 $(\nu\>\>\>e_L)$ & $(\>{\bf 1},\>{\bf 2}\>,\>-{1\over 2})$
\\
$\hat E$
&${\stilde e}^*_R$ & $e^\dagger_R$ & $(\>{\bf 1},\> {\bf 1},\>1)$
\\  \hline
$\hat H_2$ &$({\cal H}_2^+\>\>\>{\cal H}_2^0 )$&
$(\stilde {\cal H}_2^+ \>\>\> \stilde {\cal H}_2^0)$& 
$(\>{\bf 1},\>{\bf 2}\>,\>+{1\over 2})$
\\ $\hat H_1$ & $({\cal H}_1^0 \>\>\> {\cal H}_1^-)$ & $(\stilde {\cal H}_1^0 \>\>\> \stilde {\cal H}_1^-)$& 
$(\>{\bf 1},\>{\bf 2}\>,\>-{1\over 2})$
\\  \hline
\end{tabular}
\caption[Chiral supermultiplets in the MSSM]{Chiral supermultiplets in the MSSM, their field content, and their representations in the gauge groups. 
Here $u=u,c,t$; $d=d,s,b$; $e=e,\mu,\tau$ and $\nu=\nu_{e},\nu_{\mu},\nu_{\tau}$.}
\vspace{-0.6cm}
\label{tab:chiral}
\end{center}
\end{table}

\renewcommand{\arraystretch}{1.55}
\begin{table}[t]
\begin{center}
\vspace{1 cm}
\begin{tabular}{|c|c|c|c|c|}
\hline
Superfields & spin 1/2 & spin 1 & $(SU(3)_C, \> SU(2)_L,\> U(1)_Y)$\\
\hline
\hline
$\hat G^a$ &$ \stilde g$& $g$ & $(\>{\bf 8},\>{\bf 1}\>,\> 0)$
\\
\hline
$\hat W^i$ & $ \stilde W^\pm\>\>\> \stilde W^0 $&
 $W^\pm\>\>\> W^0$ & $(\>{\bf 1},\>{\bf 3}\>,\> 0)$
\\
\hline
$\hat B$ & $\stilde B^0$&
 $B^0$ & $(\>{\bf 1},\>{\bf 1}\>,\> 0)$
\\
\hline
\end{tabular}
\caption[Gauge supermultiplets in
the MSSM]{Gauge supermultiplets in
the MSSM, their field content, and their representations in the gauge groups.}
\vspace{-0.45cm}
\label{tab:gauge}
\end{center}
\end{table}

In order to keep anomaly cancellation, contrary to the SM a second
Higgs doublet is needed~\cite{glawei}. One Higgs doublet, ${\cal H}_{1}$,
gives mass to the $d$-type fermions (with weak isospin -1/2), the
other doublet, ${\cal H}_{2}$, gives mass to the $u$-type fermions (with weak
isospin +1/2). 
All SM multiplets, including the two Higgs doublets, are extended to
supersymmetric multiplets, resulting in scalar partners for quarks and
leptons (``squarks'' and ``sleptons'') and fermionic partners for the
SM gauge boson and the Higgs bosons (``gauginos'' and ``gluinos'') as shown in 
\refta{tab:chiral} and \refta{tab:gauge}.   

The mass eigenstates of the gauginos are linear combinations of these
fields, denoted as ``neutralinos'' and ``charginos''. Also the left- and
right-handed squarks (and sleptons) can mix, yielding the mass
eigenstates (denoted by the indices
$1,2$ instead of $L,R$). The EW interaction eigenstates and mass eigenstates 
of the MSSM particle spectrum are given in \refta{tab:MSSMparticles}. Here 
mass eigenstates are written with the assumption of no flavor violation. If 
flavor violation is assumed, all six up- and down-type squarks and all six charged sleptons mix separately
to give six mass eigenstates (see \refse{sec:sfermions}).

\begin{table}[htb]
\begin{center}
\renewcommand{\arraystretch}{1.8}
\begin{tabular}{|c|c|c|} \hline
\rule[-2ex]{0mm}{5ex}%
Particle &  Electroweak eigenstate & Mass Eigenstate \\
\hline \hline  
squarks  & $\tilde u_L$,$\tilde u_R$,$\tilde d_L$,$\tilde d_R$ & $\tilde u_1$,$\tilde u_2$,$\tilde d_1$,$\tilde d_2$  \\
  & $\tilde c_L$,$\tilde c_R$, $\tilde s_L$,$\tilde s_R$ & $\tilde c_1$,$\tilde c_2$,$\tilde s_1$,$\tilde s_2$  \\
  & $\tilde t_L$,$\tilde t_R$,$\tilde b_L$,$\tilde b_R$ & $\tilde t_1$,$\tilde t_2$,$\tilde b_1$,$\tilde b_2$  \\
\hline
sleptons  & $\tilde e_L$,$\tilde e_R$,$\tilde \nu_e$  & $\tilde e_1$,$\tilde e_2$,$\tilde \nu_e$  \\
  & $\tilde \mu_L$,$\tilde \mu_R$,$\tilde \nu_{\mu}$  & $\tilde \mu_1$,$\tilde \mu_2$,$\tilde \nu_{\mu}$  \\ 
  & $\tilde \tau_L$,$\tilde \tau_R$,$\tilde \nu_{\tau}$ & $\tilde \tau_1$,$\tilde \tau_2$,$\tilde \nu_{\tau}$  \\
\hline
neutralinos  & $\tilde B$,$\tilde W$,$\tilde H_u^0$,$\tilde H_d^0$  & $\tilde \chi_1^0$,$\tilde \chi_2^0$,$\tilde \chi_3^0$,$\tilde \chi_4^0$  \\
\hline
charginos  & $\tilde W^{\pm}$,$\tilde H_u^+$,$\tilde H_d^{-}$  & $\tilde \chi_1^{\pm}$,$\tilde \chi_2^{\pm}$\\
\hline
gauge boson  & $B$,$W^1$,$W^2$,$W^3$  & $W^{\pm}$,$Z$,$\gamma$  \\
\hline
gluon and gluino  & $g$,$\tilde g$  & $g$,$\tilde g$  \\
\hline
\end{tabular}
\end{center}
\caption[The EW interaction eigenstates and mass eigenstates of the MSSM.]
{The EW interaction eigenstates and mass eigenstates of the MSSM particles. No flavor mixing is assumed here.}
\label{tab:MSSMparticles}
\end{table}

At knowing the particle content of MSSM, one can write the most general $%
SU(3)_{\rm C}\times SU(2)_{\rm L}\times U(1)_{\rm Y}$ gauge invariant and renormalizable
superpotential as \cite{mssm}
\begin{eqnarray}
W_{\rm MSSM} =\epsilon _{ab}[Y_{ij}^{e}\hat{H}_{1}^{a}\hat{L}_{i}^{b}%
\hat{E}_{j}^{C}+Y_{ij}^{d}\hat{H}_{1}^{a}\hat{Q}_{i}^{b}\hat{%
D}_{j}^{C}+Y_{ij}^{u}\hat{H}_{2}^{a}\hat{Q}_{i}^{b}\hat{U}%
_{j}^{C}-\mu \hat{H}_{1}^{a}\hat{H}_{2}^{b}]
\label{superpotential}
\end{eqnarray}
where $\hat L$ represents the chiral multiplet of a $SU(2)_{\rm L}$ doublet
lepton, $\hat E$ a $SU(2)_{\rm L}$ singlet charged lepton, $\hat H_1$ and $\hat H_2$ two Higgs multiplets with opposite hypercharge.
Similarly $\hat Q$, $\hat U$ and $\hat D$ represent chiral multiplets of quarks of a
$SU(2)_{\rm L}$ doublet and two singlets with different $U(1)_{\rm Y}$ charges
whereas $i,j=1,2,3$ are family indices and a, b are $SU(2)$ indices. The symbol $\epsilon_{ab}$ is an anti-symmetric tensor with $\epsilon_{12}=1$.

As mentioned in \refse{motivation}, SUSY is not an exact symmetry of nature. It must be a broken symmetry.
The general set-up for the SSB
parameters is given by~\cite{mssm}
\begin{eqnarray}
\label{softbreaking}
-\cL_{\rm soft}&=&(m_{\tilde Q}^2)_i^j {\tilde {\cal Q}}^{\dagger i}
{\tilde {\cal Q}}_{j}
+(m_{\tilde U}^2)^i_j {\tilde {\cal U}}_{i}^* {\tilde {\cal U}}^j
+(m_{\tilde D}^2)^i_j {\tilde {\cal D}}_{i}^* {\tilde {\cal D}}^j
\nonumber \\
& &+(m_{\tilde L}^2)_i^j {\tilde {\cal L}}^{\dagger i}{\tilde {\cal L}}_{j}
+(m_{\tilde E}^2)^i_j {\tilde {\cal E}}_{i}^* {\tilde {\cal E}}^j
\nonumber \\
& &+m^2_{H_1}\cH_1^{\dagger} \cH_1
+m^2_{H_2}\cH_2^{\dagger} \cH_2
+(B \mu \cH_1 \cH_2
+ {\rm h.c.})
\nonumber \\
& &+ ( ({\bar A}^d)_{ij}\cH_1 {\tilde {\cal D}}_{i}^*{\tilde {\cal Q}}_{j}
+({\bar A}^u)_{ij}\cH_2 {\tilde {\cal U}}_{i}^*{\tilde {\cal Q}}_{j}
+({\bar A}^e)_{ij}\cH_1 {\tilde {\cal E}}_{i}^*{\tilde {\cal E}}_{j}
\nonumber \\
& & +\frac{1}{2}M_1 {\tilde B}_L^0 {\tilde B}_L^0
+\frac{1}{2}M_2 {\tilde W}_L^a {\tilde W}_L^a
+\frac{1}{2}M_3 {\tilde G}^a {\tilde G}^a + {\rm h.c.}).
\end{eqnarray}
Here we have used calligraphic capital letters for the sfermion fields in the
interaction basis with generation indices, 
\begin{eqnarray}
\tilde {\cal U}_{1,2,3}&=&\tilde u_R,\tilde c_R,\tilde t_R  ; \quad 
\tilde {\cal D}_{1,2,3}=\tilde d_R,\tilde s_R,\tilde b_R ; \quad 
\tilde {\cal Q}_{1,2,3}=(\tilde u_L \, \tilde d_L)^T, (\tilde c_L\, \tilde s_L)^T, (\tilde t_L \, \tilde b_L)^T \notag \\ 
\tilde {\cal E}_{1,2,3}&=&\tilde e_R,\tilde \mu_R,\tilde \tau_R ; \quad  
\tilde {\cal L}_{1,2,3}=(\tilde \nu_{eL} \, \tilde e_L)^T, (\tilde \nu_{\mu L}\, \tilde \mu_L)^T, (\tilde \nu_{\tau L} \, \tilde \tau_L)^T
\end{eqnarray}
and all the gauge indices have been omitted.
Here $m_{\tilde Q}^2$ and
$m_{\tilde L}^2$ are $3 \times 3$ 
matrices in family space (with $i,j$ being the
generation indeces) for the soft masses of the
left handed squark ${\tilde {\cal Q}}$ and slepton ${\tilde {\cal L}}$
$SU(2)$ doublets, respectively. $m_{\tilde U}^2$, $m_{\tilde D}^2$ and
$m_{\tilde E}^2$ contain the soft masses for right handed up-type squark
${\tilde {\cal U}}$,  down-type squarks ${\tilde {\cal D}}$ and charged
slepton ${\tilde {\cal E}}$ $SU(2)$ singlets, respectively. $\bar A^u$, 
$\bar A^d$ and $\bar A^e$ are the $3 \times 3$ matrices for the trilinear
couplings for up-type squarks, down-type 
squarks and charged slepton, respectively.
$m_{H_1}$ and $m_{H_2}$ contain the soft
masses of the Higgs sector. In the last line $M_1$, $M_2$ and $M_3$
define the bino, wino  and gluino mass terms, respectively.

It is noteworthy that the terms in \refeq{superpotential} conserve lepton and baryon numbers,
which is neither required by gauge invariance nor by renormalization. One can add the terms of the form
\begin{eqnarray}
\epsilon _{ab}[\lambda _{ijk}%
\hat{L}_{i}^{a}\hat{L}_{j}^{b}\hat{E}_{k}^{c}+\lambda
_{ijk}^{\prime }\hat{L}_{i}^{a}\hat{Q}_{j}^{b}\hat{D}%
_{k}^{c}+\lambda _{ijk}^{\prime \prime }\hat{U}_{i}^{c}\hat{D}%
_{j}^{c}\hat{D}_{k}^{c}]
\end{eqnarray}
to \refeq{superpotential} where $\lambda _{ijk}$, $\lambda_{ijk}^{\prime }$ and $\lambda _{ijk}^{\prime \prime }$ are the R-parity violating couplings. 
However these terms violate either lepton or baryon number by one unit, and presence of
these terms have dangerous impact on matter i.e. these terms
lead to fast proton decay, which is in contradiction to experimental observations.
So in order to avoid this situation we have to introduce ad-hoc symmetry,
known as R-parity, defined as \cite{fayet,Ferr}
\begin{equation}
R_{p}=(-1)^{3(B-L)+2s}  
\label{1.46}
\end{equation}
where $B$ represents baryon number, $L$ the lepton number and $s$ the
intrinsic spin of the particle. Invariance of Lagrangian under R-parity
implies $-1$ phase for sparticle and $+1$ for SM particles. Under this condition the $L$ and $B$ number violating processes are
prohibited, this prevents the proton from decaying rapidly.

Though R-parity is introduced by hand just to safe proton decay, but it has
large impact on particle physics phenomenology.  The
conservation of R-parity demands that the sparticles are always produced in
pairs. e.g. the LSP must be stable and
is assumed as the excellent candidate for DM. To detect a LSP, collider experiments
search for missing transverse energy that would arise if one of these particles were created
during a collision process and escaped undetected. For example, at the LHC, the major SUSY production processes are gluinos $\tilde g$ and
squarks $\tilde q$ e.g., $ p + p \rightarrow \tilde g + \tilde q$. These then decay into lighter SUSY particles. The final states involve two lightest neutralinos $\tilde \chi^{0}_{1}$ (giving rise to missing transverse energy $E_{T}^{\rm miss}$), quarks (jets) and leptons. The signal is
thus $E_{T}^{\rm miss}$+ jets +leptons, which should be observable at the LHC detectors.


\subsection{The scalar fermion sector}
\label{subsec:sfermions}

The squarks and charged sleptons mass term (sneutrinos being treated differently) of the MSSM Lagrangian is given by
\BE
{\cal L}_{m_{\tilde{f}}} = -\frac{1}{2} 
   \Big( \tilde{f}_L^{\dag},\tilde{f}_R^{\dag} \Big)\; {\bf M_{\tilde f}} \; 
   \VL \tilde{f}_L \\[0.5ex] \tilde{f}_R \VR~,
\label{squarksmassmatrix}
\end{equation}
where
\BE
\renewcommand{\arraystretch}{1.5}
{\bf M_{\tilde f}} = \ML M_{\tilde{f}}^2 + M_Z^2\CZb(I_3^f-Q_f\sw^2) + m_f^2 
    & m_f X_f \\
    m_f X_f 
    & M_{\tilde{f}'}^2 + M_Z^2 \CZb Q_f\sw^2 + m_f^2 \VR ,
\label{squarkmassenmatrix}
\end{equation}
with $X_f=A_f - \mu \{\CTb;\Tb\}$ and $\Tb = v_2/v_1$, the ratio of the VEV's of the two Higgs doublets, corresponds to $d$-type squarks and charged sleptons whereas $\CTb$ corresponds to $u$-type squarks.
The SSB term $M_{\tilde{f}'}$ represents right handed squarks and right handed charged sleptons. 
Sneutrino mass term is given by

\BE
M_{\tilde{\nu}}^2 
  =   m_{\tilde{L}}^2 + \left(\frac{1}{2} \MZ^2 \cos 2\beta \right)    
\label{sneutrinomass}
\EE
In order to diagonalize the sfermion mass matrix and to determine the physical mass
eigenstates the following rotation has to be performed:
\BE
\VL \sfe \\ \sfz \VR = \ML \costf & \sintf \\ -\sintf & \costf \MR 
                       \VL \sfl \\ \sfr \VR .
\label{squarkrotation}
\end{equation}
The mixing angle $\tsf$ is given for $\Tb > 1$ by:
\BEA
\costf &=& \frac{\sqrt {M_{\tilde f}^2 + M_Z^2\CZb(I_3^f-Q_f\sw^2) + m_f^2 - m_{\tilde f_2}^2}}
          {\sqrt{m_{\tilde f_1}^2-m_{\tilde f_2}^2 }} \\
\sintf &=& \frac{m_f X_f}
    {
\sqrt {M_{\tilde f}^2 + M_Z^2\CZb(I_3^f-Q_f\sw^2) + m_f^2 - m_{\tilde f_2}^2} \sqrt{m_{\tilde f_1}^2-m_{\tilde f_2}^2 }
} ~. 
\label{stt}
\EEA
The masses are given by the eigenvalues of the mass matrix:
\BEA
\label{Squarkmasse} 
m_{\tilde{f}_{1,2}}^2 &=& m_f^2+\frac{1}{2} \Big[M_{\tilde{f}}^2 + M_{\tilde{f}'}^2 + M_Z^2 \CZb I^f_3 \\
& & \mp \sqrt {[M_{\tilde f}^2-M_{\tilde{f}'}^2+ M_Z^2\CZb(I_3^f-Q_f\sw^2)]^2+4m_f^2|X_f|^2}~\Big].
\EEA
Since the non-diagonal entry of the mass matrix
\refeq{squarkmassenmatrix} is proportional to the fermion mass, 
mixing becomes particularly important for $\tilde f = \Stop$, 
in the case of $\Tb \gg 1$ also for $\tilde f = \Sbot$.

\subsection{The Higgs sector of the MSSM}
\label{subsec:higgssector}

The two Higgs doublets form the Higgs potential~\cite{hhg}
\BEA
V &=& (m_1^2 + |\mu|^2) |\cHe|^2 + (m_2^2 + |\mu|^2) |\cHz|^2 
      - m_{12}^2 (\epsilon_{ab} \cHe^a\cHz^b + \hc)  \non \\
  & & + \frac{1}{8}({g_1}^2+{g_2}^2) \left[ |\cHe|^2 - |\cHz|^2 \right]^2
        + \frac{1}{2} {g_2}^2|\cHe^{\dag} \cHz|^2~,
\label{higgspot}
\EEA
which contains $m_1, m_2, m_{12}$ as SSB parameters.
The doublet fields $\cHe$ and $\cHz$ are decomposed  in the following way:
\BEA
\cHe &=& \VL \cHe^0 \\[0.5ex] \cHe^- \VR \; = \; \VL v_1 
        + \frac{1}{\sqrt2}(\phi_1^0 - i\chi_1^0) \\[0.5ex] -\phi_1^- \VR  
        \non \\
\cHz &=& \VL \cHz^+ \\[0.5ex] \cHz^0 \VR \; = \; \VL \phi_2^+ \\[0.5ex] 
        v_2 + \frac{1}{\sqrt2}(\phi_2^0 + i\chi_2^0) \VR~.
\label{higgsfeldunrot}
\EEA
The potential (\ref{higgspot}) can be described with the help of two  
independent parameters (besides $g_1$ and $g_2$): 
$\Tb$ and $M_A^2 = -m_{12}^2(\Tb+\CTb)$,
where $M_A$ is the mass of the $\cp$-odd $A$ boson.

The diagonalization of the bilinear part of the Higgs potential,
i.e.\ the Higgs mass matrices, is performed via the orthogonal
transformations 
\BEA
\label{hHdiag}
\VL H^0 \\[0.5ex] h^0 \VR &=& \ML \Ca & \Sa \\[0.5ex] -\Sa & \Ca \MR 
\VL \phi_1^0 \\[0.5ex] \phi_2^0 \VR  \\
\label{AGdiag}
\VL G^0 \\[0.5ex] A^0 \VR &=& \ML \Cb & \Sbe \\[0.5ex] -\Sbe & \Cb \MR 
\VL \chi_1^0 \\[0.5ex] \chi_2^0 \VR  \\
\label{Hpmdiag}
\VL G^{\pm} \\[0.5ex] H^{\pm} \VR &=& \ML \Cb & \Sbe \\[0.5ex] -\Sbe & 
\Cb \MR \VL \phi_1^{\pm} \\[0.5ex] \phi_2^{\pm} \VR~.
\EEA
The mixing angle $\al$ is determined through
\BE
\tan 2\al = \tan 2\be \; \frac{\MA^2 + M_Z^2}{\MA^2 - M_Z^2} ;
\qquad  -\frac{\pi}{2} < \al < 0~.
\label{alphaborn}
\end{equation}

One gets the following Higgs spectrum:
\BEA
\mbox{2 neutral bosons},\, {\cal CP} = +1 &:& h^0, H^0 \non \\
\mbox{1 neutral boson},\, {\cal CP} = -1  &:& A^0 \non \\
\mbox{2 charged bosons}                   &:& H^+, H^- \non \\
\mbox{3 unphysical Goldstone bosons}      &:& G^0, G^+, G^- .
\EEA

The masses of the gauge bosons are given in analogy to the SM:
\BE
M_W^2 = \frac{1}{2} g_2^2 (v_1^2+v_2^2) ;\qquad
M_Z^2 = \frac{1}{2}(g_1^2+g_2^2)(v_1^2+v_2^2) ;\qquad M_\gamma=0.
\end{equation}

\bigskip
At tree level the mass matrix of the neutral $\cp$-even Higgs bosons
is given in the $\Pe$-$\Pz$-basis 
in terms of $\MZ$, $\MA$, and $\Tb$ by
\BEA
M_{\rm Higgs}^{2, {\rm tree}} &=& \ML \mpe^2 & \mpez^2 \\ 
                           \mpez^2 & \mpz^2 \MR \non\\
&=& \ML \MA^2 \SQb + \MZ^2 \CQb & -(\MA^2 + \MZ^2) \Sbe \Cb \\
    -(\MA^2 + \MZ^2) \Sbe \Cb & \MA^2 \CQb + \MZ^2 \SQb \MR,
\label{higgsmassmatrixtree}
\EEA
which by diagonalization according to \refeq{hHdiag} yields the
tree-level Higgs boson masses
\BE
M_{\rm Higgs}^{2, {\rm tree}} 
   \stackrel{\al}{\longrightarrow}
   \ML m_{H,{\rm tree}}^2 & 0 \\ 0 &  m_{h,{\rm tree}}^2 \MR~.
\end{equation}
The mixing angle $\alpha$ satisfies
\BE
\tan 2\alpha = \tan 2\beta \frac{\MA^2 + \MZ^2}{\MA^2 - \MZ^2},
\quad - \frac{\pi}{2} < \alpha < 0 .
\label{alpha}
\end{equation}
Since we treat all MSSM parameters as real there is no mixing between
$\cp$-even and $\cp$-odd Higgs bosons.

The tree-level results for the neutral 
$\cp$-even Higgs-boson masses of the MSSM read
\BE
m^2_{(H, h),{\rm tree}} =
 \frac{1}{2} \KKL \MA^2 + \MZ^2
         \pm \sqrt{(\MA^2 + \MZ^2)^2 - 4 \MZ^2 \MA^2 \CQZb} \KKR~.
\label{eq:mhtree}
\end{equation}
This implies an upper bound of $m_{h, {\rm tree}} \leq \MZ$ for the
light $\cp$-even Higgs-boson mass of the MSSM. The
direct prediction of an upper bound for the mass of the light $\cp$-even
Higgs-boson mass is one of the most striking phenomenological
predictions of the MSSM. The existence of such a bound, which does not
occur in the case of the SM Higgs boson, can be related to the fact that
the quartic term in the Higgs potential of the MSSM is given in terms of
the gauge couplings, while the quartic coupling is a free parameter in
the SM.


\subsection{Charginos}
\label{subsec:charginos}

The charginos $\tilde{\chi}_i^+\; (i=1,2)$ are four component Dirac
fermions. The mass eigenstates are obtained from the winos
$\tilde{W}^\pm$ and the charged higgsinos $\tilde{H}^-_1$,
$\tilde{H}^+_2$:
\BE
\tilde{W}^+ = \VL -i \la^+ \\[0.5ex] i \bar{\la}^- \VR 
              \quad;\quad
\tilde{W}^- = \VL -i \la^- \\[0.5ex] i \bar{\la}^+ \VR 
              \quad;\quad
\tilde{H}^+_2 = \VL \psi^+_{H_2} \\[0.5ex] \bar{\psi}^-_{H_1} \VR
                \quad;\quad
\tilde{H}^-_1 = \VL \psi^-_{H_1} \\[0.5ex] \bar{\psi}^+_{H_2} \VR~.
\end{equation}
The chargino masses are defined as mass eigenvalues of the
diagonalized mass matrix,
\BE
{\cal L}_{\tilde{\chi}^+,{\rm mass}} = -\frac{1}{2}\,
         \Big( \psi^+,\psi^- \Big) \ML 0 & {\bf X}^T \\ {\bf X} & 0 \MR
         \VL \psi^+ \\ \psi^- \VR + \hc~,
\end{equation}
or given in terms of two-component fields
\BE
\left. \begin{array}{c} 
       \psi^+ = (-i\la^+, \psi^+_{H_2}) \\[1.5ex] 
       \psi^- = (-i\la^-, \psi^-_{H_1}) 
       \end{array} 
\right.~,
\end{equation}
where {\bf X} is given by
\BE
{\bf X} = \ML M_2 & \sqrt2\, M_W\, \Sbe \\[1ex] \sqrt2\, M_W\, 
          \Cb & \mu \MR~.
\end{equation}

The physical (two-component) mass eigenstates are obtained via
unitary $(2 \times 2)$~matrices {\bf U} and {\bf V}:
\BE
\left. \begin{array}{c} 
       \chi_i^+ = V_{ij}\, \psi_j^+ \\[1.5ex] 
       \chi_i^- = U_{ij}\, \psi_j^- 
       \end{array} 
\right. \qquad i,j=1,2~.
\end{equation}
This results in a four-component Dirac spinor
\BE
\tilde{\chi}_i^+ = \VL \chi_i^+ \\[0.5ex] \bar{\chi}_i^- \VR 
                   \qquad i=1,2~,
\end{equation}
where {\bf U} and {\bf V} are given by
\BE
{\bf U} = {\bf O}_- \qquad;\qquad
{\bf V} = \left\{\begin{array}{cc} {\bf O}_+ & 
          \quad\det {\bf X}>0 \\[1ex] 
          \sigma_3\, {\bf O}_+ & \quad\det {\bf X}<0 \end{array} \right.
\end{equation}
with
\BE
{\bf O}_\pm = \ML \cos\phi_\pm & \sin\phi_\pm \\[0.5ex] 
              -\sin\phi_\pm & \cos\phi_\pm \MR~;
\end{equation}
$\cos\phi_\pm$ and $\sin\phi_\pm$ are given by
$(\epsilon = \mbox{sgn} [\det {\bf X}])$
\BEA
\tan\phi_+ = \frac{\sqrt{2}\, M_W(\Sbe m_{\tilde{\chi}_1^+} 
             + \epsilon\, \Cb m_{\tilde{\chi}_2^+})}
             {(M_2\, m_{\tilde{\chi}_1^+} 
             + \epsilon\, \mu\, m_{\tilde{\chi}_2^+} )}  \non \\ 
\tan\phi_- = \frac{-\mu\, m_{\tilde{\chi}_1^+} 
             - \epsilon\, M_2\, m_{\tilde{\chi}_2^+}} 
             {\sqrt{2}\, M_W (\Sbe m_{\tilde{\chi}_1^+} 
             + \epsilon\, \Cb m_{\tilde{\chi}_2^+})}~. 
\EEA
(If $\phi_+ < 0$ it has to be replaced by $\phi{_+} + \pi$.)
$m_{\tilde{\chi}_1^+}$ and $m_{\tilde{\chi}_2^+}$ are the eigenvalues
of the diagonalized matrix
\BEA
{\bf M}^2_{{\rm diag},\tilde{\chi}^+} &=& 
{\bf V\, X^{\dagger}\, X\, V}^{-1} \; = \; 
{\bf U^*\, X\, X^{\dagger}\, (U^*)}^{-1} \non \\
{\bf M}_{{\rm diag},\tilde{\chi}^+} &=& 
{\bf U^*\, X\, V}^{-1} \; =\; 
\ML m_{\tilde{\chi}_1^+} & 0 \\[0.5ex] 0 & m_{\tilde{\chi}_2^+} \MR .
\EEA
They are given by
\BEA
m^2_{\tilde{\chi}_{1,2}^+} &=& \frac{1}{2}\, \bigg\{ 
    M_2^2 + \mu^2 + 2M_W^2 \mp \Big[ (M_2^2-\mu^2)^2 \non \\ 
& & +\; 4M_W^4\CQZb + 4M_W^2(M_2^2+\mu^2+2\,\mu\, M_2\, \SZb) 
    \Big]^{\frac{1}{2}} \bigg\}~.
\label{Charmasse}
\EEA


\subsection{Neutralinos}
\label{subsec:neutralinos}

Neutralinos $\tilde{\chi}_i^0\; (i=1,2,3,4)$ are four-component
Majorana fermions. They are the mass eigenstates of the
photino,~$\tilde{\gamma}$, the zino,~$\tilde Z$, and the neutral higgsinos,
$\tilde{H}^0_1$ and $\tilde{H}^0_2$, with
\BE
\tilde{\gamma} = \VL -i \la_\gamma \\[0.5ex] 
                      i \bar{\la}_\gamma \VR  \quad;\quad
\tilde{Z} = \VL -i \la_Z \\[0.5ex] i \bar{\la}_Z \VR
            \quad;\quad
\tilde{H}^0_1 = \VL \psi^0_{H_1} \\[0.5ex] \bar{\psi}^0_{H_1} \VR
              \quad;\quad
\tilde{H}^0_2 = \VL \psi^0_{H_2} \\[0.5ex] \bar{\psi}^0_{H_2} \VR~.
\end{equation}
Analogously to the SM, the photino and zino are mixed states from the
bino,~$\tilde B$, and the wino,~$\tilde W$,
\BE
\tilde{B} = \VL -i \la^\prime \\[0.5ex] i \bar{\la}^\prime \VR 
\qquad;\qquad
\tilde{W}^3 = \VL -i \la^3 \\[0.5ex] i \bar{\la}^3 \VR~,
\end{equation}
with
\BEA
\tilde{\gamma} &=& \tilde{W}^3\, \sw + \tilde{B}\, \cw \non \\
\tilde{Z} &=& \tilde{W}^3\, \cw - \tilde{B}\, \sw~.
\EEA
The mass term in the Lagrange density is given by
\BE
{\cal L}_{\tilde{\chi}^0,{\rm mass}} = -\frac{1}{2}(\psi^0)^T\, {\bf Y}\, 
                                  \psi^0 + \hc~,
\end{equation}
with the two-component fermion fields
\BE
(\psi^0)^T = (-i\la^\prime , -i\la^3 , \psi_{H_1}^0 , 
              \psi_{H_2}^0)~.
\end{equation}
The mass matrix {\bf Y} is given by
\BE
\renewcommand{\arraystretch}{1.2}
{\bf Y} = \MLv M_1 & 0 & -M_Z\sw\Cb & M_Z\sw\Sbe \\ 0 & 
          M_2 & M_Z\cw\Cb & -M_Z\cw\Sbe \\
          -M_Z\sw\Cb & M_Z\cw\Cb & 0 & -\mu \\ M_Z\sw\Sbe & 
          -M_Z\cw\Sbe & -\mu & 0 \MR~.
\label{Y}
\end{equation}
The physical neutralino mass eigenstates are obtained with the
unitary transformation matrix~{\bf N}:
\BE
\chi_i^0 = N_{ij}\, \psi_j^0 \qquad i,j=1,\ldots,4,
\end{equation}
resulting in the four-component spinor (representing the mass
eigenstate) 
\BE
\tilde{\chi}_i^0 = \VL \chi_i^0 \\[0.5ex] \bar{\chi}_i^0 \VR 
\qquad i=1,\ldots,4~.
\end{equation} 
The diagonal mass matrix is then given by
\BE
{\bf M}_{{\rm diag},\tilde{\chi}^0} = {\bf N^*\, Y\, N}^{-1}~.
\end{equation}


\subsection{Gluinos}
\label{subsec:gluinos}

The gluino, $\gl$, is the spin~1/2 superpartner (Majorana fermion) of
the gluon. According to the 8 generators of $SU(3)_C$ (colour octet),
there are 8 gluinos, all having the same Majorana mass
\BE
\mgl = |M_3|~. 
\end{equation}
In SUSY GUTs $M_1$, $M_2$ and $M_3$ are not independent but connected
via 
\BE
\mgl = M_3 = \frac{g_3^2}{g_2^2}\, M_2 \; = \; 
      \frac{\al_s}{\al_{\rm em}}\, \sw^2\, M_2, \;\;
M_1 = \frac{5}{3} \frac{\sw^2}{\cw^2}\, M_2~.
\label{G-GUT}
\end{equation}

\subsection{Scalar fermion sector with flavor mixing}
\label{sec:sfermions}

In \refse{Sec:CKM} we saw how quarks are rotated from the EW interaction eigenstate basis to the mass eigenstate basis. Since squarks belong to the same supermultiplet, they need to be rotated parallel to the quarks. The rotation is performed via same matrix i.e. the CKM matrix and the relavent terms in the SSB Lagrangian given in \refeq{softbreaking} get rotated from the interaction eigenstate basis to what is known as the Super-CKM basis
\begin{eqnarray}
\label{eq:lsoft-superCKM}
   \mathcal{L}_{\rm soft}
 &=& -  {\tilde U}_{Ri}^* m_{\tilde U_R ij }^2 \tilde {U}_{Rj} 
   - {\tilde D}_{Ri}^* m_{\tilde D_R ij}^2 {\tilde D}_{Rj}
-  {\tilde U}_{Li}^* m_{\tilde U_L ij}^2 {\tilde U}_{Lj} 
   - {\tilde D}_{Li}^* m_{\tilde D_L ij}^2 {\tilde D}_{Lj} \nonumber \\
&&-   {\tilde U}_{Li} {\cal A}^u_{ij} {\tilde U}^*_{Rj} {\cal H}_2^0
- {\tilde D}_{Li} (\VCKM)_{ki} {\cal A}^u_{kj} {\tilde U}^*_{Rj} 
{\cal H}_2^+  
 -  {\tilde U}_{Li} (\VCKM^*)_{ik} {\cal A}^d_{kj} {\tilde D}^*_{Rj} 
{\cal H}_1^- \nonumber \\
&&+ {\tilde D}_{Li} {\cal A}^d_{ij} {\tilde D}^*_{Rj} {\cal H}_1^0 +  \text{h.c.},  
\end{eqnarray}
where ${\tilde U}_{L,R}$ with $U = u, c, t$ represents up-type squarks, ${\tilde D}_{L,R}$ with $D = d, s, b$ represents down-type squarks in Super-CKM basis. The soft masses $m_{\tilde U_L}$, $m_{\tilde U_R}$, $m_{\tilde D_L}$, $m_{\tilde D_R}$ and trilinear couplings ${\cal A}^{q}$ with $q=u,d$ in Super-CKM basis are related to the EW interaction eigenstate basis by
\begin{eqnarray}
{\cal A}^{q} &=& V^{q}_L {\bar A}^q V^{q \dagger}_R, \quad
 m_{\tilde U_R}^2 = V_R^u  m_{\tilde U}^2 V_R^{u \dagger}, \nonumber \\
 m_{\tilde D_R}^2 &=& V_R^d  m_{\tilde D}^2 V_R^{d \dagger}, \quad
 m_{\tilde U_L}^2 = V_L^u  m_{\tilde Q}^2 V_L^{u \dagger}, \nonumber \\
 m_{\tilde D_L}^2 &=& V_L^d  m_{\tilde Q}^2 V_L^{d \dagger}. 
\label{eq:su2}
\end{eqnarray}

In the Super-CKM basis, not only squarks with different flavor can mix among themselves but we will have left-right mixing also. This will results in $6 \times 6$
mass matrices for up-type and down-type squarks. The same arguments hold for the sleptons but in this case flavor mixing will be induced by the PMNS matrix of
the neutrino sector and transmitted by the (tiny) neutrino Yukawa couplings. Thus we will have $6 \times 6$ mass matrix for the charged sleptons in the so called Super-PMNS basis, however for the sneutrinos we have a $3 \times 3$ mass matrix, since within the MSSM even with type~I
seesaw (to be defined below), we have only three EW interaction eigenstates, ${\tilde \nu}_{L}$
with $\nu=\nu_e, \nu_\mu, \nu_\tau$ (right handed neutrinos decouple below their respective mass scale). 

The non-diagonal entries in this $6 \times 6$ general matrix for sfermions
can be described in terms of a set of 
dimensionless parameters $\deFABij$ ($F=Q,U,D,L,E; A,B=L,R$; $i,j=1,2,3$, 
$i \neq j$) where  $F$ identifies the sfermion type, $L,R$ refer to the 
``left-'' and ``right-handed'' SUSY partners of the corresponding
fermionic degrees of freedom, and $i,j$
indices run over the three generations. 
(Non-zero values for the $\deFABij$ are generated via the processes
discussed in the introduction.)

One usually writes the $6\times 6$ non-diagonal mass matrices,  
${\mathcal M}_{\tilde u}^2$ and ${\mathcal M}_{\tilde d}^2$ being ordered respectively as $(\SupL, \SchaL,
\StopL, \SupR, \SchaR, \StopR)$,  $(\SdownL, \SstrL, \SbotL, \SdownR,
\SstrR, \SbotR)$ in the Super-CKM basis, ${\mathcal M}_{\tilde l}^2$ 
being ordered as $(\SelL, \SmuL, \StauL, \SelR, \SmuR,
\StauR)$ in the Super-PMNS basis and write them in terms of left- and right-handed blocks
$M^2_{\tilde q \, AB}$, $M^2_{\tilde l \, AB}$ ($q=u,d$, $A,B=L,R$),
which are non-diagonal $3\times 3$ matrices, 
\begin{equation}
\cM_{\tilde q}^2 =\left( \begin{array}{cc}
M^2_{\tilde q \, LL} & M^2_{\tilde q \, LR} \\[.3em] 
M_{\tilde q \, LR}^{2 \, \dagger} & M^2_{\tilde q \,RR}
\end{array} \right), \qquad \tilde q= \tilde u, \tilde d~,
\label{eq:blocks-matrix}
\end{equation} 
 where:
 \begin{alignat}{5}
 M_{\tilde u \, LL \, ij}^2 
  = &  m_{\tilde U_L \, ij}^2 + \left( m_{u_i}^2
     + (T_3^u-Q_u\sw^2 ) M_Z^2 \cos 2\beta \right) \delta_{ij},  \notag\\
 M^2_{\tilde u \, RR \, ij}
  = &  m_{\tilde U_R \, ij}^2 + \left( m_{u_i}^2
     + Q_u\sw^2 M_Z^2 \cos 2\beta \right) \delta_{ij} \notag, \\
  M^2_{\tilde u \, LR \, ij}
  = &  \left< \cH_2^0 \right> {\cal A}_{ij}^u- m_{u_{i}} \mu \cot \beta \, \delta_{ij},
 \notag, \\
 M_{\tilde d \, LL \, ij}^2 
  = &  m_{\tilde D_L \, ij}^2 + \left( m_{d_i}^2
     + (T_3^d-Q_d \sw^2 ) M_Z^2 \cos 2\beta \right) \delta_{ij},  \notag\\
 M^2_{\tilde d \, RR \, ij}
  = &  m_{\tilde D_R \, ij}^2 + \left( m_{d_i}^2
     + Q_d\sw^2 M_Z^2 \cos 2\beta \right) \delta_{ij} \notag, \\
  M^2_{\tilde d \, LR \, ij}
  = &  \left< \cH_1^0 \right> {\cal A}_{ij}^d- m_{d_{i}} \mu \tb \, \delta_{ij}~,
\label{eq:SCKM-entries}
\end{alignat}
and
\begin{equation}
{\mathcal M}_{\tilde l}^2 =\left( \begin{array}{cc}
M^2_{\tilde l \, LL} & M^2_{\tilde l \, LR} \\[.3em]
M_{\tilde l \, LR}^{2 \, \dagger} & M^2_{\tilde l \,RR}
\end{array} \right),
\label{eq:slep-6x6}
\end{equation} 
 where:
 \begin{alignat}{5}
M_{\tilde l \, LL \, ij}^2 
  = &  m_{\tilde L \, ij}^2 + \left( m_{l_i}^2
     + (-\edz + \sw^2 ) \MZ^2 \cos 2\beta \right) \delta_{ij},  \notag\\
 M^2_{\tilde l \, RR \, ij}
  = &  m_{\tilde E \, ij}^2 + \left( m_{l_i}^2
     -\sw^2 \MZ^2 \cos 2\beta \right) \delta_{ij} \notag, \\
  M^2_{\tilde l \, LR \, ij}
  = &  \left< \cH_1^0 \right> {\cal A}_{ij}^e- m_{l_{i}} \mu \tb \, \delta_{ij},
\label{eq:slep-matrix}
\end{alignat}
with, $i,j=1,2,3$, $Q_u=2/3$, $Q_d=-1/3$, $T_3^u=1/2$ and
$T_3^d=-1/2$. $(m_{u_1},m_{u_2}, m_{u_3})=(m_u,m_c,m_t)$, $(m_{d_1},m_{d_2},
m_{d_3})=(m_d,m_s,m_b)$ are the quark masses and $(m_{l_1},m_{l_2},
m_{l_3})=(m_e,m_\mu,m_\tau)$ are the lepton masses.

It should be noted that the non-diagonality in flavor comes
exclusively from the SSB parameters, that could be
non-vanishing for $i \neq j$, namely: the masses $m_{\tilde U_L \, ij}^2$,
$m_{\tilde U_R \, ij}^2$, $m_{\tilde D_L \, ij}^2$, 
$m_{\tilde D_R \, ij}^2$,  $m_{\tilde L \, ij}$, $m_{\tilde E \, ij}$ 
and the trilinear couplings, ${\cal A}_{ij}^f$.   

In the sneutrino sector there is, correspondingly, a one-block $3\times
3$ mass matrix, that is referred to the $(\tinu_{eL}, \tinu_{\mu L},
\tinu_{\tau L})$ Super-PMNS basis: 
\begin{equation}
{\mathcal M}_{\tilde \nu}^2 =\left( \begin{array}{c}
M^2_{\tilde \nu \, LL}   
\end{array} \right),
\label{eq:sneu-3x3}
\end{equation} 
 where:
\begin{equation} 
  M_{\tilde \nu \, LL \, ij}^2 
  =   m_{\tilde L \, ij}^2 + \left( 
      \frac{1}{2} \MZ^2 \cos 2\beta \right) \delta_{ij},   
\label{eq:sneu-matrix}
\end{equation} 
 
It is important to note that due to $SU(2)_{\rm L}$ gauge invariance
the same soft masses $m_{\tilde Q \, ij}$ enter in both up-type and
down-type squarks mass matrices similarly $m_{\tilde L \, ij}$ enter in
both the slepton and sneutrino $LL$ mass matrices. 
The SSB parameters for the up-type squarks differ from
corresponding ones for down-type squarks by a rotation with CKM
matrix. The same would hold for sleptons i.e.\ the soft SUSY-breaking
parameters of the sneutrinos would differ from the corresponding ones
for charged sleptons by a rotation with the PMNS matrix. However, taking
the neutrino masses and oscillations 
into account in the SM leads to LFV effects that are extremely small. 
For instance, in $\mu \to e \gamma$  they are of \order{10^{-47}} in case
of Dirac neutrinos with mass around 1~eV and maximal
mixing~\cite{Kuno:1999jp,DiracNu,MajoranaNu}, and of \order{10^{-40}} in case
of Majorana neutrinos~\cite{Kuno:1999jp,MajoranaNu}. Consequently we do not
expect large effects from the inclusion of neutrino mass effects here
and neglect a rotation with the PMNS matrix. 
The sfermion mass matrices in terms of the $\deFABij$ are given as
\begin{equation}  
m^2_{\tilde U_L}= \left(\begin{array}{ccc}
 m^2_{\tilde Q_{1}} & \de_{12}^{QLL} m_{\tilde Q_{1}}m_{\tilde Q_{2}} & 
 \de_{13}^{QLL} m_{\tilde Q_{1}}m_{\tilde Q_{3}} \\
 \de_{21}^{QLL} m_{\tilde Q_{2}}m_{\tilde Q_{1}} & m^2_{\tilde Q_{2}}  & 
 \de_{23}^{QLL} m_{\tilde Q_{2}}m_{\tilde Q_{3}}\\
 \de_{31}^{QLL} m_{\tilde Q_{3}}m_{\tilde Q_{1}} & 
 \de_{32}^{QLL} m_{\tilde Q_{3}}m_{\tilde Q_{2}}& m^2_{\tilde Q_{3}} 
\end{array}\right)~,
\label{mUL}
\end{equation}
 
\noindent
\begin{equation}
m^2_{\tilde D_L}= V_{\rm CKM}^\dagger \, m^2_{\tilde U_L} \, V_{\rm CKM}~,
\label{mDL}
\end{equation}
 
\noindent 
\begin{equation}  
m^2_{\tilde U_R}= \left(\begin{array}{ccc}
 m^2_{\tilde U_{1}} & \de_{12}^{URR} m_{\tilde U_{1}}m_{\tilde U_{2}} & 
 \de_{13}^{URR} m_{\tilde U_{1}}m_{\tilde U_{3}}\\
 \de_{{21}}^{URR} m_{\tilde U_{2}}m_{\tilde U_{1}} & m^2_{\tilde U_{2}}  & 
 \de_{23}^{URR} m_{\tilde U_{2}}m_{\tilde U_{3}}\\
 \de_{{31}}^{URR}  m_{\tilde U_{3}} m_{\tilde U_{1}}& 
 \de_{{32}}^{URR} m_{\tilde U_{3}}m_{\tilde U_{2}}& m^2_{\tilde U_{3}} 
\end{array}\right)~,
\end{equation}

\noindent 
\begin{equation}  
m^2_{\tilde D_R}= \left(\begin{array}{ccc}
 m^2_{\tilde D_{1}} & \de_{12}^{DRR} m_{\tilde D_{1}}m_{\tilde D_{2}} & 
 \de_{13}^{DRR} m_{\tilde D_{1}}m_{\tilde D_{3}}\\
 \de_{{21}}^{DRR} m_{\tilde D_{2}}m_{\tilde D_{1}} & m^2_{\tilde D_{2}}  & 
 \de_{23}^{DRR} m_{\tilde D_{2}}m_{\tilde D_{3}}\\
 \de_{{31}}^{DRR}  m_{\tilde D_{3}} m_{\tilde D_{1}}& 
 \de_{{32}}^{DRR} m_{\tilde D_{3}}m_{\tilde D_{2}}& m^2_{\tilde D_{3}} 
\end{array}\right)~,
\end{equation}

\noindent 
\begin{equation}
v_2 {\cal A}^u  =\left(\begin{array}{ccc}
 m_u A_u & \de_{12}^{ULR} m_{\tilde Q_{1}}m_{\tilde U_{2}} & 
 \de_{13}^{ULR} m_{\tilde Q_{1}}m_{\tilde U_{3}}\\
 \de_{{21}}^{ULR}  m_{\tilde Q_{2}}m_{\tilde U_{1}} & 
 m_c A_c & \de_{23}^{ULR} m_{\tilde Q_{2}}m_{\tilde U_{3}}\\
 \de_{{31}}^{ULR}  m_{\tilde Q_{3}}m_{\tilde U_{1}} & 
 \de_{{32}}^{ULR} m_{\tilde Q_{3}} m_{\tilde U_{2}}& m_t A_t 
\end{array}\right)~,
\label{v2Au}
\end{equation}

\noindent 
\begin{equation}
v_1 {\cal A}^d  =\left(\begin{array}{ccc}
 m_d A_d & \de_{12}^{DLR} m_{\tilde Q_{1}}m_{\tilde D_{2}} & 
 \de_{13}^{DLR} m_{\tilde Q_{1}}m_{\tilde D_{3}}\\
 \de_{{21}}^{DLR}  m_{\tilde Q_{2}}m_{\tilde D_{1}} & m_s A_s & 
 \de_{23}^{DLR} m_{\tilde Q_{2}}m_{\tilde D_{3}}\\
 \de_{{31}}^{DLR}  m_{\tilde Q_{3}}m_{\tilde D_{1}} & 
 \de_{{32}}^{DLR} m_{\tilde Q_{3}} m_{\tilde D_{2}}& m_b A_b 
\end{array}\right)~.
\label{v1Ad}
\end{equation}

\noindent \begin{equation}  
m^2_{\tilde L}= \left(\begin{array}{ccc}
 m^2_{\tilde L_{1}} & \delta_{12}^{LLL} m_{\tilde L_{1}}m_{\tilde L_{2}} & \delta_{13}^{LLL} m_{\tilde L_{1}}m_{\tilde L_{3}} \\
 \delta_{21}^{LLL} m_{\tilde L_{2}}m_{\tilde L_{1}} & m^2_{\tilde L_{2}}  & \delta_{23}^{LLL} m_{\tilde L_{2}}m_{\tilde L_{3}}\\
\delta_{31}^{LLL} m_{\tilde L_{3}}m_{\tilde L_{1}} & \delta_{32}^{LLL} m_{\tilde L_{3}}m_{\tilde L_{2}}& m^2_{\tilde L_{3}} \end{array}\right)\end{equation}

\noindent \begin{equation}
v_1 {\cal A}^e  =\left(\begin{array}{ccc}
m_e A_e & \delta_{12}^{ELR} m_{\tilde L_{1}}m_{\tilde E_{2}} & \delta_{13}^{ELR} m_{\tilde L_{1}}m_{\tilde E_{3}}\\
\delta_{21}^{ELR}  m_{\tilde L_{2}}m_{\tilde E_{1}} & m_\mu A_\mu & \delta_{23}^{ELR} m_{\tilde L_{2}}m_{\tilde E_{3}}\\
\delta_{31}^{ELR}  m_{\tilde L_{3}}m_{\tilde E_{1}} & \delta_{32}^{ELR}  m_{\tilde L_{3}} m_{\tilde E_{2}}& m_{\tau}A_{\tau}\end{array}\right)\label{v1Al}\end{equation}

\noindent \begin{equation}  
m^2_{\tilde E}= \left(\begin{array}{ccc}
 m^2_{\tilde E_{1}} & \delta_{12}^{ERR} m_{\tilde E_{1}}m_{\tilde E_{2}} & \delta_{13}^{ERR} m_{\tilde E_{1}}m_{\tilde E_{3}}\\
 \delta_{21}^{ERR} m_{\tilde E_{2}}m_{\tilde E_{1}} & m^2_{\tilde E_{2}}  & \delta_{23}^{ERR} m_{\tilde E_{2}}m_{\tilde E_{3}}\\
\delta_{31}^{ERR}  m_{\tilde E_{3}} m_{\tilde E_{1}}& \delta_{32}^{ERR} m_{\tilde E_{3}}m_{\tilde E_{2}}& m^2_{\tilde E_{3}} \end{array}\right)\end{equation}

In this thesis, for simplicity, we are assuming that all $\deFABij$
parameters are real, therefore, hermiticity of 
${\mathcal M}_{\tilde q}^2$, ${\mathcal M}_{\tilde l}^2$ and 
${\mathcal M}_{\tilde \nu}^2$ implies $\delta_{ij}^{FAB}= \delta_{ji}^{FBA}$.

The next step is to rotate the squark states from the Super-CKM basis, 
${\tilde q}_{L,R}$, to the physical basis. 
If we set the order in the Super-CKM basis as above, 
$(\SupL, \SchaL, \StopL, \SupR, \SchaR, \StopR)$ and  
$(\SdownL, \SstrL, \SbotL, \SdownR, \SstrR, \SbotR)$, 
and in the physical basis as
${\tilde u}_{1,..6}$ and ${\tilde d}_{1,..6}$, respectively, these last
rotations are given by two $6 \times 6$ matrices, $R^{\tilde u}$ and
$R^{\tilde d}$,  
\BE
\VL  \tiu_{1} \\ \tiu_{2}  \\ \tiu_{3} \\
                                    \tiu_{4}   \\ \tiu_{5}  \\\tiu_{6}   \VR
  \; = \; R^{\tiu}  \VL \SupL \\ \SchaL \\\StopL \\ 
  \SupR \\ \SchaR \\ \StopR \VR ~,~~~~
\VL  \tid_{1} \\ \tid_{2}  \\  \tid_{3} \\
                                   \tid_{4}     \\ \tid_{5} \\ \tid_{6}  \VR             \; = \; R^{\tid}  \VL \SdownL \\ \SstrL \\ \SbotL \\
                                      \SdownR \\ \SstrR \\ \SbotR \VR ~,
\label{newsquarks}
\end{equation} 
yielding the diagonal mass-squared matrices for squarks as follows,
\BEA
{\rm diag}\{m_{\tiu_1}^2, m_{\tiu_2}^2, 
          m_{\tiu_3}^2, m_{\tiu_4}^2, m_{\tiu_5}^2, m_{\tiu_6}^2 
           \}  & = &
R^{\tiu}  \;  {\cal M}_{\tiu}^2   \; 
 R^{\tiu \dagger}    ~,\\
{\rm diag}\{m_{\tid_1}^2, m_{\tid_2}^2, 
          m_{\tid_3}^2, m_{\tid_4}^2, m_{\tid_5}^2, m_{\tid_6}^2 
          \}  & = &
R^{\tid}  \;   {\cal M}_{\tid}^2   \; 
 R^{\tid \dagger}    ~.
\EEA 

Similarly we need to rotate the sleptons and sneutrinos from the Super-PMNS basis to the physical mass eigenstate basis, 
\BE
\VL  \til_{1} \\ \til_{2}  \\ \til_{3} \\
                                    \til_{4}   \\ \til_{5}  \\\til_{6}   \VR
  \; = \; R^{\til}  \VL \SelL \\ \SmuL \\\StauL \\ 
  \SelR \\ \SmuR \\ \StauR \VR ~,~~~~
\VL  \tinu_{1} \\ \tinu_{2}  \\  \tinu_{3}  \VR             \; = \; R^{\tinu}  \VL \tinu_{eL} \\ \tinu_{\mu L}  \\  \tinu_{\tau L}   \VR ~,
\label{rotsquarks}
\end{equation} 
with $R^{\til}$ and $R^{\tinu}$ being the respective $6\times 6$ and
$3\times 3$ unitary rotating matrices that yield the diagonal
mass-squared matrices as follows, 
\BEA
{\rm diag}\{m_{\til_1}^2, m_{\til_2}^2, 
          m_{\til_3}^2, m_{\til_4}^2, m_{\til_5}^2, m_{\til_6}^2 
           \}  & = &
R^{\til}  \;  {\cal M}_{\til}^2   \; 
 R^{\til \dagger}    ~,\\
{\rm diag}\{m_{\tinu_1}^2, m_{\tinu_2}^2, 
          m_{\tinu_3}^2  
          \}  & = &
R^{\tinu}  \;   {\cal M}_{\tinu}^2   \; 
 R^{\tinu \dagger}    ~.
\EEA 

\section{Minimal Flavor Violation}

The SM has been very successfully tested by
low-energy flavor observables both from the kaon and $B_d$
sectors. In particular, the two $B$~factories have established that
$B_d$ flavor and $\cp$-violating processes are well described
by the SM up to an accuracy of the $\sim 10\%$
level~\cite{HFAgroup}.
This immediately implies a tension between the solution
of the hierarchy problem, calling for a New Physics (NP) scale at or below
the TeV scale, and the explanation of the Flavor Physics data
require a multi-TeV NP scale, if the new flavor-violating
couplings are of generic size.

An elegant way to simultaneously solve the above problems
is provided by the MFV
hypothesis~\cite{MFV1,MFV2}, where flavor and $\cp$-violation in quark sector
is assumed to entirely originate from the CKM matrix, even
in theories beyond the SM. For example in the MSSM
the off-diagonality in the sfermion mass matrix reflects the   
misalignment (in flavor space) between fermions and sfermions mass
matrices, that cannot be diagonalized simultaneously. 
This misalignment can be produced from various
origins. For instance, off-diagonal sfermion
mass matrix entries can be generated by RGE running. Going from a high energy scale, where no flavor violation is
assumed, down to the EW scale, such entries can be generated due to
presence of non diagonal Yukawa matrices in RGE's. 
For instance, in the CMSSM (see \citere{AbdusSalam:2011fc} and
references therein), the RGE effects on non-diagonal sfermion SSB
parameters are affected only by non-diagonal elements on
the Yukawa couplings and the trilinear terms which are taken as
proportional to the Yukawas at the GUT scale.  
We choose the following form of the Yukawa matrices (working in the
Super-CKM basis~\cite{slha2}), 
\begin{align}
Y^{d} = {\rm diag}(y_{d},y_{s},y_{b}), \quad
Y^{u} = V_{\rm CKM}^{\dag} {\rm diag}(y_{u},y_{c},y_{t})~.
\end{align}
Hence, all flavor violation in the quark and squark  sector is generated
by the RGE's and controlled by the CKM matrix, i.e.\ the Yukawa
couplings have a strong impact on the size of the induced off-diagonal
entries in the squark mass matrices.

The situation is somewhat different in the slepton sector where
neutrinos are strictly massless (in the SM and the MSSM).
Consequently, there is no slepton mixing, which would induce 
LFV in the charged sector, allowing not yet observed processes like $l_i \to l_j \ga$ 
($i > j$; $l_{3,2,1} = \tau, \mu, e$)~\cite{bm}. 
However in the neutral sector, we have strong experimental
evidence that shows that the neutrinos are massive and mix among
themselves~\cite{Neutrino-Osc}. 
In order to incorporate this, a seesaw mechanism (to be defined below) is
used to generate neutrino masses, and the PMNS matrix plays the role of
the CKM matrix in the lepton sector. Extending the MFV hypothesis for
leptons~\cite{MFVinlepton} we can assume that the flavor mixing in the
lepton and slepton sector is induced and controlled by the seesaw mechanism. 

\section{The Constrained MSSM}

Within the CMSSM the SSB
parameters are assumed to be universal at the Grand Unification scale
$M_{\rm GUT} \sim 2 \times 10^{16} \gev$,
\begin{eqnarray}
\label{soft}
& (m_Q^2)_{i j} = (m_U^2)_{i j} = (m_D^2)_{i j} = (m_L^2)_{i j}
 = (m_E^2)_{i j} = m_0^2\  \delta_{i j}, & \nonumber \\
& m_{H_1}^2 = m_{H_2}^2 = m_0^2, &\\
& m_{\tilde{g}}\ =\ m_{\tilde{W}}\ =\ m_{\tilde{B}}\ =\ m_{1/2}, &  
\nonumber \\
& ({\bar A}^u)_{i j}= A_0 e^{i \phi_A} (Y^u)_{i j},\ \ \ ({\bar A}^d)_{i j}= A_0 e^{i \phi_A}
(Y^d)_{i j},\ \ \ 
({\bar A}^e)_{i j}= A_0e^{i \phi_A} (Y^e)_{i j}. & \nonumber 
\end{eqnarray}
There is a common mass (square) for all the scalars, $m_0^2$, a single gaugino 
mass, $m_{1/2}$, and all the trilinear SSB terms are directly 
proportional to the corresponding Yukawa couplings in the superpotential 
with a proportionality constant $A_0 e^{i \phi_A}$, containing a
potential non-trivial complex phase.

With the use of the RGE of the MSSM,
one can obtain the SUSY spectrum at the EW scale.
All the SUSY masses and mixings are then given as 
a function of $m_0^2$, $m_{1/2}$, $A_0$, and 
$\tb$. We require radiative symmetry breaking to fix $|\mu|$ and 
$|B \mu|$ \cite{rge,bertolini} with the tree--level Higgs potential.

By definition, this model fulfills the MFV hypothesis, since the only
flavor violating terms stem from the CKM matrix. 
The important point is that, even in a model with universal SSB terms at some high energy scale as the CMSSM, some
off-diagonality in the squark mass matrices appears at the EW scale. 
Working in the basis where the squarks are rotated parallel to the
quarks i.e. the Super-CKM basis, the squark mass
matrices are not flavor diagonal at the EW scale.
This is due to the fact that at $M_{\rm GUT}$ there exist two non-trivial 
flavor structures, namely the two Yukawa matrices for the up and down quarks, 
which are not simultaneously diagonalizable. This implies that 
through RGE evolution some flavor mixing leaks into the sfermion mass matrices.
In a general SUSY model the presence of new flavor structures
in the SSB terms would generate large flavor mixing in
the sfermion mass matrices. However, in the CMSSM, the two Yukawa matrices are the only 
source of flavor change. As always in the Super-CKM basis, any off-diagonal entry 
in the sfermion mass matrices at the EW scale will be necessarily
proportional to a product of Yukawa couplings. This will play a crucial role in the analysis in chapter 6.

\section{Seesaw extensions of the MSSM}
As already mentioned in the introduction, the neutrino masses can be generated through dimension 5 operator.
There are many possible ways to form a dimension-5 gauge singlet term at low energy through the tree-level exchange 
of a heavy particle at the high energy: 
(i) each $L_L$-$\phi$ pair forms a fermion singlet, 
(ii) each of the $L_L$-$L_L$ and $\phi$-$\phi$ pair forms a scalar triplet, 
(iii) each $L_L$-$\phi$ pair forms a fermion triplet,
and (iv) each of the $L_L$-$L_L$ and $\phi$-$\phi$ pair forms a scalar singlet. 
Case (i) can arise from the tree-level exchange of a right handed fermion singlet 
and this corresponds to the Type-I seesaw mechanism
\cite{seesaw:I}. 
Case (ii) arises when the heavy particle is a Higgs triplet giving rise to the
Type-II seesaw mechanism \cite{Magg:1980ut,Lazarides:1980nt}. 
For case (iii) the exchanged particle 
should be a right-handed fermion triplet, which corresponds to 
the Type-III seesaw mechanism \cite{Foot-Type-III,Ma:2002pf}. 
The last scenario gives terms only of the form $\overline{\nu_L^C} e_L$, which 
cannot generate a neutrino mass. We describe Type-I seesaw mechanisms 
in \refse{sec:mssmI} in detail.  
\subsection{Supersymmetric Type-I seesaw model}
\label{sec:mssmI}

In order to provide an explanation for the (small) neutrino masses, the
MSSM can be extended by the type-I seesaw mechanism~\cite{seesaw:I}.
The superpotential for MSSM-seesaw I can be written as

\begin{eqnarray}
\label{superpotentialSeesaw1}
W_{\rm SI}&=&W_{\rm MSSM}+ Y_{\nu}^{ij}\epsilon_{\alpha \beta} {\hat H}_2^{\alpha} {\hat N}_i^c {\hat L}_j^{\beta}
+ \frac{1}{2} M_{N}^{ij} {\hat N}_i^c {\hat N}_j^c,
\end{eqnarray}
where $W_{\rm MSSM}$ is given in \refeq{superpotential} and ${\hat N}_i^c$
is the additional superfield that contains the three right-handed neutrinos,
$\nu_{Ri}$, and 
their scalar partners, $\tilde \nu_{Ri}$. $M_N^{ij}$ denotes the $3\times3$ Majorana mass matrix for heavy right handed neutrino.
The full set of SSB terms is given by,
\begin{eqnarray}
\label{softbreakingSeesaw1}
-\cL_{\rm soft,SI} &=& - \cL_{\rm soft}
+(m_{\tilde \nu}^2)^i_j {\tilde \nu}_{Ri}^* {\tilde \nu}_{R}^j
+ (\frac{1}{2}B_{\nu}^{ij} M_{N}^{ij} {\tilde \nu}_{Ri}^* {\tilde \nu}_{Rj}^*
+A_{\nu}^{ij}h_2 {\tilde \nu}_{Ri}^* {\tilde l}_{Lj}+ {\rm h.c.}),
\end{eqnarray}
with $\cL_{\rm soft}$ given by \refeq{softbreaking}, 
$(m_{\tilde \nu}^2)^i_j$,  $A_{\nu}^{ij}$ and $B_{\nu}^{ij}$
are the new SSB parameters.

By the seesaw mechanism three of the neutral fields acquire heavy masses and
decouple at high energy scale that we will denote as $M_N$, below this scale the effective theory
contains the MSSM plus an operator that provides masses to the neutrinos.
\begin{equation}
W_{\rm EW,SI}=W_{\rm MSSM}+ \frac{1}{2}(Y_{\nu} L  H_2)^{T}  M_{N}^{-1} (Y_{\nu} L  H_2),
\end{equation}
where $W_{\rm EW,SI}$ represent the MSSM seesaw I superpotential at EW scale. This framework naturally explains neutrino oscillations in agreement with
experimental data~\cite{Neutrino-Osc}. At the electroweak scale an
effective Majorana mass matrix for light neutrinos, 
\begin{equation}
\label{meff}
m_{\rm eff}=-\frac{1}{2}v_u^2 Y_{\nu}\cdot M_{N}^{-1}\cdot Y^{ T}_{\nu}, 
\end{equation}
arises from Dirac neutrino Yukawa $Y_{\nu}$ (that can be assumed of
the same order as the charged-lepton and quark Yukawas), and heavy Majorana
masses $M_N$.  The smallness of the neutrino masses implies that the scale
$M_N$ is very high, \order{10^{14} \gev}. 

From \refeqs{superpotentialSeesaw1} and (\ref{softbreakingSeesaw1}) 
we can observe that 
one can choose a basis such that the Yukawa coupling matrix,
$Y^e_{ij}$, and the mass matrix of the right-handed neutrinos, $M_N^{ij}$, are
diagonalized as $Y^e_\delta$ and $M_R^\delta$, respectively. In this case
the neutrino Yukawa couplings $Y_{\nu}^{ij}$ are not generally diagonal,
giving rise to LFV. Here it is important to note that the lepton-flavor
conservation is not a consequence of the SM gauge symmetry, even in the absence
of the right-handed neutrinos. 
Consequently, slepton mass terms can violate
the lepton-flavor conservation in a manner consistent with the gauge
symmetry.  Thus the scale of LFV can be identified with the
EW scale, much lower than the right-handed neutrino scale $M_N$, leading
to potentially observable rates.

In the SM augmented by 
right-handed neutrinos, the flavor violating processes such as
$\mu \to e \gamma$, $\tau \to \mu \gamma$ etc., 
whose rates are proportional to inverse powers of $M_R^\delta$, would be
highly suppressed with such a large $M_N$ scale, and hence are far
beyond current experimental bounds. However, in SUSY theories, the
neutrino Dirac couplings $Y_\nu$ 
enter in the RGE's of
the SSB sneutrino and slepton masses, generating LFV. In
the basis where the charged-lepton Yukawa couplings matrix $Y^{e}$ is diagonal, the soft
slepton-mass matrix acquires corrections that contain off-diagonal
contributions from the RGE running from $M_{\rm GUT}$ down to the 
Majorana mass scale $M_N$,
of the following form (in the leading-log approximation)~\cite{LFVhisano}: 
\begin{align}
(m_{\tilde L}^2)_{ij} &\sim \frac 1{16\pi^2} (6m^2_0 + 2A^2_0)
\left({Y_{\nu}}^{\dagger} Y_{\nu}\right)_{ij}  
\log \KL \frac{M_{\rm GUT}}{M_N} \KR \, \nonumber\\
(m_{\tilde E}^2)_{ij} &\sim 0  \, \nonumber\\
({\bar A}^e)_{ij} &\sim  \frac 3{8\pi^2} {A_0 Y^{e}}_i
\left({Y_{\nu}}^{\dagger} Y_{\nu}\right)_{ij}  
\log \KL \frac{M_{\rm GUT}}{M_N} \KR \,
\label{offdiagonal}
\end{align}
Consequently, even if the soft scalar masses were universal  at the
unification scale, quantum corrections between the GUT scale and 
the seesaw scale $M_N$
would modify this structure via renormalization-group 
running, which generates off-diagonal 
contributions~\cite{Cannoni:2013gq,gllv,Mismatch,Antusch,EGL,casas-ibarra} at 
$M_N$ in a basis such that $Y^{e}$ is diagonal. Below this  
scale, the off-diagonal contributions remain almost unchanged. 
   
Therefore the seesaw mechanism induces non trivial values for slepton
$\deFABij$ resulting in a prediction for LFV decays  
$l_i \to l_j \ga$, $(i >  j)$ that can be much larger than
the non-SUSY case. These rates depend on the structure of $Y_\nu$ at a seesaw
scale $M_N$ in a basis where $Y^e$ and $M_N$ are diagonal.  
By using the approach of \citere{casas-ibarra} a general form of $Y_\nu$
containing all neutrino experimental information can be written as: 
\begin{equation}
Y_\nu = \frac{\sqrt{2}} {v_2} \sqrt{M_R^\delta} R  \sqrt{m_\nu^\delta} U^\dagger~,
\label{eq:casas} 
\end{equation}
where $R$ is a general orthogonal matrix and $m_\nu^\de$ denotes the
diagonalized neutrino mass matrix. In this basis the matrix~$U$ can be
identified with the $U_{\rm PMNS}$ matrix obtained as: 
\begin{equation}
m_\nu^\delta=U^T m_{\rm eff} U~.
\end{equation}

In order to find values for the slepton generation mixing
parameters we need a
specific form of the product $Y_\nu^\dagger Y_\nu$ as shown in
\refeq{offdiagonal}. The 
simple consideration of direct hierarchical neutrinos with a common
scale for right handed neutrinos provides a representative reference
value. In this case using \refeq{eq:casas} we find 
\begin{equation}
Y_\nu^\dagger Y_\nu= \frac{2}{v_u^2}M_R U m_\nu^\delta U^\dagger~.
\label{eq:ynu2}
\end{equation}
Here $M_R$ is the common mass assigned to the $\nu_R$'s. In the conditions
considered here, LFV effects are independent of the matrix $R$. 

For the forthcoming numerical analysis the values of the Yukawa couplings
etc.\ have to be set to yield values in agreement with the
experimental data for neutrino masses and mixings.
In our computation, by considering a normal  hierarchy among the neutrino 
masses, we fix 
$m_{\nu_3} \sim \sqrt{\Delta m^2_{\text{atm}}} \sim 0.05 \ev$ and
require $m_{\nu_2}/m_{\nu_3}=0.17$, 
$m_{\nu_2} \sim  100 \cdot m_{\nu_1}$ consistent with the measured values of 
$\Delta m^2_{\text{sol}}$ and $\Delta m^2_{\text{atm}}$~\cite{neu-fits}. 
The matrix $U$ is identified with $U_{\rm PMNS}$ with the $\cp$-phases set to
zero and neutrino mixing angles set to the center of their
experimental values.  

One can observe that $m_{\rm eff}$ remains unchanged by consistent
changes on the scales of $M_N$ and $Y_\nu$. This is no longer correct
for the off-diagonal entries in the slepton mass matrices
(parameterized by slepton $\deFABij$). These quantities have quadratic dependence
on $Y_\nu$ and logarithmic dependence on $M_N$, see \refeq{offdiagonal}. 
Therefore larger values of $M_N$ imply larger LFV effects. 
By setting $M_N=10^{14} \gev$, the largest values of
$Y_\nu$ are of about 0.29, this
implies an important restriction on the parameters space arising from the
$\br(\mu\to e \ga)$.
An example of models  with almost degenerate $\nu_R$ can be found in
\cite{Cannoni:2013gq}. For our numerical analysis we tested several scenarios
and we found that the one defined here is the simplest and also the
one with larger LFV prediction. 


\chapter{Precision Observables}
\label{precision-observables}
In this chapter we will present the calculational details and experimetal status of the various low energy observables considered in this thesis.

\section{Higher order corrections to EWPO}
\label{sec:EWPO-calc}

EWPO that are known with an accuracy at the per-mille level or better 
have the potential to allow a discrimination between quantum effects of 
the SM and SUSY models, see \citere{PomssmRep} for a review.  Examples 
are the $W$-boson mass $\MW$ and the $Z$-boson observables, such as the 
effective leptonic weak missxing angle $\sweff$.

The $W$-boson mass can be evaluated from
\begin{equation}
\MW^2 \KL 1 - \frac{\MW^2}{\MZ^2} \KR = 
\frac{\pi \al}{\wz \Gmu} (1 + \De r)
\end{equation}
where $\al$ is the fine-structure constant and $\Gmu$ the Fermi 
constant.  This relation arises from comparing the prediction for muon 
decay with the experimentally precisely known Fermi constant.  
The one-loop contributions to $\De r$ can be written as
\begin{equation}
\De r = \De\al - \frac{\cw^2}{\sw^2}\De\rho + (\De r)_{\text{rem}},
\end{equation}
where $\De\al$ is the shift in the fine-structure constant due to the 
light fermions of the SM, $\De\al \propto \log(\MZ/m_f)$, and $\De\rho$ 
is the leading contribution to the $\rho$ parameter~\cite{rho} from 
(certain) fermion and sfermion loops (see below).  The remainder part 
$(\De r)_{\text{rem}}$ contains in particular the contributions from the 
Higgs sector.
The effective leptonic weak mixing angle at the $Z$-boson resonance, 
$\sweff$, is defined through the vector and axial-vector couplings 
($g_{\text{V}}^\ell$ and $g_{\text{A}}^\ell$) of leptons ($\ell$) to the 
$Z$~boson, measured at the $Z$-boson pole.  If this vertex is written as 
$i\bar \ell\ga^\mu (g_{\text{V}}^\ell - g_{\text{A}}^\ell \ga_5) \ell 
Z_\mu$ then
\begin{equation}
\sweff = \frac 14 \KL 1 -
  \re\frac{g_{\text{V}}^\ell}{g_{\text{A}}^\ell}\KR\,.
\end{equation}
Loop corrections enter through higher-order contributions to 
$g_{\text{V}}^\ell$ and $g_{\text{A}}^\ell$.
Both of these (pseudo-)observables are affected by shifts in the 
quantity $\De\rho$ according to
\begin{equation}
\label{eq:precobs}
\De\MW \approx \frac{\MW}{2}\frac{\cw^2}{\cw^2 - \sw^2} \De\rho\,, \quad
\De\sweff \approx - \frac{\cw^2 \sw^2}{\cw^2 - \sw^2} \De\rho\,.
\end{equation}
The quantity $\De\rho$ is defined by the relation
\begin{equation}
\De\rho = \frac{\Si_Z^{\text{T}}(0)}{\MZ^2} -
          \frac{\Si_W^{\text{T}}(0)}{\MW^2}
\label{eq:drho}
\end{equation} 
with the unrenormalized transverse parts of the $Z$- and $W$-boson 
self-energies at zero momentum, $\Si_{Z,W}^{\text{T}}(0)$.  It 
represents the leading universal corrections to the EWPO
induced by mass splitting between partners in 
isospin doublets~\cite{rho}. Consequently, it is sensitive to the 
mass-splitting effects induced by flavor mixing.

Within the SM the corrections to $\De\rho$ stem from the
splitting in one $SU(2)$ doublet. Due to the mixing of various scalar
fermion states the picture is slightly more involved in the MSSM.
In MSSM without flavor violation the well known results for the third
generation squark contribution to $\De\rho$ (without flavor mixing) can
be written as  
\begin{eqnarray}
\De\rho = \frac{3G_{\mu}}{8 \sqrt{2}\pi^2} \big[
-\sin^2 \tst \cos^2 \tst F_{0}(\mste^2,\mstz^2)
-\sin^2 \tsb \cos^2 \tsb F_{0}(\msbe^2,\msbz^2)
\nonumber \\
+\cos^2 \tst \cos^2 \tsb F_{0}(\mste^2,\msbe^2)
+\sin^2 \tsb \cos^2 \tst F_{0}(\mste^2,\msbz^2)
\nonumber \\
+\sin^2 \tst \cos^2 \tsb F_{0}(\mstz^2,\msbe^2)
+\sin^2 \tst \sin^2 \tsb F_{0}(\mstz^2,\msbz^2) \big] 
\label{drhoMSSM}
\end{eqnarray}
with
\begin{align}
F_{0}(m^2_1, m^2_2) &=
m^2_1+m^2_2-\frac{2m^2_1 m^2_2}{m^2_1-m^2_2} \ln\KL\frac{m^2_1}{m^2_2}\KR~.
\label{Fun:f0}
\end{align}
In the absence of intergenerational mixing there are only $2 \times 2$
mixing matrices to be taken into account, here parametrized by $\tst$
($\tsb$) in the scalar top (bottom) case.
Here one can see that squarks do not need to be the
$SU(2)$ partners to give contribution to $\De\rho$. In particular the
first two terms of \refeq{drhoMSSM} describe contributions from the 
same type (up type or down type) of scalar quarks.
Going from this simple case to the one with generation mixing one 
finds contribution from all three generations, including two $6 \times 6$
mixing matrices (which are difficult to analyze analytically).
The two gauge boson self-energies are then
  given by (see also \citere{delrhoNMFV}), 
\begin{align*}
\Sigma_{ZZ}(0)  & =\frac{e^{2}}{288\pi^{2}\sw^{2}\cw^{2}}(-%
{\displaystyle\sum\limits_{s,t=1}^{6}}
{\displaystyle\sum\limits_{i,j=1}^{3}}
2[\frac{1}{8}F_{0}(m_{\tilde{u}_{s}}^{2},m_{\tilde{u}_{t}}^{2})+\frac{1}%
{4}(A_0^{\rm fin}(m_{\tilde{u}_{s}}^{2})+A_0^{\rm fin}(m_{\tilde{u}_{t}}^{2}))]\\
& \{3R_{t,j}^{\tilde{u}}R_{t,j}^{\tilde{u}^{\ast}}-4\sw^{2}(R_{t,j}%
^{\tilde{u}}R_{t,j}^{\tilde{u}^{\ast}}+R_{t,3+j}^{\tilde{u}}R_{t,3+j}%
^{\tilde{u}^{\ast}})\}\\
& \{3R_{s,i}^{\tilde{u}}R_{s,i}^{\tilde{u}^{\ast}}-4\sw^{2}(R_{s,i}%
^{\tilde{u}}R_{s,i}^{\tilde{u}^{\ast}}+R_{s,3+i}^{\tilde{u}}R_{s,3+i}%
^{\tilde{u}^{\ast}})\}\\
& -%
{\displaystyle\sum\limits_{s,t=1}^{6}}
{\displaystyle\sum\limits_{i,j=1}^{3}}
2[\frac{1}{8}F_{0}(m_{\tilde{d}_{s}}^{2},m_{\tilde{d}_{t}}^{2})+\frac{1}%
{4}(A_0^{\rm fin}(m_{\tilde{d}_{s}}^{2})+A_0^{\rm fin}(m_{\tilde{d}_{t}}^{2}))]\\
& \{3R_{t,j}^{\tilde{d}}R_{t,j}^{\tilde{d}^{\ast}}-2\sw^{2}(R_{t,j}%
^{\tilde{d}}R_{t,j}^{\tilde{d}^{\ast}}+R_{t,3+j}^{\tilde{d}}R_{t,3+j}%
^{\tilde{d}^{\ast}})\}\\
& \{3R_{s,i}^{\tilde{u}}R_{s,i}^{\tilde{u}^{\ast}}-2\sw^{2}(R_{s,i}%
^{\tilde{d}}R_{s,i}^{\tilde{d}^{\ast}}+R_{s,3+i}^{\tilde{d}}R_{s,3+i}%
^{\tilde{d}^{\ast}}\}\\
& +%
{\displaystyle\sum\limits_{s=1}^{6}}
{\displaystyle\sum\limits_{i=1}^{3}}
A_0^{\rm fin}(m_{\tilde{u}_{s}}^{2})[(3-4\sw^{2})^{2}R_{s,i}^{\tilde{u}}%
R_{s,i}^{\tilde{u}^{\ast}}+16\sw^{4}R_{s,3+i}^{\tilde{u}}R_{s,3+i}%
^{\tilde{u}^{\ast}}]\\
& +%
{\displaystyle\sum\limits_{s=1}^{6}}
{\displaystyle\sum\limits_{i=1}^{3}}
A_0^{\rm fin}(m_{\tilde{d}_{s}}^{2})[(3-2\sw^{2})^{2}R_{s,i}^{\tilde{d}}%
R_{s,i}^{\tilde{d}^{\ast}}+4\sw^{4}R_{s,3+i}^{\tilde{d}}R_{s,3+i}^{\tilde
{d}^{\ast}}])
\end{align*}

\begin{align*}
\Sigma_{WW}(0)  & =\frac{e^{2}}{32\pi^{2}\sw^{2}}(-%
{\displaystyle\sum\limits_{s,t=1}^{6}}
{\displaystyle\sum\limits_{i,j=1}^{3}}
4[\frac{1}{8}F_{0}(m_{\tilde{u}_{s}}^{2},m_{\tilde{d}_{t}}^{2})+\frac{1}%
{4}(A_0^{\rm fin}(m_{\tilde{u}_{s}}^{2})+A_0^{\rm fin}(m_{\tilde{d}_{t}}^{2}))]\\
& R_{s,i}^{\tilde{u}}R_{t,j}^{\tilde{d}}R_{s,j}^{\tilde{u}^{\ast}}%
R_{t,i}^{\tilde{d}^{\ast}}\\
& +%
{\displaystyle\sum\limits_{s=1}^{6}}
{\displaystyle\sum\limits_{i=1}^{3}}
A_0^{\rm fin}(m_{\tilde{u}_{s}}^{2})R_{s,i}^{\tilde{u}}R_{s,i}^{\tilde{u}^{\ast}}+%
{\displaystyle\sum\limits_{s=1}^{6}}
{\displaystyle\sum\limits_{i=1}^{3}}
A_0^{\rm fin}(m_{\tilde{d}_{s}}^{2})R_{s,i}^{\tilde{d}}R_{s,i}^{\tilde{d}^{\ast}}%
\end{align*}
Here 
$R^{\tilde{u}}$ and $R^{\tilde{d}}$ are the $6\times6$ rotation matrices for
the up and down-type squarks respectively, see \refeq{newsquarks}.
The finite part of the one point integral function is given by
\begin{align}
A_{0}^{\rm fin}(m^{2})=m^{2}(1-\log\frac{m^{2}}{\mu^{2}})~.
\end{align}

Here it is important to note that
the corrections will come, as in \refeq{drhoMSSM}, from states
connected via $SU(2)$ as well as from ``same flavor'' contributions
stemming from the $Z$~boson self-energy, see \refeq{eq:drho}. 
Larger splitting between ``same flavor'' states due to the
intergenerational mixing thus leads to the expectation of increasing
contributions to $\De\rho$ from flavor violation effects.

The effects from flavor violation in the squark 
entering via $\De\rho$ were already evaluated in Ref.
\cite{delrhoNMFV} and included in \fh. 
We have calculated the effects of slepton flavor mixing to $\De\rho$ via  \fa/\fc\ setup and added the results to \fh\ for our numerical evaluation. The details about the changes made to \fa, \fc\ and \fh\ will be discussed in \refse{sec:feynhiggs}.
In \reffi{FeynDiagZSelf} and \reffi{FeynDiagWSelf}, we show the generic Feynman diagrams for $Z$ and $W$ boson self energies that enter in the calculation of $\De\rho$.  
\begin{figure}[htb!]
\begin{center}

\unitlength=1.0bp%

\begin{feynartspicture}(432,190)(4,2)

\FADiagram{}
\FAProp(0.,10.)(6.,10.)(0.,){/Sine}{0}
\FALabel(3.,8.93)[t]{$Z$}
\FAProp(20.,10.)(14.,10.)(0.,){/Sine}{0}
\FALabel(17.,11.07)[b]{$Z$}
\FAProp(6.,10.)(14.,10.)(0.8,){/ScalarDash}{-1}
\FALabel(10.,5.73)[t]{$\tilde u_{t}$}
\FAProp(6.,10.)(14.,10.)(-0.8,){/ScalarDash}{1}
\FALabel(10.,14.27)[b]{$\tilde u_{s}$}
\FAVert(6.,10.){0}
\FAVert(14.,10.){0}

\FADiagram{}
\FAProp(0.,10.)(6.,10.)(0.,){/Sine}{0}
\FALabel(3.,8.93)[t]{$Z$}
\FAProp(20.,10.)(14.,10.)(0.,){/Sine}{0}
\FALabel(17.,11.07)[b]{$Z$}
\FAProp(6.,10.)(14.,10.)(0.8,){/ScalarDash}{-1}
\FALabel(10.,5.73)[t]{$\tilde d_{t}$}
\FAProp(6.,10.)(14.,10.)(-0.8,){/ScalarDash}{1}
\FALabel(10.,14.27)[b]{$\tilde d_{s}$}
\FAVert(6.,10.){0}
\FAVert(14.,10.){0}

\FADiagram{}
\FAProp(0.,10.)(6.,10.)(0.,){/Sine}{0}
\FALabel(3.,8.93)[t]{$Z$}
\FAProp(20.,10.)(14.,10.)(0.,){/Sine}{0}
\FALabel(17.,11.07)[b]{$Z$}
\FAProp(6.,10.)(14.,10.)(0.8,){/ScalarDash}{-1}
\FALabel(10.,5.73)[t]{$\tilde l_{t}$}
\FAProp(6.,10.)(14.,10.)(-0.8,){/ScalarDash}{1}
\FALabel(10.,14.27)[b]{$\tilde l_{s}$}
\FAVert(6.,10.){0}
\FAVert(14.,10.){0}

\FADiagram{}
\FAProp(0.,10.)(6.,10.)(0.,){/Sine}{0}
\FALabel(3.,8.93)[t]{$Z$}
\FAProp(20.,10.)(14.,10.)(0.,){/Sine}{0}
\FALabel(17.,11.07)[b]{$Z$}
\FAProp(6.,10.)(14.,10.)(0.8,){/ScalarDash}{-1}
\FALabel(10.,5.73)[t]{$\tilde \nu_{j}$}
\FAProp(6.,10.)(14.,10.)(-0.8,){/ScalarDash}{1}
\FALabel(10.,14.27)[b]{$\tilde \nu_{i}$}
\FAVert(6.,10.){0}
\FAVert(14.,10.){0}

\FADiagram{}
\FAProp(0.,10.)(10.,10.)(0.,){/Sine}{0}
\FALabel(5.,8.93)[t]{$Z$}
\FAProp(20.,10.)(10.,10.)(0.,){/Sine}{0}
\FALabel(15.,8.93)[t]{$Z$}
\FAProp(10.,10.)(10.,10.)(10.,15.5){/ScalarDash}{-1}
\FALabel(10.,16.57)[b]{$\tilde u_{s}$}
\FAVert(10.,10.){0}

\FADiagram{}
\FAProp(0.,10.)(10.,10.)(0.,){/Sine}{0}
\FALabel(5.,8.93)[t]{$Z$}
\FAProp(20.,10.)(10.,10.)(0.,){/Sine}{0}
\FALabel(15.,8.93)[t]{$Z$}
\FAProp(10.,10.)(10.,10.)(10.,15.5){/ScalarDash}{-1}
\FALabel(10.,16.57)[b]{$\tilde d_{s}$}
\FAVert(10.,10.){0}

\FADiagram{}
\FAProp(0.,10.)(10.,10.)(0.,){/Sine}{0}
\FALabel(5.,8.93)[t]{$Z$}
\FAProp(20.,10.)(10.,10.)(0.,){/Sine}{0}
\FALabel(15.,8.93)[t]{$Z$}
\FAProp(10.,10.)(10.,10.)(10.,15.5){/ScalarDash}{-1}
\FALabel(10.,16.57)[b]{$\tilde l_{s}$}
\FAVert(10.,10.){0}

\FADiagram{}
\FAProp(0.,10.)(10.,10.)(0.,){/Sine}{0}
\FALabel(5.,8.93)[t]{$Z$}
\FAProp(20.,10.)(10.,10.)(0.,){/Sine}{0}
\FALabel(15.,8.93)[t]{$Z$}
\FAProp(10.,10.)(10.,10.)(10.,15.5){/ScalarDash}{-1}
\FALabel(10.,16.57)[b]{$\tilde \nu_{i}$}
\FAVert(10.,10.){0}

\end{feynartspicture}
\end{center}
\caption[Generic Feynman diagrams for $Z$ boson self-energies]{Generic Feynman diagrams for $Z$~boson self-energies
containing squarks and sleptons in loops. 
$\tilde u_{s,t}$,$\tilde d_{s,t}$ and $\tilde l_{s,t}$ denote the six mass eigenstates of up-type, down-type and charged sleptons respectively.
$\tilde \nu_{i,j}$ are the three sneutrinos states $\tilde \nu_{e}$,
$\tilde \nu_{\mu}$ and $\tilde \nu_{\tau}$.}   
\label{FeynDiagZSelf}
\end{figure}

\begin{figure}[htb!]
\begin{center}

\unitlength=1.0bp%

\begin{feynartspicture}(432,190)(4,2)

\FADiagram{}
\FAProp(0.,10.)(6.,10.)(0.,){/Sine}{0}
\FALabel(3.,8.93)[t]{$W$}
\FAProp(20.,10.)(14.,10.)(0.,){/Sine}{0}
\FALabel(17.,11.07)[b]{$W$}
\FAProp(6.,10.)(14.,10.)(0.8,){/ScalarDash}{-1}
\FALabel(10.,5.73)[t]{$\tilde u_{t}$}
\FAProp(6.,10.)(14.,10.)(-0.8,){/ScalarDash}{1}
\FALabel(10.,14.27)[b]{$\tilde d_{s}$}
\FAVert(6.,10.){0}
\FAVert(14.,10.){0}

\FADiagram{}
\FAProp(0.,10.)(6.,10.)(0.,){/Sine}{0}
\FALabel(3.,8.93)[t]{$W$}
\FAProp(20.,10.)(14.,10.)(0.,){/Sine}{0}
\FALabel(17.,11.07)[b]{$W$}
\FAProp(6.,10.)(14.,10.)(0.8,){/ScalarDash}{-1}
\FALabel(10.,5.73)[t]{$\tilde l_{t}$}
\FAProp(6.,10.)(14.,10.)(-0.8,){/ScalarDash}{1}
\FALabel(10.,14.27)[b]{$\tilde \nu_{i}$}
\FAVert(6.,10.){0}
\FAVert(14.,10.){0}

\FADiagram{}
\FAProp(0.,10.)(10.,10.)(0.,){/Sine}{0}
\FALabel(5.,8.93)[t]{$W$}
\FAProp(20.,10.)(10.,10.)(0.,){/Sine}{0}
\FALabel(15.,8.93)[t]{$W$}
\FAProp(10.,10.)(10.,10.)(10.,15.5){/ScalarDash}{-1}
\FALabel(10.,16.57)[b]{$\tilde u_{s}$}
\FAVert(10.,10.){0}

\FADiagram{}
\FAProp(0.,10.)(10.,10.)(0.,){/Sine}{0}
\FALabel(5.,8.93)[t]{$W$}
\FAProp(20.,10.)(10.,10.)(0.,){/Sine}{0}
\FALabel(15.,8.93)[t]{$W$}
\FAProp(10.,10.)(10.,10.)(10.,15.5){/ScalarDash}{-1}
\FALabel(10.,16.57)[b]{$\tilde d_{s}$}
\FAVert(10.,10.){0}

\FADiagram{}
\FAProp(0.,10.)(10.,10.)(0.,){/Sine}{0}
\FALabel(5.,8.93)[t]{$W$}
\FAProp(20.,10.)(10.,10.)(0.,){/Sine}{0}
\FALabel(15.,8.93)[t]{$W$}
\FAProp(10.,10.)(10.,10.)(10.,15.5){/ScalarDash}{-1}
\FALabel(10.,16.57)[b]{$\tilde l_{s}$}
\FAVert(10.,10.){0}

\FADiagram{}
\FAProp(0.,10.)(10.,10.)(0.,){/Sine}{0}
\FALabel(5.,8.93)[t]{$W$}
\FAProp(20.,10.)(10.,10.)(0.,){/Sine}{0}
\FALabel(15.,8.93)[t]{$W$}
\FAProp(10.,10.)(10.,10.)(10.,15.5){/ScalarDash}{-1}
\FALabel(10.,16.57)[b]{$\tilde \nu_{i}$}
\FAVert(10.,10.){0}
\end{feynartspicture}
\end{center}
\caption[Generic Feynman diagrams for $W$ boson self-energies]{Generic Feynman diagrams for $W$ boson self-energies
containing squarks and sleptons in loops. 
$\tilde u_{s,t}$,$\tilde d_{s,t}$ and $\tilde l_{s,t}$ denote the six mass eigenstates of up-type, down-type and charged sleptons respectively.
$\tilde \nu_{i,j}$ are the three sneutrinos states $\tilde \nu_{e}$,
$\tilde \nu_{\mu}$ and $\tilde \nu_{\tau}$.}   
\label{FeynDiagWSelf}
\end{figure}

The present experimental uncertainties for the EWPO are \cite{LEPEWWG}
\begin{equation}
\label{EWPO-today}
\de\MW^{\text{exp,today}} \sim 15 \mev, \quad
\de\sweff^{\text{exp,today}} \sim 15 \times 10^{-5}\,, 
\end{equation}
which will further be reduced~\cite{Baak:2013fwa} to
\begin{equation}
\label{EWPO-future}
\de\MW^{\text{exp,future}} \sim 4\mev, \quad
\de\sweff^{\text{exp,future}} \sim 1.3 \times 10^{-5}\,,
\end{equation}
at the ILC and at the GigaZ option of the ILC, respectively. 
Even higher precision could be expected from the FCC-ee, see, e.g.,
\citere{fcc-ee-paris}.

The prediction of $\MW$ also suffers from various kinds of
theoretical uncertainties, parametric and intrinsic. Starting with the
parametric uncertainties, an 
experimental error of $1 \gev$ on $\mt$ yields a parametric uncertainty on
$\MW$ of about $6 \mev$, while the parametric uncertainties induced by the
current experimental error of the hadronic contribution to the shift in
the fine-structure constant, $\Delta\alpha_{\rm had}$, and by the
experimental error of $\MZ$ amount to about $2 \mev$ and $2.5 \mev$,
respectively. The uncertainty of the $\MW$ prediction caused by the
experimental uncertainty of the Higgs mass 
$\delta \Mh^{\rm exp} \lsim 0.3 \gev$ is
signifcantly smaller ($\approx 0.2 \mev$). The intrinsic 
uncertainties from unknown higher-order corrections in the case of no
flavor mixing have been estimated to be around (4.7-9.4)~MeV in the
MSSM \cite{Heinemeyer:2006px,Haestier:2005ja} depending on the SUSY mass
scale. For our forthcoming numerical analysis, we have added the parameteric uncertanities in quadrature and add
the result linearly to the uncertanity from 
the unknown higher order corrections in the case of no flavor mixing.
We assume an additional 10\% uncertanity from the flavor mixing
contribution to $\De\rho^{\rm MSSM}$ and (via \refeq{eq:precobs}) 
add it linearly to the other uncertainties.

  
\section{Higher-order corrections in the Higgs sector}
\label{sec:higgs-ho}

In order to calculate one-loop corrections to the Higgs boson
masses, the renormalized Higgs boson
self-energies are needed. Here we follow the procedure used in
\citeres{mhcMSSMlong,mhiggsf1lC} (and references therein). The parameters appearing in the Higgs
potential, see \refeq{higgspot}, are renormalized as follows:
\begin{align}
\label{rMSSM:PhysParamRenorm}
  \MZ^2 &\to \MZ^2 + \dMZsq,  & \tadh &\to \tadh +
  \dtadh, \\ 
  \MW^2 &\to \MW^2 + \dMWsq,  & \tadH &\to \tadH +
  \dtadH, \notag \\ 
  M_{\rm Higgs}^2 &\to M_{\rm Higgs}^2 + \de M_{\rm Higgs}^2, & 
  \tanb &\to \tanb (1+\dtanb). \notag 
\end{align}
$M_{\rm Higgs}^2$ denotes the tree-level Higgs boson mass matrix given
in \refeq{higgsmassmatrixtree}. $\tadh$ and $\tadH$ are the tree-level
tadpoles, i.e.\ the terms linear in $h$ and $H$ in the Higgs potential.

The field renormalization matrices of both Higgs multiplets
can be set up symmetrically, 
\begin{align}
\label{rMSSM:higgsfeldren}
  \begin{pmatrix} h \\[.5em] H \end{pmatrix} \to
  \begin{pmatrix}
    1+\tfrac{1}{2} \dZ{hh} & \tfrac{1}{2} \dZ{hH} \\[.5em]
    \tfrac{1}{2} \dZ{hH} & 1+\tfrac{1}{2} \dZ{HH} 
  \end{pmatrix} \cdot
  \begin{pmatrix} h \\[.5em] H \end{pmatrix}~.
\end{align}

\noindent
For the mass counter term matrices we use the definitions
\begin{align}
  \delta M_{\rm Higgs}^2 =
  \begin{pmatrix}
    \dmhsq  & \dmhHsq \\[.5em]
    \dmhHsq & \dmHsq  
  \end{pmatrix}~.
\end{align}
The renormalized self-energies, $\hSi(p^2)$, can now be expressed
through the unrenormalized self-energies, $\Si(p^2)$, the field
renormalization constants and the mass counter terms.
This reads for the $\cp$-even part,
\begin{subequations}
\label{rMSSM:renses_higgssector}
\begin{align}
\ser{hh}(p^2)  &= \se{hh}(p^2) + \dZ{hh} (p^2-\mhtree^2) - \dmhsq, \\
\ser{hH}(p^2)  &= \se{hH}(p^2) + \dZ{hH}
(p^2-\tfrac{1}{2}(\mhtree^2+\mHtree^2)) - \dmhHsq, \\ 
\ser{HH}(p^2)  &= \se{HH}(p^2) + \dZ{HH} (p^2-\mHtree^2) - \dmHsq~.
\end{align}
\end{subequations}

Inserting the renormalization transformation into the Higgs mass terms
leads to expressions for their counter terms which consequently depend
on the other counter terms introduced in~(\ref{rMSSM:PhysParamRenorm}). 

For the $\cp$-even part of the Higgs sectors, these counter terms are:
\begin{subequations}
\label{rMSSM:HiggsMassenCTs}
\begin{align}
\dmhsq &= \de\MA^2 \cos^2(\alpha-\beta) + \delta \MZ^2 \sin^2(\alpha+\beta) \\
&\quad + \tfrac{e}{2 \MZ \sw \cw} (\dtadH \cos(\alpha-\beta)
\sin^2(\alpha-\beta) + \dtadh \sin(\alpha-\beta)
(1+\cos^2(\alpha-\beta))) \notag \\ 
&\quad + \dtanb \sinb \cosb (\MA^2 \sin 2 (\alpha-\beta) + \MZ^2 \sin
2 (\alpha+\beta)), \notag \\ 
\dmhHsq &= \tfrac{1}{2} (\de\MA^2 \sin 2(\alpha-\beta) - \dMZsq \sin
2(\alpha+\beta)) \\ 
&\quad + \tfrac{e}{2 \MZ \sw \cw} (\dtadH \sin^3(\alpha-\beta) -
\dtadh \cos^3(\alpha-\beta)) \notag \\ 
&\quad - \dtanb \sinb \cosb (\MA^2 \cos 2 (\alpha-\beta) + \MZ^2 \cos
2 (\alpha+\beta)), \notag \\ 
\dmHsq &= \de\MA^2 \sin^2(\alpha-\beta) + \dMZsq \cos^2(\alpha+\beta) \\
&\quad - \tfrac{e}{2 \MZ \sw \cw} (\dtadH \cos(\alpha-\beta)
(1+\sin^2(\alpha-\beta)) + \dtadh \sin(\alpha-\beta)
\cos^2(\alpha-\beta)) \notag \\ 
&\quad - \dtanb \sinb \cosb (\MA^2 \sin 2 (\alpha-\beta) + \MZ^2 \sin
2 (\alpha+\beta))~. \notag 
\end{align}
\end{subequations}

For the field renormalization we chose to give each Higgs doublet one
renormalization constant,
\begin{align}
\label{rMSSM:HiggsDublettFeldren}
  \cHe \to (1 + \tfrac{1}{2} \dZ{\cHe}) \cHe, \quad
  \cHz \to (1 + \tfrac{1}{2} \dZ{\cHz}) \cHz~.
\end{align}
This leads to the following expressions for the various field
renormalization constants in \refeq{rMSSM:higgsfeldren}:
\begin{subequations}
\label{rMSSM:FeldrenI_H1H2}
\begin{align}
  \dZ{hh} &= \sinasq \dZ{\cHe} + \cosasq \dZ{\cHz}, \\[.2em]
  \dZ{hH} &= \sina \cosa (\dZ{\cHz} - \dZ{\cHe}), \\[.2em]
  \dZ{HH} &= \cosasq \dZ{\cHe} + \sinasq \dZ{\cHz}~.
\end{align}
\end{subequations}
The counter term for $\tb$ can be expressed in terms of the vacuum
expectation values as
\begin{equation}
\de\tb = \frac{1}{2} \KL \dZ{\cHz} - \dZ{\cHe} \KR +
\frac{\de v_2}{v_2} - \frac{\de v_1}{v_1}~,
\end{equation}
where the $\de v_i$ are the renormalization constants of the $v_i$:
\begin{equation}
v_1 \to \KL 1 + \dZ{\cHe} \KR \KL v_1 + \de v_1 \KR, \quad
v_2 \to \KL 1 + \dZ{\cHz} \KR \KL v_2 + \de v_2 \KR~.
\end{equation}
It can be shown that the divergent parts of $\de v_1/v_1$ and 
$\de v_2/v_2$ are equal~\cite{mhiggsf1l,mhiggsf1lC}.
Consequently, one can set $\de v_2/v_2 - \de v_1/v_1$ to zero.

Similarly for the charged Higgs sector, the renormalized self-energy is expressed in terms of the unrenormalized one and the corresponding counter-terms as:
\noindent \begin{equation}
\hat{\Sigma}_{H^{-}H^{+}}\left(p^{2}\right)=\Sigma_{H^{-}H^{+}}\left(p^{2}\right)+\delta Z_{H^{-}H^{+}}\left(p^{2}-m^{2}_{H^{\pm},{\rm tree}} \right)-\delta m_{H^{\pm}}^{2},\end{equation}
where, 
\noindent \begin{equation}
\delta m_{H^{\pm}}^{2}=\delta M_{A}^{2}+\delta M_{W}^{2}\end{equation}
and,
\noindent \begin{equation}
\delta Z_{H^{-}H^{+}}=\sin^{2}\beta \, \,\dZ{\cHe}  
+\cos^{2}\beta \,\,\dZ{\cHz}. \end{equation}

The renormalization conditions are fixed by an appropriate
renormalization scheme. For the mass counter terms on-shell conditions
are used, resulting in:
\begin{align}
\label{rMSSM:mass_osdefinition}
  \dMZsq = \re \se{ZZ}(\MZ^2), \quad \dMWsq = \re \se{WW}(\MW^2),
  \quad \de\MA^2 = \re \se{AA}(\MA^2). 
\end{align}
For the gauge bosons $\Si$ denotes the transverse part of the self-energy. 
Since the tadpole coefficients are chosen to vanish in all orders,
their counter terms follow from $T_{\{h,H\}} + \de T_{\{h,H\}} = 0$: 
\begin{align}
  \dtadh = -{\tadh}, \quad \dtadH = -{\tadH}~. 
\end{align}
For the remaining renormalization constants for $\de\tb$, $\dZ{\cHe}$
and $\dZ{\cHz}$ the most convenient
choice is a \drbar\ renormalization of $\de\tb$, $\dZ{\cHe}$
and $\dZ{\cHz}$, 
\begin{subequations}
\label{rMSSM:deltaZHiggsTB}
\begin{align}
  \dZ{\cHe} &= \dZ{\cHe}^{\drbarm}
       \; = \; - \KKL \re \Sip_{HH \; |\al = 0} \KKR^{\rm div}, \\[.5em]
  \dZ{\cHz} &= \dZ{\cHz}^{\drbarm} 
       \; = \; - \KKL \re \Sip_{hh \; |\al = 0} \KKR^{\rm div}, \\[.5em]
  \dtanb &= \edz (\dZ{\cHz} - \dZ{\cHe}) = \dtanb^{\drbarm}~.
\end{align}
\end{subequations}
The corresponding renormalization scale, $\mudim$, is set to 
$\mudim = \mt$ in all numerical evaluations. 

Finally, in the 
  Feynman diagrammatic (FD) approach that we are following here, the higher-order corrected 
$\cp$-even Higgs boson masses are derived by finding the
poles of the $(h,H)$-propagator 
matrix. The inverse of this matrix is given by
\BE
\left(\Delta_{\rm Higgs}\right)^{-1}
= - i \ML p^2 -  \mHtree^2 + \hSi_{HH}(p^2) &  \hSi_{hH}(p^2) \\
     \hSi_{hH}(p^2) & p^2 -  \mhtree^2 + \hSi_{hh}(p^2) \MR~.
\label{higgsmassmatrixnondiag}
\end{equation}
Determining the poles of the matrix $\Delta_{\rm Higgs}$ in
\refeq{higgsmassmatrixnondiag} is equivalent to solving
the equation
\begin{equation}
\left[p^2 - \mhtree^2 + \hSi_{hh}(p^2) \right]
\left[p^2 - \mHtree^2 + \hSi_{HH}(p^2) \right] -
\left[\hSi_{hH}(p^2)\right]^2 = 0\,.
\label{eq:proppole}
\end{equation}
Similarly, in the case of the charged Higgs sector, the corrected Higgs mass is derived by the position of the pole in the charged Higgs propagator, which is defined by: 
\noindent \begin{equation}
p^{2}-m^{2}_{H^{\pm},{\rm tree}} +
\hat{\Sigma}_{H^{-}H^{+}}\left(p^{2}\right)=0.
\label{eq:proppolech}
\end{equation}

The present experimental uncertanity at the LHC for the mass of
light neutral higgs $\Mh$ is
$\leq 300 \mev$~\cite{ATLAS:2013mma,CMS:yva}. This can possibly be reduced by
about 50\% at the LHC and below the level of $\sim 50 \mev$ at the
ILC~\cite{dbd}. 
Similarly, for the mass of heavy neutral higgs $\MH$
and charged higgs boson $\MHp$ an uncertainity at the $1\%$ level could
be expected at the LHC~\cite{cmsHiggs}.
This sets the goal for the theoretical uncertainty, which should be
reduced to the same (or higher) level of accuracy.

\begin{figure}[htb!]
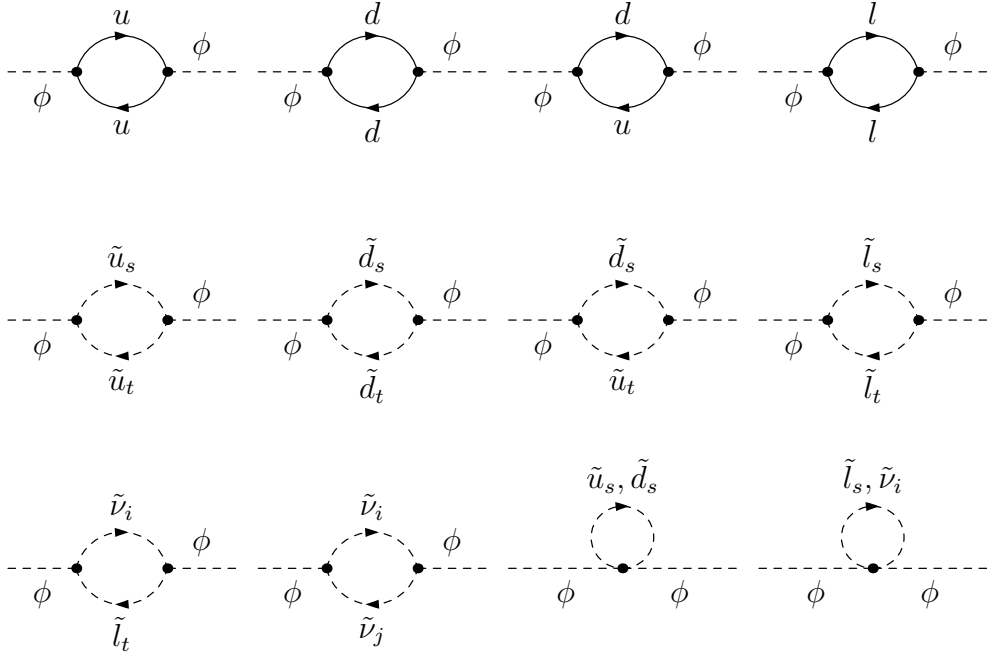

\begin{center}

\unitlength=1.0bp%

\begin{feynartspicture}(432,280)(4,3)

\FADiagram{}
\FAProp(0.,10.)(6.,10.)(0.,){/ScalarDash}{0}
\FALabel(3.,8.93)[t]{$\phi$}
\FAProp(20.,10.)(14.,10.)(0.,){/ScalarDash}{0}
\FALabel(17.,11.07)[b]{$\phi$}
\FAProp(6.,10.)(14.,10.)(0.8,){/Straight}{-1}
\FALabel(10.,5.73)[t]{$u$}
\FAProp(6.,10.)(14.,10.)(-0.8,){/Straight}{1}
\FALabel(10.,14.27)[b]{$u$}
\FAVert(6.,10.){0}
\FAVert(14.,10.){0}

\FADiagram{}
\FAProp(0.,10.)(6.,10.)(0.,){/ScalarDash}{0}
\FALabel(3.,8.93)[t]{$\phi$}
\FAProp(20.,10.)(14.,10.)(0.,){/ScalarDash}{0}
\FALabel(17.,11.07)[b]{$\phi$}
\FAProp(6.,10.)(14.,10.)(0.8,){/Straight}{-1}
\FALabel(10.,5.73)[t]{$d$}
\FAProp(6.,10.)(14.,10.)(-0.8,){/Straight}{1}
\FALabel(10.,14.27)[b]{$d$}
\FAVert(6.,10.){0}
\FAVert(14.,10.){0}

\FADiagram{}
\FAProp(0.,10.)(6.,10.)(0.,){/ScalarDash}{0}
\FALabel(3.,8.93)[t]{$\phi$}
\FAProp(20.,10.)(14.,10.)(0.,){/ScalarDash}{0}
\FALabel(17.,11.07)[b]{$\phi$}
\FAProp(6.,10.)(14.,10.)(0.8,){/Straight}{-1}
\FALabel(10.,5.73)[t]{$u$}
\FAProp(6.,10.)(14.,10.)(-0.8,){/Straight}{1}
\FALabel(10.,14.27)[b]{$d$}
\FAVert(6.,10.){0}
\FAVert(14.,10.){0}

\FADiagram{}
\FAProp(0.,10.)(6.,10.)(0.,){/ScalarDash}{0}
\FALabel(3.,8.93)[t]{$\phi$}
\FAProp(20.,10.)(14.,10.)(0.,){/ScalarDash}{0}
\FALabel(17.,11.07)[b]{$\phi$}
\FAProp(6.,10.)(14.,10.)(0.8,){/Straight}{-1}
\FALabel(10.,5.73)[t]{$l$}
\FAProp(6.,10.)(14.,10.)(-0.8,){/Straight}{1}
\FALabel(10.,14.27)[b]{$l$}
\FAVert(6.,10.){0}
\FAVert(14.,10.){0}

\FADiagram{}
\FAProp(0.,10.)(6.,10.)(0.,){/ScalarDash}{0}
\FALabel(3.,8.93)[t]{$\phi$}
\FAProp(20.,10.)(14.,10.)(0.,){/ScalarDash}{0}
\FALabel(17.,11.07)[b]{$\phi$}
\FAProp(6.,10.)(14.,10.)(0.8,){/ScalarDash}{-1}
\FALabel(10.,5.73)[t]{$\tilde u_{t}$}
\FAProp(6.,10.)(14.,10.)(-0.8,){/ScalarDash}{1}
\FALabel(10.,14.27)[b]{$\tilde u_{s}$}
\FAVert(6.,10.){0}
\FAVert(14.,10.){0}

\FADiagram{}
\FAProp(0.,10.)(6.,10.)(0.,){/ScalarDash}{0}
\FALabel(3.,8.93)[t]{$\phi$}
\FAProp(20.,10.)(14.,10.)(0.,){/ScalarDash}{0}
\FALabel(17.,11.07)[b]{$\phi$}
\FAProp(6.,10.)(14.,10.)(0.8,){/ScalarDash}{-1}
\FALabel(10.,5.73)[t]{$\tilde d_{t}$}
\FAProp(6.,10.)(14.,10.)(-0.8,){/ScalarDash}{1}
\FALabel(10.,14.27)[b]{$\tilde d_{s}$}
\FAVert(6.,10.){0}
\FAVert(14.,10.){0}

\FADiagram{}
\FAProp(0.,10.)(6.,10.)(0.,){/ScalarDash}{0}
\FALabel(3.,8.93)[t]{$\phi$}
\FAProp(20.,10.)(14.,10.)(0.,){/ScalarDash}{0}
\FALabel(17.,11.07)[b]{$\phi$}
\FAProp(6.,10.)(14.,10.)(0.8,){/ScalarDash}{-1}
\FALabel(10.,5.73)[t]{$\tilde u_{t}$}
\FAProp(6.,10.)(14.,10.)(-0.8,){/ScalarDash}{1}
\FALabel(10.,14.27)[b]{$\tilde d_{s}$}
\FAVert(6.,10.){0}
\FAVert(14.,10.){0}

\FADiagram{}
\FAProp(0.,10.)(6.,10.)(0.,){/ScalarDash}{0}
\FALabel(3.,8.93)[t]{$\phi$}
\FAProp(20.,10.)(14.,10.)(0.,){/ScalarDash}{0}
\FALabel(17.,11.07)[b]{$\phi$}
\FAProp(6.,10.)(14.,10.)(0.8,){/ScalarDash}{-1}
\FALabel(10.,5.73)[t]{$\tilde l_{t}$}
\FAProp(6.,10.)(14.,10.)(-0.8,){/ScalarDash}{1}
\FALabel(10.,14.27)[b]{$\tilde l_{s}$}
\FAVert(6.,10.){0}
\FAVert(14.,10.){0}

\FADiagram{}
\FAProp(0.,10.)(6.,10.)(0.,){/ScalarDash}{0}
\FALabel(3.,8.93)[t]{$\phi$}
\FAProp(20.,10.)(14.,10.)(0.,){/ScalarDash}{0}
\FALabel(17.,11.07)[b]{$\phi$}
\FAProp(6.,10.)(14.,10.)(0.8,){/ScalarDash}{-1}
\FALabel(10.,5.73)[t]{$\tilde l_{t}$}
\FAProp(6.,10.)(14.,10.)(-0.8,){/ScalarDash}{1}
\FALabel(10.,14.27)[b]{$\tilde \nu_{i}$}
\FAVert(6.,10.){0}
\FAVert(14.,10.){0}

\FADiagram{}
\FAProp(0.,10.)(6.,10.)(0.,){/ScalarDash}{0}
\FALabel(3.,8.93)[t]{$\phi$}
\FAProp(20.,10.)(14.,10.)(0.,){/ScalarDash}{0}
\FALabel(17.,11.07)[b]{$\phi$}
\FAProp(6.,10.)(14.,10.)(0.8,){/ScalarDash}{-1}
\FALabel(10.,5.73)[t]{$\tilde \nu_{j}$}
\FAProp(6.,10.)(14.,10.)(-0.8,){/ScalarDash}{1}
\FALabel(10.,14.27)[b]{$\tilde \nu_{i}$}
\FAVert(6.,10.){0}
\FAVert(14.,10.){0}

\FADiagram{}
\FAProp(0.,10.)(10.,10.)(0.,){/ScalarDash}{0}
\FALabel(5.,8.93)[t]{$\phi$}
\FAProp(20.,10.)(10.,10.)(0.,){/ScalarDash}{0}
\FALabel(15.,8.93)[t]{$\phi$}
\FAProp(10.,10.)(10.,10.)(10.,15.5){/ScalarDash}{-1}
\FALabel(10.,16.57)[b]{$\tilde u_{s}, \tilde d_{s}$}
\FAVert(10.,10.){0}

\FADiagram{}
\FAProp(0.,10.)(10.,10.)(0.,){/ScalarDash}{0}
\FALabel(5.,8.93)[t]{$\phi$}
\FAProp(20.,10.)(10.,10.)(0.,){/ScalarDash}{0}
\FALabel(15.,8.93)[t]{$\phi$}
\FAProp(10.,10.)(10.,10.)(10.,15.5){/ScalarDash}{-1}
\FALabel(10.,16.57)[b]{$\tilde l_{s}, \tilde \nu_{i}$}
\FAVert(10.,10.){0}

\end{feynartspicture}
\end{center}
\caption[Generic Feynman diagrams for the Higgs boson self-energies]{
Generic Feynman diagrams for the Higgs boson self-energies. $\phi$ denotes any of the Higgs bosons, $h$, $H$, $A$ or
$H^\pm$; $u$ stand for $u,c,t$; $d$ stand for $d,s,b$; $l$ stand for $e,\mu,\tau$; $\tilde u_{s,t}$, $\tilde d_{s,t}$ and $\tilde l_{s,t}$ are the six mass
eigenstates of up-type, down-type squarks and charged sleptons respectively and $\tilde \nu_{i,j}$ are the three
sneutrinos states $\tilde \nu_{e}$, $\tilde \nu_{\mu}$ and 
$\tilde \nu_{\tau}$.} 
\label{FeynDiagHSelf}
\end{figure}

\begin{figure}[htb!]
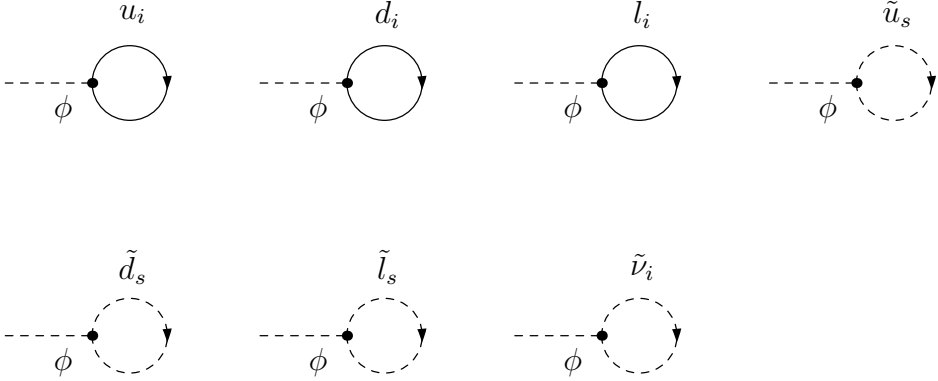

\begin{center}

\unitlength=1.0bp%

\begin{feynartspicture}(432,190)(4,2)

\FADiagram{}
\FAProp(0.,10.)(7.5,10.)(0.,){/ScalarDash}{0}
\FALabel(5.,8.93)[t]{$\phi$}
\FAProp(7.5,10.)(7.5,10.)(14.,10.){/Straight}{-1}
\FALabel(11.,16.)[]{$u_{i}$}
\FAVert(7.5,10.){0}

\FADiagram{}
\FAProp(0.,10.)(7.5,10.)(0.,){/ScalarDash}{0}
\FALabel(5.,8.93)[t]{$\phi$}
\FAProp(7.5,10.)(7.5,10.)(14.,10.){/Straight}{-1}
\FALabel(11.,16.)[]{$d_{i}$}
\FAVert(7.5,10.){0}

\FADiagram{}
\FAProp(0.,10.)(7.5,10.)(0.,){/ScalarDash}{0}
\FALabel(5.,8.93)[t]{$\phi$}
\FAProp(7.5,10.)(7.5,10.)(14.,10.){/Straight}{-1}
\FALabel(11.,16.)[]{$l_{i}$}
\FAVert(7.5,10.){0}

\FADiagram{}
\FAProp(0.,10.)(7.5,10.)(0.,){/ScalarDash}{0}
\FALabel(5.,8.93)[t]{$\phi$}
\FAProp(7.5,10.)(7.5,10.)(14.,10.){/ScalarDash}{-1}
\FALabel(11.,16.)[]{$\tilde u_{s}$}
\FAVert(7.5,10.){0}

\FADiagram{}
\FAProp(0.,10.)(7.5,10.)(0.,){/ScalarDash}{0}
\FALabel(5.,8.93)[t]{$\phi$}
\FAProp(7.5,10.)(7.5,10.)(14.,10.){/ScalarDash}{-1}
\FALabel(11.,16.)[]{$\tilde d_{s}$}
\FAVert(7.5,10.){0}

\FADiagram{}
\FAProp(0.,10.)(7.5,10.)(0.,){/ScalarDash}{0}
\FALabel(5.,8.93)[t]{$\phi$}
\FAProp(7.5,10.)(7.5,10.)(14.,10.){/ScalarDash}{-1}
\FALabel(11.,16.)[]{$\tilde l_{s}$}
\FAVert(7.5,10.){0}

\FADiagram{}
\FAProp(0.,10.)(7.5,10.)(0.,){/ScalarDash}{0}
\FALabel(5.,8.93)[t]{$\phi$}
\FAProp(7.5,10.)(7.5,10.)(14.,10.){/ScalarDash}{-1}
\FALabel(11.,16.)[]{$\tilde \nu_{i}$}
\FAVert(7.5,10.){0}

\end{feynartspicture}
\end{center}
\caption[Generic Feynman diagrams for the Higgs boson tadpoles]{
Generic Feynman diagrams for the Higgs boson tadpoles. 
$\phi$ denotes any of the Higgs bosons, $h$ or $H$; $u$ stand for $u,c,t$; $d$ stand for $d,s,b$; $l$ stand for $e,\mu,\tau$; $\tilde u_{s,t}$, $\tilde d_{s,t}$ and $\tilde l_{s,t}$ are the six mass
eigenstates of up-type, down-type squarks and charged sleptons respectively and $\tilde \nu_{i,j}$ are the three
sneutrinos states $\tilde \nu_{e}$, $\tilde \nu_{\mu}$ and 
$\tilde \nu_{\tau}$.} 
\label{FeynDiagHTad}
\end{figure}

Higher order corrections to the masses and mixing angles of the Higgs bosons in the MSSM have already been calculated in the literature. For the light and heavy $\cp$-even Higgs boson masses, complete one-loop contributions exist~\cite{ERZ,mhiggsf1lA,mhiggsf1lB,mhiggsf1lC}. Almost all the dominant contributions at two-loop level are also known~\cite{mhiggsletter,mhiggslong,mhiggslle,mhiggsFD2,bse,mhiggsEP0,mhiggsEP1,mhiggsEP1b,mhiggsEP2,mhiggsEP3,mhiggsEP3b,mhiggsEP4,mhiggsEP4b,mhiggsRG1,mhiggsRG1a}. For example, with the assumption of vanishing external momenta, the \order{\alt\als} contributions have been calculated in the Feynman diagrammatic (FD) approach and effective potential (EP) approach and the \order{\alt^2}, \order{\alb\als}, \order{\alt\alb} and \order{\alb^2} contributions are calculated in the EP approach. The momentum dependence at two-loop level was evaluated in \citeres{mhiggs2lp2, Mh-p2-BH4,Borowka:2015ura,Mh-p2-DDVS} and in ~\citere{mhiggsEP5}, a nearly full two-loop calculation in EP approach that also include the leading three-loop contributions has been presented. The code {\tt H3m}~\cite{mhiggsFD3l} adds the leading three-loop corrections of \order{\alt\als^2} to the \fh\ results. In the very recent work~\cite{Mh-logresum}, a combination of full one-loop results supplimented with leading and subleading two-loop contributions and a resummation of the leading and subleading logarithmic contributions from scalar-top sector is presented. This combination reduce the theoretical uncertainty from about 3 $\gev$ to about 2 $\gev$, for scalar-top masses at or below the $\tev$ scale, for the light $\cp$-even Higgs boson mass.
 Flavor violation effects for the case of squarks in MI approach were calculated in \cite{arana,arana-NMFV2}. We have calculated the effects of slepton mixing to the Higgs boson masses in \fa/\fc\ setup and added the result to the \fh. The details about the changes in \fa, \fc\ and \fh\ can be found in \refse{sec:feynhiggs}. We also calculate the effects of squark mixing in MFV CMSSM and MFV \CMSSMI.
In \reffi{FeynDiagHSelf}, we show generic Feynman diagrams for Higg self energy while Feynman diagrams for tadpoles are shown in \reffi{FeynDiagHTad}.

\section{\boldmath{$B$}-physics observables}
\label{sec:bpo}
In this thesis, we also consider several $B$-physics observables (BPO):
\bsg, \bmm\ and \dmbs. 
Concerning \bsg\ included in the calculation are the most 
relevant loop contributions to the Wilson coefficients:
(i)~loops with Higgs bosons (including the resummation of large $\tb$
effects~\cite{Isidori:2002qe}),  
(ii)~loops with charginos and 
(iii)~loops with gluinos. 
For \bmm\ there are three types of relevant one-loop corrections
contributing to the relevant Wilson coefficients:
(i)~Box diagrams, 
(ii)~$Z$-penguin diagrams and 
(iii)~neutral Higgs boson $\phi$-penguin diagrams, where $\phi$ denotes the
three neutral MSSM Higgs bosons, $\phi = h, H, A$ (again large resummed
$\tb$ effects have been taken into account).
In our numerical evaluation there are included
what are known to be the dominant contributions to
these three types of diagrams \cite{Chankowski:2000ng}: chargino
contributions to box and $Z$-penguin diagrams and chargino and gluino
contributions to $\phi$-penguin diagrams.   
Concerning \dmbs, in the MSSM 
there are in general three types of one-loop diagrams that contribute:
(i)~Box diagrams, 
(ii)~$Z$-penguin diagrams and 
(iii)~double Higgs-penguin diagrams (again including the resummation of
large $\tb$ enhanced effects).
In our numerical evaluation there are included again what are known to be the
dominant contributions to these three types of diagrams in scenarios
with non-minimal flavor violation (for a review see, for instance,
\cite{Foster:2005wb}): gluino contributions to box 
diagrams, chargino contributions to box and $Z$-penguin diagrams, and 
chargino and gluino contributions to double $\phi$-penguin diagrams. 
More details about the calculations employed can be
found in \citeres{arana,arana-NMFV2}.
We perform our numerical calculation with the {\tt BPHYSICS} subroutine
taken from the {\tt SuFla} code~\cite{sufla} (with some additions and
improvements as detailed in \citeres{arana,arana-NMFV2}), which has
been implemented as a subroutine into (a private version of) \fh.
The experimental values used in the numerical analysis\footnote{
Using the most up-to-date value of 
$\bmm = 2.9\pm 0.7 \times 10^{-9}$~\cite{bmm-CMS-LHCb}
would have had a minor impact on our analysis.
}
and SM prediction of these observables
is given in the 
\refta{tab:ExpStatus-BPO}~\cite{hfag:rad,Misiak:2009nr,Chatrchyan:2013bka,Aaij:2013aka,Buras:2012ru,hfag:pdg,Buras:1990fn,Golowich:2011cx}.
\begin{table}[htb!]
\BC
\renewcommand{\arraystretch}{1.5}
\begin{tabular}{|c|c|c|}
\hline
Observable & Experimental Value & SM Prediction  \\\hline
\bsg & $3.43\pm 0.22 \times 10^{-4}$ & $3.15\pm 0.23 \times 10^{-4}$ \\
\bmm & $(3.0)^{+1.0}_{-0.9} \times 10^{-9}$ & $3.23\pm 0.27 \times 10^{-9}$ \\
\dmbs & $116.4\pm 0.5 \times 10^{-10} \mev $ 
      & $(117.1)^{+17.2}_{-16.4} \times 10^{-10} \mev$   \\
\hline
\end{tabular}
\caption[Experimental values of BPO with their SM prediction.]
{Experimental values (used in our numerical analysis) of $B$-physics observables  with
their SM prediction.}  
\label{tab:ExpStatus-BPO}
\renewcommand{\arraystretch}{1.0}
\EC
\end{table}

\section{\boldmath{\hbs}}
\label{sec:hbs-calc}
\begin{figure}[ht!]
\begin{center}
\psfig{file=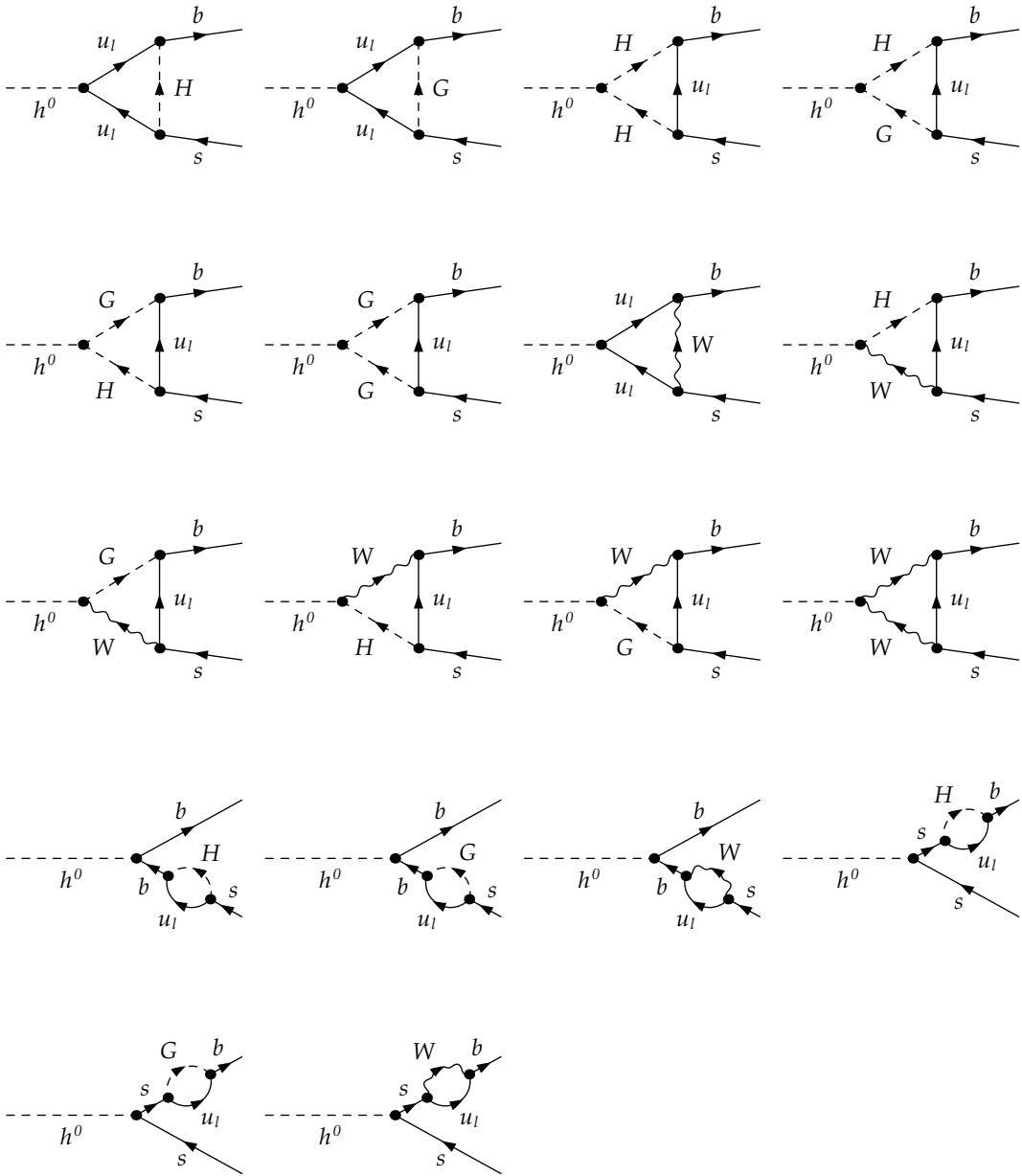}
\end{center}
\caption[Feynman diagrams for the decay $h\rightarrow b\bar{s}+\bar{b}s$]{Feynman diagrams showing SUSY-EW contributions (except neutralino-chargino) to the decay process $h\rightarrow b\bar{s}+\bar{b}s$.}   
\label{QFVHD-Rem}
\end{figure}

\begin{figure}[ht!]
\begin{center}
\psfig{file=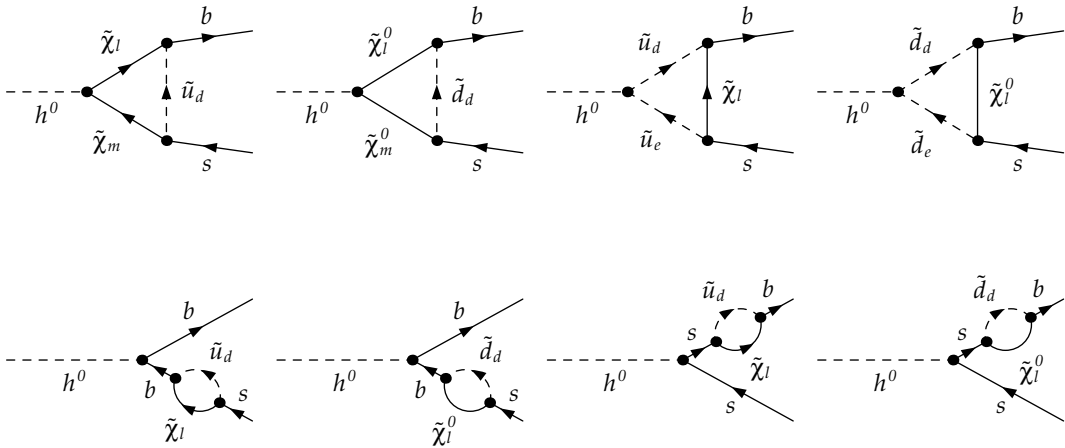}
\vspace{0.2cm}
\end{center}
\caption[Feynman diagrams for the decay $h\rightarrow b\bar{s}+\bar{b}s$]{Feynman diagrams showing neutralino-chargino contributions to the decay process $h\rightarrow b\bar{s}+\bar{b}s$.}   
\label{QFVHD-EW}
\end{figure}

\begin{figure}[ht!]
\begin{center}
\psfig{file=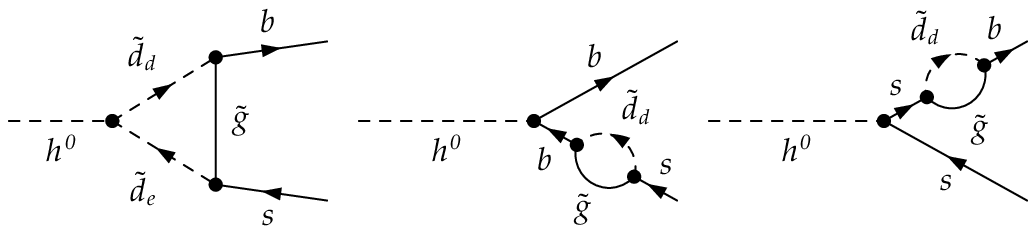}
\end{center}
\caption[Feynman diagrams for the decay $h\rightarrow b\bar{s}+\bar{b}s$]{Feynman diagrams showing SUSY-QCD contributions to the decay process $h\rightarrow b\bar{s}+\bar{b}s$.}   
\label{QFVHD-QCD}
\end{figure}

In SM the branching ratio \brhbs\ can be
at most of \order{10^{-7}}~\cite{HdecNMFV}, too small to have a
chance of detection at the LHC. But because of the strong FCNC gluino
couplings and the $\tb$-enhancement inherent to the MSSM Yukawa
couplings, we may expect several orders of magnitude increase of the
branching ratio as compared to the SM result, see 
\citere{HdecNMFV, SUSY-QCD}~. This decay in the framework of the MSSM has been analyzed in the
literature: the SUSY-QCD contributions for this decay were calculated 
in~\cite{HdecNMFV,SUSY-QCD}, and the SUSY-EW contributions using the mass
insertion approximation were calculated in~\cite{Demir}. Later
in~\cite{SUSY-EW} the SUSY-EW contributions and their interference
effects with the SUSY-QCD contribution were calculated using exact
diagonalization of the squark mass matrices. In all these
analysis, only LL mixing in the
squarks mass matrix was considered, and experimental constraints were
imposed only from \bsg. Most recently in~\cite{SUSY-EW-RR} also RR
mixing has been included. 
However mixing of the LR or RL elements 
of the mass matrix and constraints from other BPO or  potential other constraints were not taken into account 
(except in the most recent analysis in~\cite{SUSY-EW-RR}).
We (re-)calculate full one-loop 
contributions from SUSY-QCD as well as SUSY-EW loops with the help of
the \fa~\cite{feynarts,famssm} and \fc~\cite{formcalc} packages.
The lengthy analytical results are not shown here. We take into account the experimental constraints
not only from BPO but also from the EWPO. In the scalar quark sector we not only
consider the LL mixing, but also include the LR-RL and RR mixing for
our analysis of \brhbs. For our  numerical analysis we define 
\begin{align}
\brhbs = \frac{\Ga(\hbs)}{\Ga_{h, {\rm tot}}^{\rm MSSM}}
\end{align}
where $\Gamma_{h, {\rm tot}}^{\rm MSSM}$ is the total decay width of the
light Higgs boson $h$ of the MSSM, as evaluated with
\fh~\cite{feynhiggs,mhiggslong,mhiggsAEC,mhcMSSMlong,Mh-logresum}. 
The contributing Feynman diagrams for the decay \hbs\ are shown in
\reffi{QFVHD-QCD}-\ref{QFVHD-Rem}.
Which BR might be detectable at the LHC or an $e^+e^-$ collider such as
the ILC can only be established by means of specific experimental
analyses, which, to our knowledge, do not exist yet. However, in the
literature it is expected to measure BR's at the level of $10^{-3}$ at
the LHC~\cite{HdecNMFV}.
In the clean ILC environment in general Higgs boson branching ratios
below the level of $10^{-4}$ can be observed, see
e.g.\ \citere{ILCreview} for a recent review. We will take this as a
rough guideline down to which level the decay \hbs\ could be observable.
Feynman diagram for SUSY-EW contributions to the decay $\hbs$ are shown in \reffi{QFVHD-Rem} and \reffi{QFVHD-EW}, and SUSY-QCD contributions are shown in \reffi{QFVHD-QCD}.

 
\section{\boldmath{$l_i\rightarrow l_j \gamma$}}

Neutrino oscillation experiments\cite{Neutrino-Osc} have established the existence of lepton
flavor violation. So, as a natural consequence of neutrino
oscillations, one would expect flavour mixing in the charged lepton sector as well. This
mixing can be manifested in rare decay processes such as $\mu\to e\gamma$, $\tau\to e\gamma$, and
$\tau\to \mu\gamma$. However, if only the lepton Yukawa couplings carry this information on
flavour mixing, as in the SM with massive neutrinos, the expected
rates of these processes are extremely tiny\cite{Kuno:1999jp,DiracNu,MajoranaNu} being proportional to the ratio of
masses of neutrinos over the masses of the $W$ bosons. These values are very
far from the present experimental upper bounds \cite{Adam:2013mnn,Aubert:2009ag} that can be
read from \refta{tab:ExpStatus-LFV}.
The situation in the MSSM (extended by the seesaw mechanism) is completely different. Here lepton-slepton misallignment (generated by the presence of seesaw parameters in the RGE's) can dominate the SM contribution by several orders of magnitude. Thus making the study of rare LFV processes very attractive. 

We analyze these processes in the framework of CMSSM (extended by Type I seesaw mechanism). MSSM contributions to these decays originate from lepton-slepton-neutralino and lepton-slepton-chargino couplings. The predictions for $\br(l_i \to l_j \gamma)$ 
are obtained with {\tt SPheno 3.2.4}. We checked that the use of this code produces results similar to the ones obtained by our private codes used
in \citere{Cannoni:2013gq}. 
\begin{table}[h!]
\BC
\renewcommand{\arraystretch}{1.5}
\begin{tabular}{|c|c|}
\hline
Observable & Experimental value \\\hline
$\br(\mu \rightarrow e \gamma)$ & $ < 5.7 \times 10^{-13}$ \\
$\br(\tau \rightarrow e \gamma)$ & $ < 3.3 \times 10^{-8}$ \\
$\br(\tau \rightarrow \mu \gamma)$ & $ < 4.4 \times 10^{-8}$ \\
\hline
\end{tabular}
\caption{Present experimental status of LFV processes; their SM
  prediction is zero.}
\label{tab:ExpStatus-LFV}
\renewcommand{\arraystretch}{1.0}
\EC
\end{table}
\clearpage
\section{\boldmath $h \rightarrow l_i^{\pm} l_j^{\mp}$}
\label{sec:LFVHD-calc}
 
Since the discovery of Higgs boson, special effort has been made to determine its properties. The motivation
for such an effort resides on understanding the mechanism for EWSB. At present, several
aspects of the Higgs boson are to some extent well known, in particular those related with some of its expected
“standard” decay modes, namely: $WW^{*}$, $ZZ^{*}$, $\gamma \gamma$, $b \bar{b}$ and $\tau \bar{\tau}$ . Currently, measurements of these decay modes have
shown compatibility with the SM expectations, although with large associated uncertainties \cite{CMS:2014ega}.
Indeed, it is due to these large uncertainties that there is still room for nonstandard decay properties, something
that has encouraged such searches at the LHC as well. Searches for invisible Higgs decays have been published in
\cite{CMSAad2014,CMSChatrchyan2014}. Recently CMS collaboration using the 2012 dataset taken at $\sqrt{s} = 8 \rm TeV$ with an integrated luminosity of 19.7 $\rm fb^{-1}$, has found a 2.5 $\sigma$ excess in the $h\rightarrow \mu \tau$ channel, which translates into ${\rm BR}(h\rightarrow \mu \tau) \approx 0.89^{+40}_{-37} \%$ \cite{CMSLFVHD}. However there is no statistically significant excess in the ATLAS results\cite{ATLAS-LFVHD}. 

One needs to find the theoretical framework which can accomodate larger rates for LFVHD to explain CMS excess while still respecting the upper bounds on cLFV's. Efforts in such direction have been done in different contexts, with pioneer works in Refs. \cite{Pilaftsis1992,Diaz2000}. More recenty, Ref. \cite{Herrero2013} studied the problem in the MSSM, while \cite{Arhrib2013} in the R-parity violating MSSM. These decays have been considered as well in the inverse seesaw model in \cite{Herrero2014}. Possible effects due to vectorlike leptons have been investigated in \cite{Falkowski2014}. Extended scalar sectors involving several Higgs doublets and flavor symmetries (Yukawa textures) have been examined too \cite{Bhattacharyya2011,Bhattacharyya2012, Arroyo2013, Campos2014}. Finally, the Type-III Two Higgs Doublet Model has been considered in Refs. \cite{Davidson2010,Kopp2014}. Basically, the bottom line of these analyses is that unless one deals with extra Higgs doublets, LFVHD are below the LHC reach.  

\begin{figure}[htb!]
\begin{center}
\psfig{file=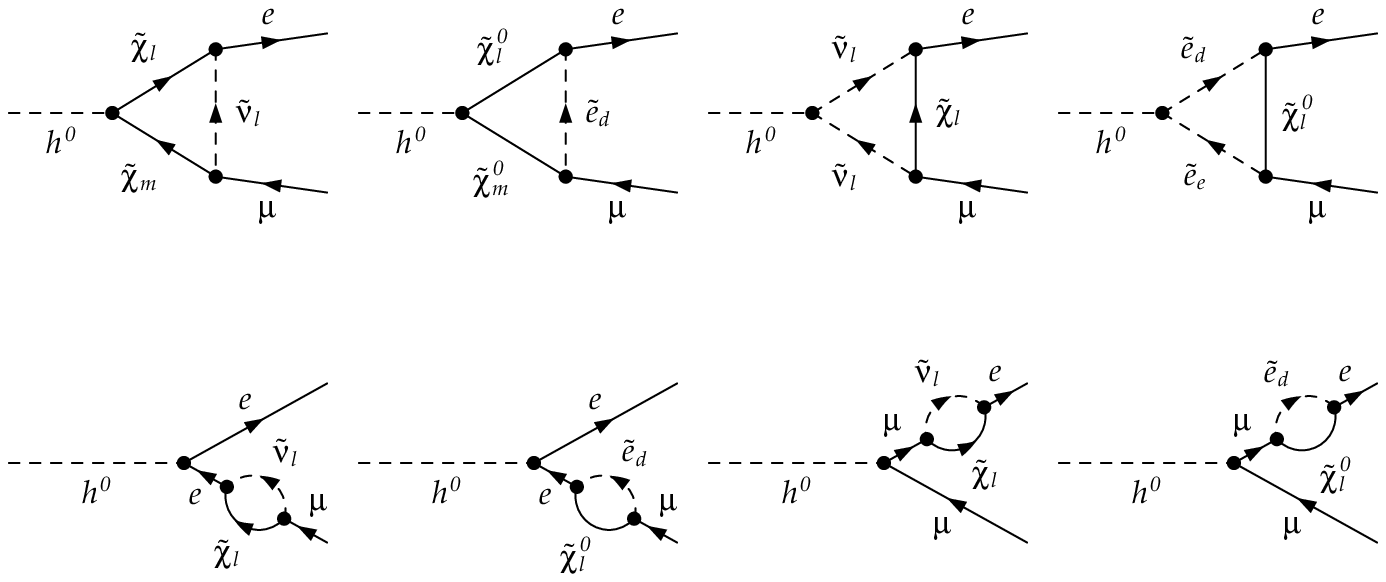}
\vspace{1.0cm}
\psfig{file=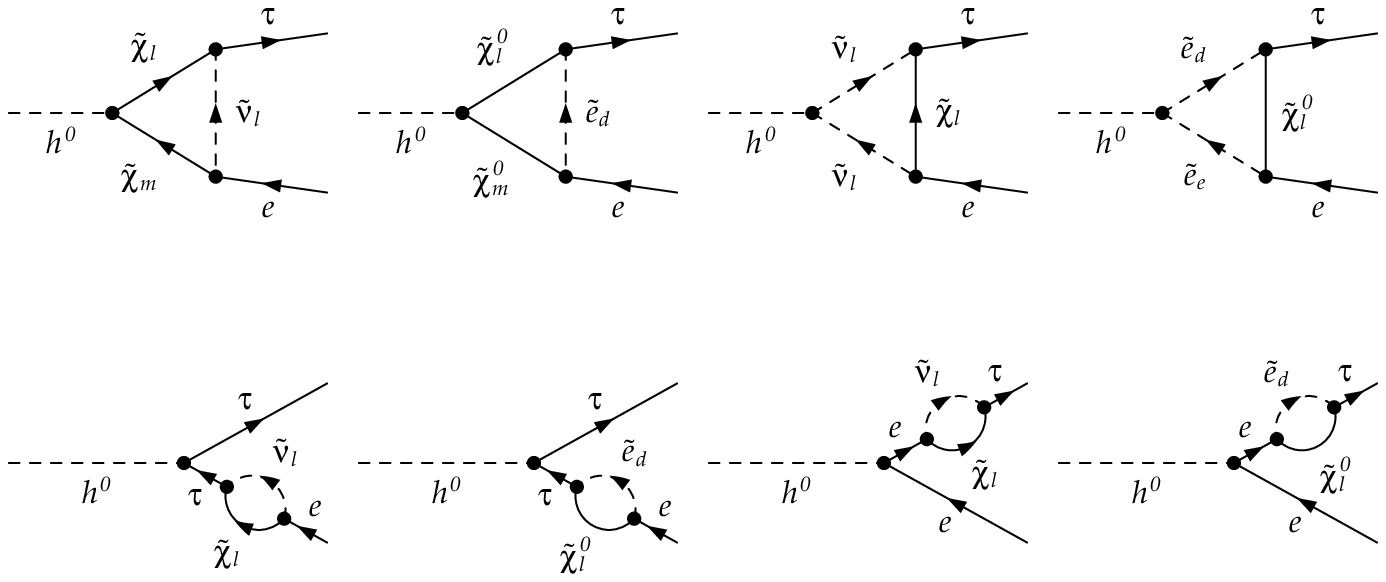}
\vspace{0.7cm}
\psfig{file=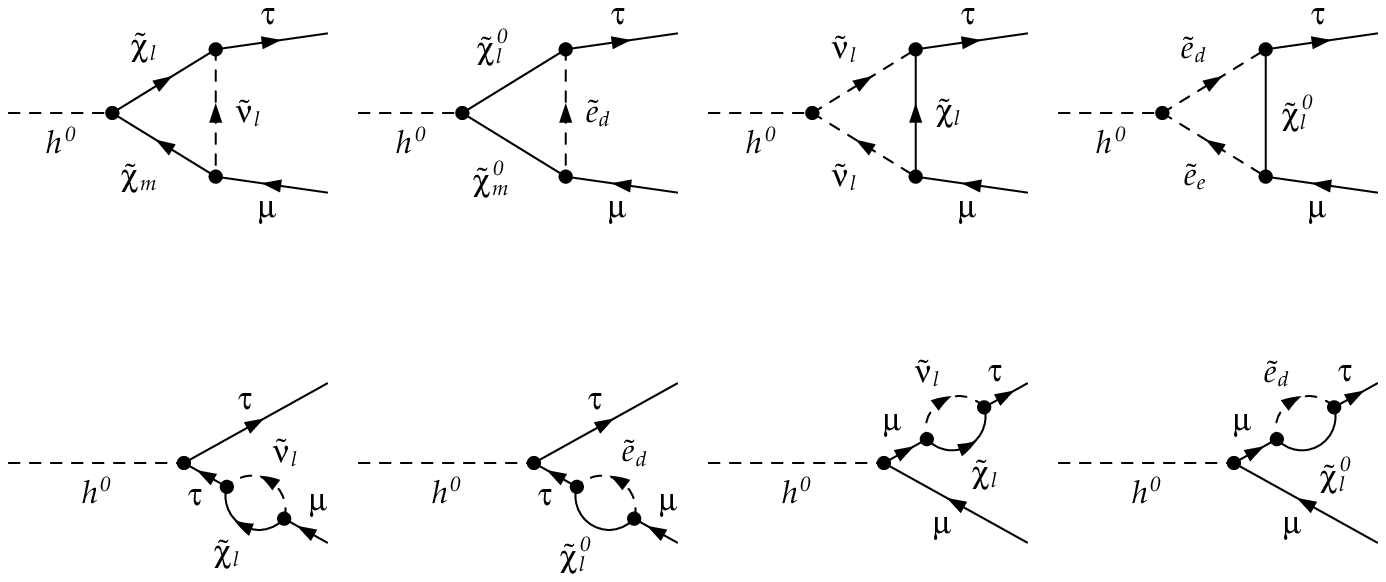}
\end{center}
\caption{Feynman diagrams for LFV decays $h \rightarrow l_i^{\pm} l_j^{\mp}$.}
\label{Diag_LFVHD}
\end{figure}

In this thesis we calculate the LFVHD in SUSY using FD approach.
We study the lepton-slepton misalignment effects to LFVHD, both in the MI approach and in MFV \CMSSMI. We do not use mass
insertion approximation and exact diagonalization of the
slepton mass matrix is performed.  Feynman diagrams entering our calculation are shown in \reffi{Diag_LFVHD} where first two rows correspond to the decay $h \rightarrow e^{\pm} \mu^{\mp}$, middle two rows correspond to $h \rightarrow e^{\pm} \tau^{\mp}$ and last two rows correspond to $h \rightarrow \mu^{\pm} \tau^{\mp}$. 
For the analytical calculation we used \fa/\fc\ setup. For this purpose, we implimented LFV Feynman rules for the MSSM in these packages (see \refse{sec:feynhiggs} for details). 

For numerical analysis we define the branching ratios of LFVHD as
\BE
{\rm BR}(h \rightarrow l_i^{\pm} l_j^{\mp})= \frac{\Gamma(h \rightarrow l_i^{\pm} l_j^{\mp})}{\Gamma(h \rightarrow l_i^{\pm} l_j^{\mp})+\Gamma_h^{\rm MSSM}} 
\EE 
where $i,j=e, \mu, \tau$ and $\Gamma_h^{\rm MSSM}$ is total decay width of $\cp$-even light Higgs boson $h$. 
\section{Changes in {\tt FeynArts}, {\tt FormCalc} and {\tt FeynHiggs}}
\label{sec:feynhiggs}
{\tt FeynArts}\cite{feynarts} and {\tt FormCalc}\cite{formcalc} provide
a high level of automation for perturbative calculations up to one loop.  This is 
particularly important for models with a large particle content such as 
the MSSM \cite{famssm}.  Here we briefly describe the recent extension 
of the implementation of the MSSM in these packages to include LFV.  
Details on the previous inclusion of NMFV can be found in 
\citeres{feynarts,interplay}. This involves firstly the modification of 
the slepton couplings in the existing \fa\ model file for the MSSM 
and secondly the corresponding initialization routines for the slepton 
masses and mixings, i.e.\ the $6\times 6$ and $3\times3$ diagonalization 
of the mass matrices in \fc.

\subsection{{\tt FeynArts} Model File}

\fa' add-on model file \texttt{FV.mod} applies algebraic 
substitutions to the Feynman rules of \texttt{MSSM.mod} to upgrade 
minimal to non-minimal flavor mixing in the sfermion sector.  The 
original version modified only the squark sector, i.e.\ NMFV, and needed 
to be generalized to include LFV.  We solved this by allowing the user 
to choose which sfermion types to introduce non-minimal mixing for 
through the variable \texttt{\$FV} (set before model initialization, of 
course).  For example,
\begin{verbatim}
  $FV = {11, 12, 13, 14};
  InsertFields[..., Model -> {MSSM, FV}]
\end{verbatim}
sets non-minimal mixing for all four sfermion types, with 11 = $\tinu$, 
12 = $\til$, 13 = $\tiu$, and 14 = $\tid$ as usual in \texttt{MSSM.mod}.  
For compatibility with the old NMFV-only version, the default is 
\verb|$FV = {13, 14}|.

\texttt{FV.mod} introduces the following new quantities:
\begin{center}
\begin{tabular}{ll}
\texttt{UASf[$s_1$,$s_2$,$t$]} &
	the slepton mixing matrix $R$, where \\
&	$s_1, s_2 = 1\dots 6$, \\
&	$t = 1\,(\tinu\/), 2\,(\til\/), 3\,(\tiu\/), 4\,(\tid\/)$, \\[1ex]
\texttt{MASf[$s$,$t$]} &
	the slepton masses, where \\
&	$s = 1\dots 6$, \\
&	$t = 1\,(\tinu\/), 2\,(\til\/), 3\,(\tiu\/), 4\,(\tid\/)$.
\end{tabular}
\end{center}
Entries $4\dots 6$ are unused for the sneutrino.


\subsection{Model initialization in {\tt FormCalc}}

The initialization of the generalized slepton-mixing parameters 
\texttt{MASf} and \texttt{UASf} is already built into \fc's regular 
MSSM model-initialization file \texttt{model\_mssm.F} but not turned on 
by default.  It must be enabled by adjusting the \texttt{FV} 
preprocessor flag in \texttt{run.F}:
\begin{verbatim}
   #define FV 2
\end{verbatim}
where 2 is the lowest sfermion type $t$ for which flavor violation is 
enabled, i.e.\ $\til$.

The flavor-violating parameters $\deFABij$ are represented in \fc\ 
by the \texttt{deltaSf} matrix:
\begin{center}
\begin{tabular}{ll}
\texttt{double complex deltaSf($s_1$,$s_2$,$t$)} &
	the matrix $(\delta_t)_{s_1s_2}$, where \\
&	$s_1, s_2 = 1\dots 6$ ($1\dots 3$ for $\tinu$), \\
&	$t = 2\,(\til\/), 3\,(\tiu\/), 4\,(\tid\/)$.
\end{tabular}
\end{center}
Since $\delta$ is an Hermitian matrix, only the entries above the
diagonal are considered.  The $\deFABij$ are located at the following 
places in the matrix $\delta$:
$$
\left(\begin{array}{ccc|ccc}
\noent & \delta^{LLL}_{12} & \delta^{LLL}_{13} &
\noent & \delta^{ELR}_{12} & \delta^{ELR}_{13} \\
\noent & \noent & \delta^{LLL}_{23} &
\delta^{ERL*}_{12} & \noent & \delta^{ELR}_{23} \\
\noent & \noent & \noent &
\delta^{ERL*}_{13} & \delta^{ERL*}_{23} & \noent \\ \hline
\noent & \noent & \noent &
\noent & \delta^{ERR}_{12} & \delta^{ERR}_{13} \\
\noent & \noent & \noent &
\noent & \noent & \delta^{ERR}_{23} \\
\noent & \noent & \noent &
\noent & \noent & \noent
\end{array}\right)
$$
The trilinear couplings $A_f$ acquire non-zero off-diagonal entries in 
the presence of LFV through the relations
\begin{equation}
m_{f,i} (A_f)_{ij} = (M_{\tif,LR}^2)_{ij}\,,
\quad
i, j = 1\dots 3\, ,
\end{equation}
see \refeq{eq:slep-matrix}.  These off-diagonal trilinear couplings (and 
hence the $\delta$'s) appear directly in the Higgs--slepton--slepton 
couplings, whereas all other effects are mediated through the masses 
and mixings.

The described changes are contained in the \fa\ 3.9 and {\tt FormCalc} 8.4 
packages which are publicly available from {\tt www.feynarts.de.}


\subsection{Inclusion of LFV into {\tt FeynHiggs}}

As discussed above, the new corrections to the (renormalized) 
Higgs-boson self-energies (and thus to the Higgs-boson masses), as well 
as to $\De\rho$ (and thus to $\MW$ and $\sweff$) have been included in 
\fh~\cite{feynhiggs,mhiggslong,mhiggsAEC,mhcMSSMlong,Mh-logresum}.

The corrections are activated by setting one or more of the $\deFABij$ to 
non-zero values.  All $\deFABij$ that are not set are assumed to be zero. 
The non-zero value can be set in three ways:
\begin{itemize}
\item
by including them in the input file, e.g.

\texttt{deltaLLL23~~~~0.1}

where the general format of the identifier is

\texttt{delta$F$$XY$$ij$, $F$ = L,E,Q,U,D, $XY$ = LL,LR,RL,RR, $ij$ = 12,23,13}

\item
by calling the subroutine \texttt{FHSetLFV(\ldots)} from your 
Fortran/C/C++ code.

\item
by calling the routine \texttt{FHSetLFV[\ldots]} from your Mathematica
code.

\end{itemize}
The detailed invocation of \texttt{FHSetLFV} is given in the 
corresponding man page included in the \fh\ distribution.  The LFV 
corrections are included starting from \fh\ version 2.10.2,
available from \texttt{feynhiggs.de}.


\chapter{Quark Flavor Mixing Effects in the Model Independent Approach}

MFV sceneraios put tight constraints on the possible value of the FCNC
couplings, especially for the first and second generation squarks which
are sensitive to the data on $K^0-\bar{K}^0$ and $D^0-\bar{D}^0$
mixing. However, the third generation is less constrained, since present
data on $B^0-\bar{B}^0$ mixing still leaves some room for FCNCs. This
allows some parameter space for the more general scenerios focusing on 
the mixing between second and third generation (s)quarks.
One such example is the neutral higgs decay \hbs.
The SM contribution is highly suppressed
for this process but the SUSY-QCD quark-squark-gluino loop contribution
can enhance the MSSM contribuion by several orders of magnitude. Also the
SUSY-EW one loop contribution from quark-squark-chargino and
quark-squark-neutralino loop even though subdominent, can have sizable
effects on the \brhbs, where in particular the interfrence effects of
SUSY-QCD and SUSY-EW loop corrections can be relevant.

This decay in the framework of the MSSM has been analyzed in the
literature: the SUSY-QCD contributions for this decay were calculated 
in~\cite{HdecNMFV,SUSY-QCD}, and the SUSY-EW contributions using the mass
insertion approximation were calculated in~\cite{Demir}. Later
in~\cite{SUSY-EW} the SUSY-EW contributions and their interference
effects with the SUSY-QCD contribution were calculated using exact
diagonalization of the squark mass matrices. In all these
analysis, only LL mixing in the
squarks mass matrix was considered, and experimental constraints were
imposed only from \bsg. Most recently in~\cite{SUSY-EW-RR} also RR
mixing has been included. 
However mixing of the LR or RL elements 
of the mass matrix and constraints from other BPO or  potential other constraints were not taken into account 
(except in the most recent analysis in~\cite{SUSY-EW-RR}).

In this chapter we will analyze the decay \hbs, evaluated at the full
one-loop level, by taking into account the experimental constraints
not only from BPO but also from the EWPO. In the scalar quark sector we will not only
consider the LL mixing, but also include the LR-RL and RR mixing for
our analysis of \brhbs. We will analyze this
decay in the model independent approach where flavor mixing
parameters are put in by hand without any emphasis on the origin of this
mixing (but respecting the experimental bounds from BPO and EWPO). The results 
presented in this chapter were published in \cite{EWPO-BPO-QFVHD}. 
In the next section we enlist the input parameters for our MI analysis. 

\section{Input parameters} 

Regarding our choice of MSSM parameters for our forthcoming numerical
analysis, we have chosen the framework of \cite{Arana-Catania:2013nha},
This framework is well compatible with present data.  

In this framework, six specific points in the MSSM
parameter space, have been selected. These points are allowed by
present data, including recent LHC searches and the measurements of the
muon anomalous magnetic moment. In \refta{tab:spectra} the
values of the various MSSM parameters as well as the values of the
predicted MSSM mass spectra are summarized. They were evaluated with
the program
\fh~\cite{feynhiggs,mhiggslong,mhiggsAEC,mhcMSSMlong,Mh-logresum}. For 
simplicity, and to reduce the number of 
independent MSSM input parameters we have assumed equal soft masses for
the sleptons of the first and second generations (similarly for the
squarks),  equal soft masses for the left and right slepton sectors
(similarly for the squarks, where $\tilde Q$ denotes the the
``left-handed'' squark sector, whereas $\tilde U$ and $\tilde D$ denote
the up- and down-type parts of the ``right-handed'' squark sector)
and also equal trilinear couplings for
the stop, $A_t$,  and sbottom squarks, $A_b$. In the slepton sector we
just consider the stau trilinear coupling, $A_\tau$. The other trilinear
sfermion couplings are set to zero. Regarding the SSB parameters for the gaugino
masses, $M_i$ ($i=1,2,3$),  we assume an approximate GUT relation. The
pseudoscalar Higgs mass $\MA$, and the $\mu$ parameter are also taken as
independent input parameters. In summary, the six points S1, \ldots, S6 are
defined in terms of the following subset of ten input MSSM parameters: 
\BEA
m_{\tilde L_1} &=& m_{\tilde L_2} \; ; \; m_{\tilde L_3} \; 
(\mbox{with~} m_{\tilde L_{i}} = m_{\tilde E_{i}}\,\,,\,\,i=1,2,3) \non \\
m_{\tilde Q_1} &=& m_{\tilde Q_2} \; ; \; m_{\tilde Q_3} \; 
(\mbox{with~} m_{\tilde Q_i} = m_{\tilde U_i} = m_{\tilde D_i}\,\,,\,\,i=1,2,3) 
                                                               \non \\
A_t&=&A_b\,\,;\,\,A_\tau \nonumber \\
M_2&=&2 M_1\, =\,M_3/4 \,\,;\,\,\mu \nonumber \\
\MA&\,\,;\,\, &\tb
\EEA
\begin{table}[h!]
\begin{tabular}{|c|c|c|c|c|c|c|}
\hline
 & S1 & S2 & S3 & S4 & S5 & S6 \\\hline
$m_{\tilde L_{1,2}}$& 500 & 750 & 1000 & 800 & 500 &  1500 \\
$m_{\tilde L_{3}}$ & 500 & 750 & 1000 & 500 & 500 &  1500 \\
$M_2$ & 500 & 500 & 500 & 500 & 750 &  300 \\
$A_\tau$ & 500 & 750 & 1000 & 500 & 0 & 1500  \\
$\mu$ & 400 & 400 & 400 & 400 & 800 &  300 \\
$\tb$ & 20 & 30 & 50 & 40 & 10 & 40  \\
$\MA$ & 500 & 1000 & 1000 & 1000 & 1000 & 1500  \\
$m_{\tilde Q_{1,2}}$ & 2000 & 2000 & 2000 & 2000 & 2500 & 1500  \\
$m_{\tilde Q_{3}}$  & 2000 & 2000 & 2000 & 500 & 2500 & 1500  \\
$A_t$ & 2300 & 2300 & 2300 & 1000 & 2500 &  1500 \\\hline
$m_{\tilde l_{1}}-m_{\tilde l_{6}}$ & 489-515 & 738-765 & 984-1018 & 474-802  & 488-516 & 1494-1507  \\
$m_{\tilde \nu_{1}}-m_{\tilde \nu_{3}}$& 496 & 747 & 998 & 496-797 & 496 &  1499 \\
$m_{{\tilde \chi}_1^\pm}-m_{{\tilde \chi}_2^\pm}$  & 375-531 & 376-530 & 377-530 & 377-530  & 710-844 & 247-363  \\
$m_{{\tilde \chi}_1^0}-m_{{\tilde \chi}_4^0}$& 244-531 & 245-531 & 245-530 & 245-530  & 373-844 & 145-363  \\
$M_{h}$ & 126.6 & 127.0 & 127.3 & 123.1 & 123.8 & 125.1  \\
$M_{H}$  & 500 & 1000 & 999 & 1001 & 1000 & 1499  \\
$M_{A}$ & 500 & 1000 & 1000 & 1000 & 1000 & 1500  \\
$M_{H^\pm}$ & 507 & 1003 & 1003 & 1005 & 1003 & 1502  \\
 $m_{\tilde u_{1}}-m_{\tilde u_{6}}$& 1909-2100 & 1909-2100 & 1908-2100 & 336-2000 & 2423-2585 & 1423-1589  \\
$m_{\tilde d_{1}}-m_{\tilde d_{6}}$ & 1997-2004 & 1994-2007 & 1990-2011 & 474-2001 & 2498-2503 &  1492-1509 \\
$m_{\tilde g}$ &  2000 & 2000 & 2000 & 2000 & 3000 &  1200 \\
\hline
\end{tabular}
\caption[Selected points in the MSSM parameter space.]{Selected points in the MSSM parameter space (upper part)
and their corresponding spectra (lower part). 
All mass parameters and trilinear couplings are given in GeV.} 
\label{tab:spectra}
\end{table}

The specific values of these ten MSSM parameters in \refta{tab:spectra},
to be used in the forthcoming analysis, are chosen to provide
different  
patterns in the various sparticle masses, but all leading to rather
heavy spectra, thus they are naturally in agreement with the
absence of SUSY signals at LHC. In particular  
all points lead to rather heavy squarks and gluinos above $1200\gev$ and
heavy sleptons above $500\gev$ (where the LHC limits would also
  permit substantially lighter scalar leptons). 
The values of $\MA$ within the interval
$(500,1500)\gev$, $\tb$ within the interval $(10,50)$ and a large
$A_t$ within $(1000,2500)\gev$ are fixed such that a light Higgs boson
$h$ within the LHC-favoured range $(123,127)\gev$ is obtained.
It should also be noted that the large chosen values of $\MA \ge 500$
GeV place the Higgs sector of our scenarios in the so called decoupling
regime\cite{Haber:1989xc},  
where the couplings of $h$ to gauge bosons and fermions are close to
the SM Higgs couplings, and the heavy $H$ couples like the
pseudoscalar $A$, and all heavy Higgs bosons are close in mass.
Increasing $\MA$  the heavy
Higgs bosons tend to decouple from low energy physics and the light
$h$ behaves like $H_{\rm SM}$. This type of MSSM Higgs sector seems
to be in good agreement with recent LHC data\cite{LHCHiggs}. 
We have checked with the code {\tt HiggsBounds}~\cite{higgsbounds} (but not yet taking into account the most recent update\cite{Higgsbounds-2015})
  that the Higgs sector is in agreement with the LHC searches (where S3
  is right ``at the border'').
Particularly, the so far absence of gluinos at
LHC, forbids too low $M_3$ and, therefore, given the  assumed GUT
relation, forbids also a too low $M_2$. Consequently, the
values of $M_2$ and $\mu$ are fixed as to get gaugino masses compatible
with present LHC bounds. 
Finally, we have also required that all our points lead to a prediction
of the anomalous magnetic moment of the muon in the MSSM that can fill
the present discrepancy between the SM prediction and the
experimental value (see \cite{Arana-Catania:2013nha} for more details). 

\section{Experimental constraints on $\deFABij$}

In this section we will present the present experimental constraints on
the squark mixing parameters $\deFABij$ for the above mentioned MSSM
points S1\dots S6 defined in \refta{tab:spectra}. The experimental
constraints from BPO for the same set of parameters that we are using
were already calculated in \cite{arana-NMFV2} for one $\deFABij \neq0$ ,
which we reproduce here for completeness in the \refta{tab:boundsS1S6}.   

\begin{figure}[htb!]
\begin{center}
\psfig{file=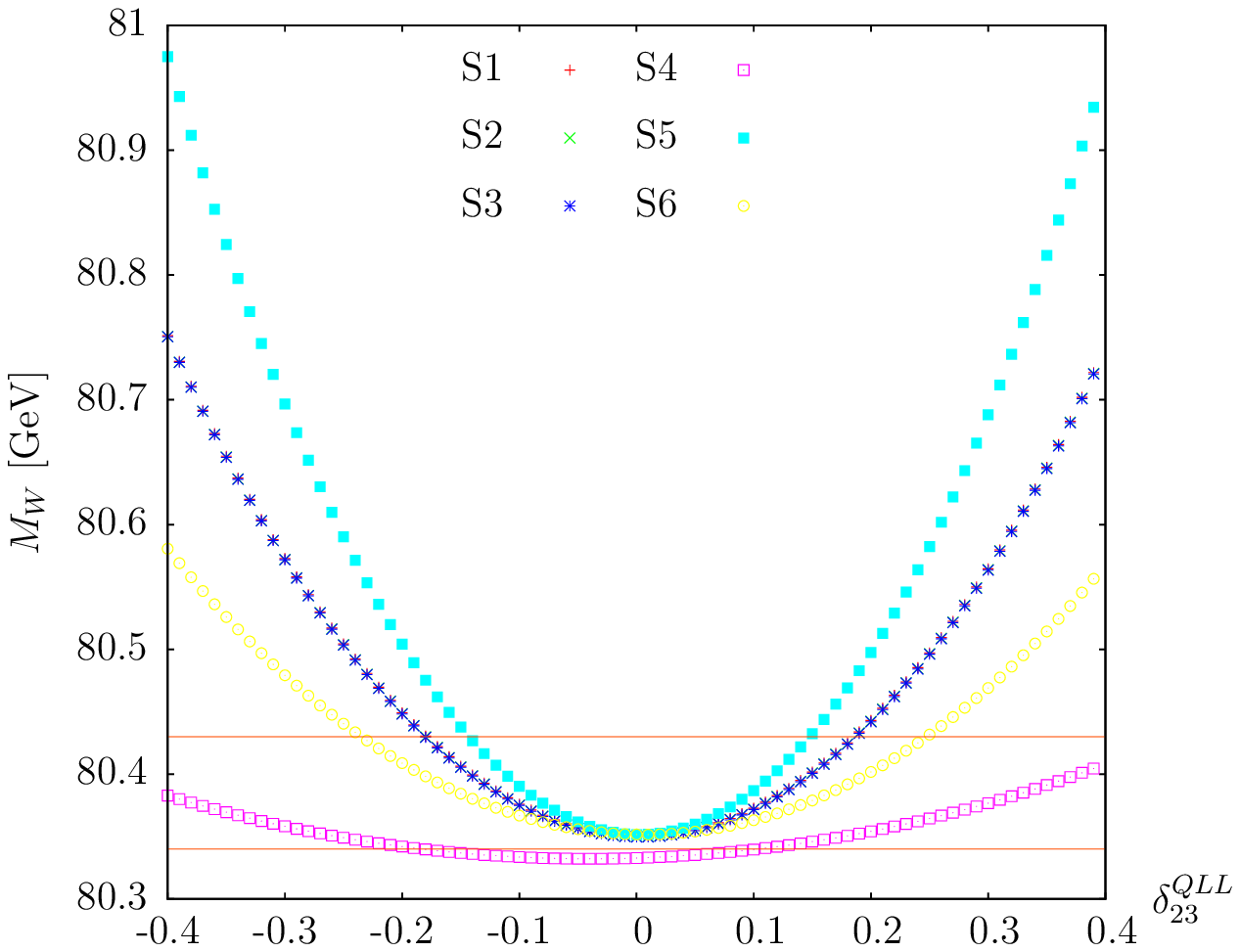  ,scale=0.52,angle=0,clip=}
\psfig{file=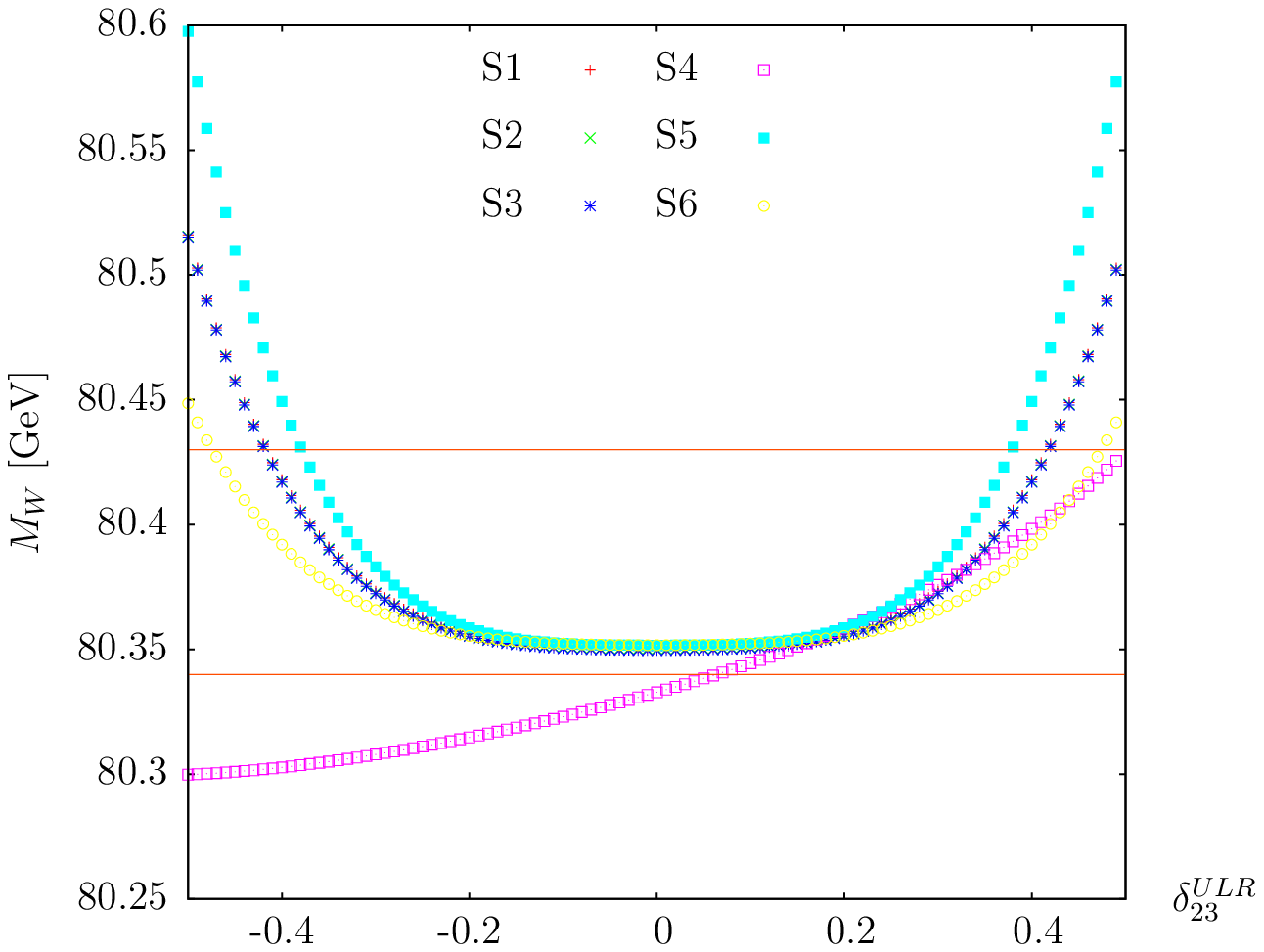  ,scale=0.52,angle=0,clip=}\\
\vspace{2.0cm}
\psfig{file=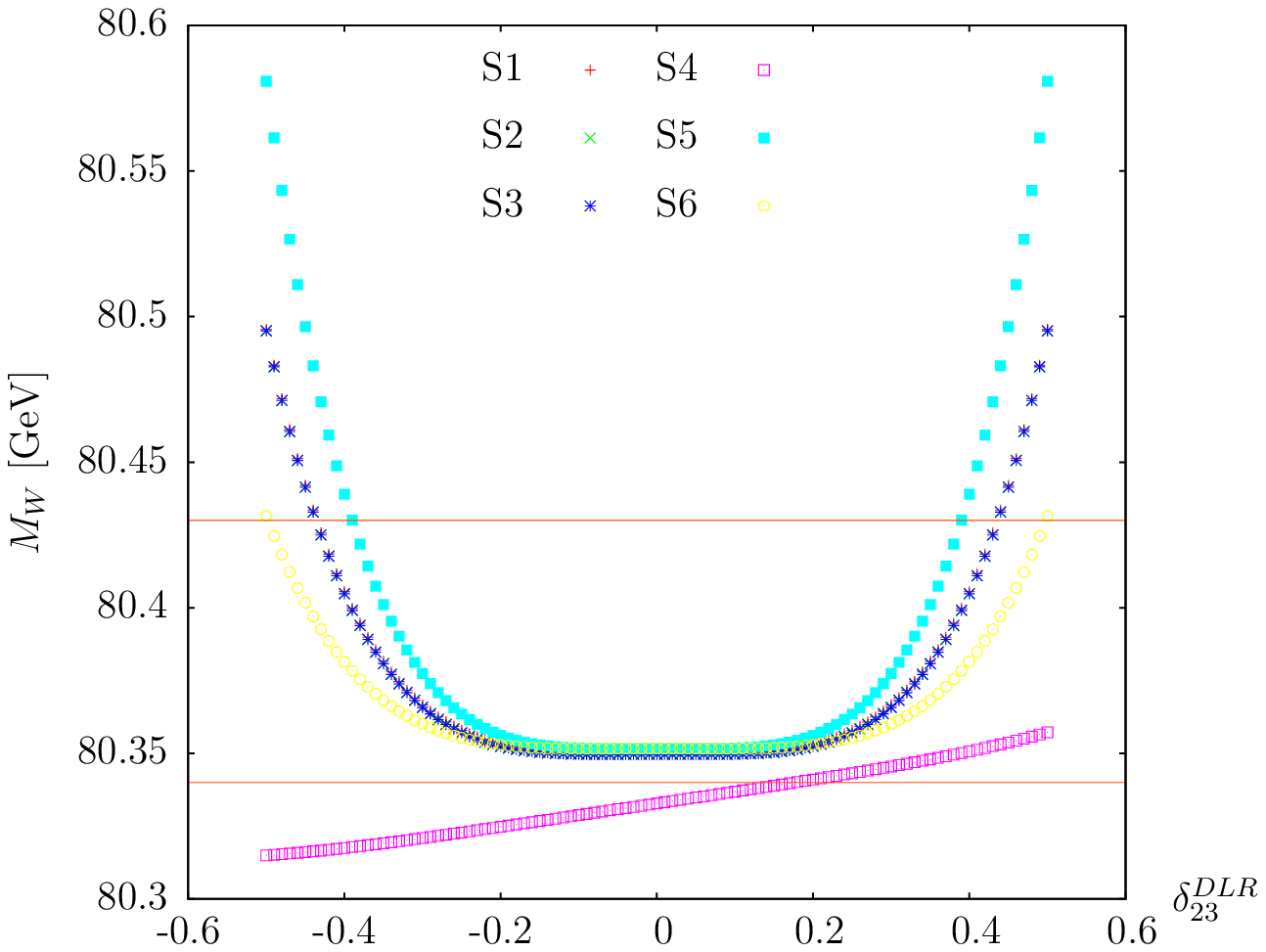 ,scale=0.52,angle=0,clip=}
\vspace{0.2cm}
\end{center}
\caption[$\MW$ as a function of squark $\deFABij$]{$\MW$ as a function of $\del{QLL}{23}$ (upper left), 
$\del{ULR}{23}$ (upper right) and $\del{DLR}{23}$ (lower). }   
\label{Fig:DMW}
\end{figure}

We now turn our attention to the constraints from $\MW$. 
In \reffi{Fig:DMW} we show the
$\MW$ as a function of $\del{QLL}{23}$, $\del{ULR}{23}$ and
$\del{DLR}{23}$ in the scenarios S1 \ldots S6. 
The area between the orange lines shows the allowed 
value of $\MW$ with $3\sigma$ experimental uncertainty.
The corresponding constraints from $\MW$ on $\deFABij$, also taking
into account the theoretical uncertainties as described at the end of
\refse{sec:EWPO-calc}, are shown in \refta{tab:EWPOboundsS1S6}. 
No constraints can be found on the $\del{RR}{ij}$, 
as their contribution to $\MW$ does not reach the MeV level, and consequently
we do not show them here. Furtheremore, the constraints for the
$\del{URL}{23}$ and $\del{DRL}{23}$ are similar to those for
$\del{ULR}{23}$ and $\del{DLR}{23}$, respectively, and not shown
here. 

On the other hand, the constraints on $\del{QLL}{23}$ are modified by the
EWPO specially the region (-0.83:-0.78) for the point S5, which was
allowed by the BPO, is now excluded. The allowed intervals for the
points S1-S3 have also shrunk. However the point S4 was already excluded
by BPO, similarly the allowed interval for S6 do not get modified by
EWPO. The constraints on $\del{ULR}{23}$ and $\del{DLR}{23}$
are less restrictive then the ones from BPO except for the point S4
where the region (0.076:0.12) is excluded for $\del{DLR}{23}$ by
EWPO.

\renewcommand{\arraystretch}{1.10}
\begin{table}[htb!]
\begin{center}
\resizebox{9.0cm}{!} {
\begin{tabular}{|c|c|c|} \hline
 & & Total allowed intervals \\ \hline
$\delta^{QLL}_{23}$ & \begin{tabular}{c}  S1 \\ S2 \\ S3 \\ S4 \\ S5 \\ S6 \end{tabular} &  
\begin{tabular}{c} 
(-0.27:0.28) \\ (-0.23:0.23) \\ (-0.12:0.06) (0.17:0.19) \\ excluded \\ (-0.83:-0.78) (-0.14:0.14) \\ (-0.076:0.14) \end{tabular} \\ \hline
$\delta^{ULR}_{23}$  & \begin{tabular}{c}  S1 \\ S2 \\ S3 \\ S4 \\ S5 \\ S6 \end{tabular}    
& \begin{tabular}{c} 
(-0.27:0.27) \\ (-0.27:0.27) \\ (-0.27:0.27) \\ excluded \\ (-0.22:0.22) \\ (-0.37:0.37) \end{tabular}   \\ \hline
$\delta^{DLR}_{23}$  & \begin{tabular}{c}  S1 \\ S2 \\ S3 \\ S4 \\ S5 \\ S6 \end{tabular}    & 
\begin{tabular}{c} 
(-0.0069:0.014) (0.12:0.13) \\ (-0.0069:0.014) (0.11:0.13) \\ (-0.0069:0.014) (0.11:0.13) \\ (0.076:0.12) (0.26:0.30) \\ (-0.014:0.021) (0.17:0.19) \\ (0:0.0069) (0.069:0.076) \end{tabular}  \\ \hline
$\delta^{URL}_{23}$  & \begin{tabular}{c}  S1 \\ S2 \\ S3 \\ S4 \\ S5 \\ S6 \end{tabular}    & 
\begin{tabular}{c}
(-0.27:0.27) \\ (-0.27:0.27) \\ (-0.27:0.27) \\ excluded \\ (-0.22:0.22) \\ (-0.37:0.37) \end{tabular} 
  \\ \hline
$\delta^{DRL}_{23}$  & \begin{tabular}{c}  S1 \\ S2 \\ S3 \\ S4 \\ S5 \\ S6 \end{tabular}    &
\begin{tabular}{c} (-0.034:0.034) \\ (-0.034:0.034) \\ (-0.034:0.034) \\ excluded \\ (-0.062:0.062) \\ (-0.021:0.021) \end{tabular} 
  \\ \hline
$\delta^{URR}_{23}$ & \begin{tabular}{c}  S1 \\ S2 \\ S3 \\ S4 \\ S5 \\ S6 \end{tabular}   & \begin{tabular}{c} 
(-0.99:0.99) \\ (-0.99:0.99) \\ (-0.98:0.97) \\ excluded \\ (-0.99:0.99) \\ (-0.96:0.94)  \end{tabular}    \\ \hline
$\delta^{DRR}_{23}$  & \begin{tabular}{c}  S1 \\ S2 \\ S3 \\ S4 \\ S5 \\ S6 \end{tabular}    &
\begin{tabular}{c}  (-0.96:0.96) \\ (-0.96:0.96) \\ (-0.96:0.94) \\ excluded \\ (-0.97:0.97) \\ (-0.97:-0.94) (-0.63:0.64) (0.93:0.97)
\end{tabular}    \\ \hline
\end{tabular}}  
\end{center}
\vspace{-1em}
\caption[Present allowed (by BPO) intervals for the $\deFABij$]
{Present allowed (by BPO) intervals for the $\deFABij$ for the MSSM points defined in
\refta{tab:spectra}\cite{arana-NMFV2}. 
}
\label{tab:boundsS1S6}
\vspace{-4em}
\end{table}
\renewcommand{\arraystretch}{1.55}

\renewcommand{\arraystretch}{1.1}
\begin{table}[htb!]
\begin{center}
\resizebox{9.0cm}{!} {
\begin{tabular}{|c|c|c|} \hline
 & & Total allowed intervals \\ \hline
$\delta^{QLL}_{23}$ & \begin{tabular}{c}  S1 \\ S2 \\ S3 \\ S4 \\ S5 \\ S6 \end{tabular} &  
\begin{tabular}{c} 
(-0.18:0.18) \\ (-0.18:0.18) \\ (-0.18:0.18) \\ (-0.53:-0.17)(0.10:0.45) \\ (-0.14:0.14) \\ (-0.23:0.23) \end{tabular} \\ \hline
$\delta^{ULR}_{23},\delta^{URL}_{23}$  & \begin{tabular}{c}  S1 \\ S2 \\ S3 \\ S4 \\ S5 \\ S6 \end{tabular}    
& \begin{tabular}{c} 
(-0.41:0.41) \\ (-0.41:0.41) \\ (-0.41:0.41) \\ (0.10:0.50) \\ (-0.39:0.39) \\ (-0.47:0.47) \end{tabular}   \\ \hline
$\delta^{DLR}_{23},\delta^{DRL}_{23}$  & \begin{tabular}{c}  S1 \\ S2 \\ S3 \\ S4 \\ S5 \\ S6 \end{tabular}    & 
\begin{tabular}{c} 
(-0.43:0.43) \\ (-0.43:0.43) \\ (-0.43:0.43) \\ (0.16:0.99) \\ (-0.39:0.39) \\ (-0.49:0.49) \end{tabular}  \\ \hline
\end{tabular}}  
\end{center}
\caption[Present allowed (by $\MW$) intervals for the $\deFABij$]
{Present allowed (by $\MW$) intervals for the squark mixing parameters
$\deFABij$ for the selected S1-S6 MSSM points defined in
\refta{tab:spectra}. 
}
\label{tab:EWPOboundsS1S6}
\end{table}
\renewcommand{\arraystretch}{1.55}
\section{\boldmath{\brhbs}}

\begin{figure}[ht!]
\begin{center}
\psfig{file=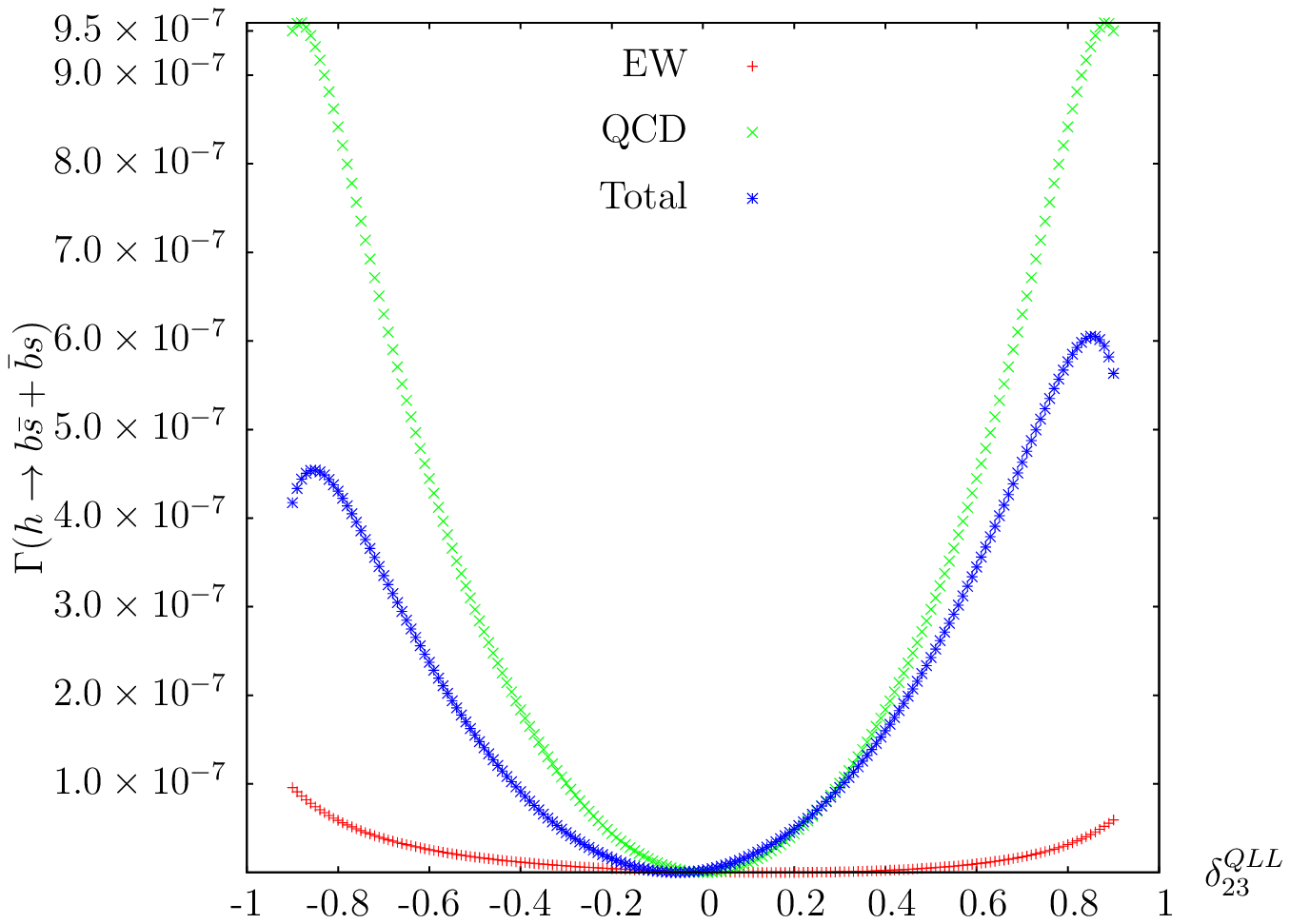  ,scale=0.52,angle=0,clip=}
\psfig{file=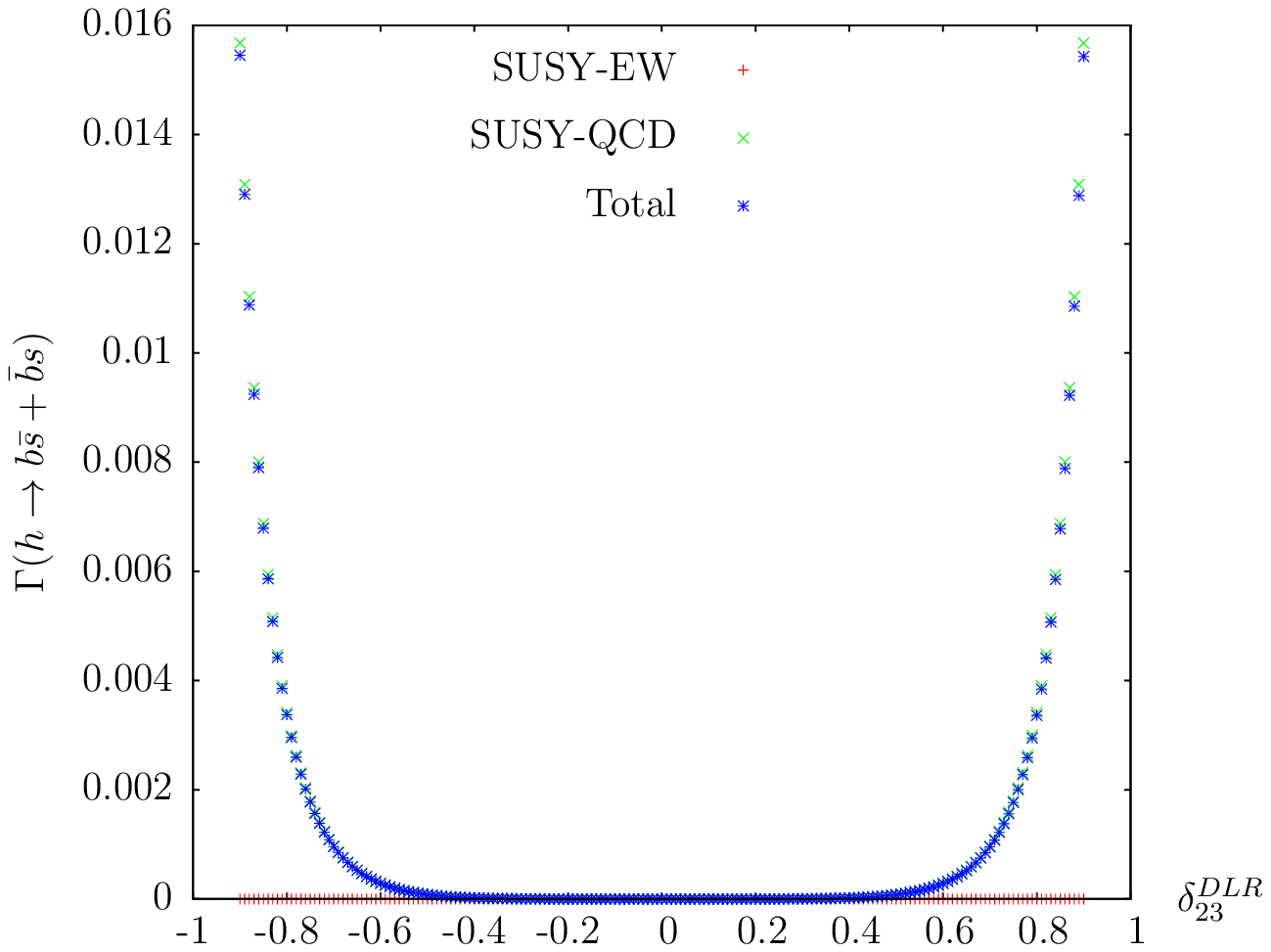  ,scale=0.52,angle=0,clip=}\\
\vspace{2.0cm}
\psfig{file=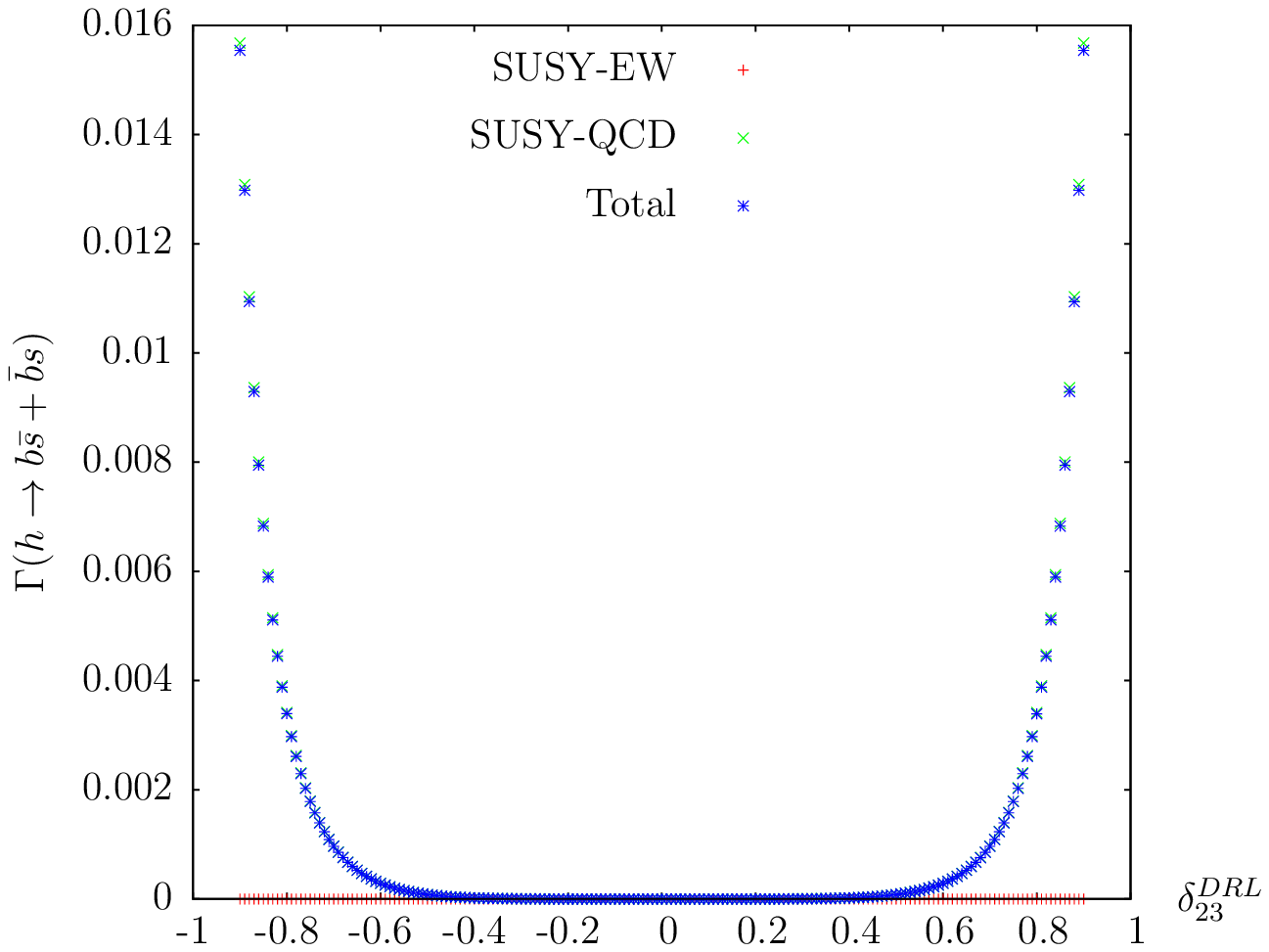 ,scale=0.52,angle=0,clip=}
\psfig{file=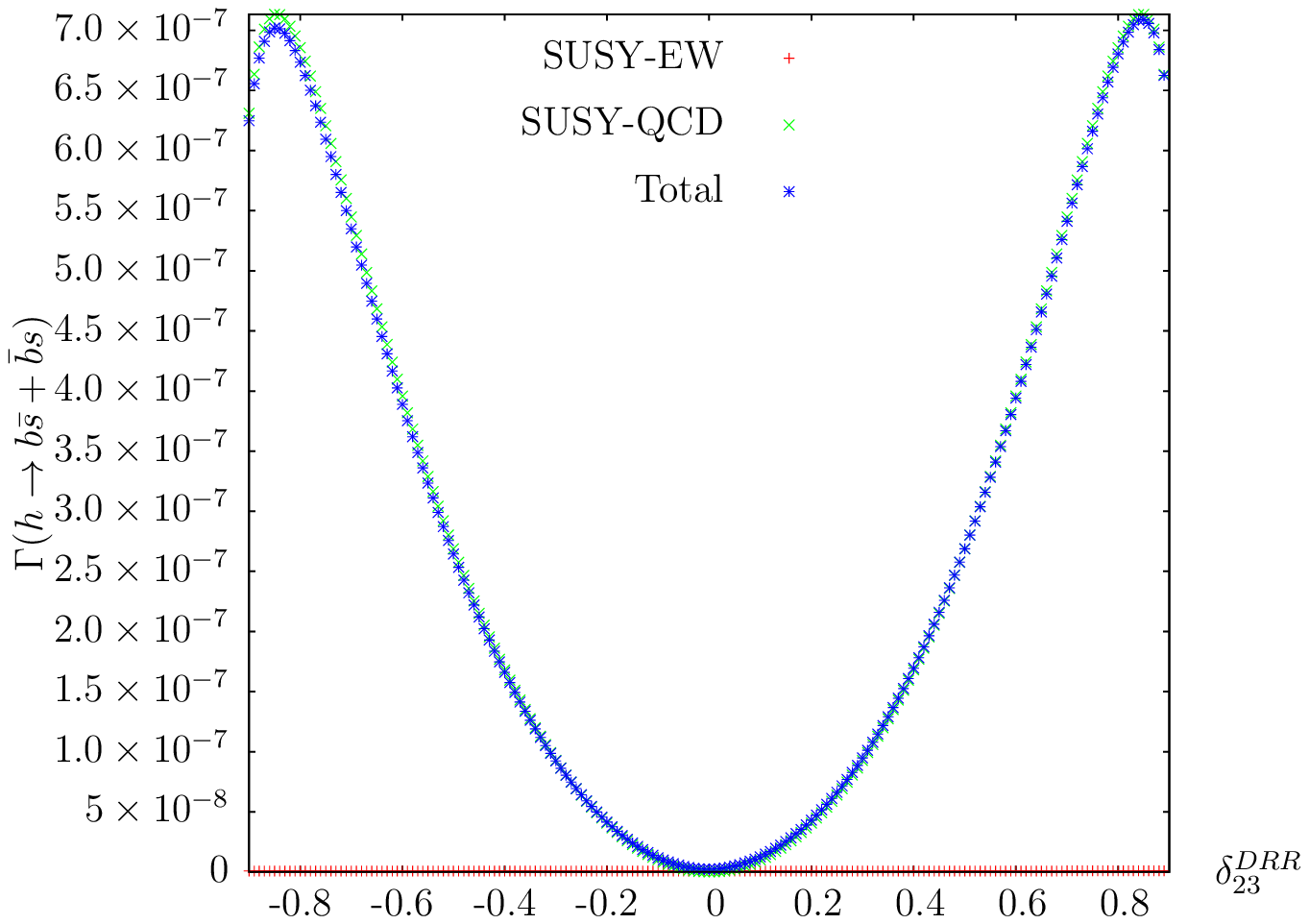 ,scale=0.52,angle=0,clip=}
\vspace{0.2cm}
\end{center}
\caption[$\Ga(\hbs)$ as a function of squark $\deFABij$]{$\Ga(\hbs)$ as a function of $\del{QLL}{23}$ (upper left),
$\del{DLR}{23}$ (upper right), $\del{DRL}{23}$ (lower left) and
$\del{DRR}{23}$ (lower right). 
}     
\label{SUSY-CONT}
\end{figure}

In order to illustrate the contributions from different diagrams we show
in \reffi{SUSY-CONT} the SUSY-EW, SUSY-QCD and total SUSY contribution
to $\Ga(\hbs)$ as a function of
$\del{QLL}{23}$ (upper left), $\del{DLR}{23}$ (upper right),
$\del{DRL}{23}$ (lower left) and $\del{DRR}{23}$ (lower
right). These four $\deFABij$ are the only relevant ones, since we
are mainly concerned with the down-type sector, and mixing with the
first generation does not play a role.
 
In order to compare our results with the literature, we have used
the same set of input parameters as in \cite{SUSY-EW}:
\begin{align}
\mu &= 800 \gev,\; \msusy = 800 \gev,\;  A_f = 500 \gev, \nonumber\\
\MA &= 400 \gev,\; M_2 = 300 \gev,\; \tb = 35 \, ,
\end{align}
where we have chosen, for simplicity, $\msusy$ as a common value for the
soft SUSY-breaking squark mass parameters, 
$\msusy = M_{\tilde Q} = M_{\tilde U} = M_{\tilde D}$,
and all the various trilinear parameters to be universal,
$A_f=A_t=A_b=A_c=A_s$. The value of the $\deFABij$'s are varied from 
-0.9~to~0.9, and GUT relations are used to calculate $M_1$ and
$M_3$. 
In \citere{SUSY-EW}, only LL mixing was considered. In this limit we find
results in qualitative agreement with \citere{SUSY-EW}. This analysis has 
been done just to illustrate the different contributions and we do not
take into account any experimental constraints. A detailed analysis for
realisitic SUSY scenerios (defined in \refta{tab:spectra}) constrained
by BPO and EWPO can be found below.

As can be seen in \reffi{SUSY-CONT}, for the decay width $\Ga(\hbs)$ the
SUSY-QCD contribution is dominant in all the cases. For LL mixing shown
in the upper left plot, the SUSY-QCD contribution reaches up
to \order{10^{-6}}, while the SUSY-EW contribution 
reach up to \order{10^{-7}}, resulting in a total contribution ``in
between'', due to the negative interference between SUSY-EW and SUSY-QCD
contribution. 
For LR and RL mixing, shown in the upper right and lower left plot,
respectively, the SUSY-QCD contribution reach up to the
maximum value of \order{10^{-2}}, while the SUSY-EW contribution reach
only up to \order{10^{-7}}. In this case total contriution is almost equal to
SUSY-QCD contribution as SUSY-EW contibution (and thus the interference)
is relatively neglible.
For RR mixing, shown in the lower right plot, the SUSY-EW
contribution of \order{10^{-10}} is again neglible compared to SUSY-QCD
contribution of \order{10^{-7}}.  

\begin{figure}[ht!]
\begin{center}
\psfig{file=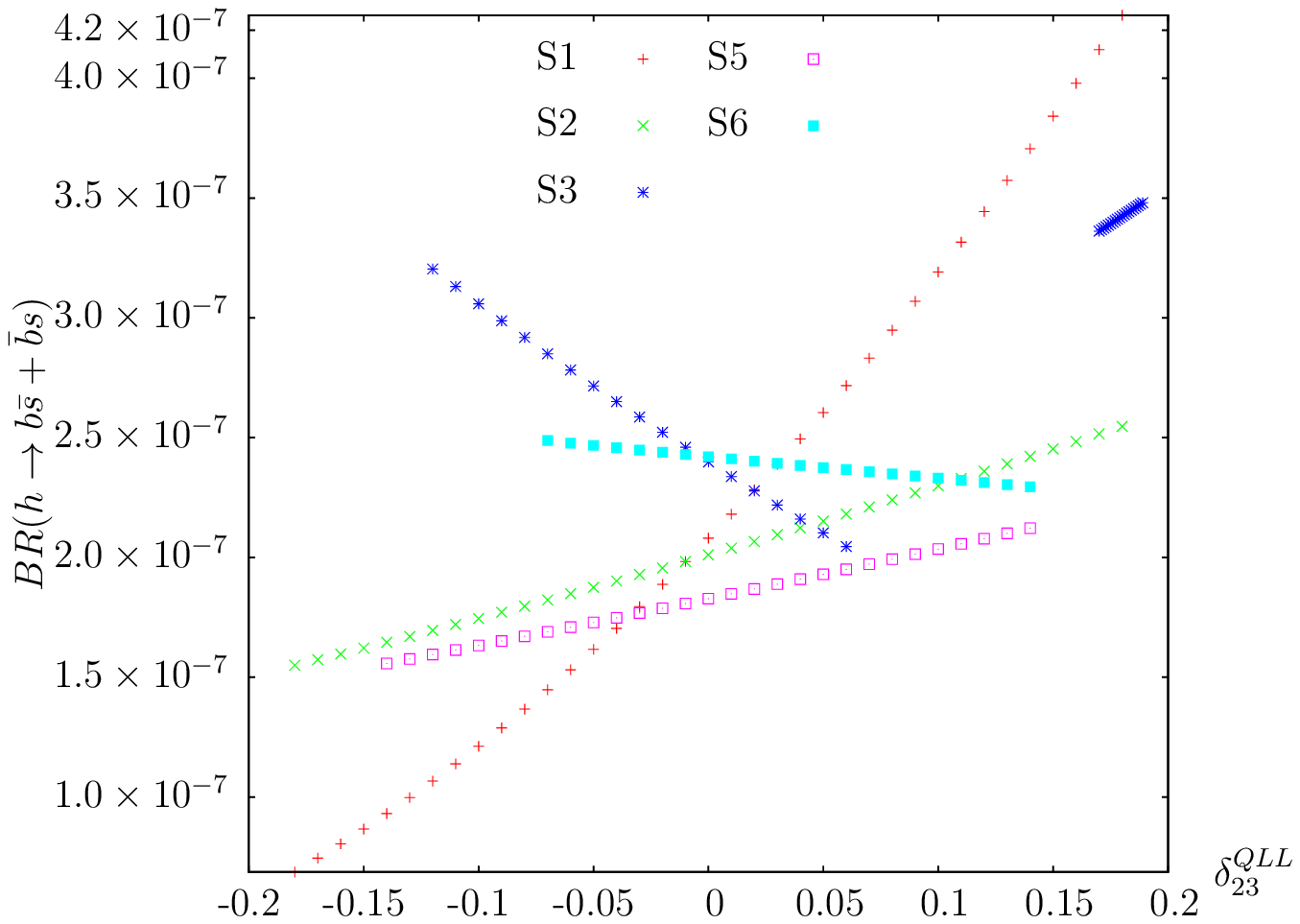  ,scale=0.51,angle=0,clip=}
\psfig{file=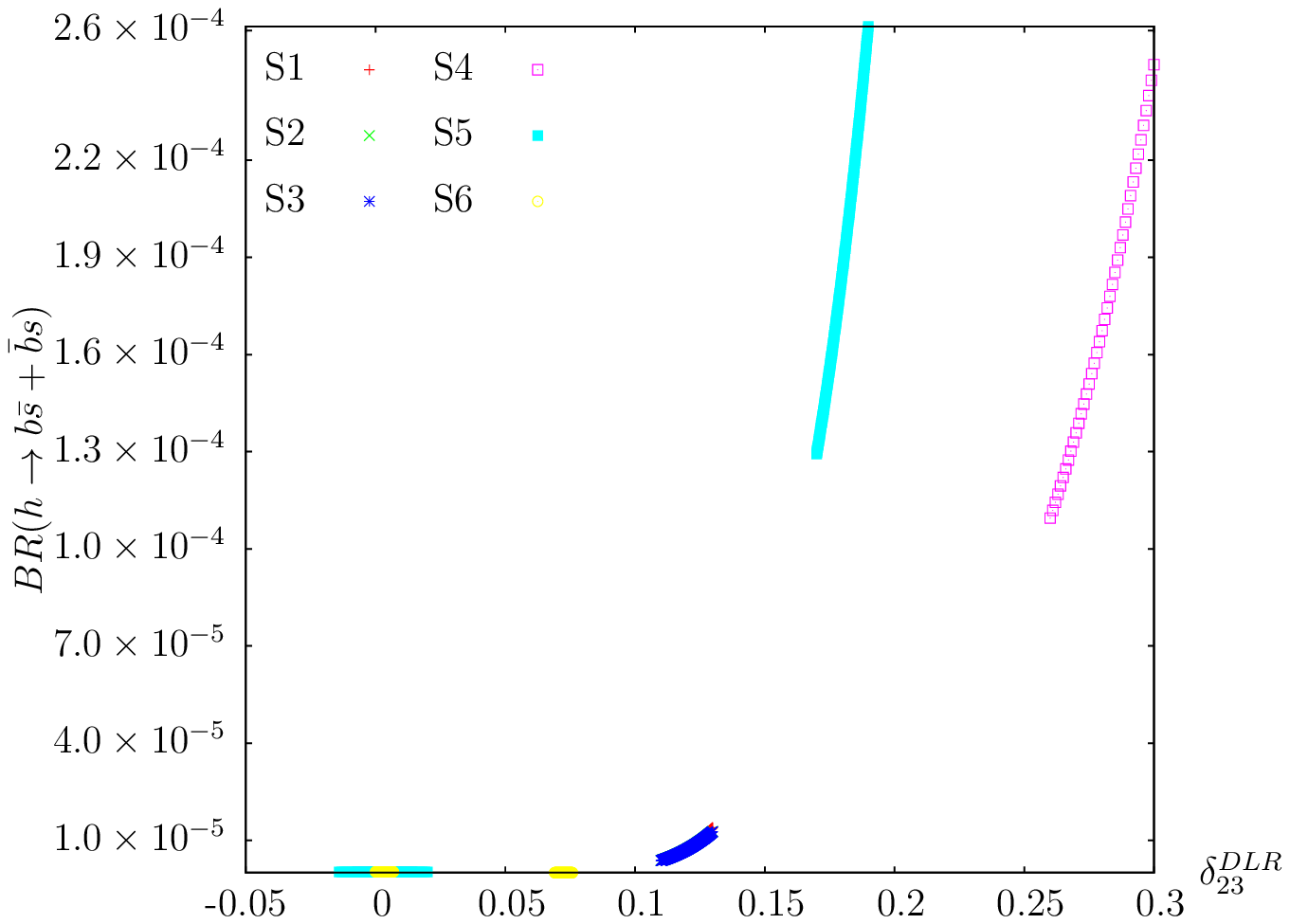  ,scale=0.51,angle=0,clip=}\\
\vspace{0.2cm}
\psfig{file=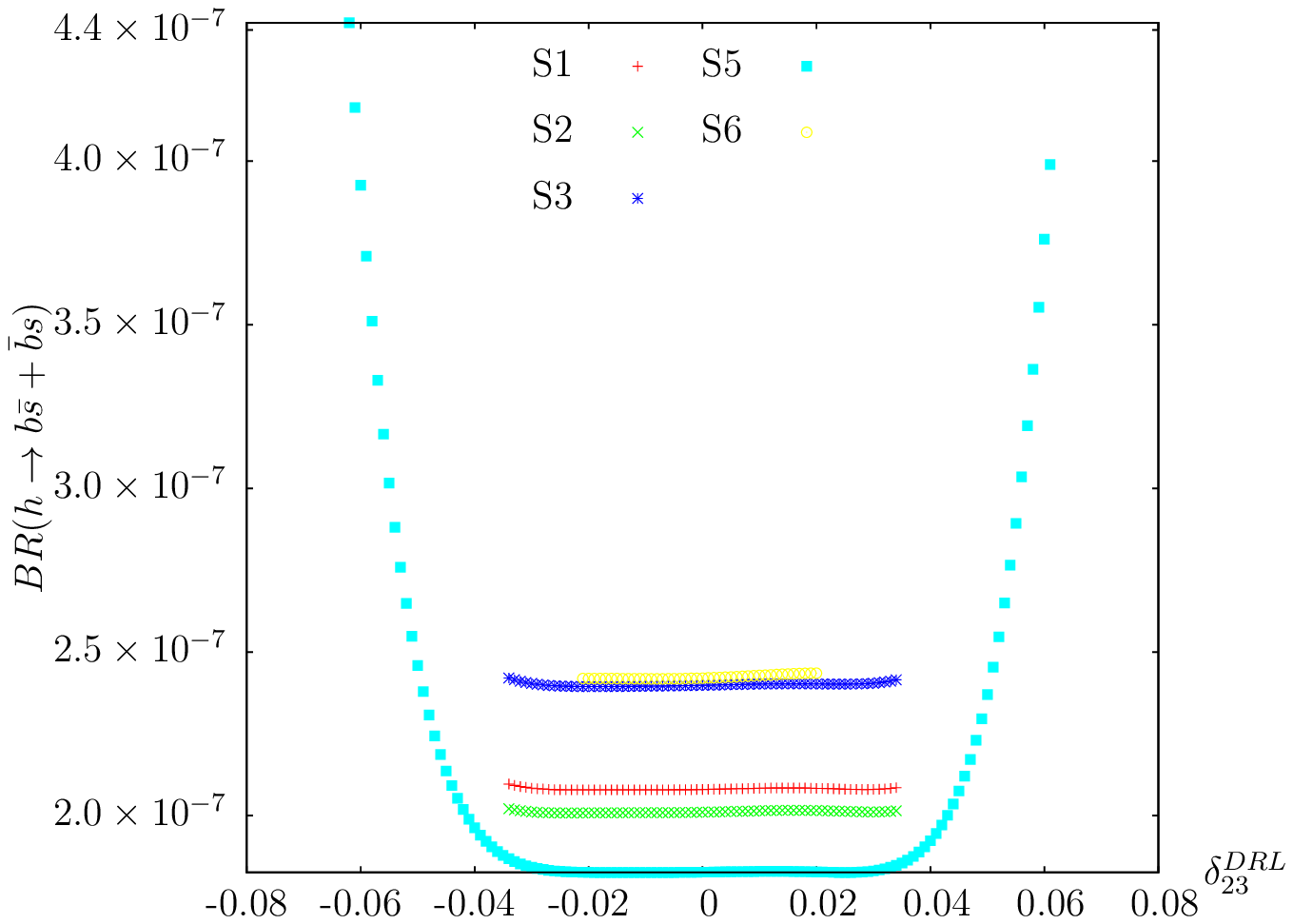  ,scale=0.51,angle=0,clip=}
\psfig{file=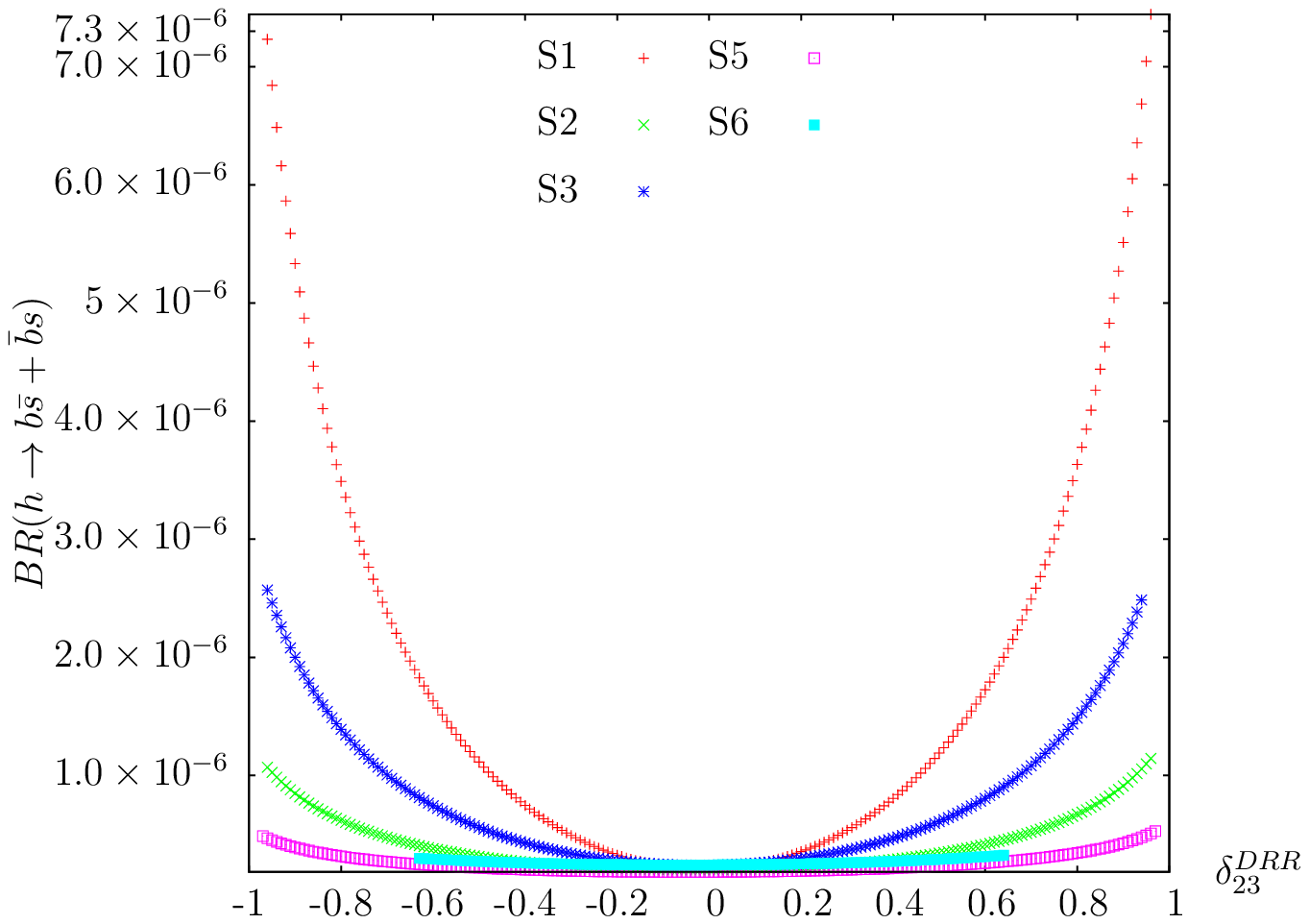  ,scale=0.51,angle=0,clip=}\\
\end{center}
\caption[\brhbs\ as a function of squark $\deFABij$]{\brhbs\ as a function of  $\del{QLL}{23}$ (upper left), 
$\del{DLR}{23}$ (upper right), $\del{DRL}{23}$ (lower left) and 
$\del{DRR}{23}$ (lower right).}
\label{Fig:QFVHD}
\end{figure}

Now we turn to realistic scenarios that are in agreement with
experimental data from BPO and EWPO. Starting point are the scenarios
S1\ldots S6 defined in \refta{tab:spectra}, where we vary the flavor
violating $\deFABij$ within the experimentally allowed ranges following
the results given in \reftas{tab:boundsS1S6}, \ref{tab:EWPOboundsS1S6}.
We start with the scenarios in which we allow one of the $\deFABij$ to
be varied, while the others are set to zero. 
In \reffi{Fig:QFVHD} we show \brhbs\ as a function of $\del{QLL}{23}$ 
(upper left), $\del{DLR}{23}$ (upper right), $\del{DRL}{23}$ (lower
left) and $\del{DRR}{23}$ (lower right), i.e.\ for the same set of
$\deFABij$ that has been analyzed in \reffi{SUSY-CONT}. 
It can be seen that allowing only one $\deFABij \neq 0$ results in
rather small values of \brhbs. LL (upper left) and RL (lower left plot)
mixing results in \order{10^{-7}} values for \brhbs. One order of
magnitude can be gained in the RR mixing case (lower right). The largest
values of \brhbs\ are obtained in the case of $\del{DLR}{23} \neq 0$
(upper right plot). Here in S4 and S5 values of 
$\brhbs \sim 2 \times 10^{-4}$ can be found, possibly in the reach of
future $e^+e^-$ colliders, see \refse{sec:hbs-calc}.

So far we have shown the effects of independent variations of
one $\deFABij$. Obviously, a realistic model would include several 
$\deFABij \neq 0$ that may interfere, increasing or decreasing the results
obtained with just the addition of independent contributions.
GUT based MFV models that induce the flavor violation
via RGE running automatically generate several $\deFABij \neq 0$ at the
EW scale.
In the following we will present results with two or three 
$\deFABij \neq 0$, where we combined the ones that showed the largest
effects. 

\begin{figure}[ht!]
\begin{center}
\psfig{file=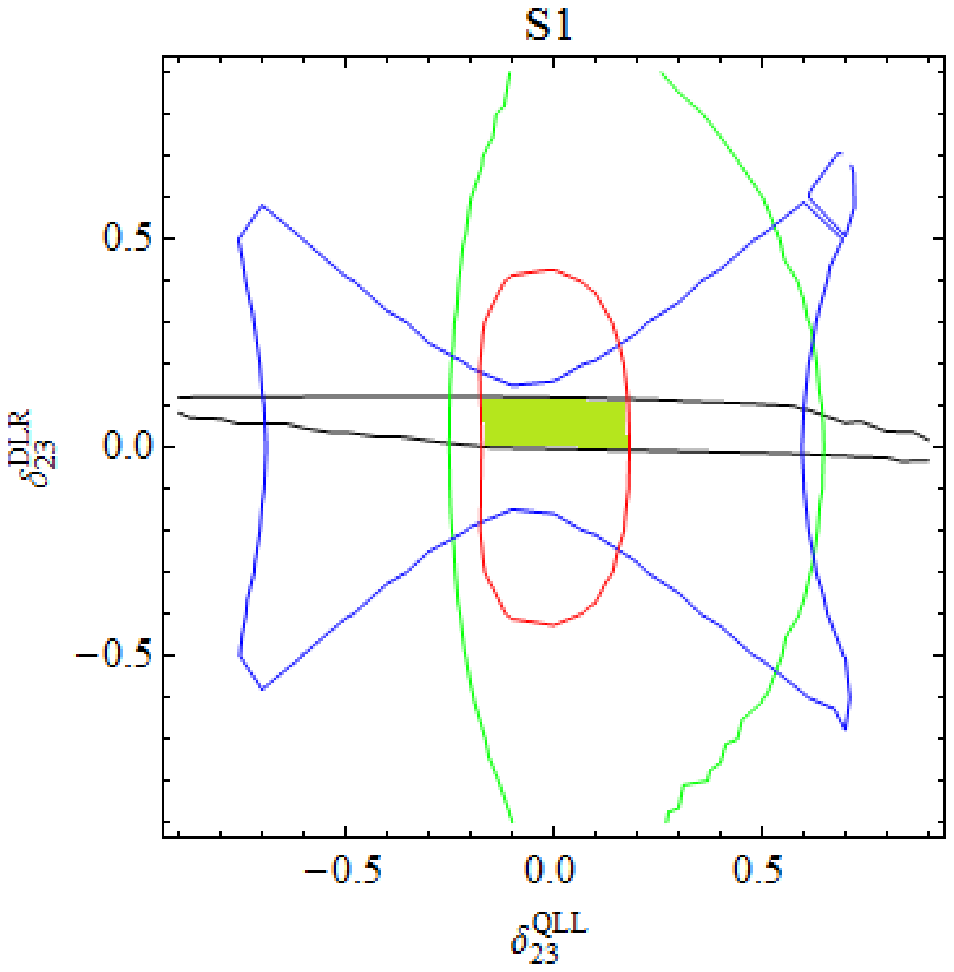  ,scale=0.70,angle=0,clip=}
\psfig{file=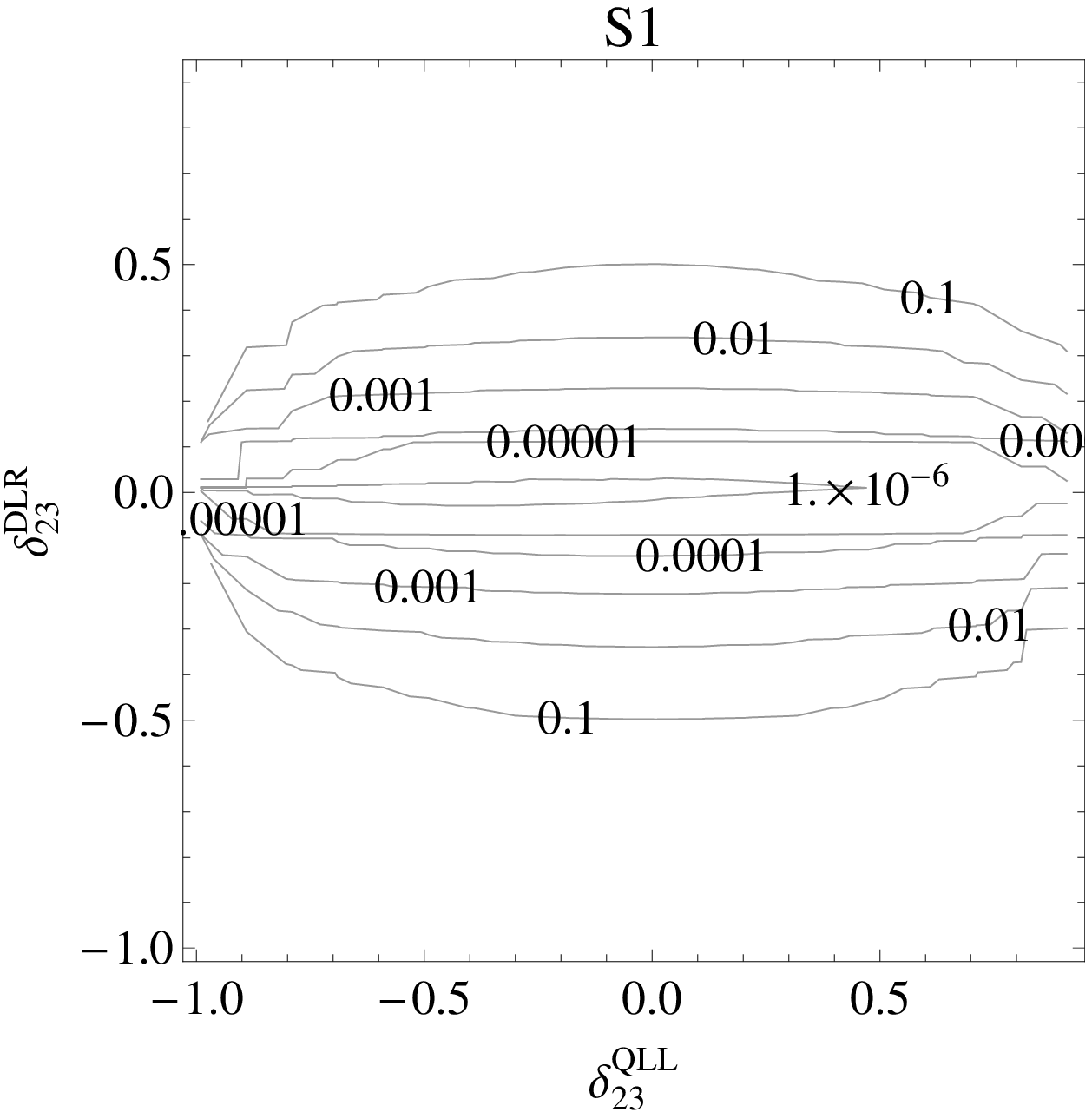  ,scale=0.52,angle=0,clip=}\\
\vspace{0.2cm}
\psfig{file=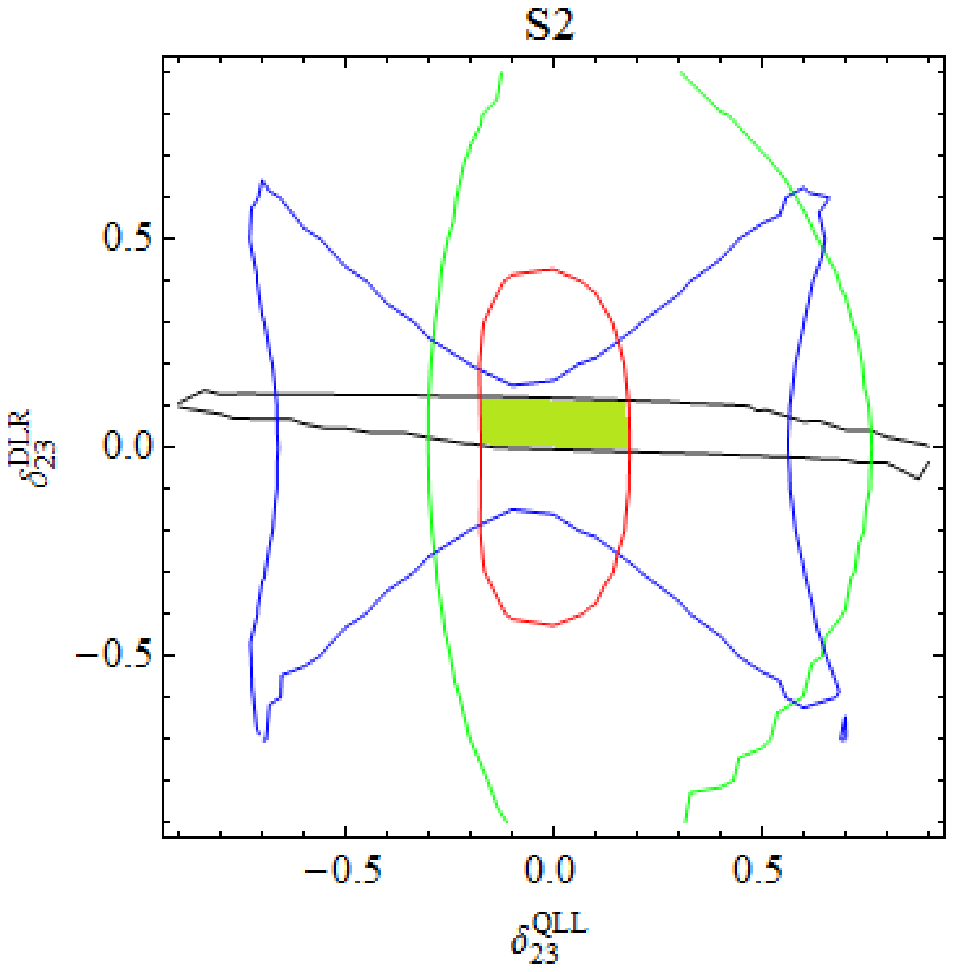 ,scale=0.70,angle=0,clip=}
\psfig{file=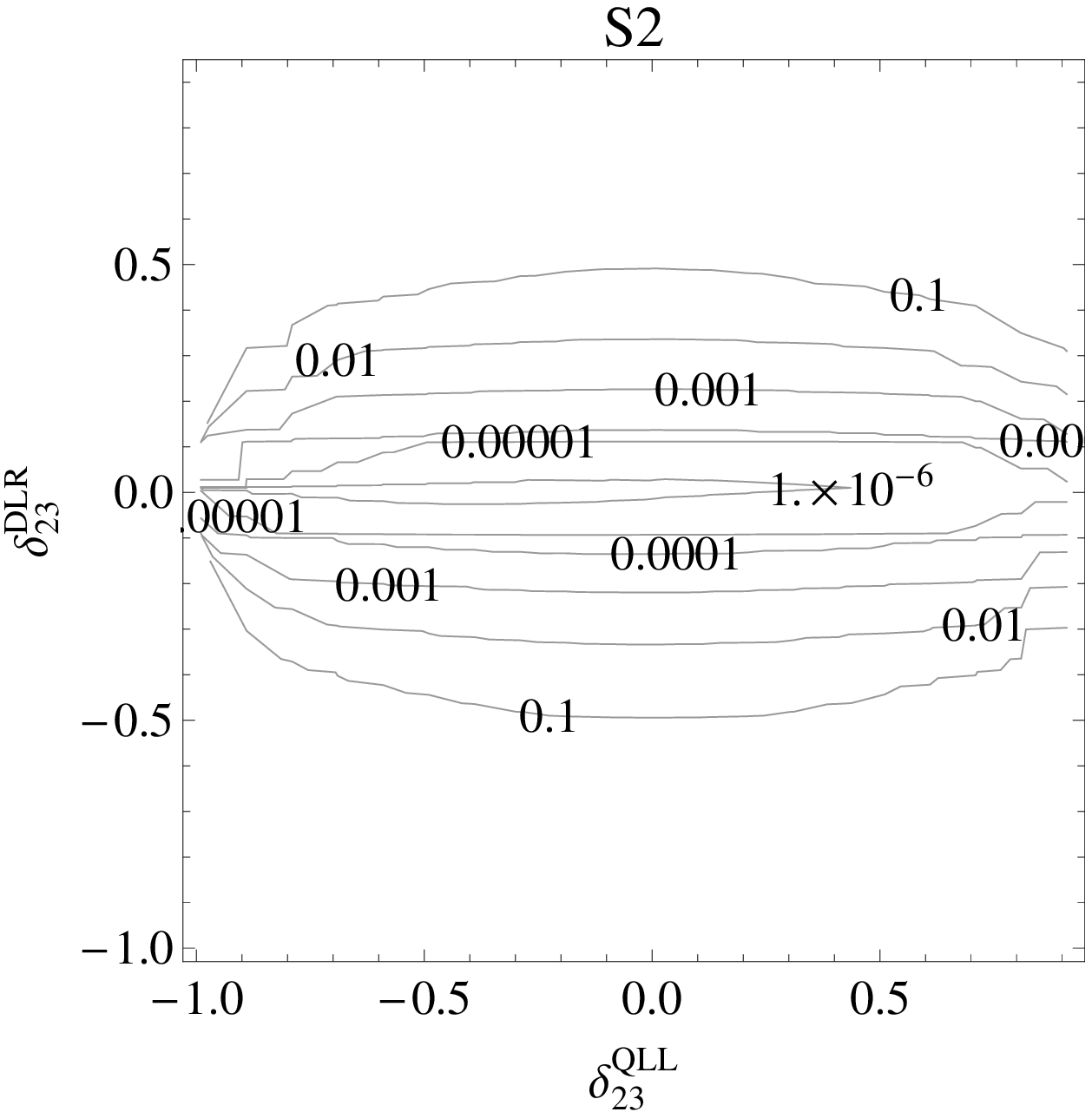 ,scale=0.52,angle=0,clip=}\\
\vspace{0.2cm}
\psfig{file=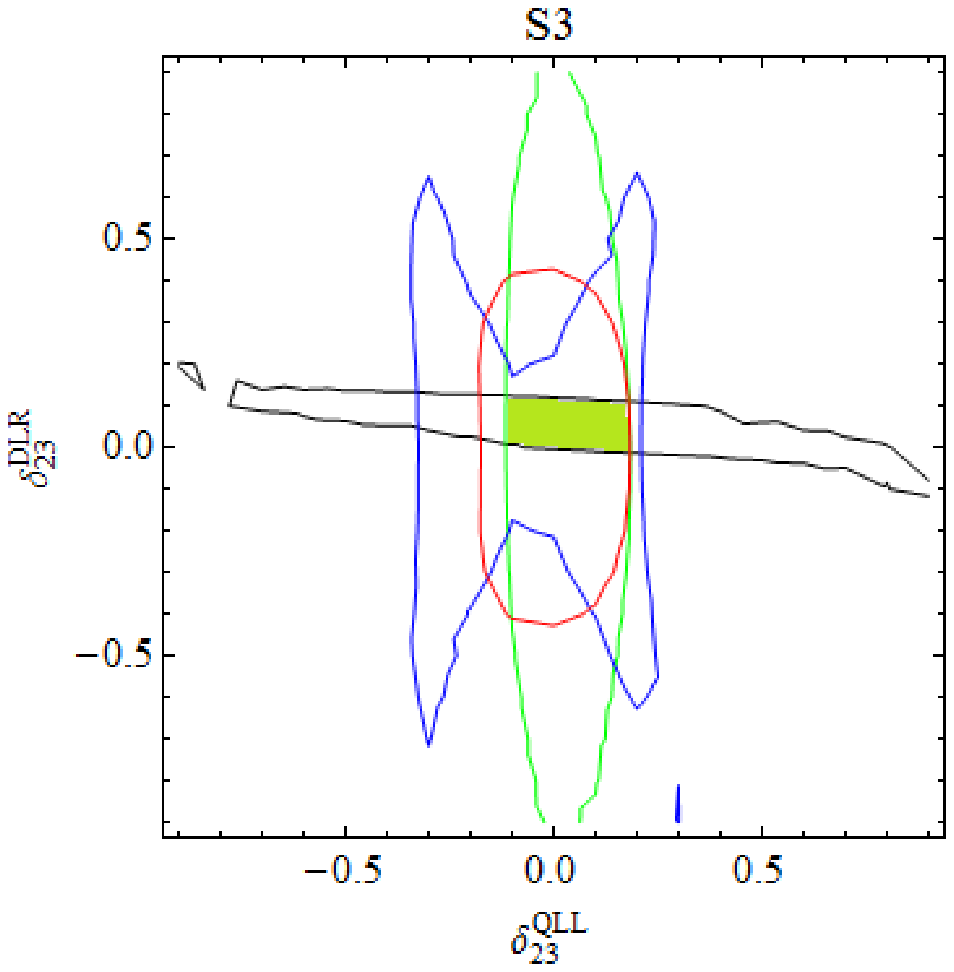 ,scale=0.70,angle=0,clip=}
\psfig{file=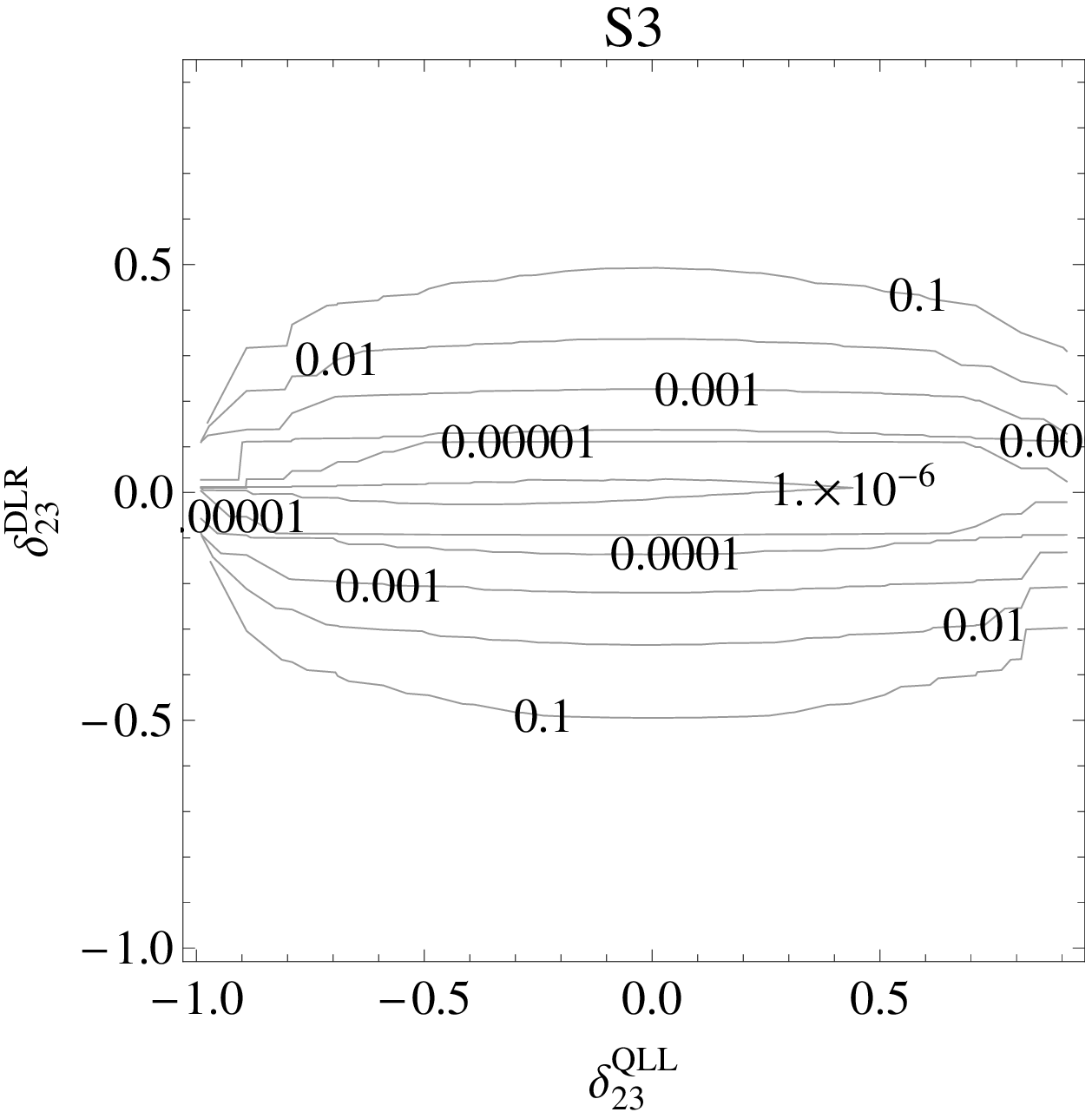 ,scale=0.52,angle=0,clip=}
\end{center}
\vspace{-2em}
\caption[Contours of EWPO, BPO and \brhbs\ in ($\del{QLL}{23}$ , $\del{DLR}{23}$) plane]{Left: Contours of \bsg\ (Black), \bmm\ (Green), \dmbs (Blue) and
$\MW$ (Red) in ($\del{QLL}{23}$ , $\del{DLR}{23}$) plane for
points S1-S3. The shaded area shows the range of values allowed by all
constraints. Right: corresponding contours for \brhbs.}     
\label{Fig:S1S3QLLDLR23}
\end{figure}

\begin{figure}[ht!]
\begin{center}
\psfig{file=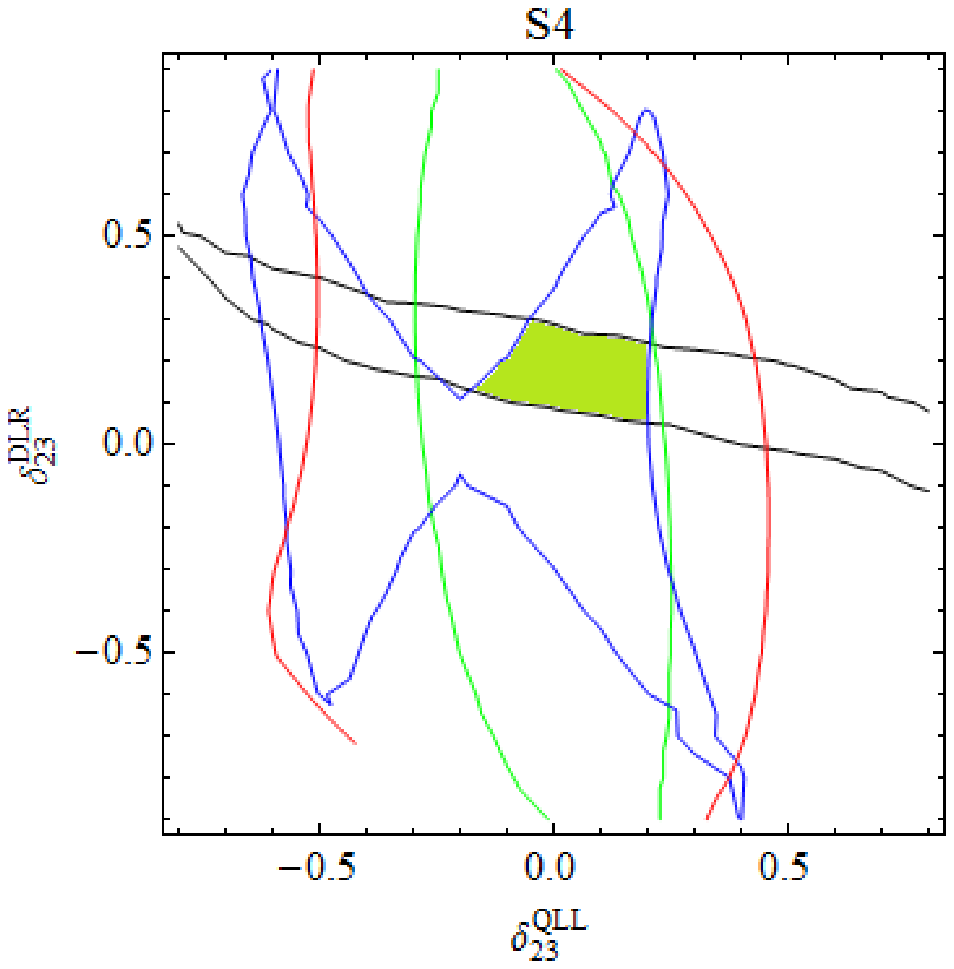  ,scale=0.69,angle=0,clip=}
\psfig{file=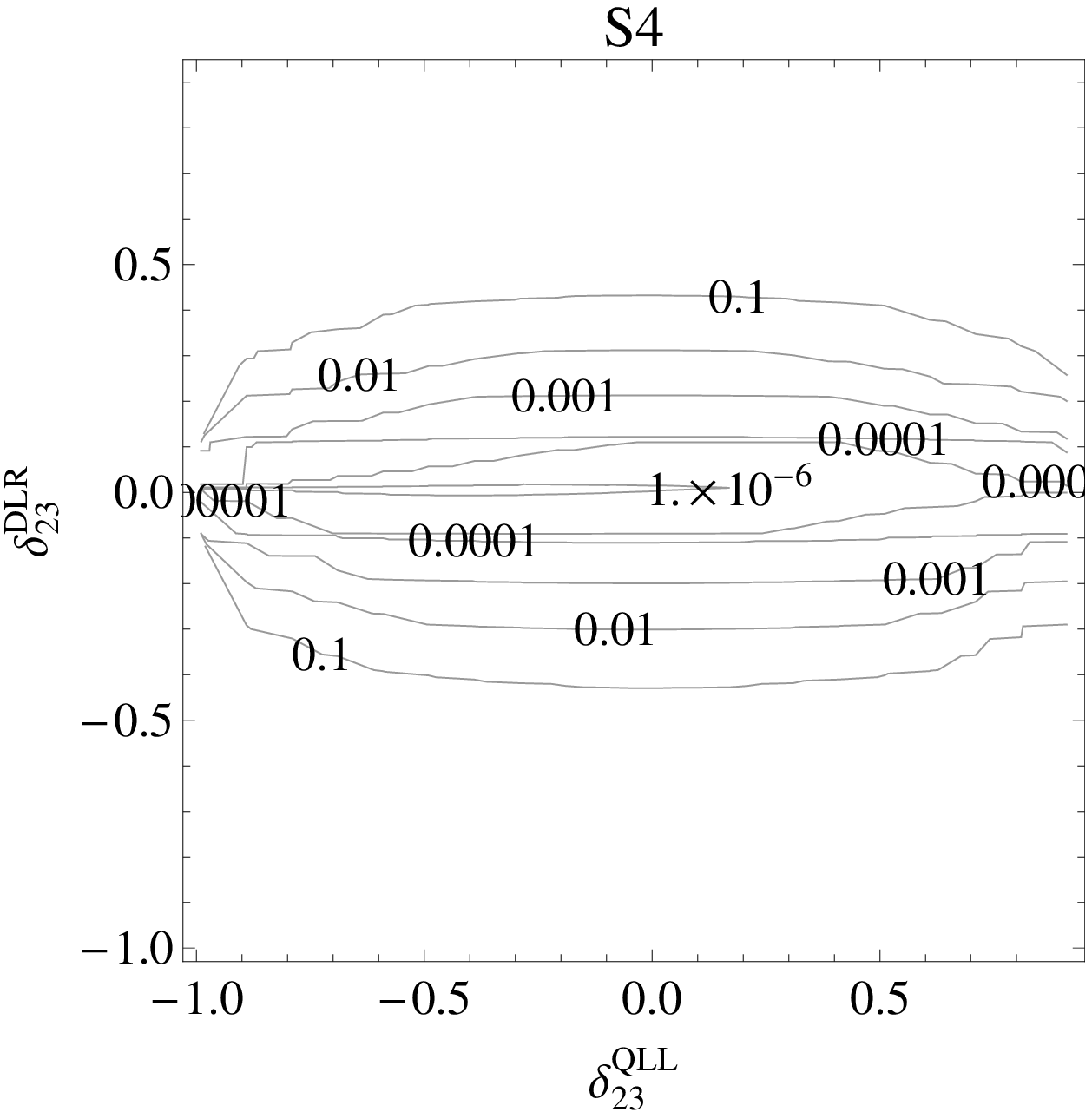  ,scale=0.52,angle=0,clip=}\\
\vspace{0.2cm}
\psfig{file=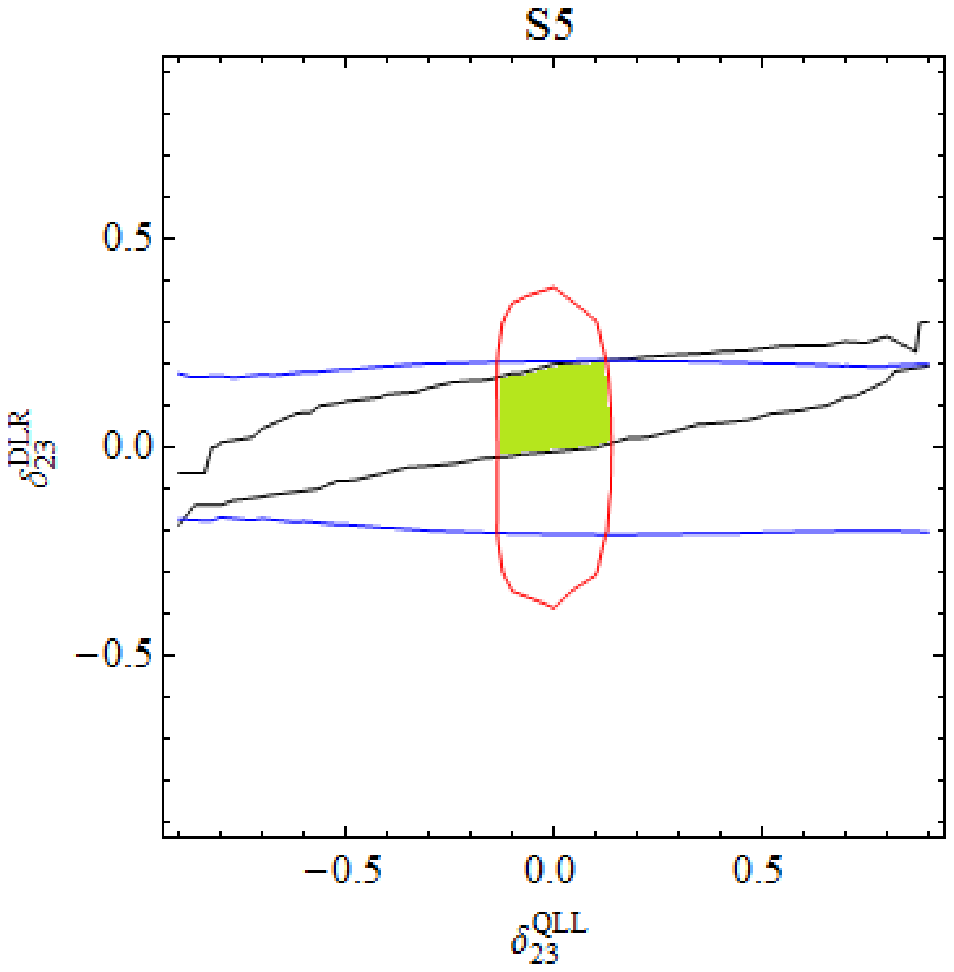 ,scale=0.69,angle=0,clip=}
\psfig{file=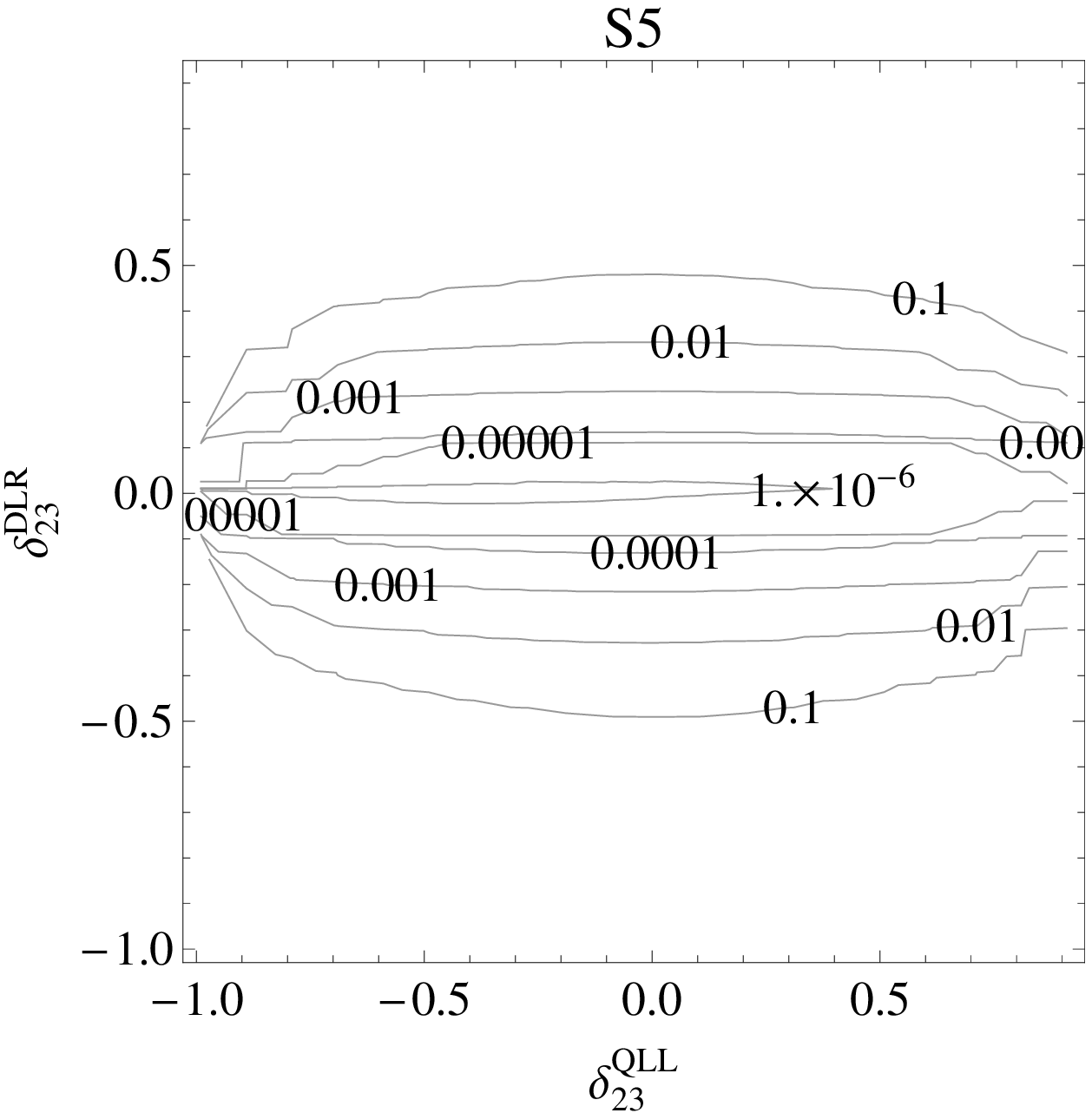 ,scale=0.52,angle=0,clip=}\\
\vspace{0.2cm}
\psfig{file=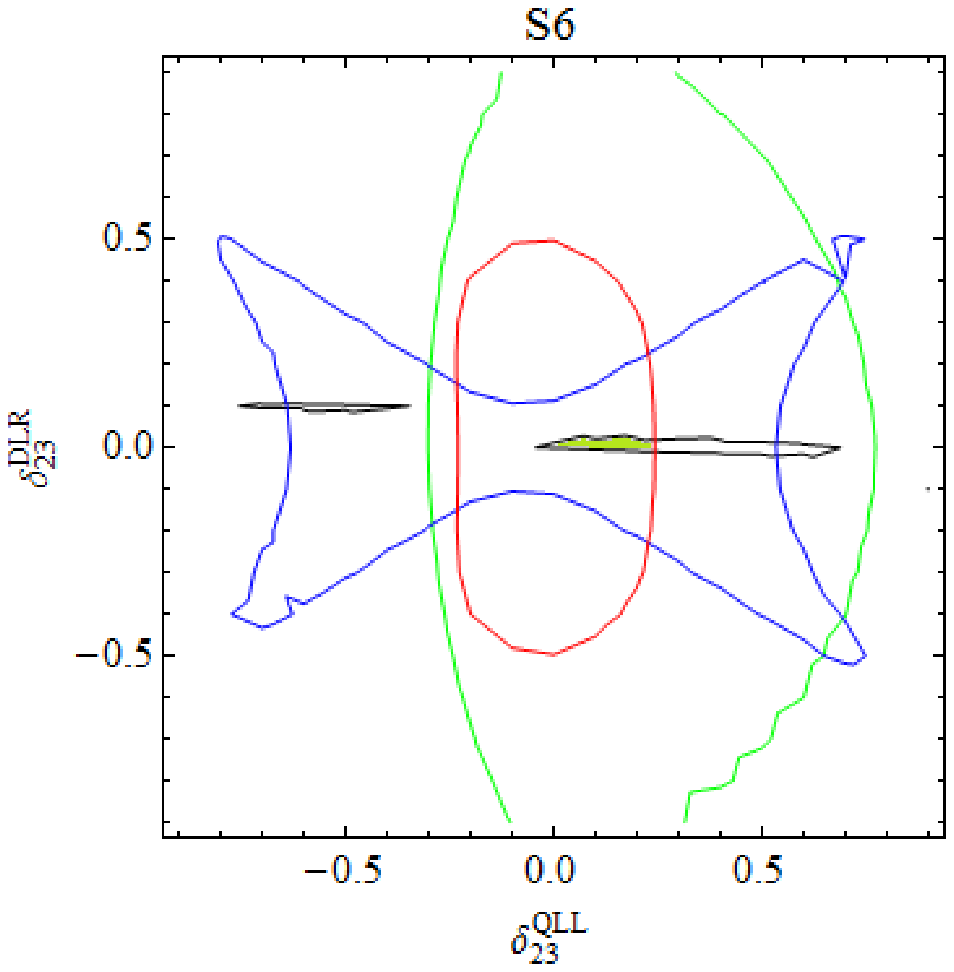 ,scale=0.69,angle=0,clip=}
\psfig{file=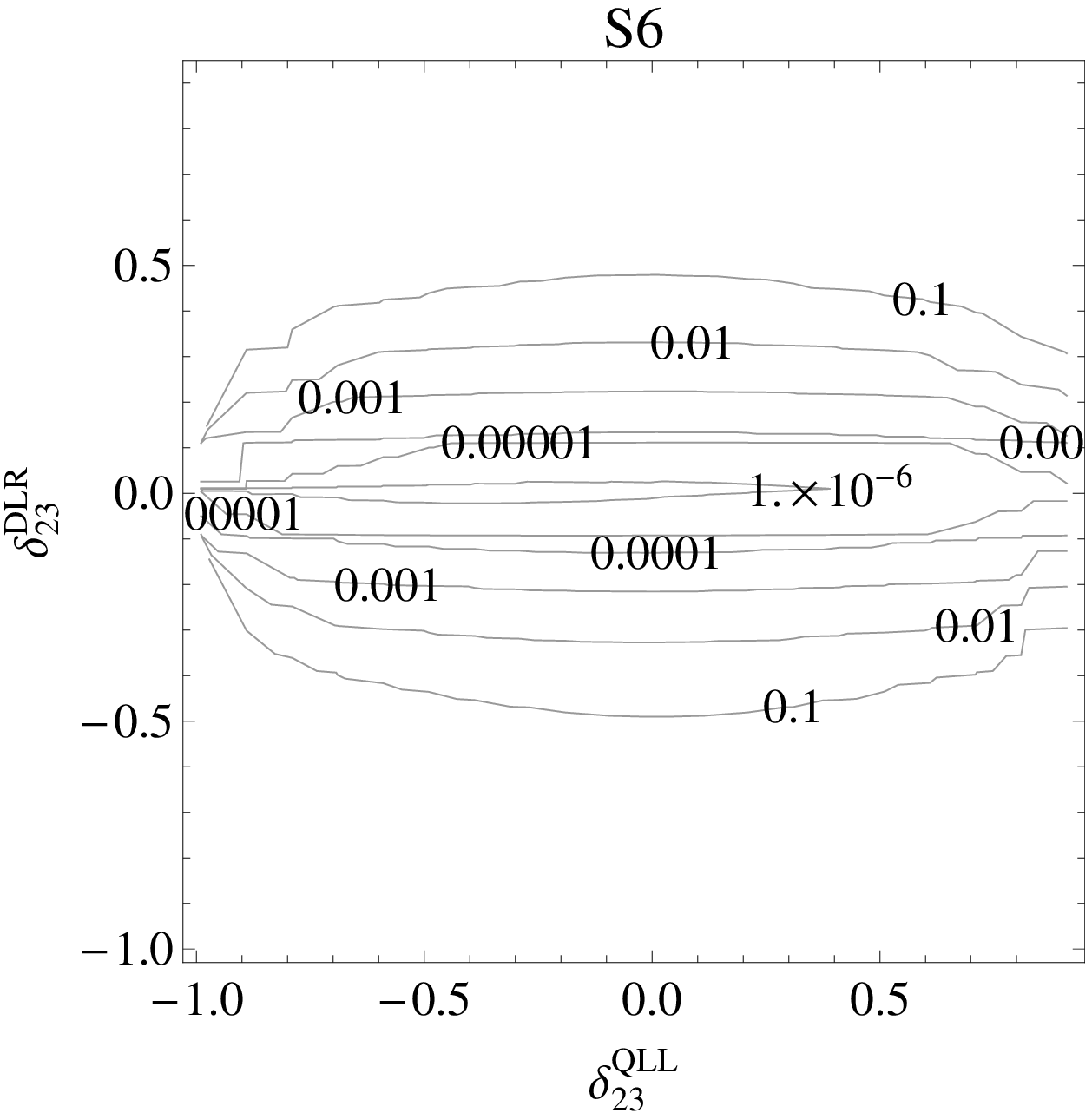 ,scale=0.52,angle=0,clip=}
\vspace{-2em}
\end{center}
\caption[Contours of EWPO, BPO and \brhbs\ in ($\del{QLL}{23}$ , $\del{DLR}{23}$) plane]{Left: Contours of \bsg\ (Black), \bmm\ (Green), \dmbs\ (Blue) and
$\MW$ (Red) in ($\del{QLL}{23}$ , $\del{DLR}{23}$) plane for
points S4-S6. The shaded area shows the range of values allowed by all
constraints. Right: corresponding contours for \brhbs.}    
\label{Fig:S4S6QLLDLR23}
\end{figure}

\begin{figure}[ht!]
\begin{center}
\psfig{file=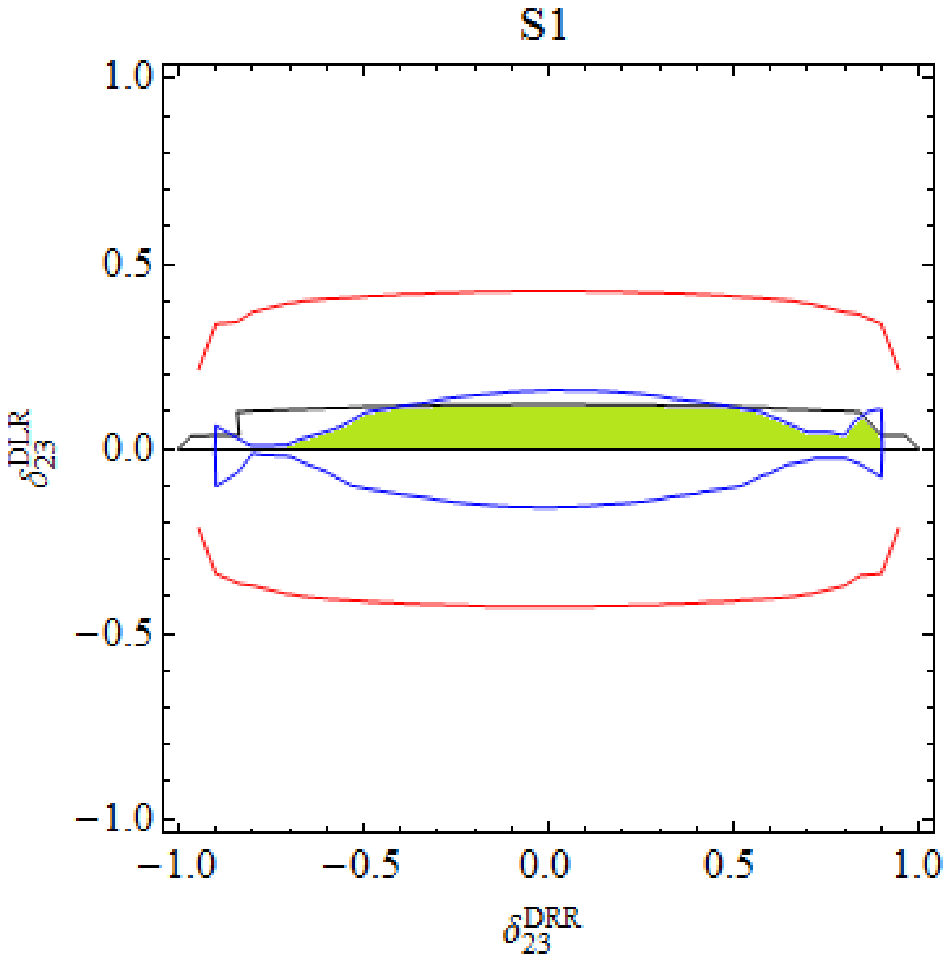  ,scale=0.69,angle=0,clip=}
\psfig{file=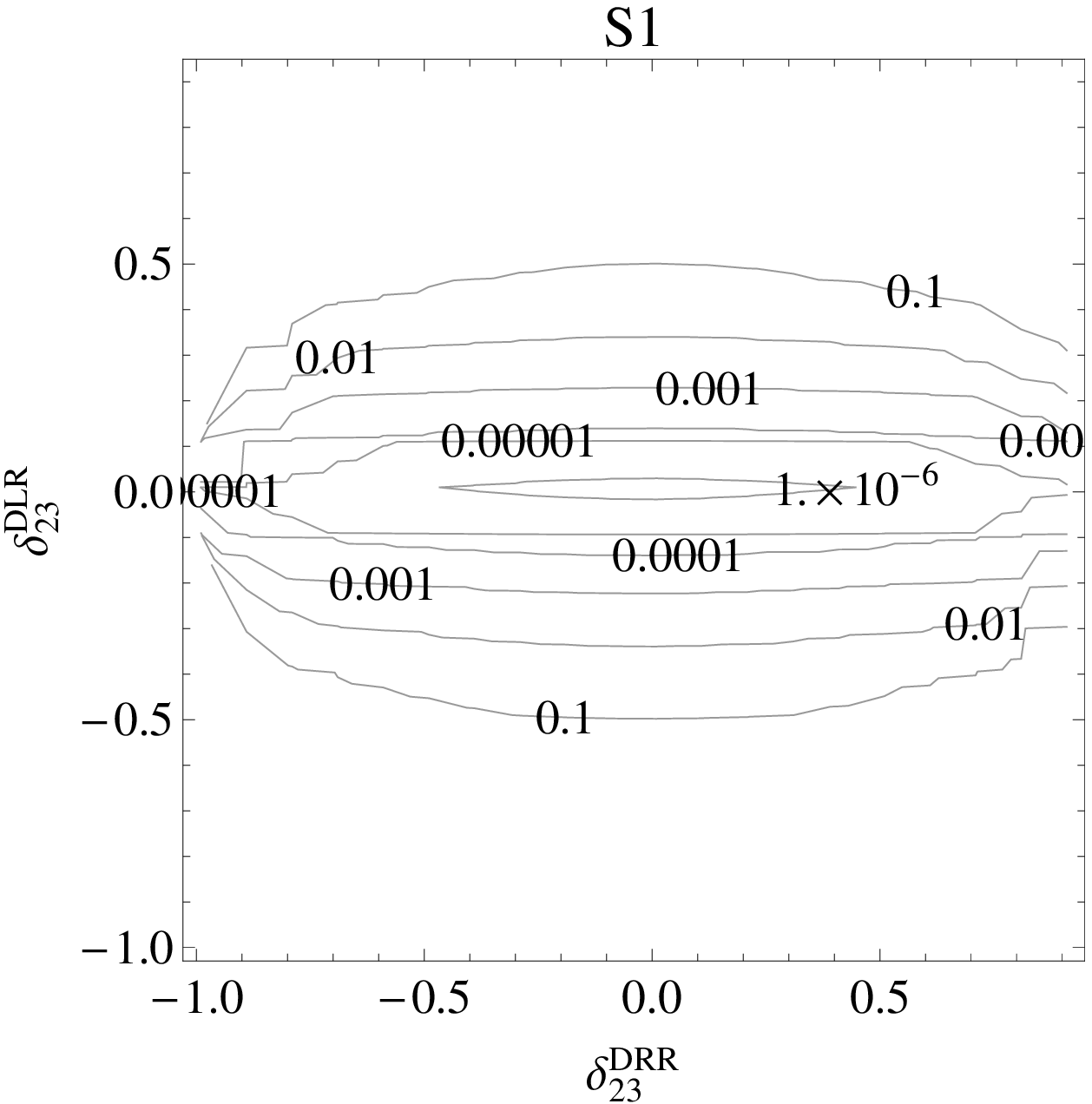  ,scale=0.52,angle=0,clip=}\\
\vspace{0.2cm}
\psfig{file=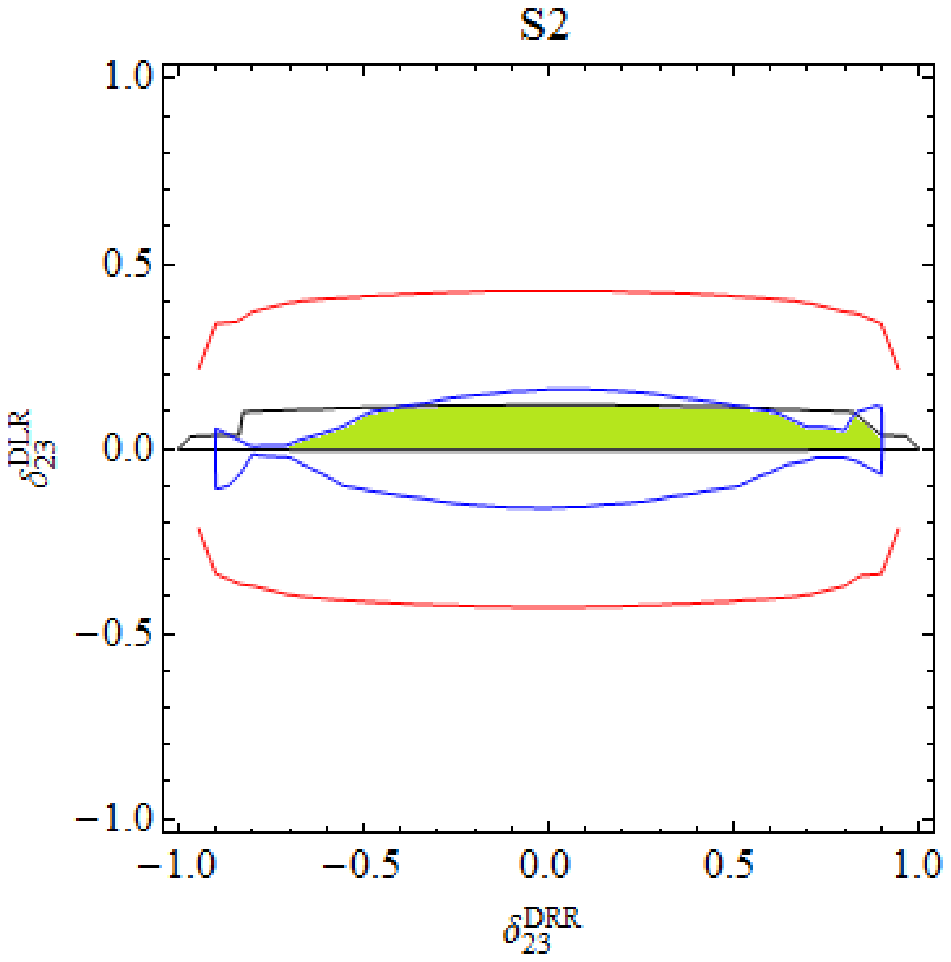 ,scale=0.69,angle=0,clip=}
\psfig{file=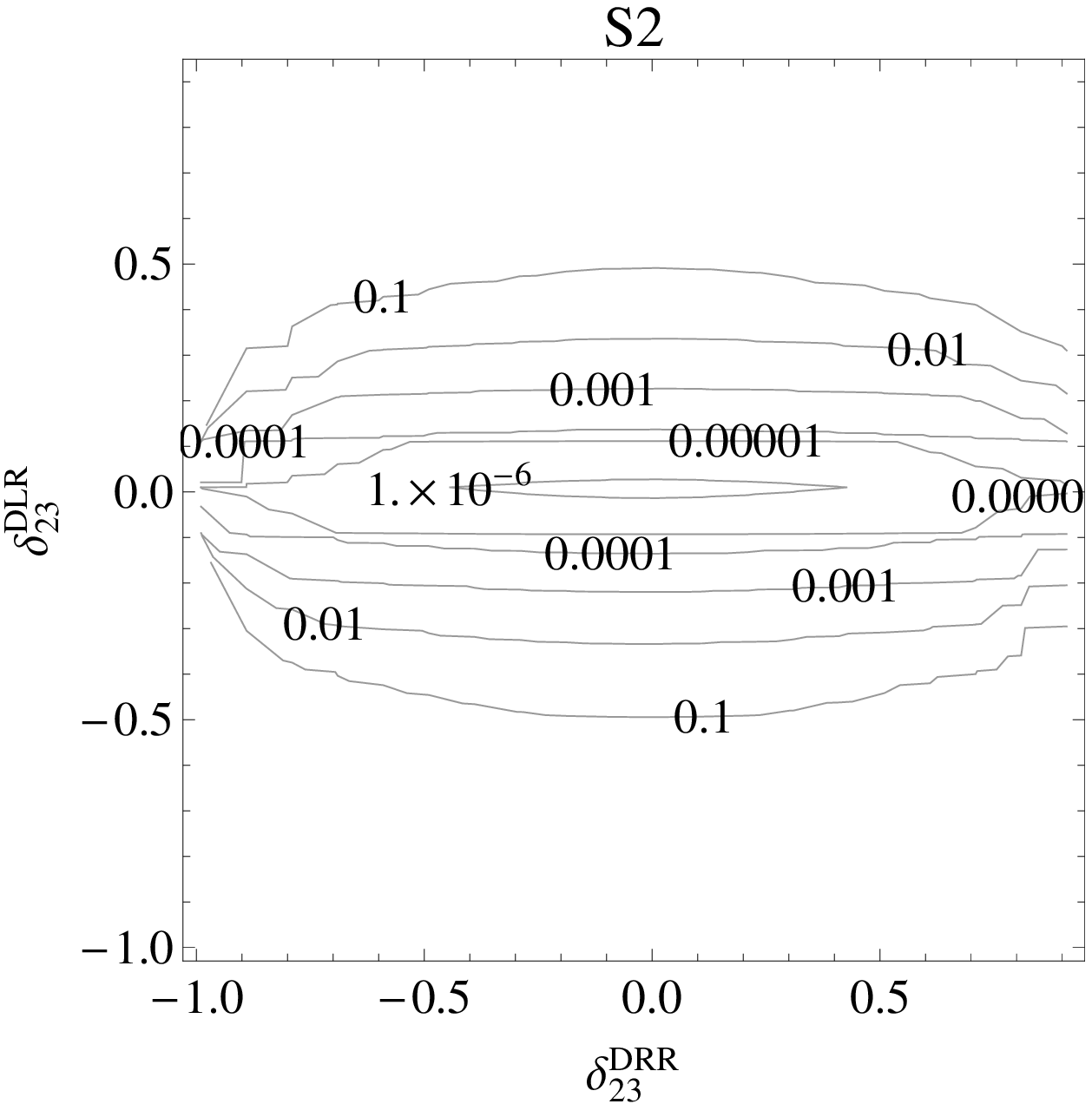 ,scale=0.52,angle=0,clip=}\\
\vspace{0.2cm}
\psfig{file=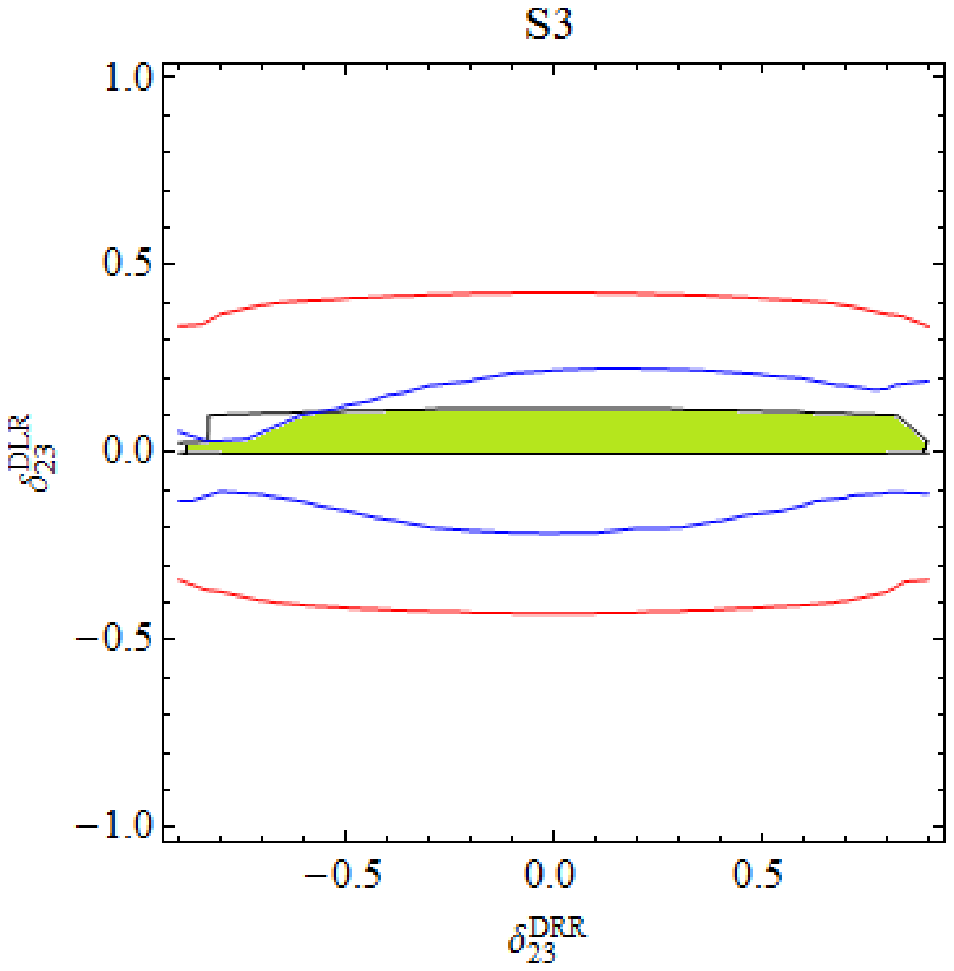 ,scale=0.69,angle=0,clip=}
\psfig{file=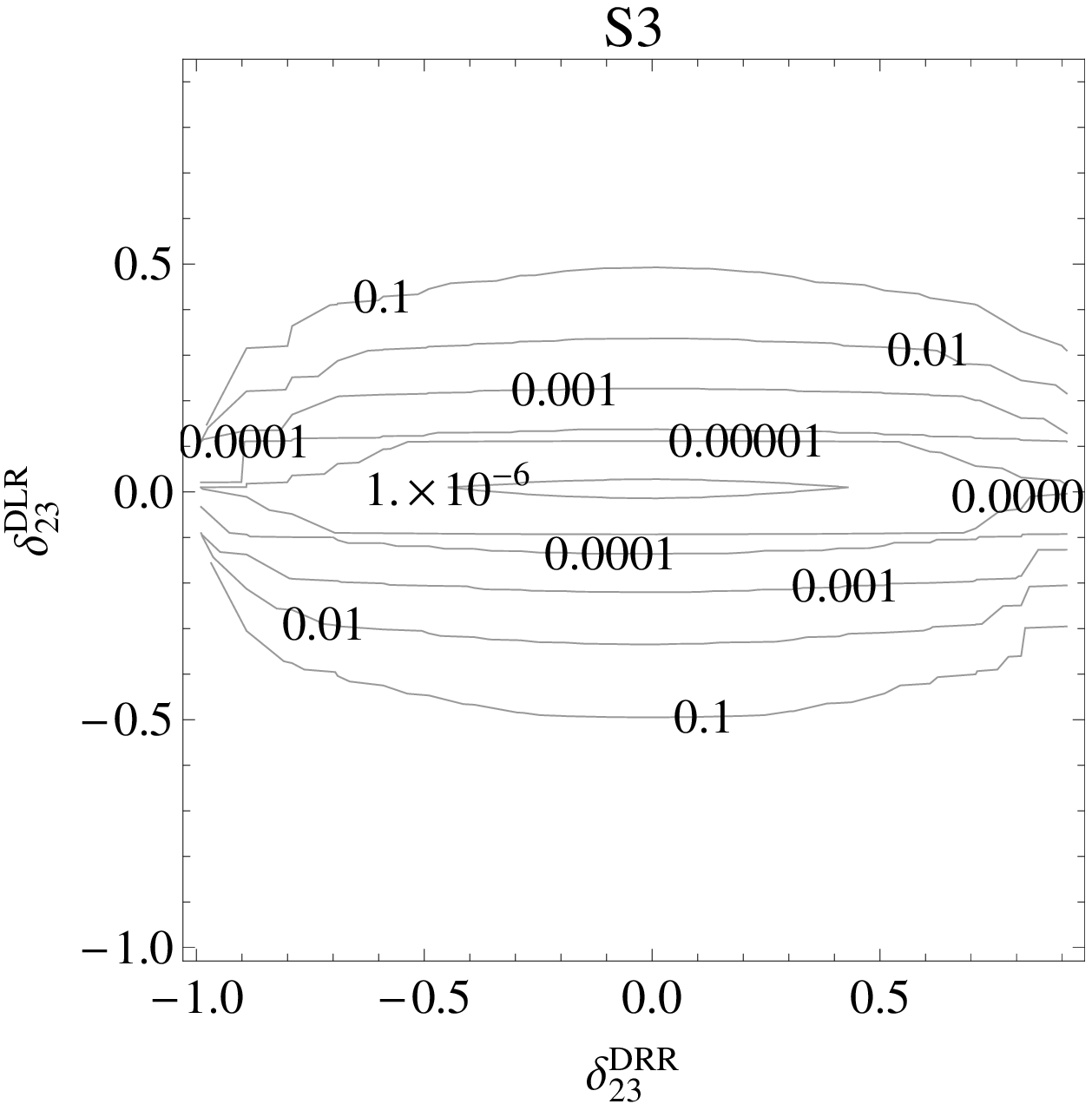 ,scale=0.52,angle=0,clip=}
\vspace{-2em}
\end{center}
\caption[Contours of EWPO, BPO and \brhbs\ in ($\del{DRR}{23}$ , $\del{DLR}{23}$) plane]
{Left: Contours of \bsg\ (Black), \bmm\ (Green), \dmbs\ (Blue) and
$\MW$ (Red) in ($\del{DRR}{23}$ , $\del{DLR}{23}$) plane for points S1-S3. 
The shaded area shows the range of values allowed by all
constraints. Right: corresponding contours for \brhbs.}   
\label{Fig:S1S3DLRDRR23}
\end{figure}

\begin{figure}[ht!]
\begin{center}
\psfig{file=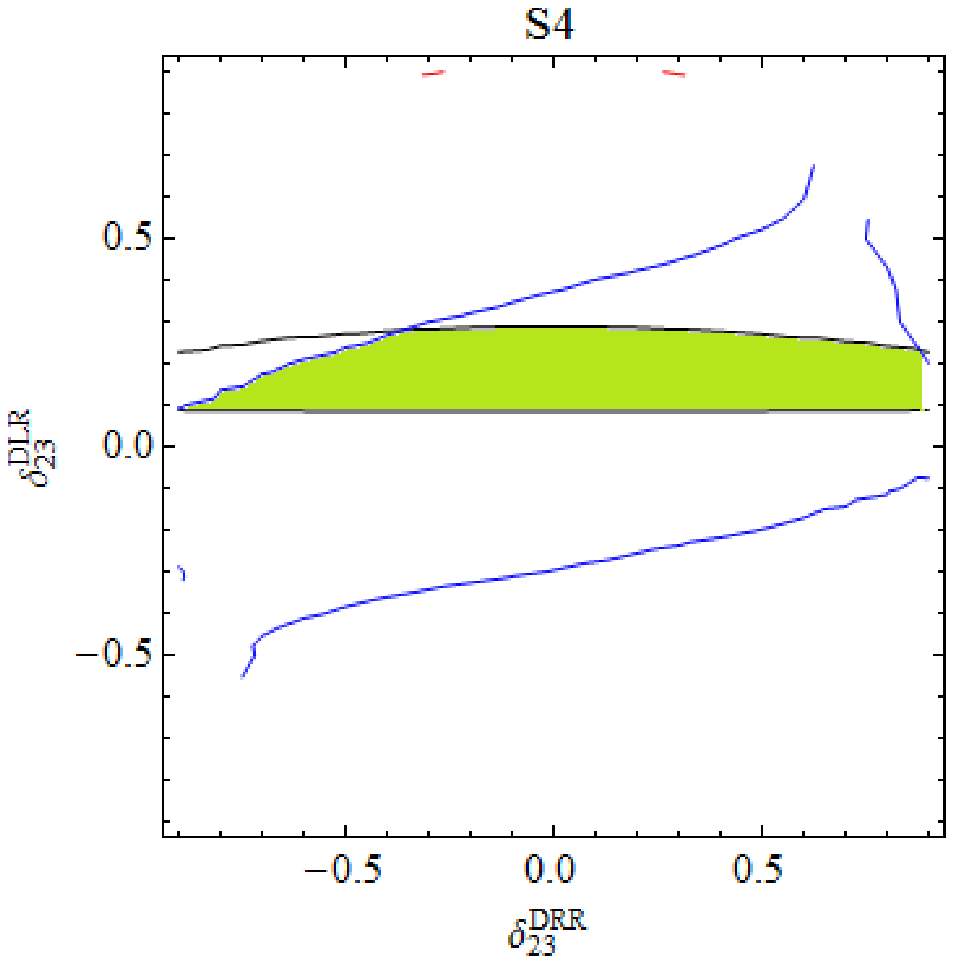  ,scale=0.69,angle=0,clip=}
\psfig{file=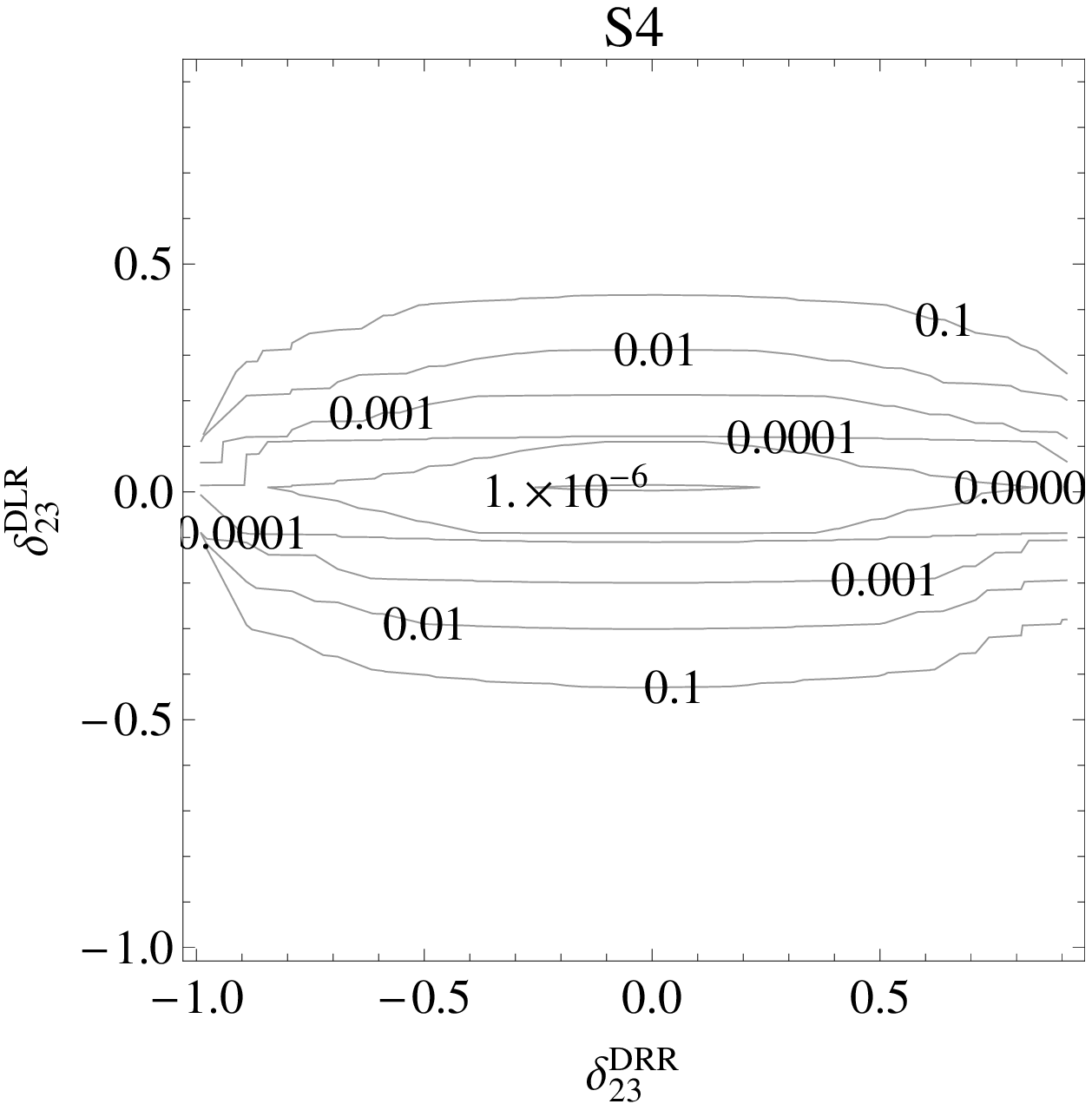  ,scale=0.52,angle=0,clip=}\\
\vspace{0.2cm}
\psfig{file=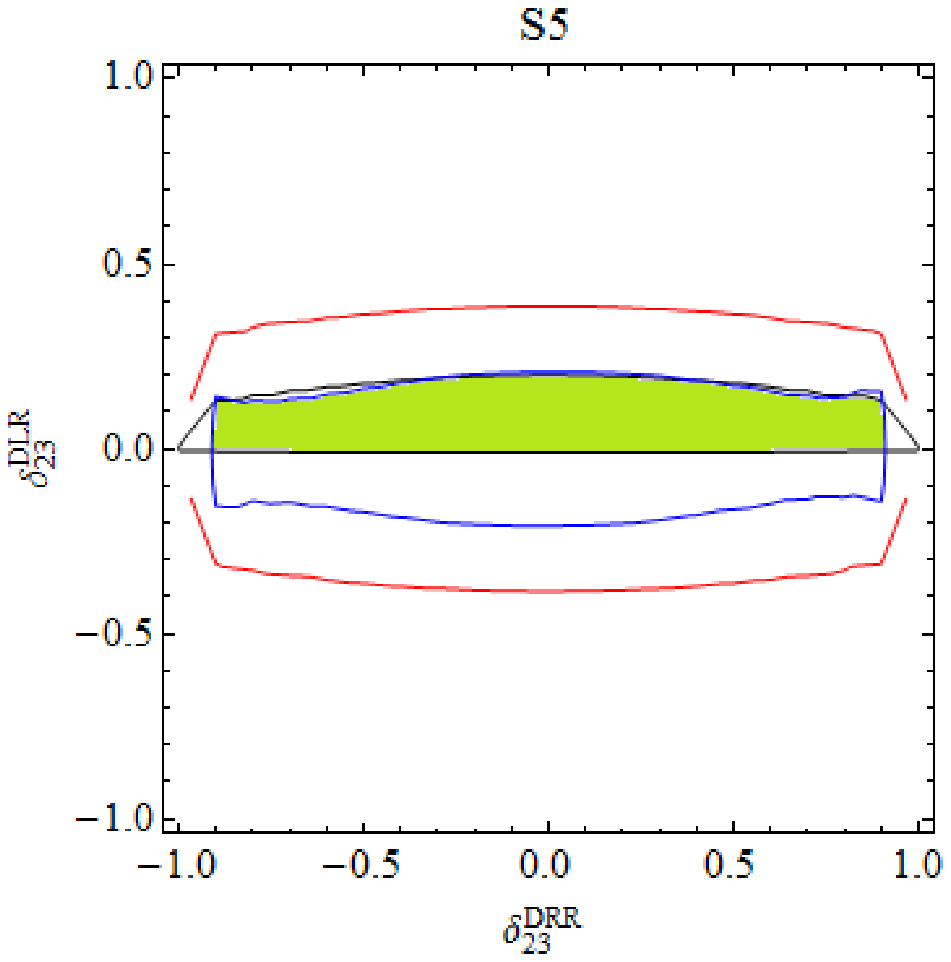 ,scale=0.69,angle=0,clip=}
\psfig{file=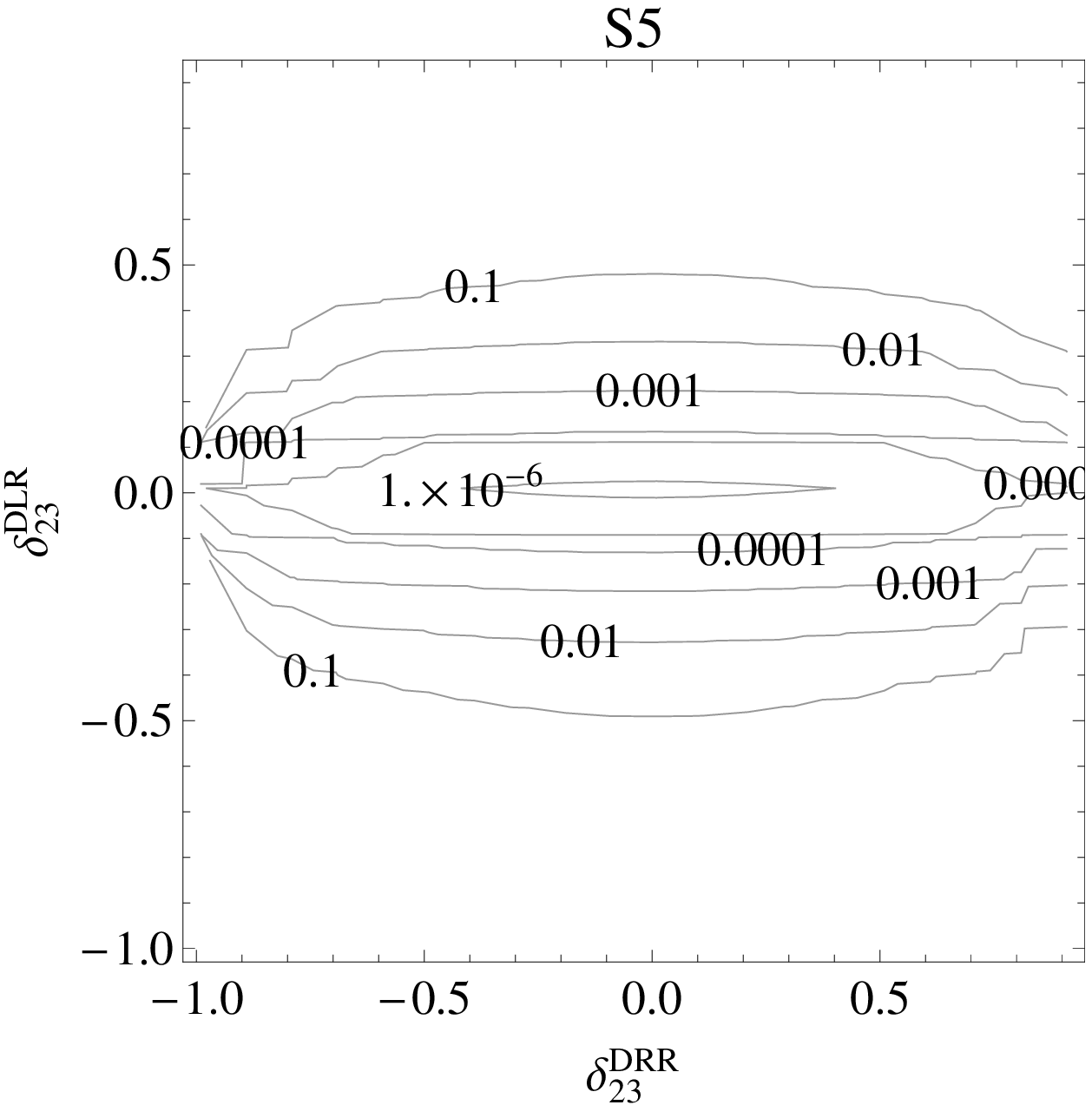 ,scale=0.52,angle=0,clip=}\\
\vspace{0.2cm}
\psfig{file=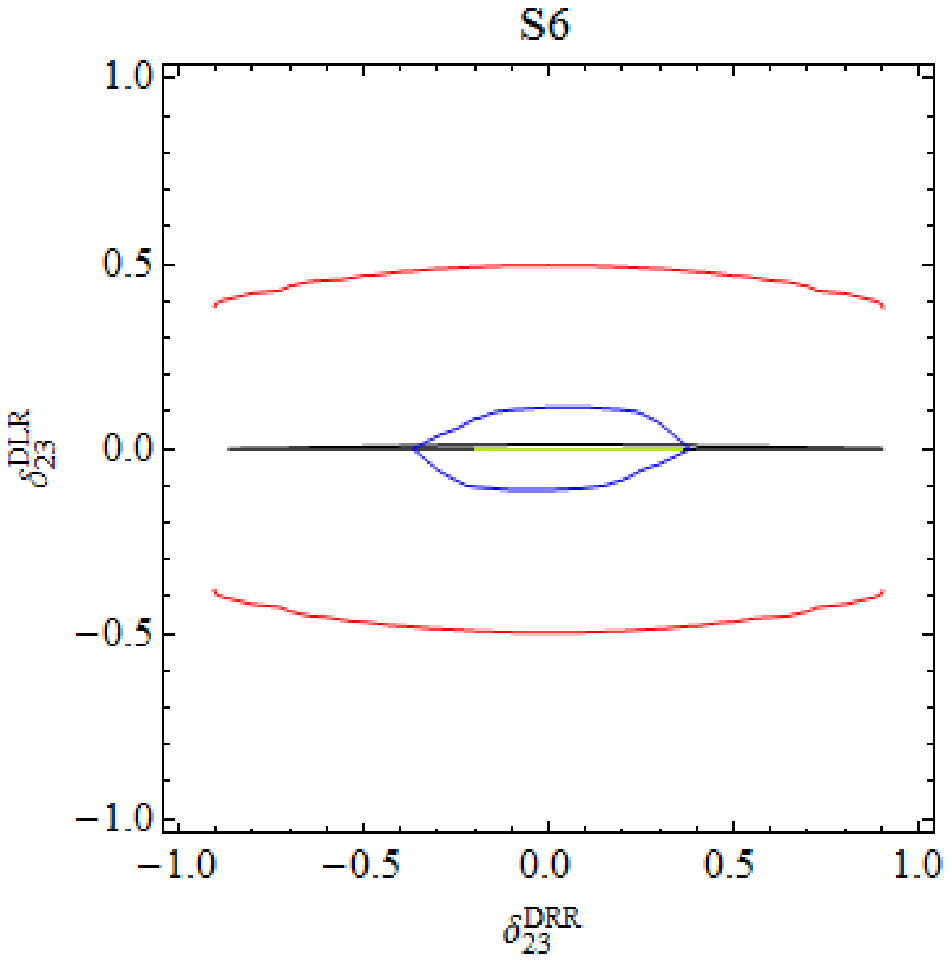 ,scale=0.69,angle=0,clip=}
\psfig{file=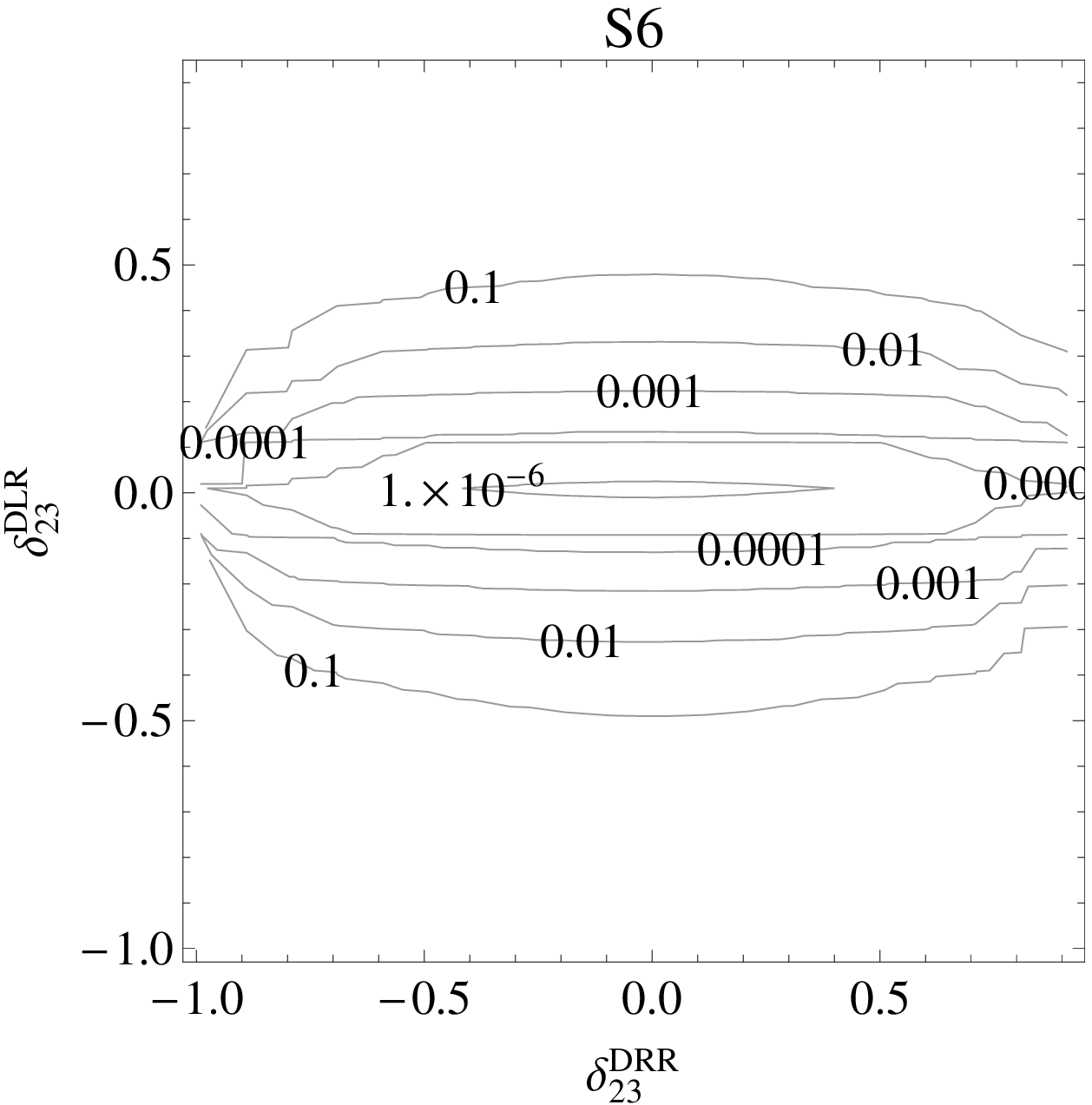 ,scale=0.52,angle=0,clip=}
\vspace{-2em}
\end{center}
\caption[Contours of EWPO, BPO and \brhbs\ in ($\del{DRR}{23}$ , $\del{DLR}{23}$) plane]
{Left: Contours of \bsg\ (Black), \bmm\ (Green), \dmbs\ (Blue) and
$\MW$ (Red) in ($\del{DRR}{23}$ , $\del{DLR}{23}$) plane for points S4-S6. 
The shaded area shows the range of values allowed by all
constraints. Right: corresponding contours for \brhbs.}   
\label{Fig:S4S6DLRDRR23}
\end{figure}

In \reffis{Fig:S1S3QLLDLR23}-\ref{Fig:S4S6DLRDRR23}, in the left columns
we show the $3\,\si$ contours (with experimental and theory
uncertainties added linearly) of 
\bsg\ (Black), \bmm\ (Green), \dmbs\ (Blue) and $\MW$ (Red). 
For non-visible contours the whole plane is allowed by that constraint. 
The right columns show, for the same parameters, the results for \brhbs. 
In \reffis{Fig:S1S3QLLDLR23} and \ref{Fig:S4S6QLLDLR23} we present the
results for the plane ($\del{QLL}{23}$,$\del{DLR}{23}$) for S1\ldots
S3 and for S4\ldots S6, respectively. Similarly,
in \reffis{Fig:S1S3DLRDRR23} and \ref{Fig:S4S6DLRDRR23} we show the
($\del{DRR}{23}$, $\del{DLR}{23}$) plane.
The shaded area in the left columns indicates the area that is allowed
by all experimental constraints. 
In the ($\del{QLL}{23}$, $\del{DLR}{23}$) planes one can see that the
large values for $\del{QLL}{23}$ are not allowed by $\MW$, on the
other hand, \bsg\ mostly restricts the value of
$\del{DLR}{23}$. 
The largest values for \brhbs\ in each plane in the arrea allowed by the
BPO and the EWPO are summarized in the upper part of \refta{tab:brhbs-2d}. 
One can see that 
in most cases we find $\br(\hbs) \sim \order{10^{-5}}$, 
which would render the observation difficult at current and future
colliders. However, in the ($\del{QLL}{23},\del{DLR}{23}$) plane in the
scenarios S4 and S5 maximum values of \order{3 \times 10^{-4}} can be
observed, which could be detectable at future ILC measurements.
In the ($\del{DRR}{23}$, $\del{DLR}{23}$) plane for these two scenarios
even values of \order{10^{-3}} are reached, which would make a
measurement of the flavor violating Higgs decay relatively easy at the
ILC.
\renewcommand{\arraystretch}{1.2}
\begin{table}[htb!]
\begin{center}
\resizebox{11.0cm}{!} {
\begin{tabular}{|c|c|c|c|} \hline
 Plane & MSSM point & Maximum possible value & Figure \\ \hline
($\del{QLL}{23},\del{DLR}{23}$) & \begin{tabular}{c}  S1 \\ S2 \\ S3 \\ S4 \\ S5 \\ S6 \end{tabular} &  
\begin{tabular}{c} 
$1.38 \times 10^{-5}$ \\ $1.39 \times 10^{-5}$ \\ $1.43 \times 10^{-5}$
  \\ $3.34 \times 10^{-4}$ \\ $2.74 \times 10^{-4}$ \\ $1.36 \times 10^{-8}$ 
\end{tabular} &
\begin{tabular}{c} 
\reffi{Fig:S1S3QLLDLR23} \\ \reffi{Fig:S1S3QLLDLR23} \\ \reffi{Fig:S1S3QLLDLR23} \\
\reffi{Fig:S4S6QLLDLR23} \\ \reffi{Fig:S4S6QLLDLR23} \\ \reffi{Fig:S4S6QLLDLR23}
\end{tabular}
\\ \hline
($\del{DRR}{23},\del{DLR}{23}$)  & \begin{tabular}{c}  S1 \\ S2 \\ S3 \\ S4 \\ S5 \\ S6 \end{tabular}    
& \begin{tabular}{c} 
$4.41 \times 10^{-6}$ \\ $3.32 \times 10^{-6}$ \\ $3.07 \times 10^{-5}$
    \\ $1.66 \times 10^{-3}$ \\ $1.97 \times 10^{-3}$ \\ $6.03 \times 10^{-8}$ 
\end{tabular} & 
\begin{tabular}{c} 
\reffi{Fig:S1S3DLRDRR23} \\ \reffi{Fig:S1S3DLRDRR23} \\ \reffi{Fig:S1S3DLRDRR23} \\
\reffi{Fig:S4S6DLRDRR23} \\ \reffi{Fig:S4S6DLRDRR23} \\ \reffi{Fig:S4S6DLRDRR23}
\end{tabular}
\\ \hline\hline
\begin{tabular}{c}($\del{QLL}{23},\del{DLR}{23}$) \\ with $\del{DRR}{23}= 0.5$ \end{tabular} & \begin{tabular}{c}  S1 \\ S2 \\ S3 \\ S4 \\ S5 \\ S6 \end{tabular}    & 
\begin{tabular}{c} 
$7.49 \times 10^{-6}$ \\ $7.33 \times 10^{-6}$ \\$3.50 \times 10^{-6}$
  \\ Excluded \\ Excluded \\ Excluded \end{tabular} &
\begin{tabular}{c} 
\reffi{Fig:S1S3QLLDLRDRR23} \\ \reffi{Fig:S1S3QLLDLRDRR23} \\ \reffi{Fig:S1S3QLLDLRDRR23} \\
\reffi{Fig:S4S6QLLDLRDRR23} \\ \reffi{Fig:S4S6QLLDLRDRR23} \\ \reffi{Fig:S4S6QLLDLRDRR23}
\end{tabular}  \\ \hline
\end{tabular}}  
\end{center}
\caption[Maximum $\brhbs$ for two and three $\deFABij \neq 0$ case]{Maximum possible value for $\brhbs$ for two and three $\deFABij \neq 0$ case for the selected S1-S6 MSSM points defined in
\refta{tab:spectra}. 
}
\label{tab:brhbs-2d}
\end{table}
\renewcommand{\arraystretch}{1.55}

\begin{figure}[ht!]
\begin{center}
\psfig{file=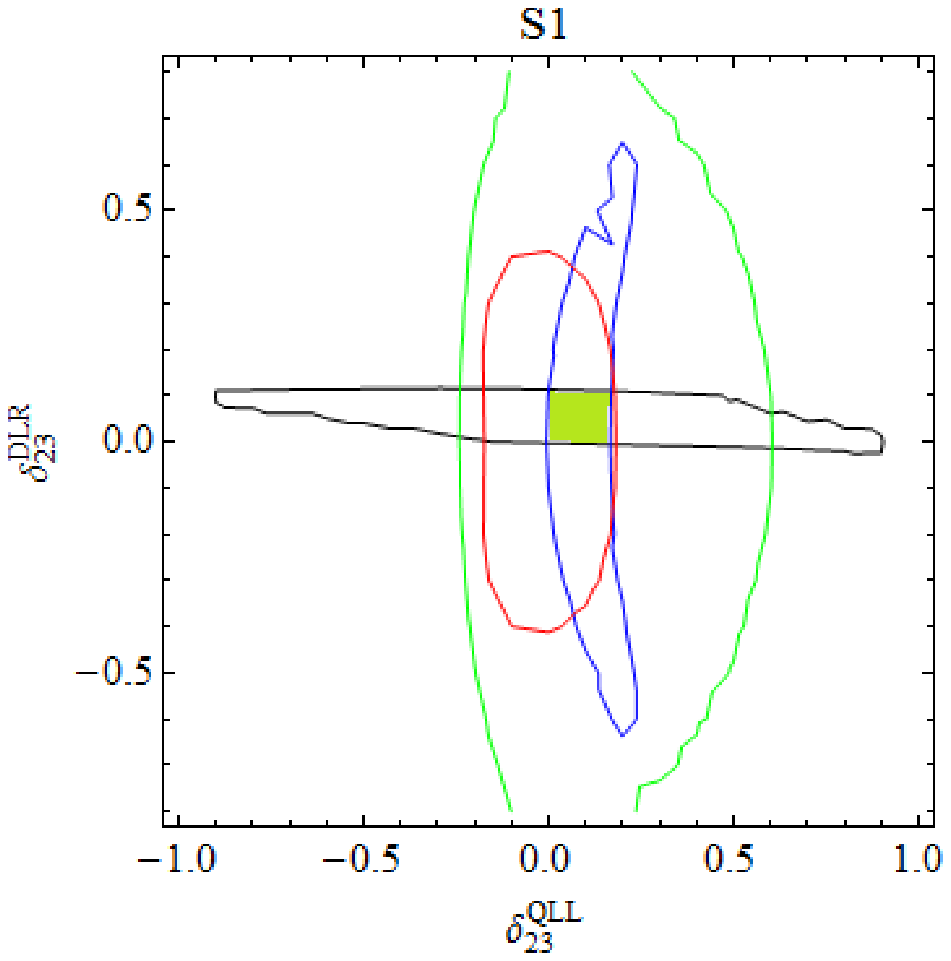  ,scale=0.69,angle=0,clip=}
\psfig{file=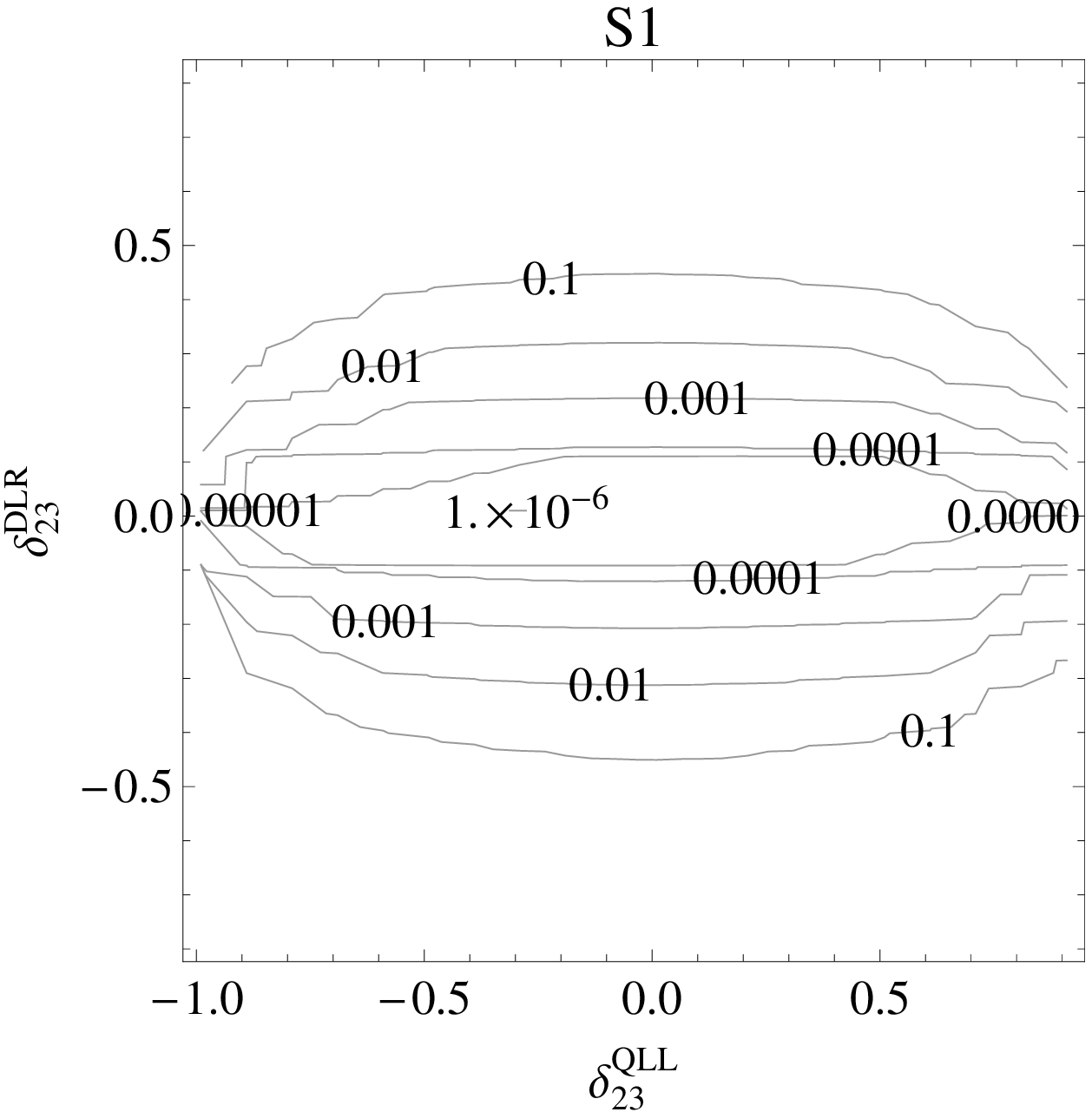  ,scale=0.52,angle=0,clip=}\\
\vspace{0.2cm}
\psfig{file=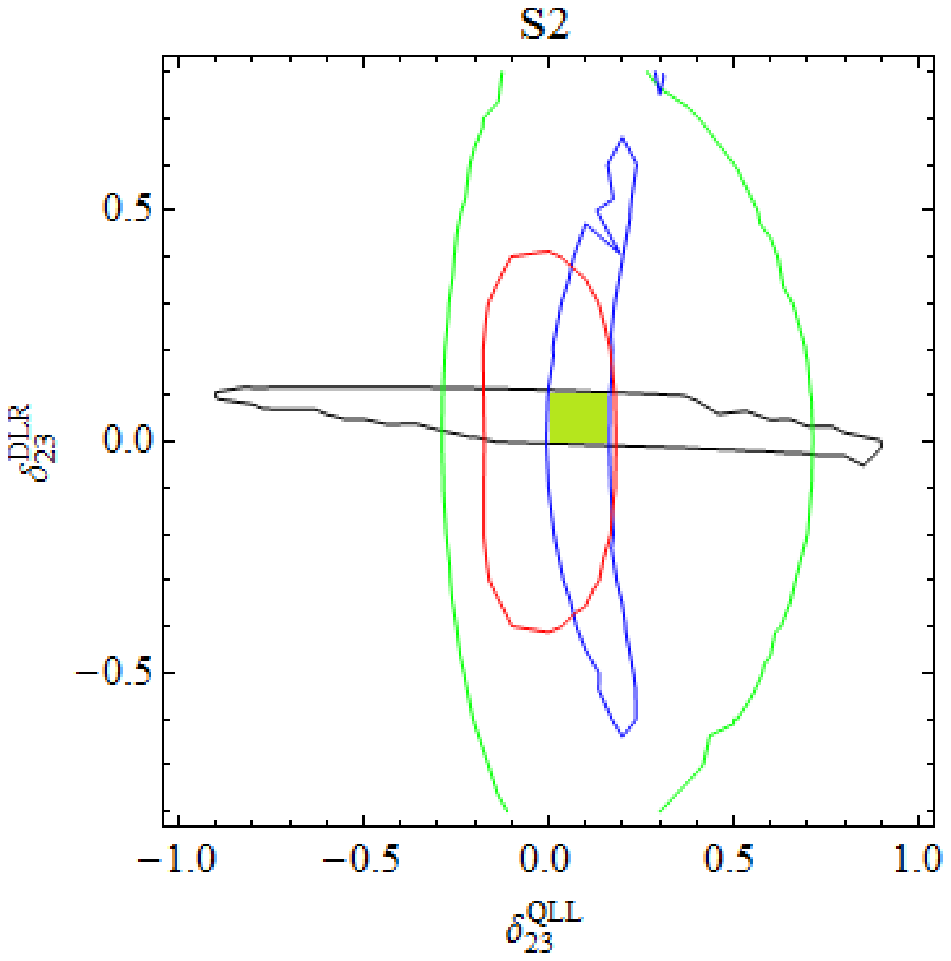 ,scale=0.69,angle=0,clip=}
\psfig{file=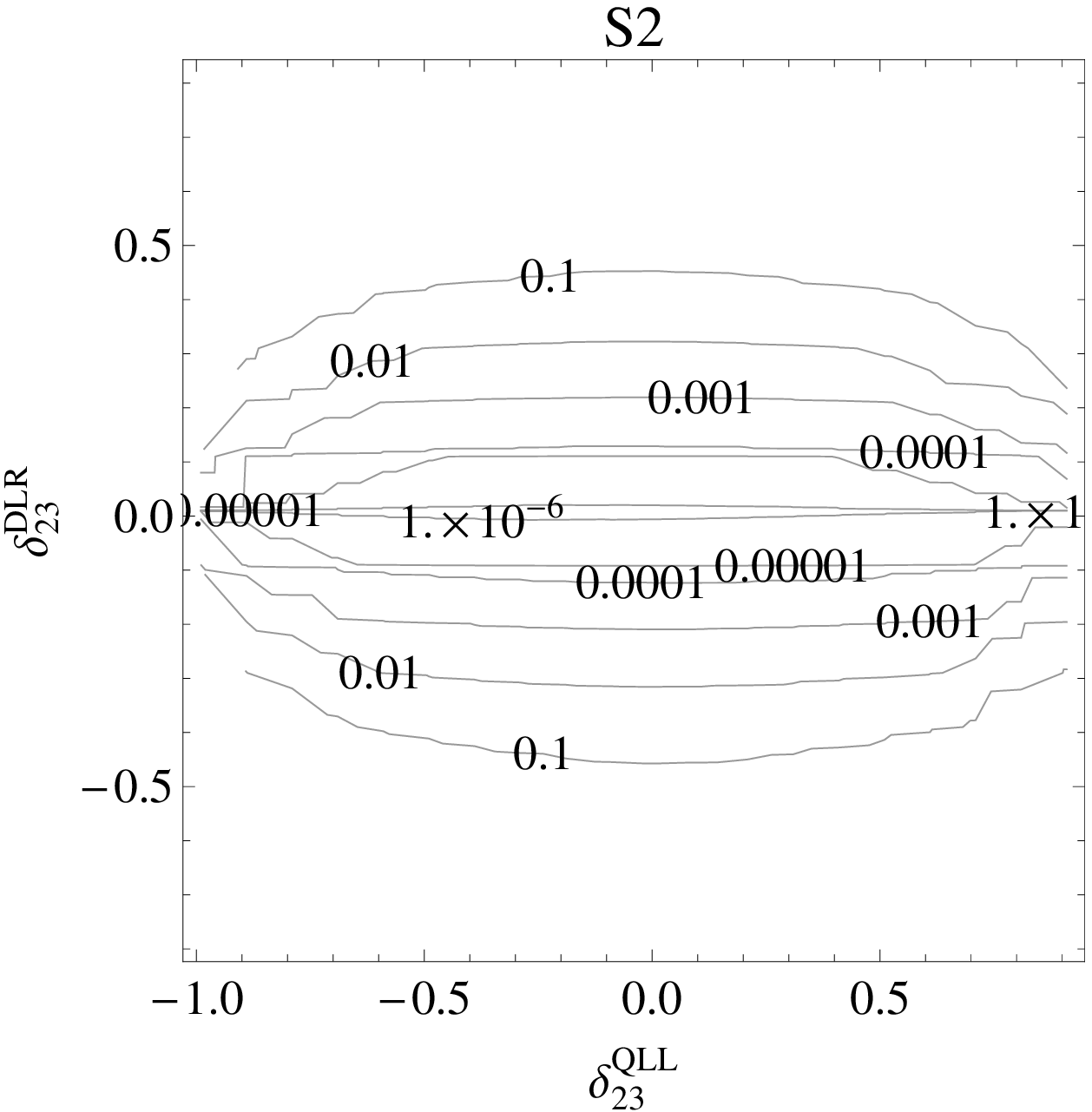 ,scale=0.52,angle=0,clip=}\\
\vspace{0.2cm}
\psfig{file=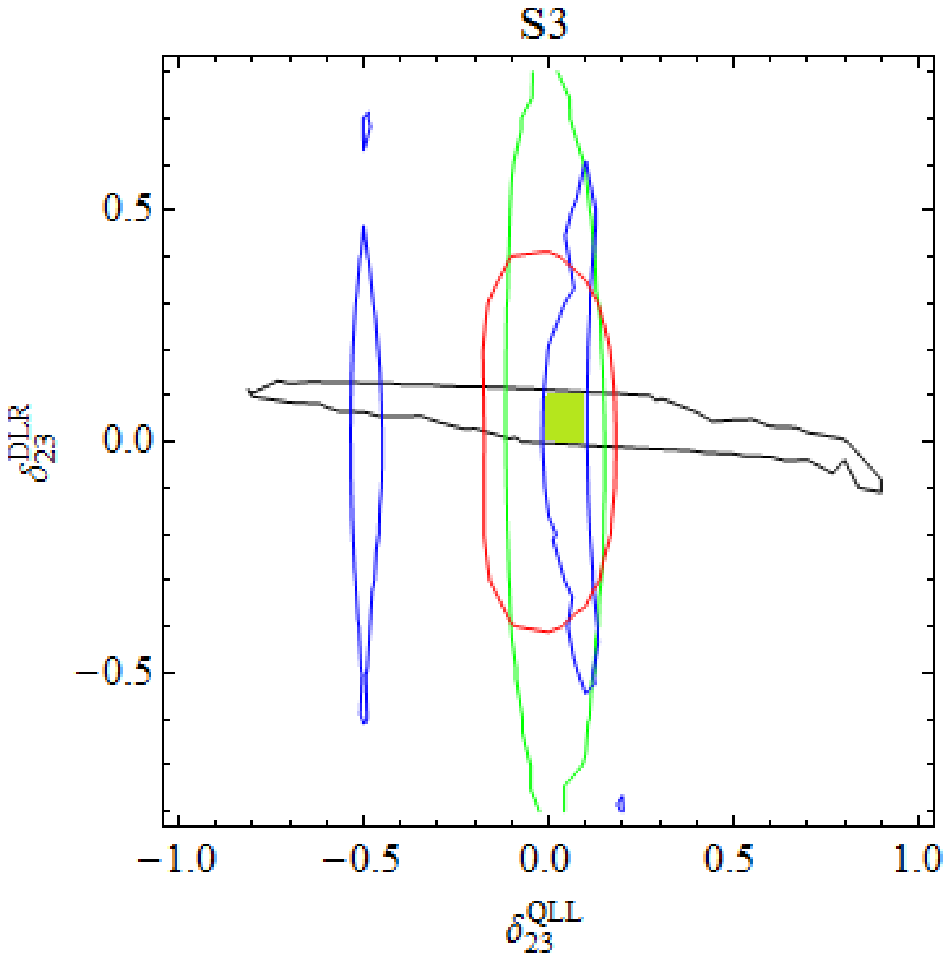 ,scale=0.69,angle=0,clip=}
\psfig{file=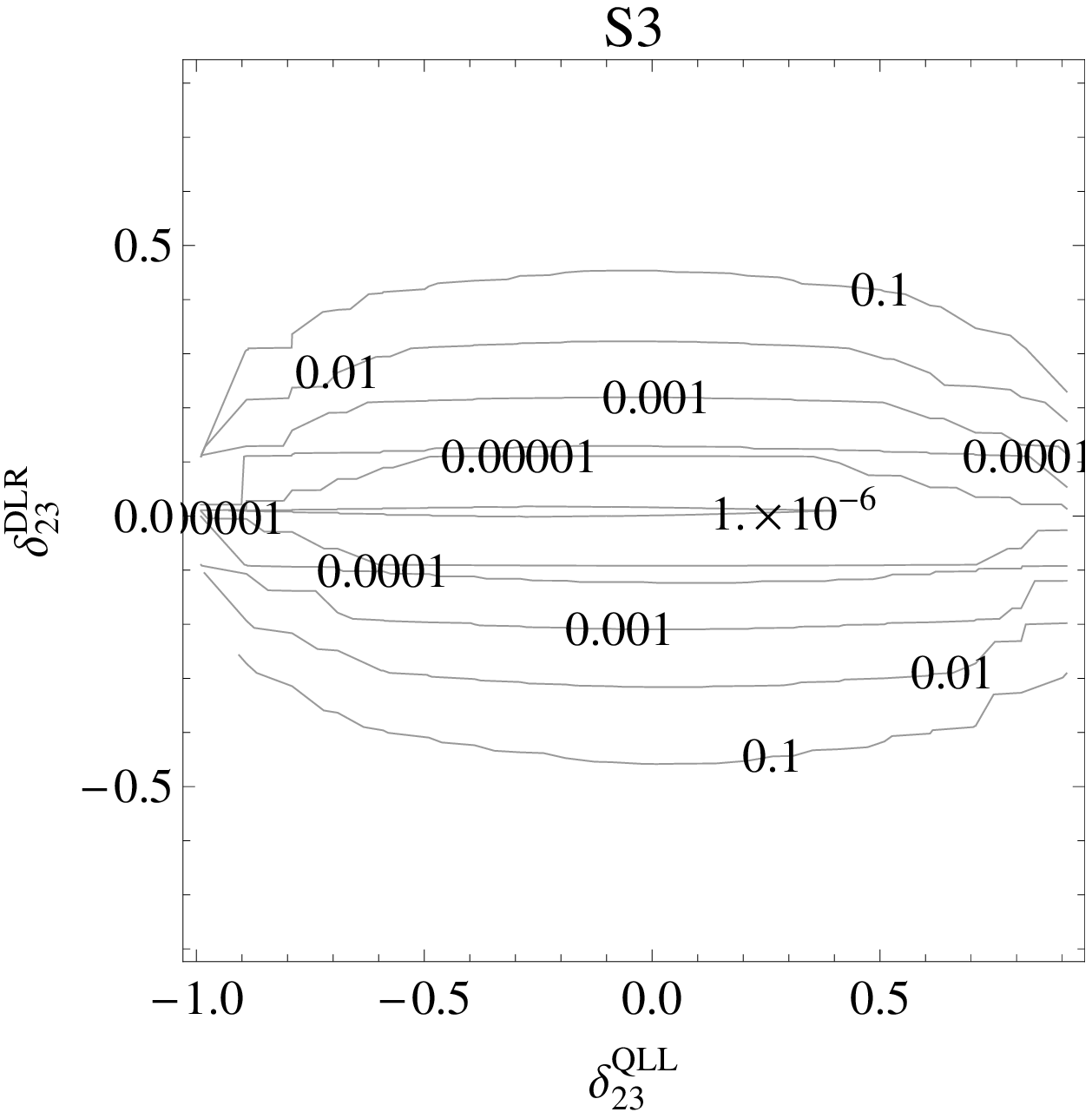 ,scale=0.52,angle=0,clip=}
\vspace{-2em}
\end{center}
\caption[Contours of EWPO, BPO and \brhbs\ in ($\del{QLL}{23}$ , $\del{DLR}{23}$) plane]
{Left: Contours of \bsg\ (Black), \bmm\ (Green), \dmbs\ (Blue) and
$\MW$ (Red) in the ($\del{QLL}{23}$ , $\del{DLR}{23}$) plane with 
$\del{DRR}{23} = 0.5$ for points S1-S3. The shaded area shows the range
    of values allowed by all 
constraints. Right: corresponding contours
for \brhbs.}    
\label{Fig:S1S3QLLDLRDRR23}
\end{figure}

\begin{figure}[ht!]
\begin{center}
\psfig{file=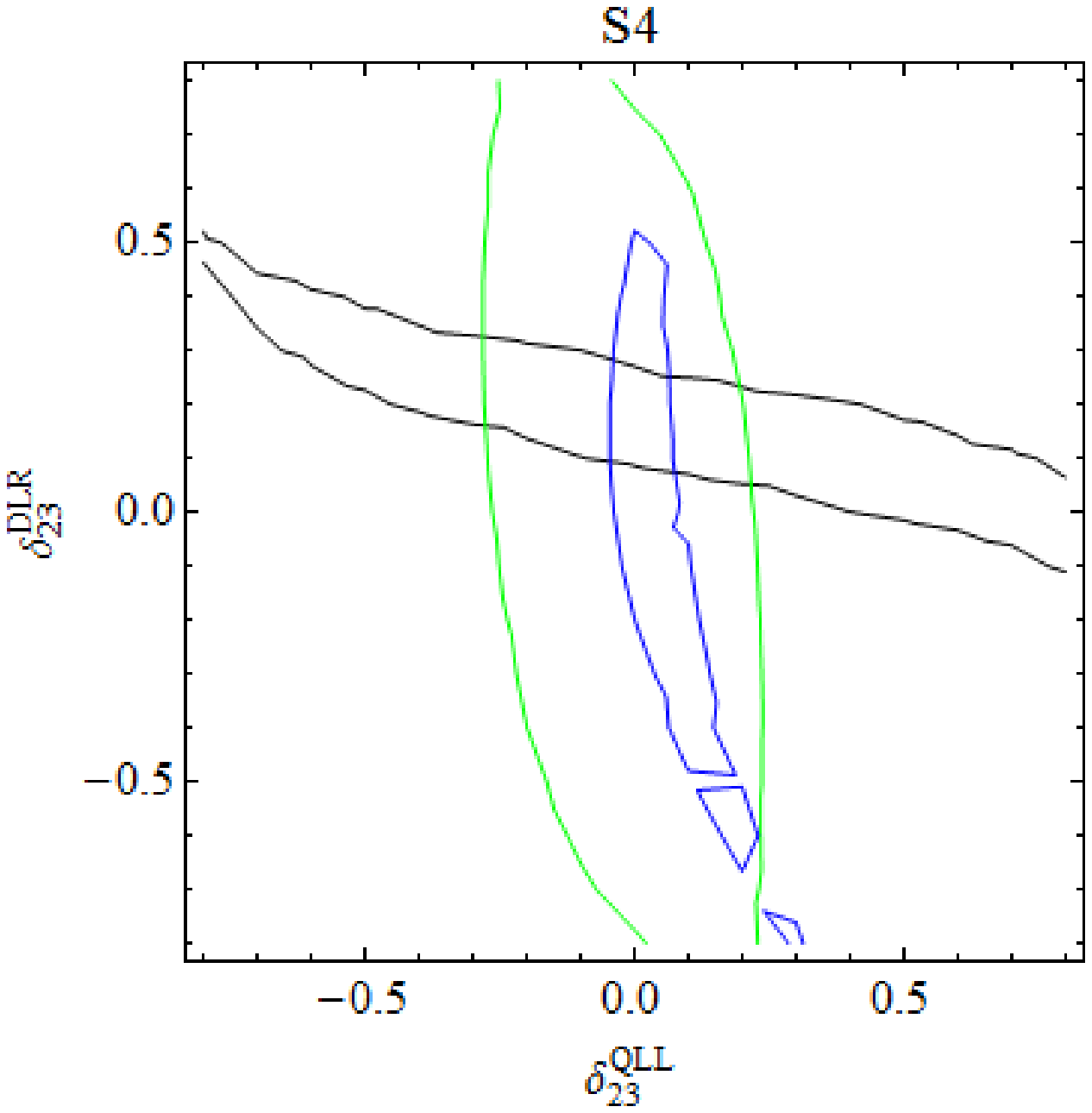  ,scale=0.52,angle=0,clip=}
\psfig{file=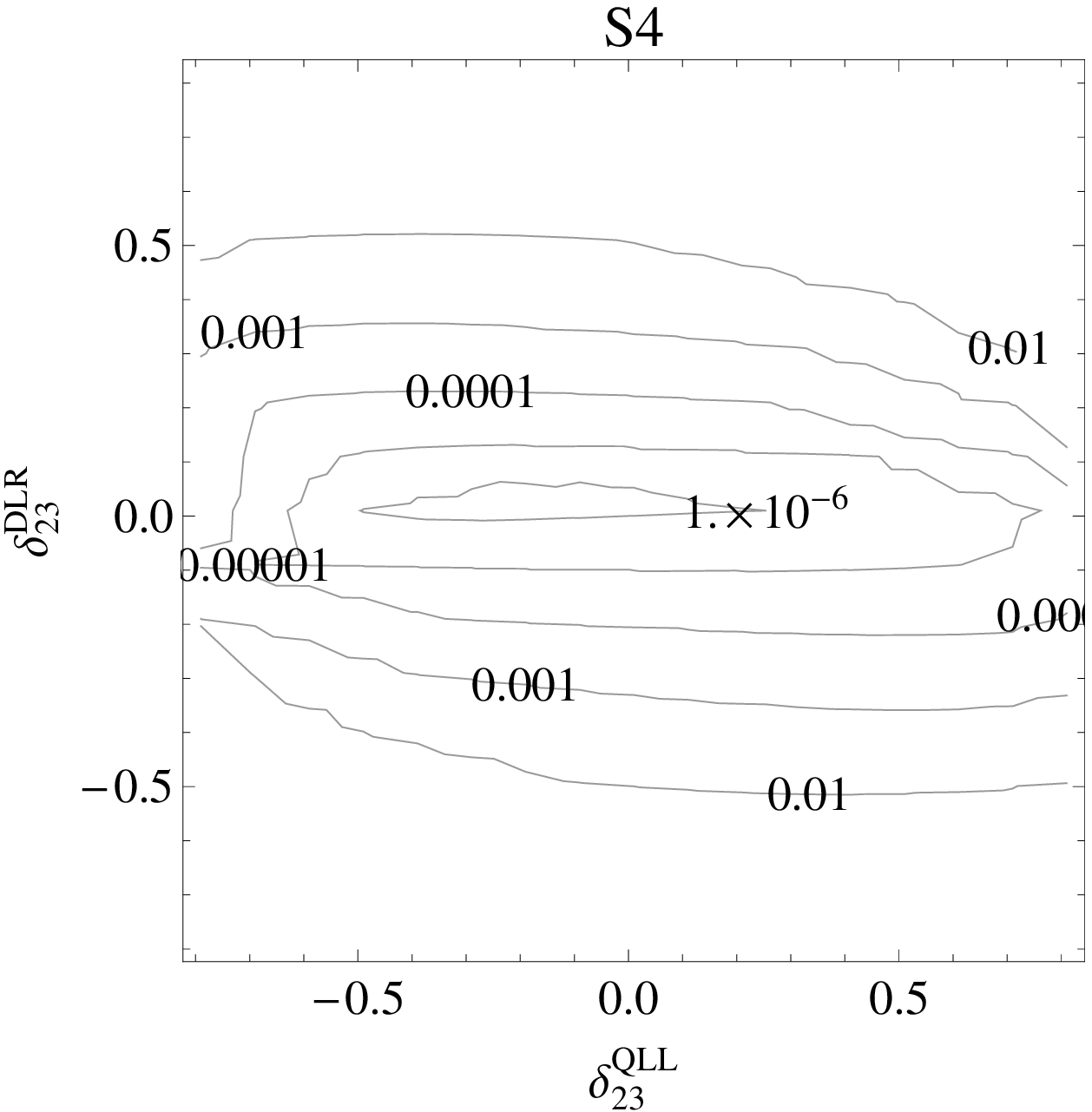  ,scale=0.52,angle=0,clip=}\\
\vspace{0.2cm}
\psfig{file=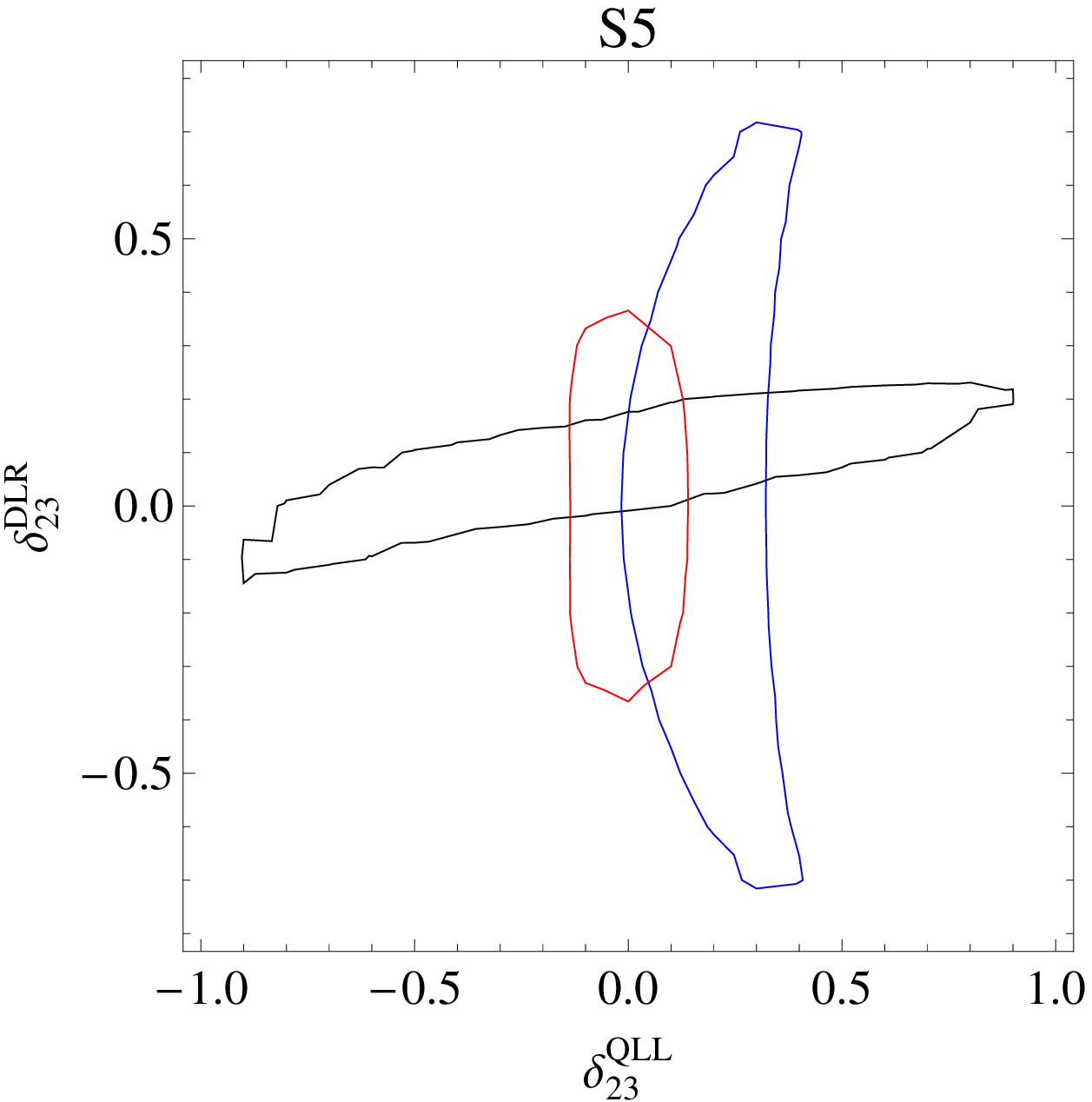 ,scale=0.52,angle=0,clip=}
\psfig{file=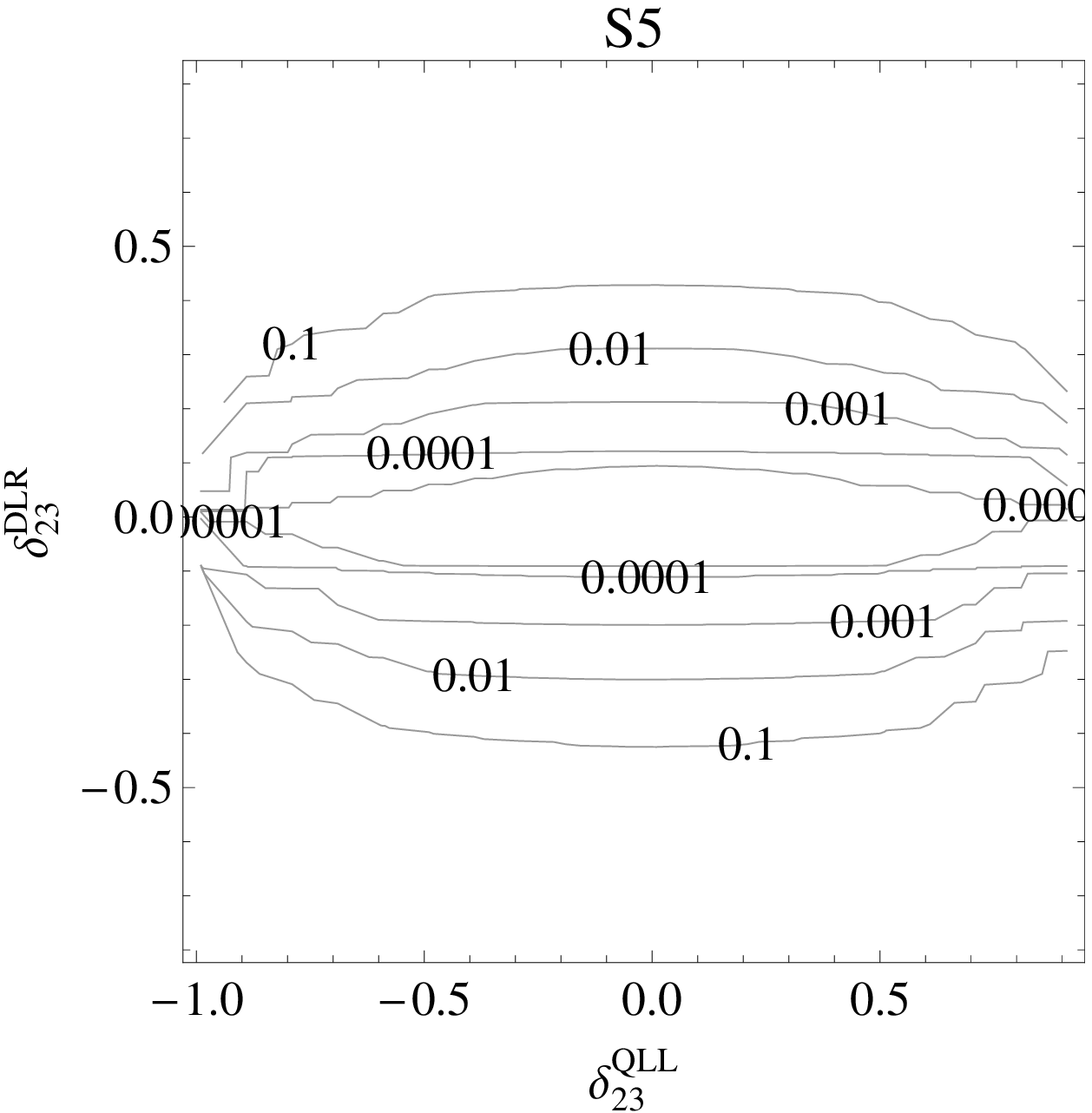 ,scale=0.52,angle=0,clip=}\\
\vspace{0.2cm}
\psfig{file=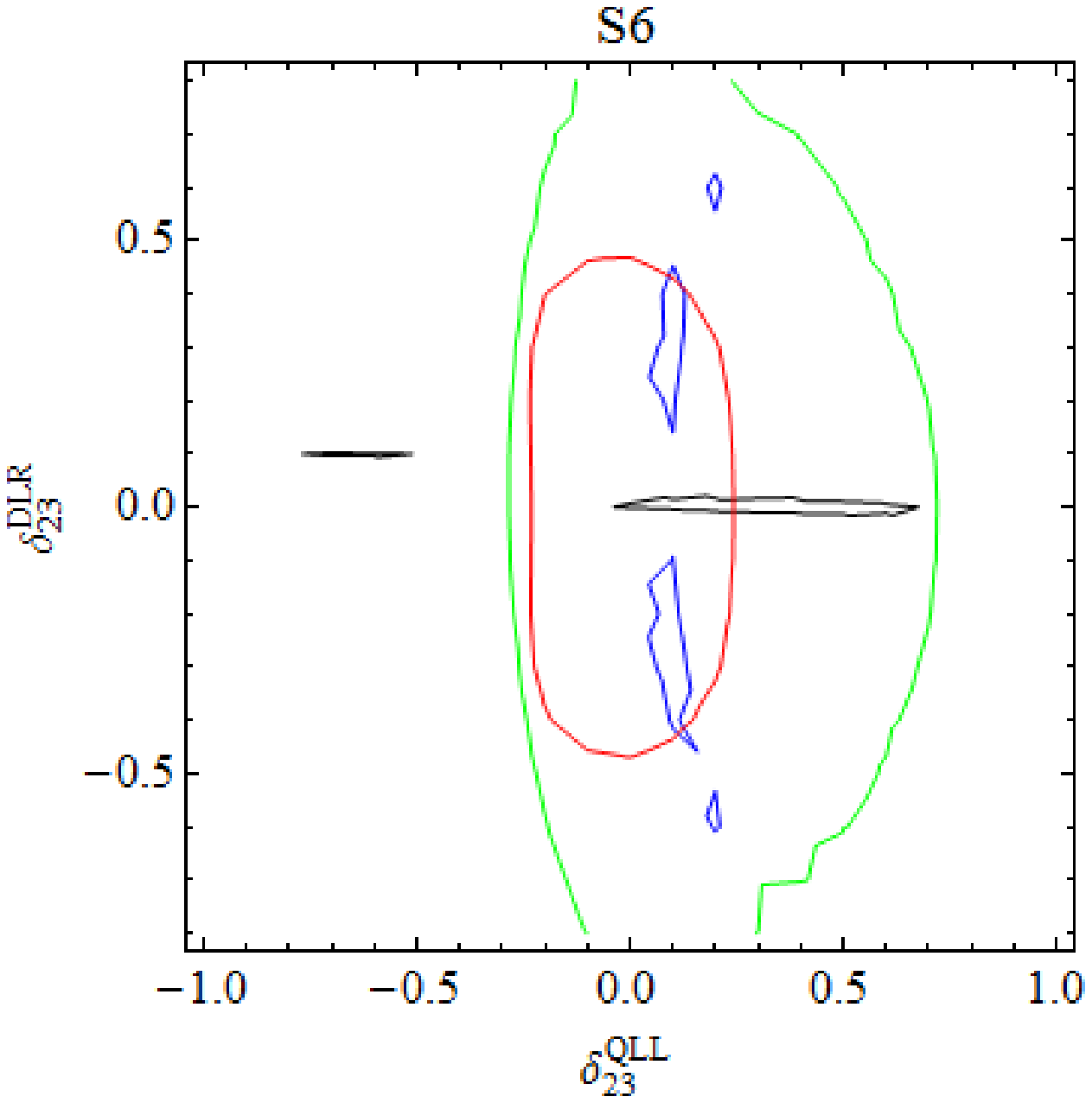 ,scale=0.52,angle=0,clip=}
\psfig{file=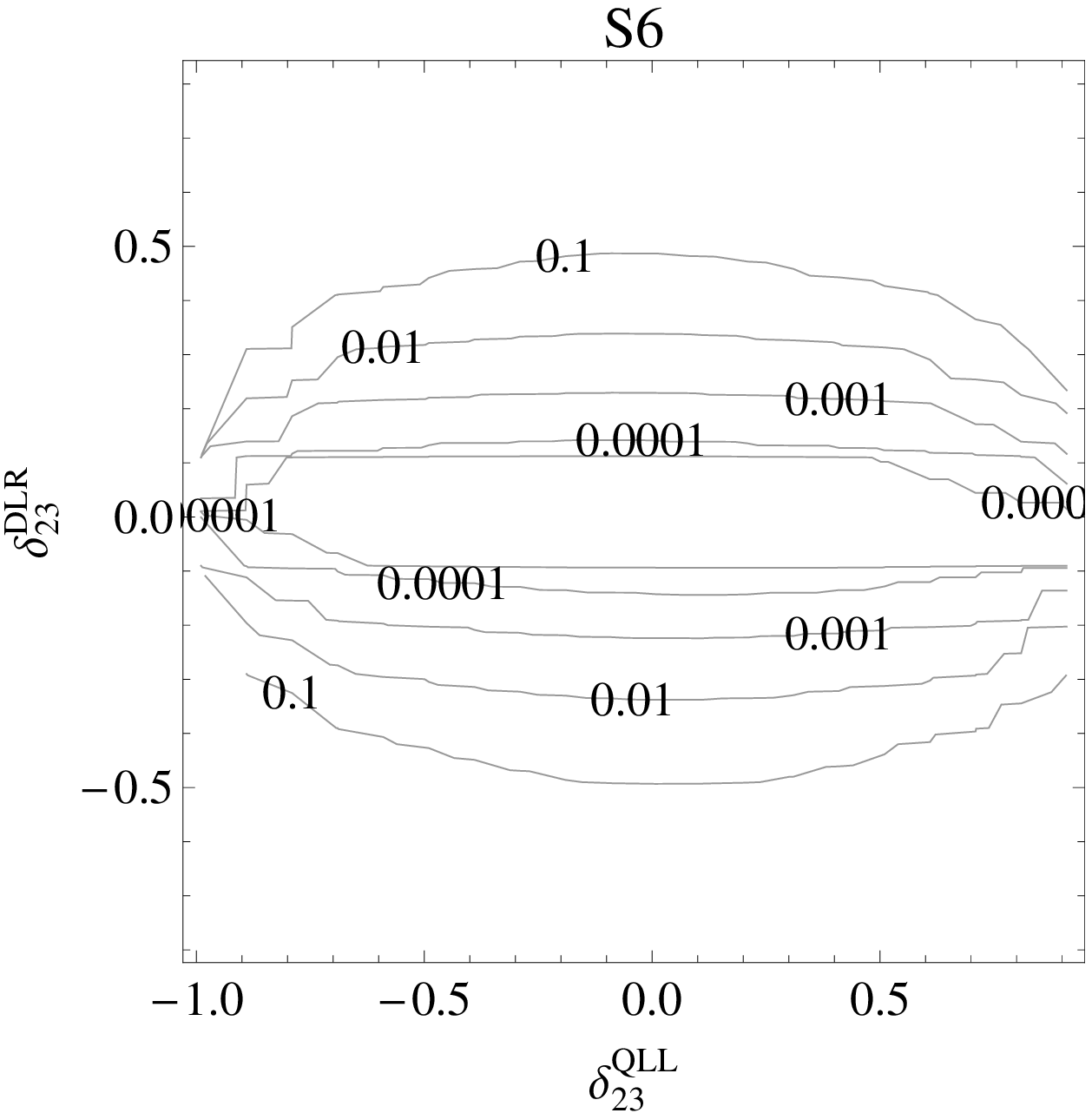 ,scale=0.52,angle=0,clip=}
\vspace{-2em}
\end{center}
\caption[Contours of EWPO, BPO and \brhbs\ in ($\del{QLL}{23}$ , $\del{DLR}{23}$) plane]
{Left: Contours of \bsg\ (Black), \bmm\ (Green), \dmbs\ (Blue) and
$\MW$ (Red) in the ($\del{QLL}{23}$ , $\del{DLR}{23}$) plane with 
$\del{DRR}{23} = 0.5$ for points S4-S6. The shaded area shows the range
    of values allowed by all 
constraints. Right: corresponding contours
for \brhbs.}    
\label{Fig:S4S6QLLDLRDRR23}
\end{figure}

\medskip
As a last step in model independent analysis, we consider the case of
three $\deFABij \neq 0 $ at a time. For this purpose we 
scan the parameters in the ($\del{QLL}{23}$, $\del{DLR}{23}$)
plane and set $\del{DRR}{23} = 0.5$. For reasons of practicability
we choose {\em one} intermediate value for $\del{DRR}{23}$; a very small
value will have no additional effect, and a very large value of
$\del{DRR}{23}$  leads to large excluded areas in the 
($\del{QLL}{23}$, $\del{DLR}{23}$) plane.
We show our results in \reffis{Fig:S1S3QLLDLRDRR23} and 
\ref{Fig:S4S6QLLDLRDRR23} in the scenarios S1-S3 and S4-S6,
respectively. Colors and shadings are chosen as in the previous
analysis.
Here it should be noted that in S4 the whole plane is excluded by
$\MW$, and in S5 by \bmm\ (both contours are not visible). In S6 no
overlap between the four constraints is found, and again this scenario
is excluded. We have checked that also a smaller value of
$\del{DRR}{23} = 0.2$ does not qualitatively change the picture for
S4, S5 and S6.
The highest values that can be reached for \brhbs\ in the three
remaining scenarios in the experimentally
allowed regions are shown in the lower part of \refta{tab:brhbs-2d}. 
One can see only very small valus or \order{5 \times 10^{-6}} are
found, i.e.\ choosing $\del{DRR}{23} \neq 0$ did not lead to
observable values of \brhbs.

To summarize, in our model independent analysis, allowing for more than
one $\deFABij \neq 0$ we find 
that the additional freedom resulted in somewhat larger values of
\brhbs\ as compared to the case of only one non-zero $\deFABij$.
In particular in the two scenarios S4 and S5 values of 
$\brhbs \sim 10^{-3} - 10^{-4}$ can be reached, allowing the
detection of the flavor violating Higgs decay at the ILC. The other
scenarios always yield values that are presumably too low for
current and future colliders.

\chapter{Lepton Flavor Mixing Effects in the Model Independent Approach}

In this chapter we analyse the lepton flavor mixing in MI approach. We use the same set of input parameter (\refta{tab:spectra}) that was used in the previous chater. As a first step, we have calculated the sensitivity of EWPO like the 
$W$-boson mass or the effective weak leptonic mixing angle to the $\deFABij$'s in slepton sector
entering in the $Z$ and $W$ boson self energies at one-loop
level through the $\rho$~parameter. Besides EWPO we also explore the effects of LFV on the MSSM Higgs 
sector. We evaluate the effects of LFV on the predictions of the masses of the light and heavy 
$\cp$-even Higgs bosons, $\Mh$ and $\MH$, as well as on the charged 
Higgs-boson mass $\MHp$. Here we do not calculate predictions for cLFV decays 
in the MI approach as they were already explored in \cite{Arana-Catania:2013nha} for the same set of input parameters that we are using. They calculated the constraints on slepton $\deFABij$'s  from cLFV decays (mentioned in the following section). We have also calculated the predictions for LFVHD which will be presented in the last section. The results presented in this chapter were published in \cite{drhoLFV}.

\section{Constraints on \boldmath{$\deFABij$} from cLFV decays}
\label{sec:limits}

We need to set the range of values for the explored
$\delta^{FAB}_{ij}$'s. We use the constraints (shown in \refta{boundsSpoints}) 
as taken from \citere{Arana-Catania:2013nha}, calculated from the following LFV processes. 
\begin{itemize}
\item[1.-] Radiative LFV decays: $\mu \to e \gamma$, $\tau \to e \gamma$ and
$\tau \to \mu \gamma$. These are sensitive to the $\deFABij$'s via the
  $(l_il_j\gamma)_{\rm 1-loop}$ vertices with a real photon. 
\item[2.-] Leptonic LFV decays: $\mu \to 3 e$, $\tau \to 3 e$ and $\tau
  \to 3 \mu$. These are sensitive to the $\deFABij$'s via the
  $(l_il_j\gamma)_{\rm 1-loop}$ vertices with a virtual photon, via
  the $(l_il_jZ)_{\rm 1-loop}$ vertices with a virtual $Z$, and via the
  $(l_il_jh)_{\rm 1-loop}$, $(l_il_jH)_{\rm 1-loop}$ and
  $(l_il_jA)_{\rm 1-loop}$ vertices with  virtual Higgs bosons. 
\item[3.-] Semileptonic LFV tau decays: $\tau \to \mu \eta$ and $\tau
  \to e \eta$. These are sensitive to the $\deFABij$'s via $(\tau \mu
  A)_{\rm 1-loop}$ and $(\tau e A)_{\rm 1-loop}$ vertices,
  respectively, with a virtual $A$, and via  
$(\tau \mu Z)_{\rm 1-loop}$ and $(\tau e Z)_{\rm 1-loop}$ vertices,
  respectively with a virtual $Z$. 
\item[4.-] Conversion of $\mu$ into $e$ in heavy nuclei: These are
  sensitive to the $\deFABij$'s via the $(\mu e\gamma)_{\rm 1-loop}$
  vertex with a virtual photon, via the $(\mu e Z)_{\rm 1-loop}$
  vertex with a virtual $Z$, and via the $(\mu e h)_{\rm 1-loop}$ and
  $(\mu e H)_{\rm 1-loop}$ vertices with a virtual $h$ and $H$
  Higgs boson, respectively.  
\end{itemize}  
\renewcommand{\arraystretch}{1.1}
\begin{table}[htb!]
\begin{center}
\resizebox{15.0cm}{!} {
\begin{tabular}{|c|c|c|c|c|c|c|}
\hline
 &  S1 &  S2 &  S3 &  S4 &  S5 & S6  
 \\ \hline
 & & & & & & \\
$|\delta^{LLL}_{12}|_{\rm max}$ & $10 \times 10^{-5}$ & $7.5\times 10^{-5}$ &   $5 \times 10^{-5}$& $6 \times 10^{-5}$ & $42\times 10^{-5}$  &  $8\times 10^{-5}$  \\ 
& & & & & & \\
\hline
& & & & & & \\
$|\delta^{ELR}_{12}|_{\rm max}$ & $2\times 10^{-6}$ & $3\times 10^{-6}$ &
$4\times 10^{-6}$  & $3\times 10^{-6}$ & $2\times 10^{-6}$  & $1.2\times 10^{-5}$   \\ 
& & & & & & \\
\hline
& & & & & & \\
$|\delta^{ERR}_{12}|_{\rm max}$ & $1.5 \times 10^{-3}$& $1.2 \times 10^{-3}$ & 
$1.1 \times 10^{-3}$ & $1 \times 10^{-3}$ & $2 \times 10^{-3}$ & $5.2 \times 10^{-3}$   \\ 
& & & & & & \\
\hline
& & & & & & \\
$|\delta^{LLL}_{13}|_{\rm max} $ &  $5 \times 10^{-2}$ & $5 \times 10^{-2}$ & 
$3 \times 10^{-2}$ &  $3 \times 10^{-2}$& $23 \times 10^{-2}$ & $5 \times 10^{-2}$   \\ 
& & & & & & \\
\hline
& & & & & & \\
 $|\delta^{ELR}_{13}|_{\rm max}$& $2\times 10^{-2}$  & $3\times 10^{-2}$ & $4\times 10^{-2}$ & $2.5\times 10^{-2}$ & $2\times 10^{-2}$ & $11\times 10^{-2}$   \\ 
& & & & & & \\
 \hline
 & & & & & & \\
$|\delta^{ERR}_{13}|_{\rm max}$ & $5.4\times 10^{-1}$  & $5\times 10^{-1}$ & 
 $4.8\times 10^{-1}$ &$5.3\times 10^{-1}$  & $7.7\times 10^{-1}$ & $7.7\times 10^{-1}$ 
  \\ 
 & & & & & & \\ 
  \hline
 & & & & & & \\ 
$|\delta^{LLL}_{23}|_{\rm max}$ & $6\times 10^{-2}$  & $6\times 10^{-2}$ & 
 $4\times 10^{-2}$& $4\times 10^{-2}$ & $27\times 10^{-2}$ & $6\times 10^{-2}$ 
  \\ 
 & & & & & & \\ 
  \hline
 & & & & & & \\ 
$|\delta^{ELR}_{23}|_{\rm max}$ & $2\times 10^{-2}$   & $3\times 10^{-2}$ & 
$4\times 10^{-2}$ & $3\times 10^{-2}$ & $2\times 10^{-2}$ & $12\times 10^{-2}$ 
  \\ 
 & & & & & & \\ 
  \hline
 & & & & & & \\ 
$|\delta^{ERR}_{23}|_{\rm max}$ & $5.7\times 10^{-1}$  & $5.2\times 10^{-1}$ & 
 $5\times 10^{-1}$& $5.6\times 10^{-1}$ & $8.3\times 10^{-1}$ & $8\times 10^{-1}$ 
  \\ 
 & & & & & & \\  
  \hline
\end{tabular}}  
\end{center}
\caption[Constraints on $|\delta^{FAB}_{ij}|$ from LFV decays.]{ Present upper bounds on the slepton mixing parameters $|\delta^{FAB}_{ij}|$ for the selected S1-S6 MSSM points defined in \refta{tab:spectra}. The bounds for $|\delta^{ERL}_{ij}|$ are similar 
to those of $|\delta^{ELR}_{ij}|$.}
\label{boundsSpoints}
\end{table}

\section{Numerical results} 
\label{sec:results}

We have implemented the full one-loop results for the $W$~and
$Z$~boson and the Higgs boson self-energies 
in $\fh$, including all LFV mixing terms (see \refse{sec:feynhiggs} for details). The analytical results are
lenghty and are not shown here. 
They can, however, be found in the latest version of our code, 
\fh\,{\tt 2.10.2}.
For the numerical investigation we have analyzed all 12 slepton $\de^{FAB}_{ij}$'s
for the MSSM scenarios defined in \refta{tab:spectra}.
In order to get a good understanding of the LFV effects to $\De\rho$ and
consequently $\MW$ and $\sweff$ we define

\begin{align}
\De\rho^{\rm LFV} &= \De\rho-\De\rho^{\rm MSSM}, \\
\de\MW^{\rm LFV} &= \MW-\MW^{\rm MSSM}, \\
\de\sweff^{\rm LFV} &= \sweff-\sweff^{\rm MSSM},
\end{align}
where $\De\rho^{\rm MSSM}$, $\MW^{\rm MSSM}$ and $\sweff^{\rm MSSM}$ are the
values of the relevant observables with all $\deFABij = 0$ (and the
latter two evaluated with the help of \refeq{eq:precobs}). Furthermore we define
\begin{align}
\De \Mh^{\rm LFV} &= \Mh - \Mh^{\rm MSSM}, \\
\De \MH^{\rm LFV} &= \MH - \MH^{\rm MSSM}, \\
\De \MHp^{\rm LFV} &= \MHp - \MHp^{\rm MSSM},
\end{align}
where $\Mh^{\rm MSSM}$, $\MH^{\rm MSSM}$ and 
$\MHp^{\rm MSSM}$ corresponds to the 
Higgs masses with all $\deFABij = 0$.
The SM results for $\MW$ and $\sweff$ are $\MW=80.361 \gev$ and
$\sweff=0.23152$ as evaluated with \fh\ (using the approximation formulas
given in \citeres{MWSMapprox,sw2effSMapprox}).
The numerical values of $\De\rho$, $\MW$, $\sweff$, 
$\Mh$, $\MH$ and $\MHp$  in the MSSM with all $\deFABij = 0$ are
summarized in \refta{absolutevalues}.  

\begin{table}[h!]
\begin{tabular}{|c|c|c|c|c|c|c|}
\hline
 &  S1 &  S2 &  S3 &  S4 &  S5 & S6  
 \\ \hline
 & & & & & & \\
$\De\rho$ & $2.66 \times 10^{-5}$ & $1.72\times 10^{-5}$ &   $1.39 \times 10^{-5}$& $2.35 \times 10^{-4}$ & $2.36\times 10^{-5}$  &  $2.14\times 10^{-5}$  \\ 
& & & & & & \\
\hline
& & & & & & \\
$\MW$ & $ 80.362$ & $ 80.362 $ & $80.361$  & $80.375$ & $80.364$  & $80.363$   \\ 
& & & & & & \\
\hline
& & & & & & \\
$\sweff$ & $0.23151$& $0.23152$ & $0.23152$ & $0.23143$ & $0.23150$ & $0.23151$   \\ 
& & & & & & \\
\hline
& & & & & & \\
$ M_{h} $ &  $126.257$ & $126.629$ & $126.916$ &  $123.205$& $123.220$ & $124.695$   \\ 
& & & & & & \\
\hline
& & & & & & \\
 $M_{H} $ & $500.187$  & $999.580$ & $999.206$ & $1001.428$ & $1000.239$ & $1499.365$   \\ 
& & & & & & \\
 \hline
 & & & & & & \\
$M_{H^{\pm}} $ & $506.888$  & $1003.182$ & $1003.005$ &$1005.605$  & $1003.454$ & $1501.553$ 
  \\ 
 & & & & & & \\ 
 \hline
\end{tabular}
\caption[The values of  $\De\rho$, $\MW$, $\sweff$, 
$\Mh$, $\MH$ and $\MHp$ with all $\deABij = 0$.]{The values of  $\De\rho$, $\MW$, $\sweff$, 
$\Mh$, $\MH$ and $\MHp$ for the selected S1-S6 MSSM points defined
in \refta{tab:spectra} (i.e.\ with all $\deFABij = 0$). Mass values are
in~GeV.}
\label{absolutevalues}
\end{table}

\begin{figure}[ht!]
\begin{center}
\psfig{file=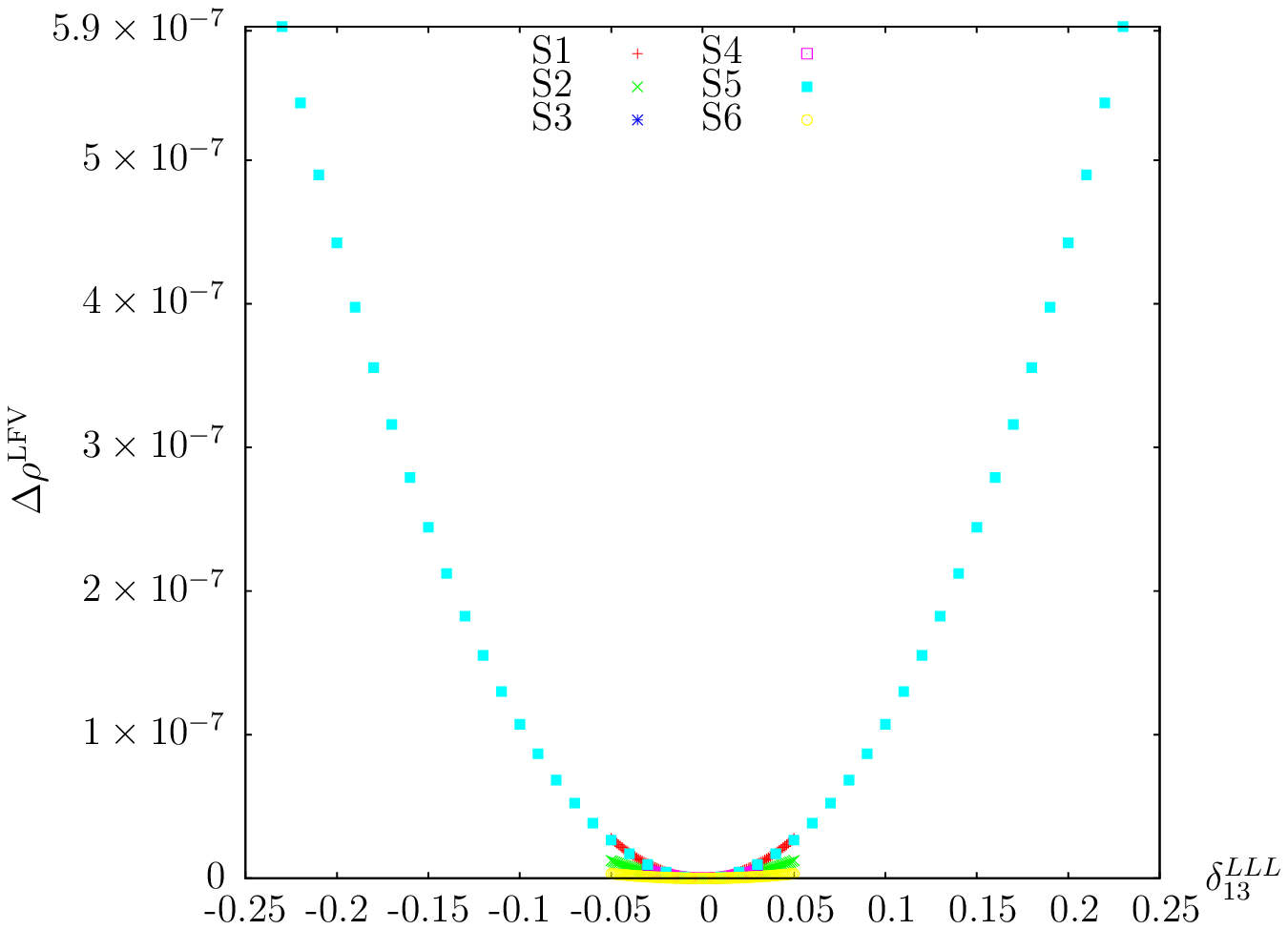  ,scale=0.53,angle=0,clip=}
\hspace{0.3cm}
\psfig{file=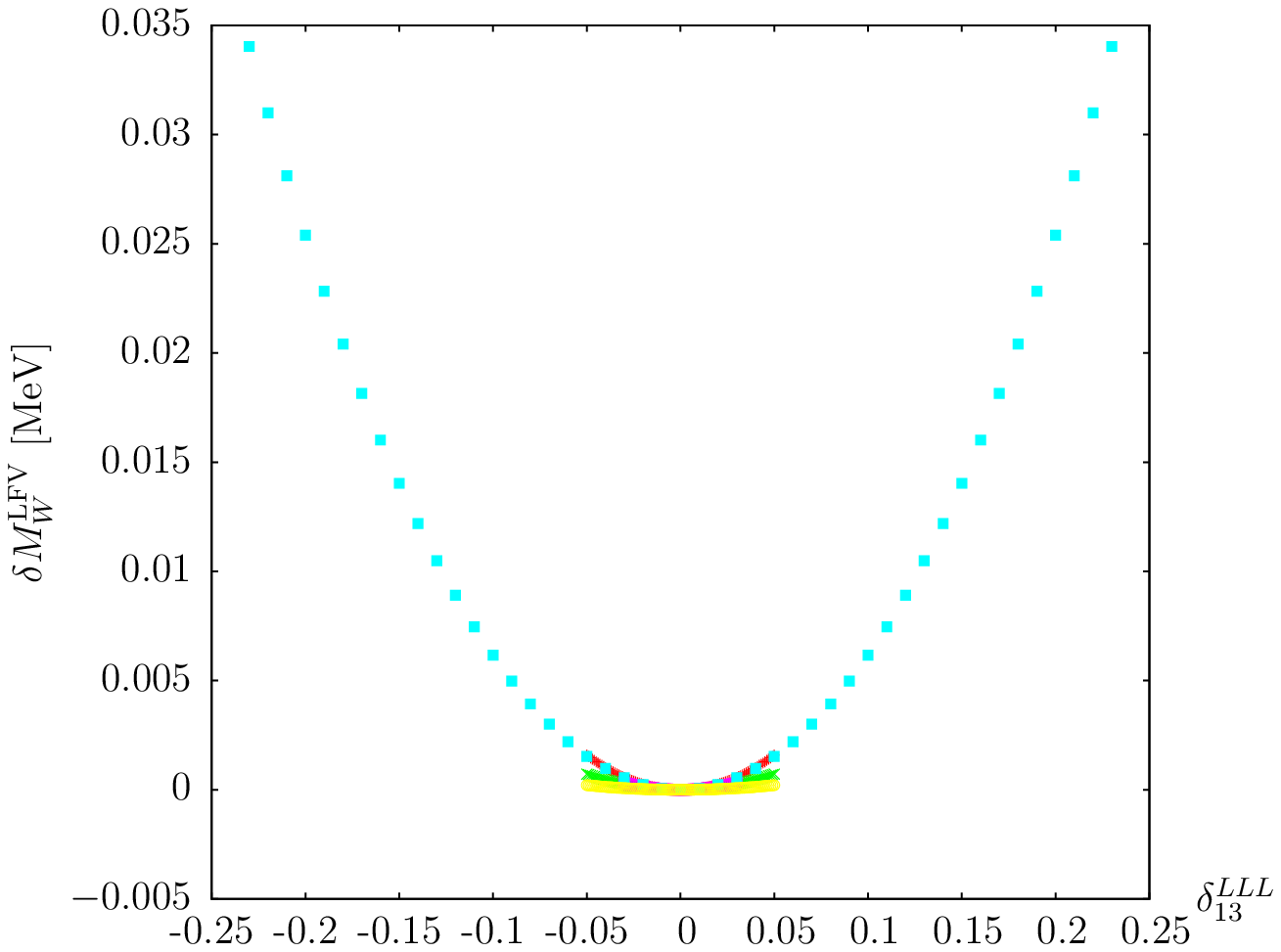  ,scale=0.53,angle=0,clip=}\\
\vspace{0.5cm}
\psfig{file=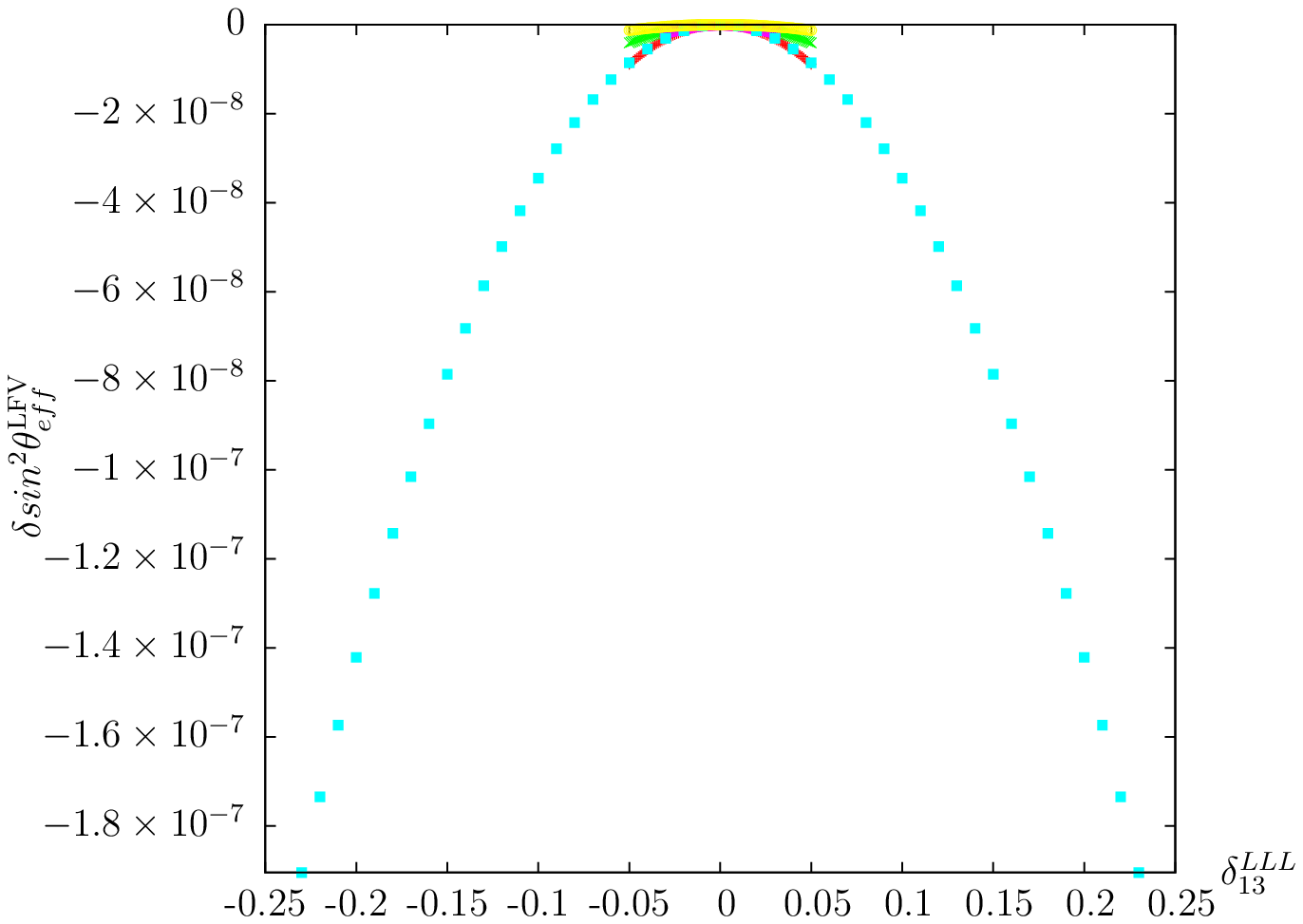 ,scale=0.53,angle=0,clip=}
\psfig{file=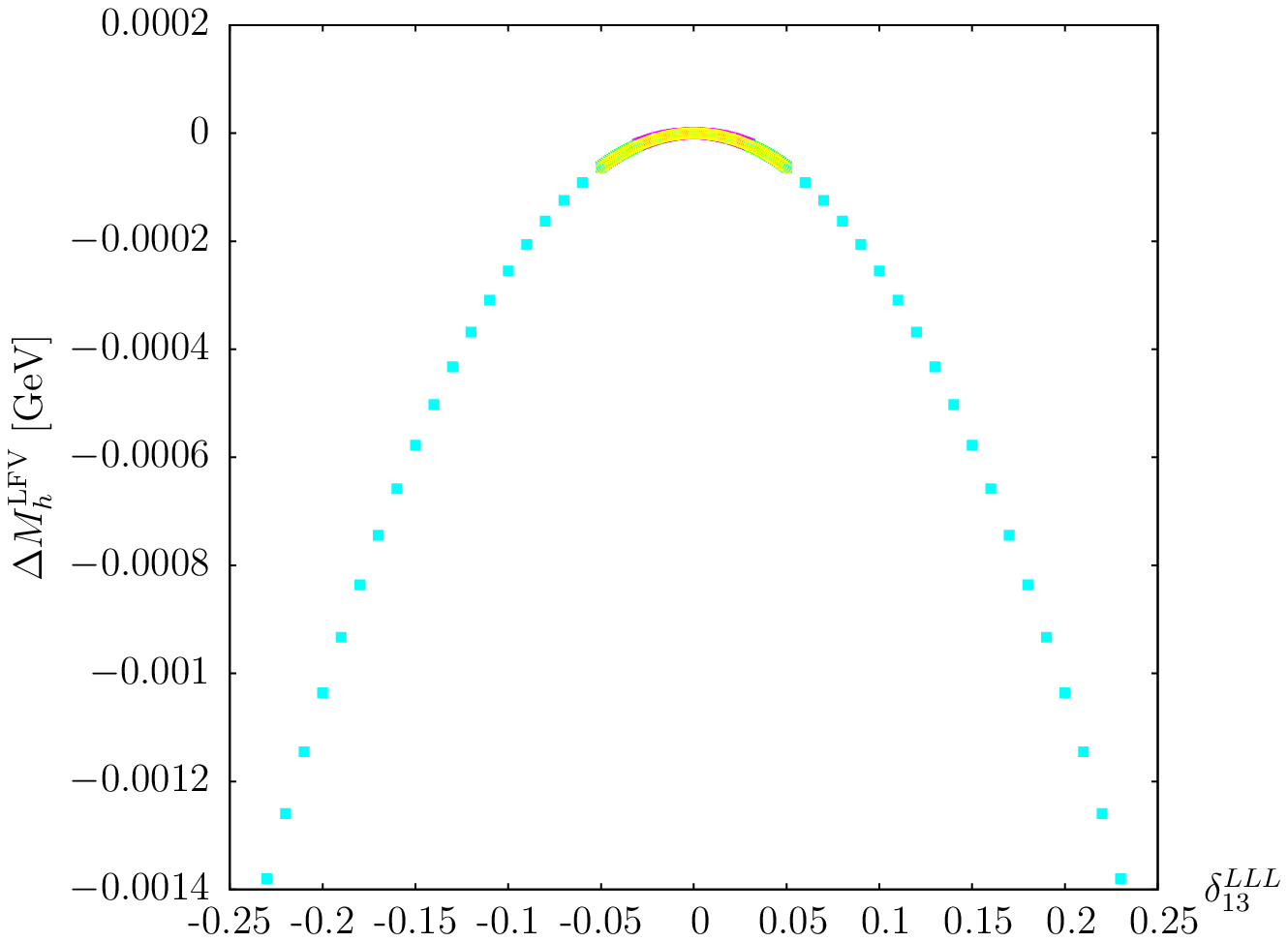   ,scale=0.53,angle=0,clip=}\\
\vspace{0.5cm}
\psfig{file=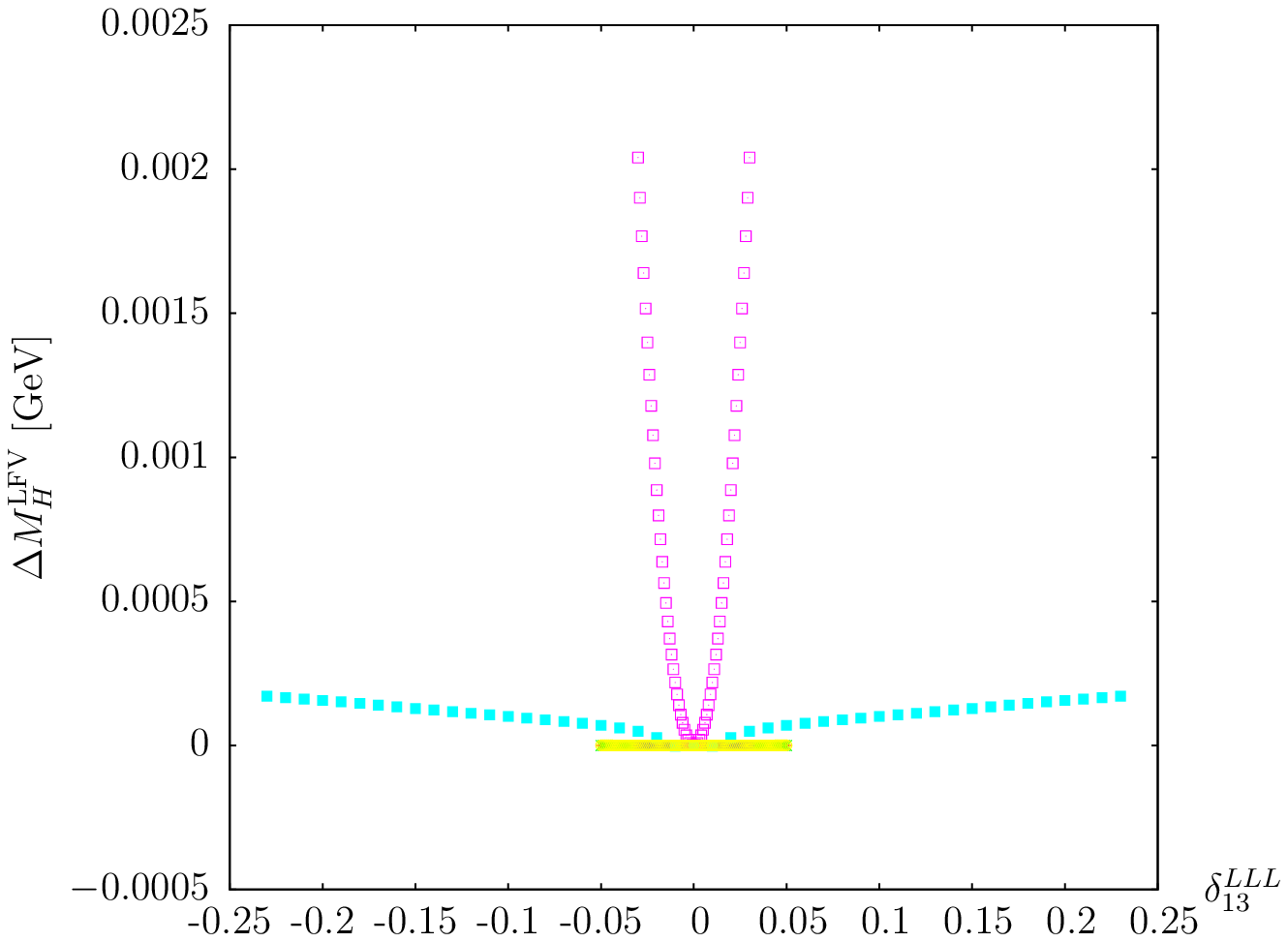  ,scale=0.53,angle=0,clip=}
\psfig{file=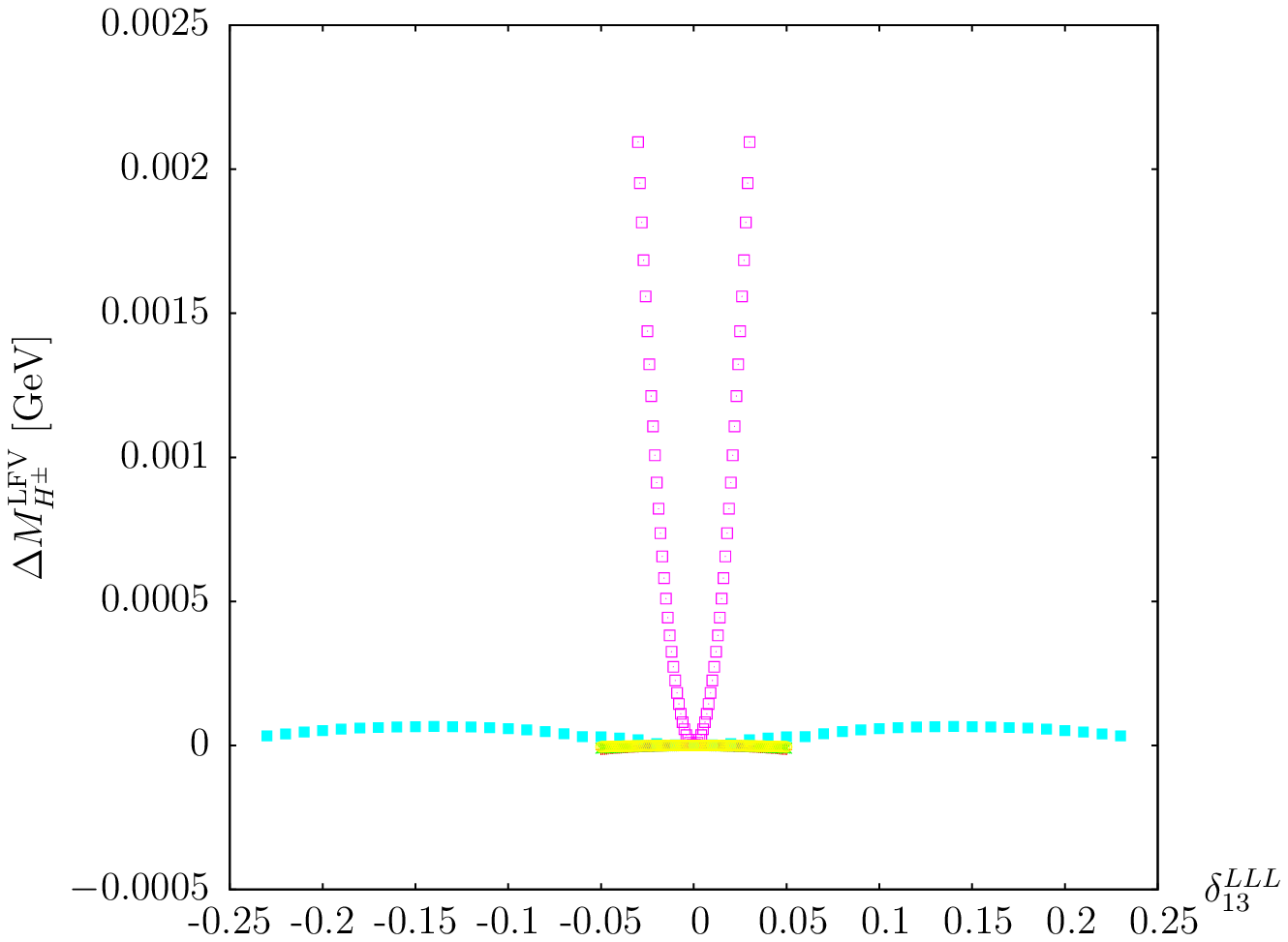  ,scale=0.53,angle=0,clip=}\\

\end{center}
\caption{EWPO and Higgs masses as a function of $\delta^{LLL}_{13}$.}  
\label{figdLL13}
\end{figure}

\begin{figure}[ht!]
\begin{center}
\psfig{file=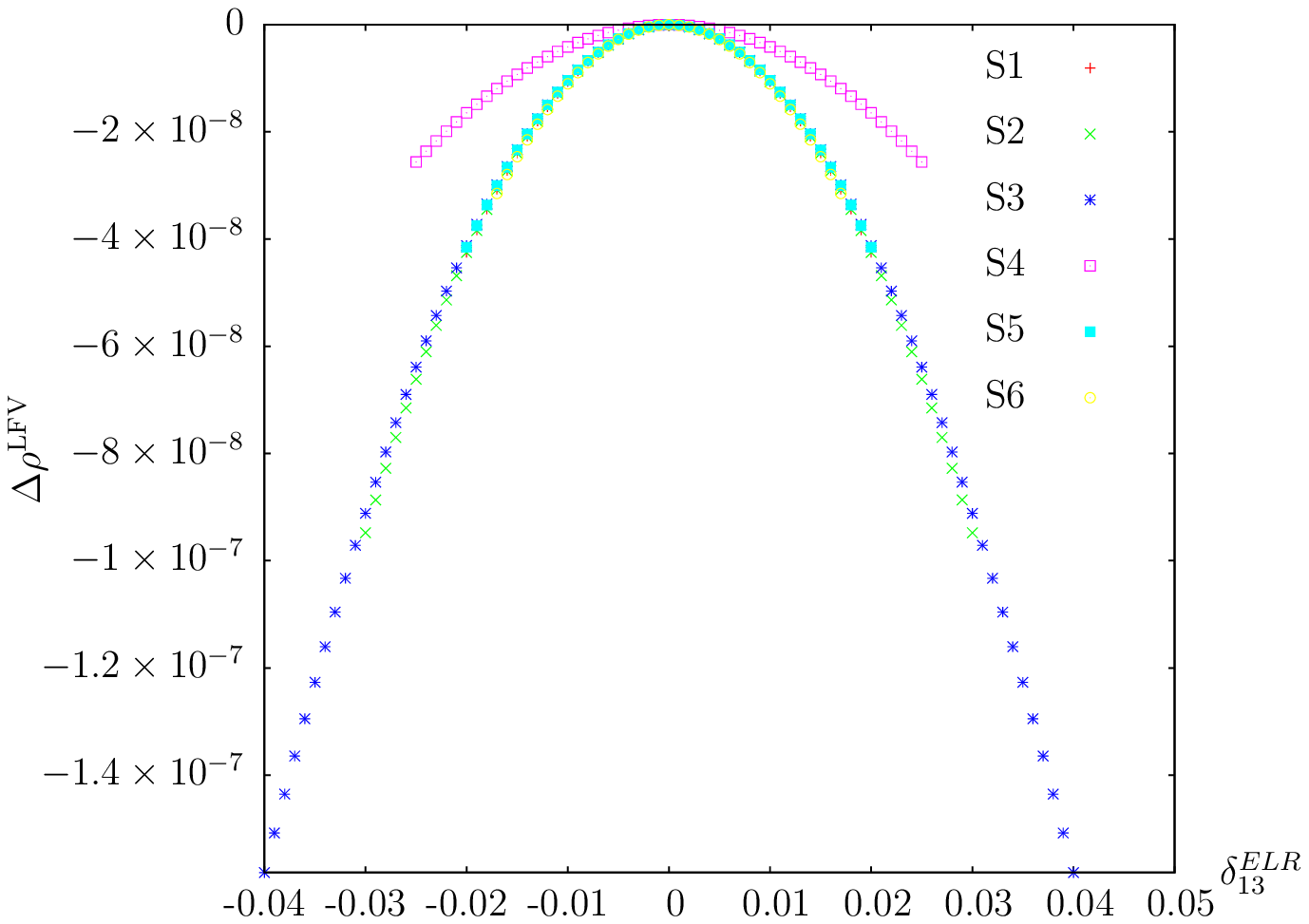  ,scale=0.53,angle=0,clip=}
\psfig{file=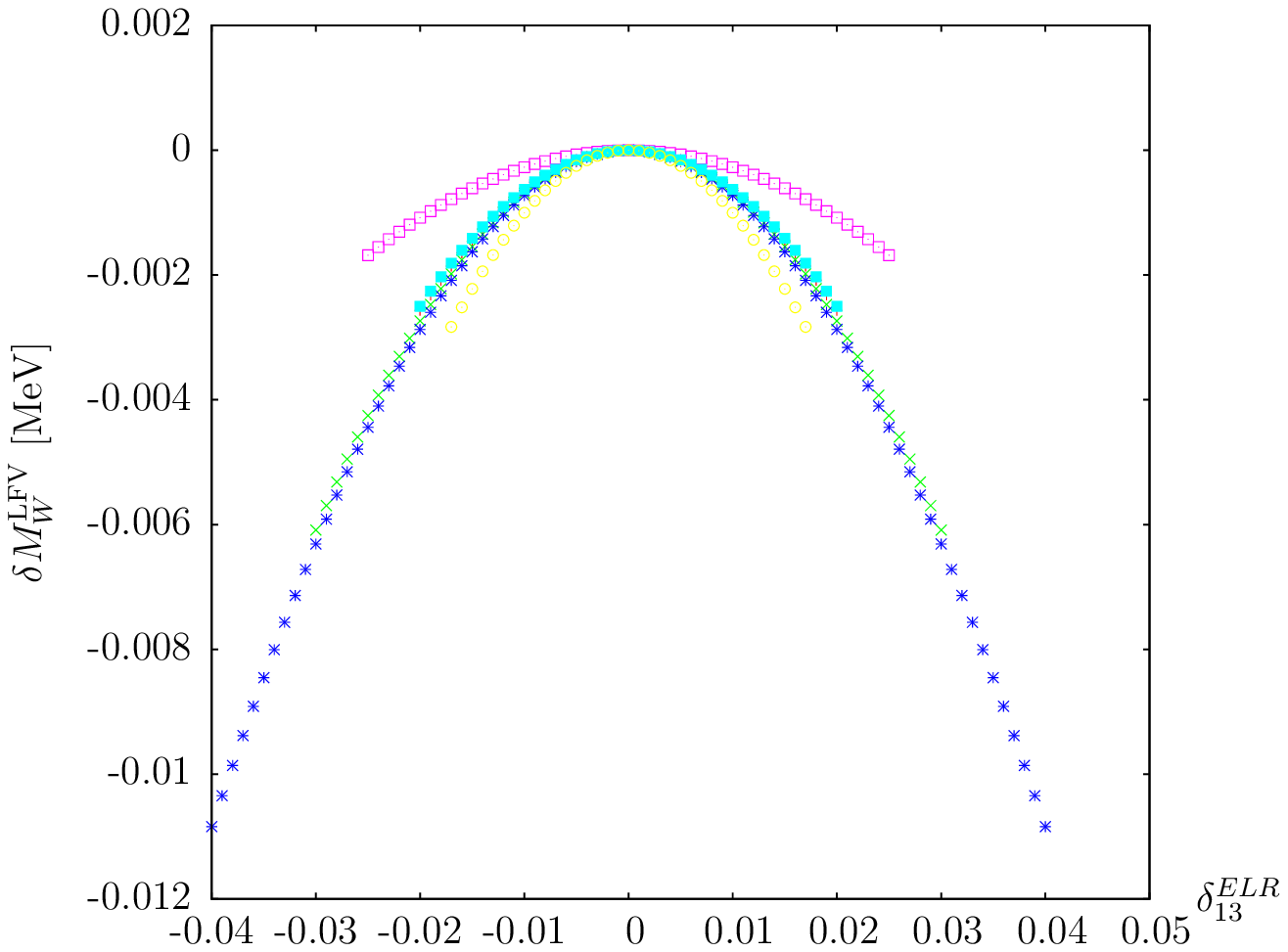  ,scale=0.53,angle=0,clip=}\\
\vspace{0.5cm}
\psfig{file=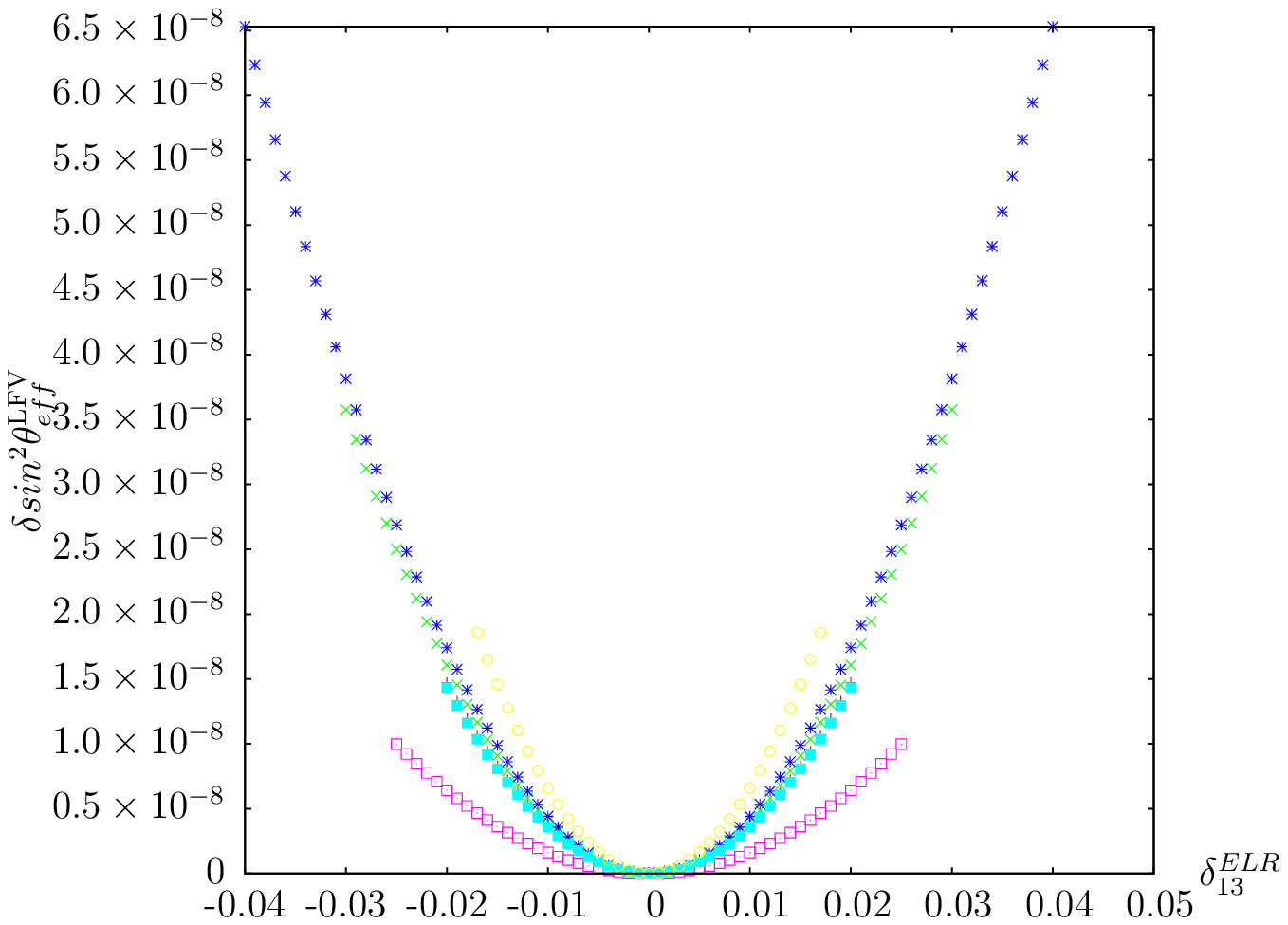 ,scale=0.53,angle=0,clip=}
\psfig{file=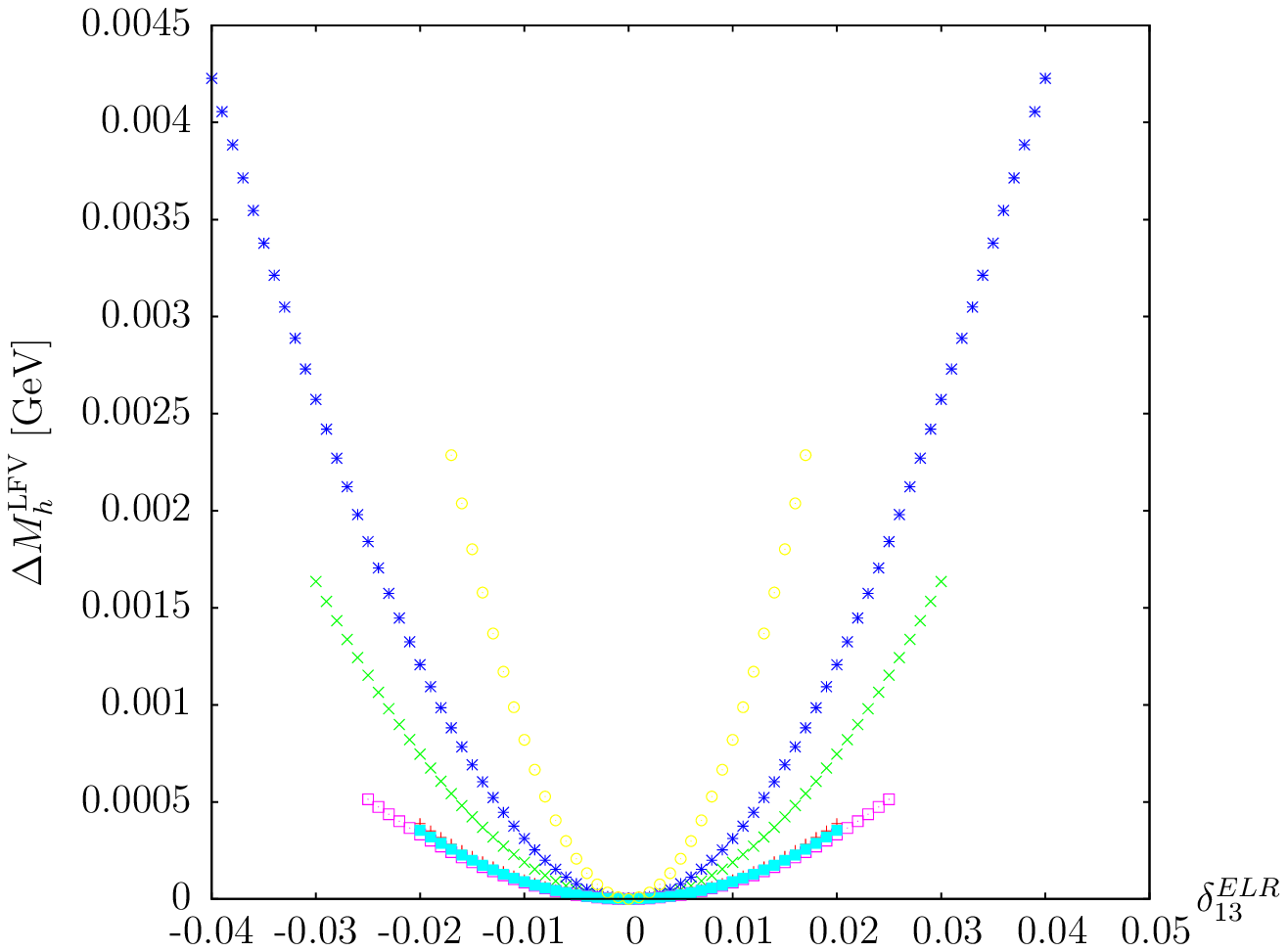   ,scale=0.53,angle=0,clip=}\\
\vspace{0.5cm}
\psfig{file=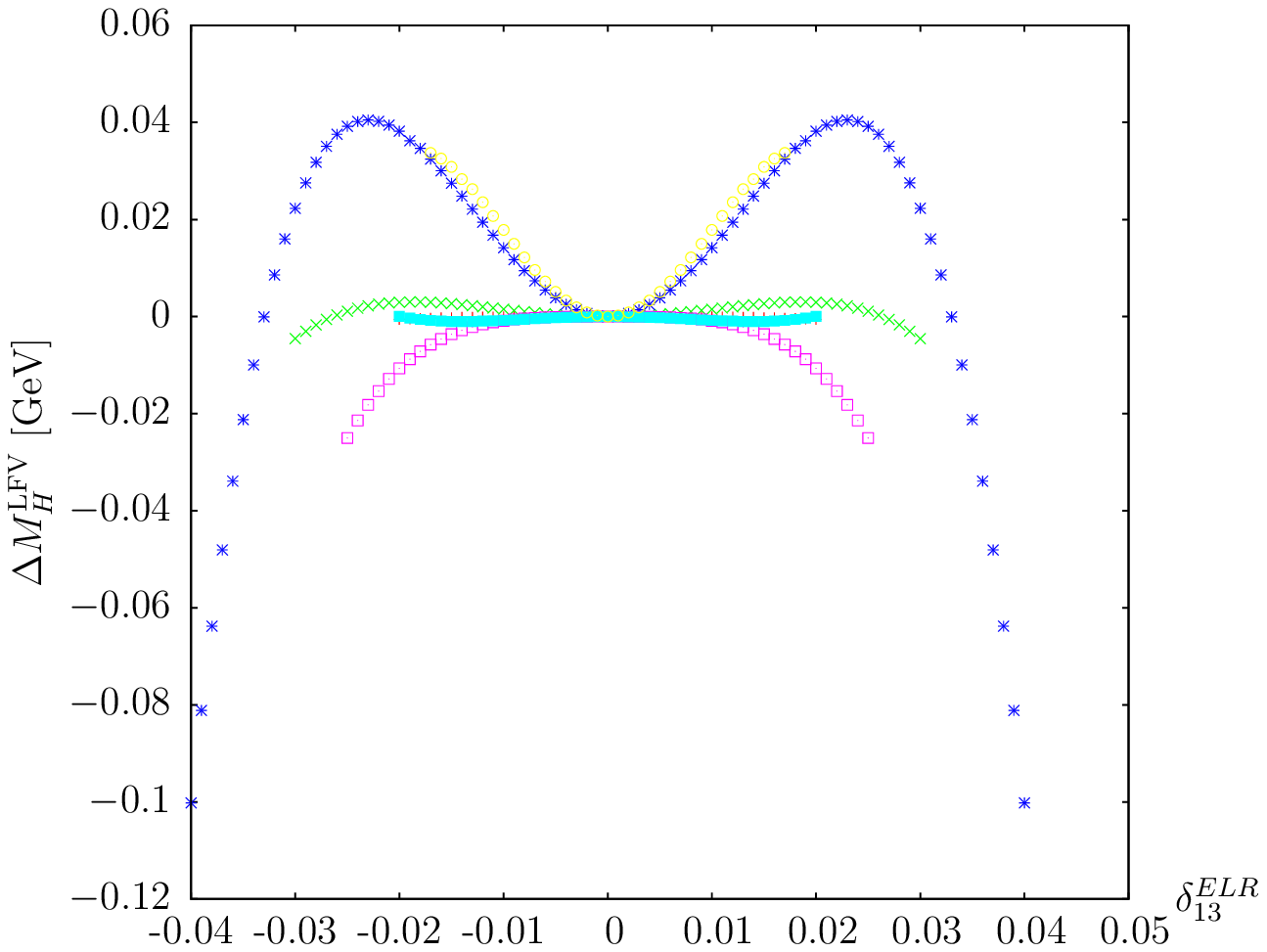  ,scale=0.53,angle=0,clip=}
\hspace{0.1cm}
\psfig{file=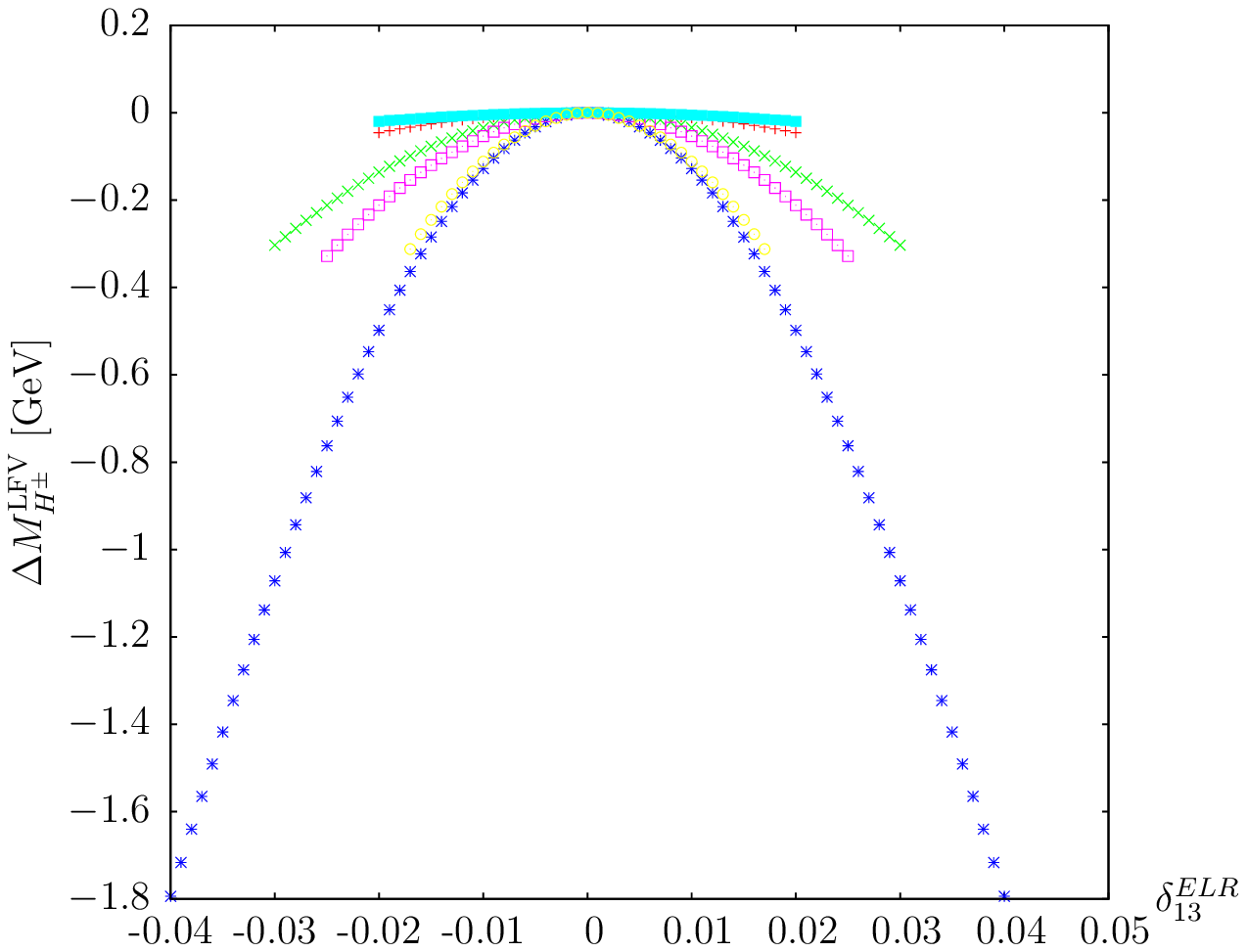  ,scale=0.53,angle=0,clip=}\\

\end{center}
\caption{EWPO and Higgs masses as a function of $\delta^{ELR}_{13}$.}  
\label{figdLR13}
\end{figure}

\begin{figure}[ht!]
\begin{center}
\psfig{file=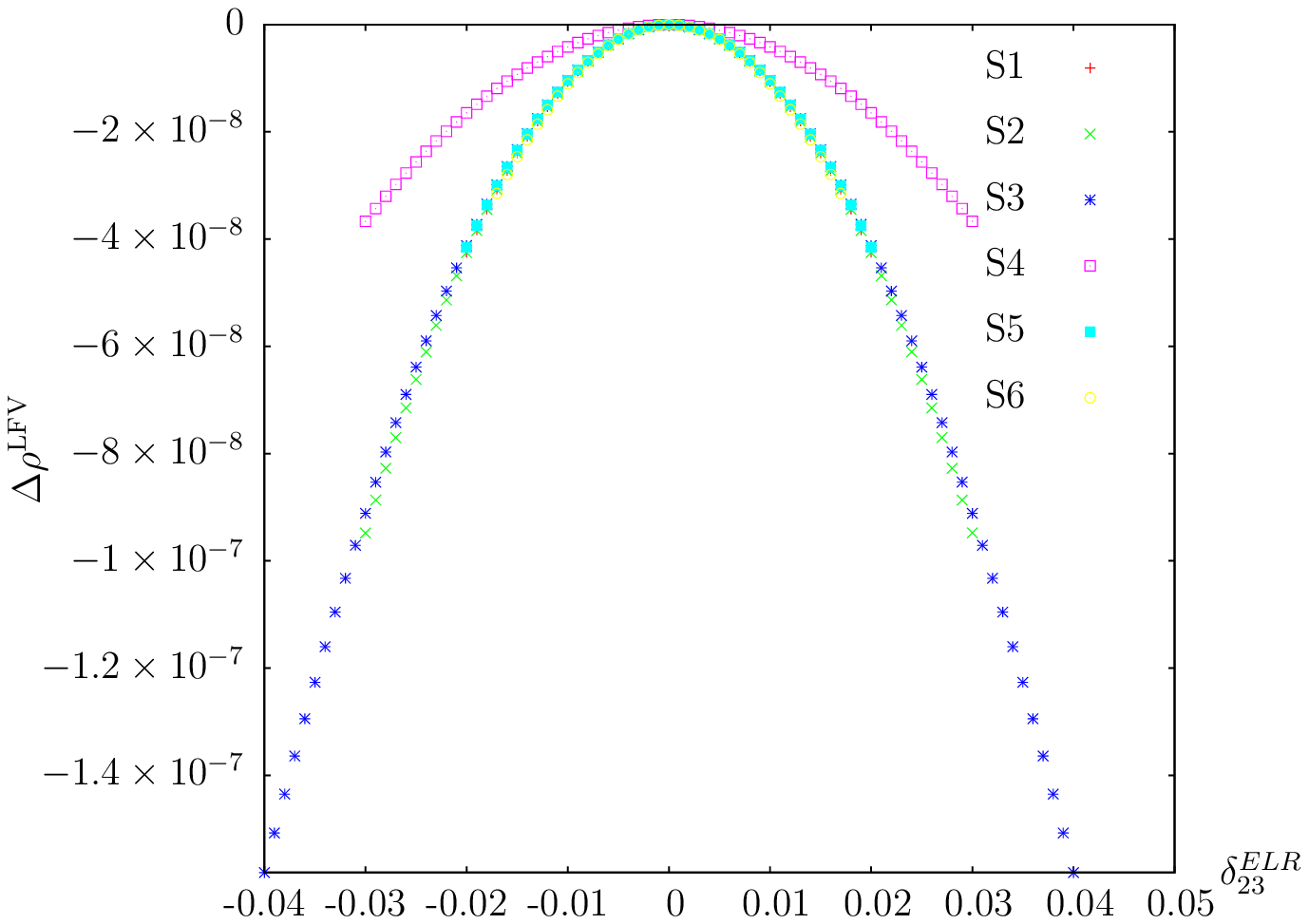  ,scale=0.53,angle=0,clip=}
\psfig{file=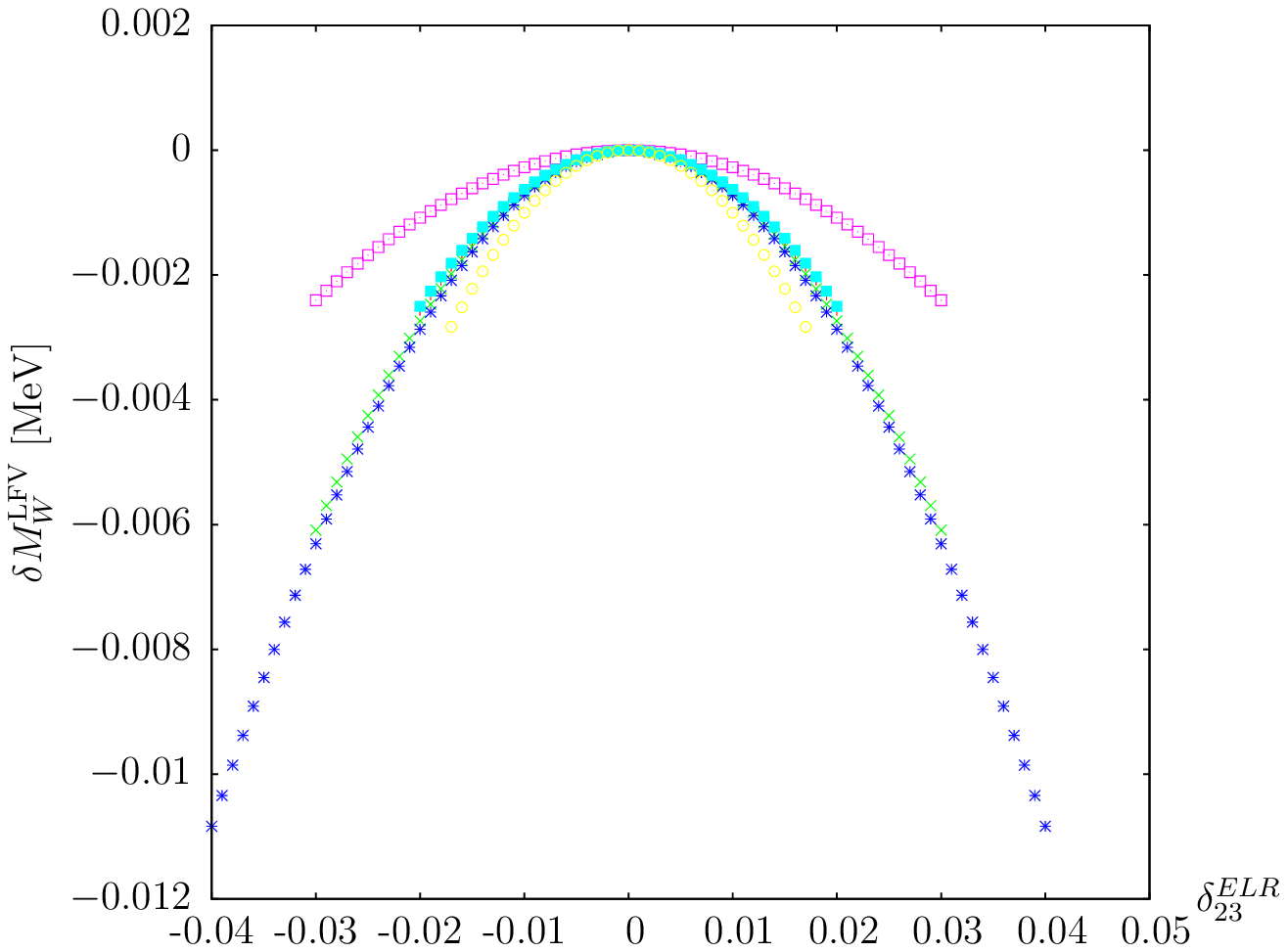  ,scale=0.53,angle=0,clip=}\\
\vspace{0.5cm}
\psfig{file=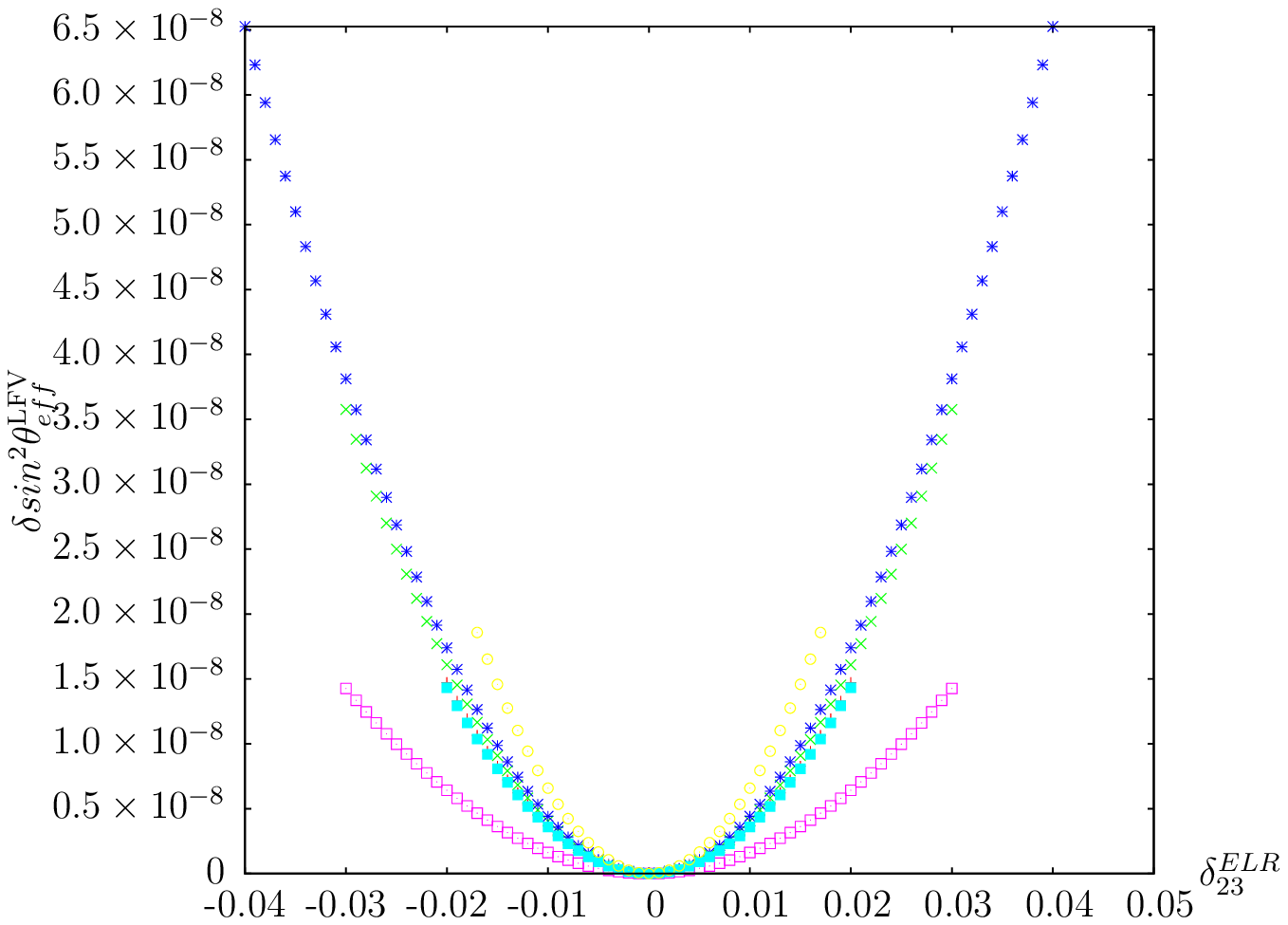 ,scale=0.53,angle=0,clip=}
\psfig{file=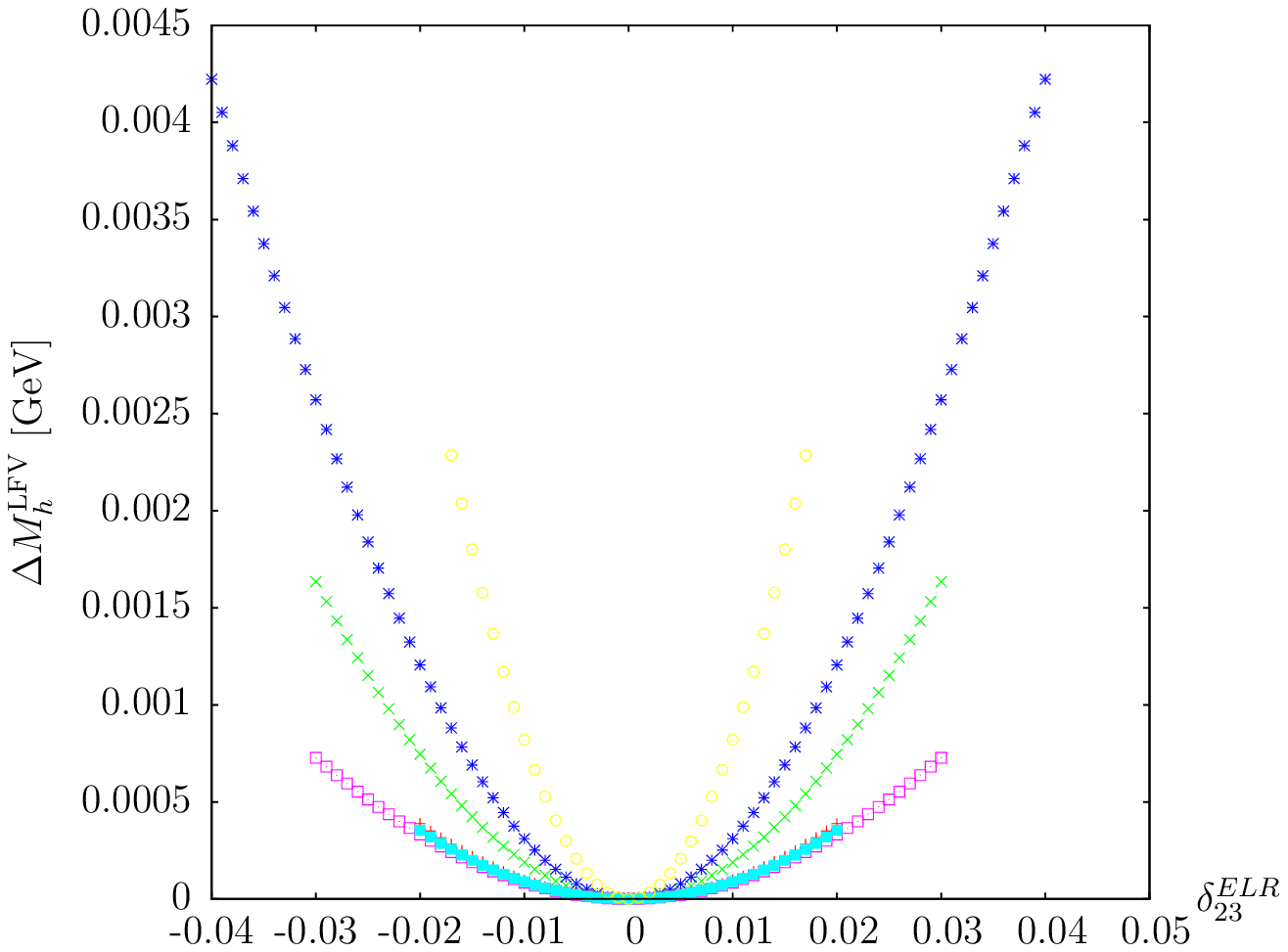   ,scale=0.53,angle=0,clip=}\\
\vspace{0.5cm}
\psfig{file=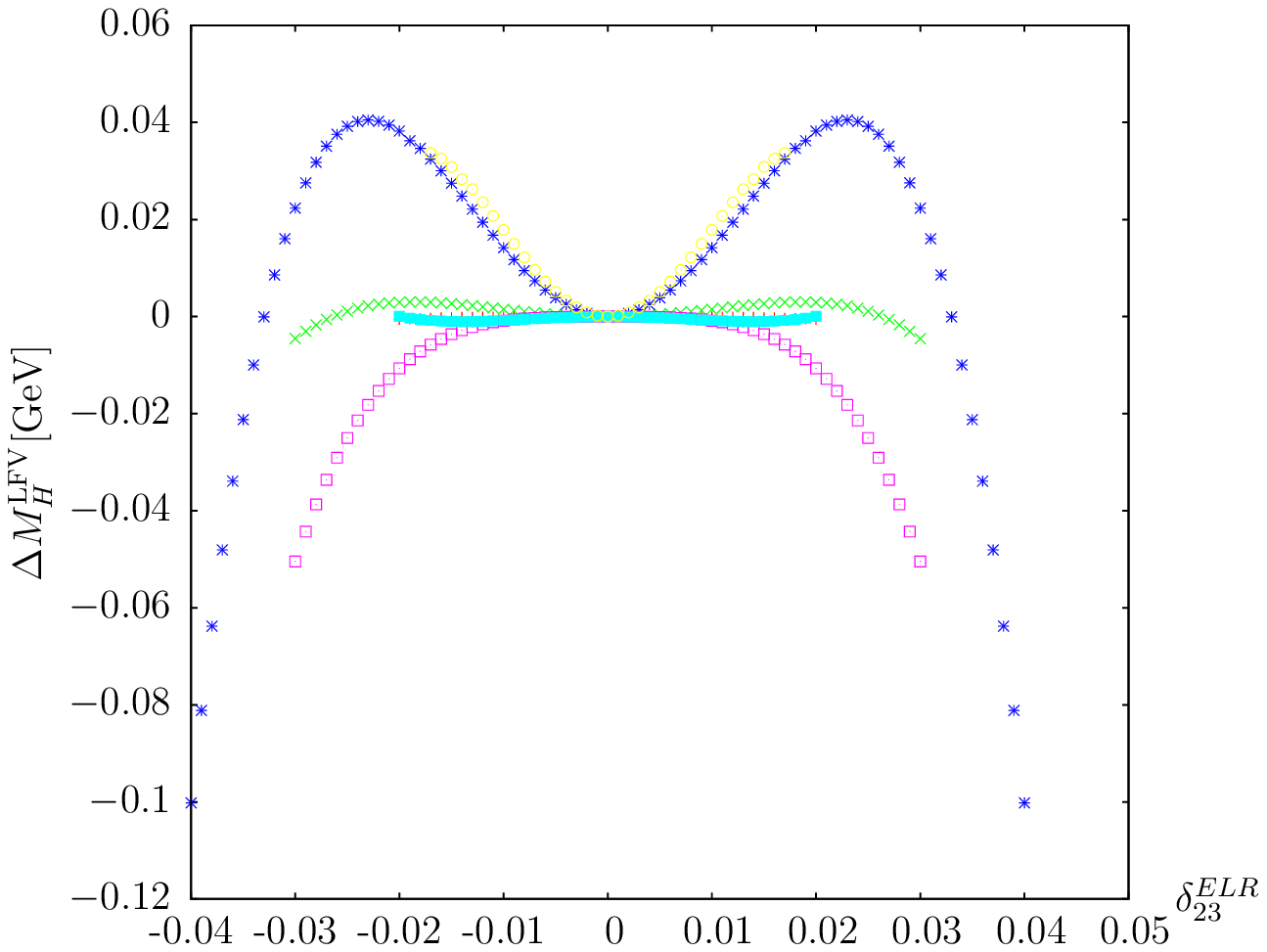  ,scale=0.53,angle=0,clip=}
\hspace{0.1cm}
\psfig{file=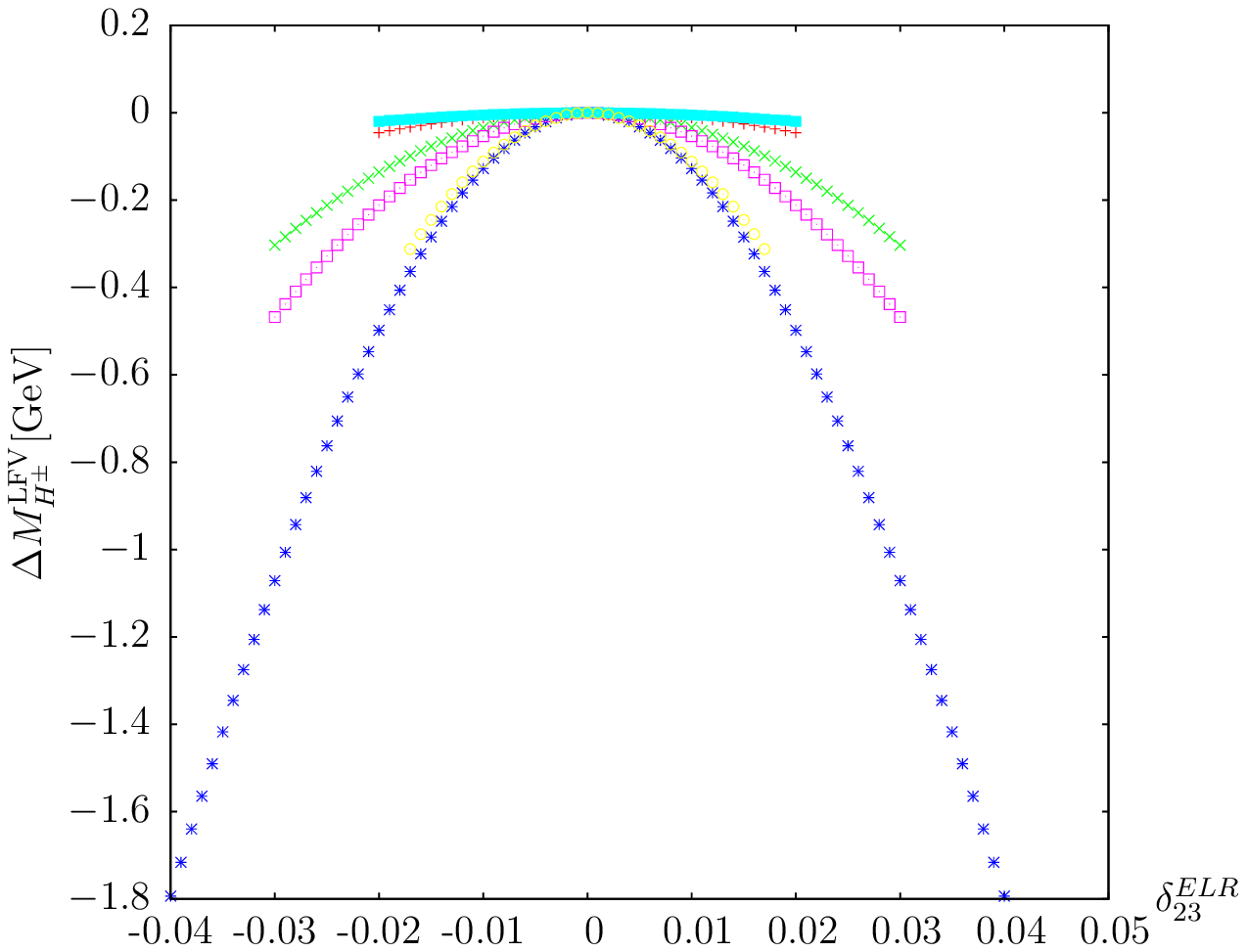  ,scale=0.53,angle=0,clip=}\\

\end{center}
\caption{EWPO and Higgs masses as a function of $\delta^{ELR}_{23}$.}  
\label{figdLR23}
\end{figure}

\begin{figure}[ht!]
\begin{center}
\psfig{file=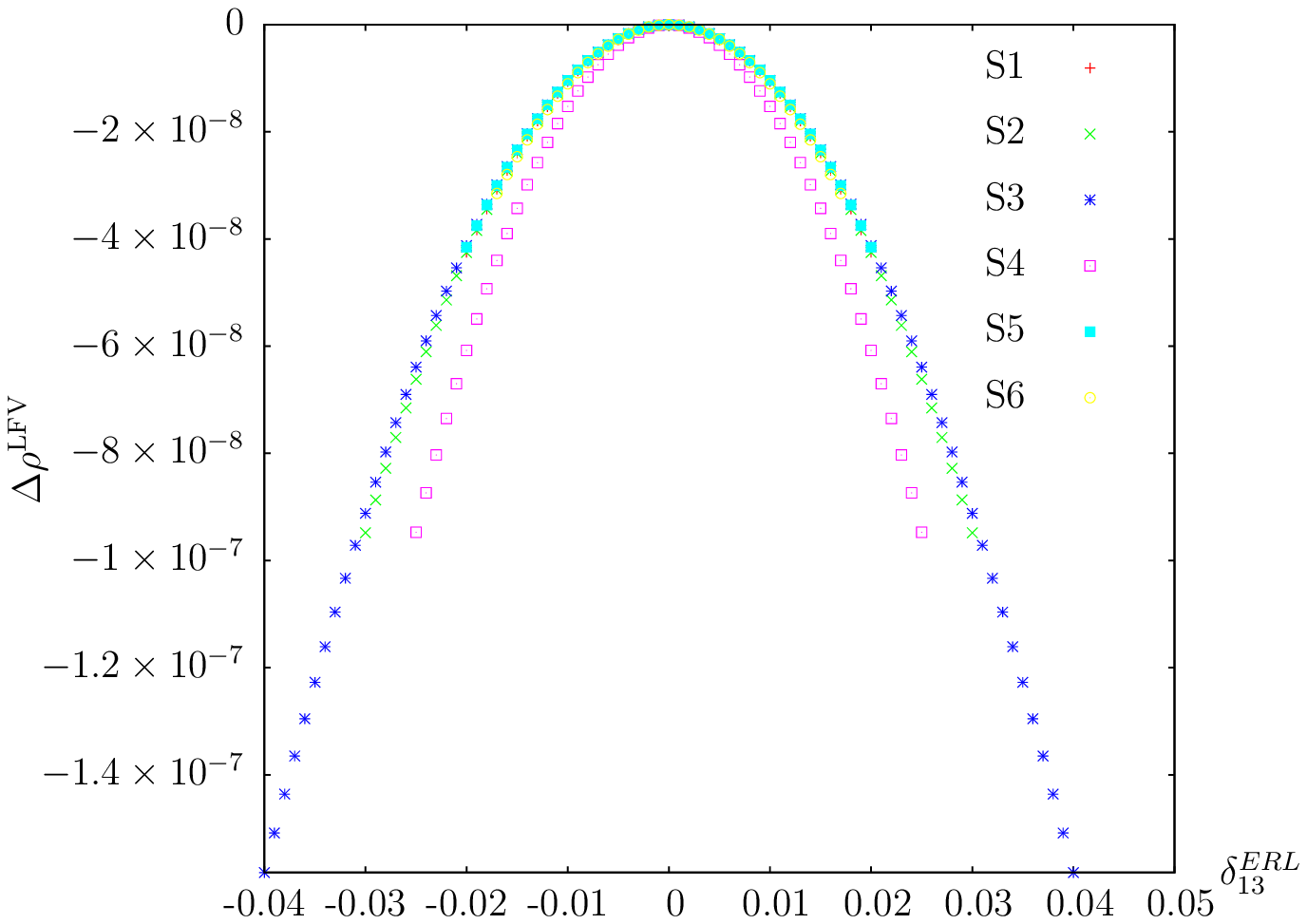  ,scale=0.53,angle=0,clip=}
\psfig{file=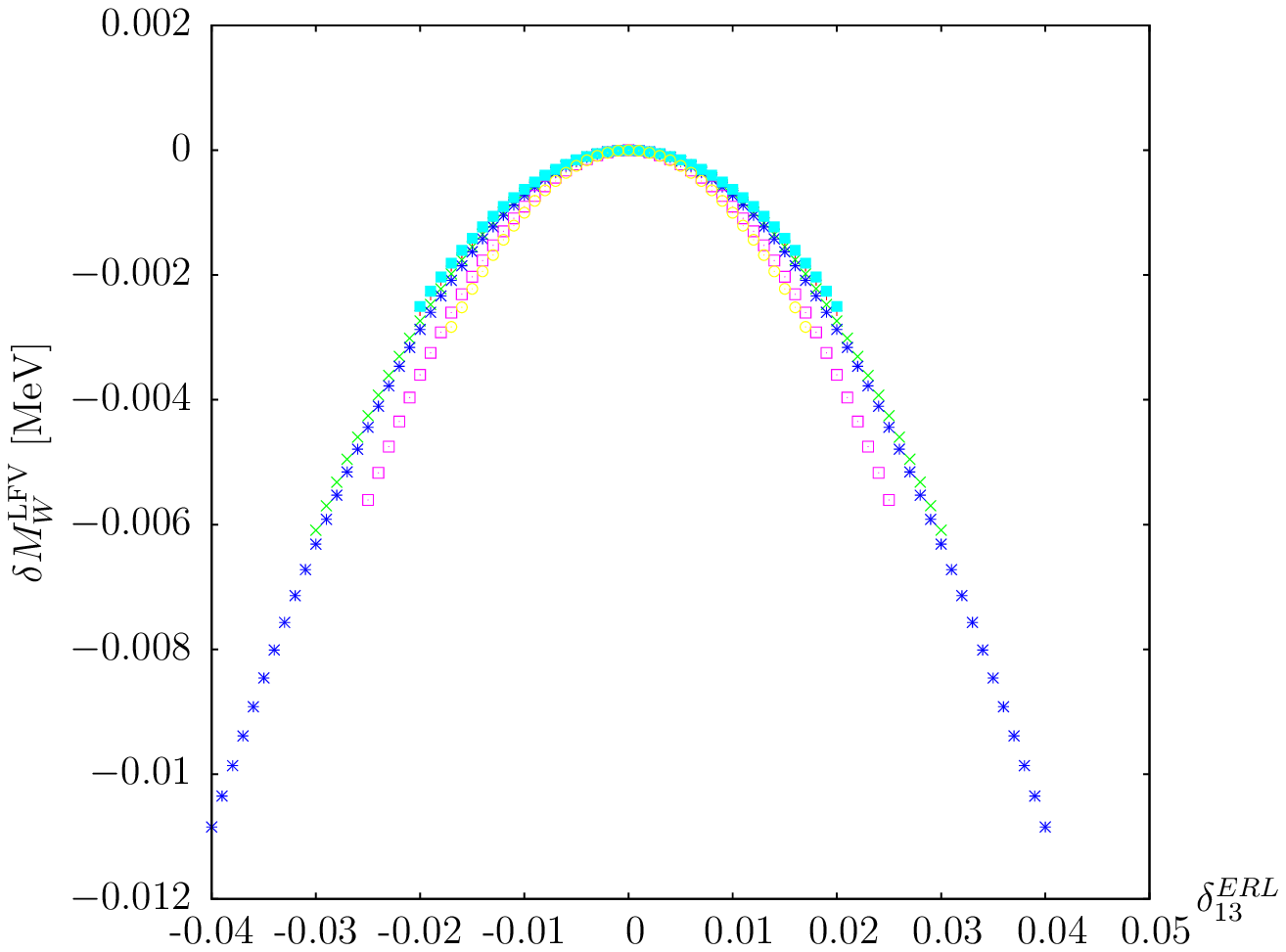  ,scale=0.53,angle=0,clip=}\\
\vspace{0.5cm}
\psfig{file=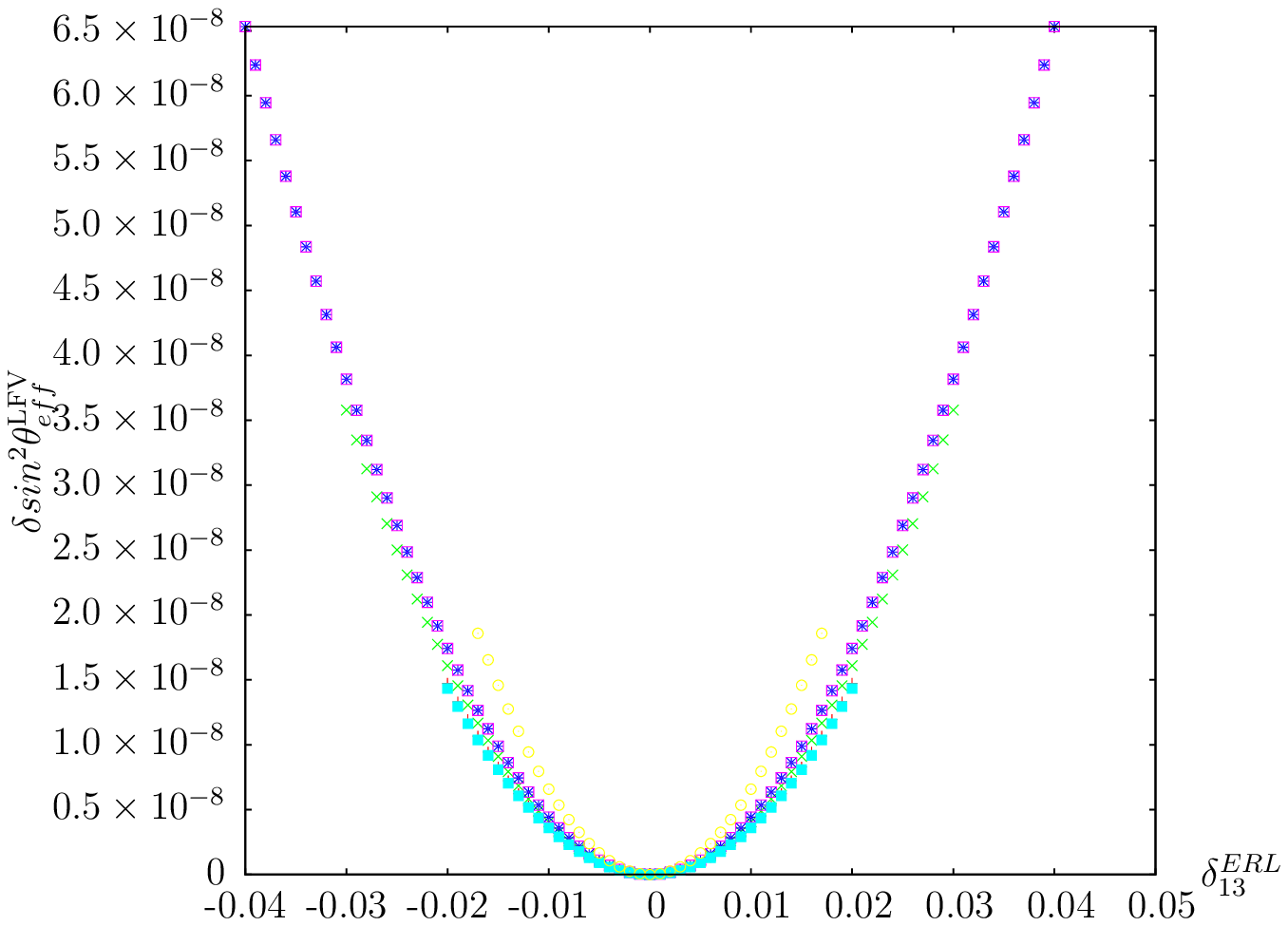 ,scale=0.53,angle=0,clip=}
\psfig{file=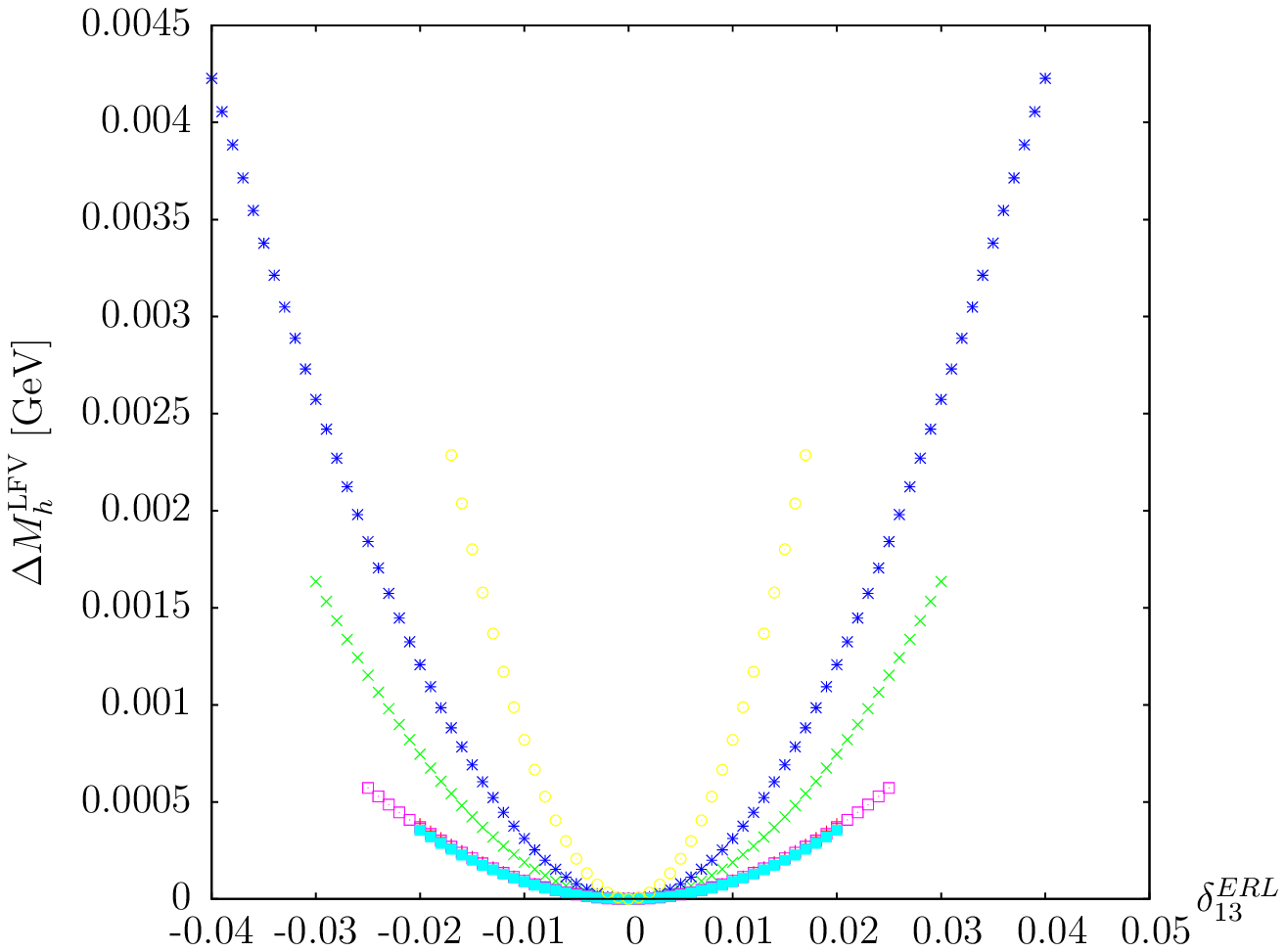   ,scale=0.53,angle=0,clip=}\\
\vspace{0.5cm}
\psfig{file=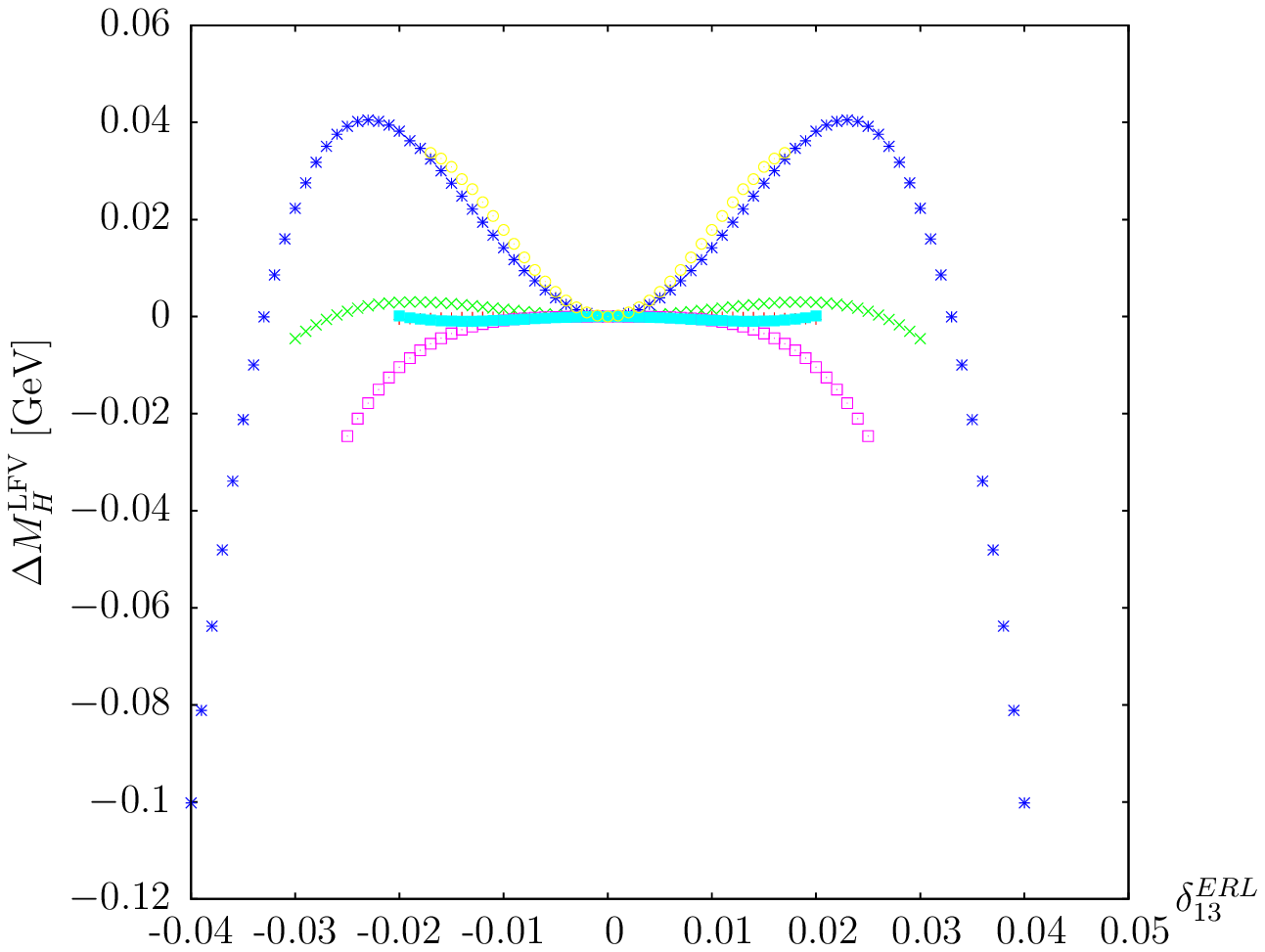  ,scale=0.53,angle=0,clip=}
\hspace{0.1cm}
\psfig{file=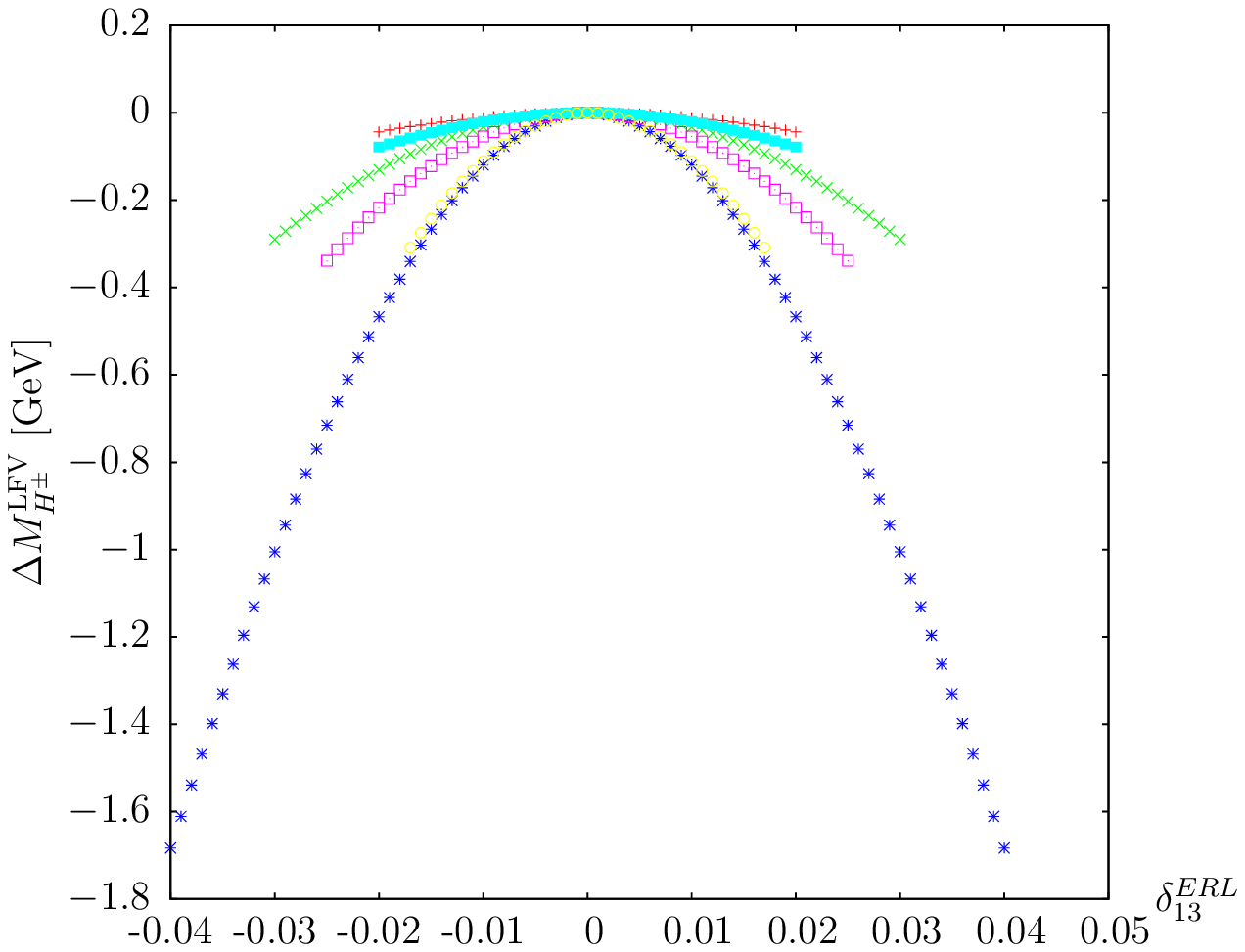  ,scale=0.53,angle=0,clip=}\\

\end{center}
\caption{EWPO and Higgs masses as a function of $\delta^{ERL}_{13}$.}  
\label{figdRL13}
\end{figure}

\begin{figure}[ht!]
\begin{center}
\psfig{file=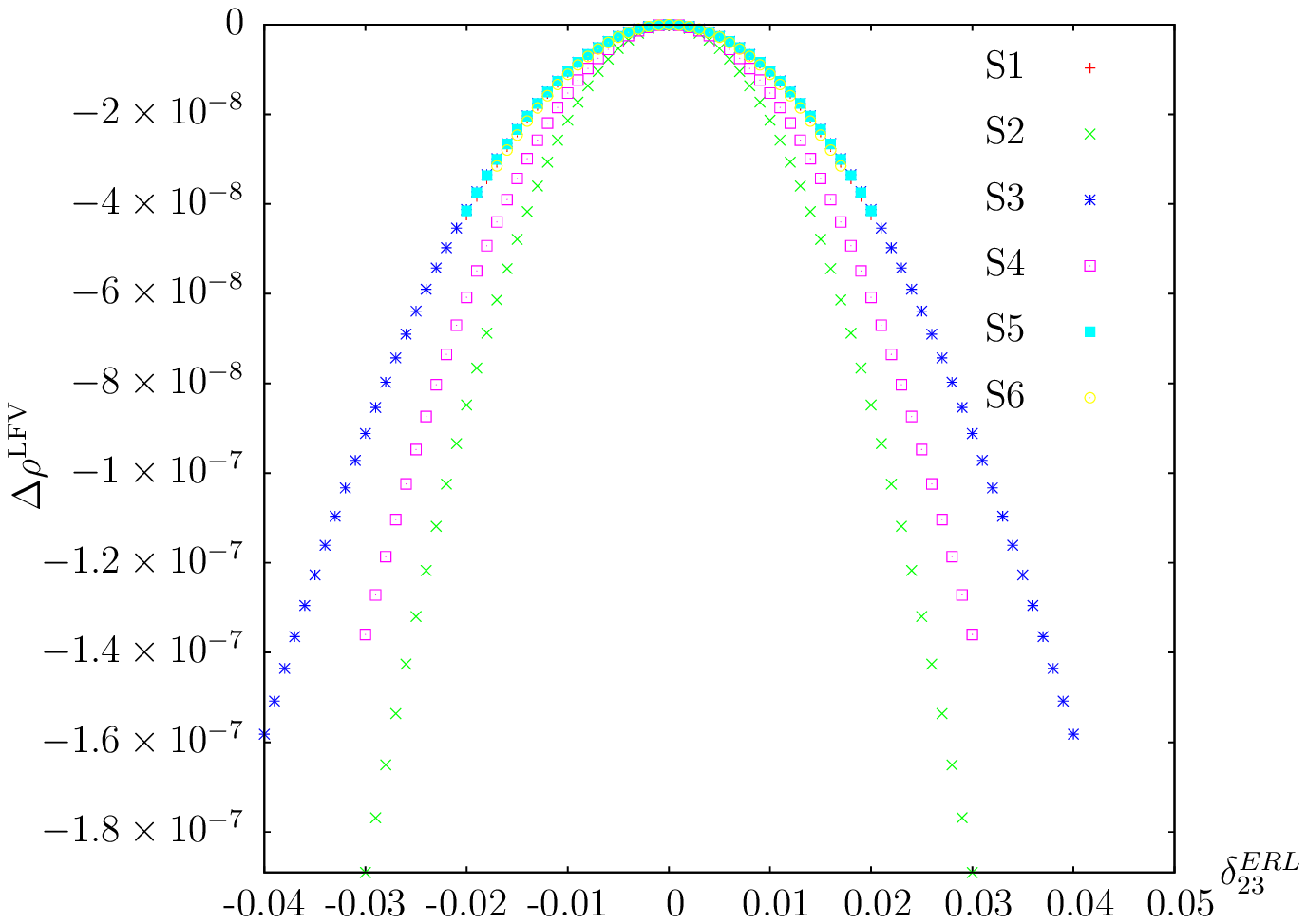  ,scale=0.53,angle=0,clip=}
\psfig{file=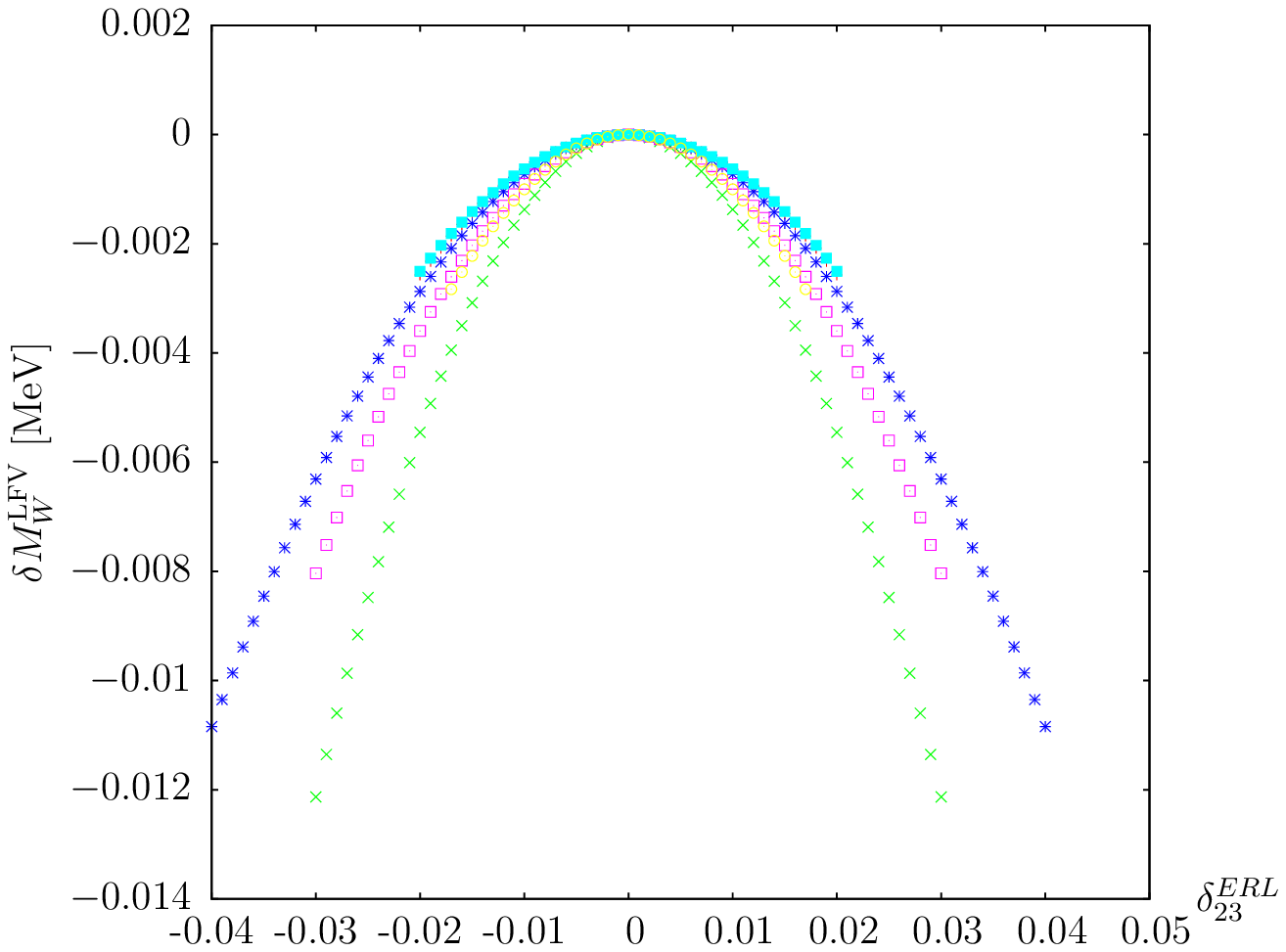  ,scale=0.53,angle=0,clip=}\\
\vspace{0.5cm}
\psfig{file=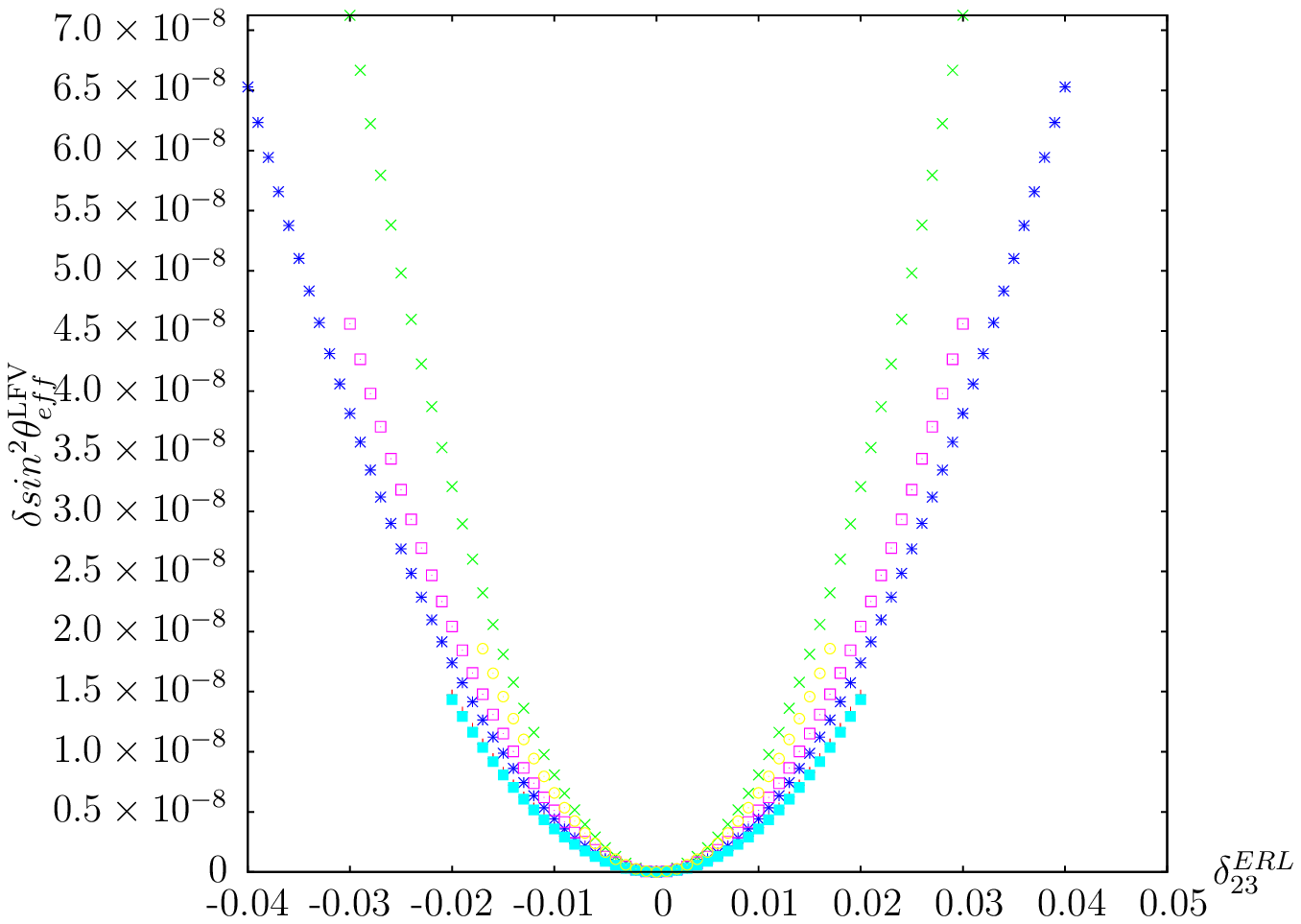 ,scale=0.53,angle=0,clip=}
\psfig{file=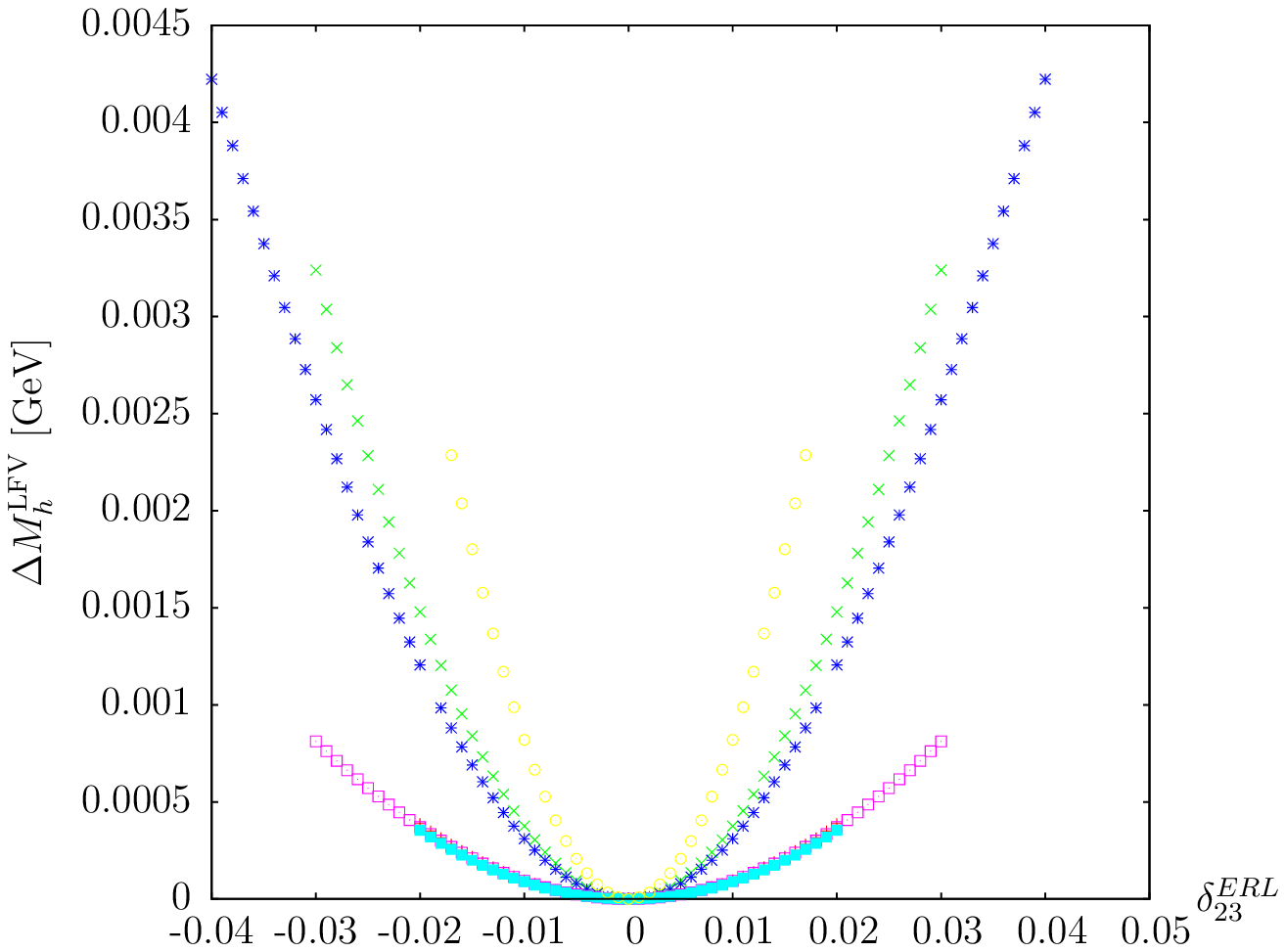   ,scale=0.53,angle=0,clip=}\\
\vspace{0.5cm}
\psfig{file=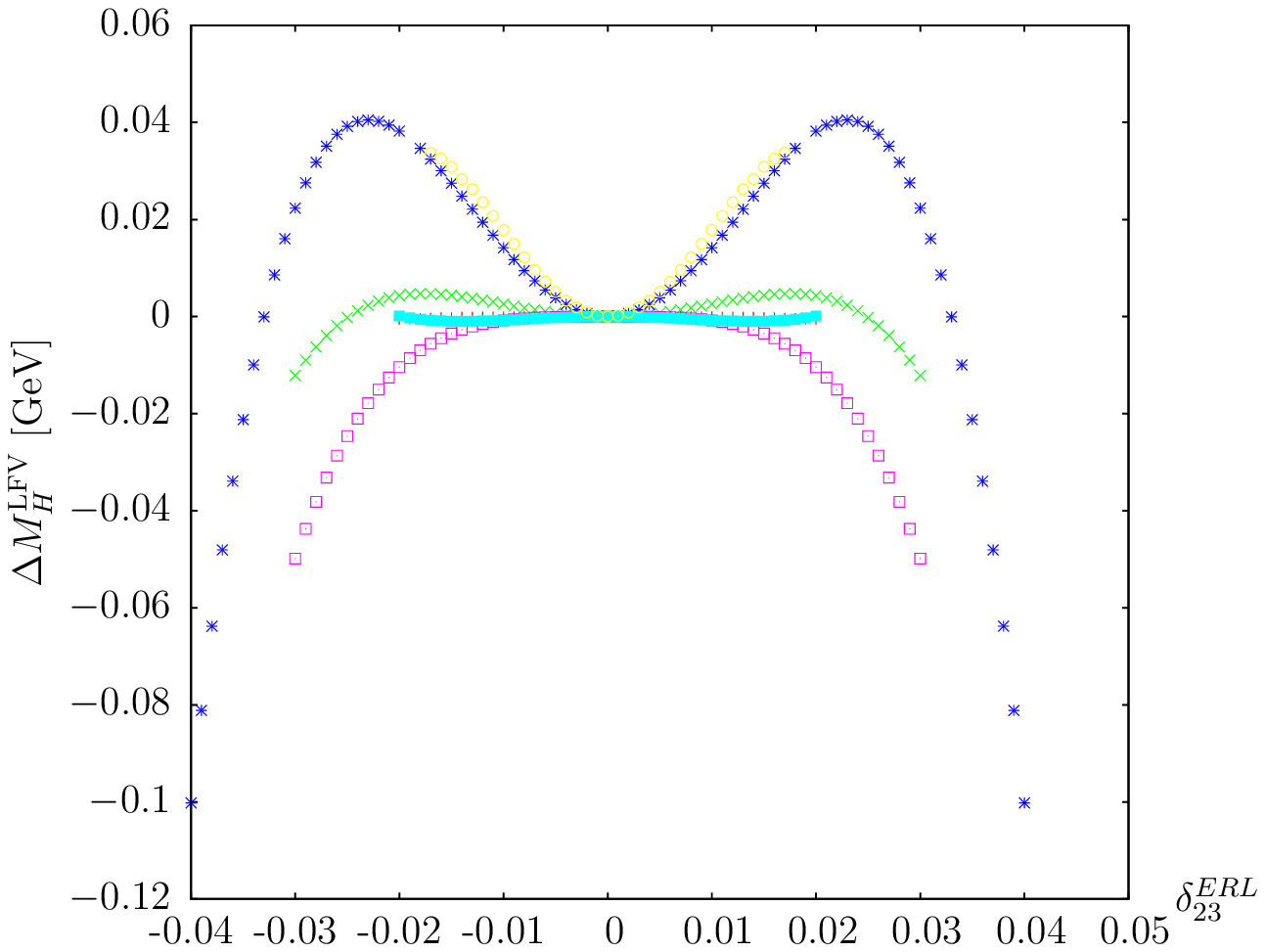  ,scale=0.53,angle=0,clip=}
\hspace{0.1cm}
\psfig{file=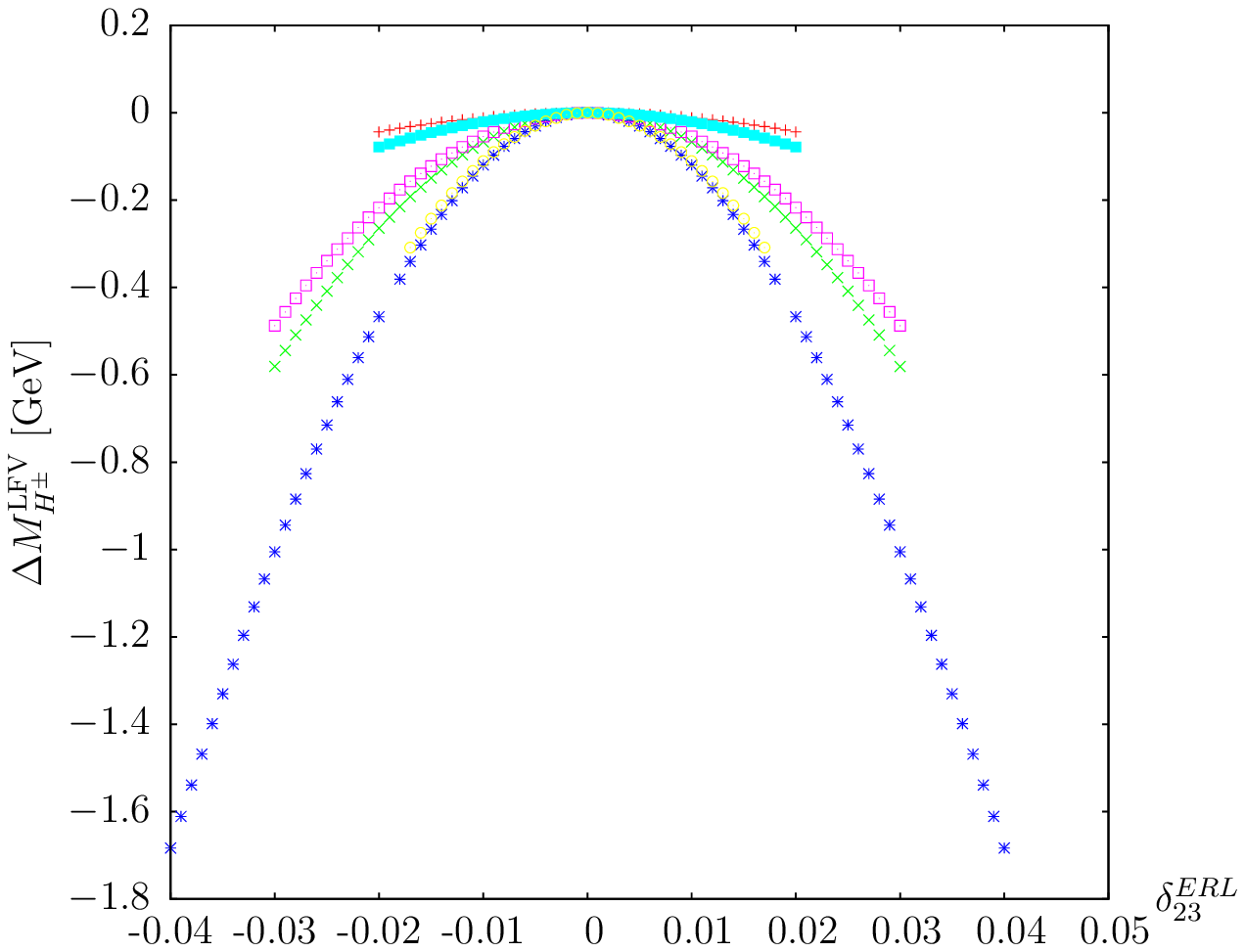  ,scale=0.53,angle=0,clip=}\\

\end{center}
\caption{EWPO and Higgs masses as a function of $\delta^{ERL}_{23}$.}  
\label{figdRL23}
\end{figure}

\begin{figure}[ht!]
\begin{center}
\psfig{file=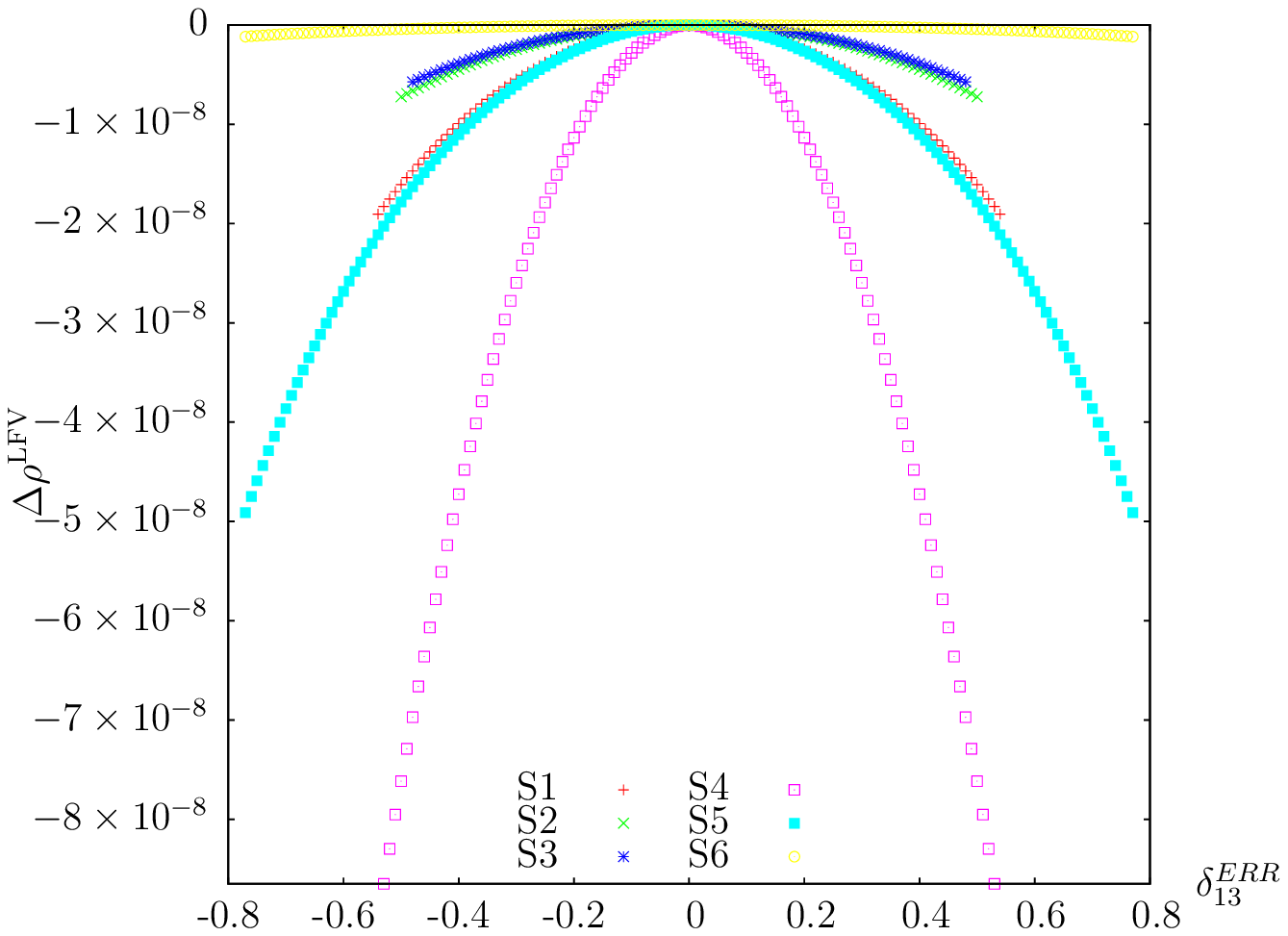  ,scale=0.53,angle=0,clip=}
\hspace{0.2cm}
\psfig{file=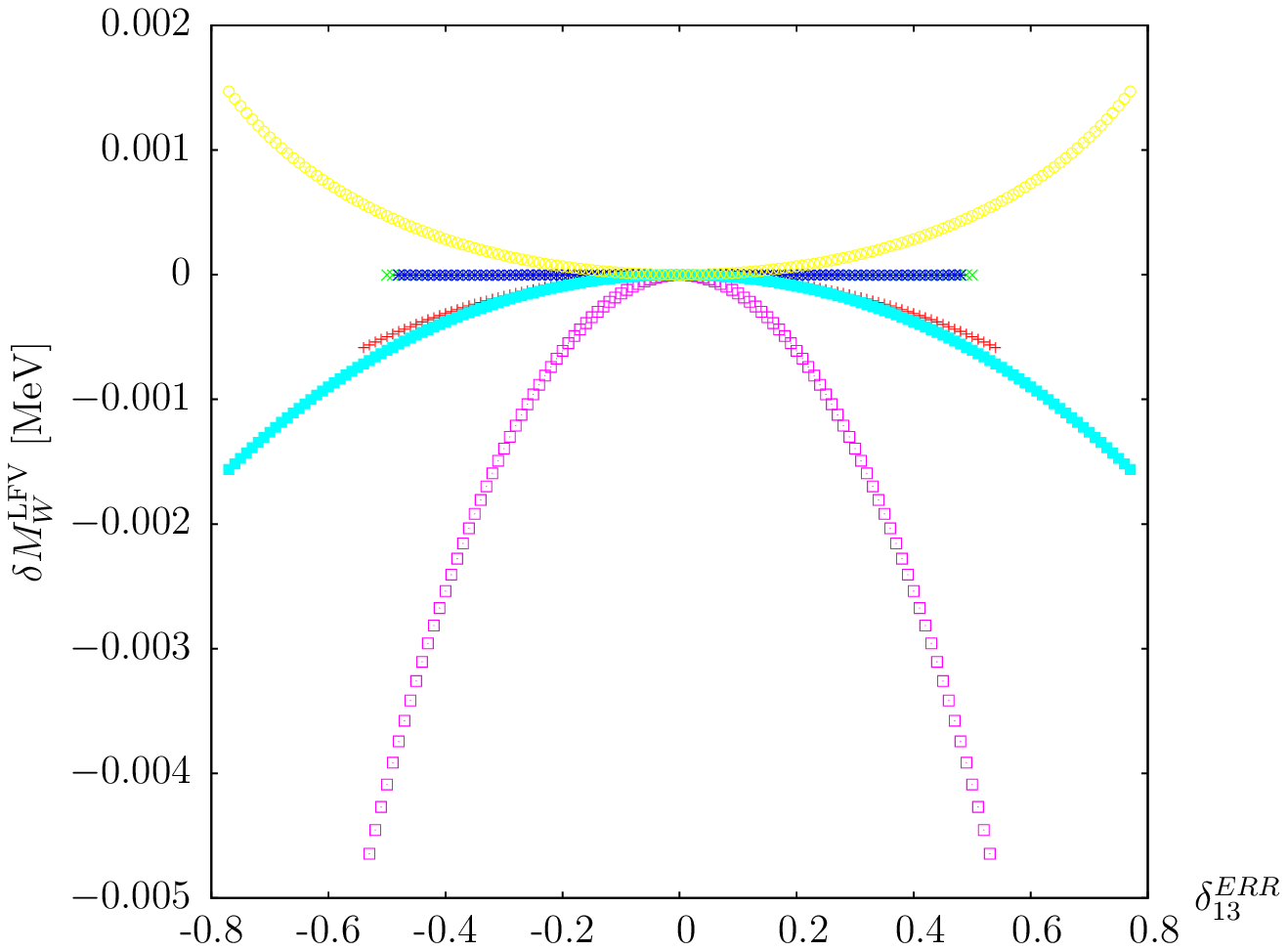  ,scale=0.53,angle=0,clip=}\\
\vspace{0.5cm}
\psfig{file=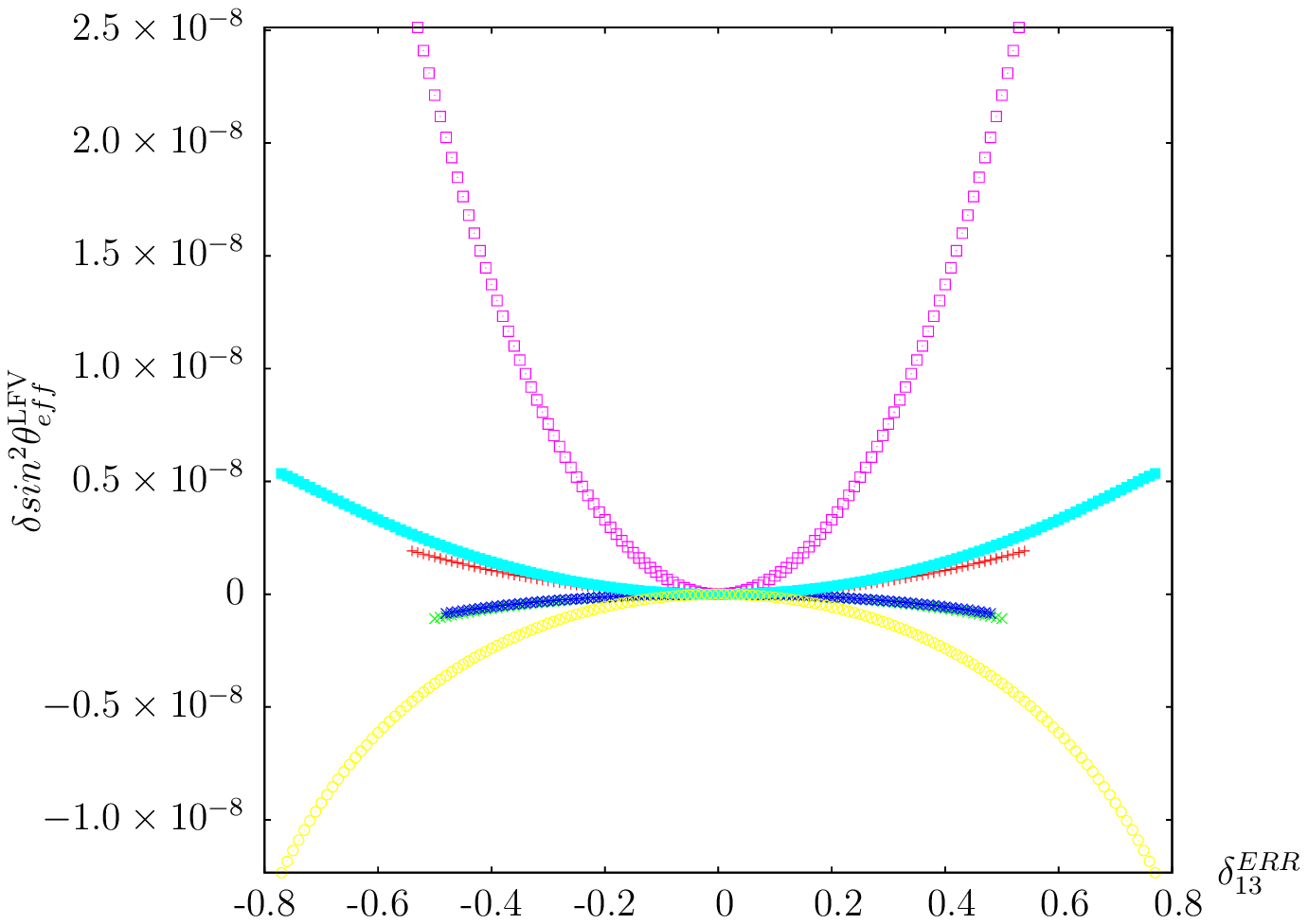 ,scale=0.53,angle=0,clip=}
\psfig{file=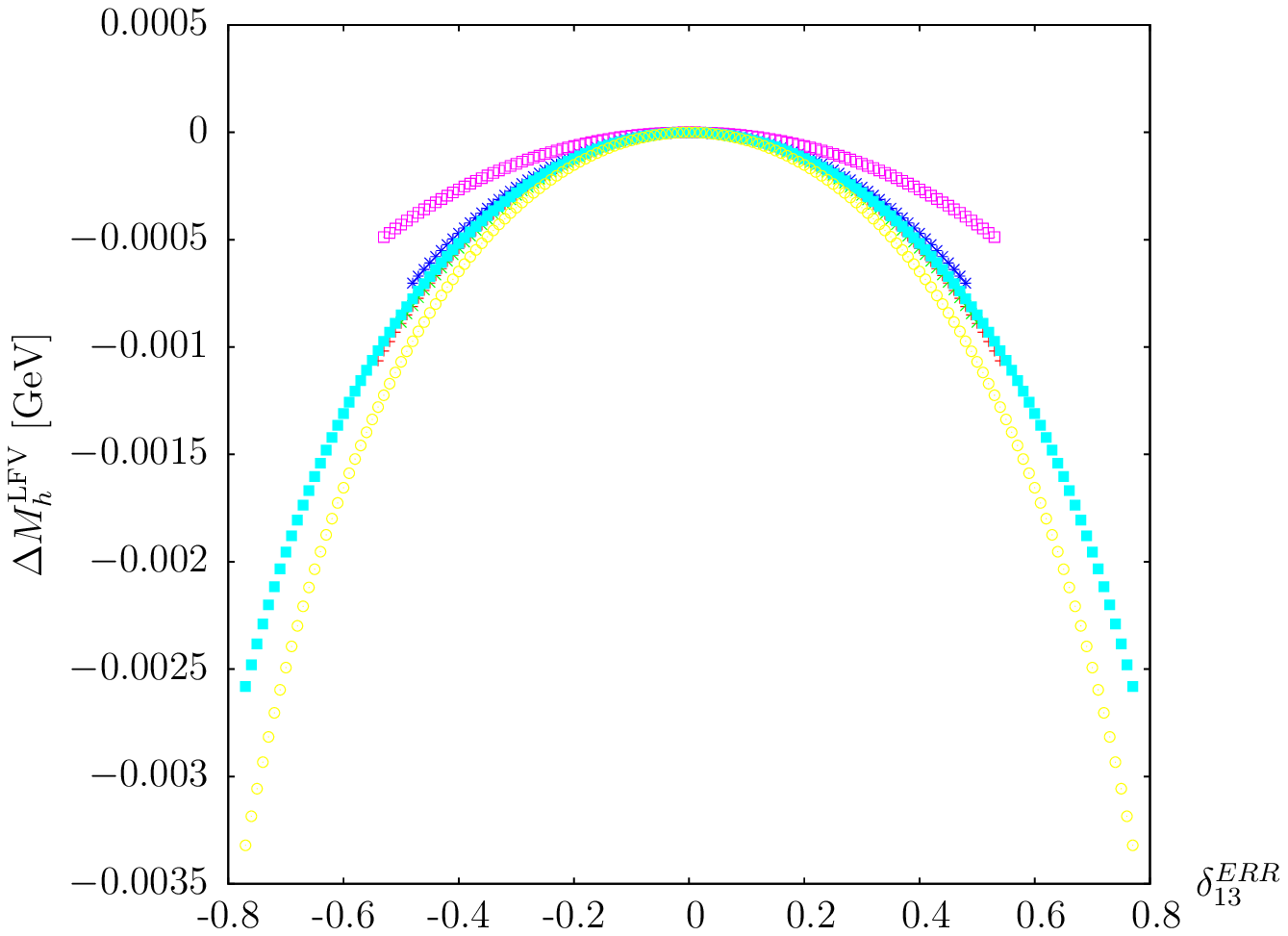   ,scale=0.53,angle=0,clip=}\\
\vspace{0.5cm}
\psfig{file=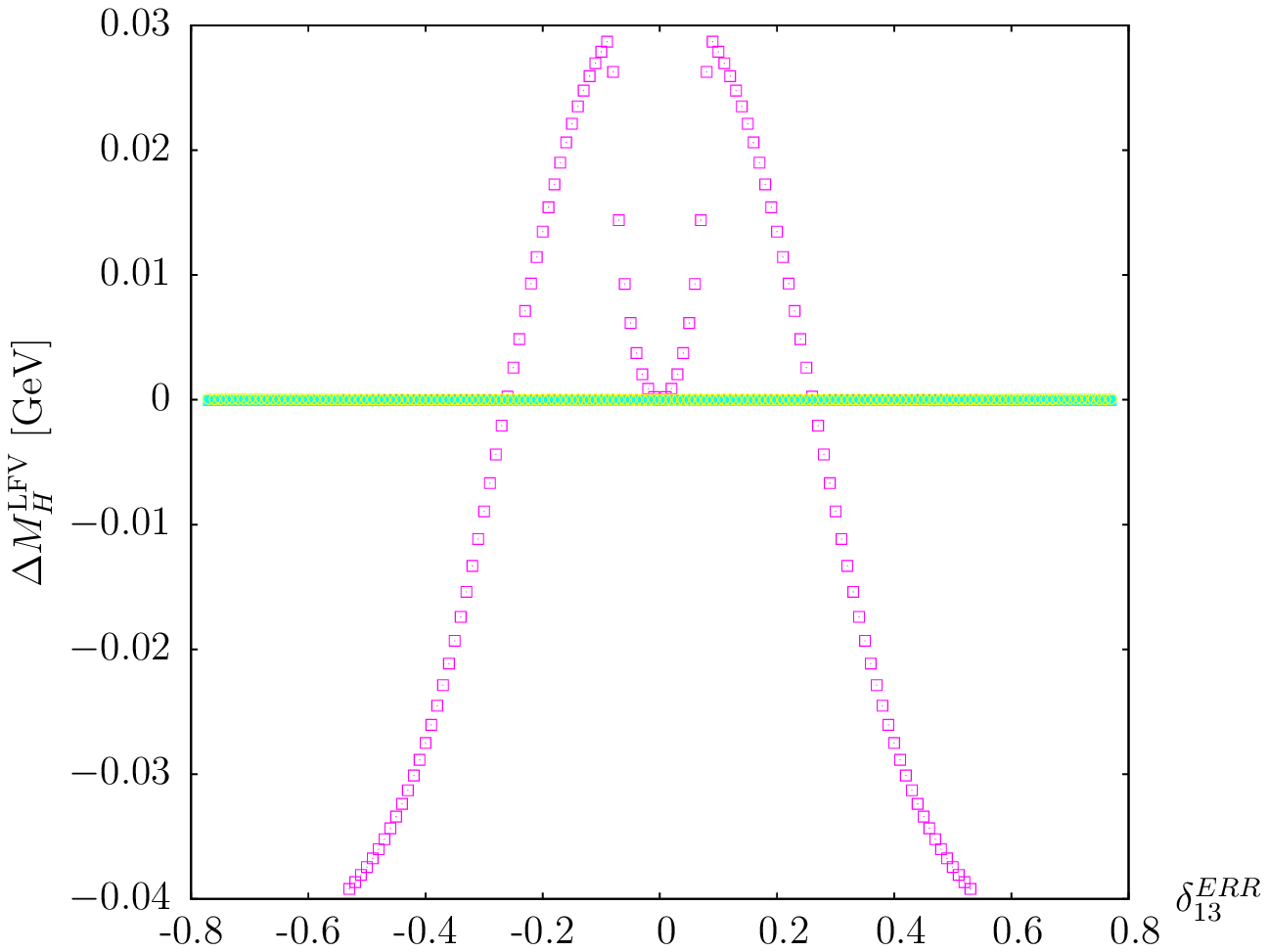  ,scale=0.53,angle=0,clip=}
\hspace{0.2cm}
\psfig{file=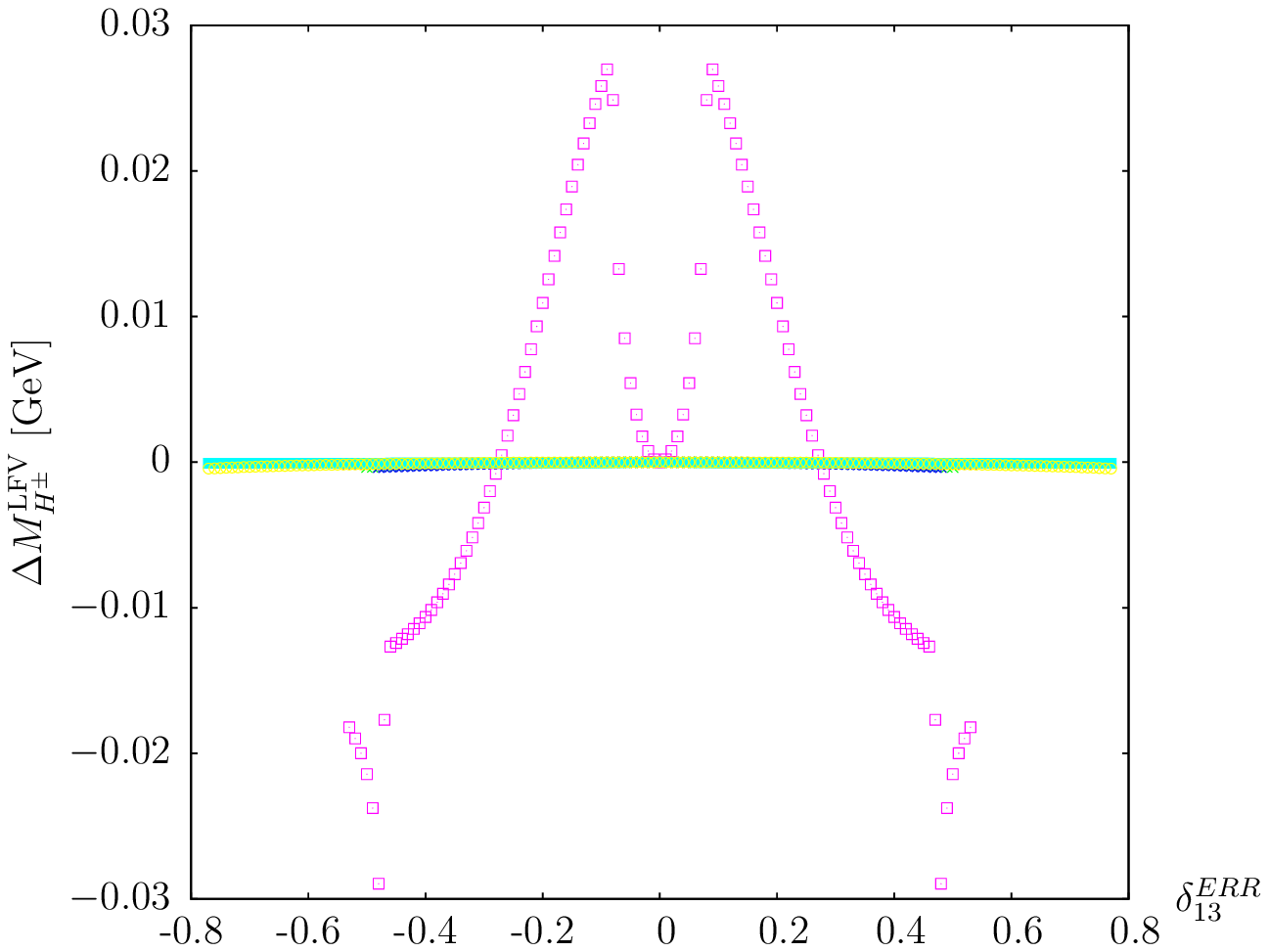  ,scale=0.53,angle=0,clip=}\\

\end{center}
\caption{EWPO and Higgs masses as a function of $\delta^{ERR}_{13}$.}  
\label{figdRR13}
\end{figure}

\begin{figure}[ht!]
\begin{center}
\psfig{file=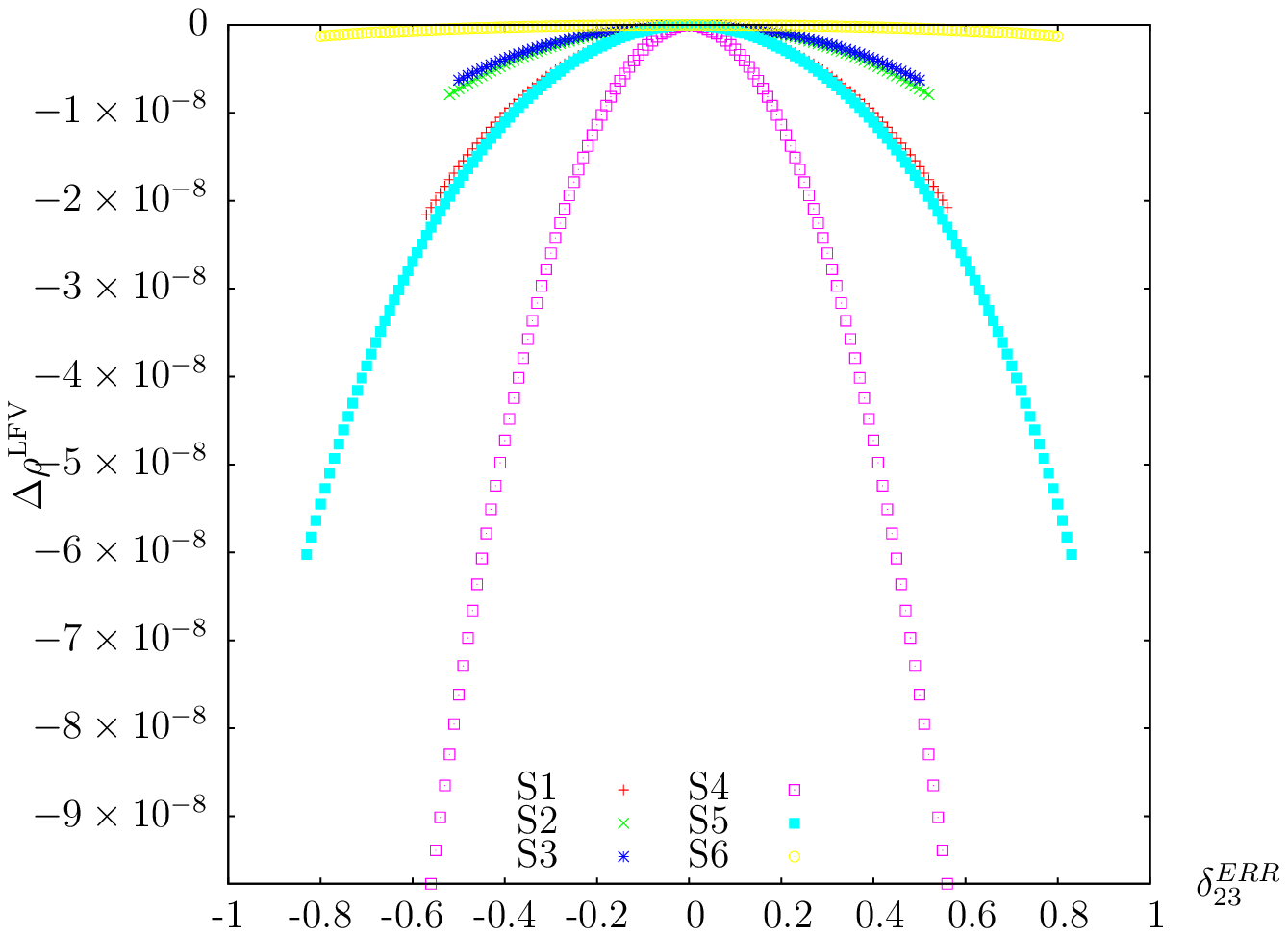  ,scale=0.53}
\hspace{0.2cm}
\psfig{file=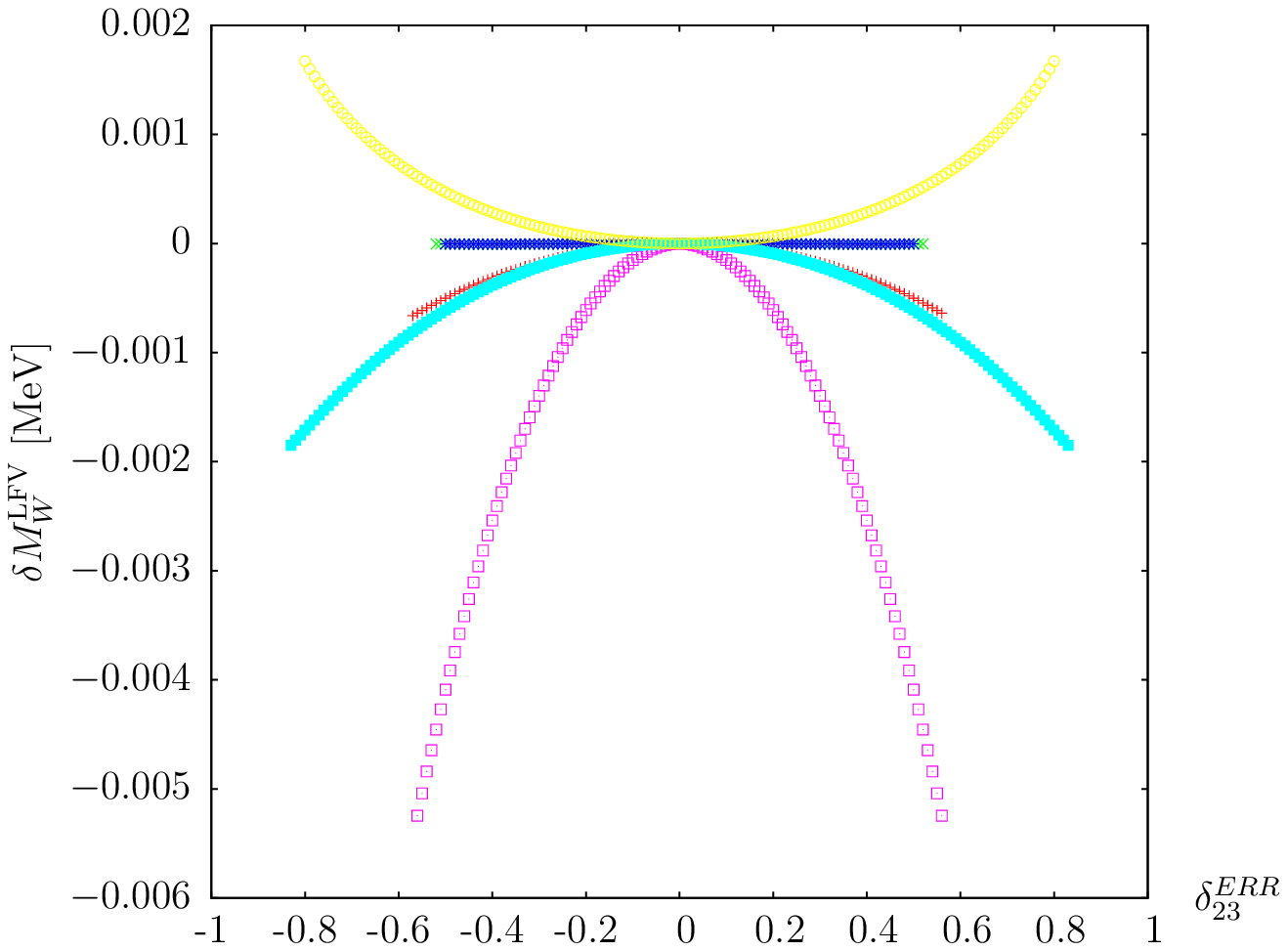  ,scale=0.53}\\
\vspace{0.5cm}
\psfig{file=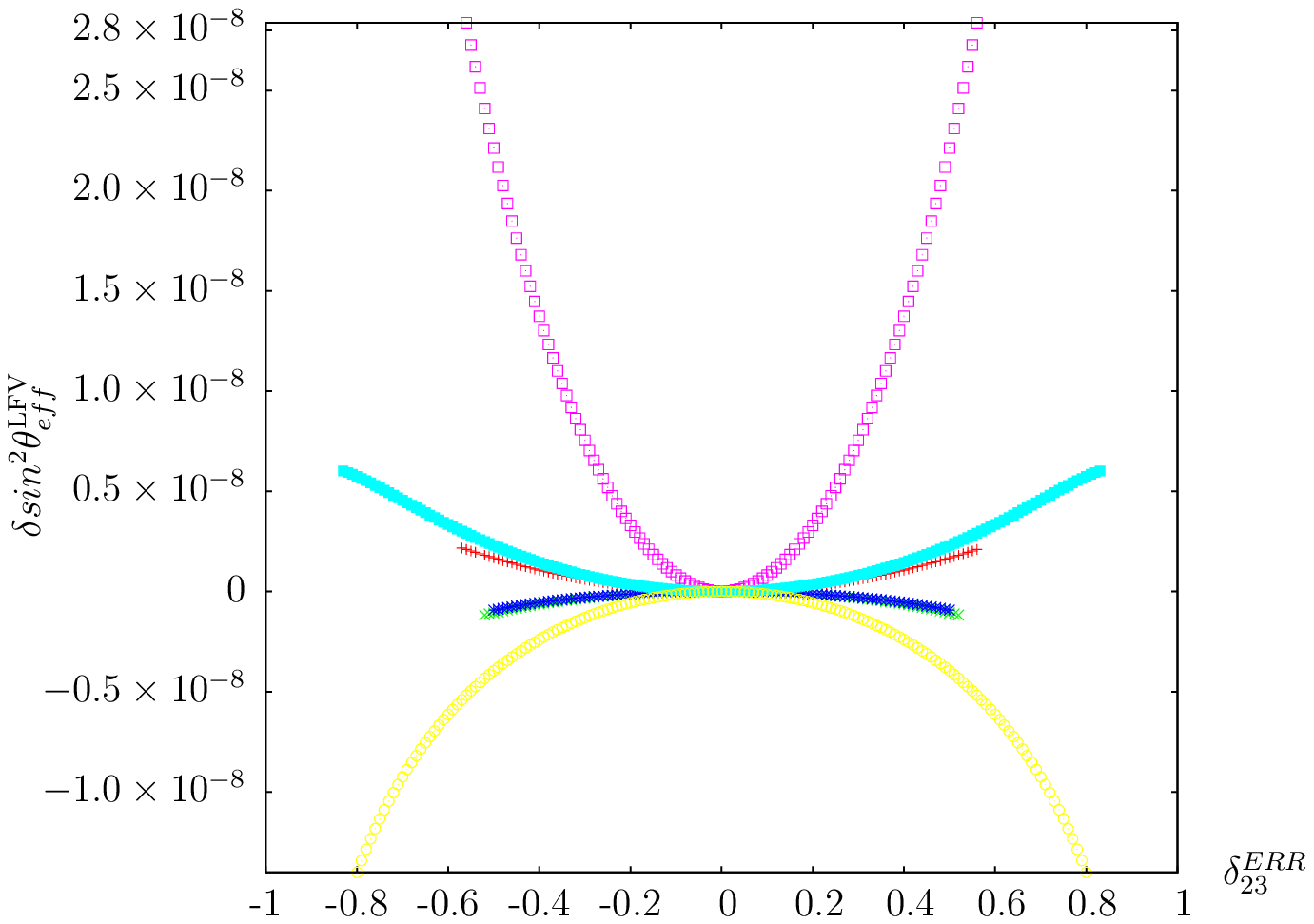 ,scale=0.53}
\psfig{file=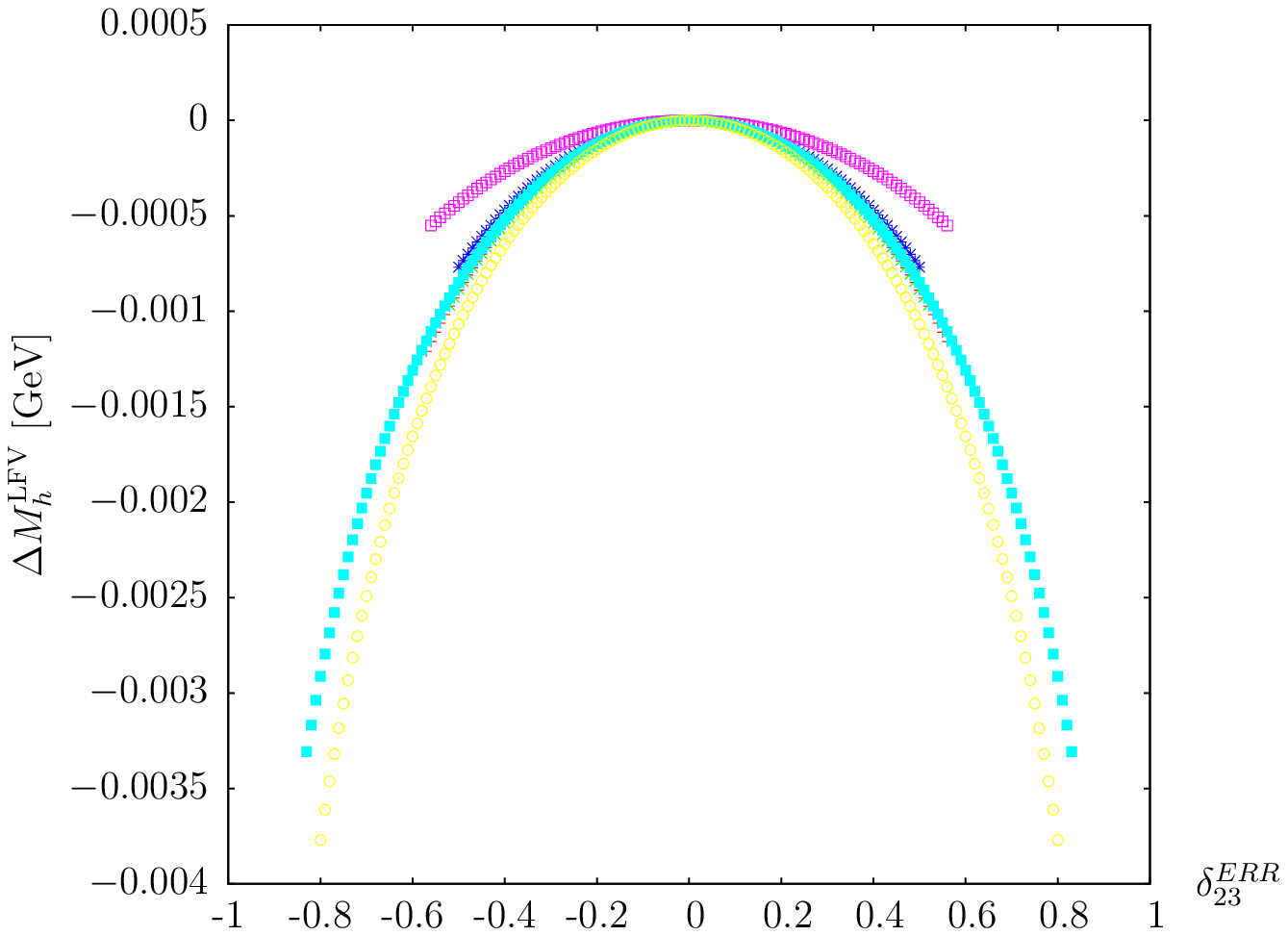   ,scale=0.53}\\
\vspace{0.5cm}
\psfig{file=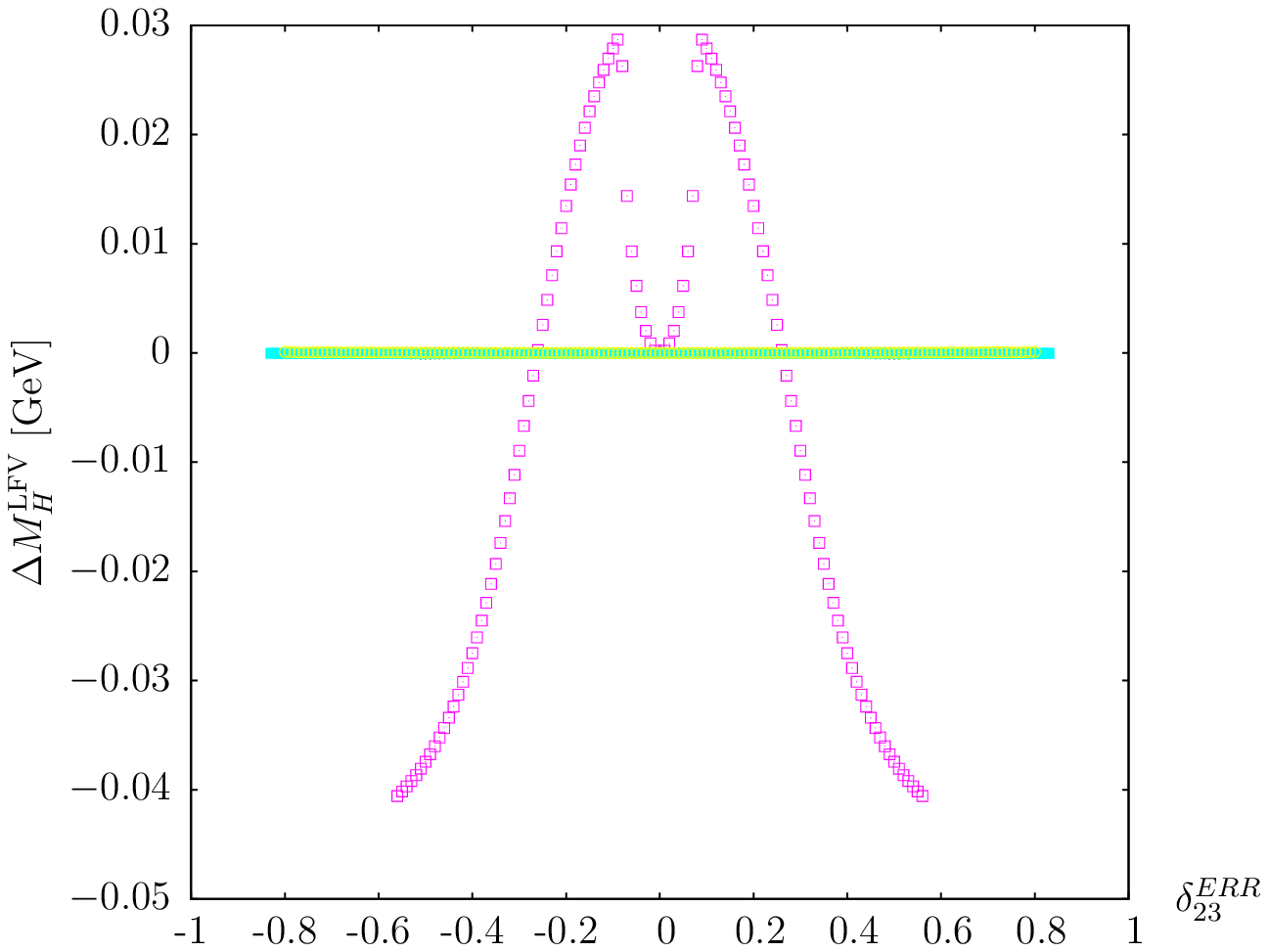  ,scale=0.53}
\hspace{0.3cm}
\psfig{file=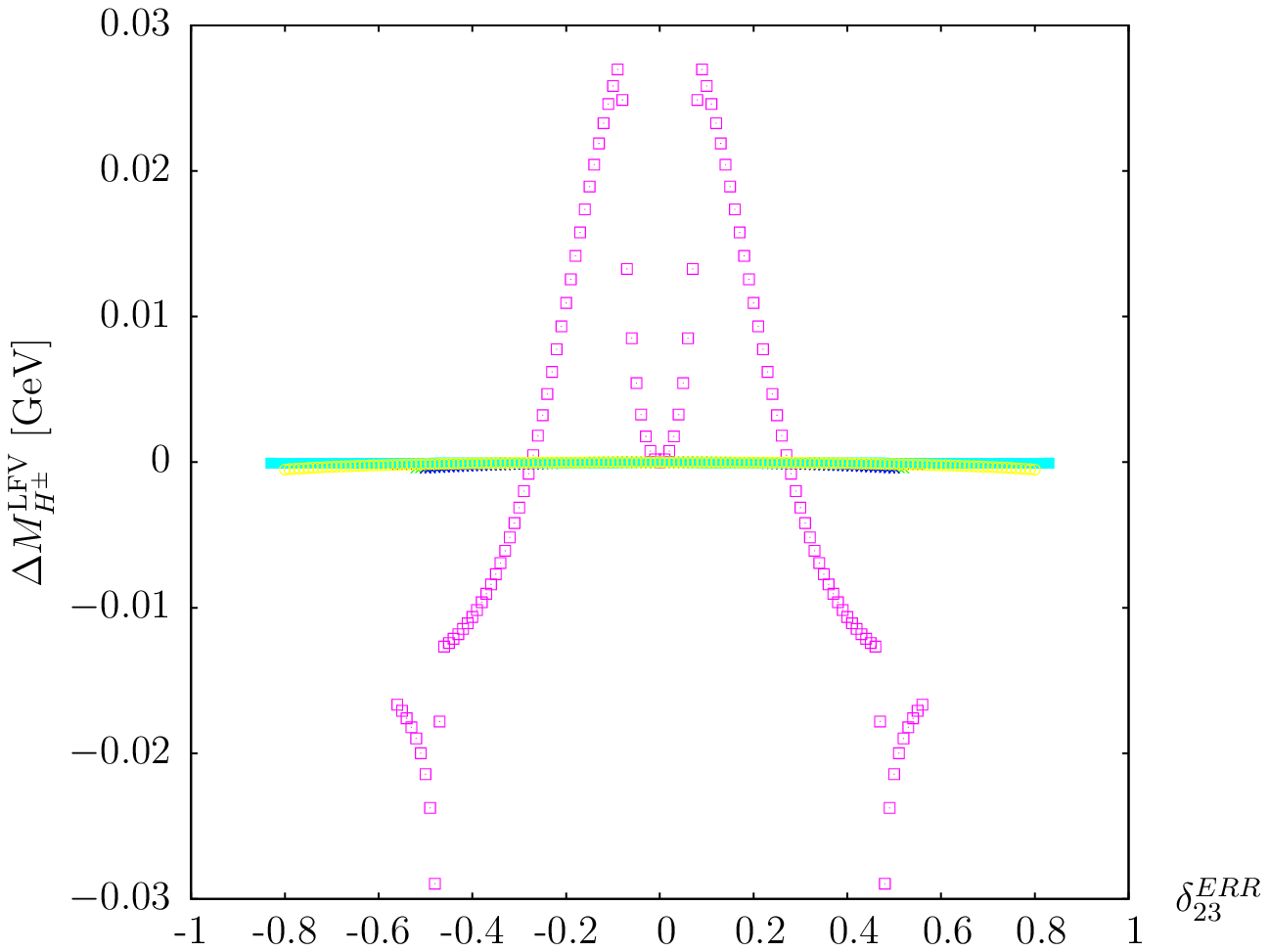  ,scale=0.53}\\

\end{center}
\caption{EWPO and Higgs masses as a function of $\delta^{ERR}_{23}$.}  
\label{figdRR23}
\end{figure}

Our numerical results are shown in \reffi{figdLL13} to \reffi{figdRR23}. 
The six plots in each figure are ordered as follows.
Upper left: $\De\rho^{\rm LFV}$, 
upper right: $\de\MW^{\rm LFV}$, 
middle left: $\de\sweff^{\rm LFV}$, 
middle right: $\De \Mh^{\rm LFV}$, 
lower left: $\De \MH^{\rm LFV}$, 
and lower right: $\De \MHp^{\rm LFV}$, 
as a function of $\de^{LL}_{13}$ (Fig.\ref{figdLL13}), 
$\de^{LLL}_{23}$ (Fig.\ref{figdLL23}),
$\de^{ELR}_{13}$ (Fig.\ref{figdLR13}), 
$\de^{ELR}_{23}$ (Fig.\ref{figdLR23}), 
$\de^{ERL}_{13}$ (Fig.\ref{figdRL13}),
$\de^{ERL}_{23}$ (Fig.\ref{figdRL23}), 
$\de^{ERR}_{13}$ (Fig.\ref{figdRR13}) 
and $\de^{ERR}_{23}$ (Fig.\ref{figdRR23}).
The legends are shown only in the first plot of each figure. 
We do not show results for LFV effects involving only the first and
second generation. While they are included for completeness in our
analytical results, they are expected to have a negligible effect on the
observables considered here. The latter is confirmed by the numerical
analysis presented in the next subsections.


Applying the most recent limits from the above listed LFV process yield
up-to-date limits on the $\deFABij$~\cite{Arana-Catania:2013nha}. 
Using the these upper bounds on $\delta^{FAB}_{ij}$, as given in the
\refta{boundsSpoints}, we calculate the corrections to the Higgs boson masses
and the EWPO. For each explored non-vanishing delta, $\delta^{FAB}_{ij}$, the 
corresponding sfermion physical masses and the sfermion rotation matrices, as
well as the EWPO and Higgs masses were numerically 
computed with {\tt FeynHiggs\,2.10.2}, where we have included the analytical
results of our calculations.


\subsection{EWPO}
\label{sec:ewpo}

\begin{figure}[ht!]
\begin{center}
\hspace{0.3cm}
\psfig{file=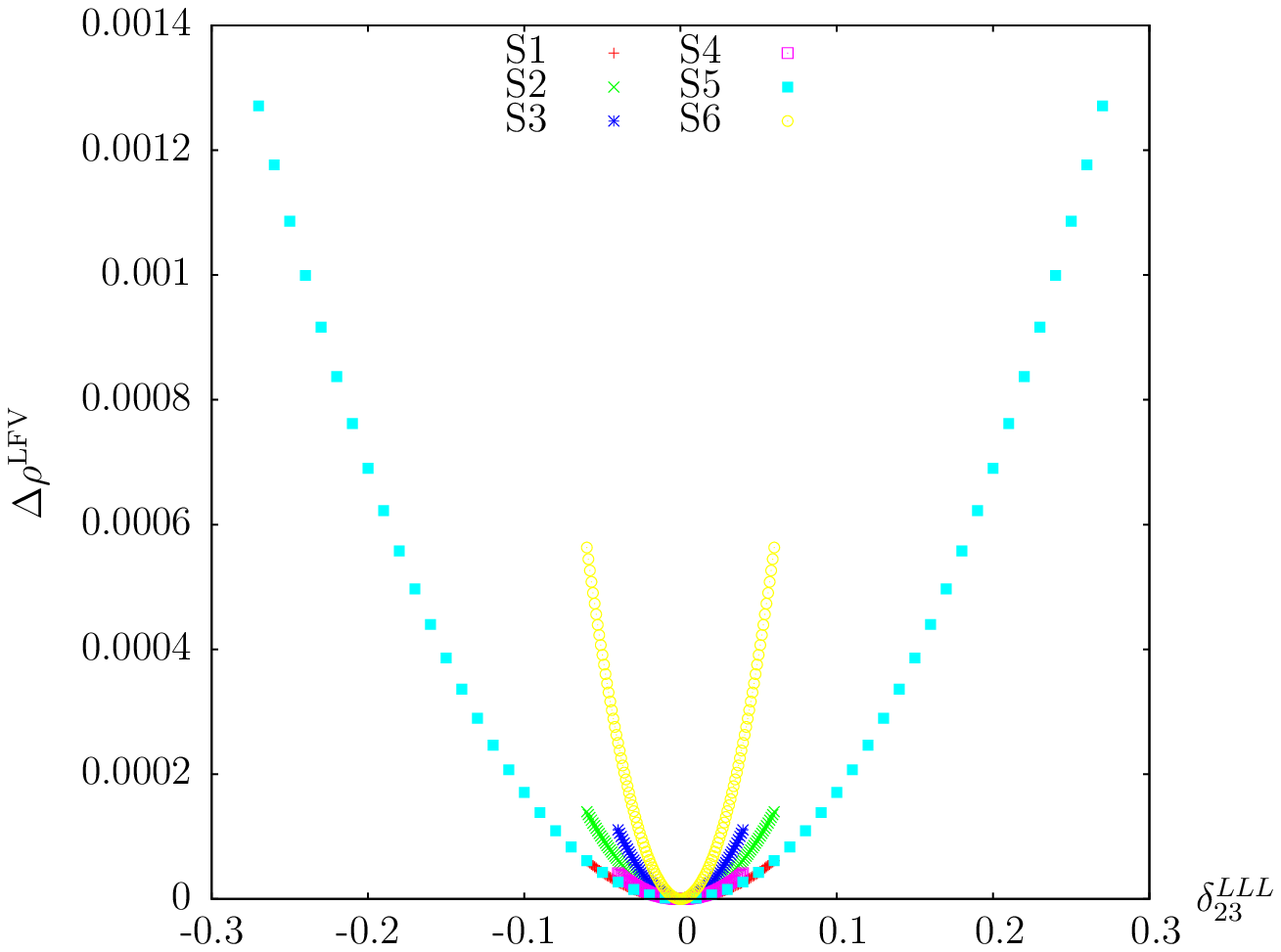  ,scale=0.53,angle=0,clip=}
\hspace{0.3cm}
\psfig{file=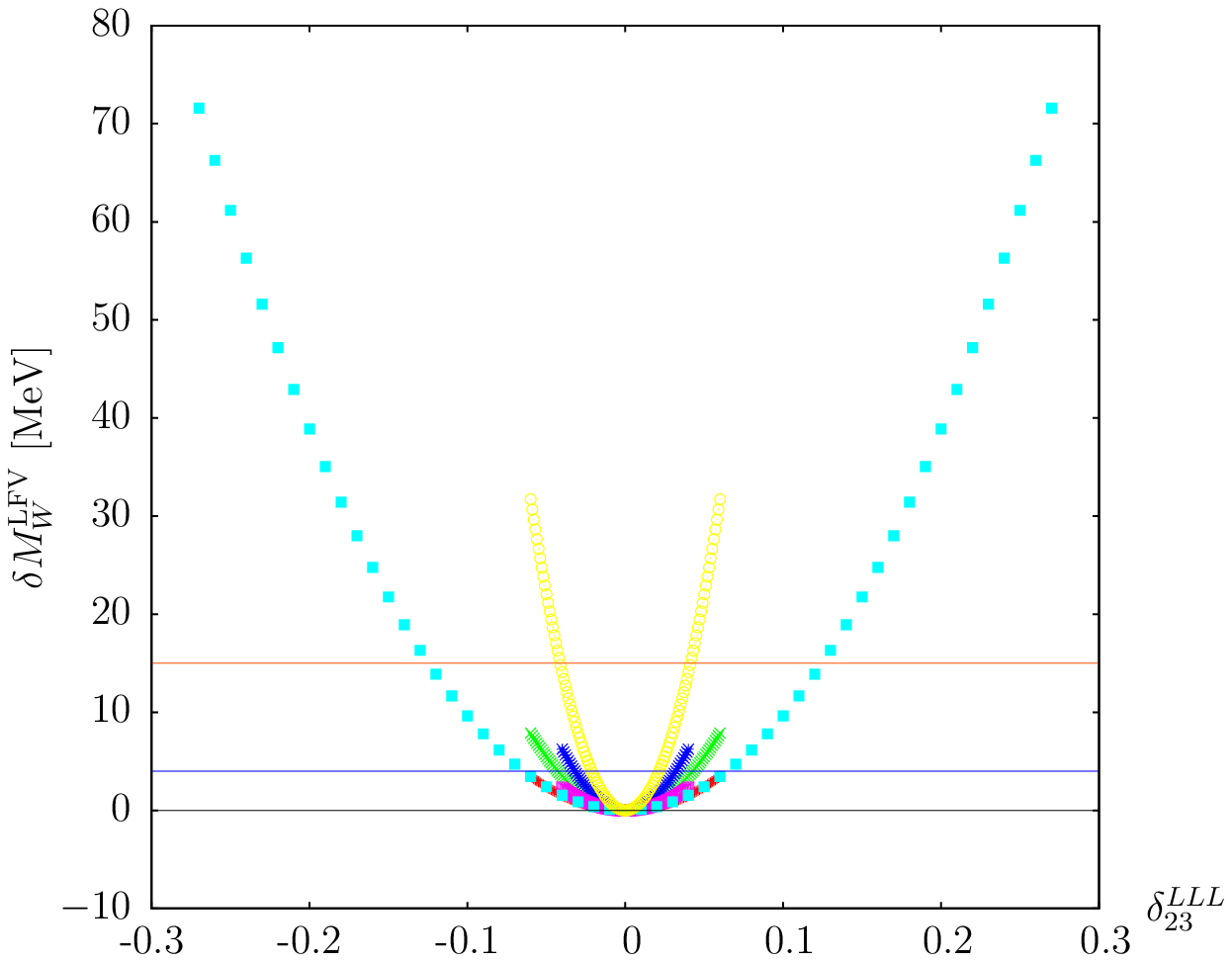  ,scale=0.53,angle=0,clip=}\\
\vspace{0.5cm}
\psfig{file=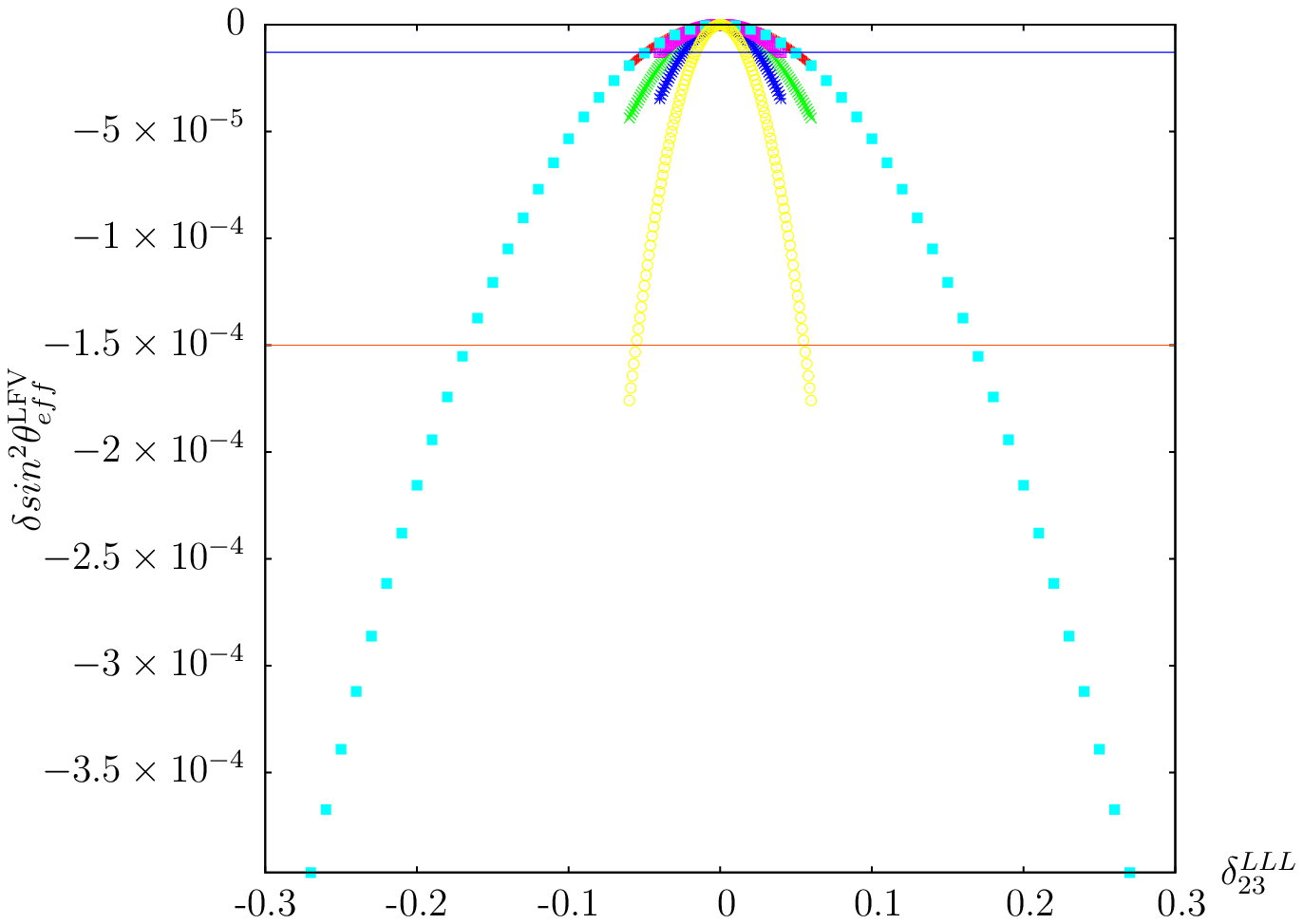 ,scale=0.53,angle=0,clip=}
\psfig{file=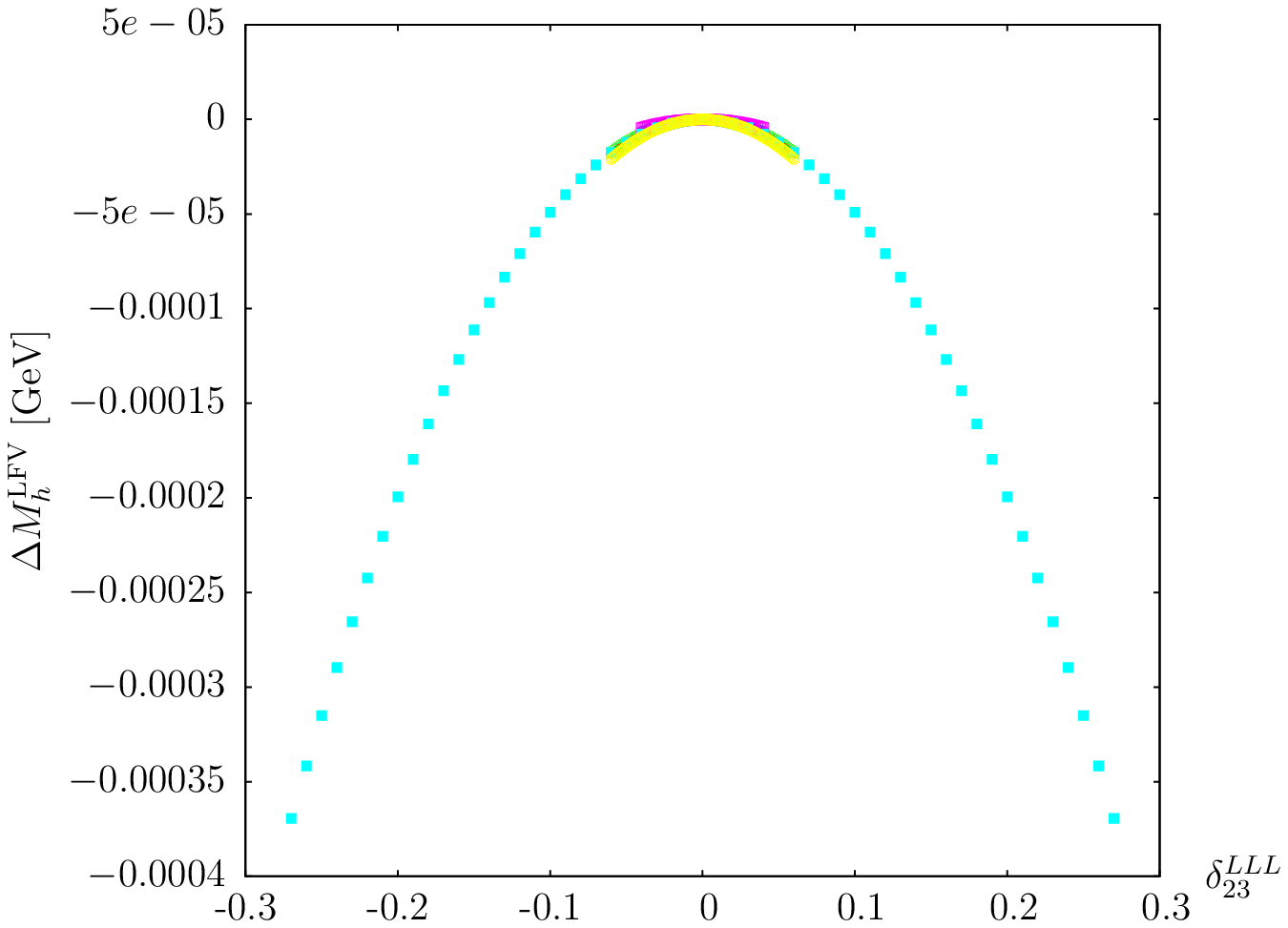   ,scale=0.53,angle=0,clip=}\\
\vspace{0.5cm}
\psfig{file=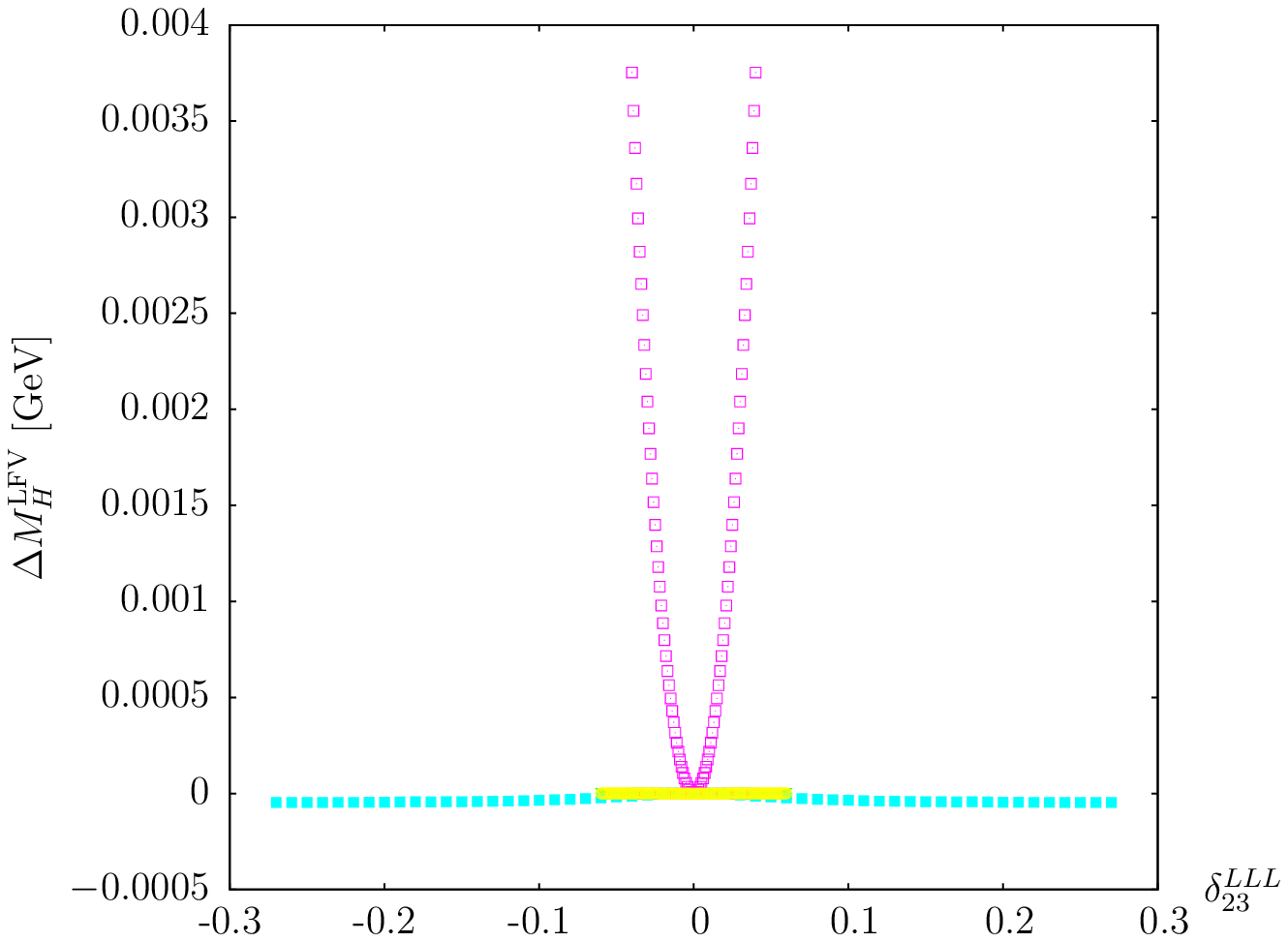  ,scale=0.53,angle=0,clip=}
\psfig{file=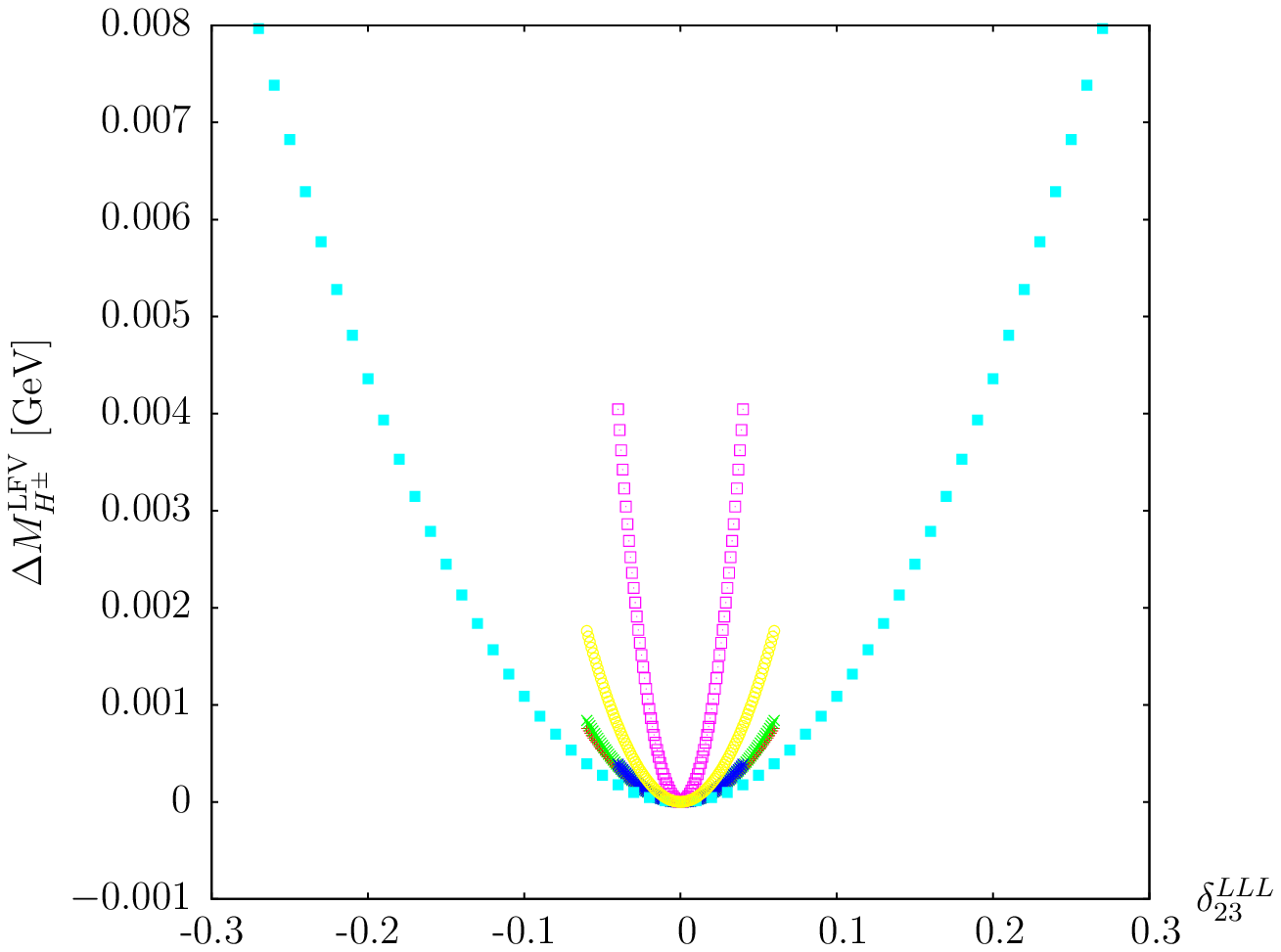  ,scale=0.53,angle=0,clip=}\\

\end{center}
\caption[EWPO and Higgs masses as a function of $\delta^{LLL}_{23}$.]{EWPO and Higgs masses as a function of $\delta^{LLL}_{23}$.
Solid red (blue) line shows the present (future) experimental uncertainty.}  
\label{figdLL23}
\end{figure}

We start with the investigation of the LFV effects on the EWPO. 
The experimental bounds on $\de^{FAB}_{12}$ where $A,B=L,R$ are very
strict (as discussed above, see \refta{boundsSpoints}) 
and does not yield sizable contribution. The bounds on the other
$\de^{FAB}_{ij}$'s are relatively less strict but still in most cases we
do not get sizable contributions for EWPO (but now can quantify their
corresponding sizes). The only sizable contribution that we get comes from
$\de^{LLL}_{23}$. The upper left plot in \reffi{figdLL23} shows our results for
$\De\rho$ as functions of $\de^{LLL}_{23}$, under the presently allowed
experimental range given in \ref{boundsSpoints}, where, depending on the
choice of the scenario (S1 \ldots S6) values of up to \order{10^{-3}} can
be reached. The largest values are found in S5, where the largest values
of $\del{LLL}{23}$ of up to $\pm 0.3$ are permitted. For the same value
of $\del{LLL}{23}$ we find the largest contributions in S6, which
possesses the relatively largest values of SSB
parameters in the slepton sector. This indicates that in general large
contributions to the EWPO are possible as soon as heavy sleptons are
involved. Consequently, while such heavy sleptons are in general
difficult to detect directly at the LHC or the ILC, their presence could
be visible in case of large LFV contributions via a shift in the EWPO. 

Turning to the (pseudo-)observables $\MW$ and $\sweff$, which are shown
in the upper right and middle left plot of \reffi{figdLL23},
respectively, we can
compare the size of the LFV contributions to the current and future
anticipated accuracies in these observables. The black line in both
plots indicates the result for $\del{LLL}{23} = 0$. The red line shows
the current level of accuracy, see \refeq{EWPO-today}, while the blue line
indicates the future ILC/GigaZ precision, see \refeq{EWPO-future}. 
We refrain from putting the absolute values of these
observables, since their values strongly depend on the choice of the
stop/sbottom sector (see \citere{PomssmRep} and references therein),
which is independent on the slepton sector under investigation here.
While the current level of accurcay only has the potential to restrict
$\del{LLL}{23}$ in~S5 and~S6, the future accuracy, in particular for $\sweff$,
can set stringent bounds in all six scenarios. 

The overall conclusion for the EWPO is that while $\del{LLL}{23}$ is most
difficult to restrict from ``conventional'' LFV observables, see
\refse{sec:limits}, it has (by far) the strongest impact on EWPO. Even with
the current precision, and even better with the (anticipated) future
accuracies, depending on the values of the scalar top/bottom sector new bounds
beyond the ``conventional'' LFV observables can be obtained.

\subsection{Higgs masses}
\label{sec:Mh}

We now turn to the effects of the LFV contribtions on the prediction of
the neutral $\cp$-even and the charged MSSM Higgs boson masses. As
discussed in \refse{sec:higgs-ho}, the theoretical accuracy should reach
a precision of $\sim 50 \mev$ in the case of $\Mh$ and about $\sim 1\%$
in the case of the heavy Higgs bosons. The calculation of $\Mh$ in the
presence of NMFV in the scalar quark
sector, as obtained in \citere{arana}, indicated that from the colored
sector corrections of \order{10 \gev} are possible (i.e.\ for NMFV
$\deFABij$ in agreement with all other precision data). Similar or even
larger corrections where found for the heavy Higgs bosons, in particular
for the mass of the charged Higgs boson. Large corrections were
connected especially to non-zero values of $\del{ULR,URL}{23}$. 
While the corrections from the scalar lepton sector are naturally much smaller
than from the scalar quark sector, it could be expected that the LFV 
contributions can exceed future and possibly even current experimental
uncertainties. In the absence of the knowledge of the exact LFV
contributions a theoretical uncertainty had to be assigned at least at
the level of \order{100 \mev} for $\Mh$ and \order{10 \gev} for $\MHp$. 
Both uncertainties are at the level (or exceeding) the future
anticipated accuracies for these Higgs-boson masses. Consequently, the
LFV have to be evaluated and analyzed in order to reach the required
level of precision.

As described above, the Higgs-boson masses are shown in the middle right
plot ($\Mh$), the lower left ($\MH$) and the lower right plot ($\MHp$)
in each figure. As expected from the NMFV analysis in the scalar quark
sector~\cite{arana}, the largest effects are found for
$\del{ELR,ERL}{23}$, but similarly for $\del{ELR,ERL}{13}$, indicating that
only the electroweak, but not the Yukawa couplings, play a relevant role
in these corrections. Contrary to the expectations, the corrections to
$\Mh$ {\em always} stay below the level of a few~MeV. While this result
eliminates the above menioned uncertainty of \order{100 \mev}, these
contributions are too small to yield a sizable numerical effect. 

Turning to the heavy Higgs bosons, the contributions to $\MH$, most
sizable again for $\del{ELR,ERL}{23,13}$, do not exceed \order{100 \mev}
and are thus effectively negligible. Substantially larger corrections
are found, in agreement with the expectations from \citere{arana} for
the charged Higgs-boson mass. They can reach the level of nearly 
$-2 \gev$, see \reffis{figdLR13} - \ref{figdRL23}. For the chosen values
of $\MA$ (or $\MHp$) this stays below the level of~1\%. However, the
absolute size of the corrections is not connected to the value of $\MHp$
in~S1-S6. Choosing starting values of $\MA$ somewhat smaller (requiering
a new evaluation of the corresponding bounds on the LFV $\deFABij$),
could yield relative corrections to $\MHp$ at the level
of~1\%. Furthremore, as in the case of the light Higgs-boson mass, the
explicit calculation of the LFV effects eliminates the theory
uncertainty associated to these effects, thus improving the theoretical
accuracy. 

\subsection{{\boldmath ${\rm BR}(h \rightarrow l_i^{\pm} l_j^{\mp})$}}
\begin{figure}[ht!]
\begin{center}
\psfig{file=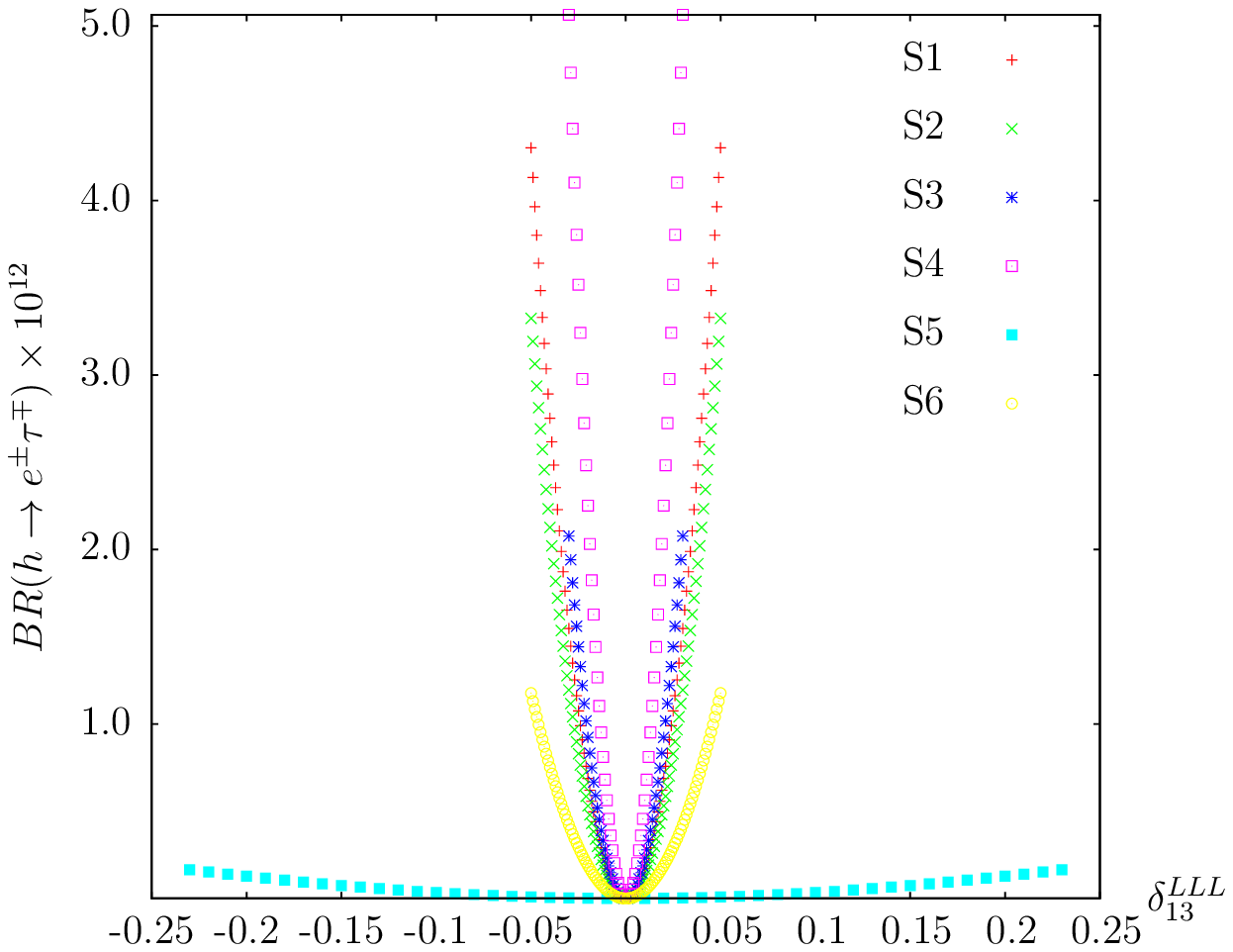  ,scale=0.52,angle=0,clip=}
\psfig{file=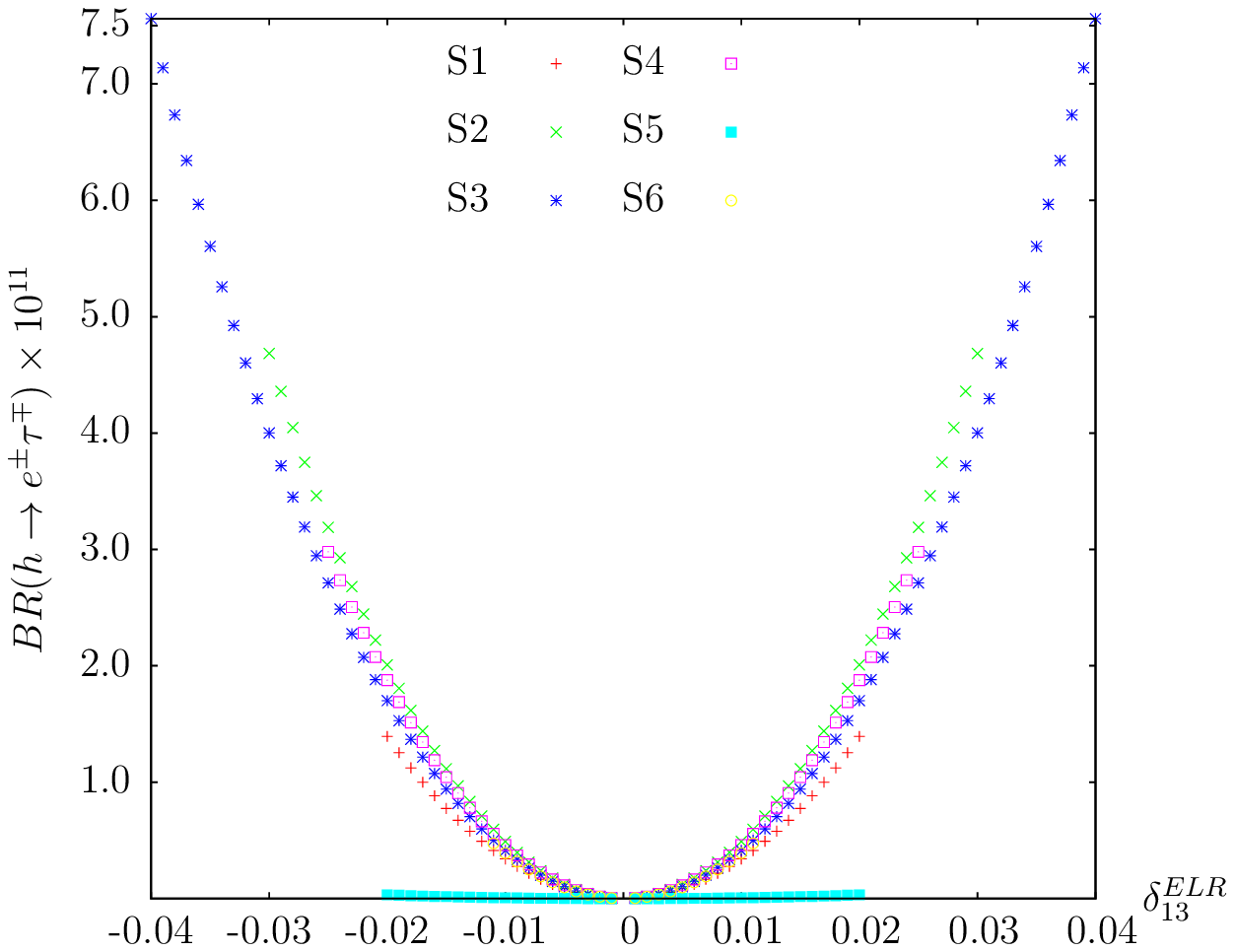  ,scale=0.52,angle=0,clip=}\\
\vspace{0.7cm}
\psfig{file=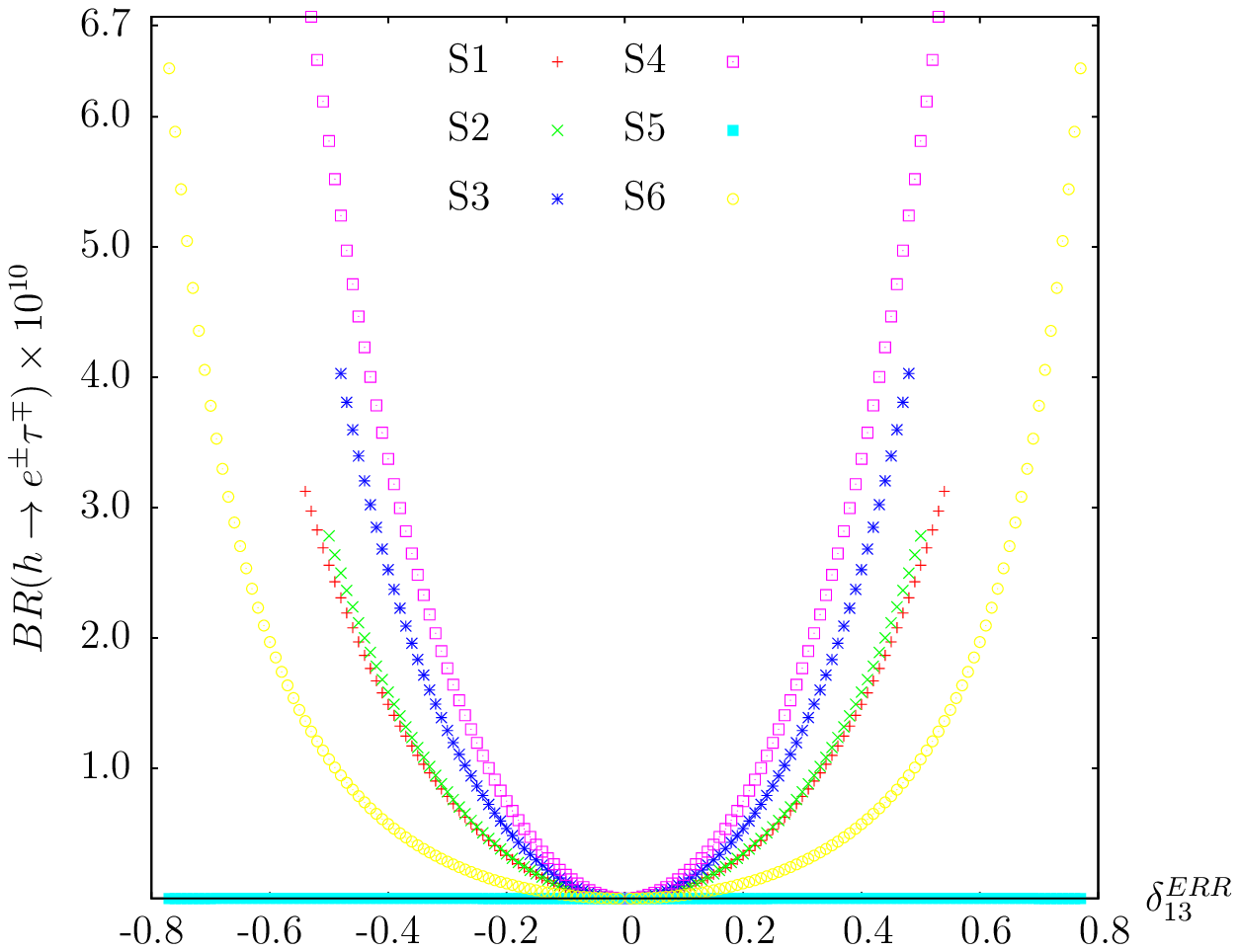 ,scale=0.52,angle=0,clip=}
\psfig{file=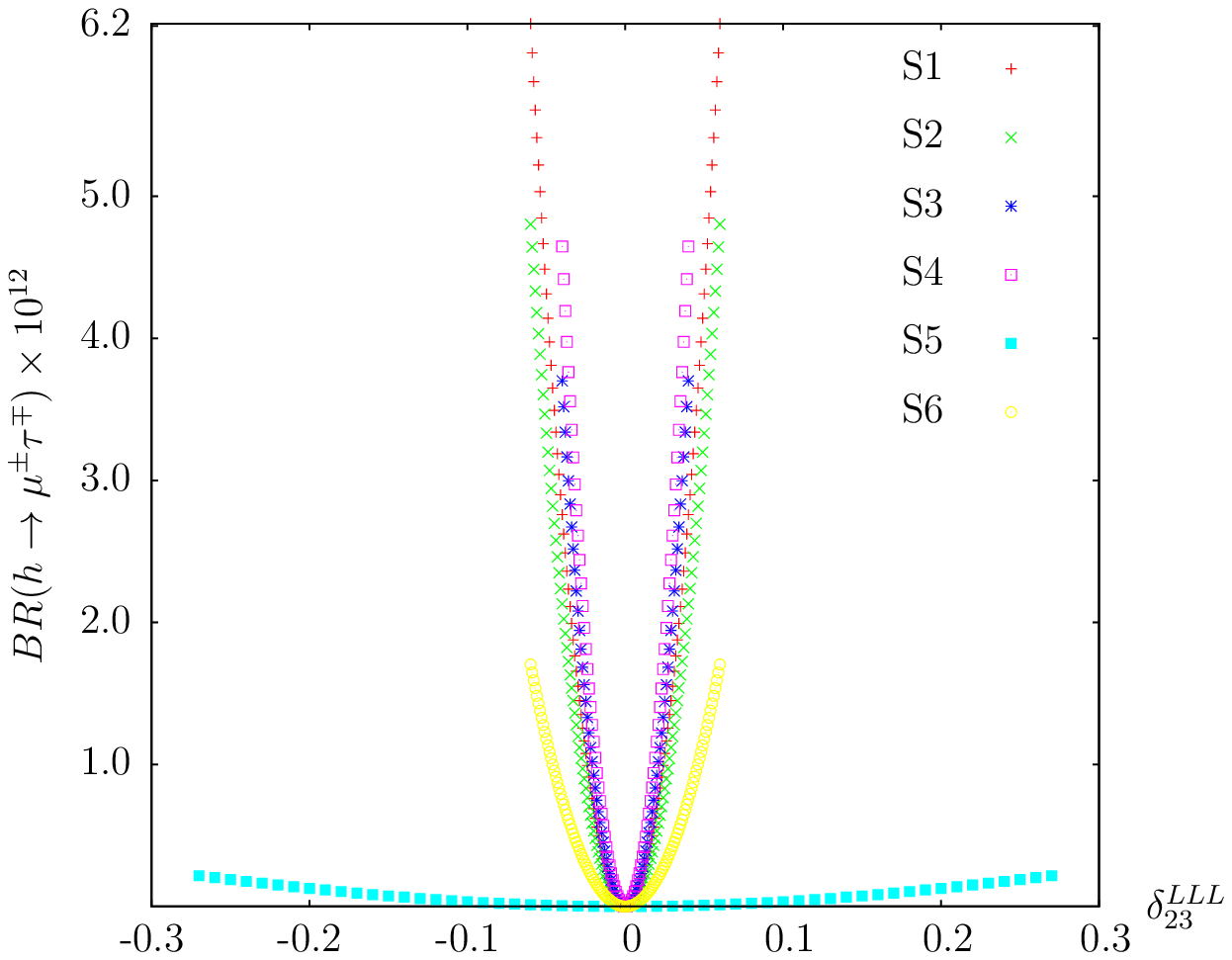   ,scale=0.52,angle=0,clip=}\\
\vspace{0.7cm}
\psfig{file=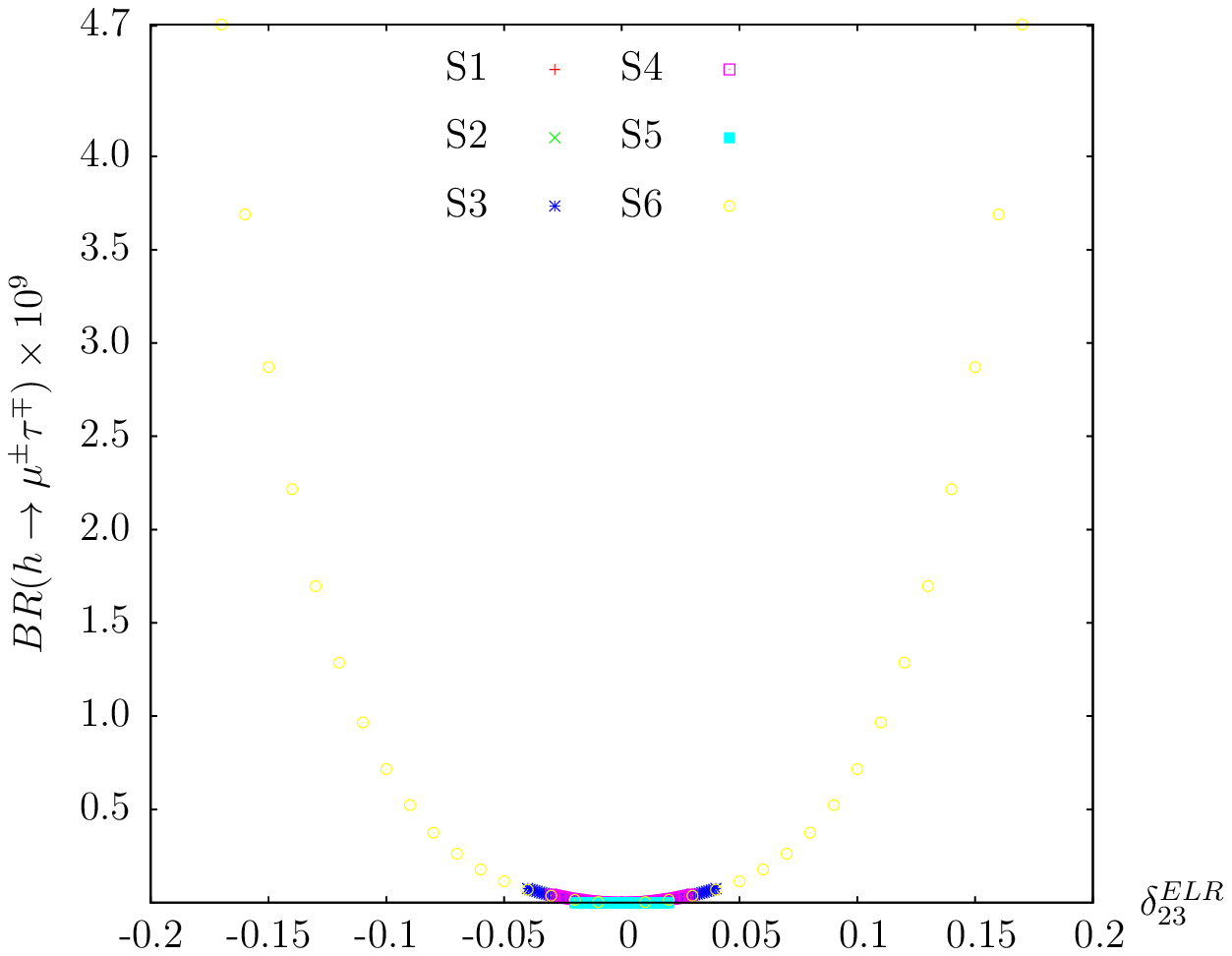  ,scale=0.52,angle=0,clip=}
\psfig{file=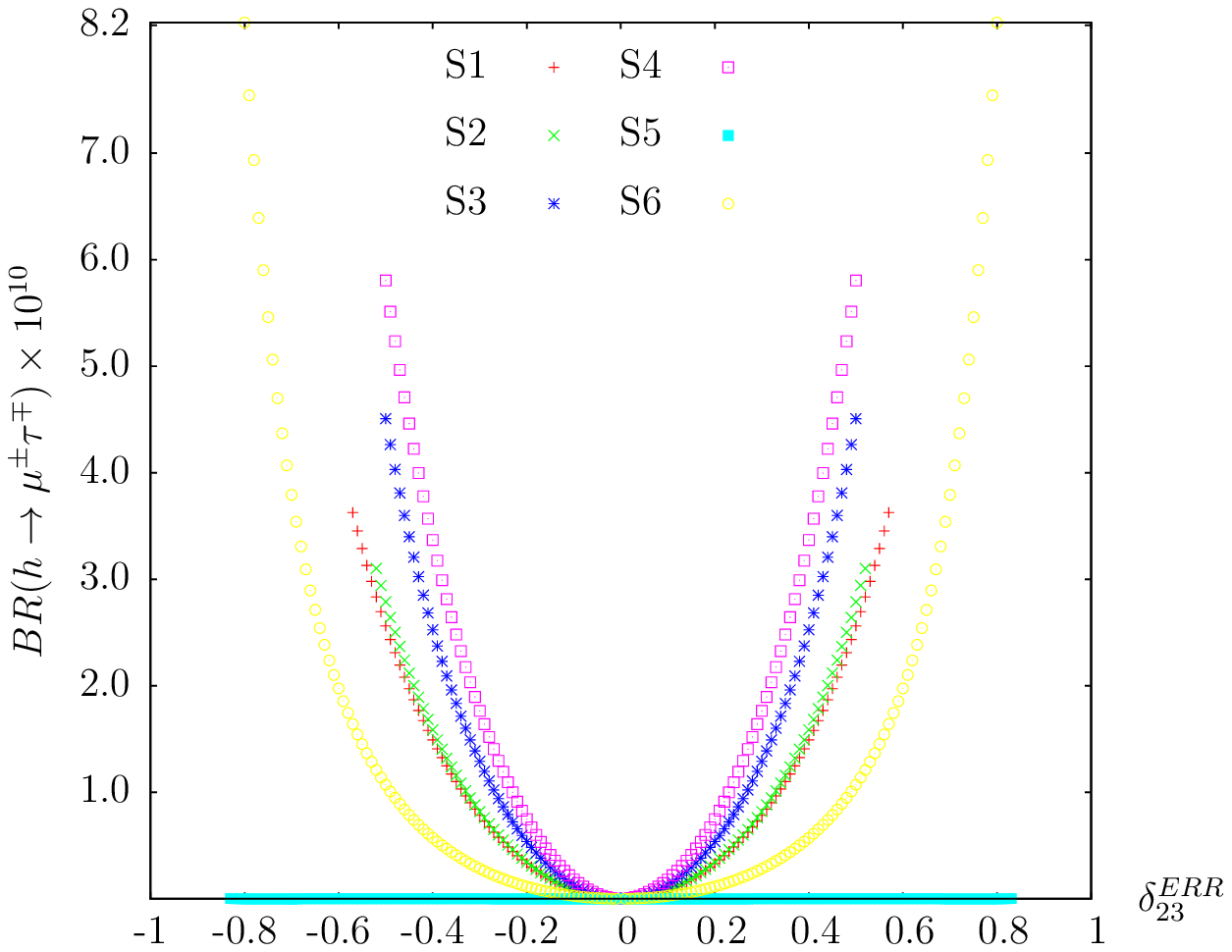  ,scale=0.52,angle=0,clip=}\\

\end{center}
\caption[LFV decays $h \rightarrow e \tau $ and $h \rightarrow \mu \tau $ as a function of slepton mixing ]{Lepton flavor violating decays $h \rightarrow e \tau $ and $h \rightarrow \mu \tau $ as a function of slepton
  mixing $\deABij$ for the six points defined in the \refta{tab:spectra}.}  
\label{fig:Hetau:Hmutau}
\end{figure}

As a last step in MI analysis, we present here the slepton mixing effects to the LFVHD. These decays were calculated using newly modified (see \refse{sec:feynhiggs}) \fa/\fc\ setup. 
The constraints from cLFV decays on slepton $\deFABij$'s are very tight and we do not expect large values for the BR's.  In \reffi{fig:Hetau:Hmutau} we present our numerical results for BR($h \rightarrow e^{\pm} \tau^{\mp} $) and BR($h \rightarrow \mu^{\pm} \tau^{\mp} $) as a function of slepton mixing $\deFABij$'s for the six points defined in the \refta{tab:spectra}. ${\rm BR}(h \rightarrow e^{\pm} \mu^{\mp})$ can only reach \order{10^{-17}} at maximum and we do not show them here. BR($h \rightarrow e^{\pm} \tau^{\mp} $) and BR($h \rightarrow \mu^{\pm} \tau^{\mp} $) can reach at most to \order{10^{-9}} for some parameter points, which is very small compared to the CMS excess \cite{CMSLFVHD}. The reason for such a small value in the experimentally allowed parameter range is the following. The same couplings namely chargino-lepton-slepton and neutralino-lepton-slepton are responsible for the cLFV decays and LFVHD, making it very difficult to find any larger values for LFVHD BR's. Our results show that if the excess shown in the CMS results\cite{CMSLFVHD} persists, we will need to find some other sources of LFV to explain CMS result. Lepton-slepton misalignment is not sufficient to explain this excess. On the other hand our results are in agreement with the ATLAS results \cite{ATLAS-LFVHD} which do not see any excess over SM background.     



\chapter{Flavor Mixing Effects in MFV CMSSM \& its Seesaw Extension}

After presenting the MI analysis in the previous chapters, here we will investigate the predictions for 
off-diagonal sfermion SSB mass terms and flavor mixing effects in 
the CMSSM and \CMSSMI.

This work is motivated by the fact that in many analyses of the CMSSM, or extensions such as the NUHM1 or NUHM2
(see \citere{AbdusSalam:2011fc} and references therein), 
the hypothesis of MFV has been used, and it has been assumed
that the contributions coming
from MFV are negligible not only for FCNC processes but for other observables like EWPO and Higgs masses as well, see, e.g.,
\citere{CMSSM-NUHM}. 
In this chapter we will analyze whether this assumption is justified, and
whether including these MFV effects could lead to additional constraints
on the CMSSM parameter space. In this respect we evaluate in the CMSSM
and in the \CMSSMI\ the following set of observables: 
\begin{itemize}
\item BPO, in particular \bsg, \bmm\ and \dmbs, 
\item EWPO, in particular $\MW$ and the
effective weak leptonic mixing angle, $\sweff$, 
\item the masses of
the neutral and charged Higgs bosons in the MSSM, 
\item QFVHD in particular \hbs, 
\item cLFV decays in particular $\mu \to e \ga$, $\tau \to e \ga$, $\tau \to \mu \ga$ as well as 
\item LFVHD in particular $h \to e^{\pm} \mu^{\mp}$, $h \to e^{\pm} \tau^{\mp}$ and $h \to \mu^{\pm} \tau^{\mp}$.
\end{itemize}
In order to perform our calculations, we used {\tt SPheno}~\cite{Porod:2003um}
to generate the CMSSM (containing also the type~I seesaw) particle
spectrum by running RGE from the GUT down to the EW scale. 
The particle spectrum was handed over in the form of an SLHA
file~\cite{SLHA} to 
\fh~\cite{feynhiggs,mhiggslong,mhiggsAEC,mhcMSSMlong,Mh-logresum} to
calculate  EWPO and Higgs boson masses. The BPO were
calculated by the {\tt BPHYSICS} subroutine included in the SuFla
code~\cite{sufla} (see also \citeres{arana,arana-NMFV2} for the improved
version used here). QFVHD and LFVHD were calculated using \fa/\fc\ setup whereas cLFV decays were calculated with {\tt SPheno 3.2.4}. The following section describes the details of our computational setup.
The results presented in this chapter were published in \cite{MFV-CMSSM}.
\section{Computational setup}
\label{sec:GUTEW}

The SUSY spectra have been generated with the code 
{\tt SPheno 3.2.4}~\cite{Porod:2003um} (for the CMSSM and the \CMSSMI). 
We defined the SLHA~\cite{SLHA} file at the GUT scale. 
In a first step within {\tt SPheno}, gauge and
Yukawa couplings at $\MZ$ scale are calculated using tree-level
formulas. Fermion masses, the $Z$~boson pole mass, the fine structure constant
$\alpha$, the Fermi constant $G_F$ and the strong coupling constant
$\alpha_s(\MZ)$ are used as input parameters. The gauge and Yukawa
couplings, calculated at $\MZ$, are then used as 
input for the one-loop
RGE's to obtain the corresponding values at the GUT scale
which is calculated from the requirement that $g_1 = g_2$. 
The CMSSM boundary
conditions are then applied to the complete set of
two-loop RGE's and are evolved to the EW scale.  
At this point the SM and SUSY radiative
corrections are applied to the gauge and Yukawa couplings, and the 
two-loop RGE's are again evolved to GUT scale. 
After applying the CMSSM  boundary conditions again
the two-loop RGE's are run down to EW scale to get SUSY spectrum. This
procedure is iterated until the required precision is achieved. 
The output is then written in the form of an SLHA, file which is used as 
input to calculate low energy observables discussed below. 

For our scans of the \CMSSMI\ parameter space we use 
{\tt SPheno 3.2.4}~\cite{Porod:2003um} with
the model ``see-saw type-I'' and apply a similar procedure to that in the CMSSM case. The neutrino
related input parameters are
included in the respective SLHA input blocks (see \citere{SLHA} for
details). The predictions for $\br(l_i \to l_j \gamma)$ 
are also obtained with {\tt SPheno 3.2.4}, see the discussion in \refse{sec:Sl}.  
We checked that the use of this code produces results similar to the ones obtained by our private codes used
in \citere{Cannoni:2013gq}.

\section{Input parameters}

In order to get an overview about the size of the effects in the CMSSM
parameter space, the relevant parameters $m_0$, $m_{1/2}$ have been
scanned as, or in case of $A_0$ and
$\tb$ have been set to all combinations of 
\begin{align}
m_0 &\eq 500 \gev \ldots 5000 \gev~, \\
m_{1/2} &\eq 1000 \gev \ldots 3000 \gev~, \\
A_0 &\eq -3000, -2000, -1000, 0 \gev~, \\
\tb &\eq 10, 20, 35, 45~,
\end{align}
with $\mu > 0$.  
Primarily we are not interested in the absolute values for EWPO BPO and Higgs masses but the
effects that comes from flavor violation within the MFV framework,
i.e.\ the effect from the off-diagonal entries in the sfermion mass matrices.
We first calculate the low-energy observables by setting all $\deFABij=0$
by hand. In a second step we evaluate the observables with the values of 
$\deFABij$ obtained through RGE running. We then evaluate the ``pure
MFV effects'', 
\begin{align}
\Dbsg &\eq \br(B \to X_s \gamma) - \br^{\rm MSSM}(B \to X_s \gamma)~, \\
\Dbmm &\eq \br(B_{s} \to \mu^+ \mu^-) 
         - \br^{\rm MSSM}(B_{s} \to \mu^+ \mu^-)~, \\
\Ddmbs &\eq \De M_{B_s} - \De M_{B_s}^{\rm MSSM}~,
\end{align}
where $\br^{\rm MSSM}(B \to X_s \gamma)$, 
$\br^{\rm MSSM}(B_s \to \mu^+ \mu^-)$ and $\De M_{B_{S}}^{\rm MSSM}$
corresponds to the  values of relevant observables with all $\deFABij =
0$. Furthermore we define 
\begin{align}
\DMh &\eq \Mh - \Mh^{\rm MSSM} \\
\DMH &\eq \MH - \MH^{\rm MSSM} \\
\DMHp &\eq \MHp - \MHp^{\rm MSSM}
\end{align}
where $\Mh^{\rm MSSM}$, $\MH^{\rm MSSM}$ and 
$\MHp^{\rm MSSM}$ corresponds to the 
Higgs masses with all $\deFABij = 0$.  Similarly we define for the EWPO
\begin{align}
\Drho &\eq \De\rho-\De\rho^{\rm MSSM} \\
\DMW &\eq \MW-\MW^{\rm MSSM} \\
\Dsweff &\eq \sweff-\sweff^{\rm MSSM}
\end{align}
where $\De\rho^{\rm MSSM}$, $\MW^{\rm MSSM}$ and $\sweff^{\rm MSSM}$ are the
values of the relavant observables with all $\deFABij = 0$. 

\section{Effects of squark mixing in the CMSSM}
\label{sec:Sq}

In this section we analyze the effects from RGE induced flavor violating
mixing in the scalar quark sector in the CMSSM (i.e.\ with no mixing in
the slepton sector). The RGE running from the GUT scale to the EW has
been performed as described in \refse{sec:GUTEW}, with the subsequent
evaluation of the low-energy observables as discussed in
Chap:~\ref{precision-observables}. 

In \reffis{fig:DelQLL13}-\ref{fig:Sq-MH-BPO} we show the results of
our CMSSM analysis in the $m_0$--$m_{1/2}$ plane for four
different combinations of $\tb = 10, 45$ (left and right column) 
and $A_0 = 0, -3000 \gev$ (upper and lower row).
This set represents four ``extreme'' cases of the parameter space and give
an overview about the possible sizes of the effects and their
dependences on $\tb$ and $A_0$ (which we verified with other, not
shown, combinations).

\subsection{Squark \boldmath{$\deFABij$'s}}
\begin{figure}[ht!]
\begin{center}
\psfig{file=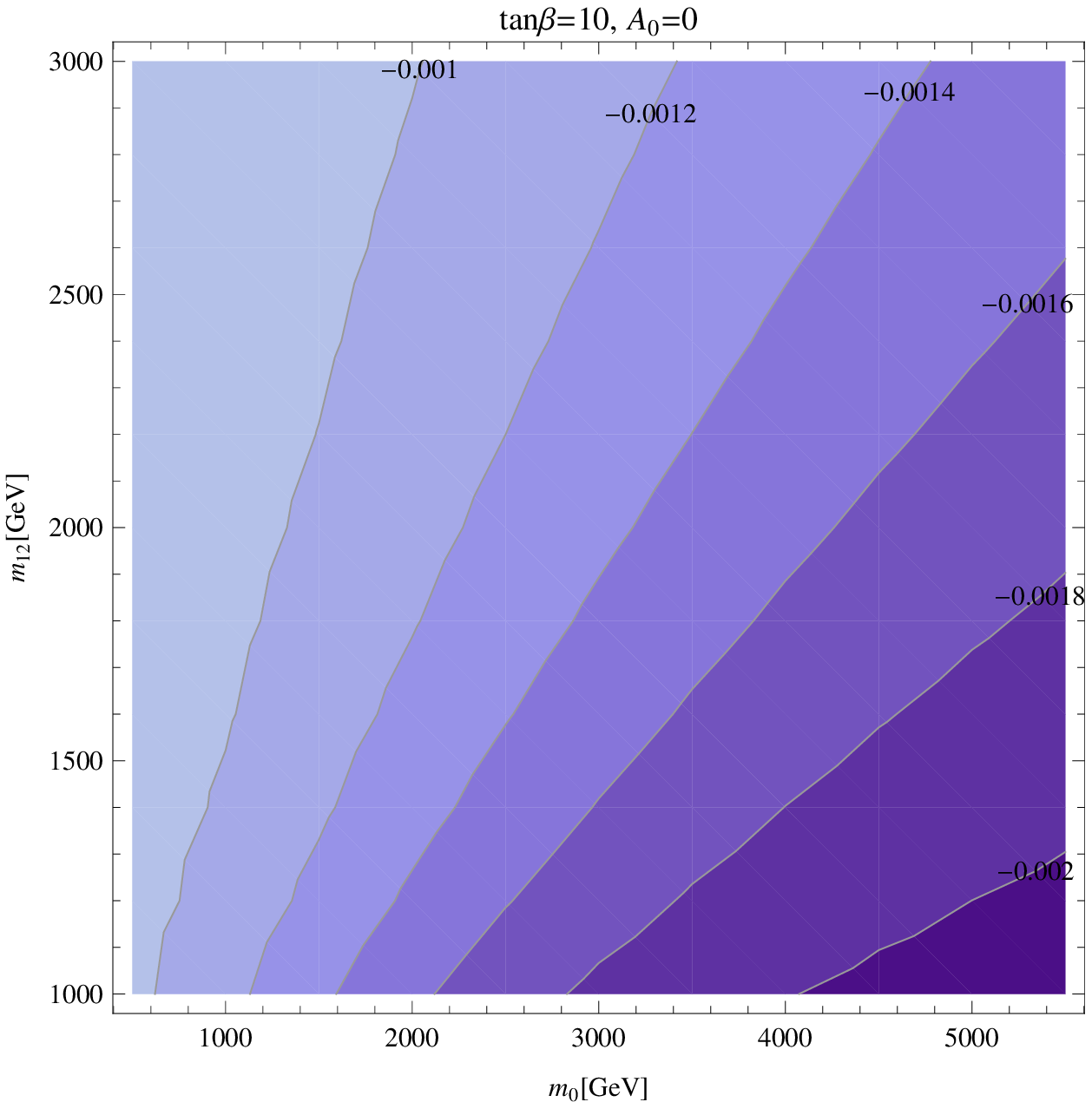  ,scale=0.51,angle=0,clip=}
\psfig{file=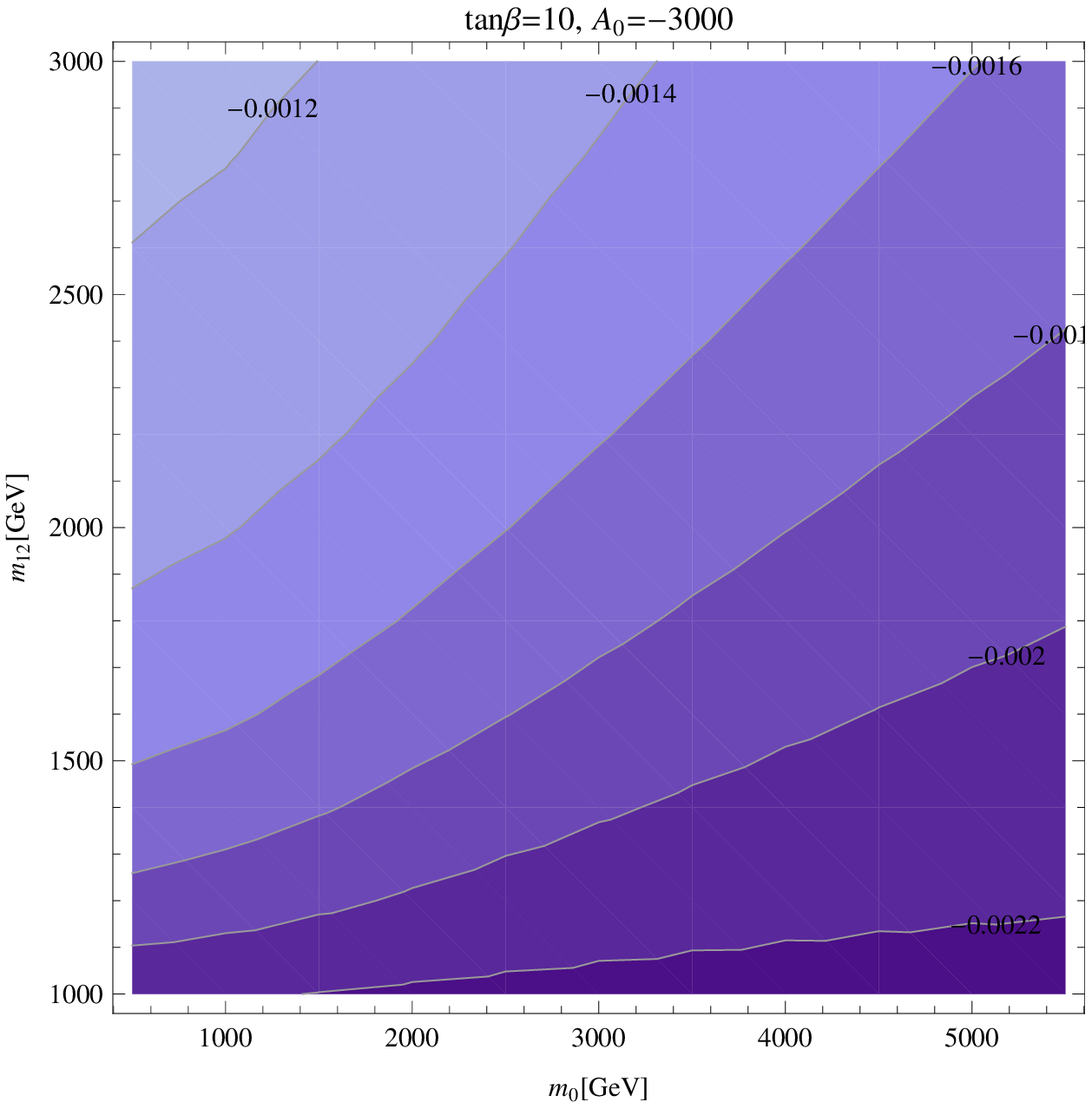  ,scale=0.51,angle=0,clip=}\\
\vspace{0.2cm}
\psfig{file=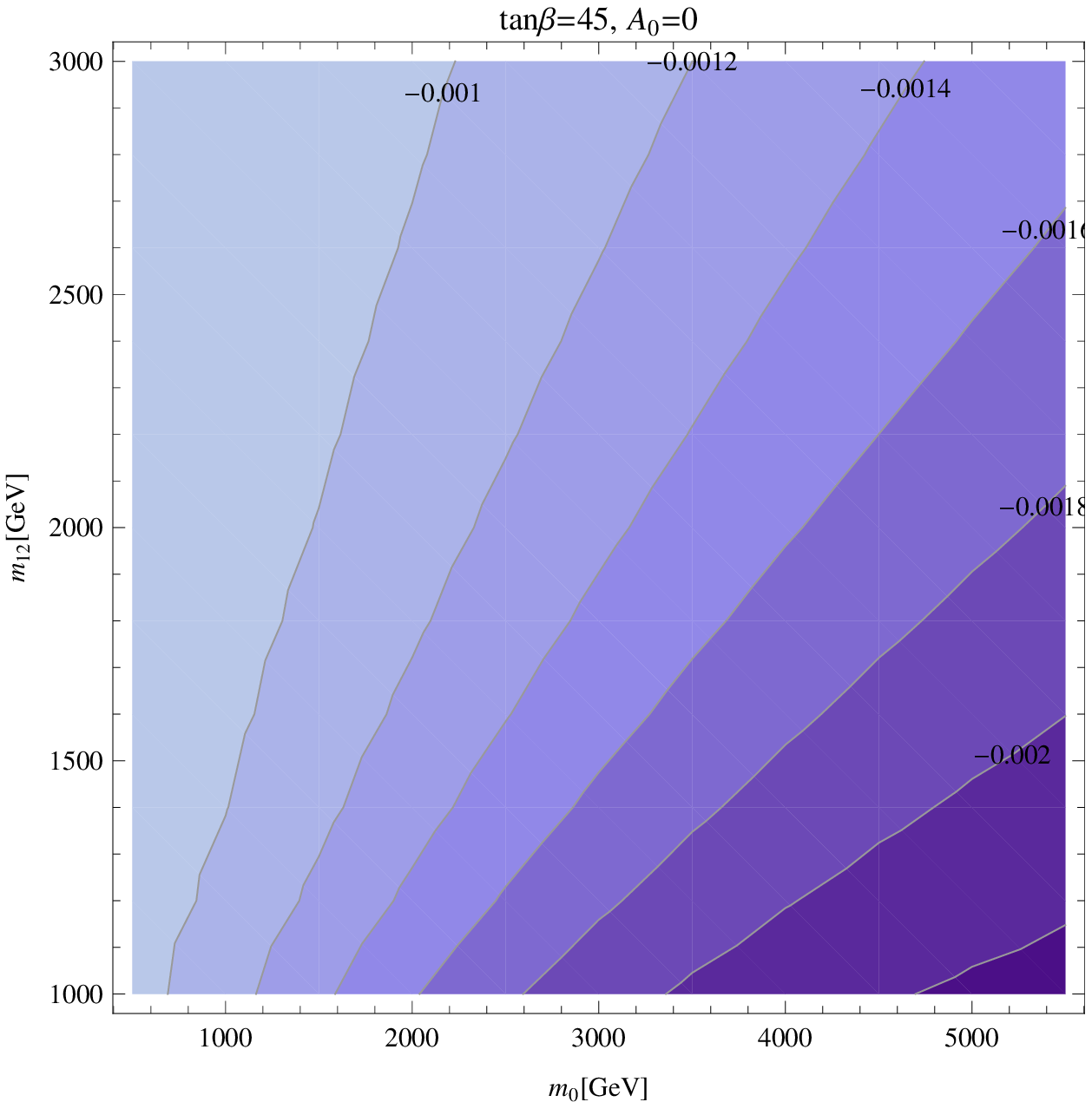 ,scale=0.51,angle=0,clip=}
\psfig{file=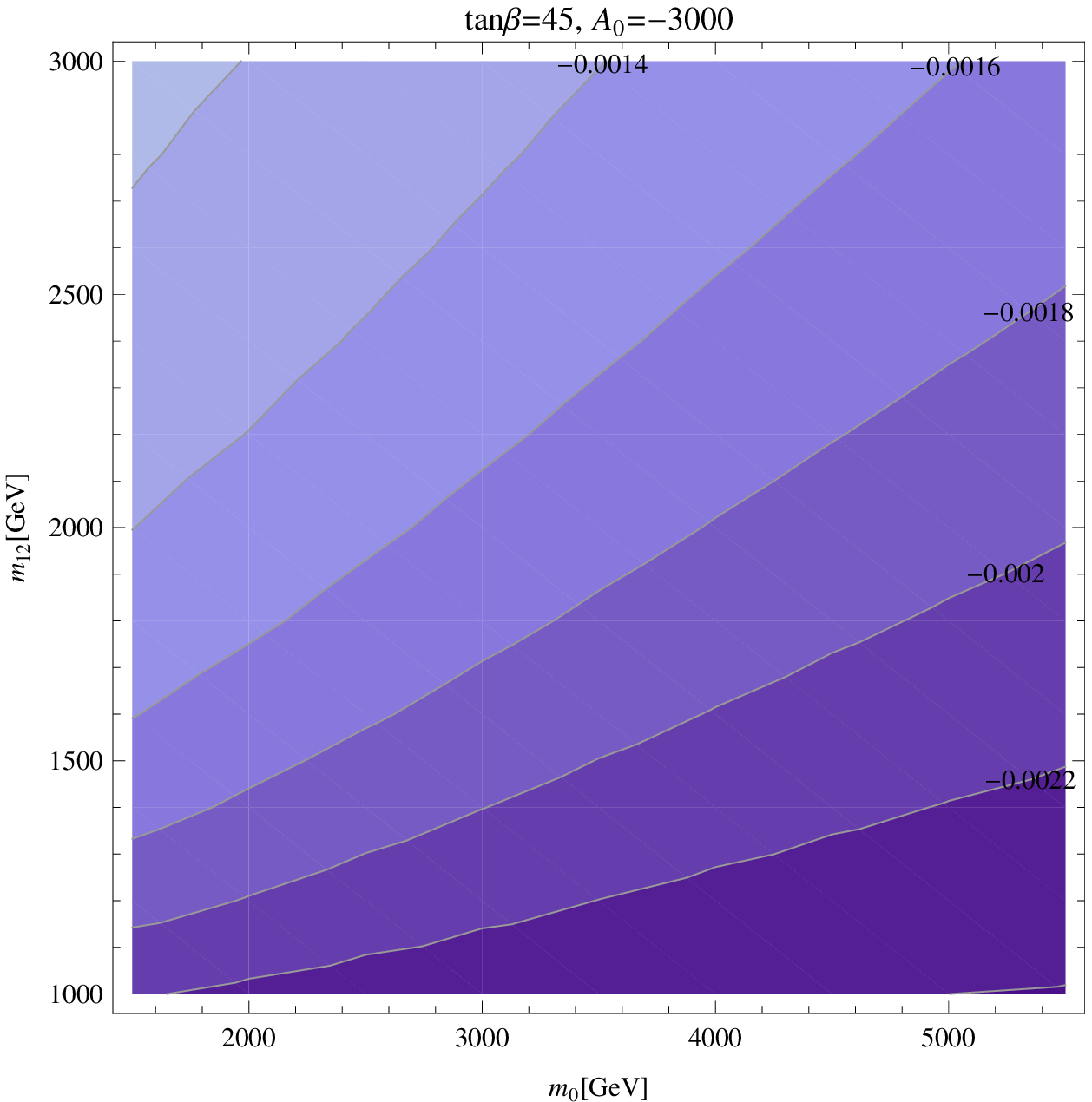   ,scale=0.51,angle=0,clip=}\\
\vspace{-0.2cm}
\end{center}
\caption[Contours of $\delta^{QLL}_{13}$  in the
  $m_0$--$m_{1/2}$ plane.]{Contours of $\delta^{QLL}_{13}$  in the
  $m_0$--$m_{1/2}$ plane for different values of $\tb$ and $A_0$ in
  the CMSSM.}   
\label{fig:DelQLL13}
\end{figure}

\begin{figure}[ht!]
\begin{center}
\psfig{file=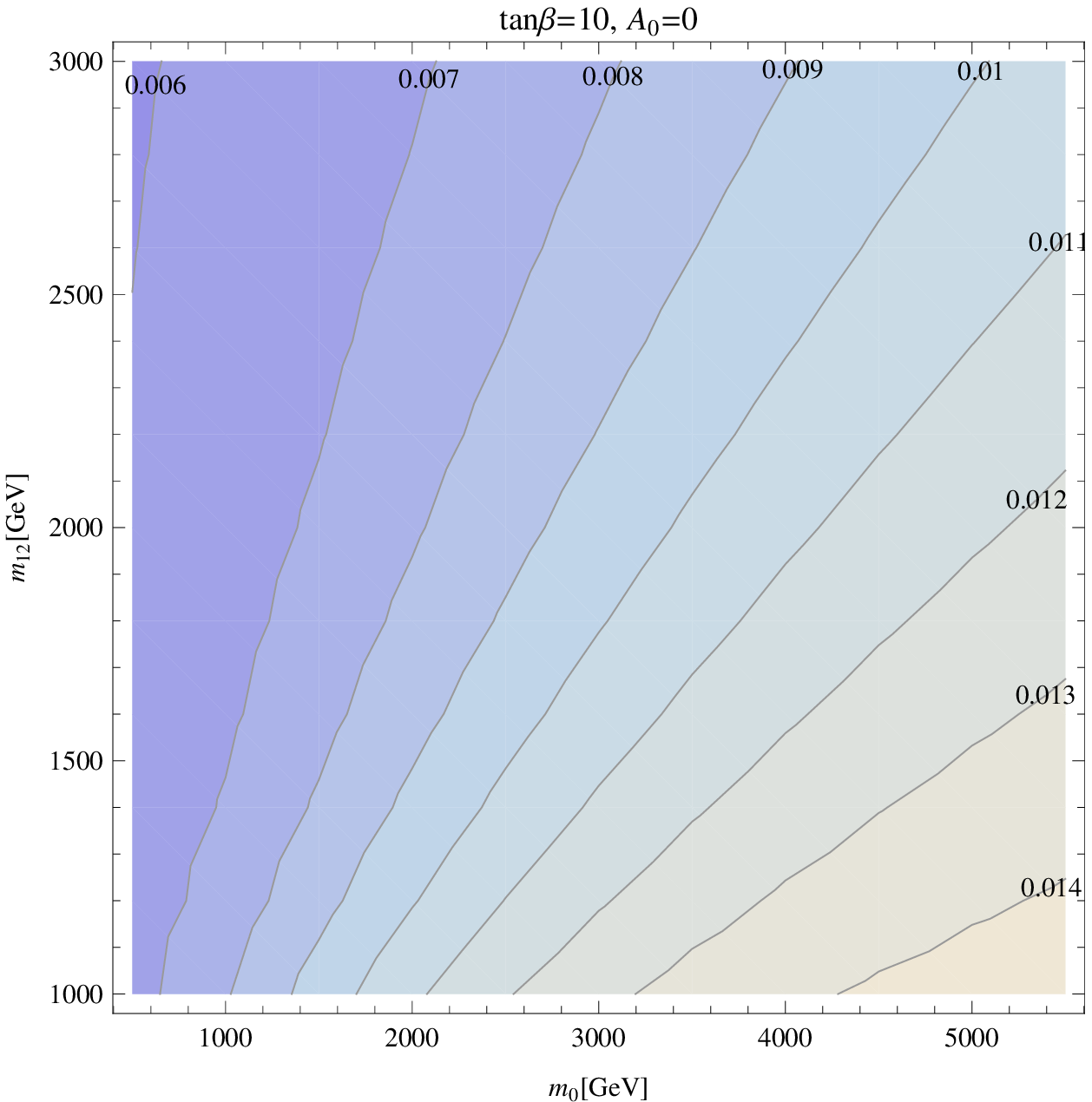  ,scale=0.51,angle=0,clip=}
\psfig{file=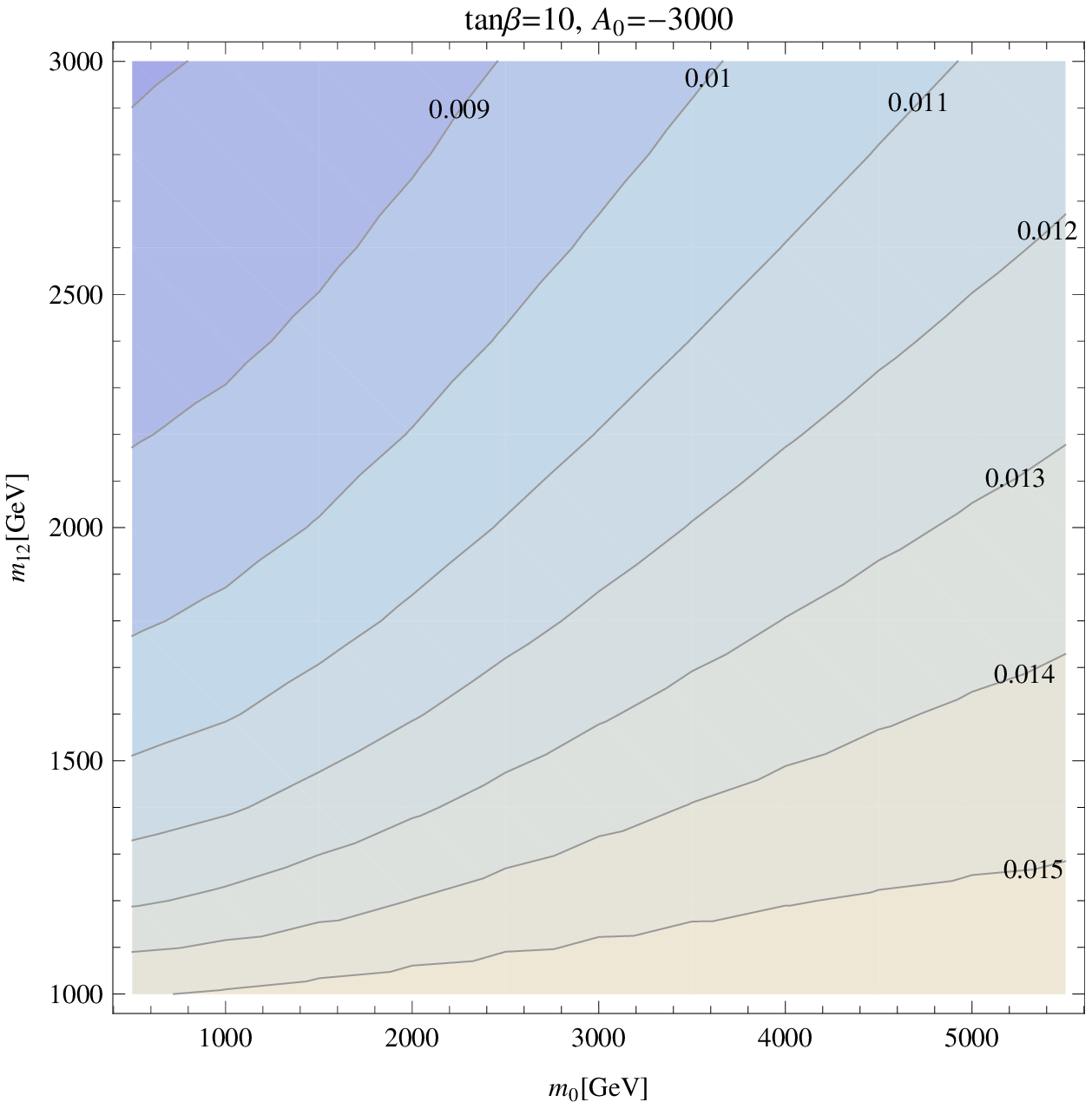  ,scale=0.51,angle=0,clip=}\\
\vspace{0.2cm}
\psfig{file=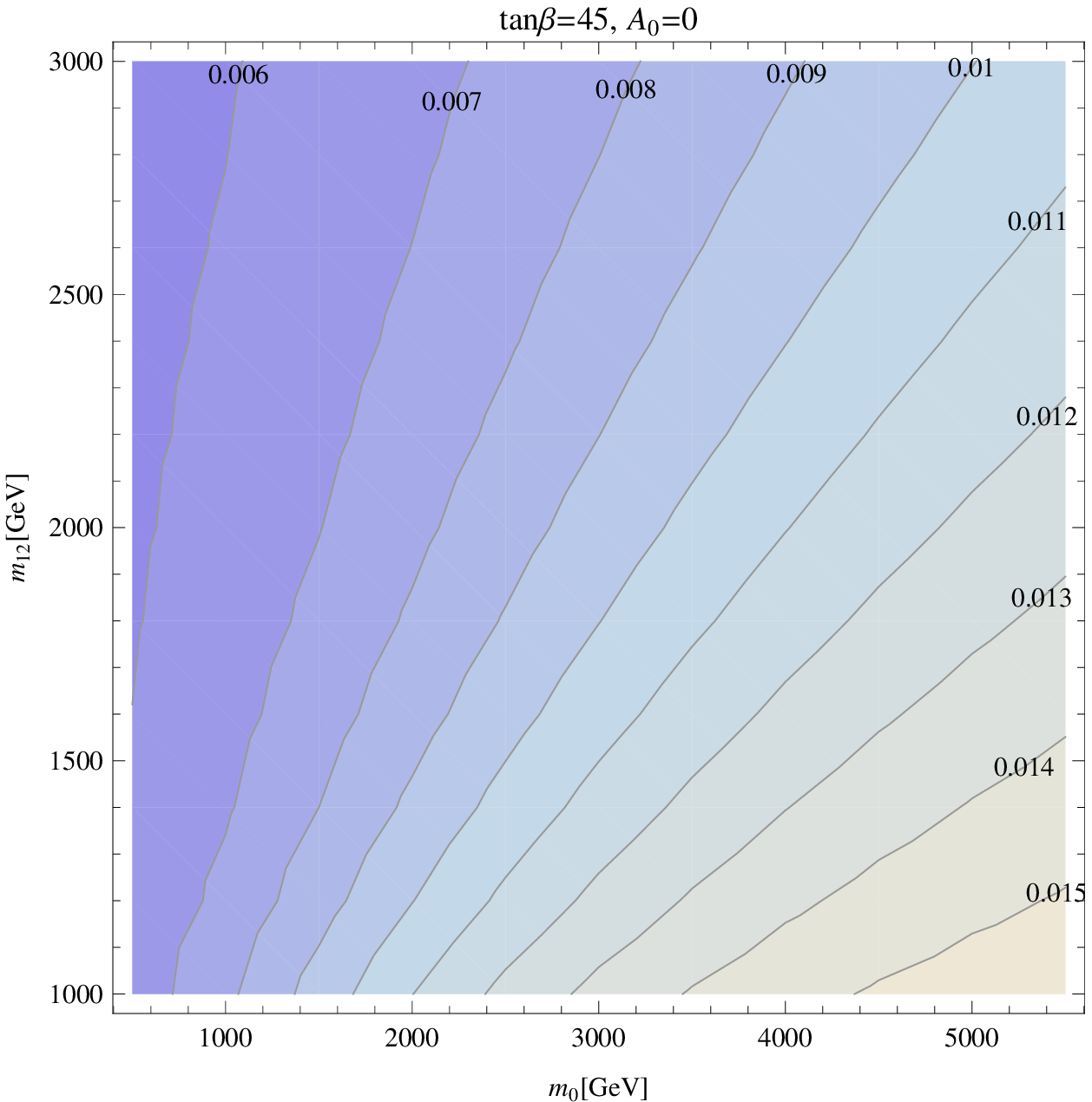 ,scale=0.51,angle=0,clip=}
\psfig{file=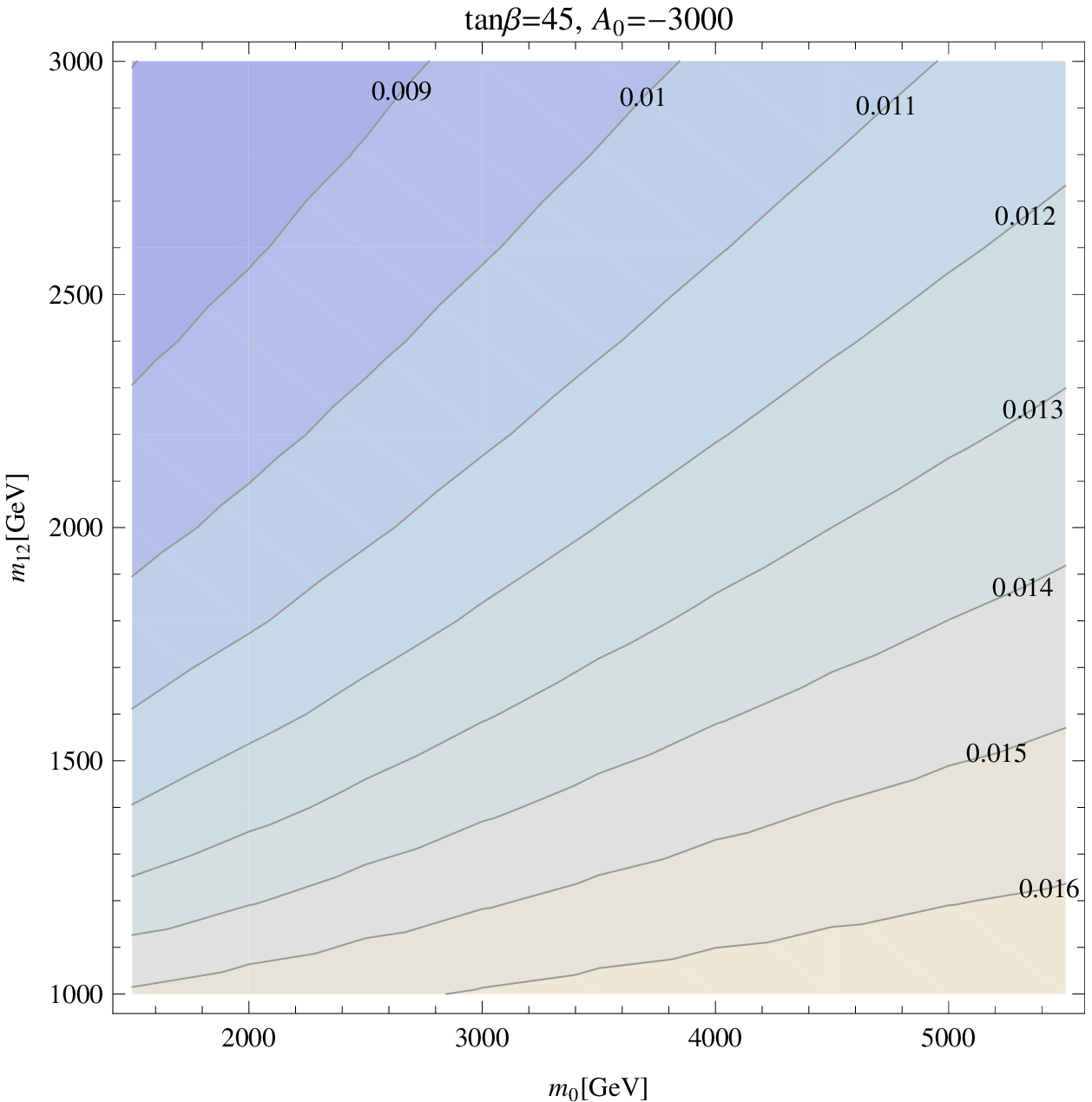   ,scale=0.51,angle=0,clip=}\\
\vspace{-0.2cm}
\end{center}
\caption[Contours of $\delta^{QLL}_{23}$ in the
  $m_0$--$m_{1/2}$ plane.]{Contours of $\delta^{QLL}_{23}$ in the
  $m_0$--$m_{1/2}$ plane for different values of $\tb$ and    
$A_0$ in the CMSSM.}  
\label{fig:DelQLL23}
\end{figure}

\begin{figure}[ht!]
\begin{center}
\psfig{file=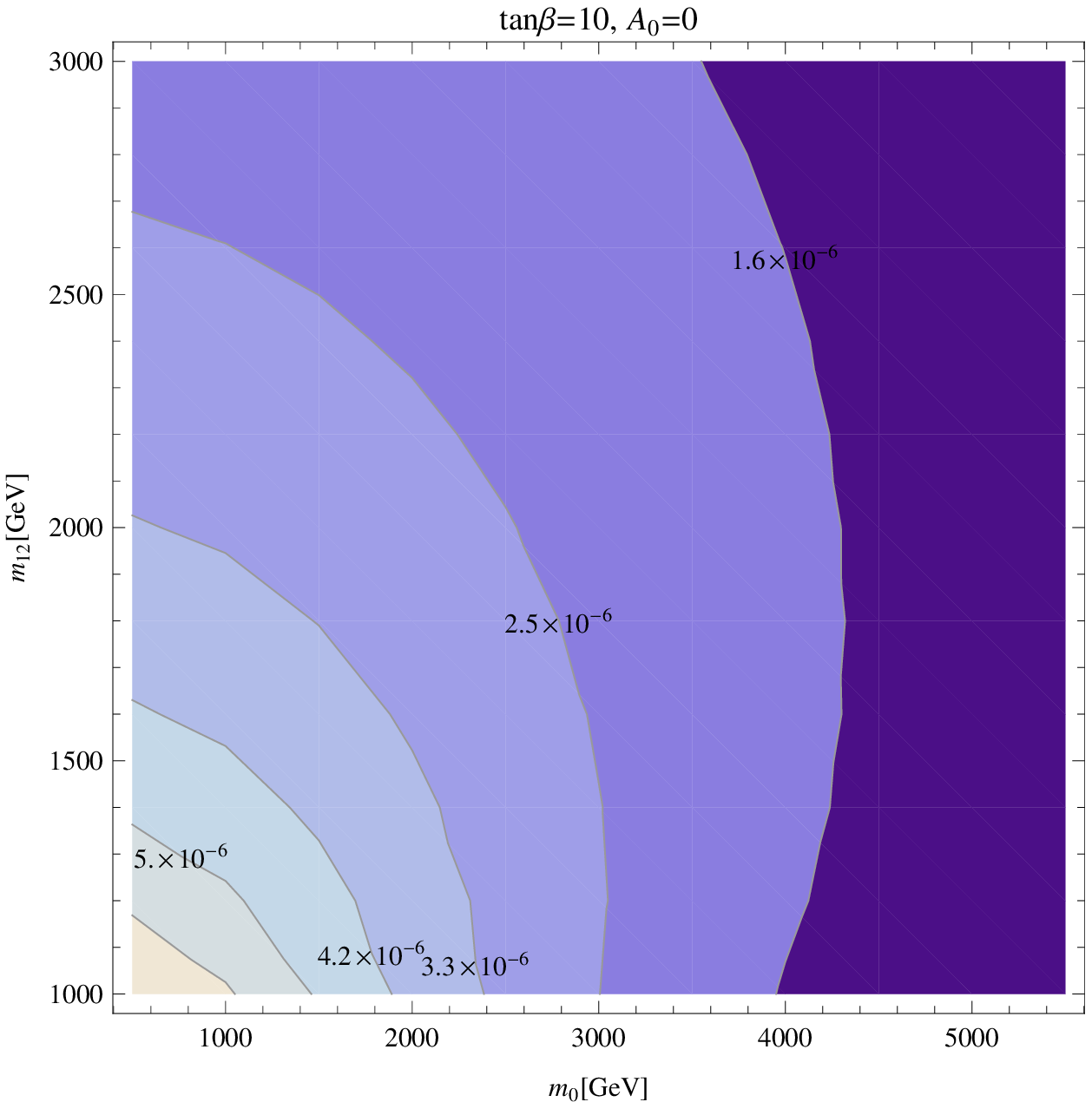  ,scale=0.51,angle=0,clip=}
\psfig{file=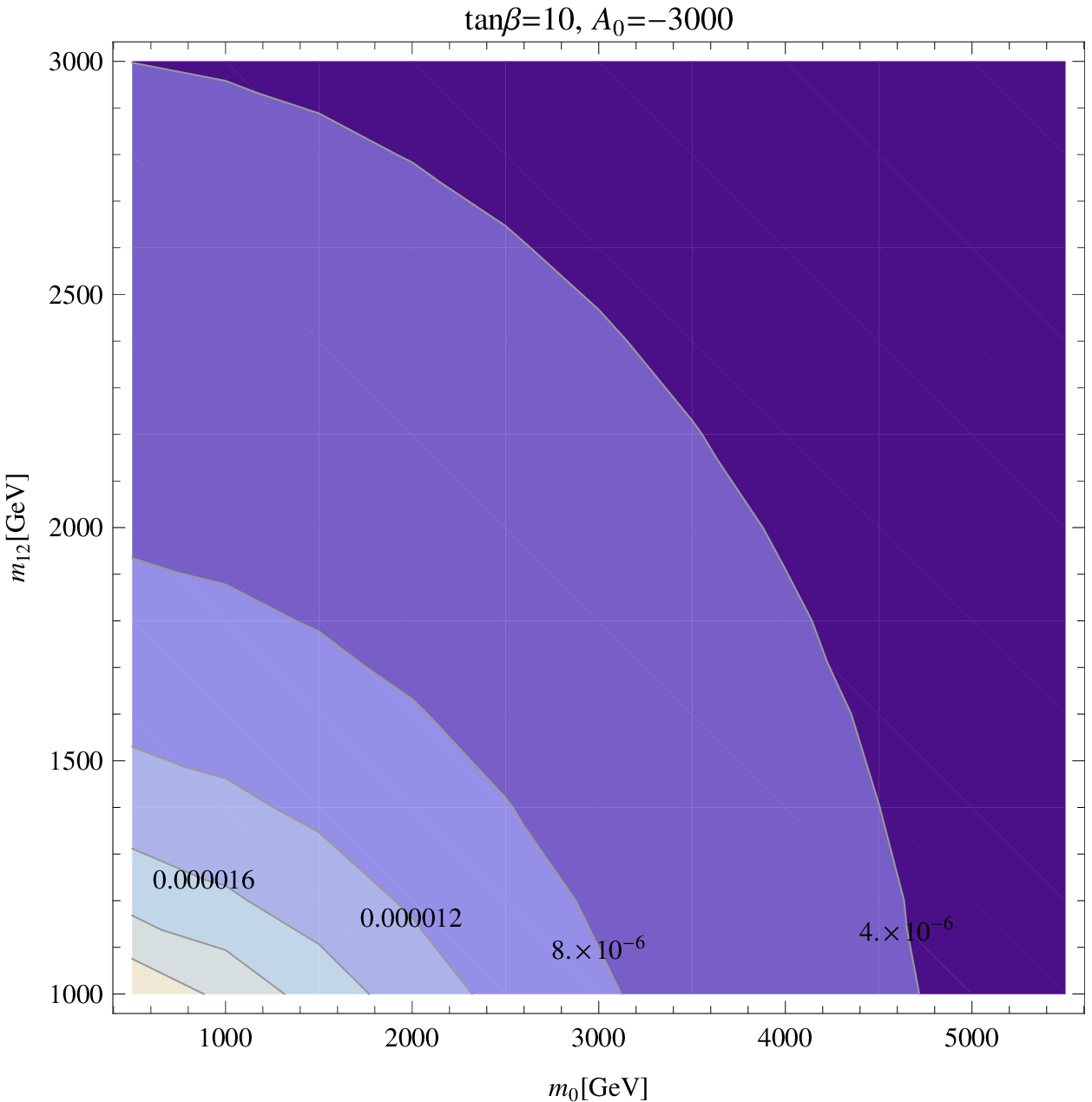  ,scale=0.51,angle=0,clip=}\\
\vspace{0.2cm}
\psfig{file=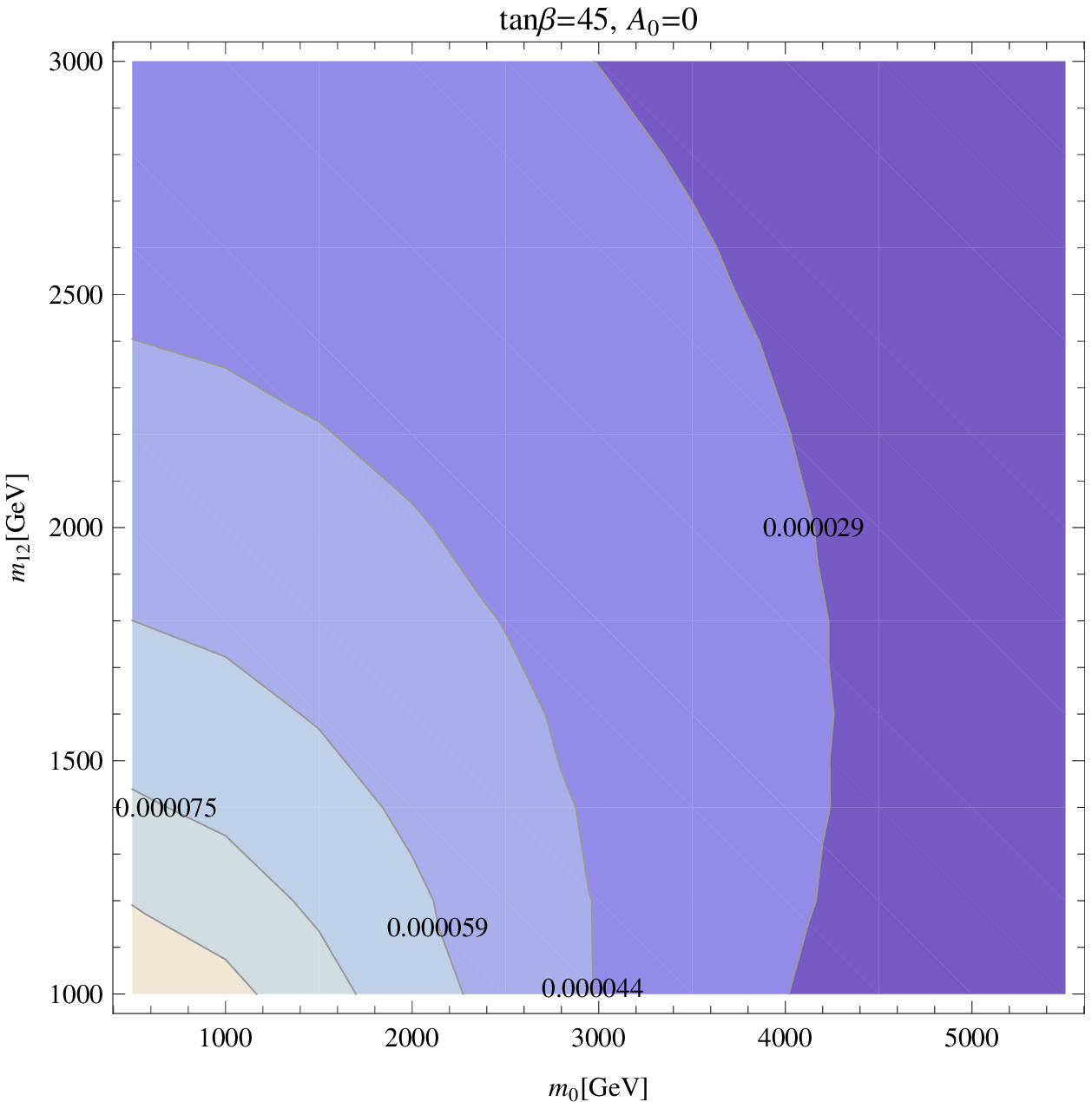 ,scale=0.51,angle=0,clip=}
\psfig{file=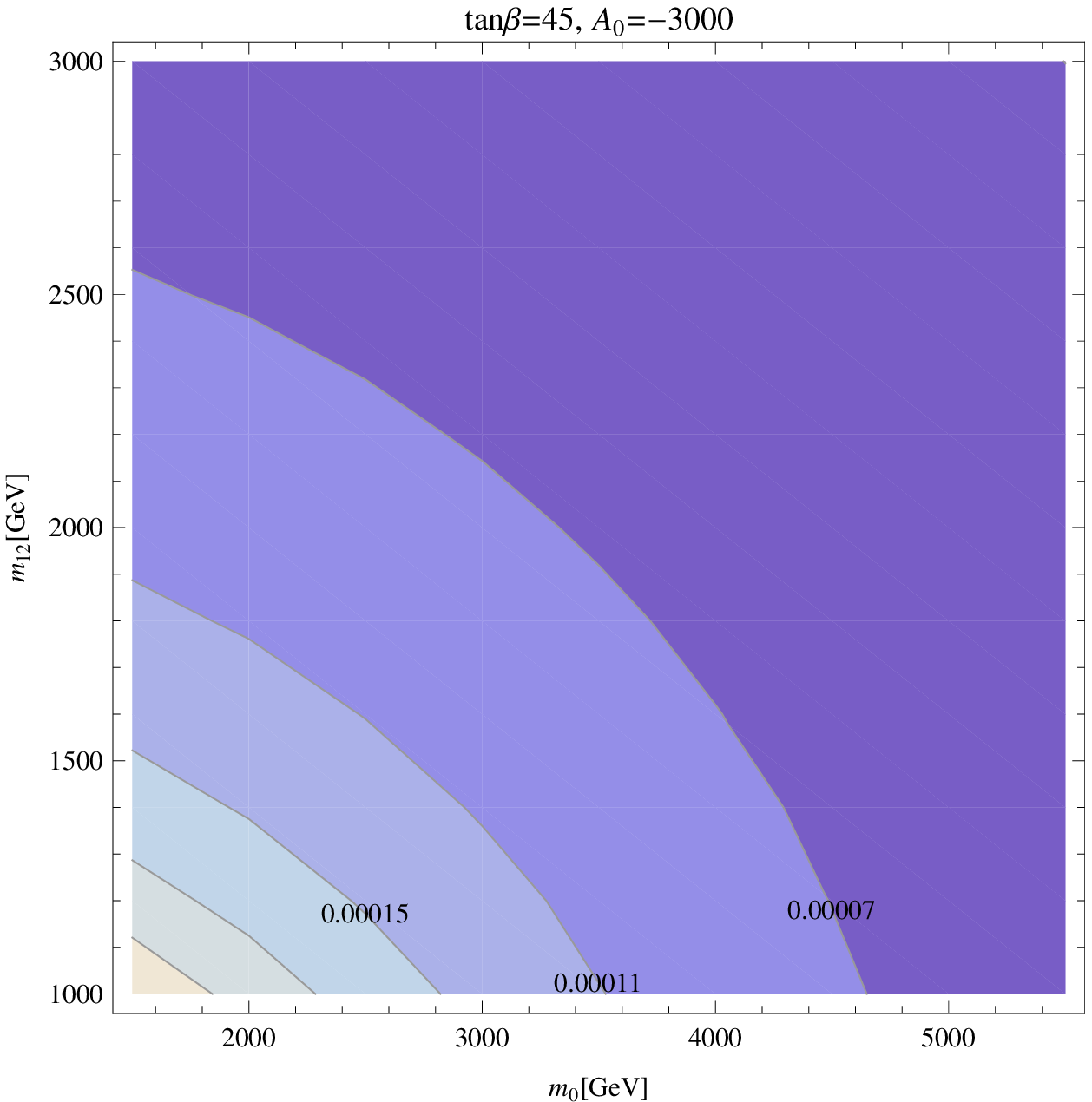   ,scale=0.51,angle=0,clip=}\\
\vspace{-0.2cm}
\end{center}
\caption[Contours of $\delta^{ULR}_{23}$ in the
  $m_0$--$m_{1/2}$ plane.]{Contours of $\delta^{ULR}_{23}$ in the
  $m_0$--$m_{1/2}$ plane for different values of $\tb$ and    
$A_0$ in the CMSSM.}  
\label{fig:DelULR23}
\end{figure}

We start with the three most relevant $\deFABij$'s. In
\reffis{fig:DelQLL13}-\ref{fig:DelULR23} we show the results for
$\del{QLL}{13}$, $\del{QLL}{23}$ and $\del{ULR}{23}$, respectively,
which are expected to yield the largest results.
The values show the expected pattern of their size with 
$\del{QLL}{23} \sim \order{10^{-2}}$ being the largest one, and
$\del{QLL}{13}$ and $\del{ULR}{23}$ about one or two orders of magnitude
smaller. All other $\deFABij$ which are not shown reach only values of
\order{10^{-5}}. 
One can observe an interesting pattern in these figures: the
values of $\deFABij$ increase with larger values of either $\tb$ or
$A_0$. 
The values for $\del{QLL}{}$ increase with $m_0$, whereas the
$\del{ULR}{}$ and $\del{DLR}{}$ decrease with $m_0$. 
This behavior can be understood for the RGE's of the non diagonal SUSY
breaking parameters (see, e.g., \citere{Martin:1993zk}), $\del{QLL}{}$'s are
defined as ratios of off-diagonal soft terms that grow with $m_0^2$ over
diagonal soft masses that also grow with $m_0$. However, 
$\del{ULR}{}$'s and $\del{DLR}{}$'s arises from the ratio of the RGE generated
off-diagonal 
trilinear terms which depend on the value of $A_0$, that is considered fixed
in our case, over $m_0$ growing diagonal soft masses.  
As discussed above, these $\deFABij \neq 0$ are often neglected
in phenomenological analyses of the CMSSM (see, e.g.,
\citere{CMSSM-NUHM}). We also emphasize that these effects are purely due to
the presenece of the CKM matrix on the RGE's, their contribution will vanish
when the mixing of the two first generation with the third generation  is
neglected (as we have checked numerically).

\subsection{EWPO}
\label{EWPO-CMSSM-Res}
\begin{figure}[ht!]
\begin{center}
\psfig{file=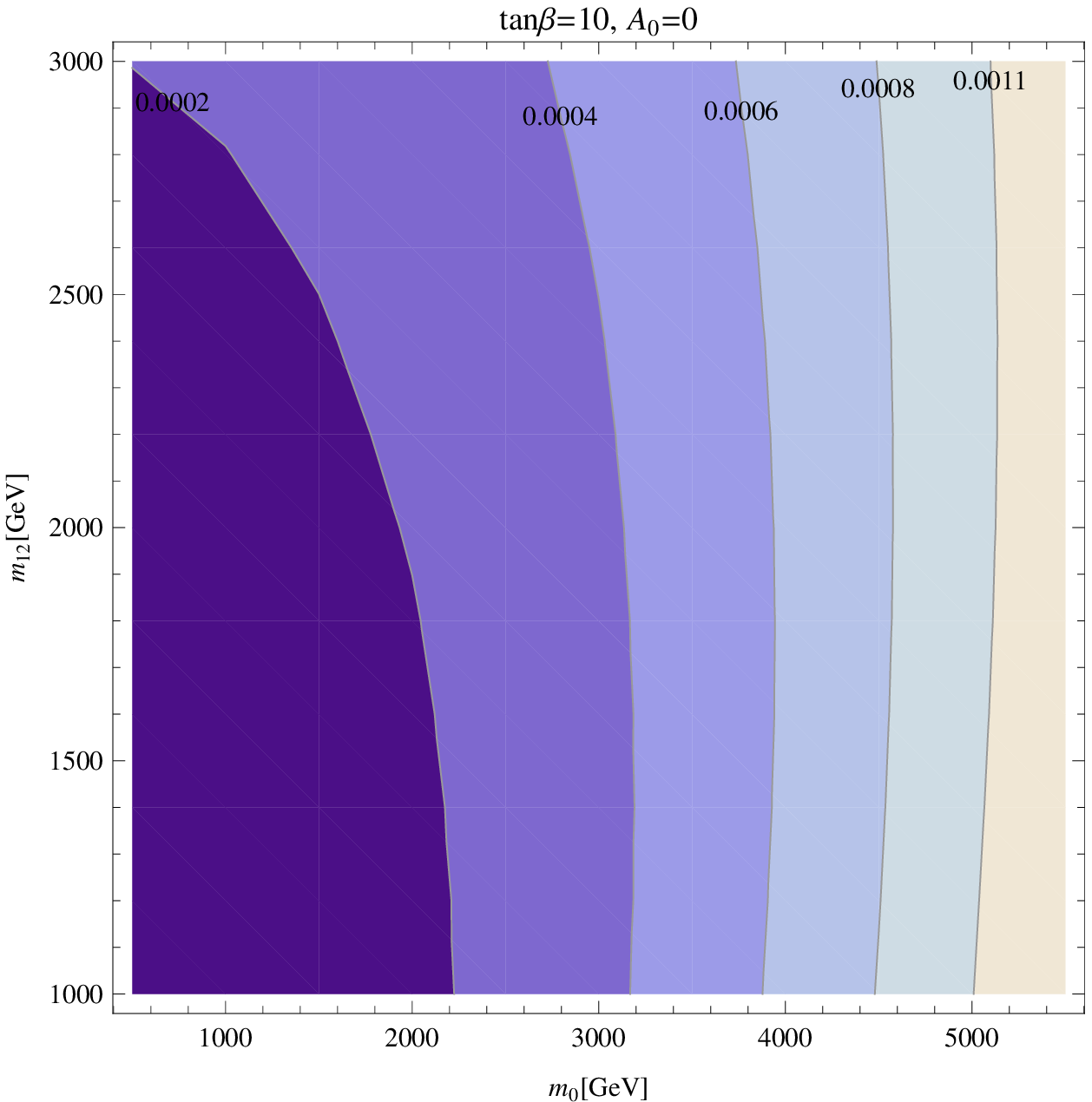  ,scale=0.51,angle=0,clip=}
\psfig{file=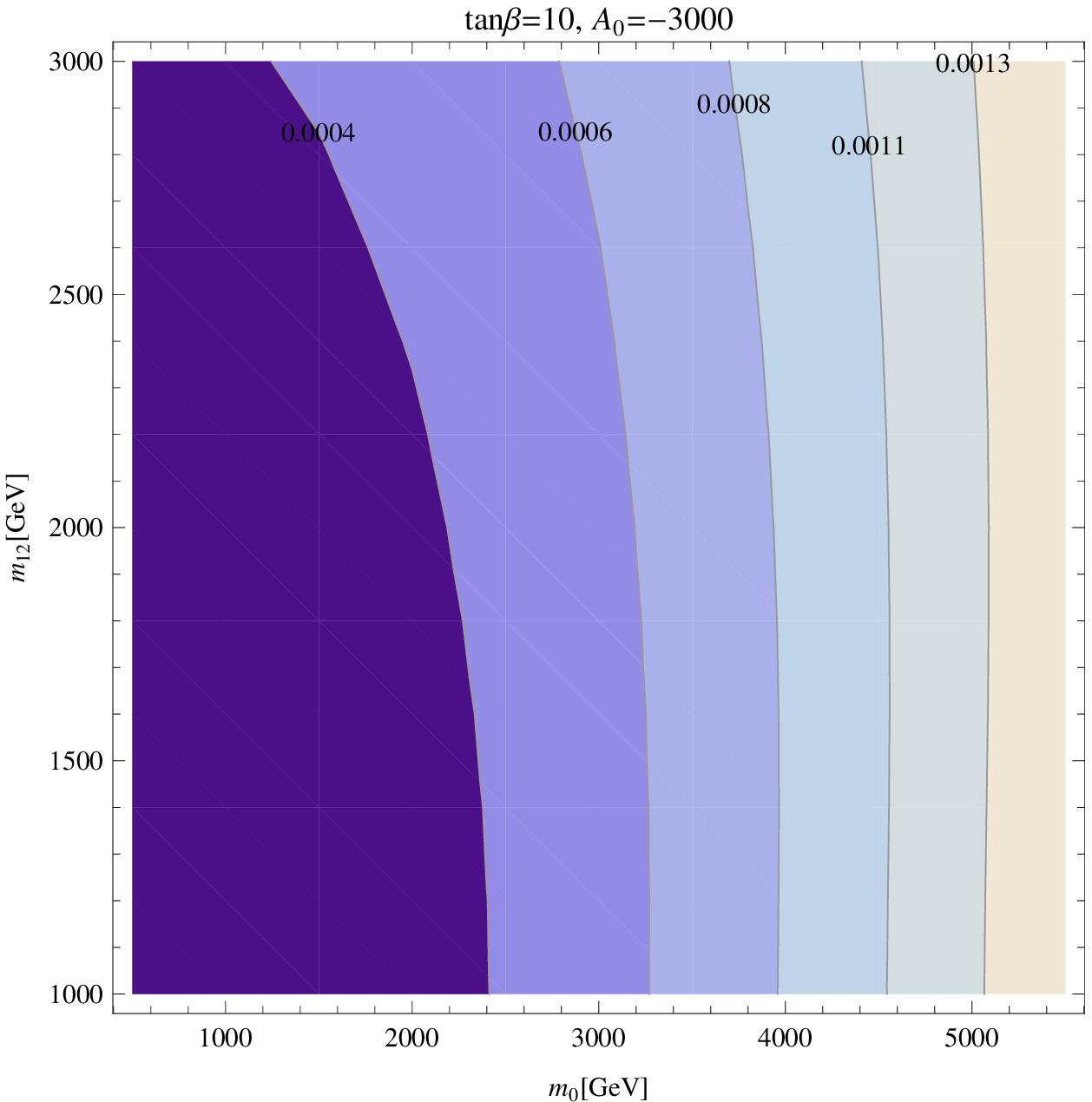  ,scale=0.51,angle=0,clip=}\\
\vspace{0.2cm}
\psfig{file=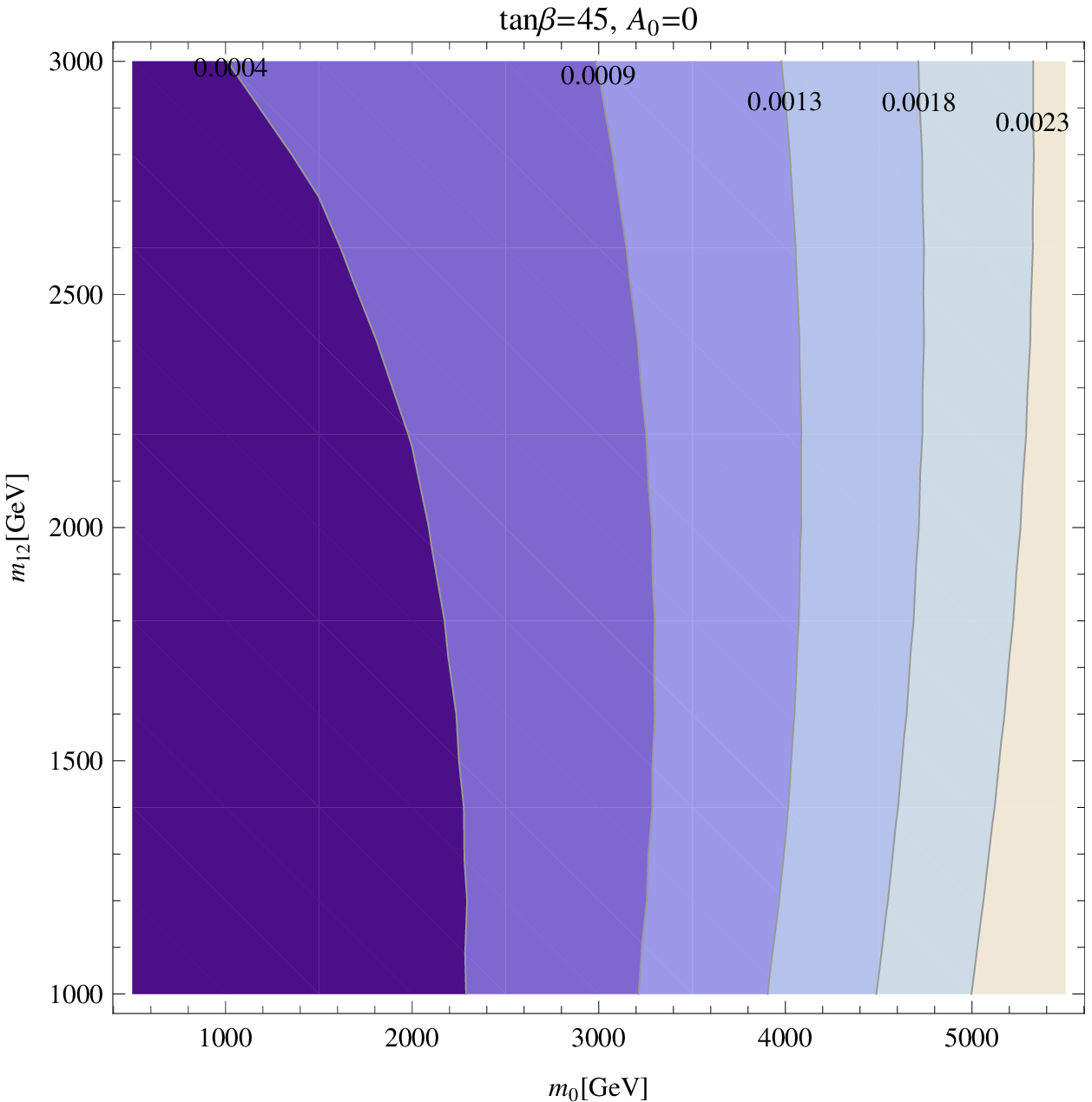 ,scale=0.51,angle=0,clip=}
\psfig{file=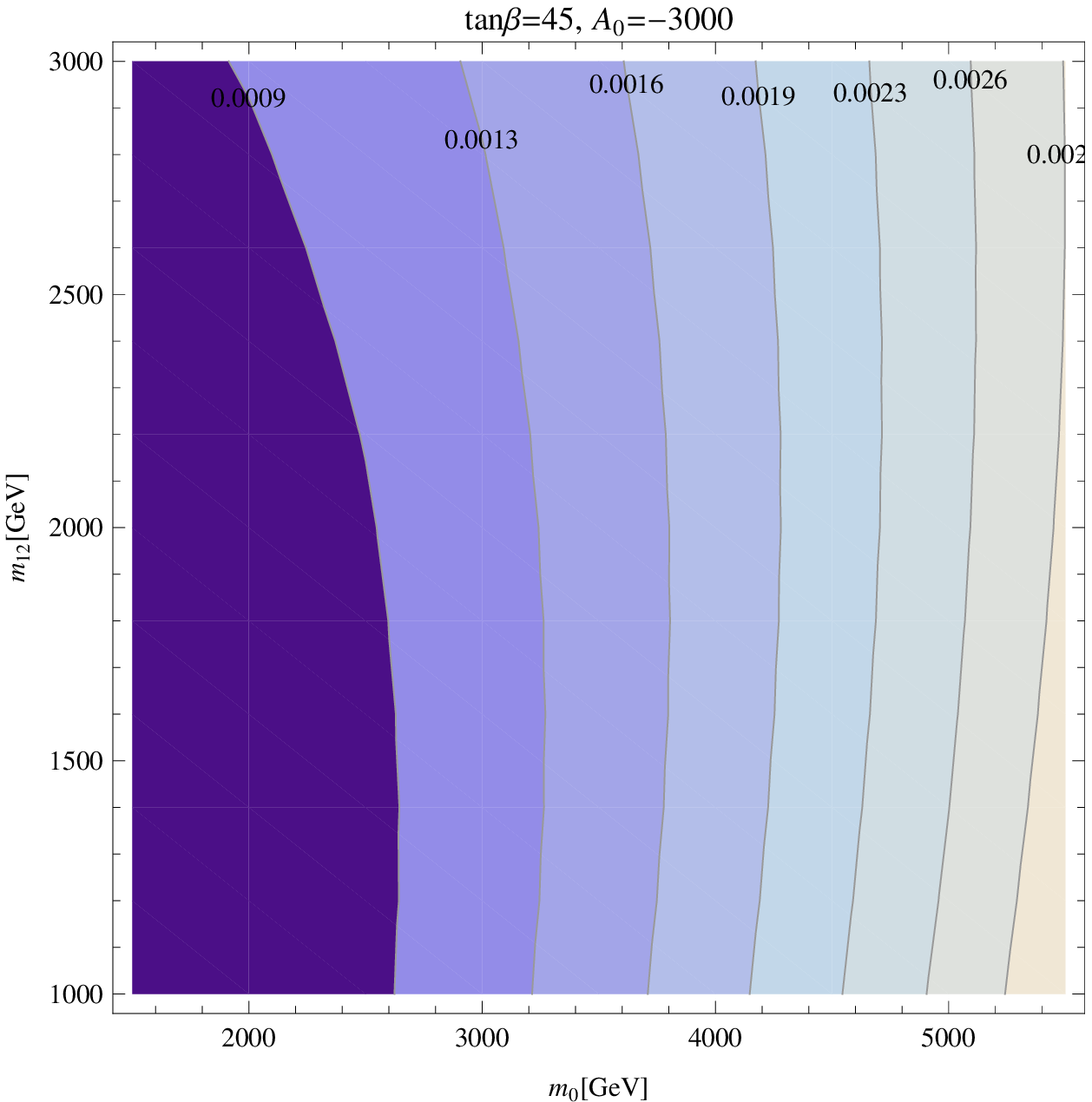   ,scale=0.51,angle=0,clip=}\\
\vspace{-0.2cm}
\end{center}
\caption[Contours of \Drho\ in the $m_0$--$m_{1/2}$ plane.]{Contours of \Drho\ in the $m_0$--$m_{1/2}$ plane for different
  values of $\tb$ and $A_0$ in the CMSSM.}  
\label{fig:SQ-delrho}
\end{figure}

\begin{figure}[ht!]
\begin{center}
\psfig{file=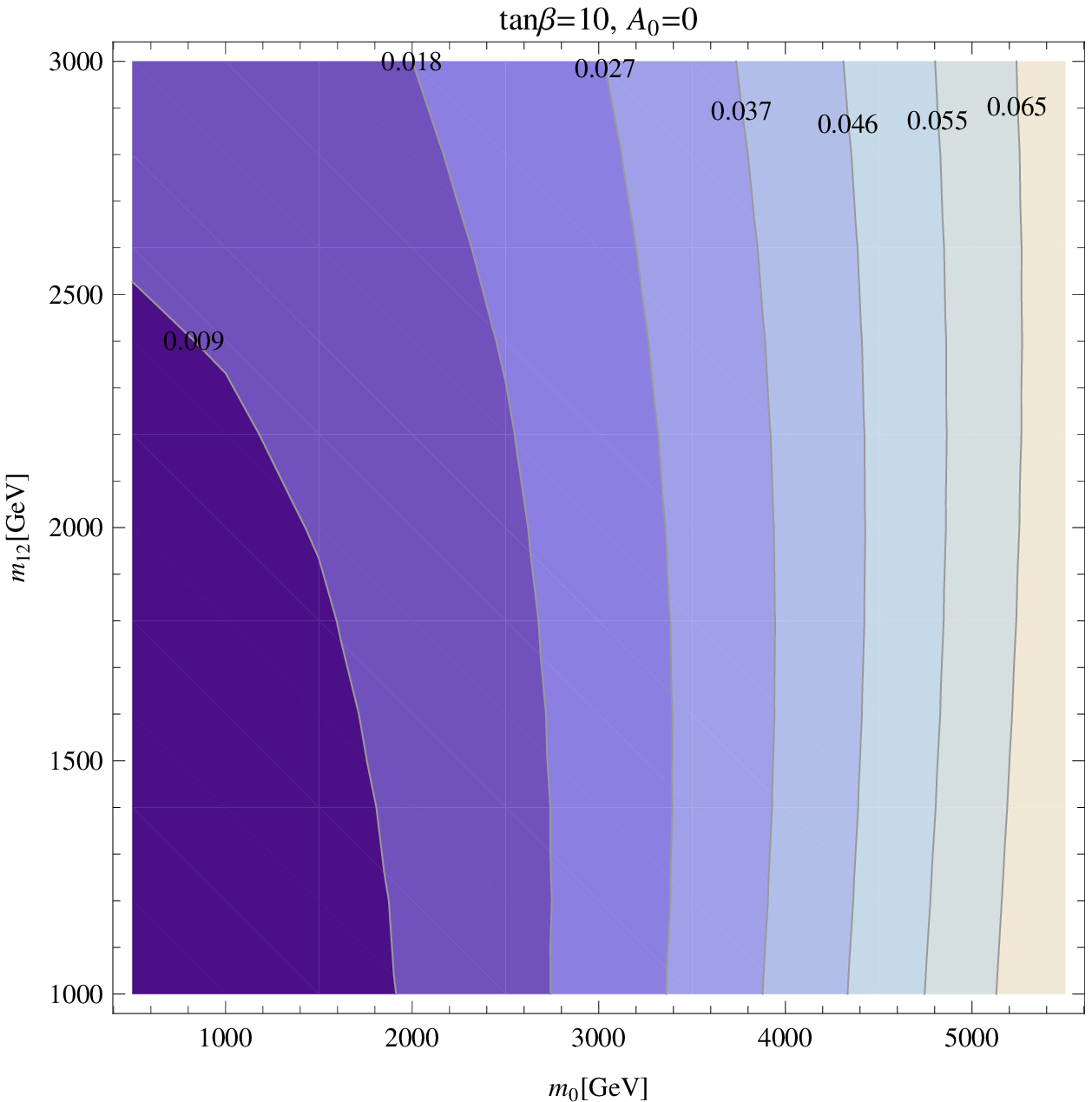  ,scale=0.51,angle=0,clip=}
\psfig{file=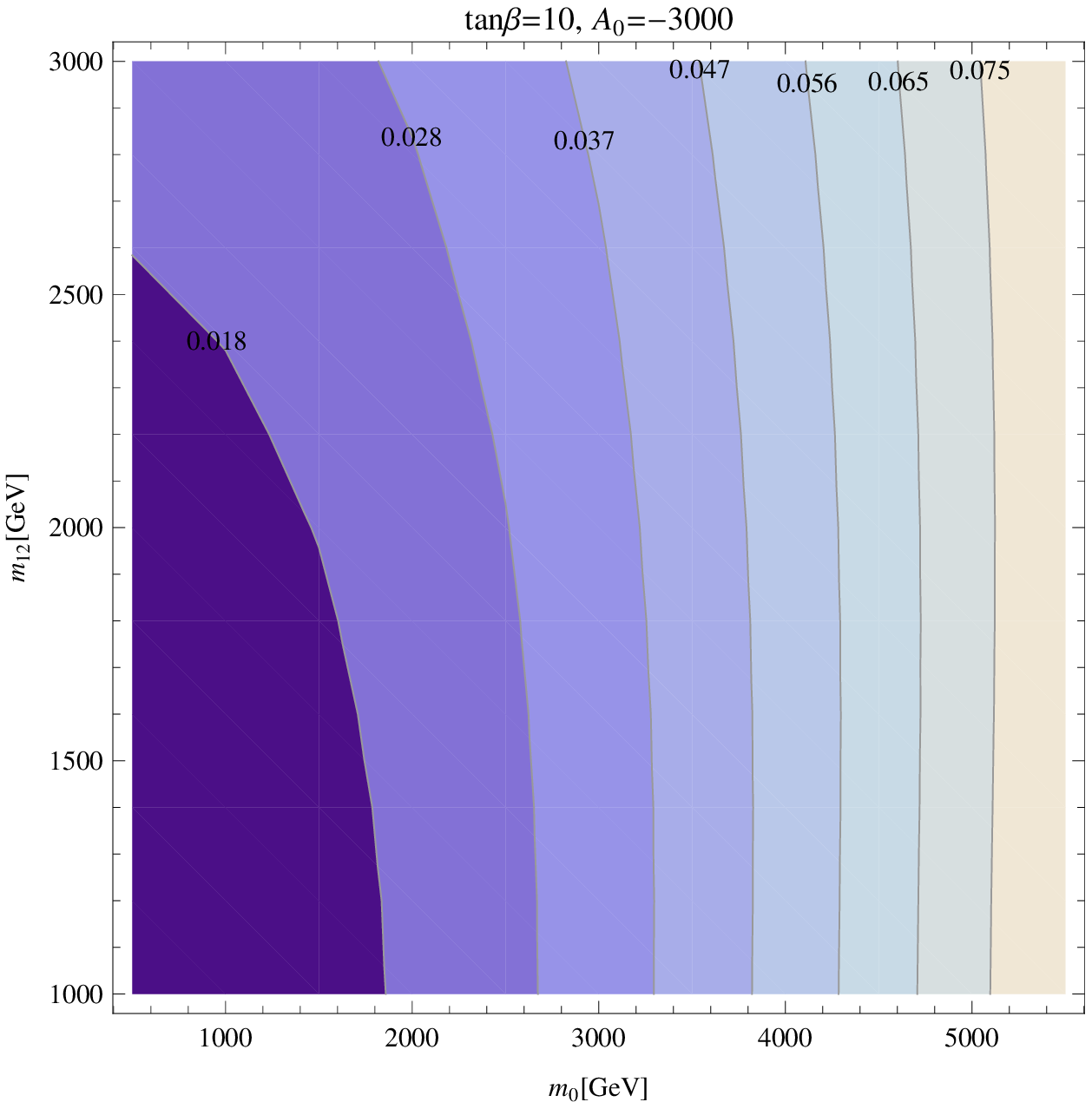  ,scale=0.51,angle=0,clip=}\\
\vspace{0.2cm}
\psfig{file=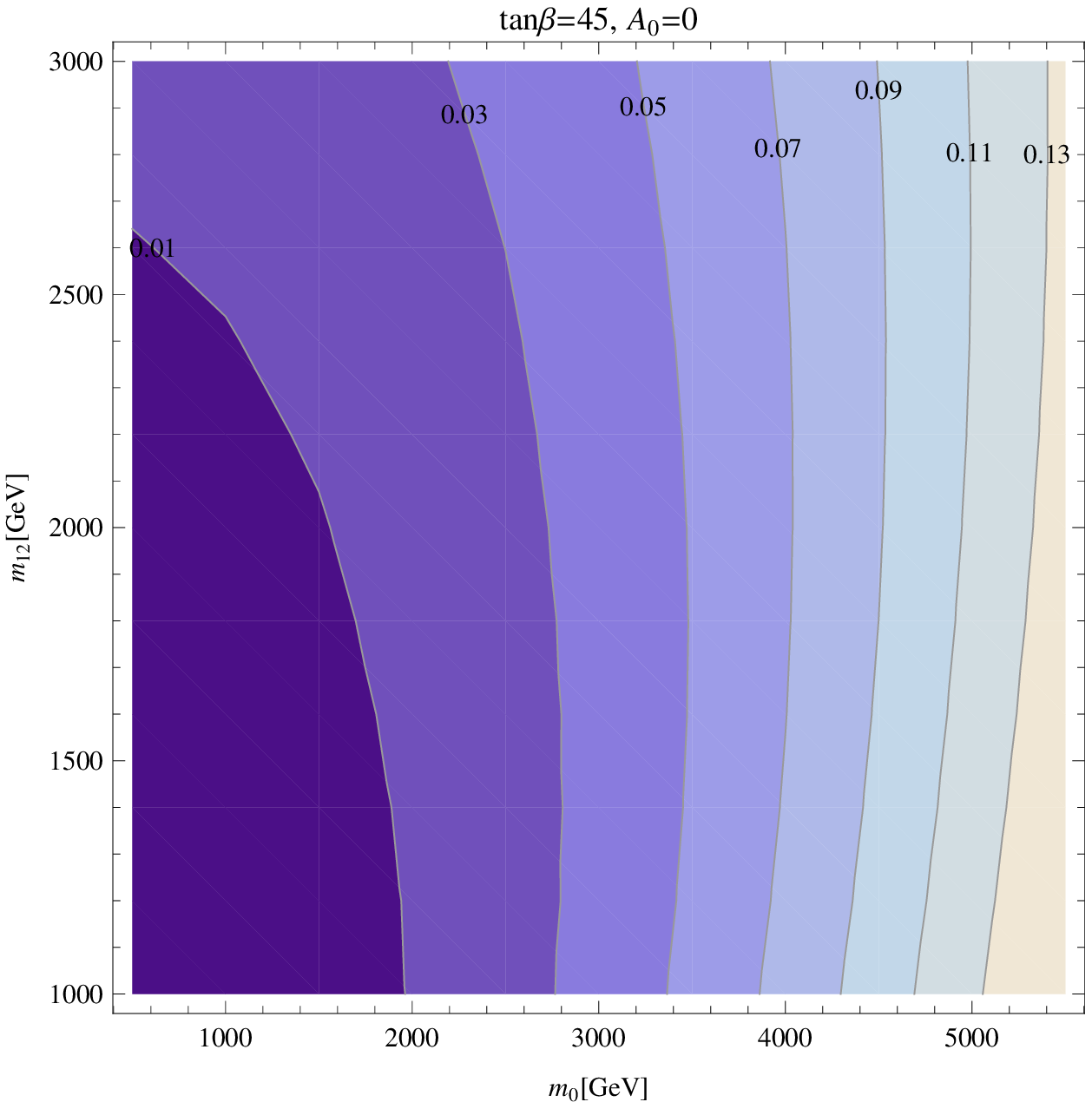 ,scale=0.51,angle=0,clip=}
\psfig{file=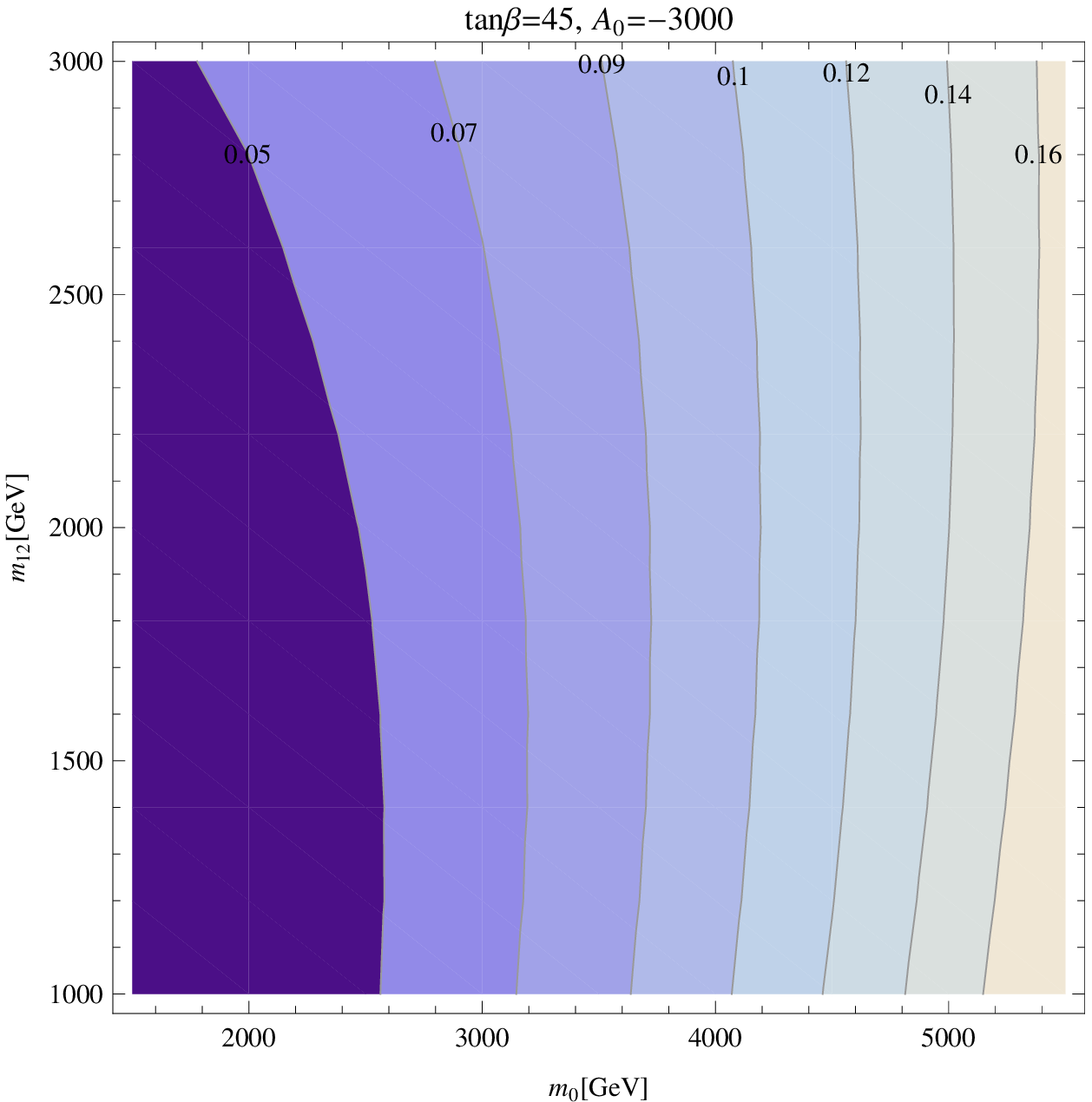   ,scale=0.51,angle=0,clip=}\\
\vspace{-0.2cm}
\end{center}
\caption[Contours of \DMW\ in GeV in the
  $m_0$--$m_{1/2}$ plane]{Contours of \DMW\ in GeV in the
  $m_0$--$m_{1/2}$ plane for different values of $\tb$ and    
$A_0$ in the CMSSM.}  
\label{fig:SQ-delMW}
\end{figure}

\begin{figure}[ht!]
\begin{center}
\psfig{file=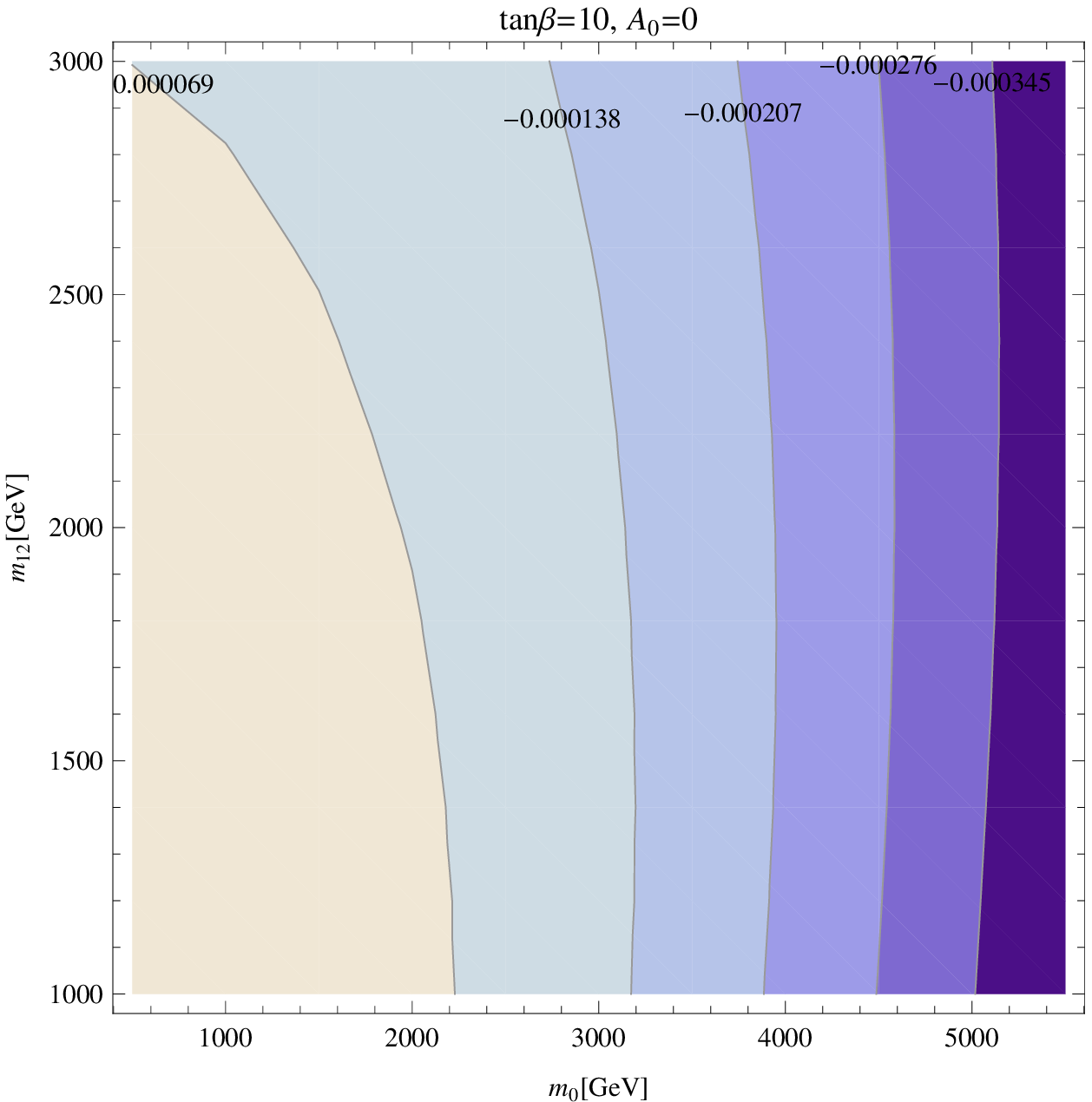  ,scale=0.51,angle=0,clip=}
\psfig{file=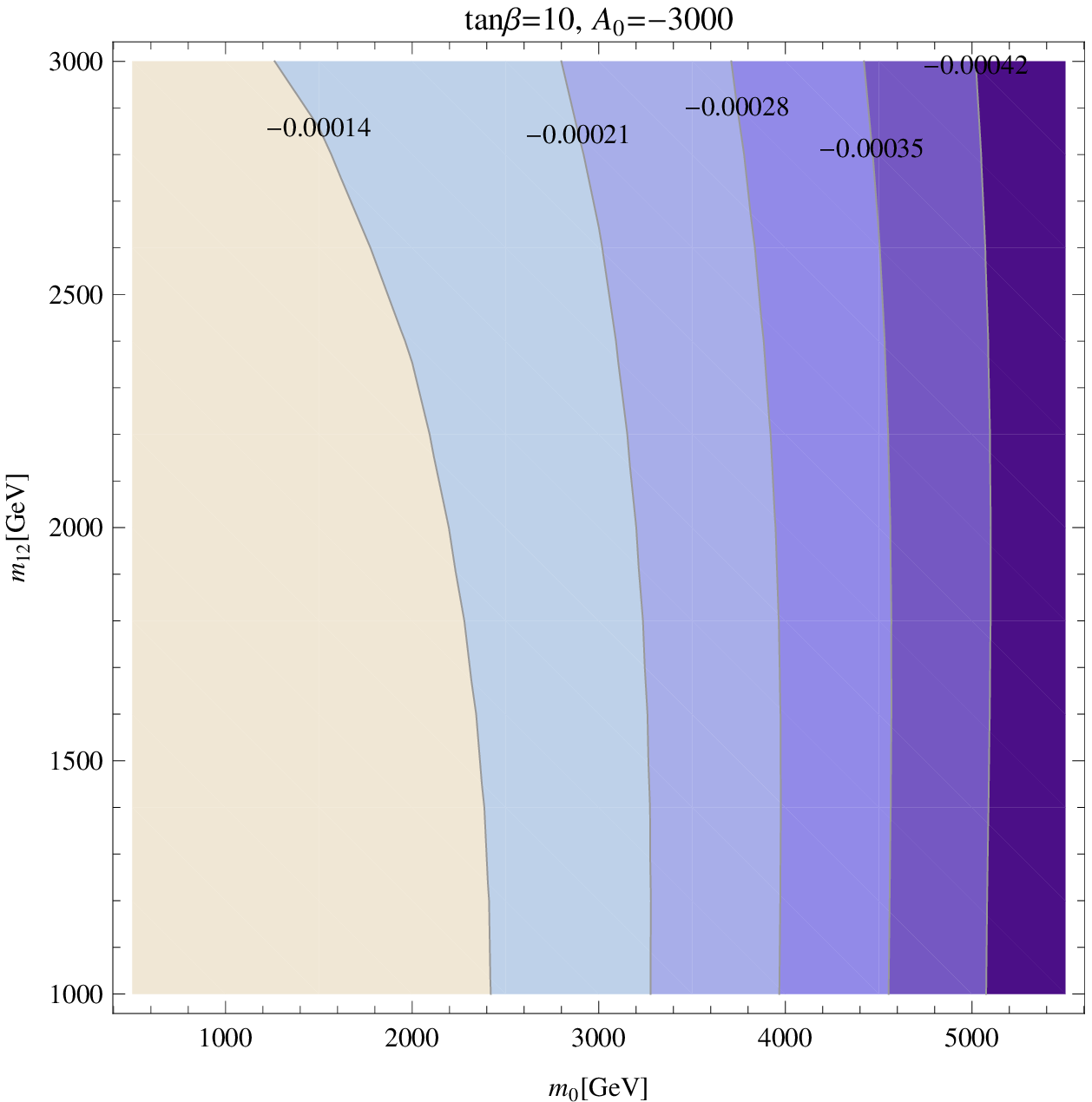  ,scale=0.51,angle=0,clip=}\\
\vspace{0.2cm}
\psfig{file=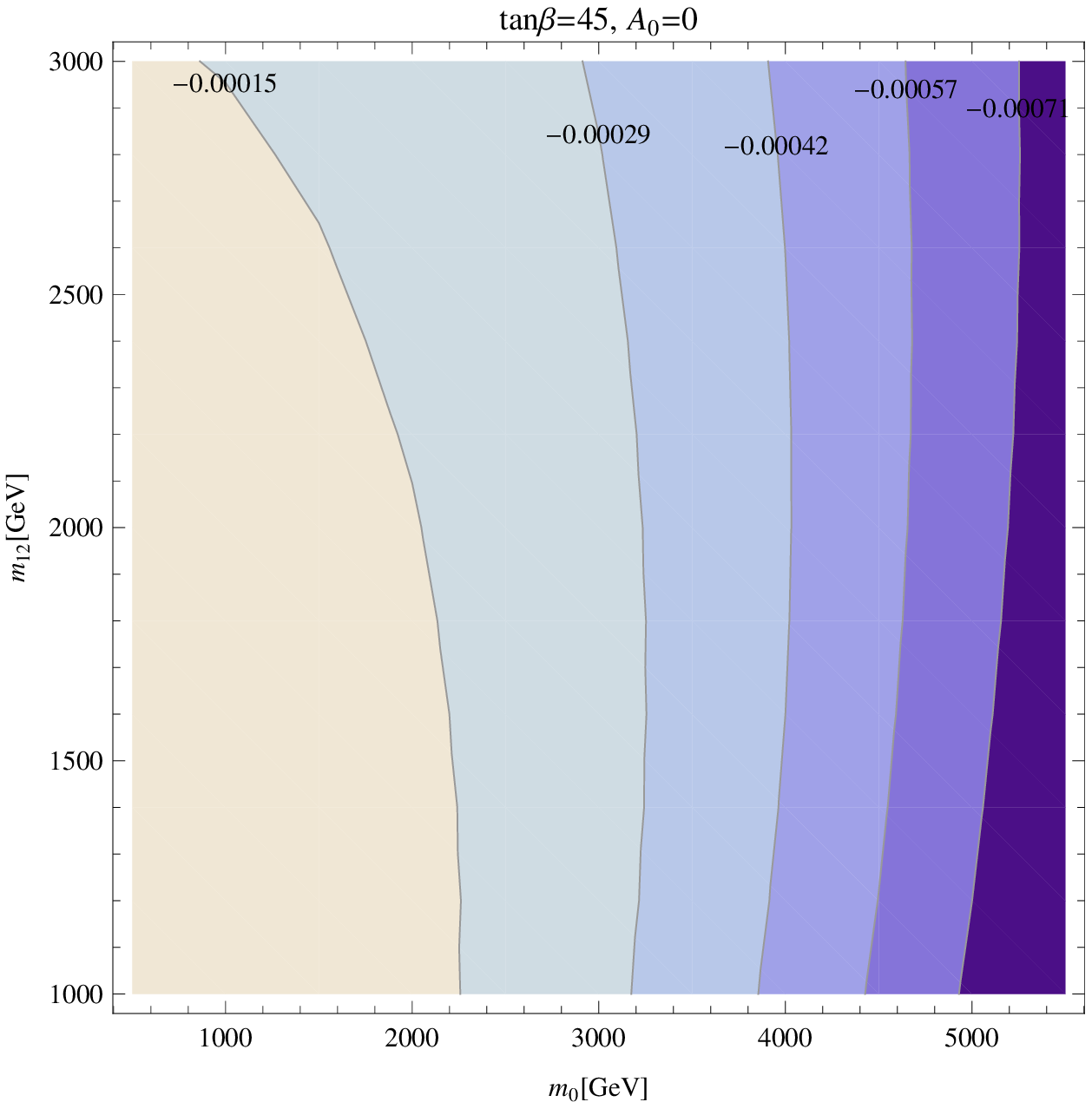 ,scale=0.51,angle=0,clip=}
\psfig{file=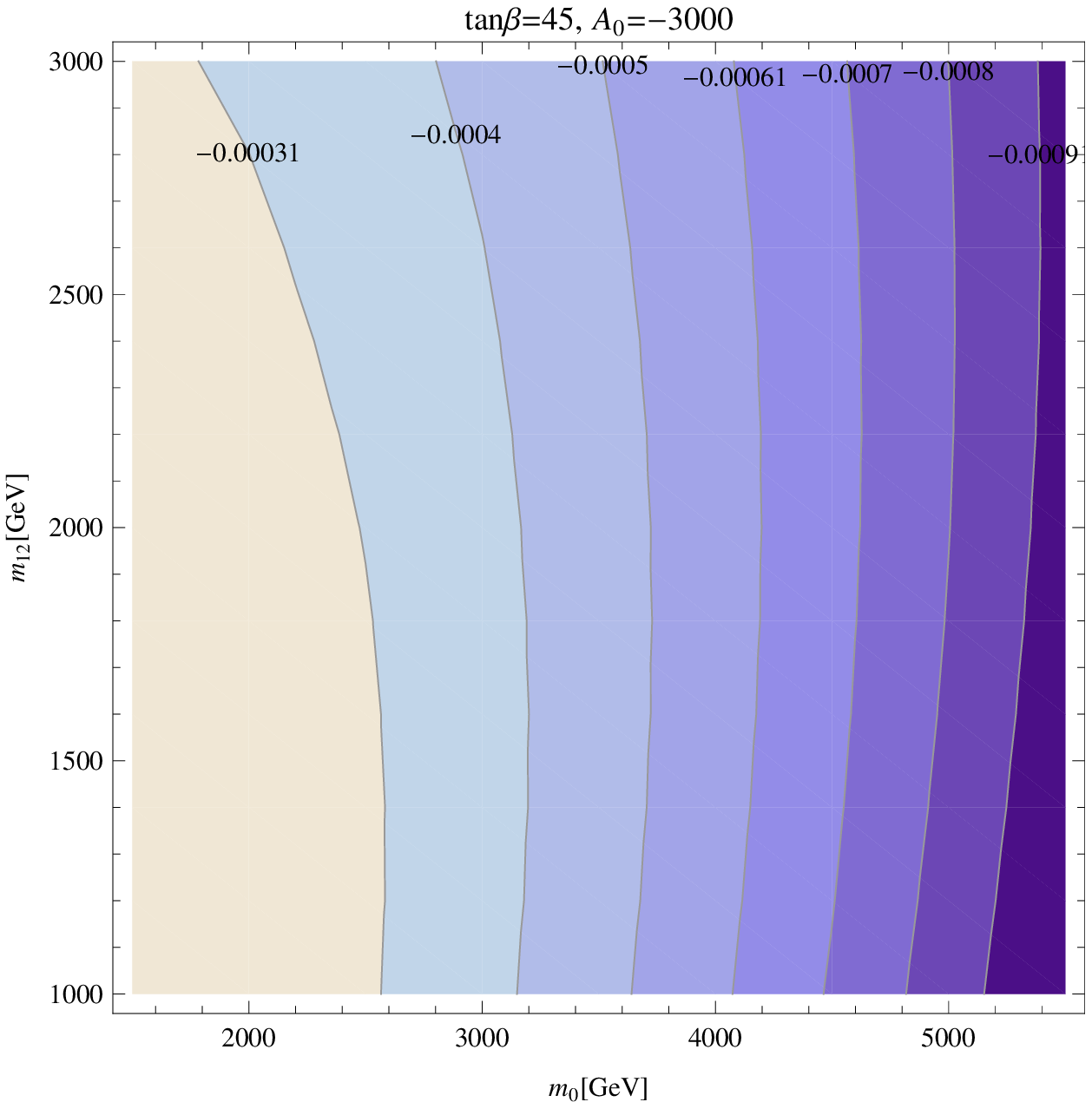   ,scale=0.51,angle=0,clip=}\\
\vspace{-0.8cm}
\end{center}
\caption[Contours of \Dsweff\ in the
  $m_0$--$m_{1/2}$ plane.]{Contours of \Dsweff\ in the
  $m_0$--$m_{1/2}$ plane for different values of $\tb$ and    
$A_0$ in the CMSSM.}  
\label{fig:SQ-delSW2}
\end{figure}

In \reffis{fig:SQ-delrho}-\ref{fig:SQ-delSW2} we analyze the effects of
the non-zero $\deFABij$ on the EWPO \Drho, \DMW\ and \Dsweff,
respectively. 
Here the same pattern is reflected for the EWPO, i.e.\ by increasing
the value of $\tb$ or $A_0$, we find larger contributions to the
EWPO. In particular one can observe a non-decoupling effect for
large values of $m_0$. Larger soft SUSY-breaking
parameters with the non-zero values in particular of $\del{QLL}{23}$,
see above, lead to an enhanced splitting in masses belonging to an
$SU(2)$ doublet, and thus to an enhanced contribution to the
$\rho$-parameter. The corresponding effects on $\MW$ and $\sweff$, 
for $m_0 \gsim 3 \tev$, exhibit corrections that are several times
larger than the current experimental accuracy (whereas the SUSY
corrections with all $\deFABij = 0$ decouple and go to
zero). Consequently, including the non-zero values of the $\deFABij$
and correctly taking these corrections into account, would yield an 
{\em upper} limit on $m_0$, which in the known analyses so far
is unconstrained from above~\cite{CMSSM-NUHM}. 
A more detailed analysis within the CMSSM will be needed to determine
the real upper bound on $m_0$, which, however, is beyond the scope of
this thesis.

\begin{figure}[ht!]
\begin{center}
\psfig{file=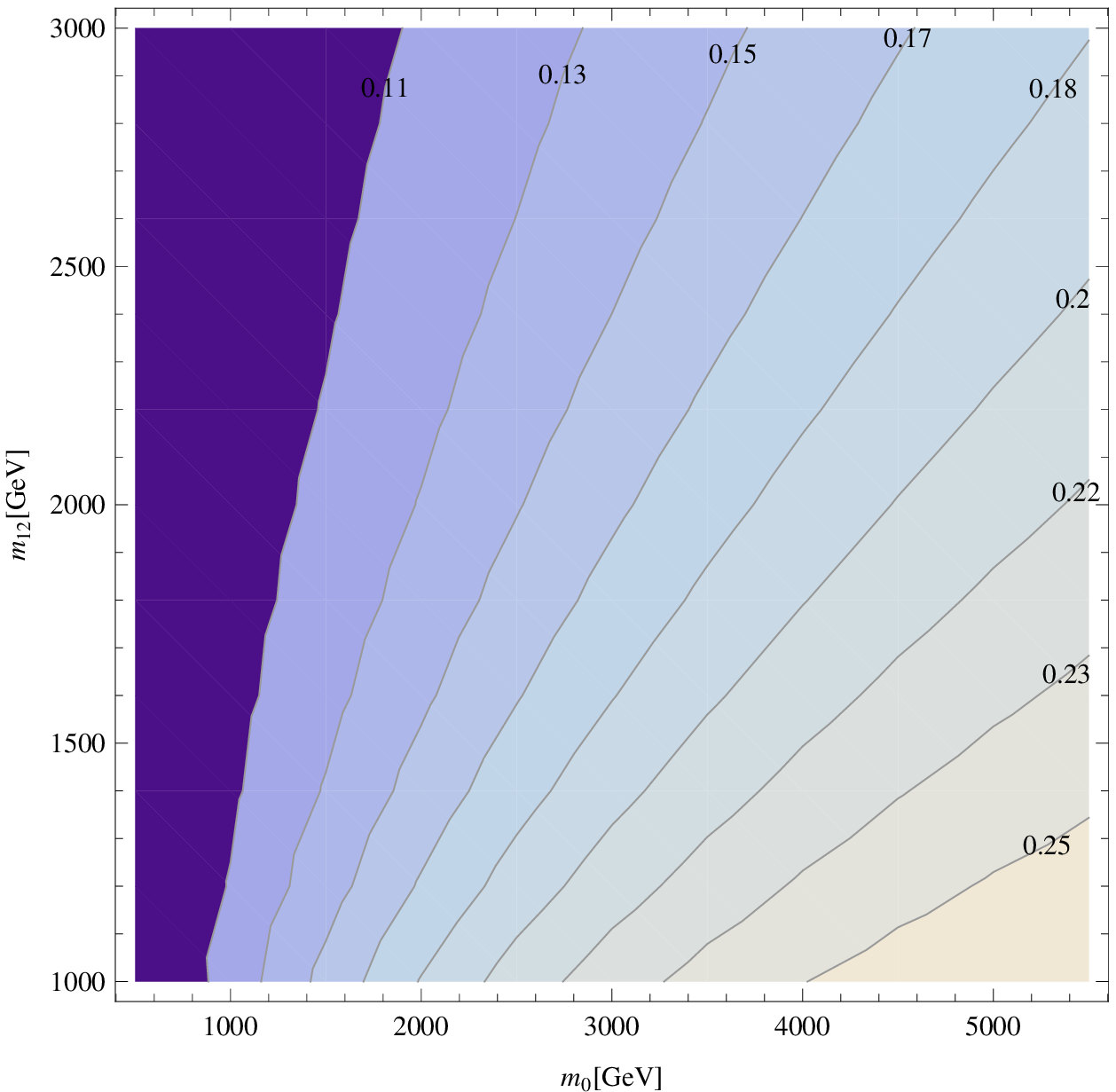,scale=0.51,angle=0,clip=}
\psfig{file=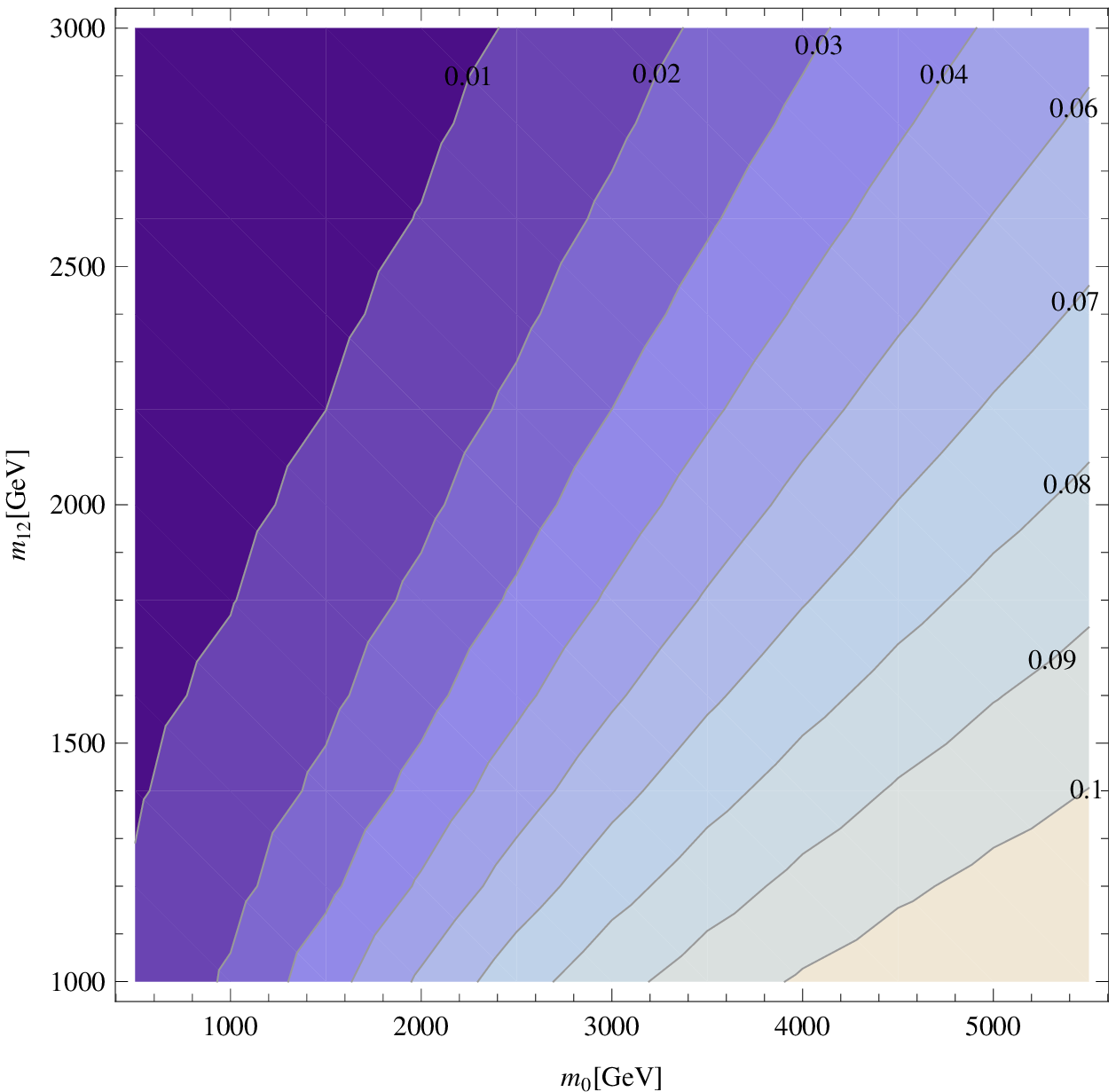,scale=0.51,angle=0,clip=} 
\vspace{-0.2cm}
\end{center}
\caption[Contours of $(m_2^2 - m_1^2)/(m_2^2 + m_1^2)$ in the $m_0$--$m_{1/2}$
plane]{Contours of $(m_2^2 - m_1^2)/(m_2^2 + m_1^2)$ in the $m_0$--$m_{1/2}$
plane for fixed values of $A_0 = 0$ and $\tb = 45$. Left: the two most
stop-like squarks (i.e.\ in the limit of zero inter-generational mixing they
coincide with the two scalar tops), right: the lightest most stop-like and most
sbottom-like squarks (see text).}
\label{fig:SQ-masses}
\end{figure}

In order to gain more insight about the source of the large
corrections to $\De\rho$ (and thus to the EWPO), we show in
\reffi{fig:SQ-masses} several relative mass (square) differences,
$(m_2^2 - m_1^2)/(m_2^2 + m_1^2)$ in the
$m_0$--$m_{1/2}$ plane for fixed $A_0 = 0$ and $\tb = 45$. 
The left plot shows the mass difference for the two most stop-like
squarks (i.e.\ in the limit of zero inter-generational mixing they
coincide with the two scalar tops). The right plot shows the relative
mass difference for the lightest most stop-like and most sbottom-like
squark. (These results are simply the {\tt Spheno} output in our
scenario.)
In both cases one can see that the relative mass differences
increase (controlled by the non-zero $\deFABij$ induced by the CKM matrix
in the RGE running) in a fashion similar as the $\del{QLL}{}$ discussed
above, i.e.\ in particular for $m_0 > m_{1/2} > 1 \tev$. These
increasing mass differences lead (together with 
contributions from the mixing matrices) to the observed increase of
$\De\rho$ as in \reffi{fig:SQ-delrho}.

Our findings can be briefly compared to the existing
literature. The EWPO in the context of flavor violation were evaluated
first in \citere{delrhoNMFV}, where correspondingly large corrections
were found for large $\del{QLL}{23}$ (in fact, that was the only
parameter dependence analyzed in that paper, and only the mixing between
the second and third generation of squarks was taken into
account). Subsequently, the EWPO were also evaluated for the full
three-generation mixing in \citere{Cao:2006xb}. The numerical analysis,
however, was restricted to a degenerate and fixed SUSY mass scale.
Correspondingly, no large effects with increasing SUSY mass scales were
analyzed and only relative small corrections were found. Due to the
different numerical set-up, however, there is no contradiction with our
results for $\De\rho$.
\subsection{Higgs masses and the BPO}
\begin{figure}[ht!]
\begin{center}
\psfig{file=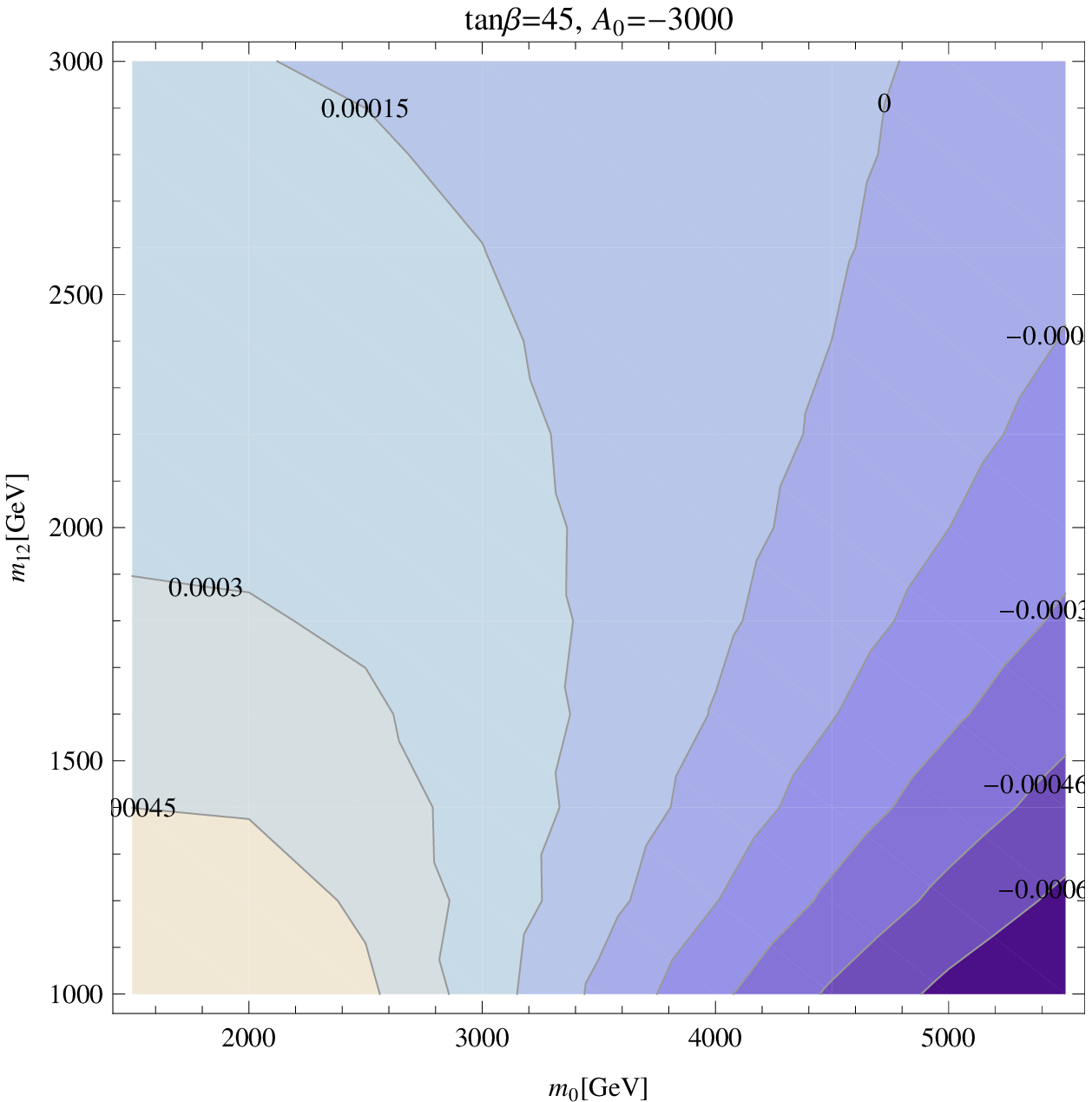,scale=0.50,angle=0,clip=}
\psfig{file=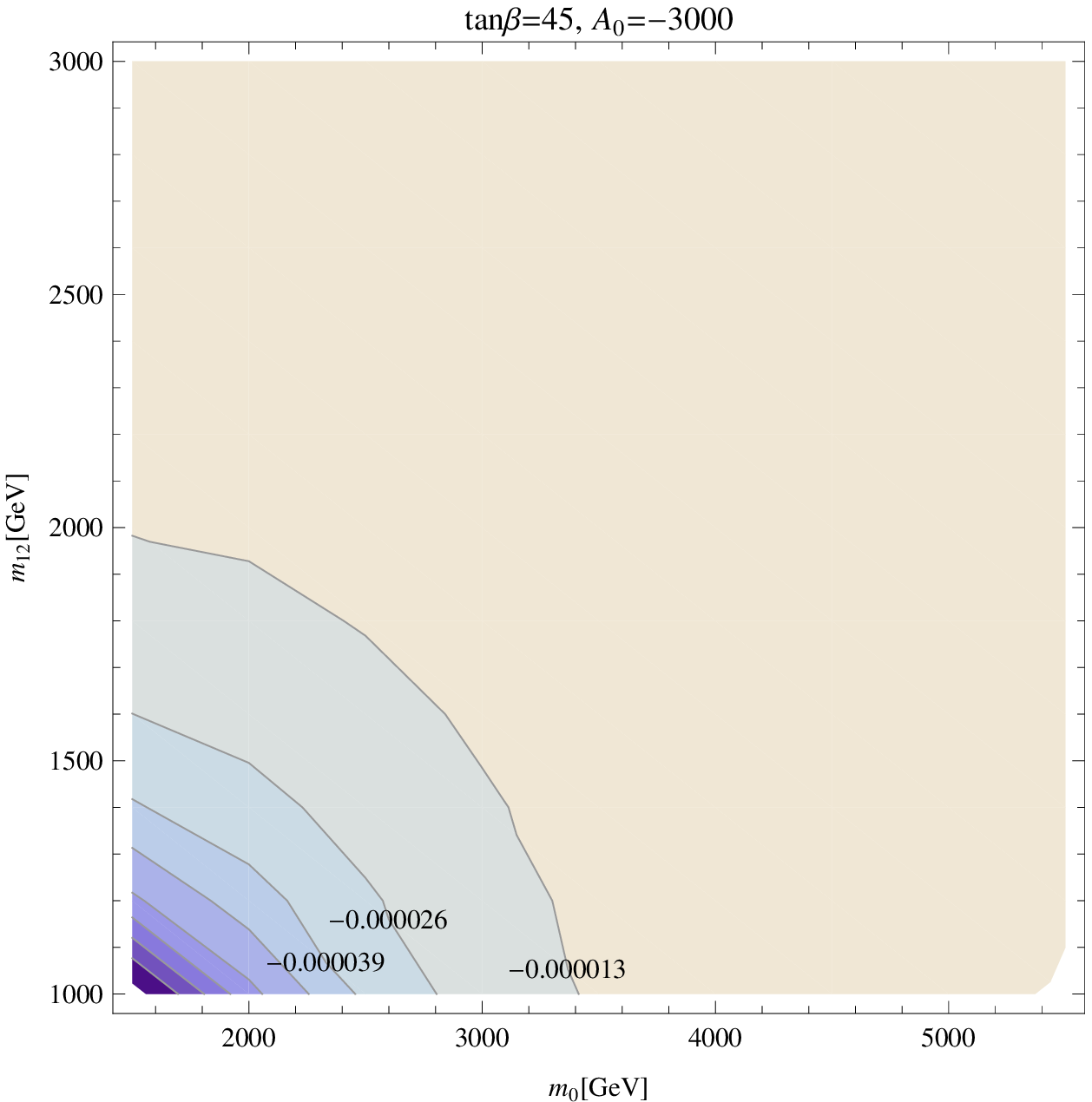,scale=0.50,angle=0,clip=}\\
\vspace{0.2cm}
\psfig{file=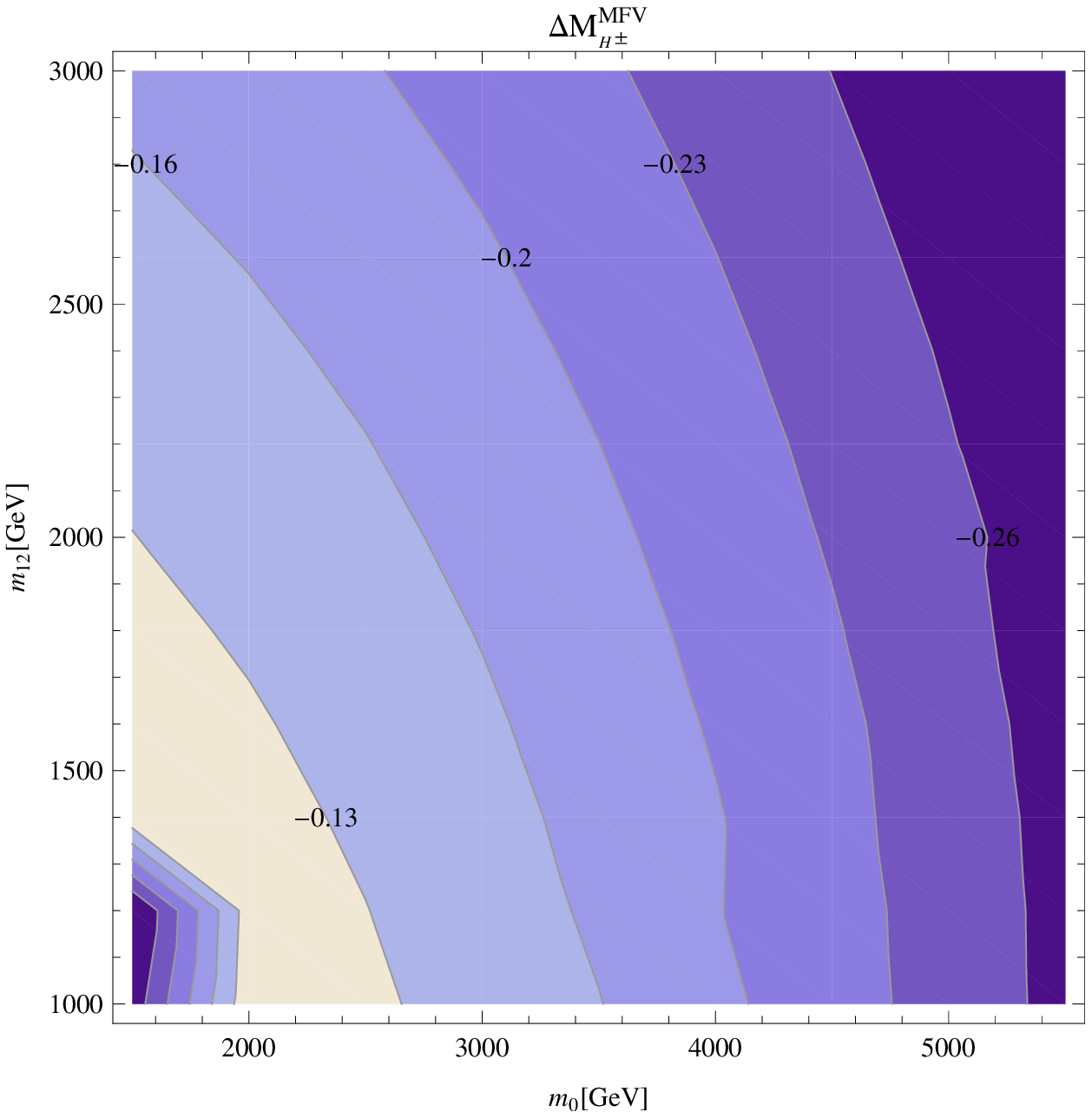,scale=0.50,angle=0,clip=}
\psfig{file=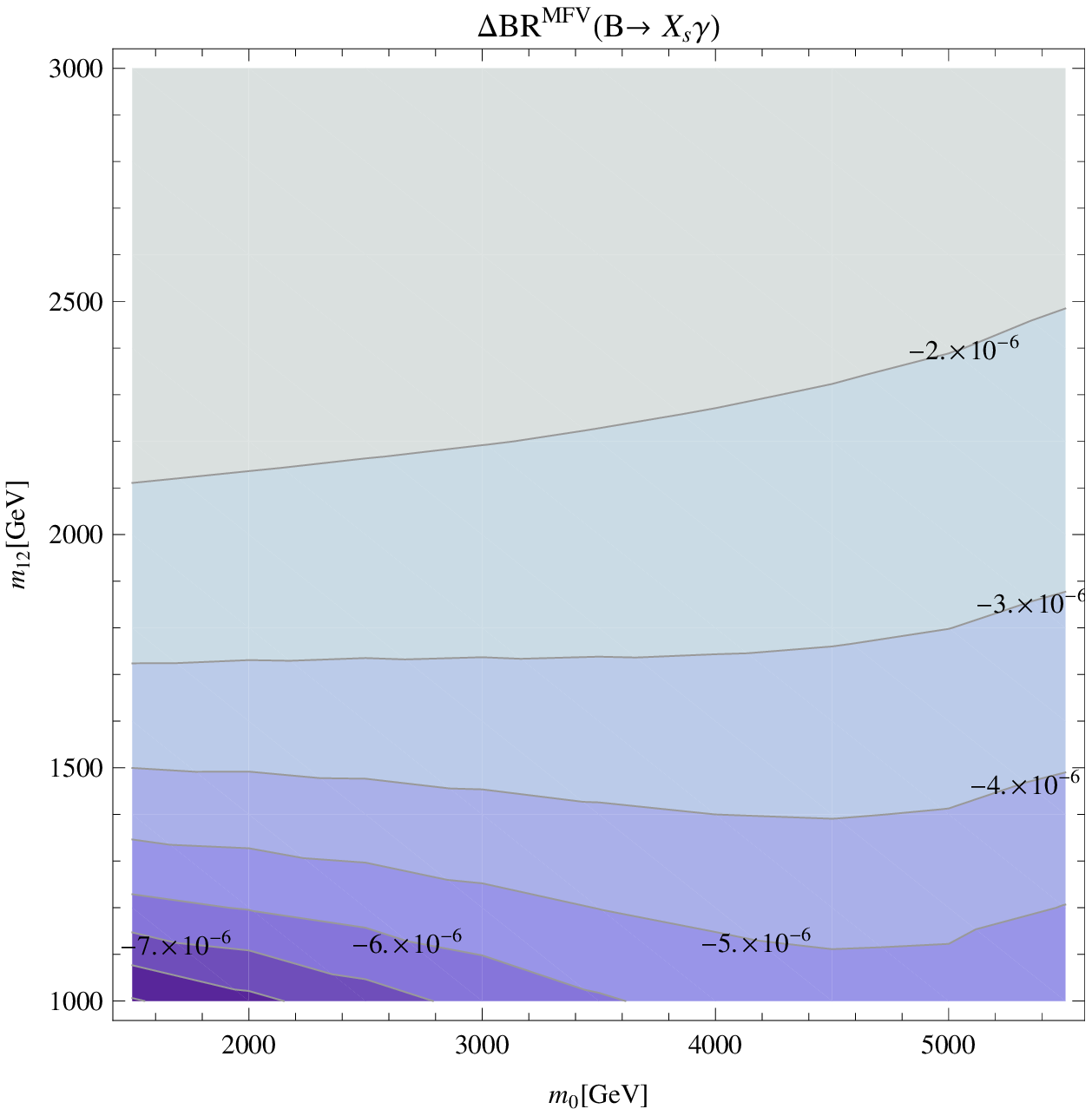,scale=0.50,angle=0,clip=}\\
\vspace{0.5cm}
\psfig{file=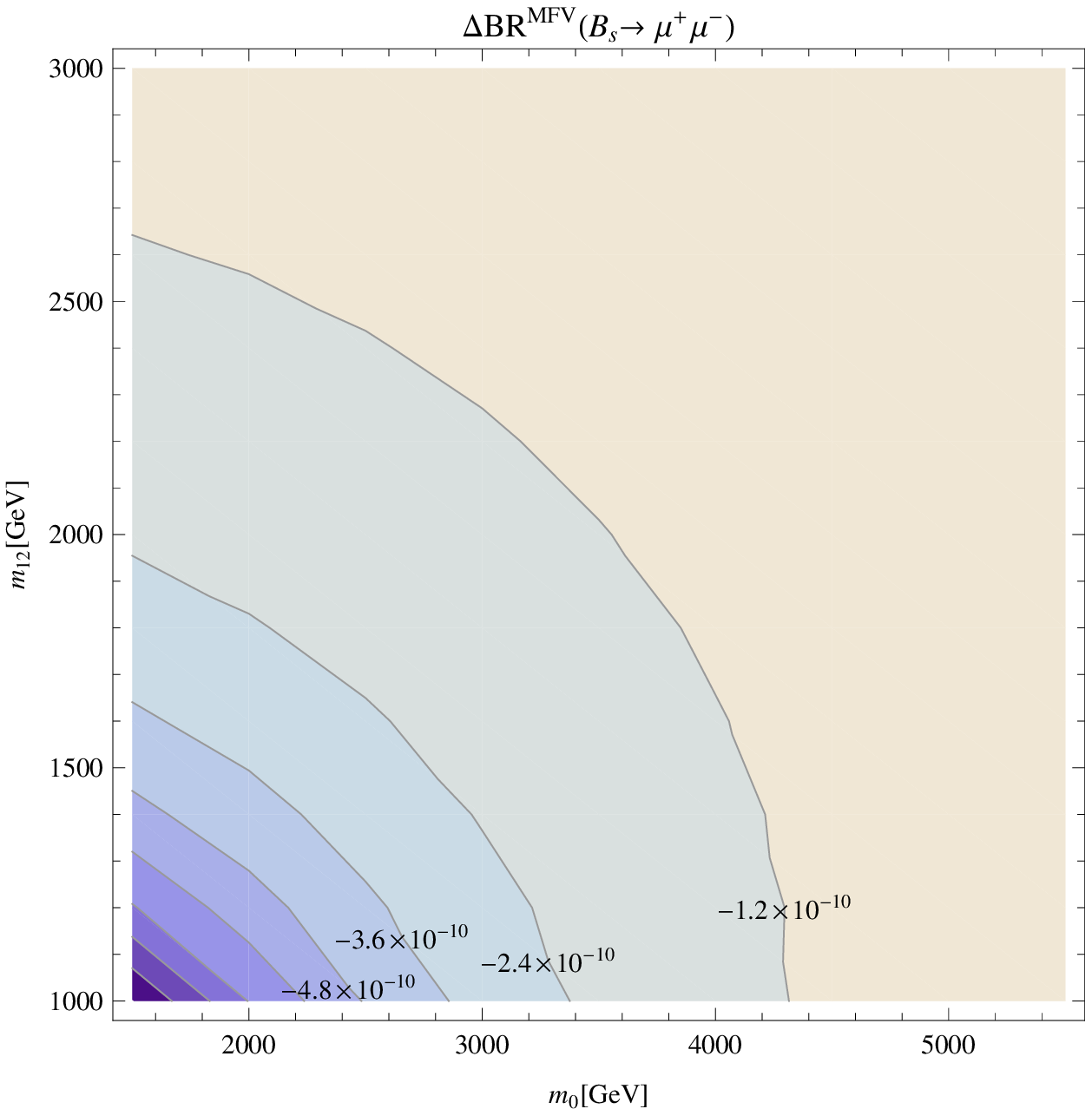,scale=0.50,angle=0,clip=}
\psfig{file=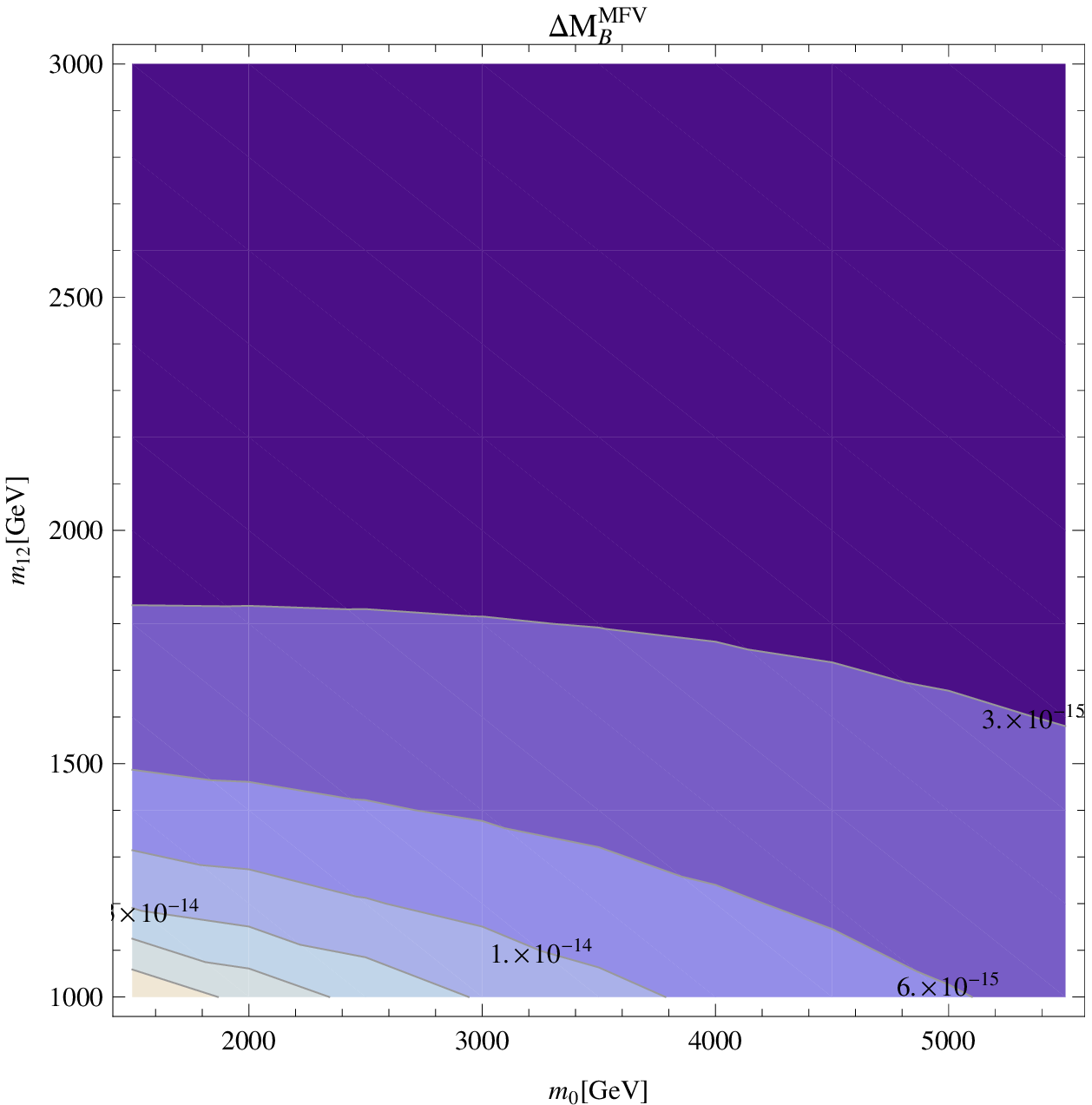,scale=0.50,angle=0,clip=}\\
\end{center}
\vspace{-1em}
\caption[Contours of Higgs mass corrections and BPO 
in the $m_0$--$m_{1/2}$ plane]{Contours of Higgs mass corrections (\DMh,
\DMH\ and \DMHp\ in GeV) and BPO (\Dbsg, \Dbmm\ and \Ddmbs) 
in the $m_0$--$m_{1/2}$ plane for $\tb = 45$ and \
$A_0 = -3000 \gev$ in the CMSSM.}
\label{fig:Sq-MH-BPO} 
\vspace{-3em}
\end{figure}

In \reffi{fig:Sq-MH-BPO} we show the results of
our CMSSM analysis with the effects of
the non-zero $\deFABij$ on the Higgs mass calculations and on the 
BPO in the $m_0$--$m_{1/2}$ plane for $\tb = 45$ and
$A_0 = -3000$. 
We only show this ``extreme'' case, where smaller values of
$\tb$ and $A_0$ would lead to smaller effects.
In the upper left, upper right and middle left plot we show \DMh,
\DMH\ and \DMHp, respectively. It can be seen that the effects on the
neutral Higgs boson masses are negligible w.r.t.\ the experimental
accuracy. The effects on $\MHp$ can reach \order{100 \mev}, 
where largest effects are found for both very small values
of $m_0$ and $m_{1/2}$ (dominated by $\del{ULR}{23}$) or very large values of 
$m_0$ and $m_{1/2}$ (dominated by $\del{QLL}{13,23}$). Corrections of up to
$-300 \mev$ are found, but still remaining
below the foreseeable future precision. Consequently,
also in the Higgs mass evaluation not taking into account the non-zero
values of the $\deFABij$ is a good approximation.
In the middle right, lower left and lower right plot of
\reffi{fig:Sq-MH-BPO} we show the results for the BPO
\Dbsg, \Dbmm\ and \Ddmbs,
respectively. The effects in \Dbsg\ are of \order{-10^{-5}} and thus one
order of magnitude smaller than the experimenal accuracay. Similarly, we
find $\Dbmm \sim \order{10^{-10}}$ and 
$\Ddmbs \sim \order{10^{-15} \gev}$, i.e.\ one or several orders of
magnitude below the experimental precision. This shows that for the BPO
neglecting the effects of non-zero $\deFABij$ in the CMSSM is a good
approximation. 
\subsection{\boldmath \brhbs}
\begin{figure}[ht!]
\begin{center}
\psfig{file=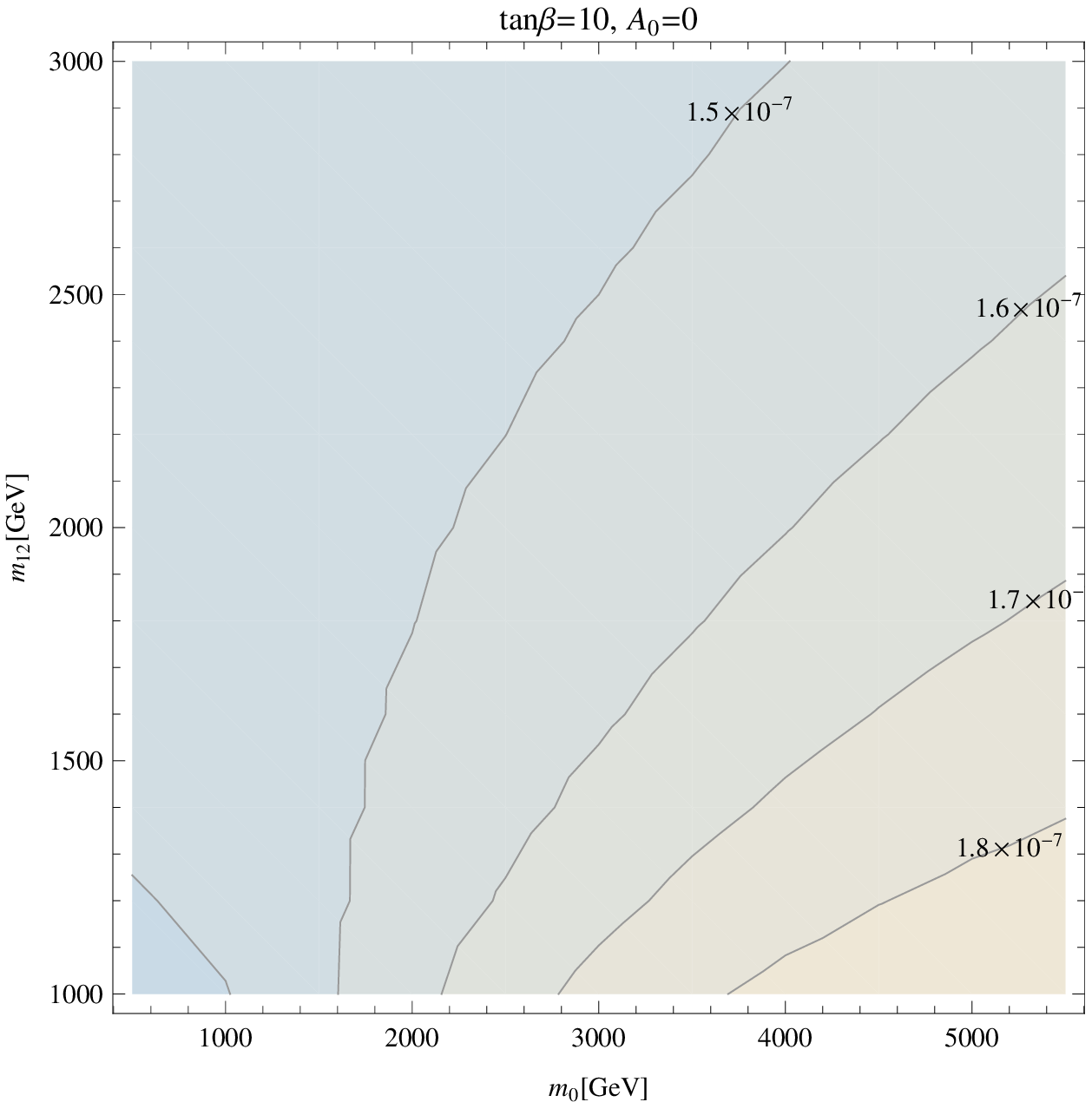  ,scale=0.51,angle=0,clip=}
\psfig{file=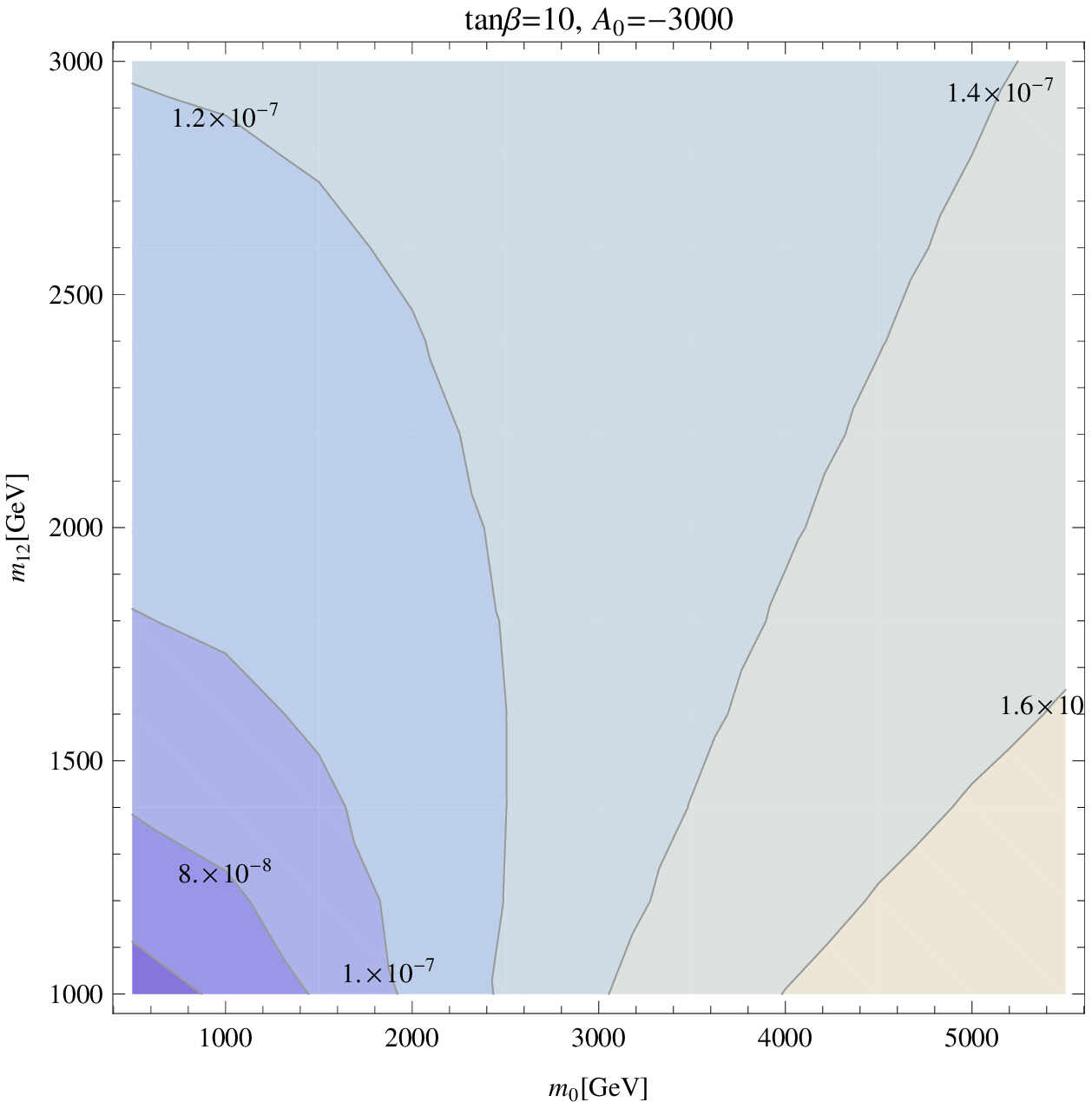  ,scale=0.51,angle=0,clip=}\\
\vspace{0.2cm}
\psfig{file=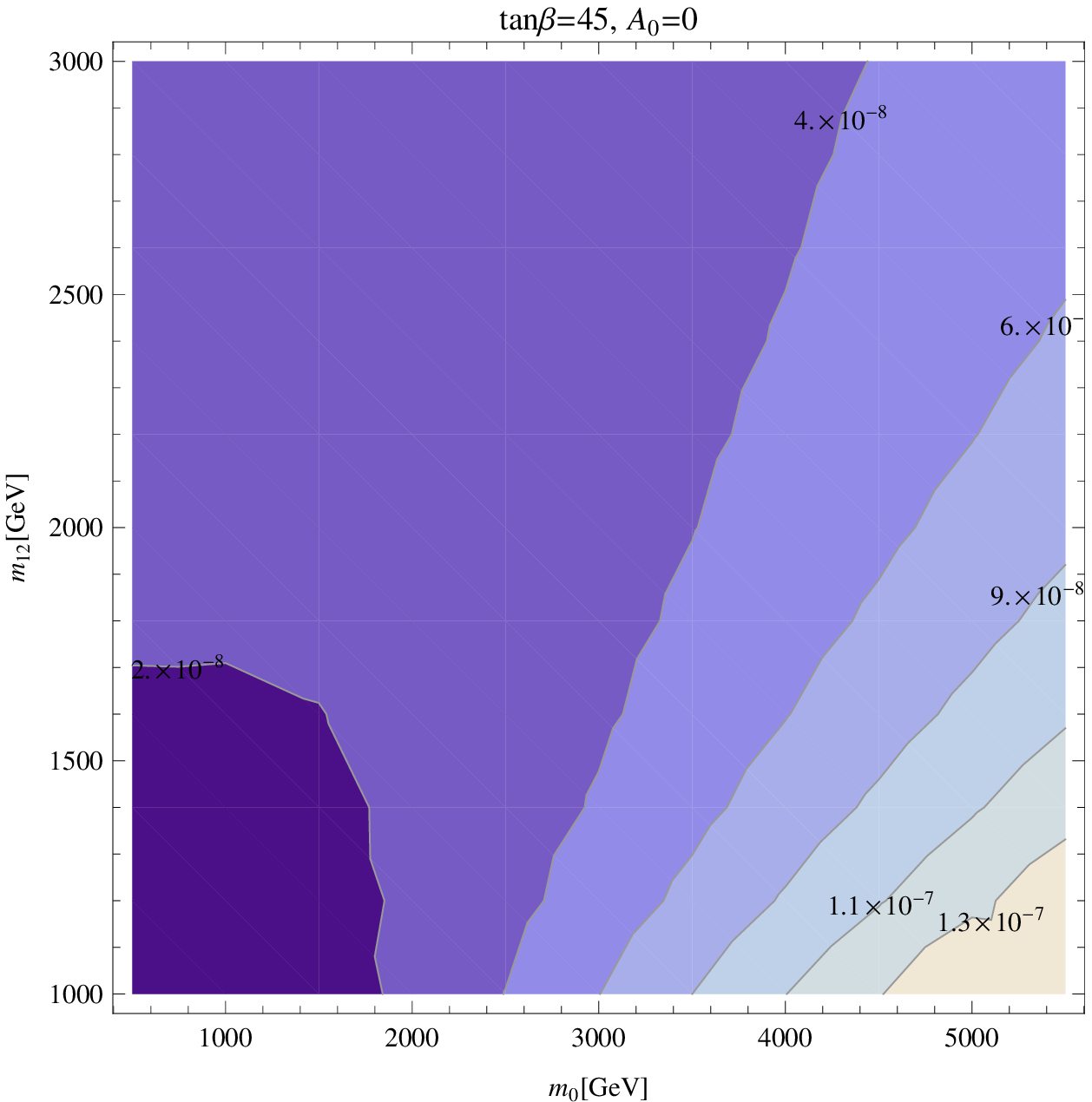 ,scale=0.51,angle=0,clip=}
\psfig{file=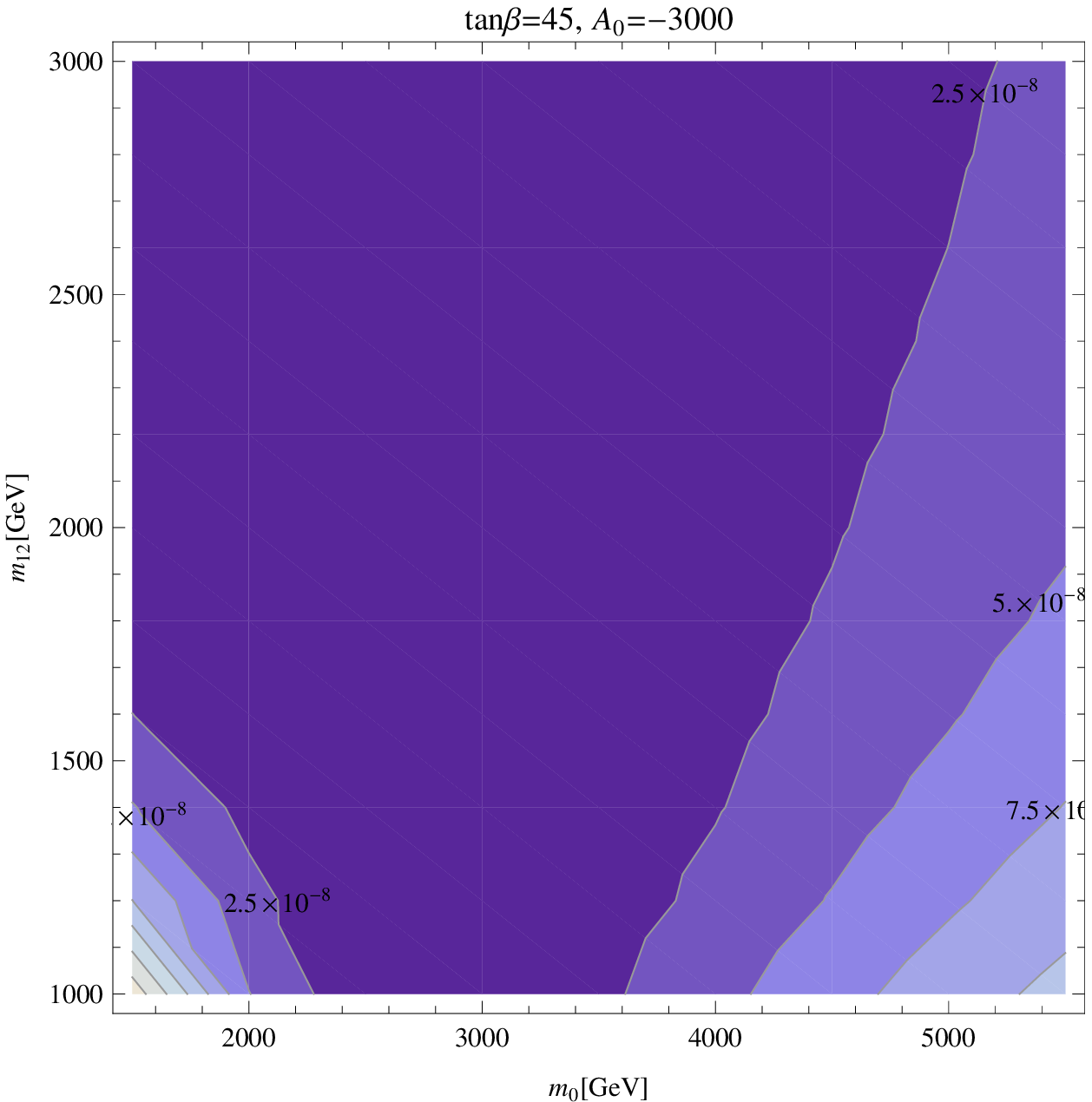   ,scale=0.51,angle=0,clip=}\\
\vspace{-0.2cm}
\end{center}
\caption[Contours of ${\rm BR}(h \rightarrow b \bar{s}+\bar{b}s)$  in the
  $m_0$--$m_{1/2}$ plane]{Contours of ${\rm BR}(h \rightarrow b \bar{s}+\bar{b}s)$  in the
  $m_0$--$m_{1/2}$ plane for different values of $\tb$ and $A_0$ in
  the CMSSM.}   
\label{fig:BRhbs}
\end{figure}

The results are shown in \reffi{fig:BRhbs}, where we display the
contours of \brhbs\ in the ($m_0$, $m_{1/2}$) plane for $\tb=10$,
$A_0=0$ (upper left), $\tb=10$, $A_0=-3000 \gev$ (upper right), 
$\tb=45$, $A_0=0$ (lower left) and $\tb=45$, $A_0=-3000 \gev$ (lower
right). By comparison with planes for other $\tb$-$A_0$ combinations we
have varyfied that these four planes constitute a representative
example. The allowed parameter space can be deduced by comparing to the
results presented above and in \citeres{mc9}. While not all the planes are in
agreement with current constraints, large parts, in particular for
larger values of $m_0$ and $m_{1/2}$ are compatible with a combination
of direct searches, flavor and electroweak precision observables as well
as astrophysical data. Upper bounds on $m_0$ at the few~TeV level
could possibly be set by including the findings of \refse{EWPO-CMSSM-Res}
into a global CMSSM analysis.

In \reffi{fig:BRhbs} one can see that for most of parameter space values
of \order{10^{-7}} are found for \brhbs, i.e.\ outside the reach of
current or future collider experiments. Even for the ``most
  extreme'' set of parameters we have analyzed, $\tb = 45$ and 
$A_0 = -3000 \gev$, no detectable rate has been found. Turning the
argument around, any observation of the decay \hbs\ at the (discussed)
future experiments would exclude the CMSSM as a possible model.
\clearpage
\section{Effects of slepton mixing in \CMSSMI.}
\label{sec:Sl}

In this section we analyze the effects of non-zero $\deFABij$ values in
the \CMSSMI.
In order to investigate the effects induced just by the mixings
in the slepton sector, such that we can compare their contribution from
the one produced by the mixings in the squak sector (and to discriminate
it from effects from mixings in the squark sector) we present here the
results with only $\deFABij$ in 
the slepton sector non-zero, i.e.\ after the RGE running with both CKM
and seesaw parameters non-zero, the $\deFABij$ from the squark
sector are set to zero by hand at the EW scale. The effects of the 
squark mixing in the \CMSSMI\ are nearly
indistinguishable from the ones analyzed in the previous subsection.

As mentioned in \refse{sec:mssmI}, the calculations in this section are
done by using the values of $Y_\nu$ constructed from \refeq{eq:casas} with
degenerate $M_R$'s. The matrix $R$ is set to the identity since it does not 
enter in \refeq{eq:ynu2} and therefore the slepton $\deFABij$'s do not depend
on it. The matrix $m_\nu^\delta$ is a diagonal mass matrix adjusted to
reproduce neutrino masses at low energy compatible with the experimental
observations and with hierarchical neutrino masses. We performed our
computation by using the seesaw scale $M_N=10^{14} \gev$. With this
choice the bound $\br(\mu \to e \gamma) < 5.7 \times 10^{-13}$ ~\cite{Adam:2013mnn}
imposes severe restrictions on the $m_0$--$m_{1/2}$ plane, excluding 
values of $m_0$  below 2--3~TeV (depending on $\tb$ and $A_0$). The values of
the slepton $\deFABij$ will increase as the scale $M_N$  increases but also
does the parameter space excluded by the $\br(\mu \to e \gamma)$
bound. For example, by increasing $M_N$ by an order of magnitude, the largest
entries in the matrix $Y_\nu$ will become of \order{1} and the bound on
$\br(\mu \to e \gamma)$ will only be satisfied if $m_0\approx 5 \tev$ (see more details below). 

\subsection{Slepton \boldmath{$\deFABij$'s}}
\begin{figure}[ht!]
\begin{center}
\psfig{file=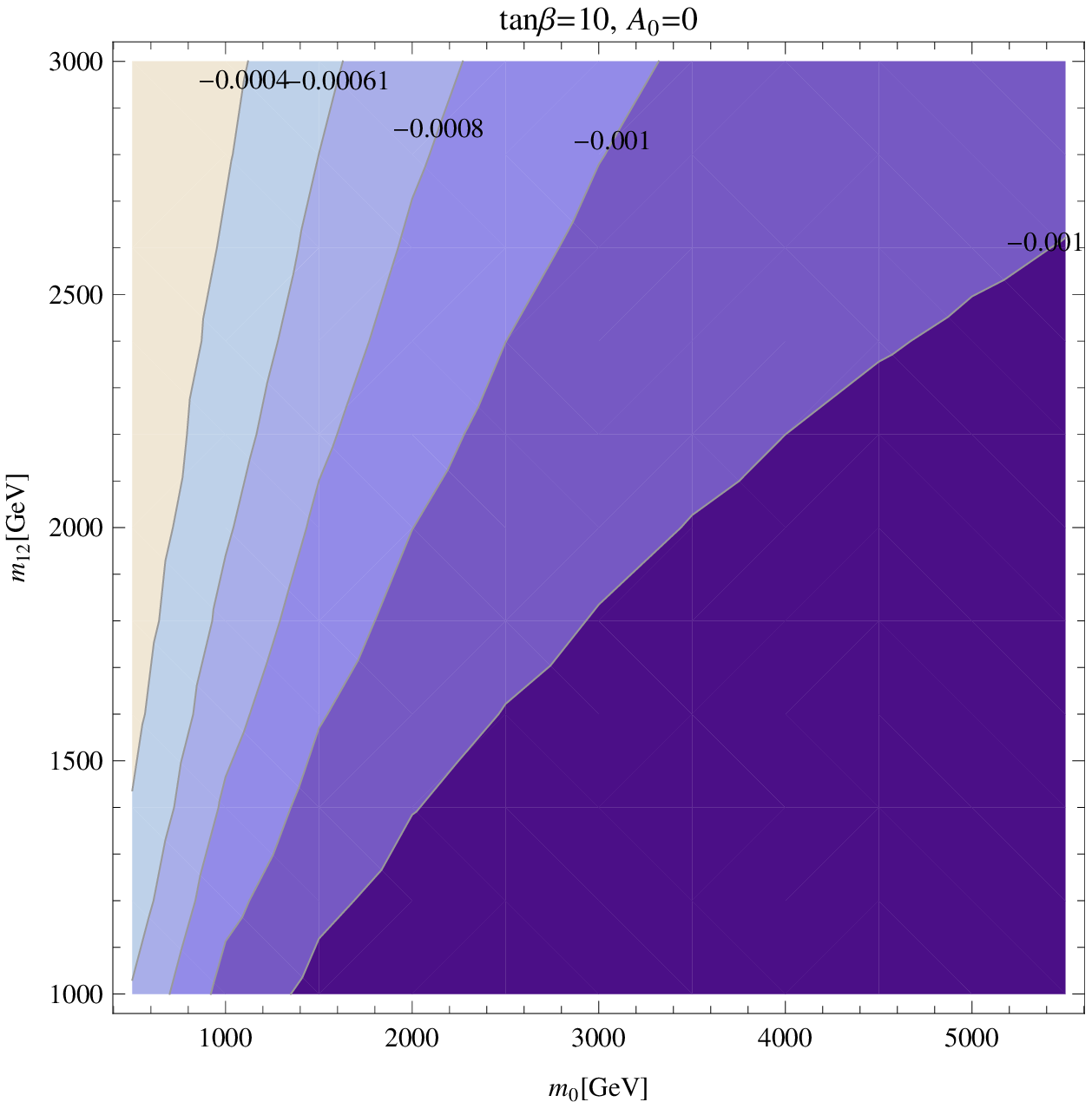  ,scale=0.51,angle=0,clip=}
\psfig{file=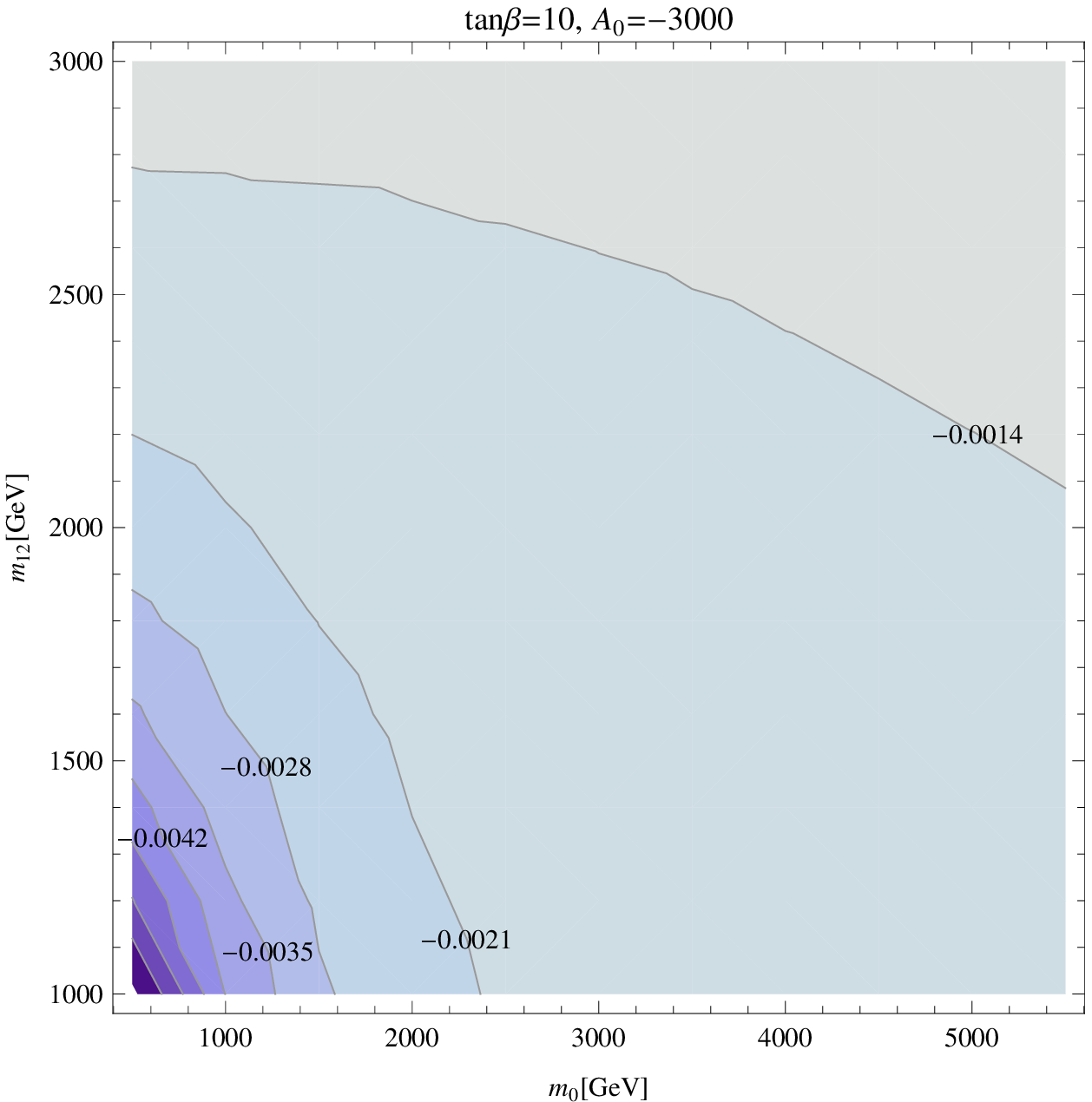  ,scale=0.51,angle=0,clip=}\\
\vspace{0.2cm}
\psfig{file=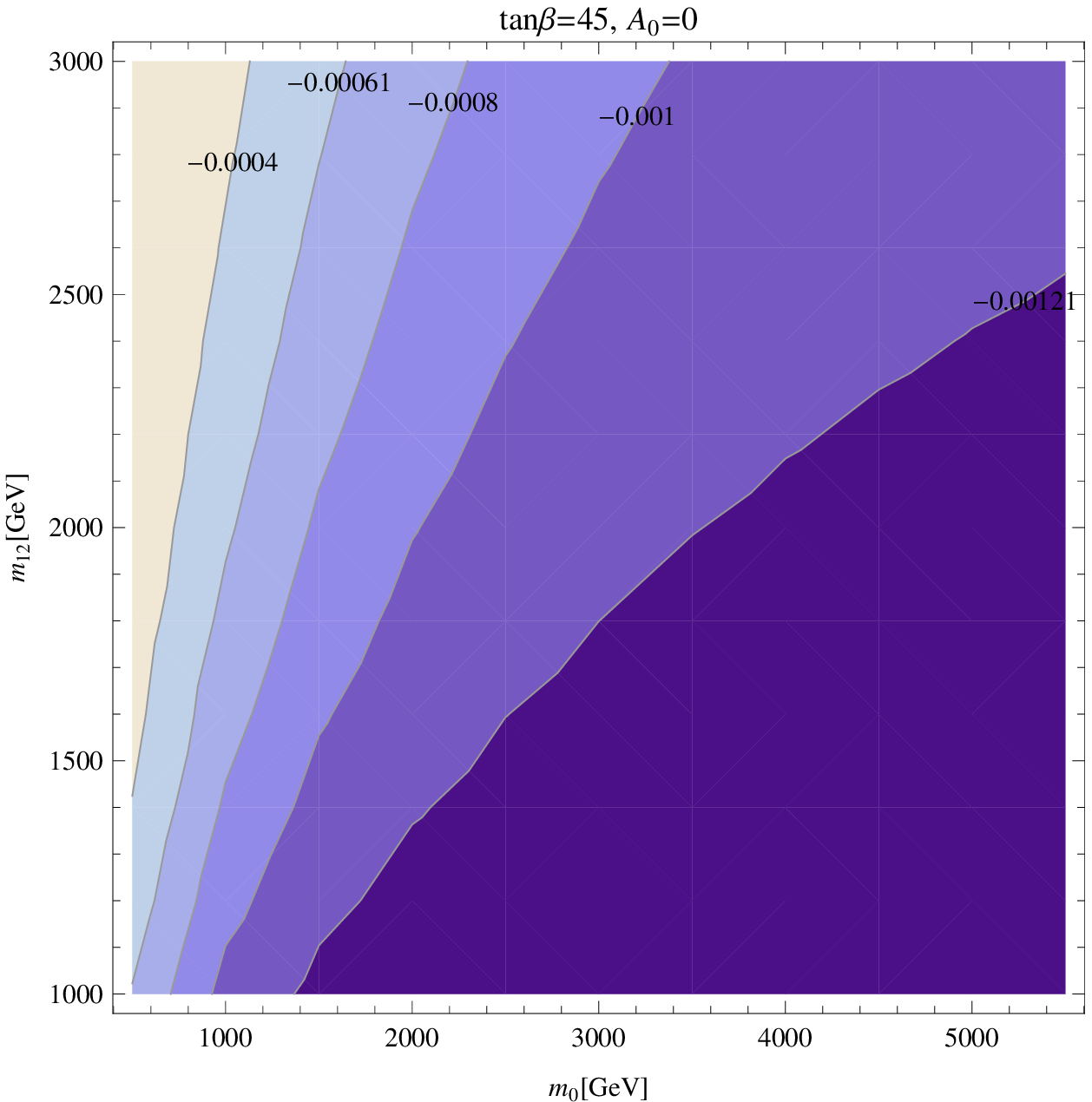 ,scale=0.51,angle=0,clip=}
\psfig{file=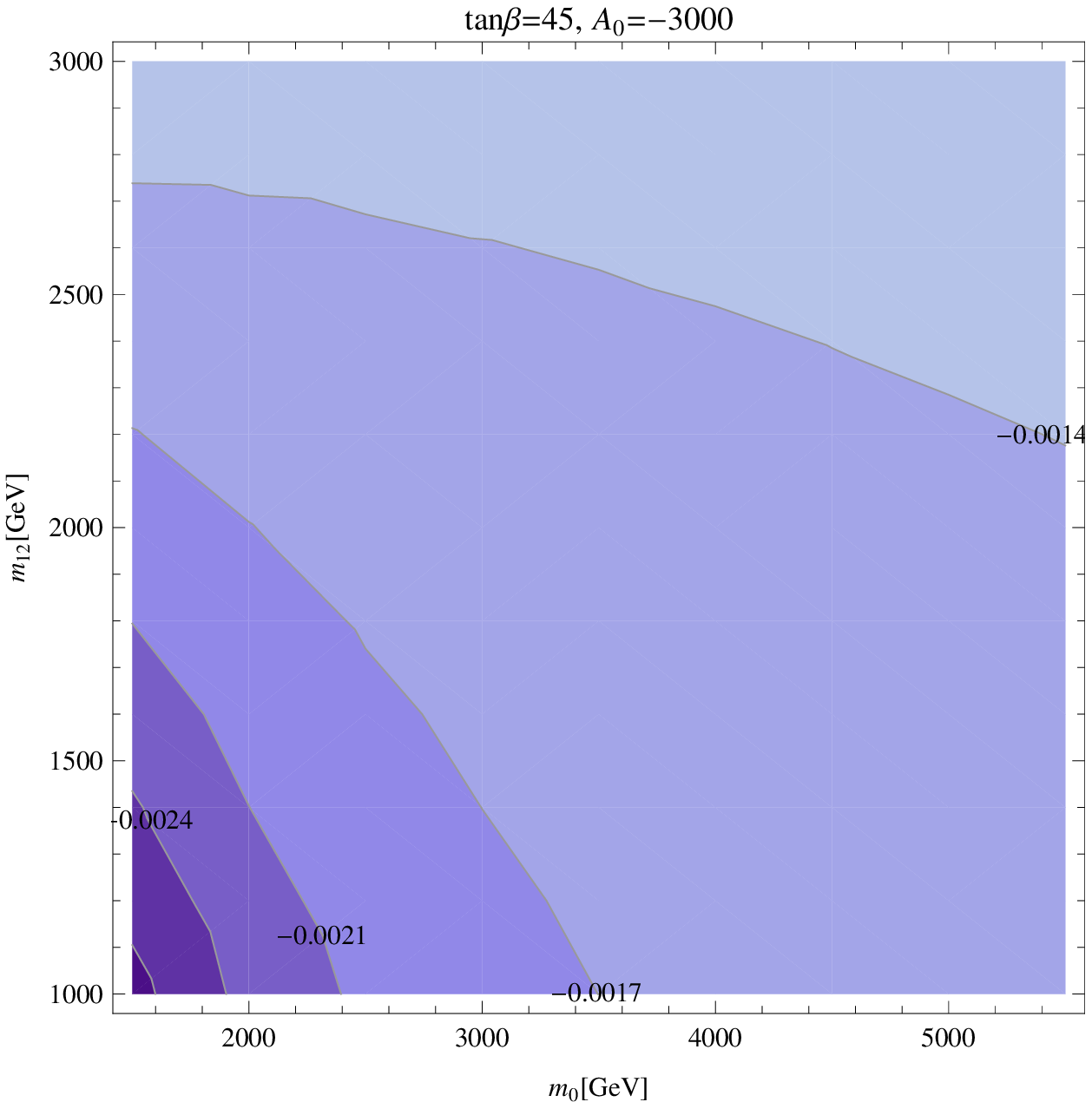   ,scale=0.51,angle=0,clip=}\\
\vspace{-0.2cm}
\end{center}
\caption[Contours of $\delta^{LLL}_{12}$  in the
  $m_0$--$m_{1/2}$ plane.]{Contours of $\delta^{LLL}_{12}$  in the
  $m_0$--$m_{1/2}$ plane for different values of $\tb$ and    
$A_0$ in the \CMSSMI. }  
\label{fig:DelLLL12}
\end{figure}

\begin{figure}[ht!]
\begin{center}
\psfig{file=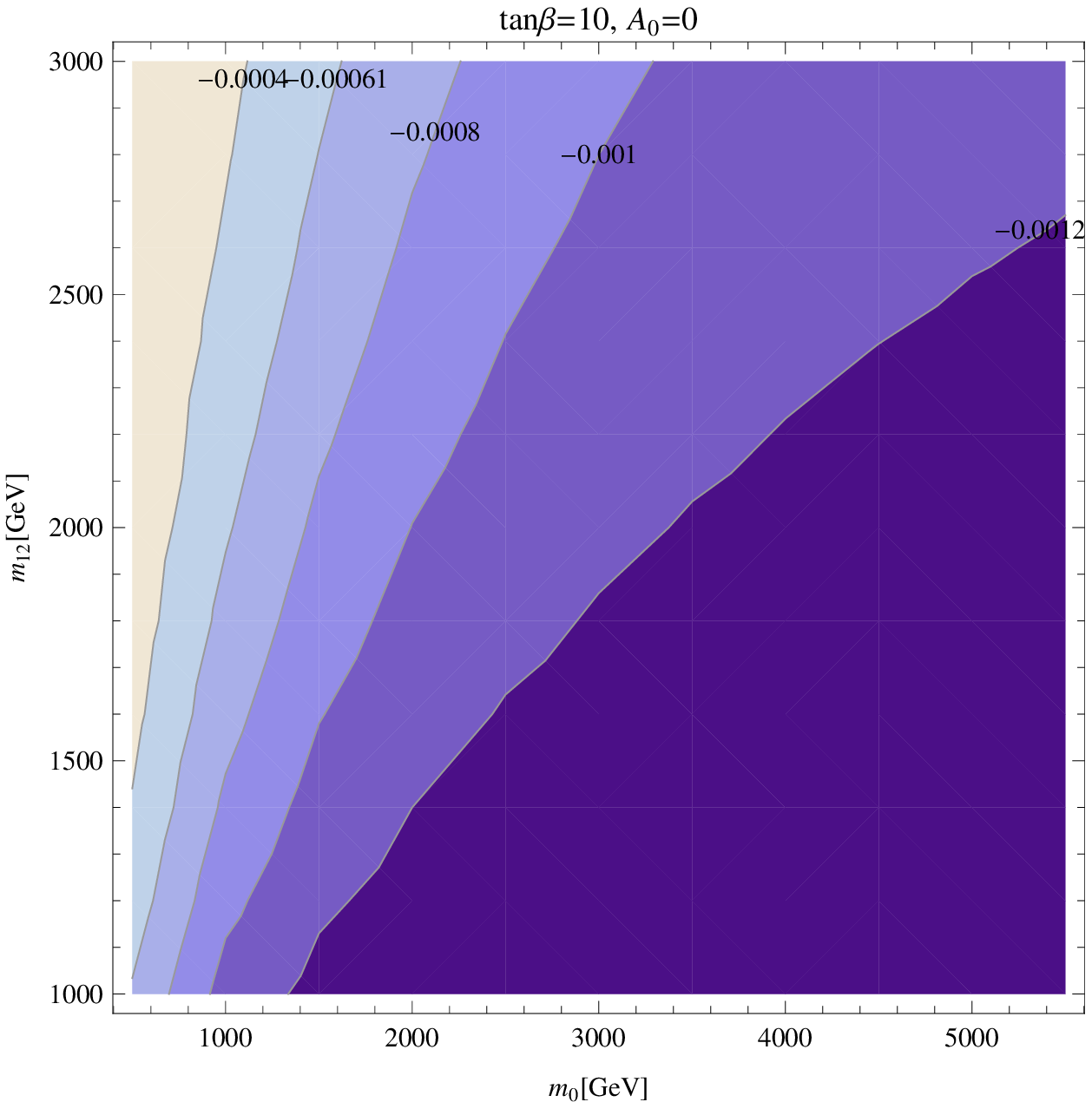  ,scale=0.51,angle=0,clip=}
\psfig{file=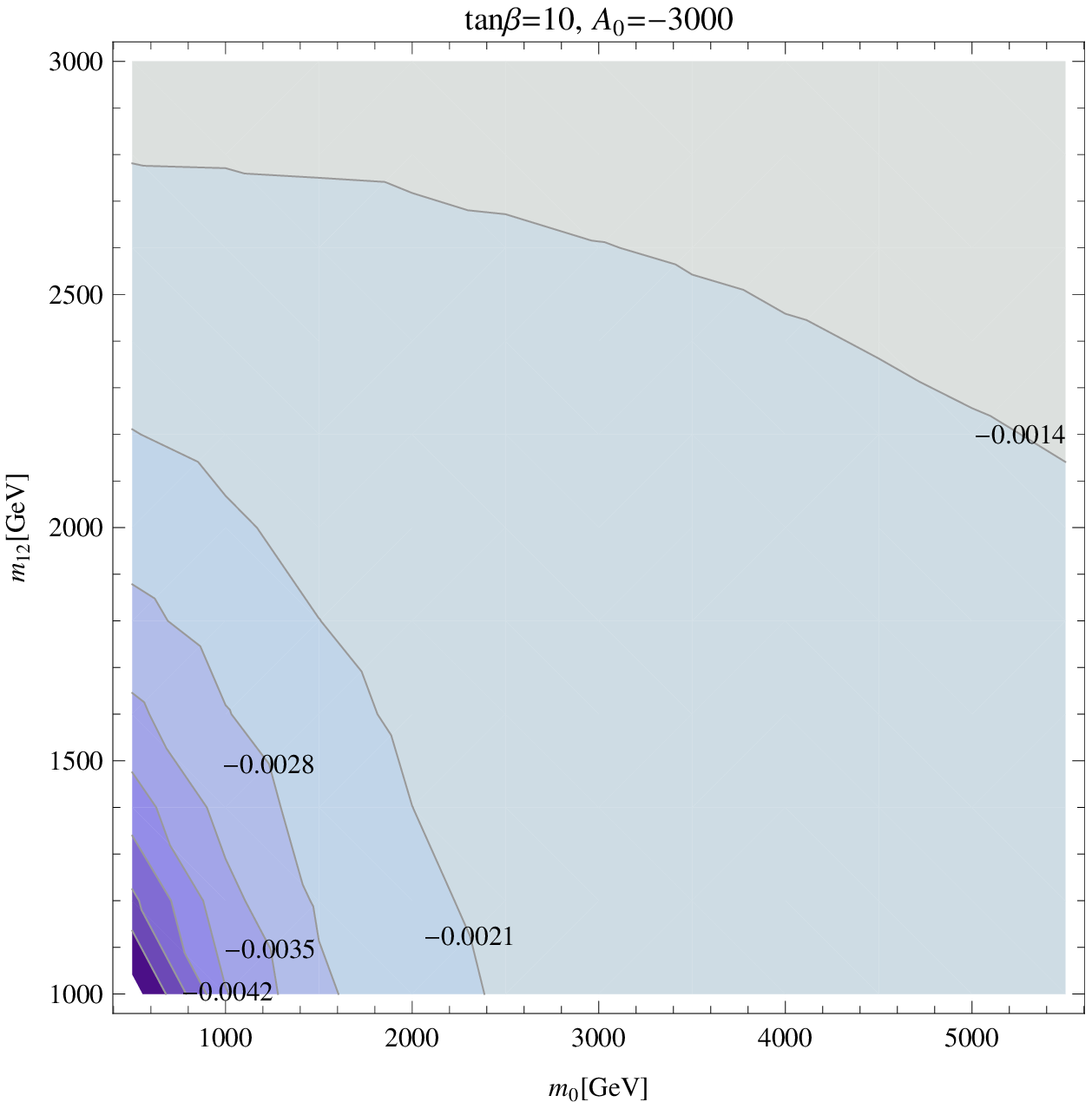  ,scale=0.51,angle=0,clip=}\\
\vspace{0.2cm}
\psfig{file=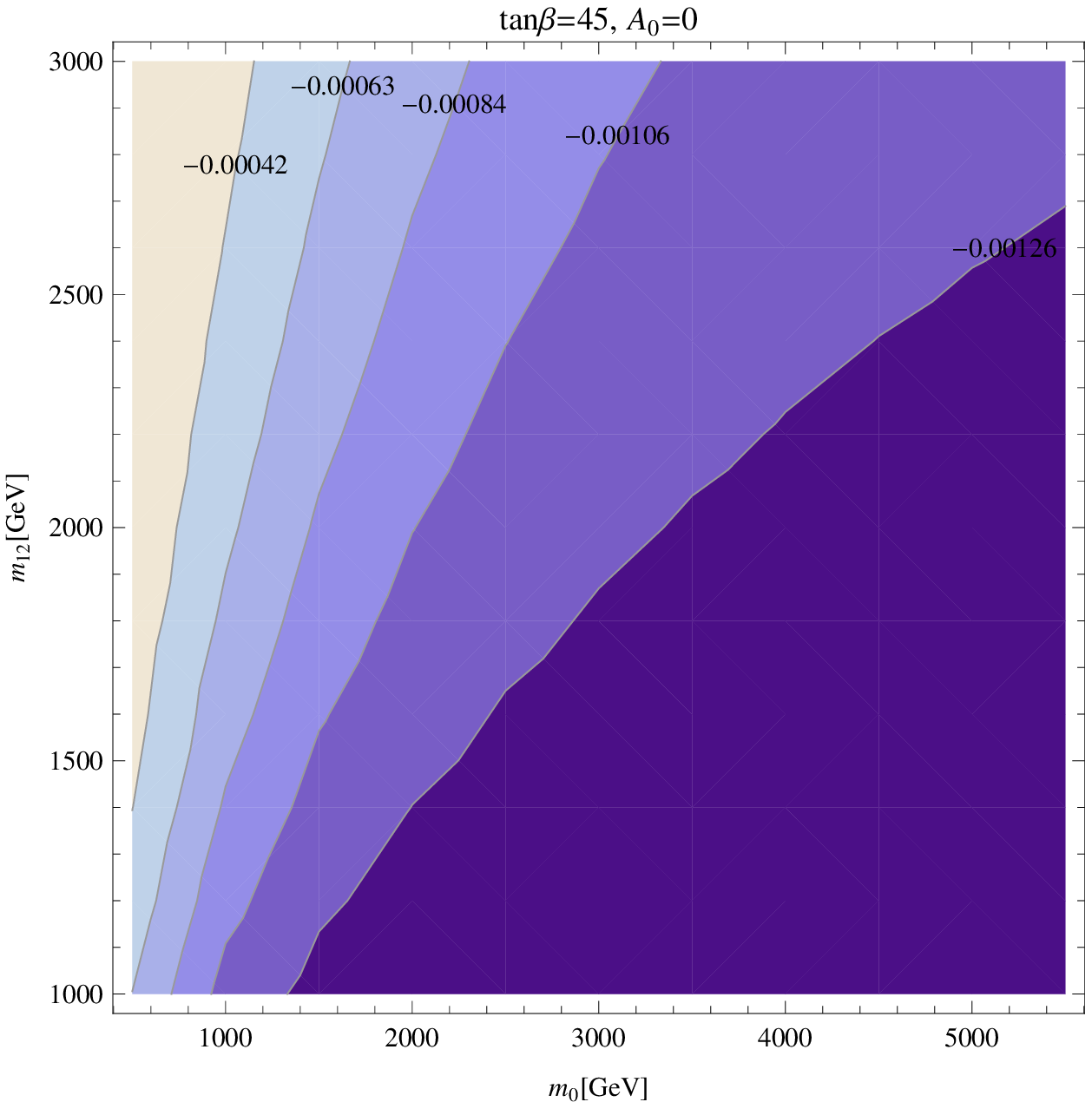 ,scale=0.51,angle=0,clip=}
\psfig{file=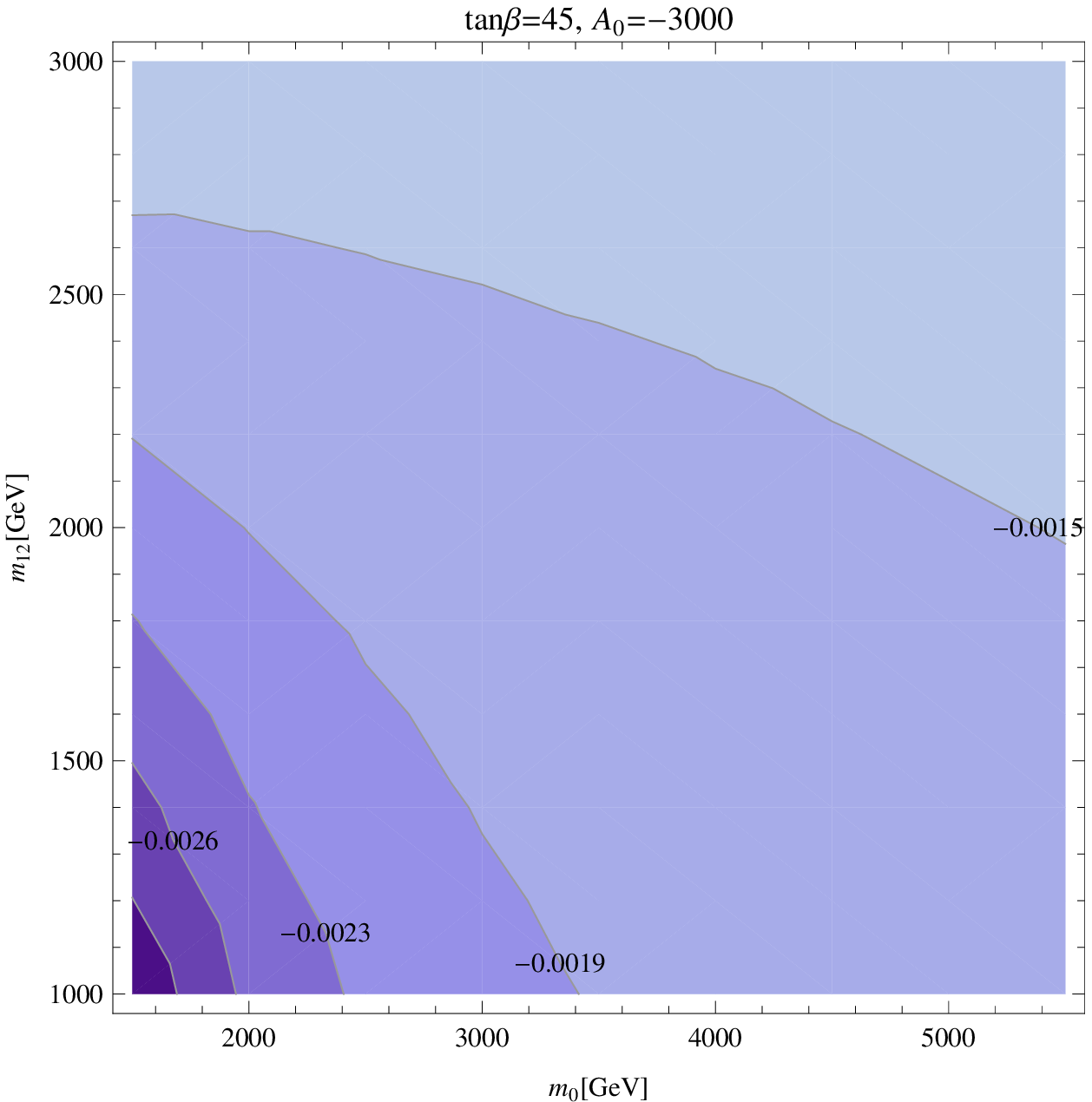   ,scale=0.51,angle=0,clip=}\\
\vspace{-0.2cm}
\end{center}
\caption[Contours of $\delta^{LLL}_{13}$ in the
  $m_0$--$m_{1/2}$ plane.]{Contours of $\delta^{LLL}_{13}$ in the
  $m_0$--$m_{1/2}$ plane for different values of $\tb$ and    
$A_0$ in the \CMSSMI. }  
\label{fig:DelLLL13}
\end{figure}

\begin{figure}[ht!]
\begin{center}
\psfig{file=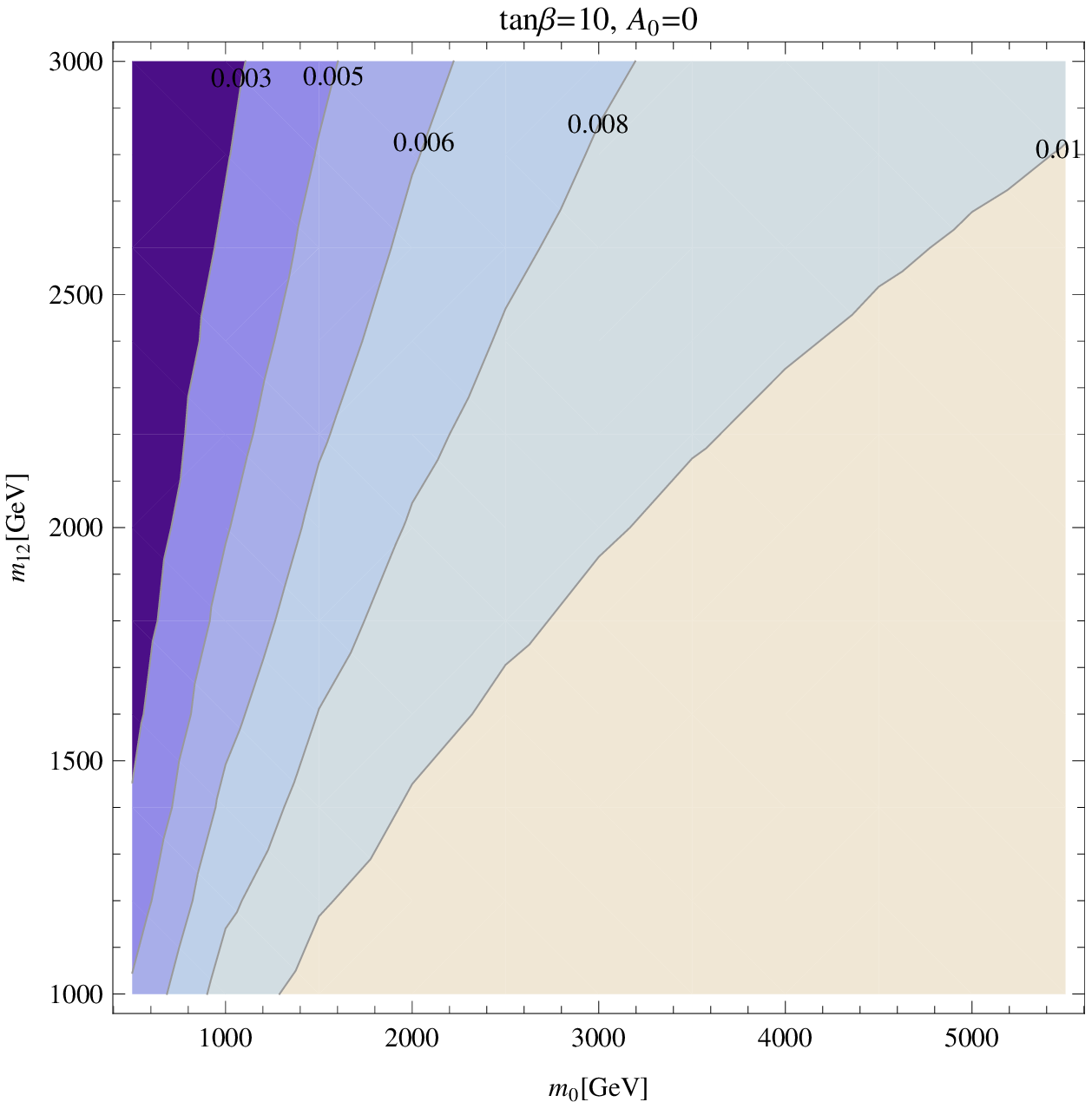  ,scale=0.51,angle=0,clip=}
\psfig{file=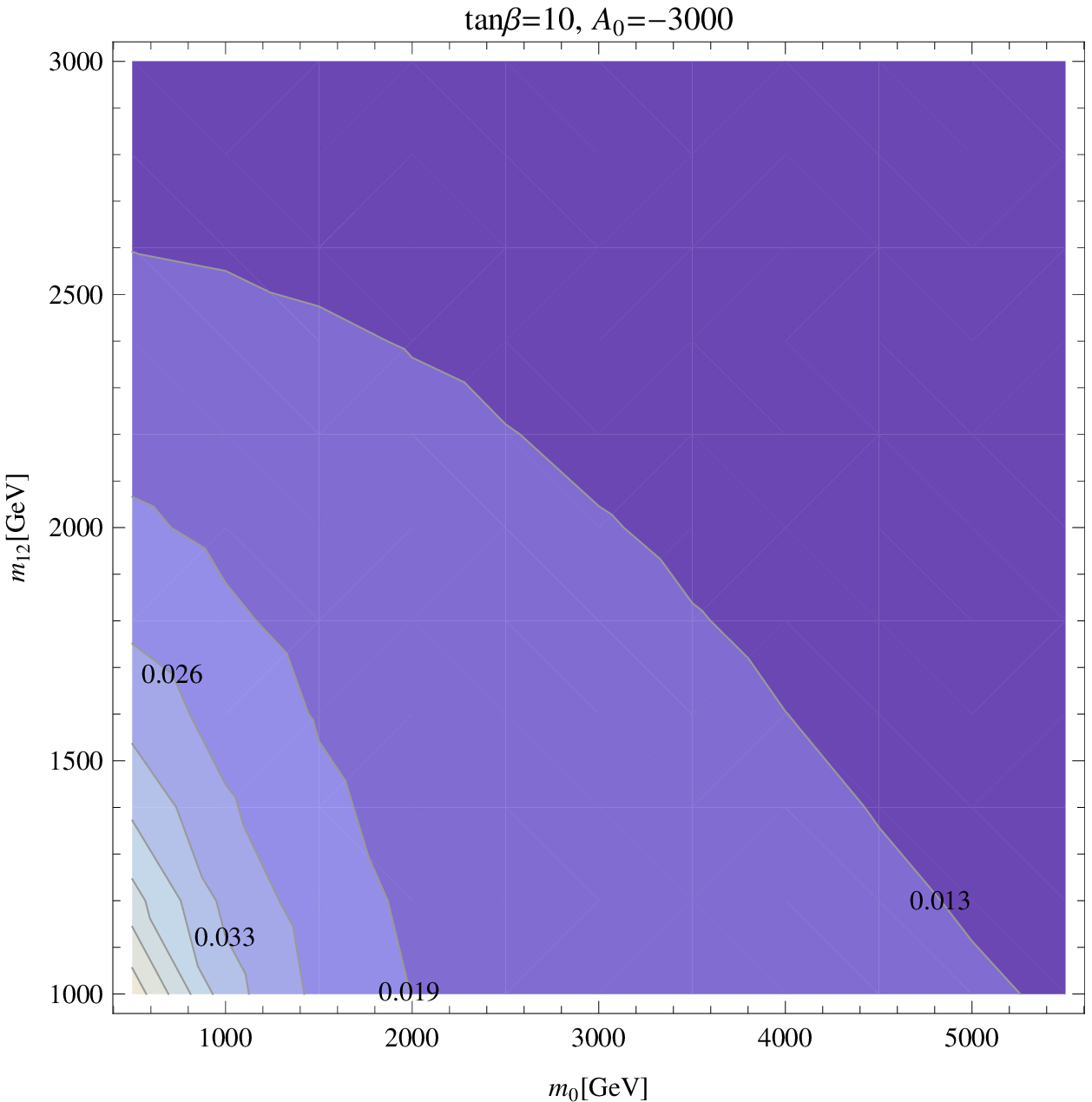  ,scale=0.51,angle=0,clip=}\\
\vspace{0.2cm}
\psfig{file=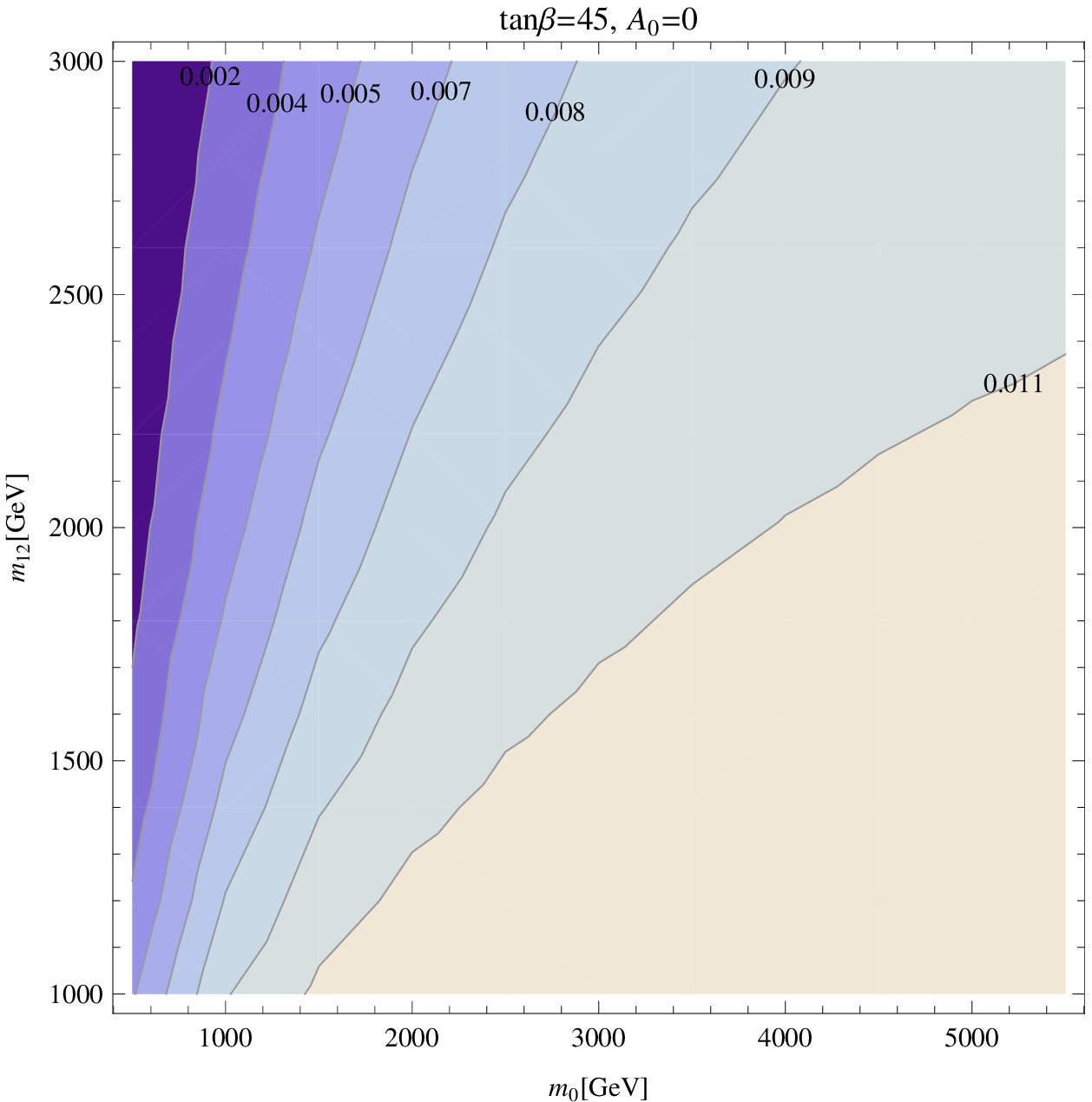 ,scale=0.51,angle=0,clip=}
\psfig{file=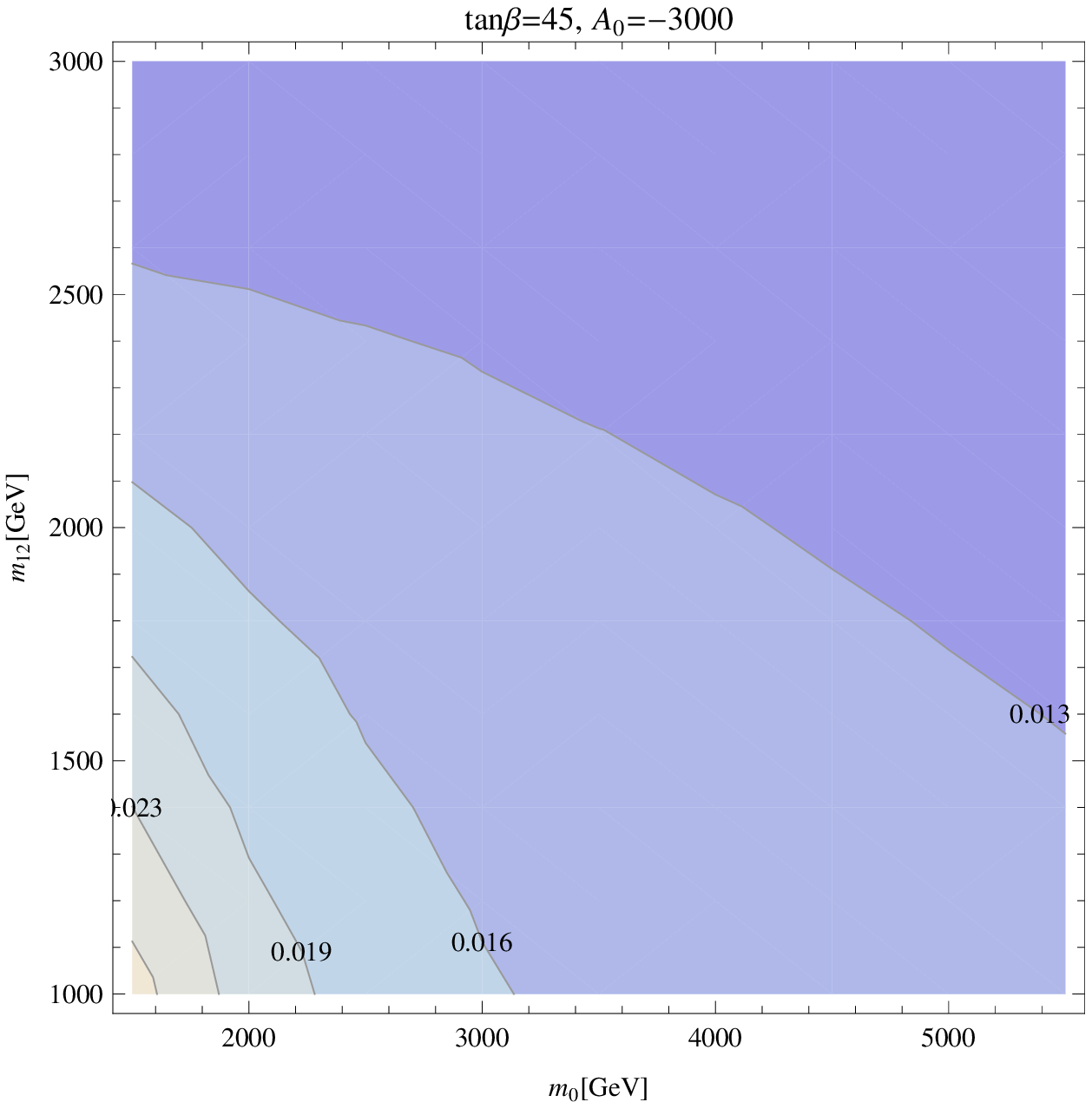   ,scale=0.51,angle=0,clip=}\\
\vspace{-0.2cm}
\end{center}
\caption[Contours of $\delta^{LLL}_{23}$ in the
  $m_0$--$m_{1/2}$ plane.]{Contours of $\delta^{LLL}_{23}$ in the
  $m_0$--$m_{1/2}$ plane for different values of $\tb$ and    
$A_0$ in the \CMSSMI.}  
\label{fig:DelLLL23}
\end{figure}

Our numerical results in the \CMSSMI\ are shown in
\reffis{fig:DelLLL12} - \ref{SL-MH}. As in the CMSSM we present the
results in the $m_0$--$m_{1/2}$ plane for four combinations of 
$\tb = 10, 45$ (upper and lower row) and $A_0 = 0, -3000 \gev$ (left
and right column), again capturing the ``extreme'' cases.
We start presenting the three most relevant $\deFABij$. 
\reffis{fig:DelLLL12}-\ref{fig:DelLLL23} show $\del{LLL}{12}$, 
$\del{LLL}{13}$ and $\del{LLL}{23}$, respectively. As expected,
$\del{LLL}{23}$ turns out to be largest of \order{0.01}, while the
other two are about one order of magnitude smaller. The dependence on
$\tb$ is not very prominent, but going from $A_0 = 0$ to $-3000 \gev$
has a strong impact on the $\deFABij$. For small $A_0$ the size of the
$\deFABij$ is increasing with larger $m_0$ and $m_{1/2}$, for 
$A_0 = -3000 \gev$ the largest values are found for small $m_0$ and
$m_{1/2}$. 
\subsection{EWPO}
\begin{figure}[ht!]
\begin{center}
\psfig{file=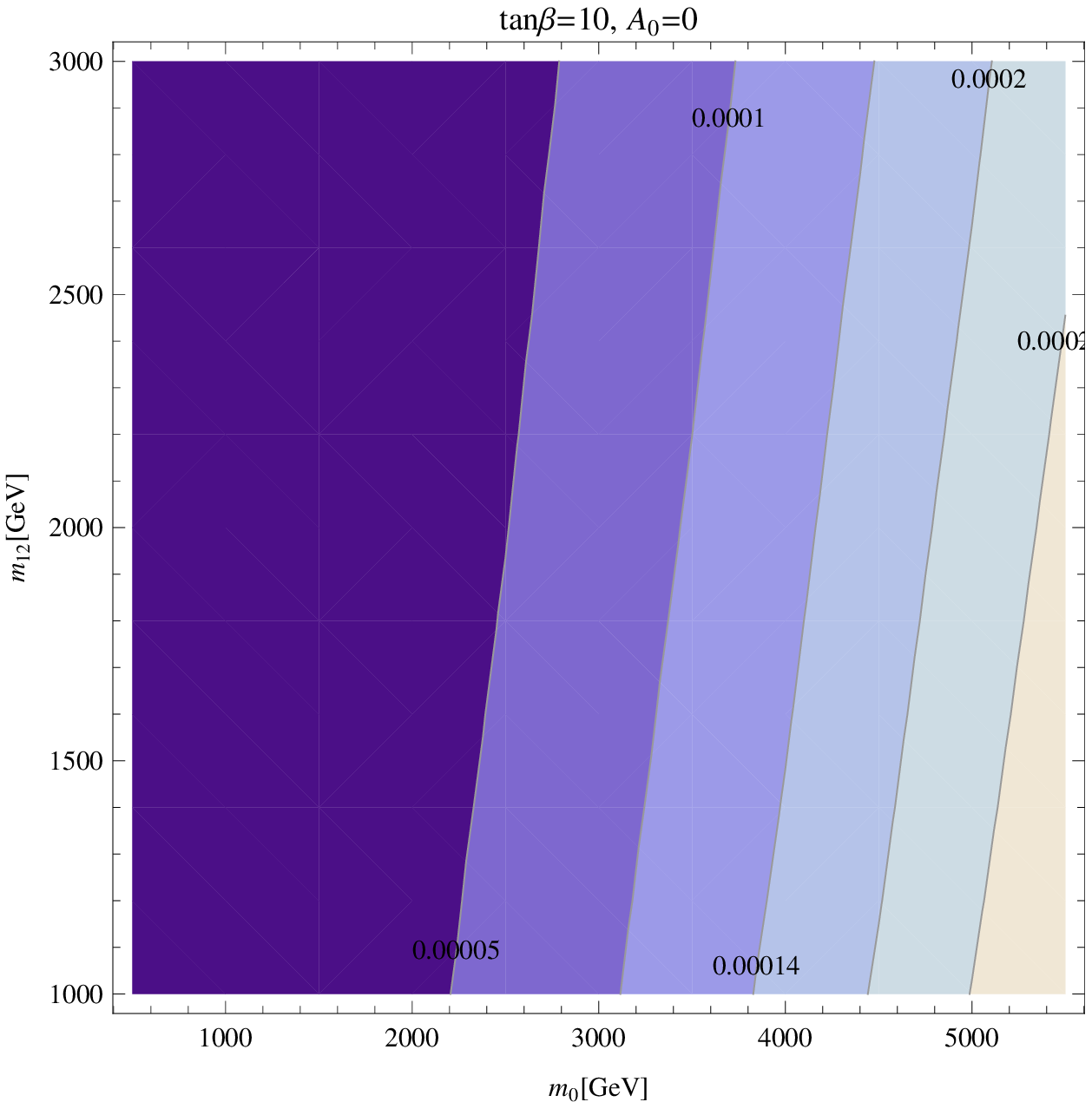  ,scale=0.51,angle=0,clip=}
\psfig{file=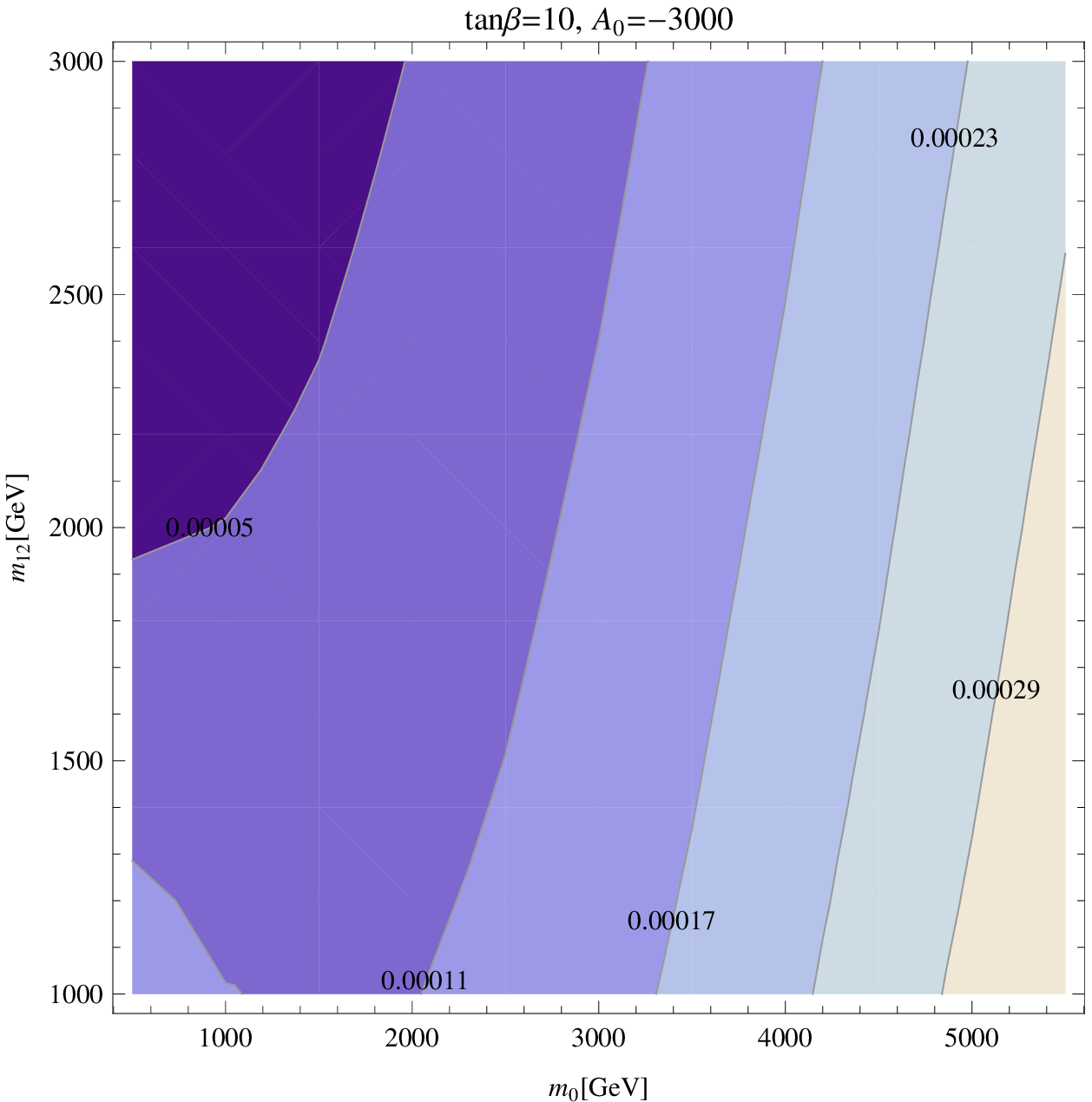  ,scale=0.51,angle=0,clip=}\\
\vspace{0.2cm}
\psfig{file=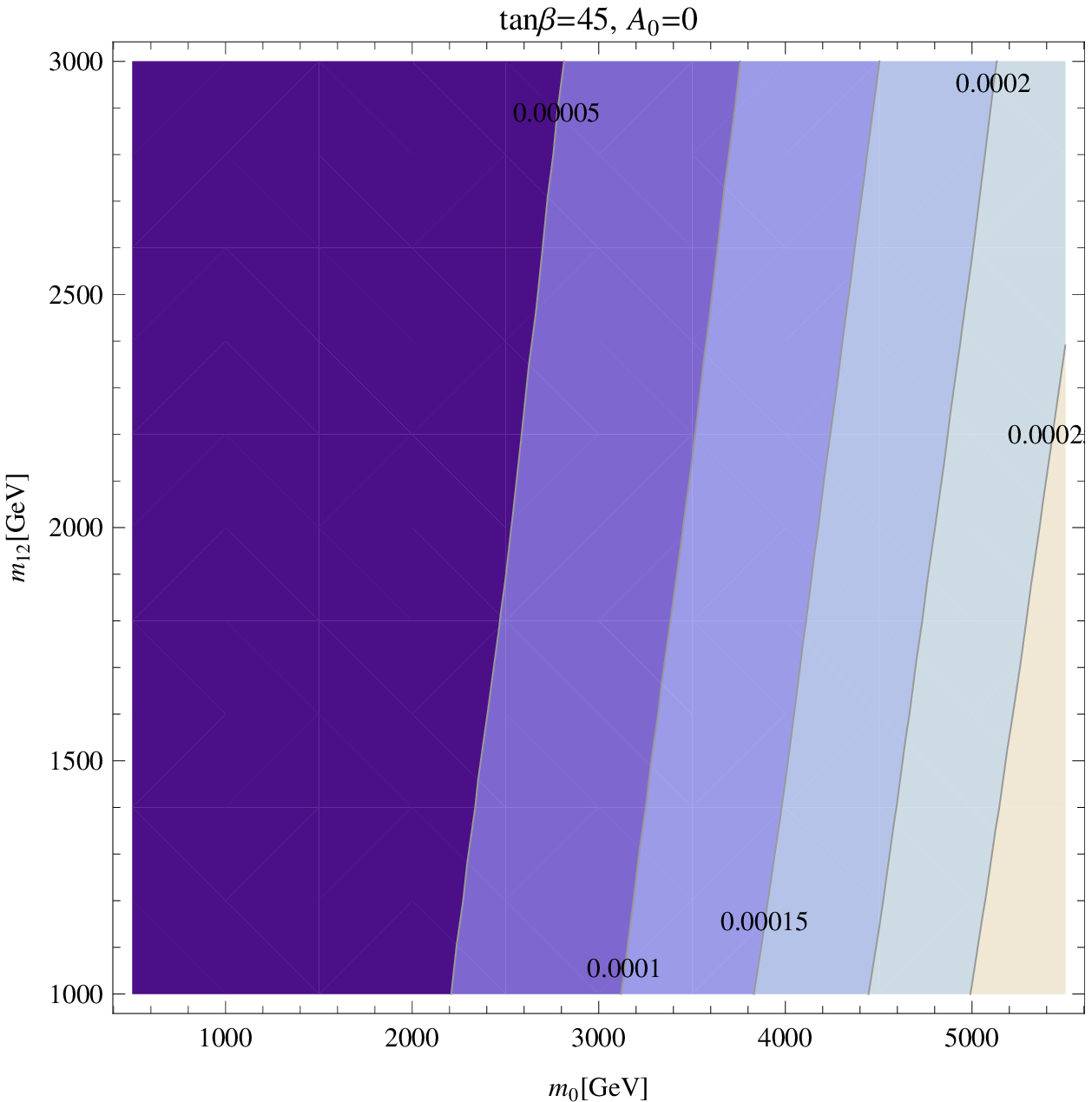 ,scale=0.51,angle=0,clip=}
\psfig{file=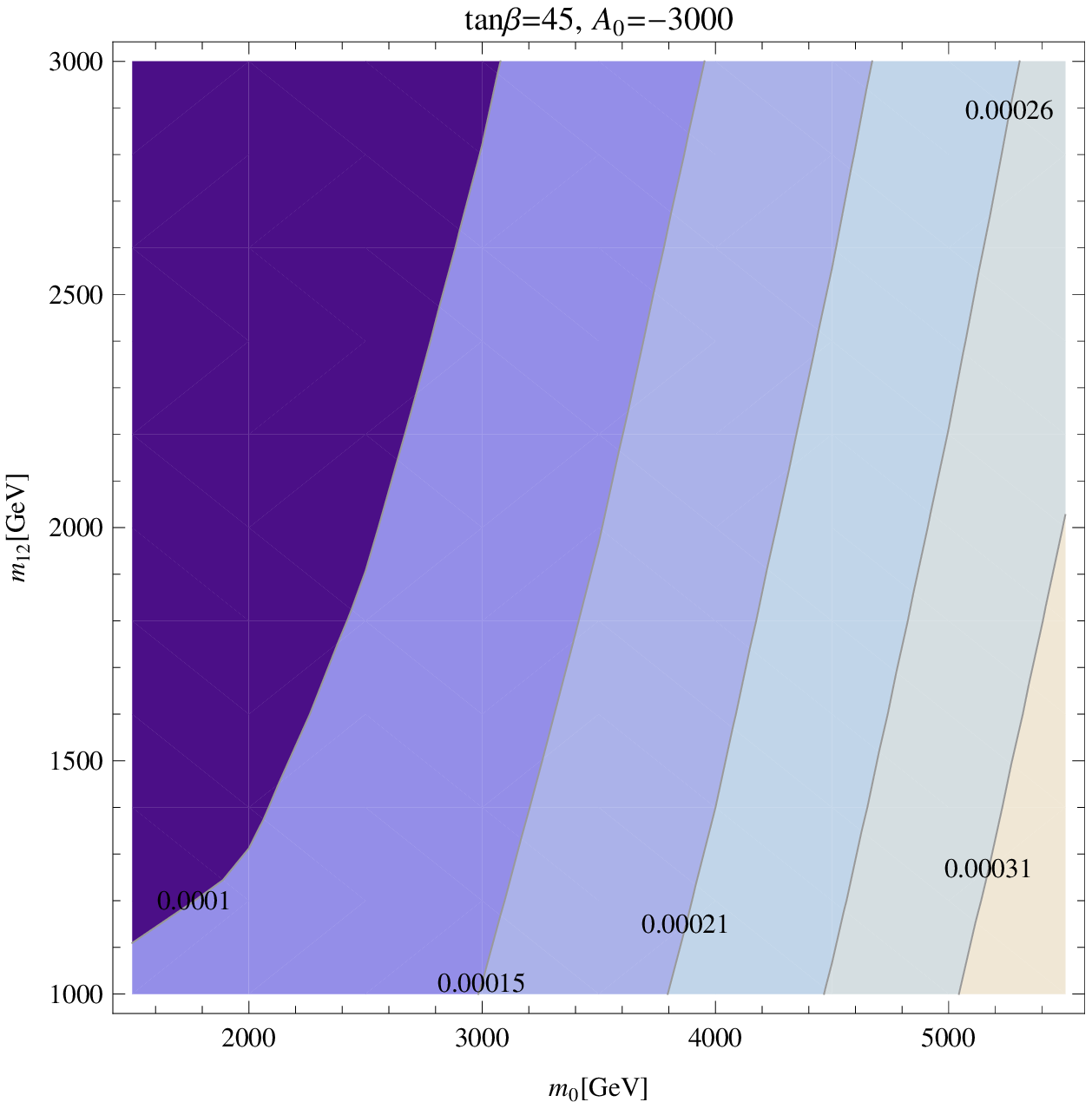   ,scale=0.51,angle=0,clip=}\\
\end{center}
\caption[Contours of \Drho\ in the
  $m_0$--$m_{1/2}$ plane.]{Contours of \Drho\ in the
  $m_0$--$m_{1/2}$ plane for different values of $\tb$ and    
$A_0$ in the \CMSSMI.}  
\label{fig:SL-delrho}
\end{figure}

\begin{figure}[ht!]
\begin{center}
\psfig{file=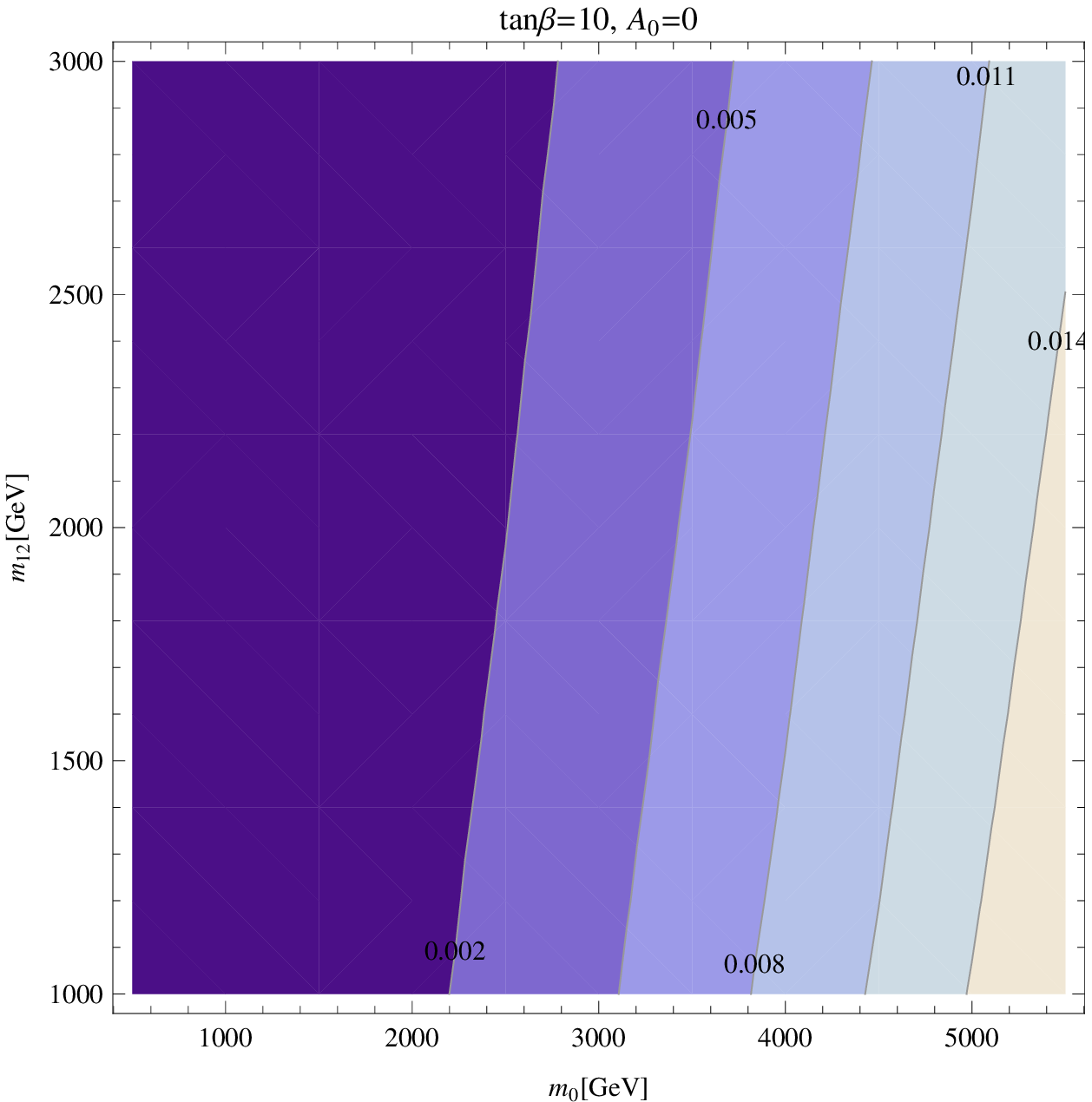  ,scale=0.51,angle=0,clip=}
\psfig{file=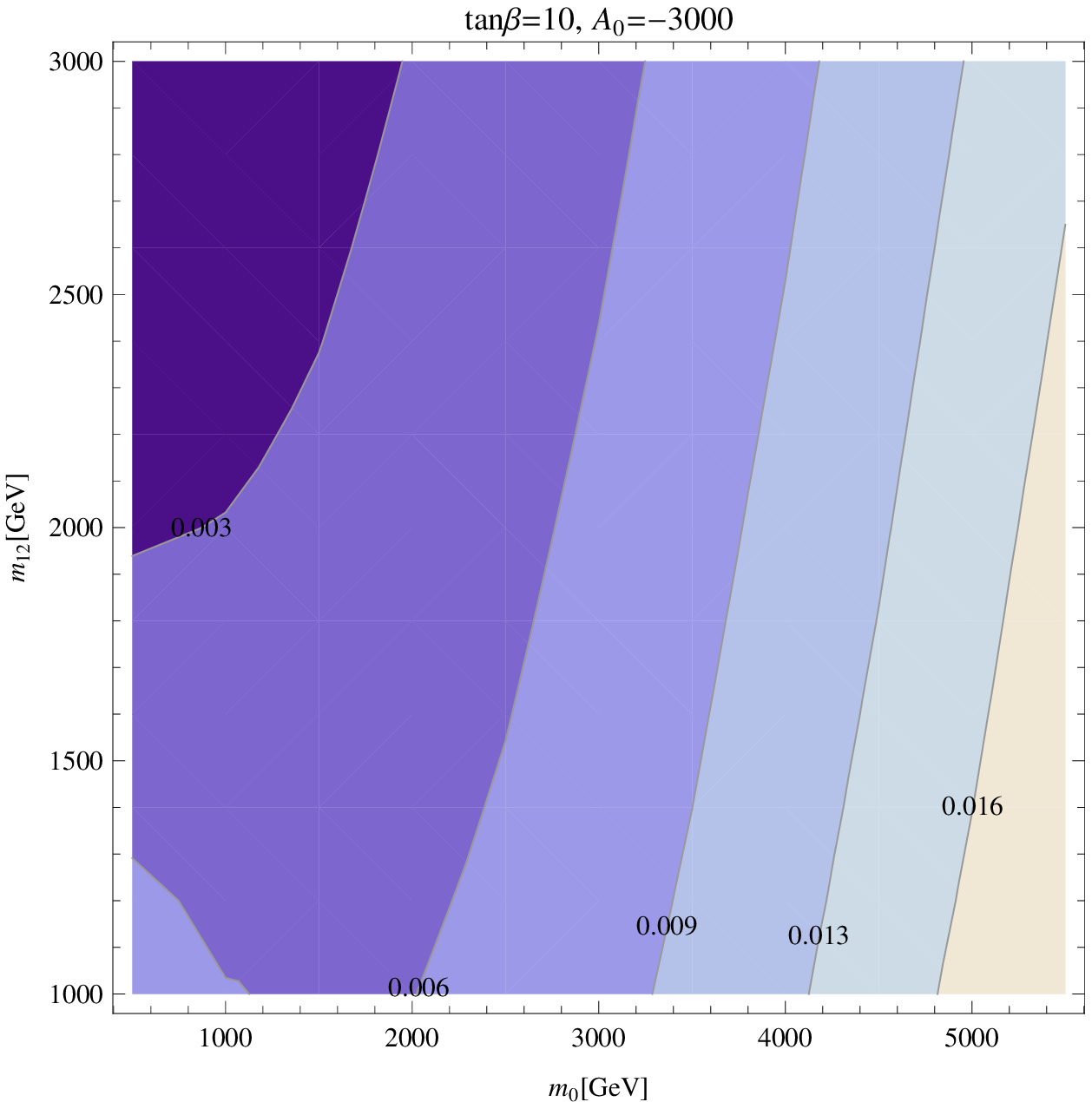  ,scale=0.51,angle=0,clip=}\\
\vspace{0.2cm}
\psfig{file=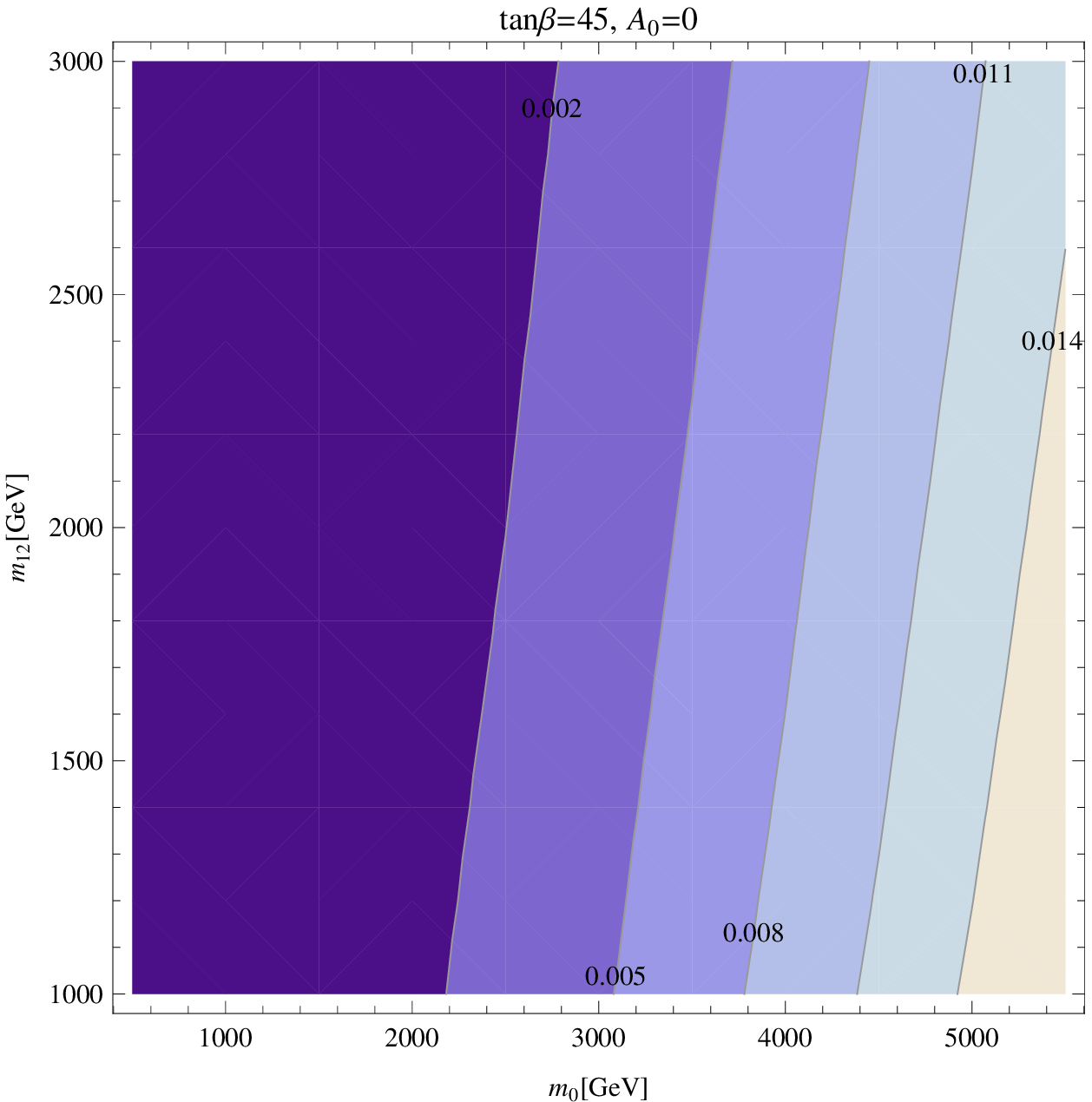 ,scale=0.51,angle=0,clip=}
\psfig{file=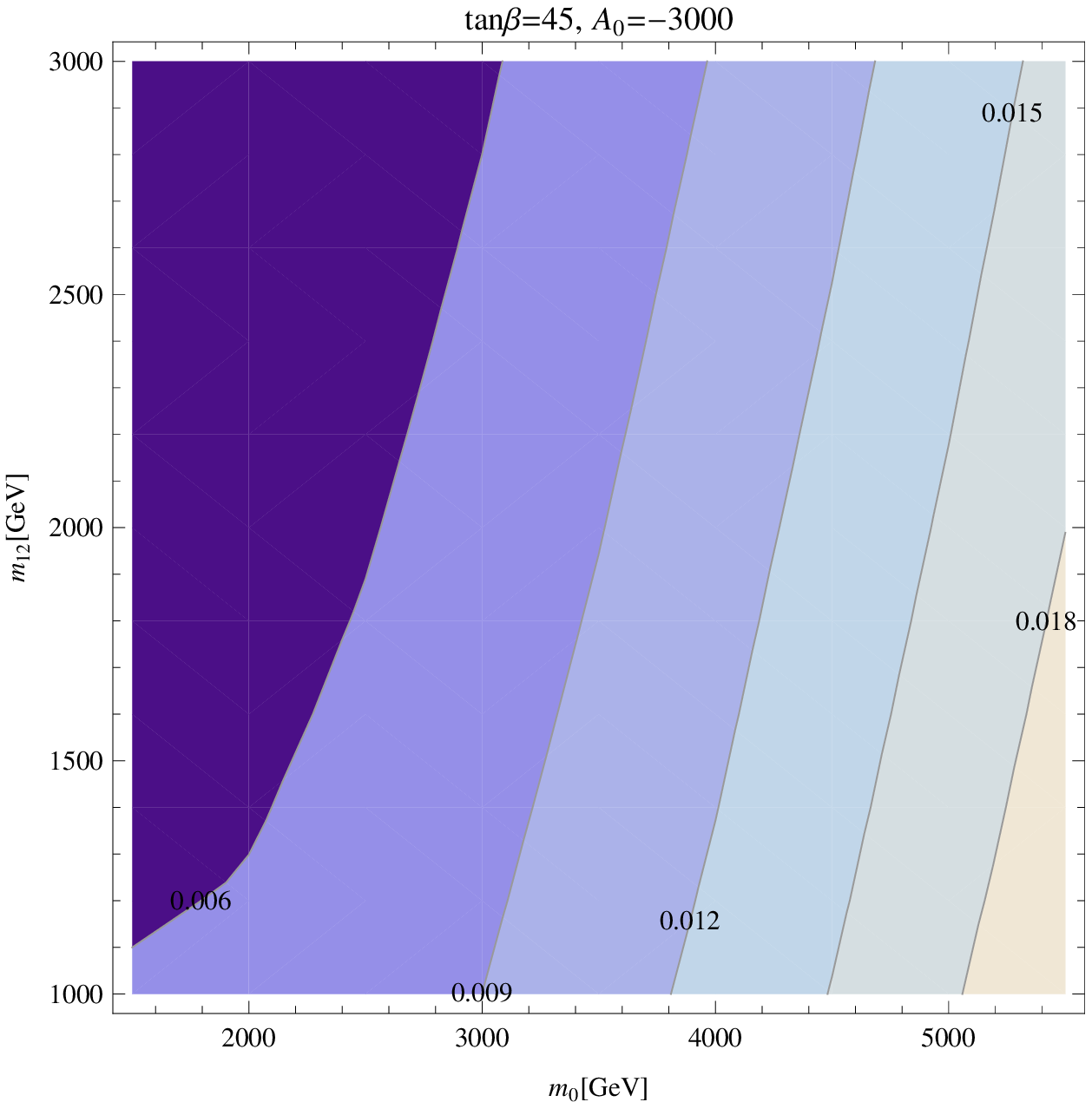   ,scale=0.51,angle=0,clip=}\\
\vspace{-0.2cm}
\end{center}
\caption[Contours of \DMW\ in GeV in the
  $m_0$--$m_{1/2}$ plane.]{Contours of \DMW\ in GeV in the
  $m_0$--$m_{1/2}$ plane for different values of $\tb$ and    
$A_0$ in the \CMSSMI.}  
\label{fig:SL-delMW}
\end{figure}

\begin{figure}[ht!]
\begin{center}
\psfig{file=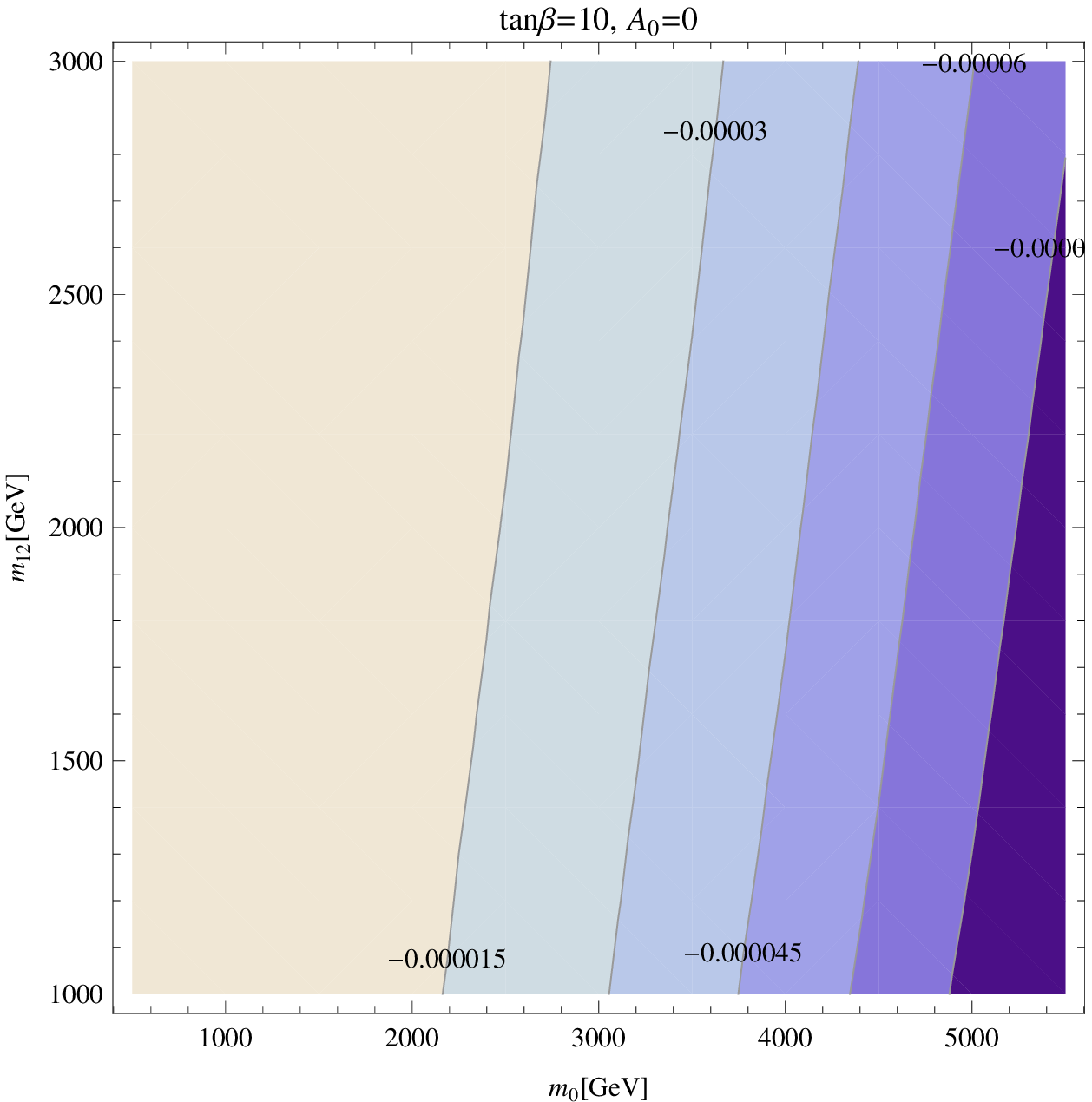  ,scale=0.51,angle=0,clip=}
\psfig{file=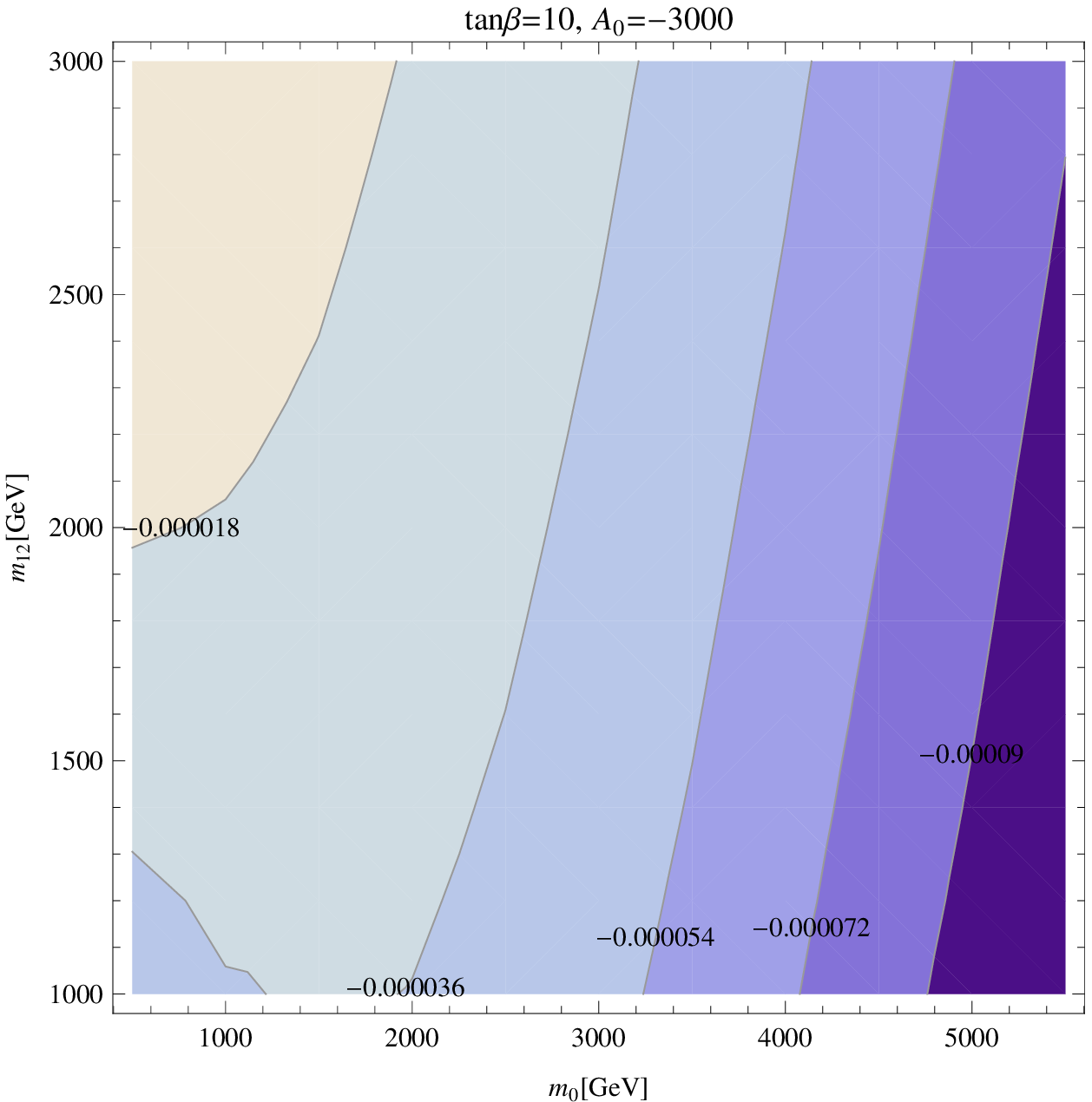  ,scale=0.51,angle=0,clip=}\\
\vspace{0.2cm}
\psfig{file=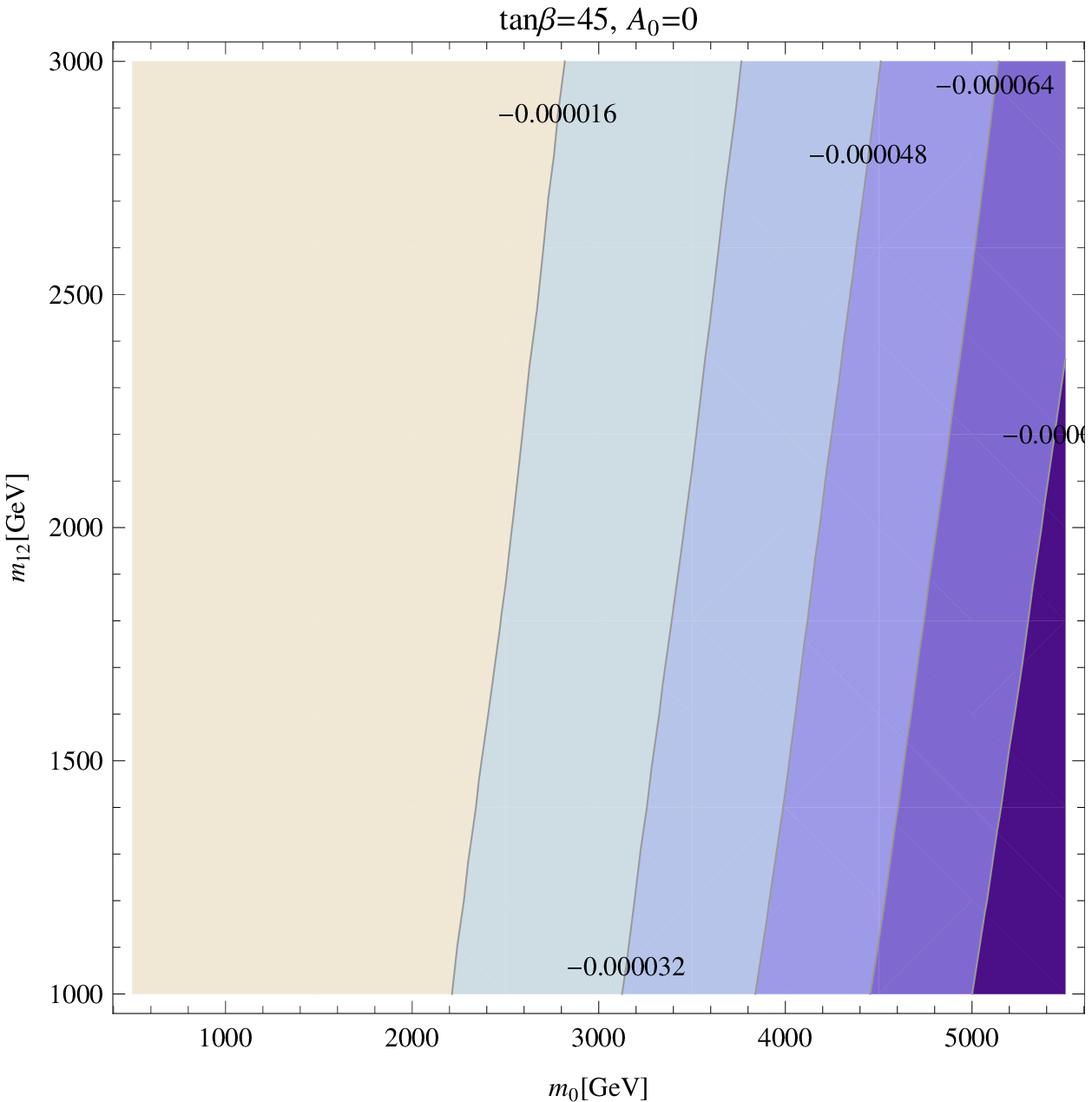 ,scale=0.51,angle=0,clip=}
\psfig{file=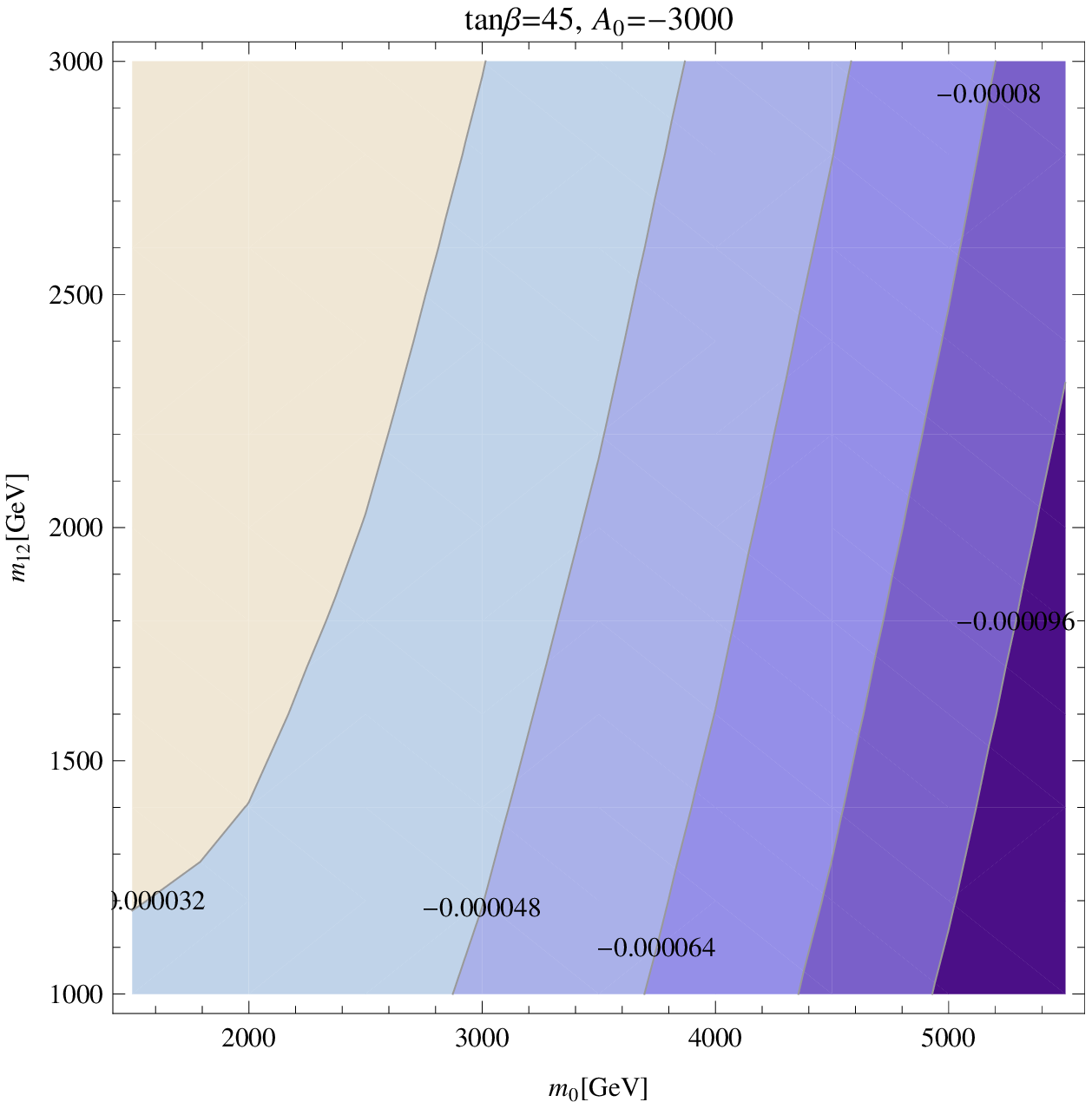   ,scale=0.51,angle=0,clip=}\\
\vspace{-0.2cm}
\end{center}
\caption[Contours of \Dsweff\ in the
  $m_0$--$m_{1/2}$ plane.]{Contours of \Dsweff\ in the
  $m_0$--$m_{1/2}$ plane for different values of $\tb$ and    
$A_0$ in the \CMSSMI. }  
\label{fig:SL-delSW2}
\end{figure}

In \reffis{fig:SL-delrho}-\ref{fig:SL-delSW2} we show the results for
the EWPO. The same pattern and non-decoupling behavior for EWPO as
in the case of CMSSM (squark $\deFABij$) can be observed. However, the
corrections induced by slepton flavor violation 
are relatively small compared to squark case. For the most extreme
cases, i.e.\ the largest values of $m_0$, the corrections to $\MW$
turn out to be of the same order of the experimental uncertainty.
For those parts of the parameter space neglecting the effects of LFV
to the EWPO could turn out to be an insufficient approximation, in
particular in view of future improved experimental accuracies.
\subsection{Higgs masses}
\begin{figure}[ht!]
\begin{center}
\psfig{file=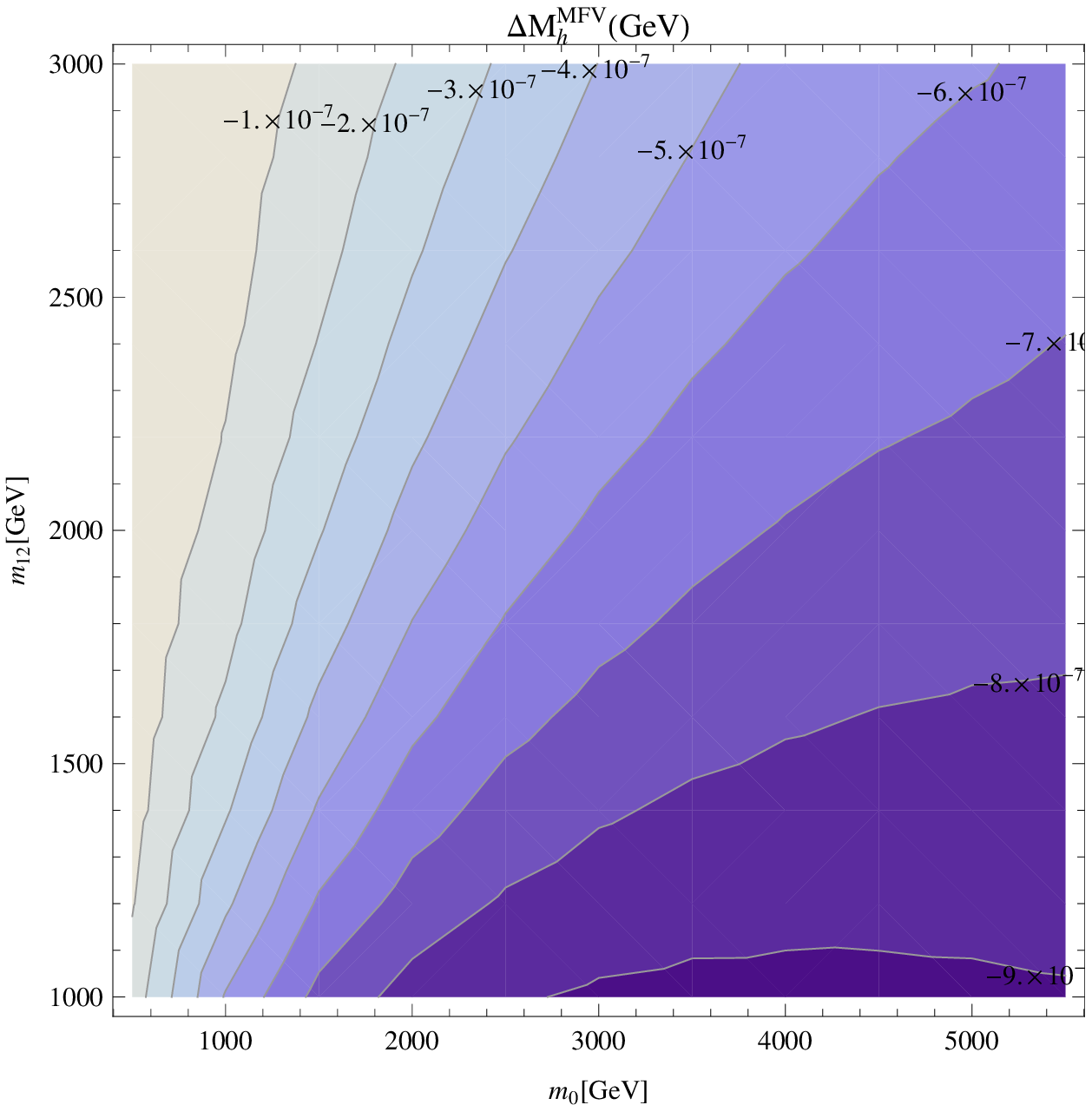 ,scale=0.51,angle=0,clip=}
\psfig{file=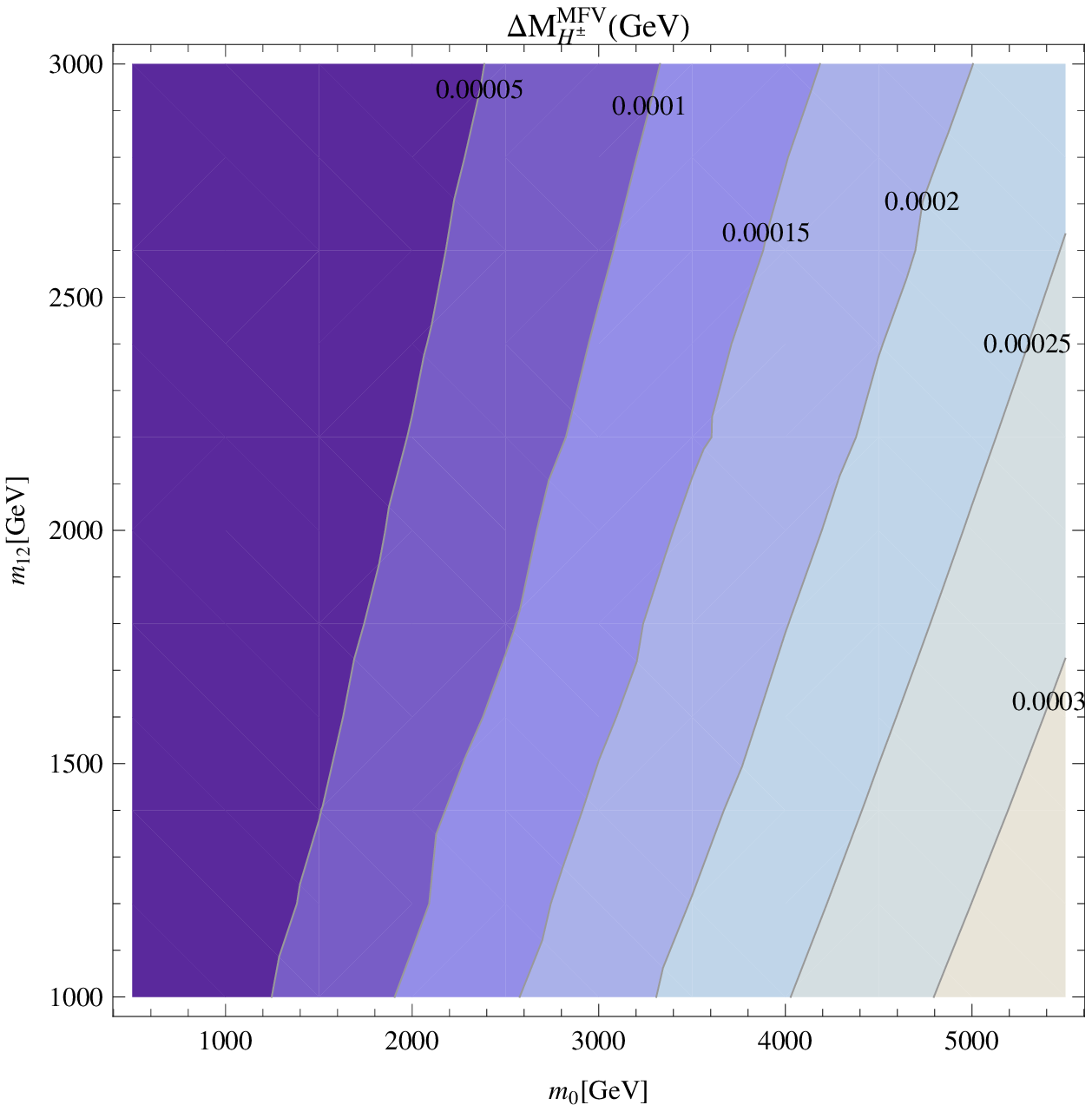 ,scale=0.51,angle=0,clip=}
\end{center}
\caption[Contours of \DMh\ (left) and \DMHp\ (right) 
in the $m_0$--$m_{1/2}$ plane.]{Contours of \DMh\ (left) and \DMHp\ (right) 
in the $m_0$--$m_{1/2}$ plane for $\tb = 10$ and $A_0 = 0$ in
the \CMSSMI.}
\label{SL-MH} 
\end{figure}

Finally, in \reffi{SL-MH} we present the corrections to the Higgs
boson masses induced by slepton flavor violation.
Here we only show \DMh\ (left) and \DMHp (right) for $\tb = 10$ and 
$A_0 = 0$. They turn out to be
negligibly small in both cases. Corrections to \DMH, which are not
shown, are even smaller. We have checked that these results hold also
for other combinations of $\tb$ and $A_0$. 
Consequently, within the Higgs sector the
approximation of neglecting the effects of the $\deFABij$ is fully
justified. 
\subsection{\boldmath{${\rm BR}(l_i \rightarrow l_j \gamma)$}}
\begin{figure}[ht!]
\begin{center}
\psfig{file=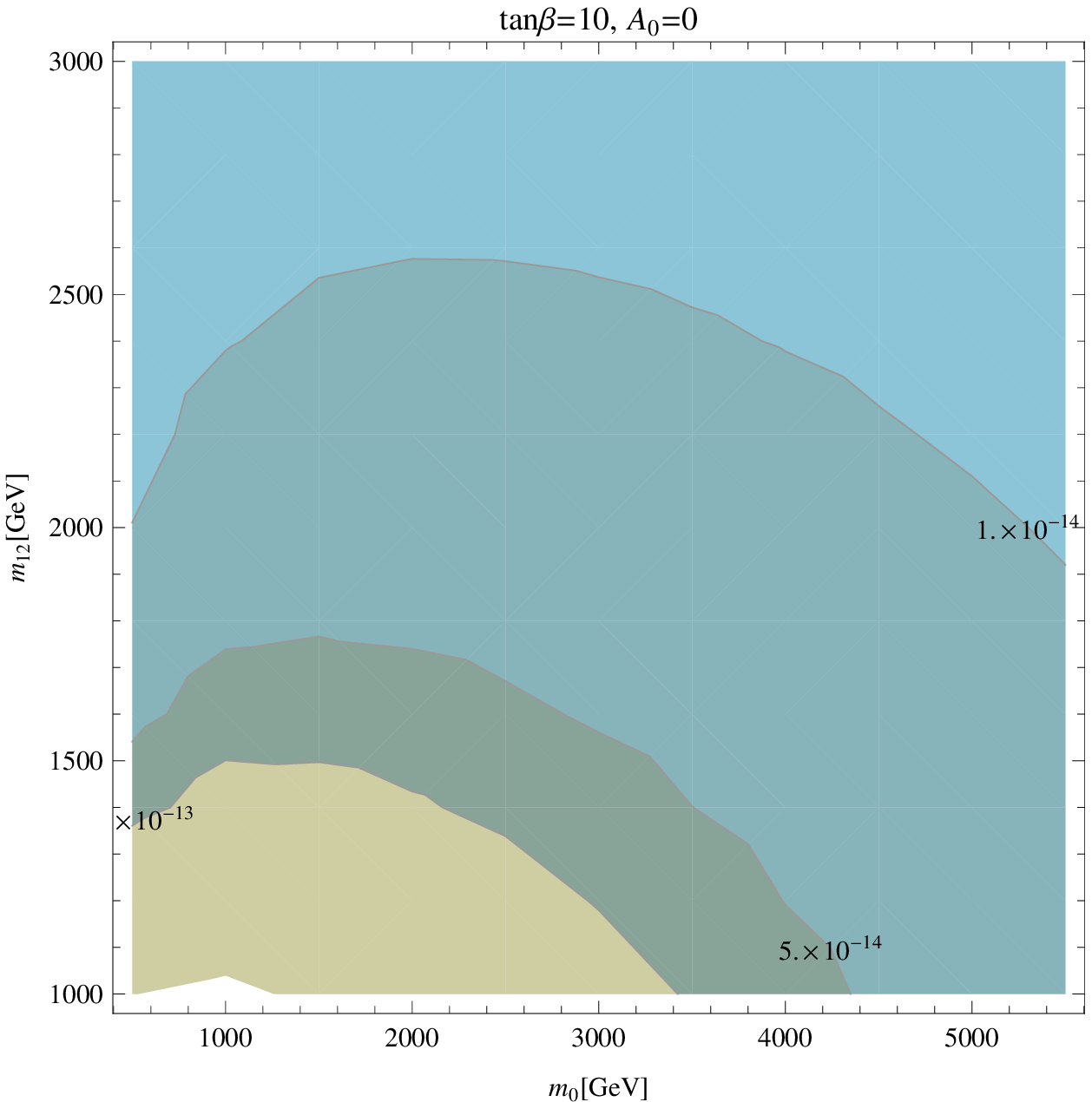  ,scale=0.51,angle=0,clip=}
\psfig{file=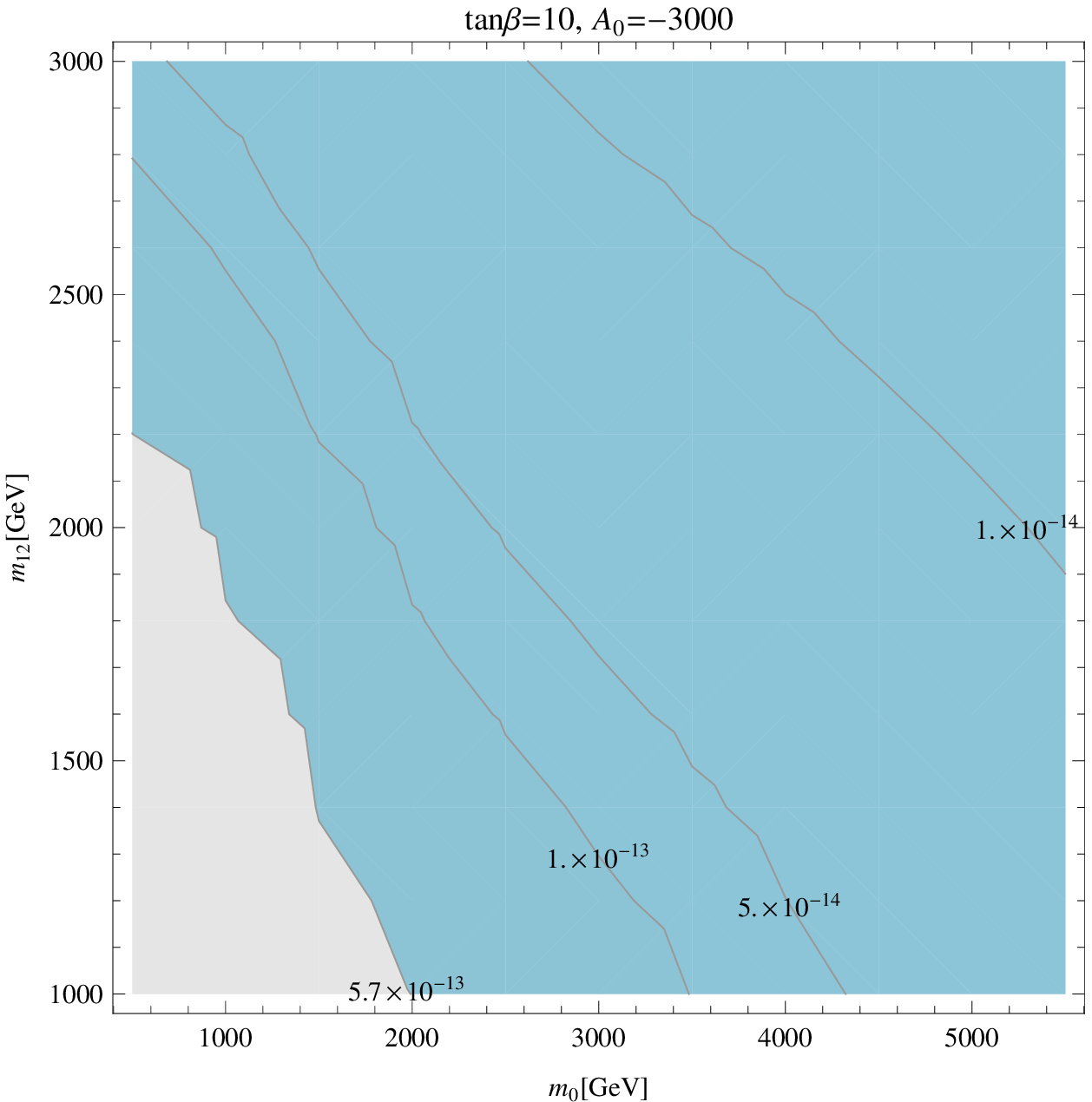  ,scale=0.51,angle=0,clip=}\\
\vspace{0.2cm}
\psfig{file=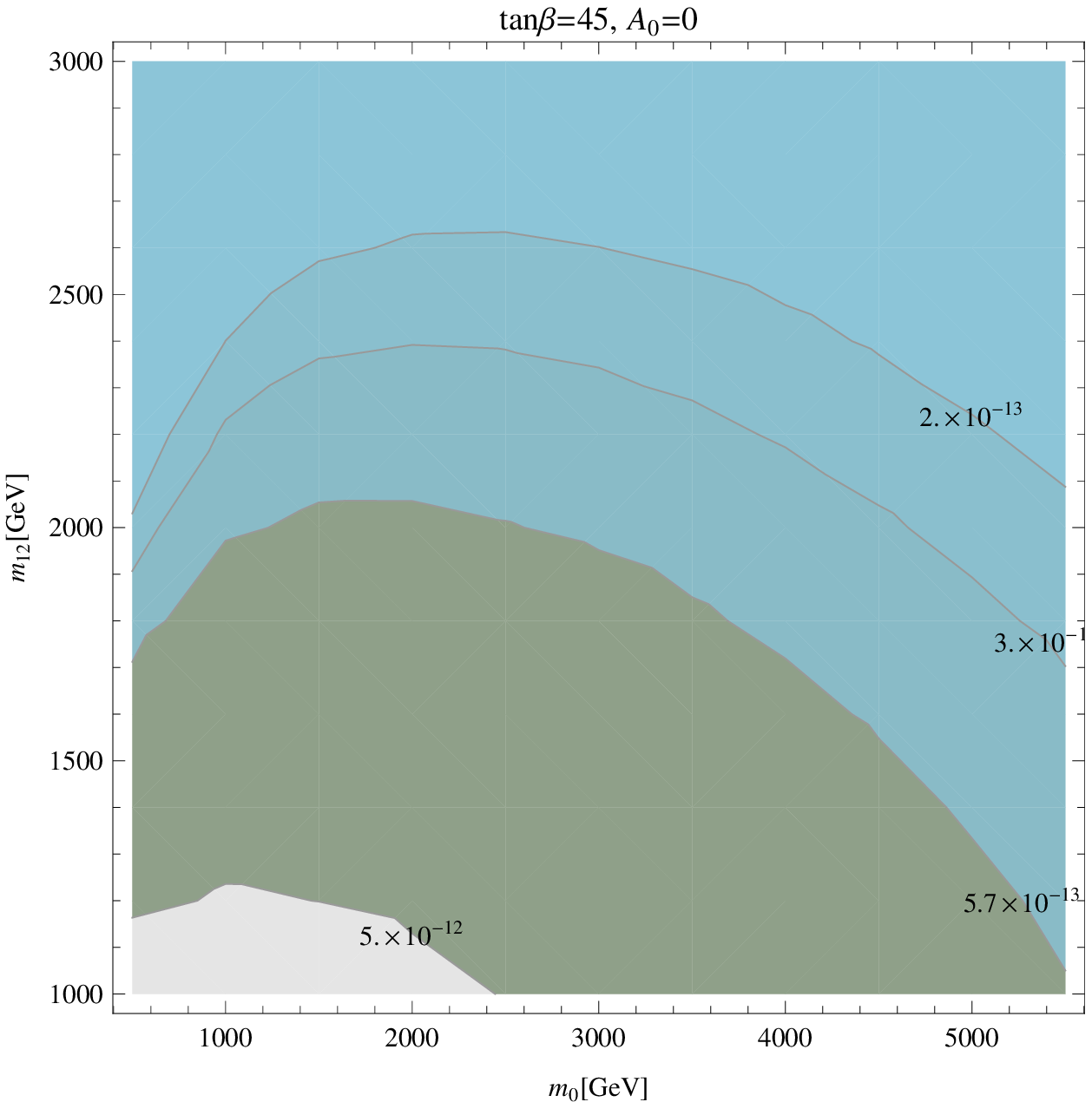 ,scale=0.51,angle=0,clip=}
\psfig{file=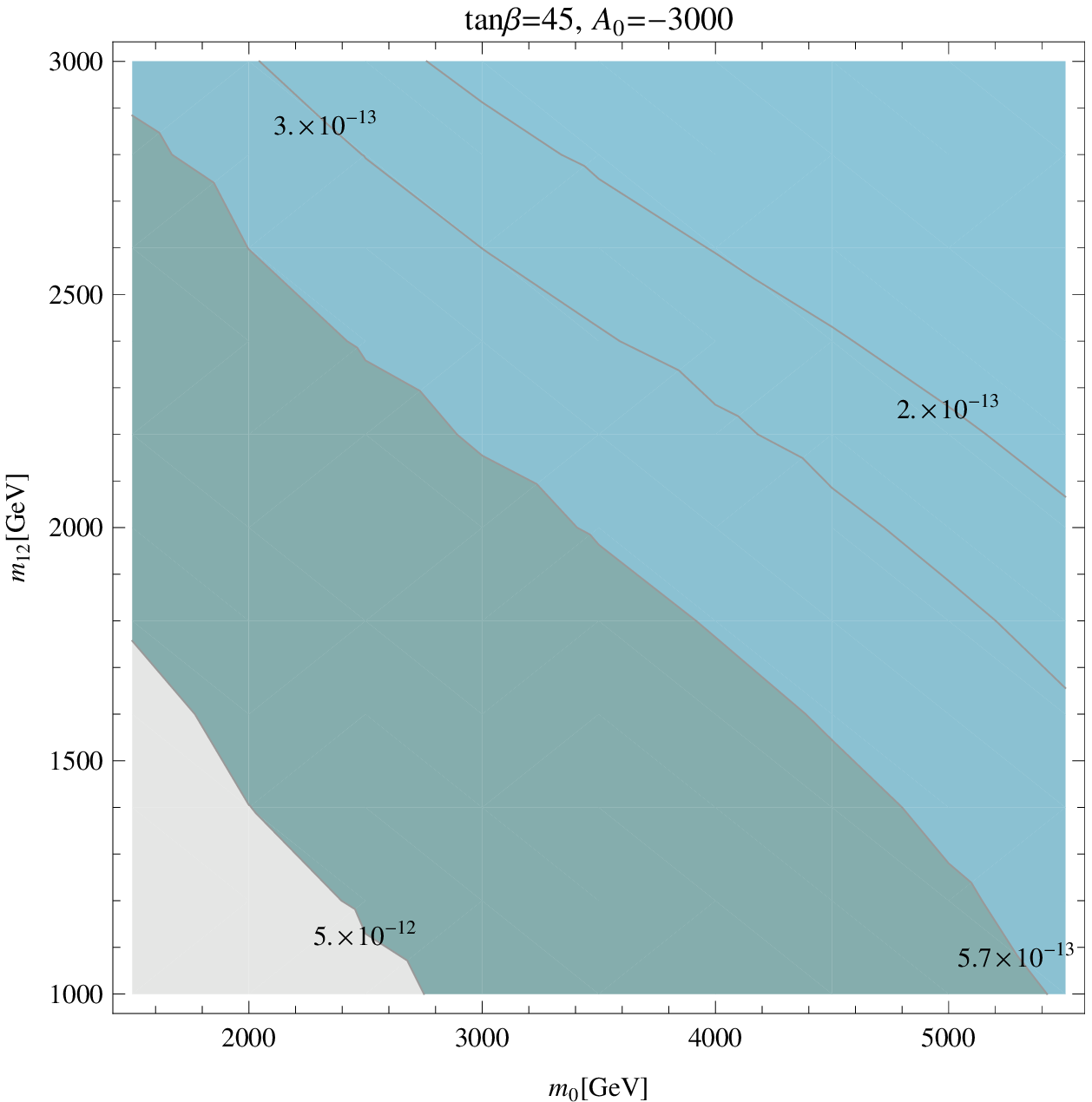   ,scale=0.51,angle=0,clip=}\\
\vspace{-0.2cm}
\end{center}
\caption[Contours of BR($\mu \to e \gamma$)  in the
  $m_0$--$m_{1/2}$ plane]{Contours of BR($\mu \to e \gamma$)  in the
  $m_0$--$m_{1/2}$ plane for different values of $\tb$ and $A_0$ in
  the \CMSSMI.}   
\label{fig:BrmegSSI}
\end{figure}

\begin{figure}[ht!]
\begin{center}
\psfig{file=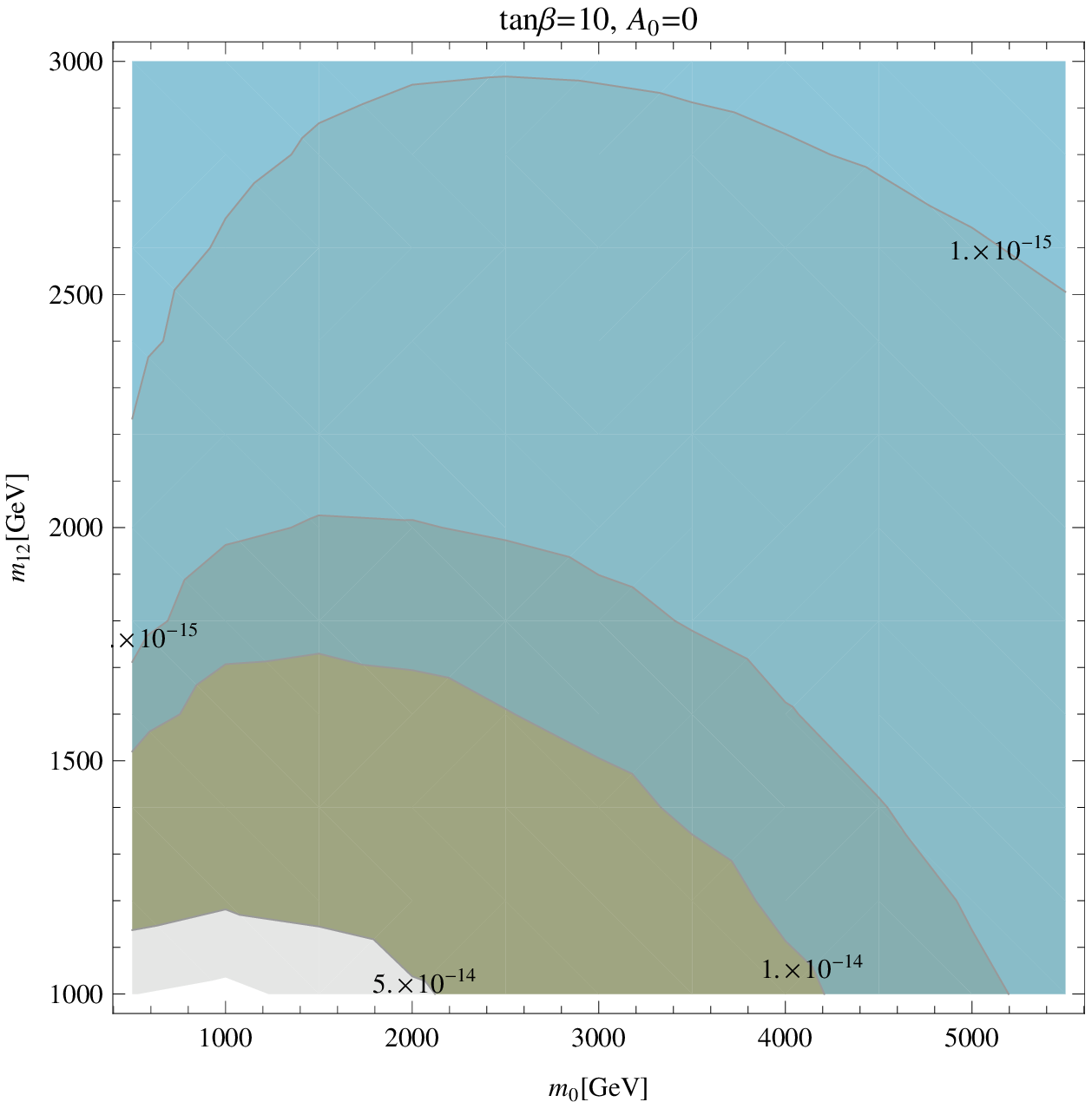  ,scale=0.51,angle=0,clip=}
\psfig{file=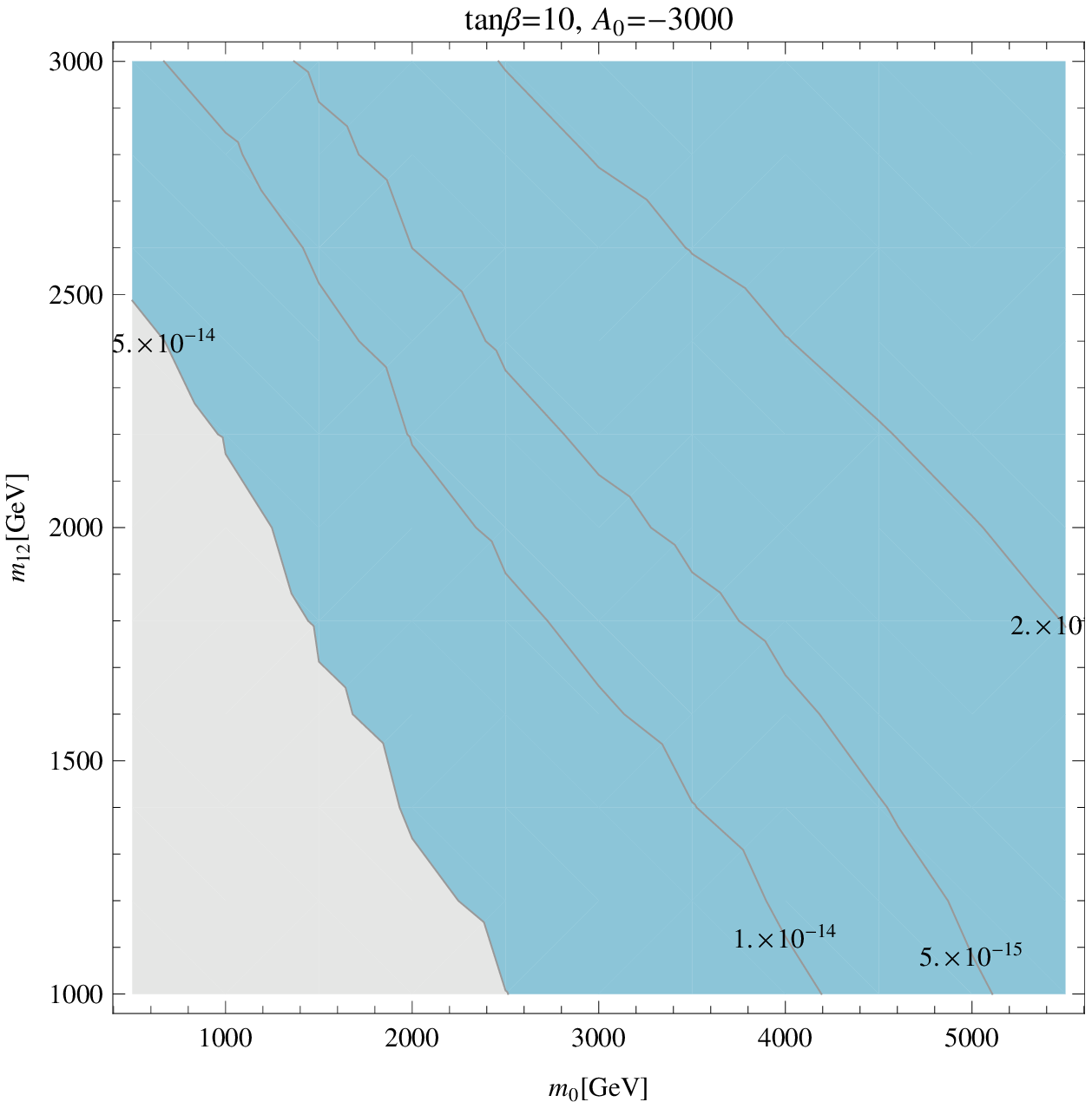  ,scale=0.51,angle=0,clip=}\\
\vspace{0.2cm}
\psfig{file=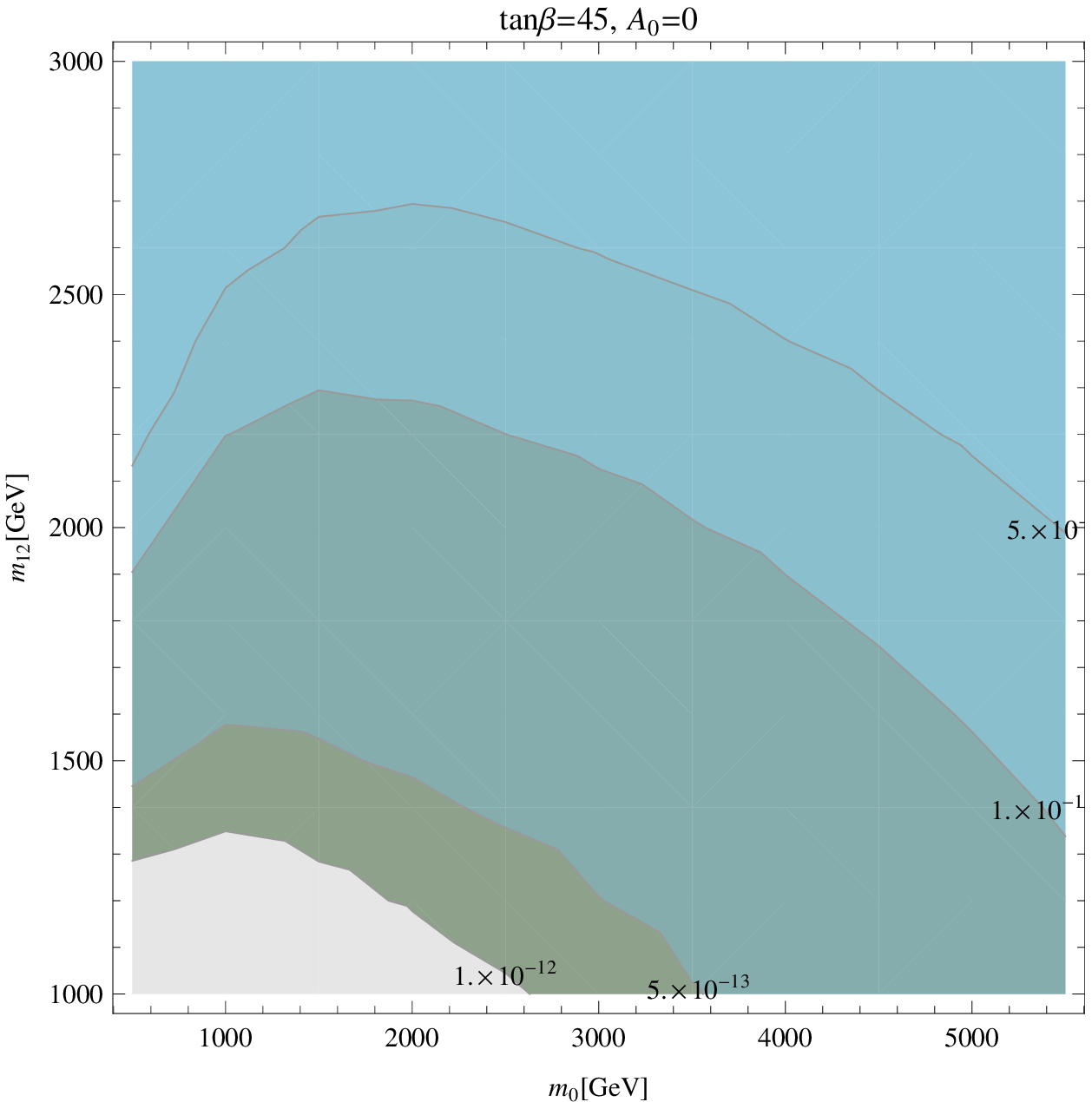 ,scale=0.51,angle=0,clip=}
\psfig{file=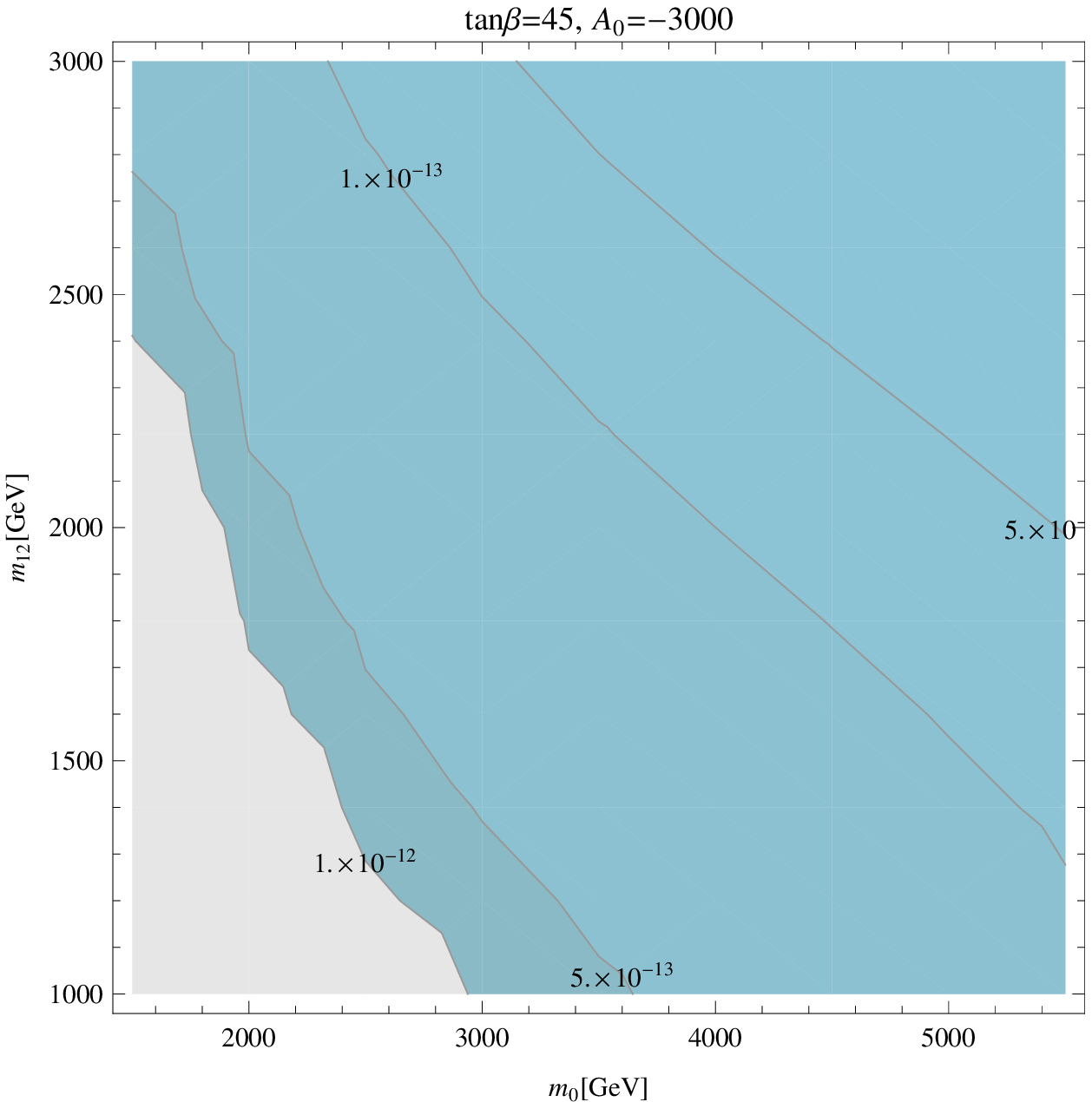   ,scale=0.51,angle=0,clip=}\\
\vspace{-0.2cm}
\end{center}
\caption[Contours of BR($\tau \to e \gamma$)  in the
  $m_0$--$m_{1/2}$ plane]{Contours of BR($\tau \to e \gamma$)  in the
  $m_0$--$m_{1/2}$ plane for different values of $\tb$ and $A_0$ in
  the \CMSSMI.}   
\label{fig:BrtegSSI}
\end{figure}

\begin{figure}[ht!]
\begin{center}
\psfig{file=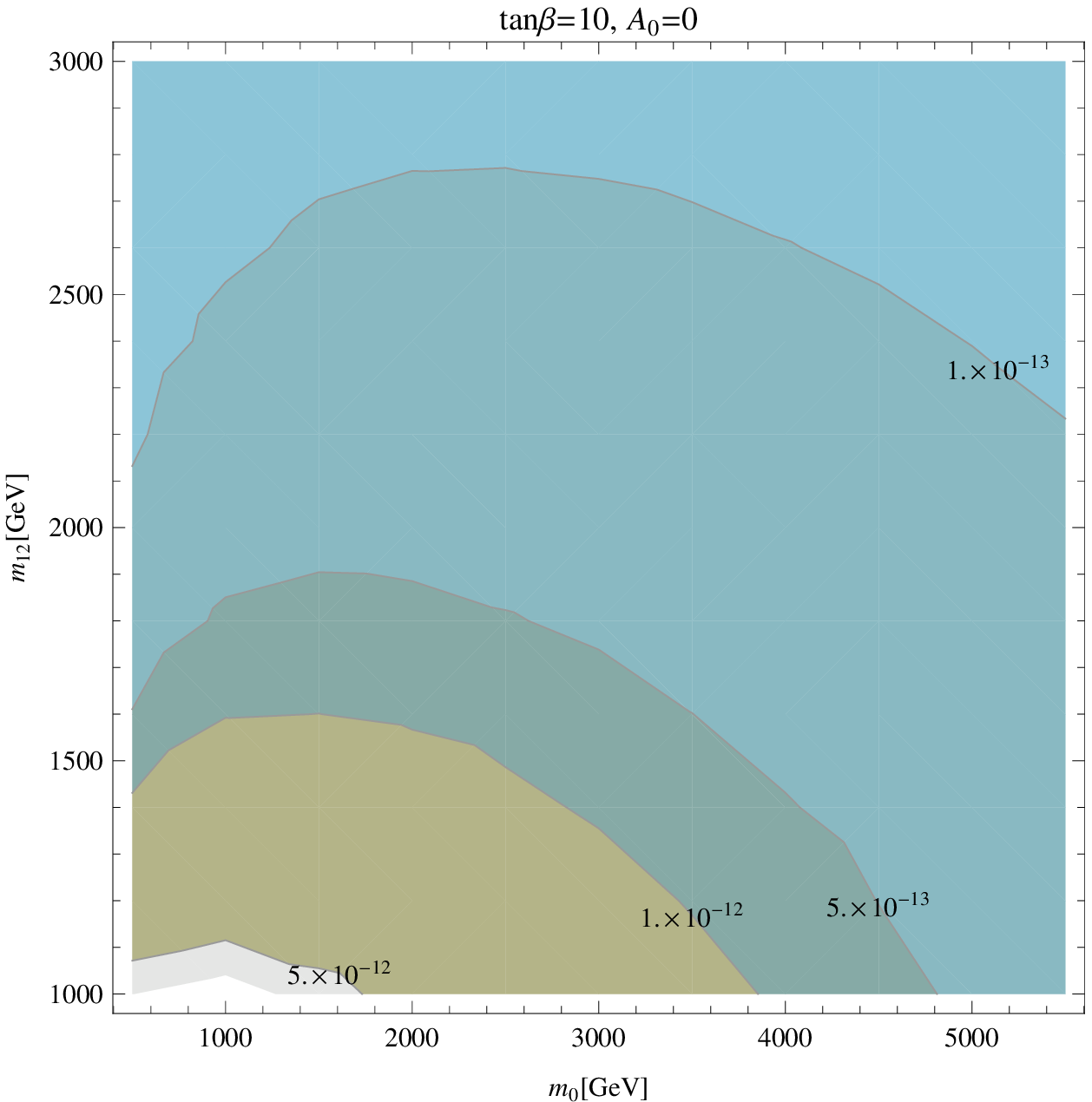  ,scale=0.51,angle=0,clip=}
\psfig{file=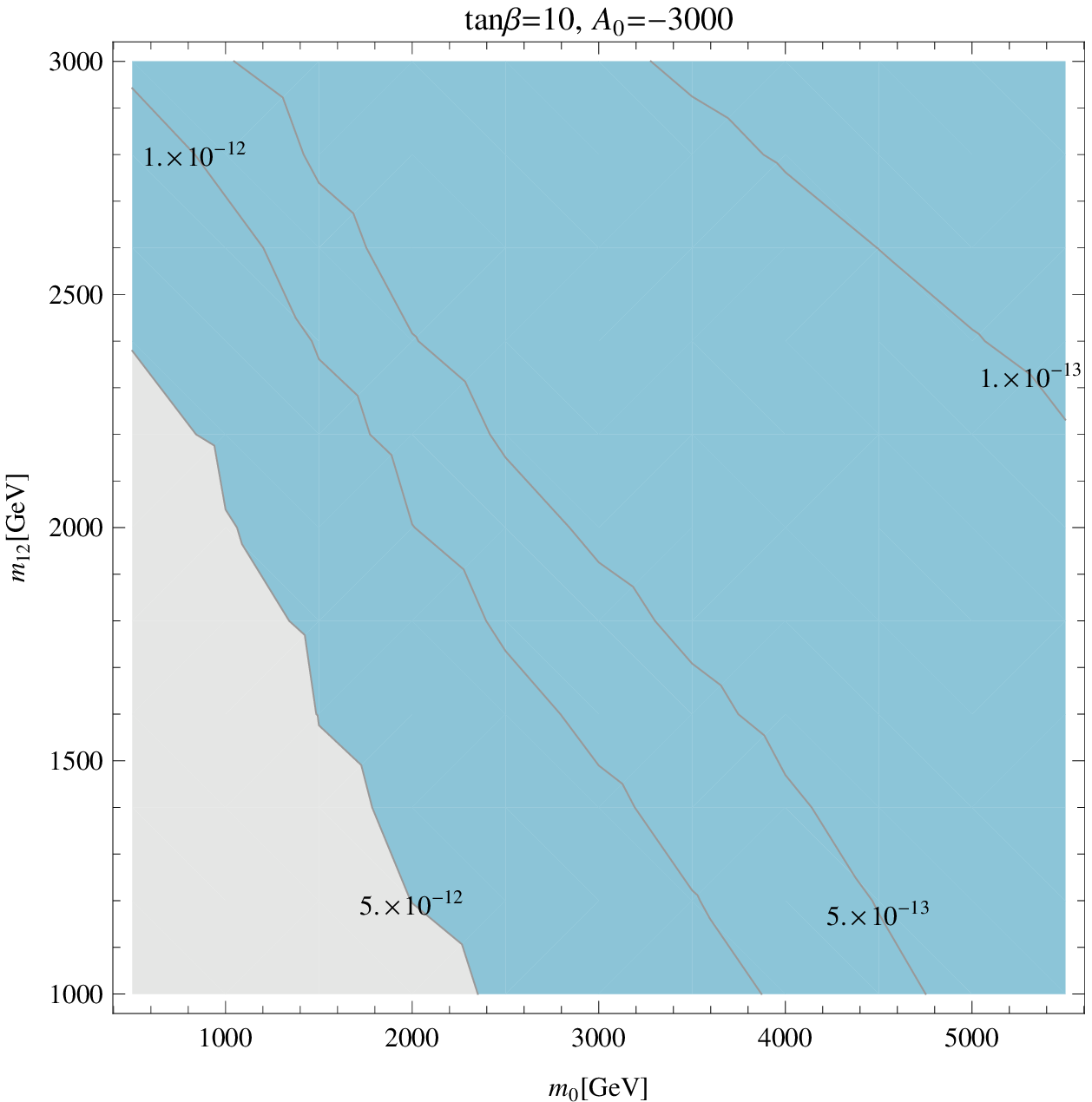  ,scale=0.51,angle=0,clip=}\\
\vspace{0.2cm}
\psfig{file=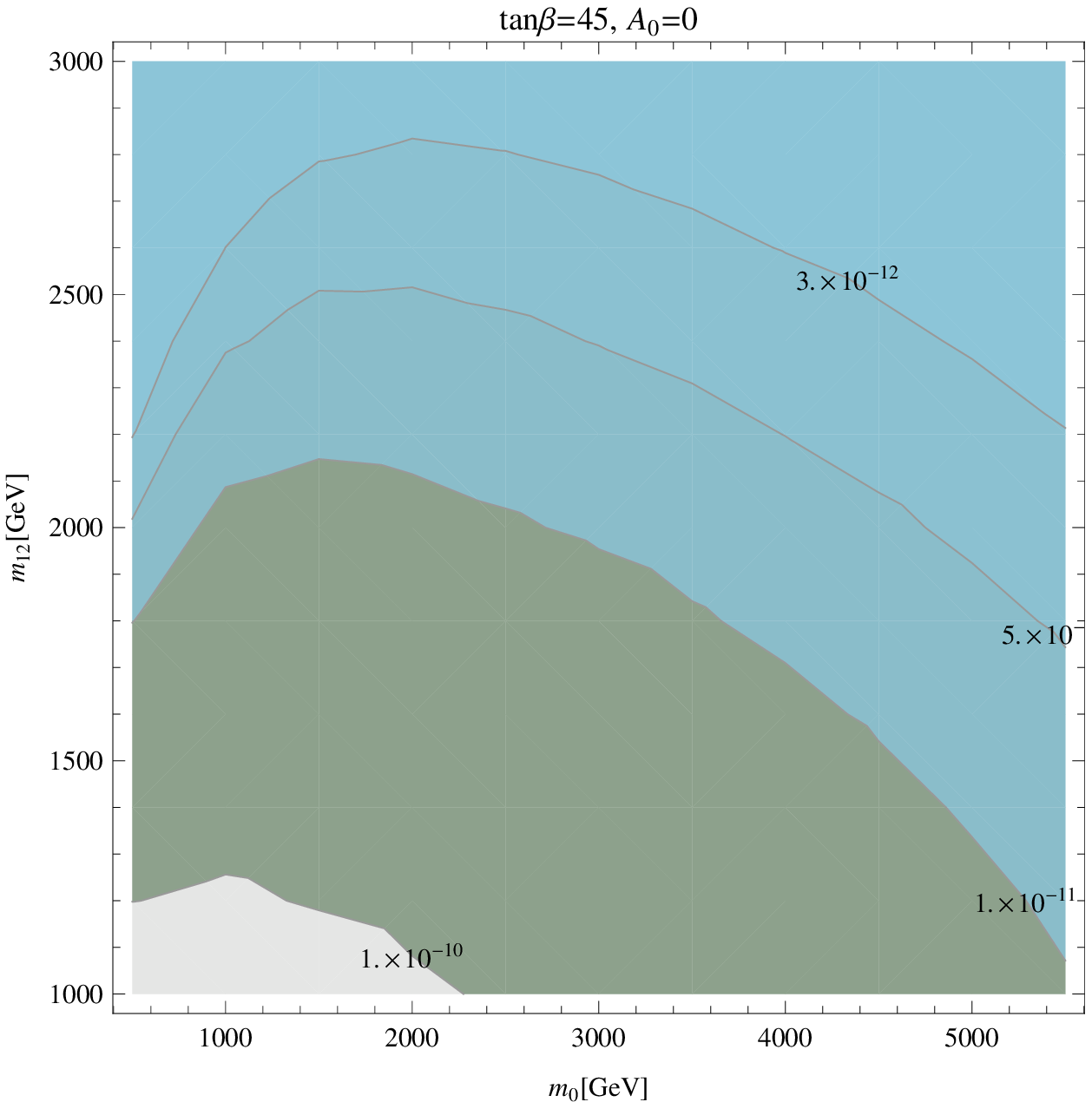 ,scale=0.51,angle=0,clip=}
\psfig{file=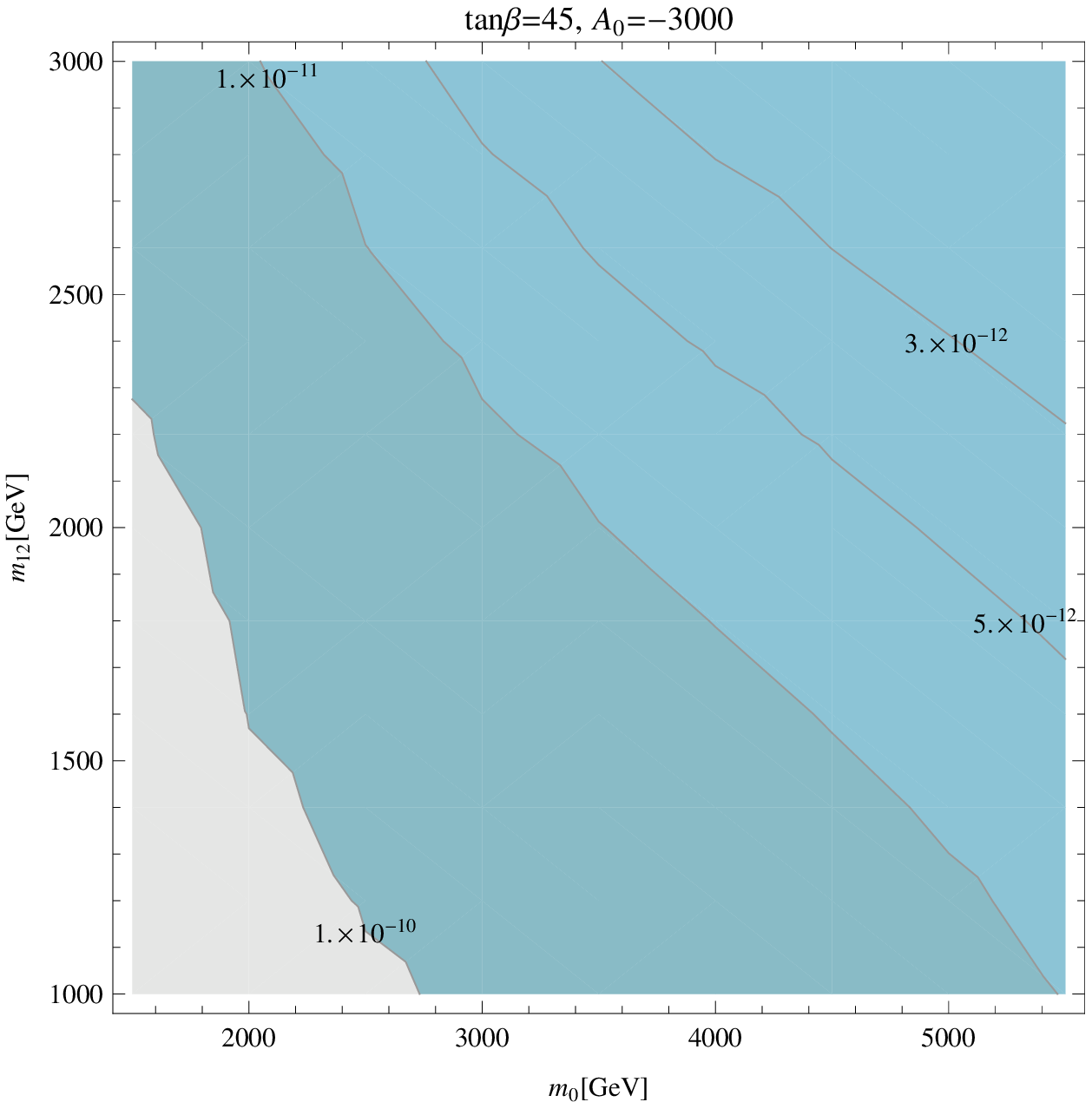   ,scale=0.51,angle=0,clip=}\\
\vspace{-0.2cm}
\end{center}
\caption[Contours of BR($\tau \to \mu \gamma$)  in the
  $m_0$--$m_{1/2}$ plane]{Contours of BR($\tau \to \mu \gamma$)  in the
  $m_0$--$m_{1/2}$ plane for different values of $\tb$ and $A_0$ in
  the \CMSSMI.}   
\label{fig:BrtmgSSI}
\end{figure}

The experimental limit BR($\mu \to e \gamma)< 5.7 \times 10^{-13}$ put severe constraints on slepton $\deFABij$'s as discussed before.
In \reffi{fig:BrmegSSI}, we show the predictions for BR($\mu \to e \gamma$)
in $m_0$--$m_{1/2}$ for different values of $A_0$ and $\tb$ in \CMSSMI.
The selected values of $Y_\nu$ result in a large prediction for, e.g.,
BR($\mu \to e \gamma$) that can eliminate some of the $m_0$--$m_{1/2}$
parameter plane, in particular combinations of low values of $m_0$ and
$m_{1/2}$. For $\tb=10$ and $A_0=0$, BR($\mu \to e \gamma$) (upper left plot of \reffi{fig:BrmegSSI}) do not exclude any region in $m_0$--$m_{1/2}$ plane, whereas with $\tb=10$ and $A_0=-3000$ lower left region below $m_0, m_{1/2}=2000 $ is excluded (see upper right plot of \reffi{fig:BrmegSSI}).  For combinations like $\tb=45,$ $A_0=0$ and $\tb=45,$ $A_0=-3000$ even larger parts of the plane are excluded by BR($\mu \to e \gamma$).
In \reffi{fig:BrtegSSI} and \reffi{fig:BrtmgSSI}, we show the predictions for BR($\tau \to e \gamma$) and BR($\tau \to \mu \gamma$) respectively.  It can be seen that these processes do not reach their respective experimental bounds ${\rm BR}(\tau \rightarrow e \gamma)< 3.3 \times 10^{-8}$, ${\rm BR}(\tau \rightarrow \mu \gamma)< 4.4 \times 10^{-8}$. Consequently they do not exclude any parameter space.
\subsection{\boldmath${\rm BR}(h \rightarrow l_i^{\pm} l_j^{\mp}$)}
\begin{figure}[ht!]
\begin{center}
\psfig{file=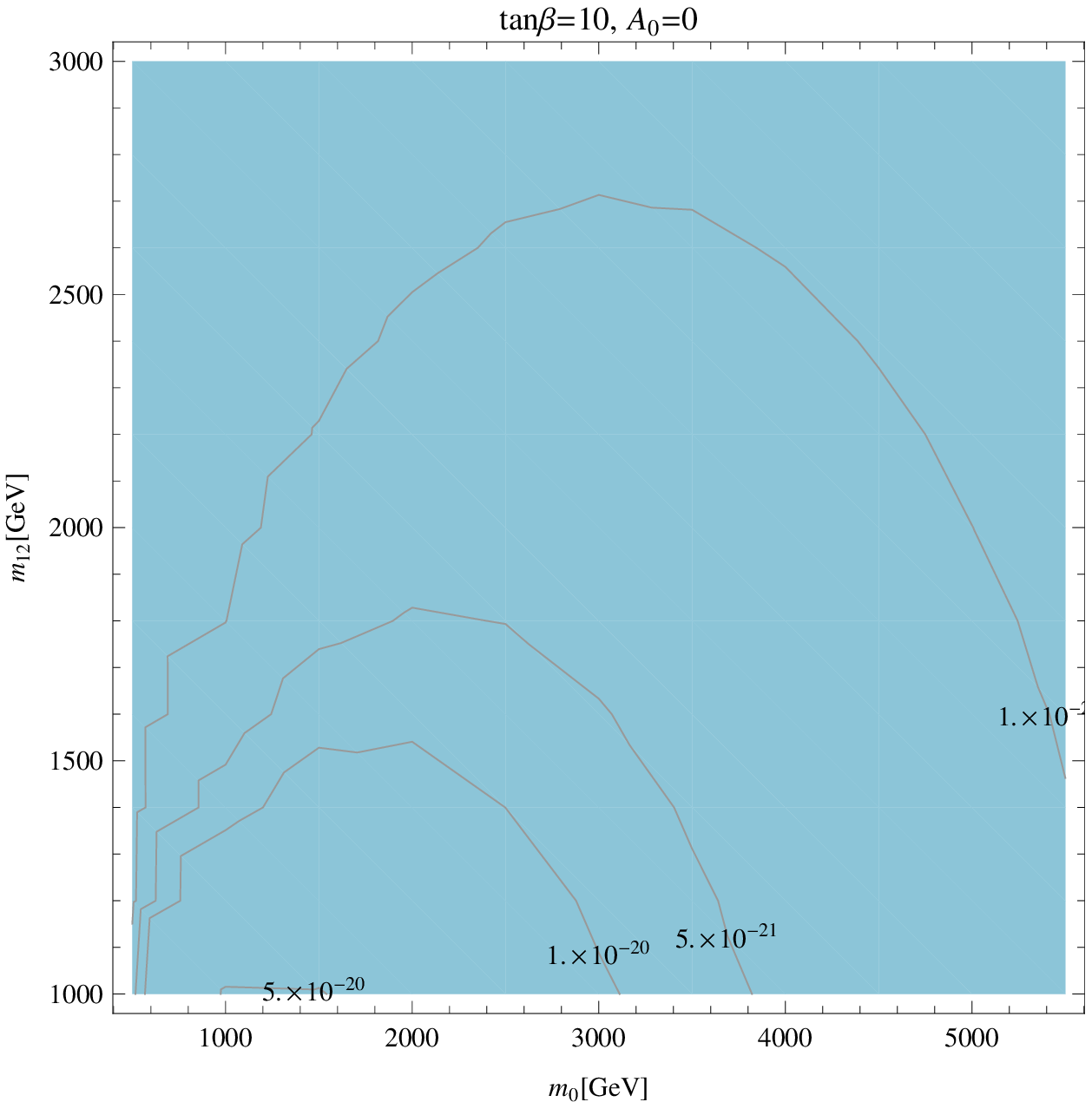  ,scale=0.51,angle=0,clip=}
\psfig{file=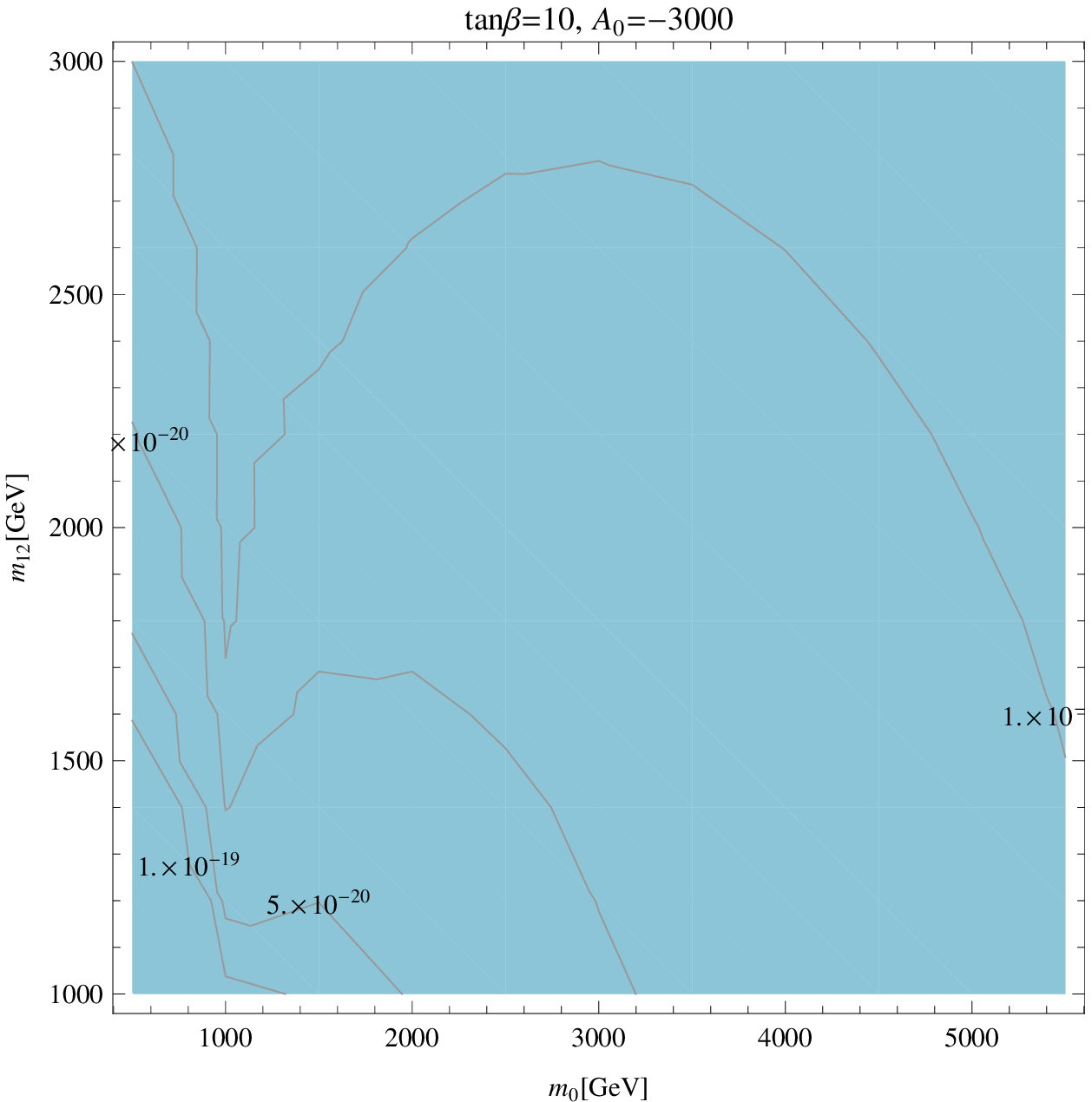  ,scale=0.51,angle=0,clip=}\\
\vspace{0.2cm}
\psfig{file=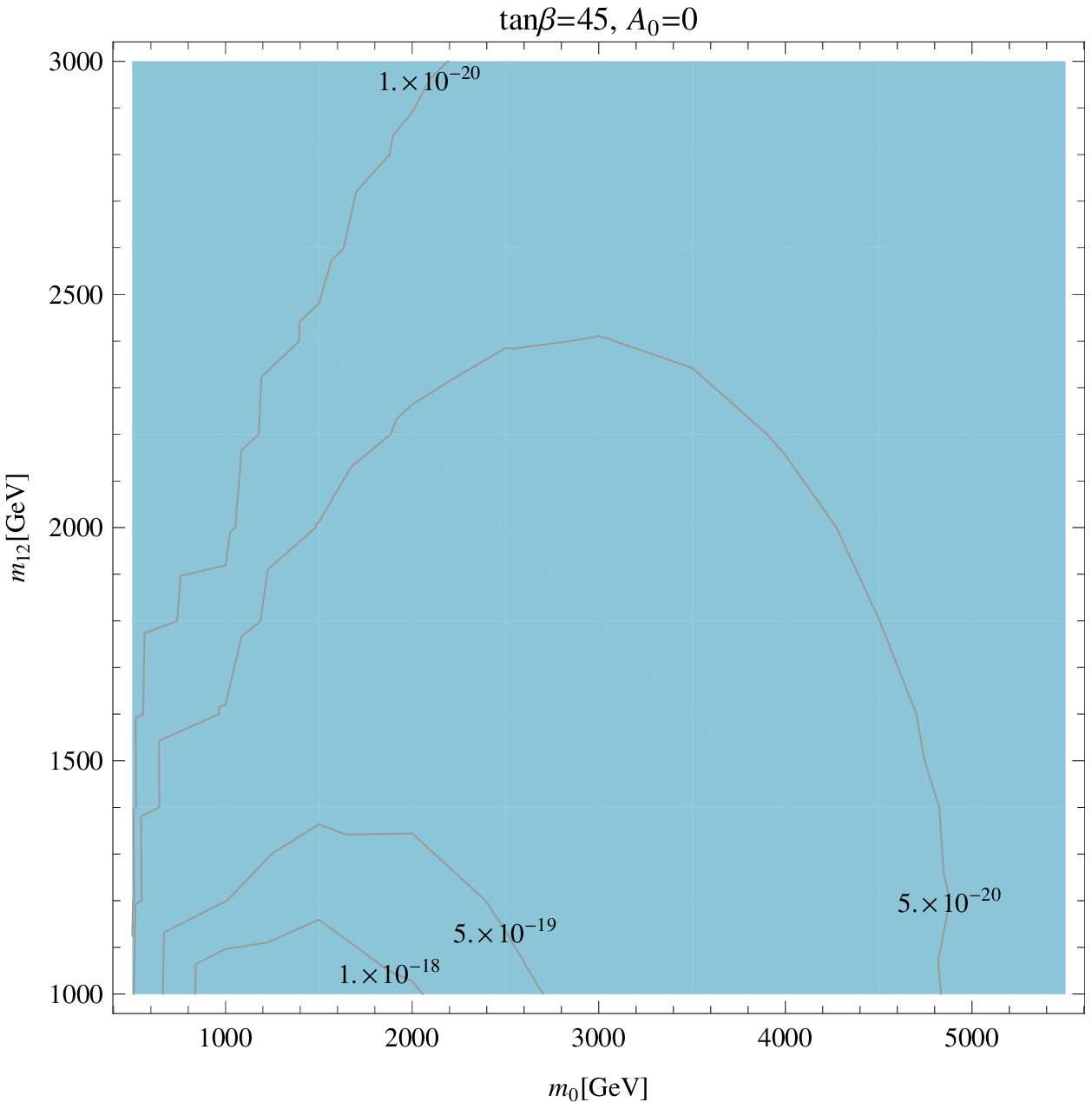 ,scale=0.51,angle=0,clip=}
\psfig{file=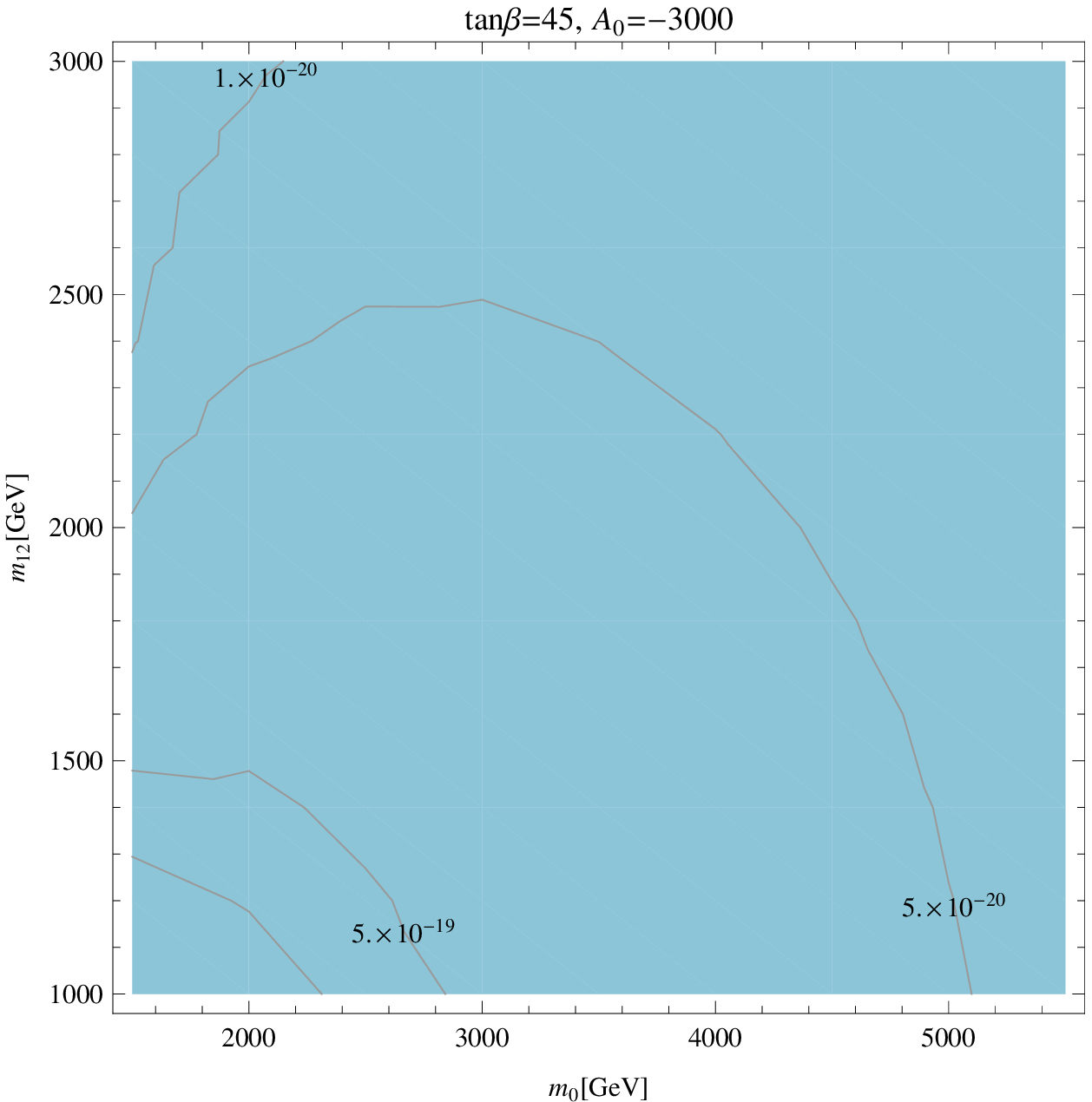   ,scale=0.51,angle=0,clip=}\\
\vspace{-0.2cm}
\end{center}
\caption[Contours of BR($h \to e \mu$)  in the
  $m_0$--$m_{1/2}$ plane]{Contours of BR($h \to e \mu$)  in the
  $m_0$--$m_{1/2}$ plane for different values of $\tb$ and $A_0$ in
  the \CMSSMI.}   
\label{fig:HMueESSI}
\end{figure}

\begin{figure}[ht!]
\begin{center}
\psfig{file=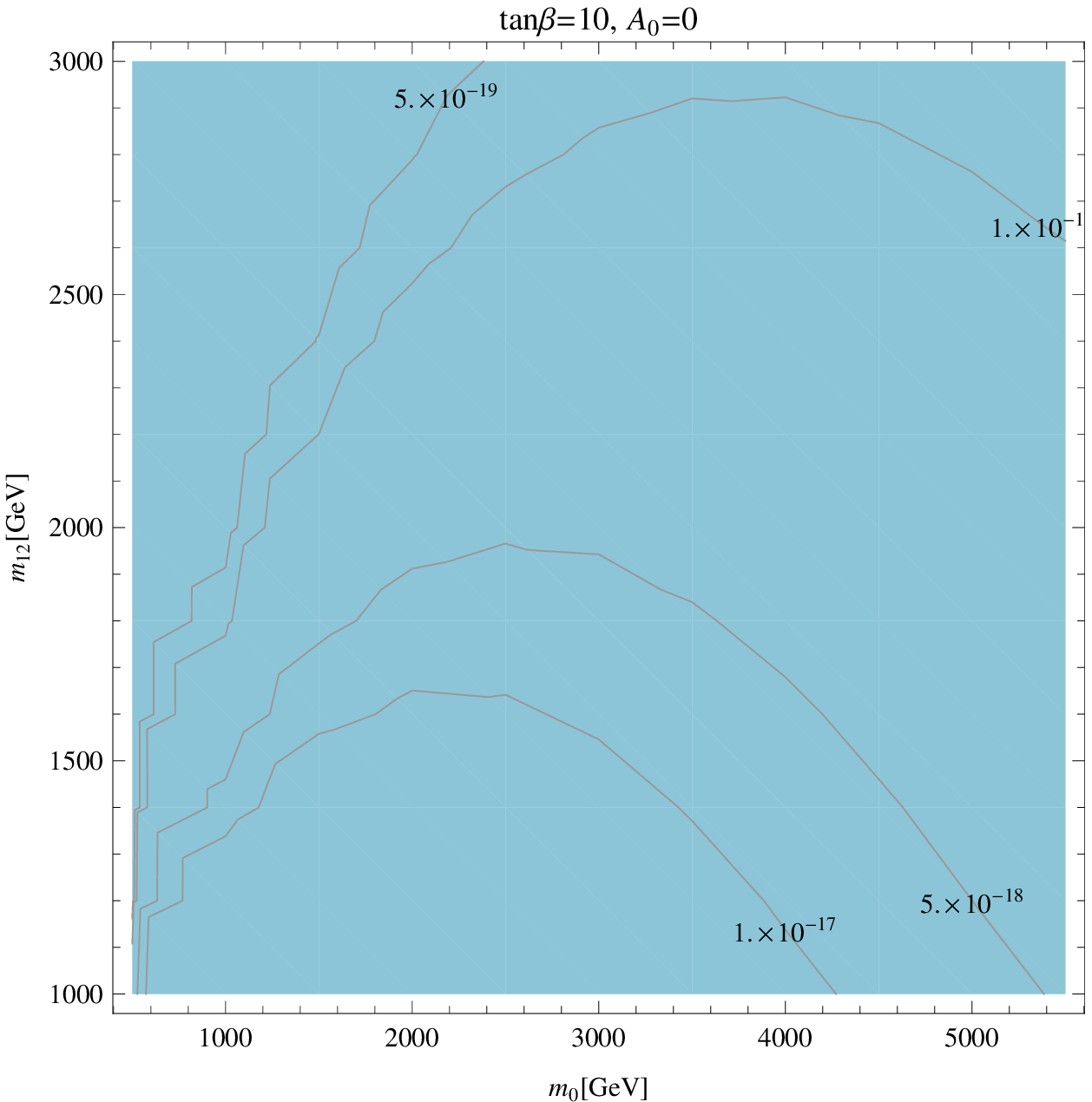  ,scale=0.51,angle=0,clip=}
\psfig{file=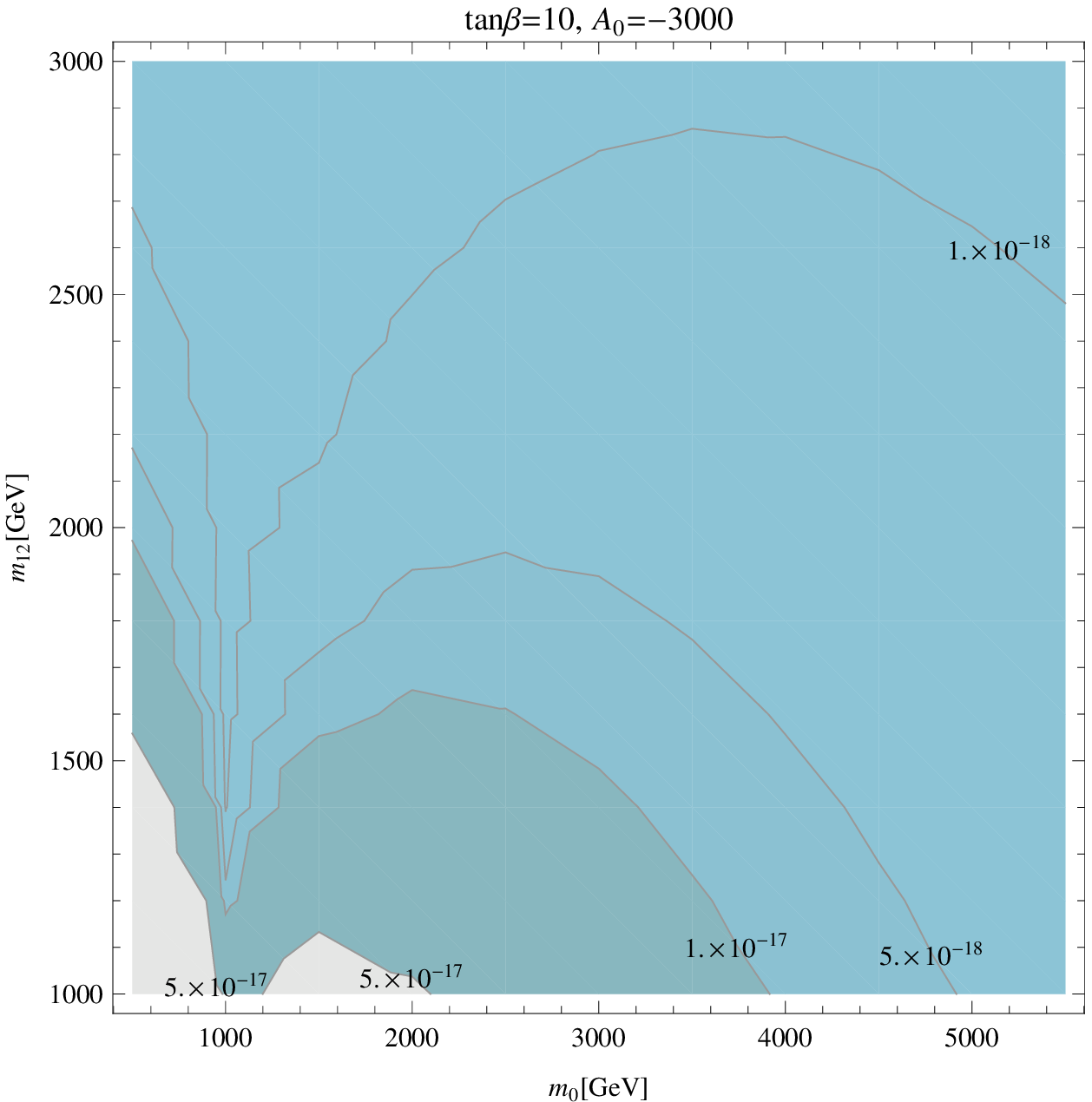  ,scale=0.51,angle=0,clip=}\\
\vspace{0.2cm}
\psfig{file=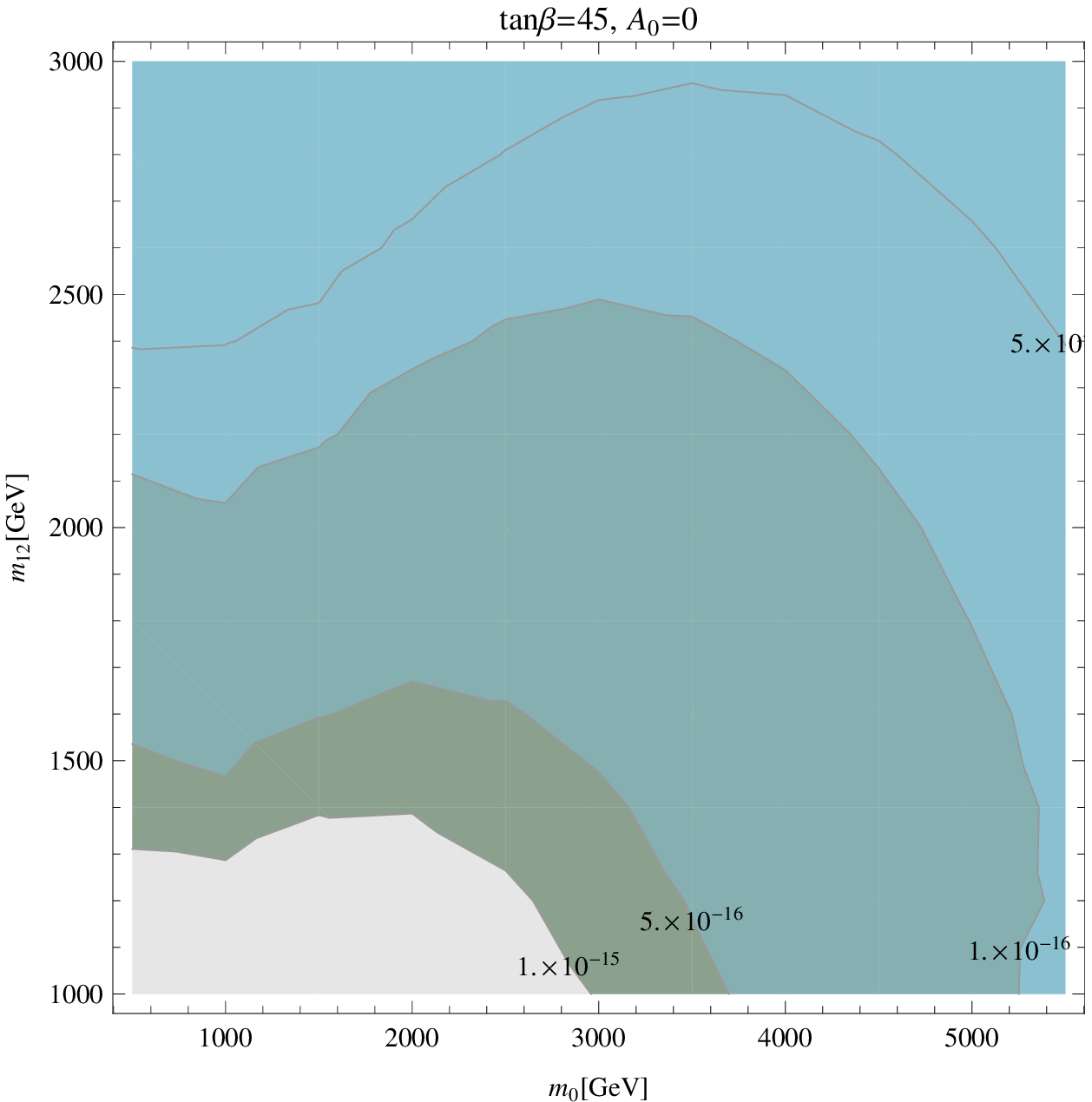 ,scale=0.51,angle=0,clip=}
\psfig{file=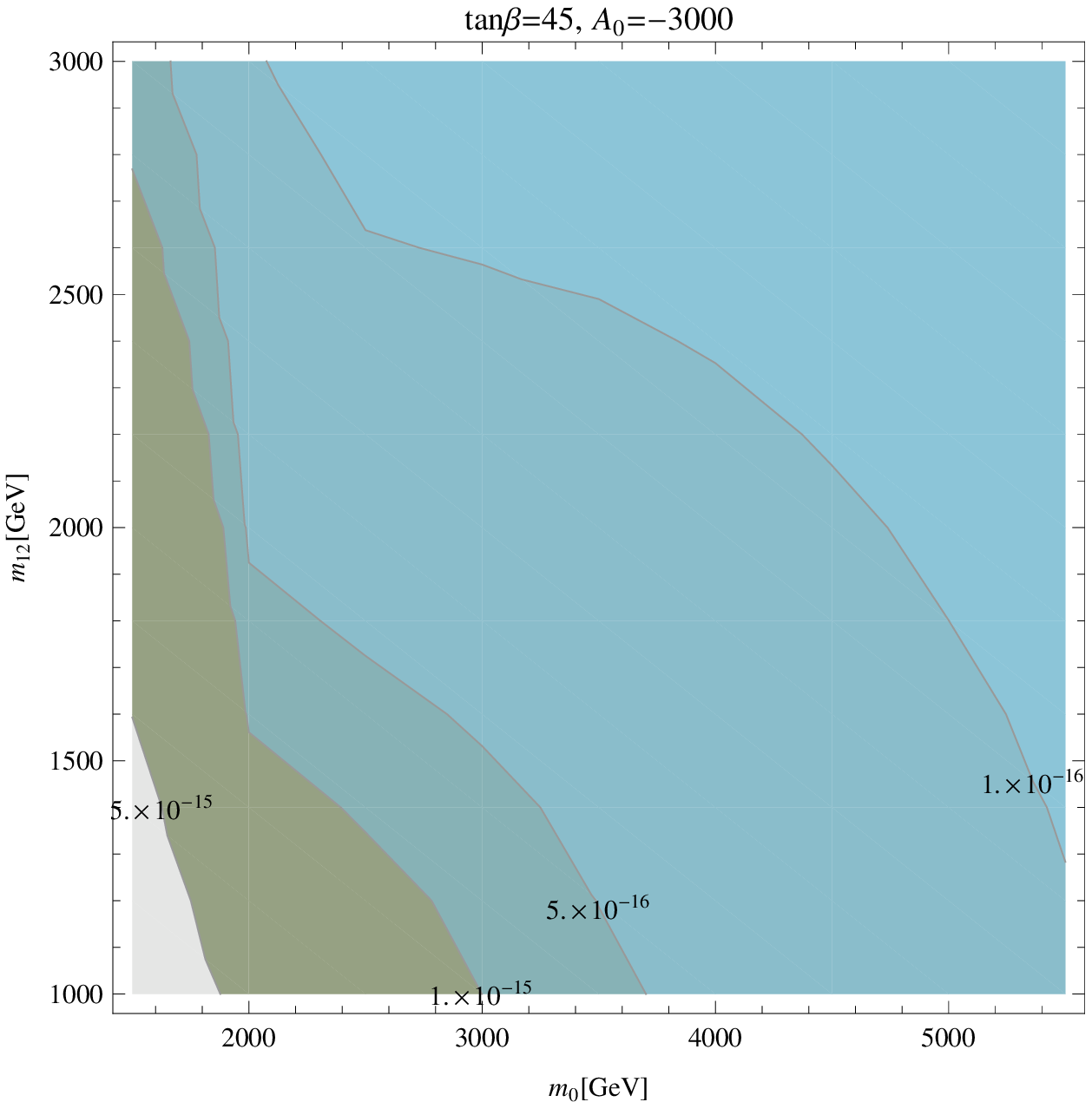   ,scale=0.51,angle=0,clip=}\\
\vspace{-0.2cm}
\end{center}
\caption[Contours of BR($h \to e \tau$)  in the
  $m_0$--$m_{1/2}$ plane]{Contours of BR($h \to e \tau$)  in the
  $m_0$--$m_{1/2}$ plane for different values of $\tb$ and $A_0$ in
  the \CMSSMI.}   
\label{fig:HTauESSI}
\end{figure}

\begin{figure}[ht!]
\begin{center}
\psfig{file=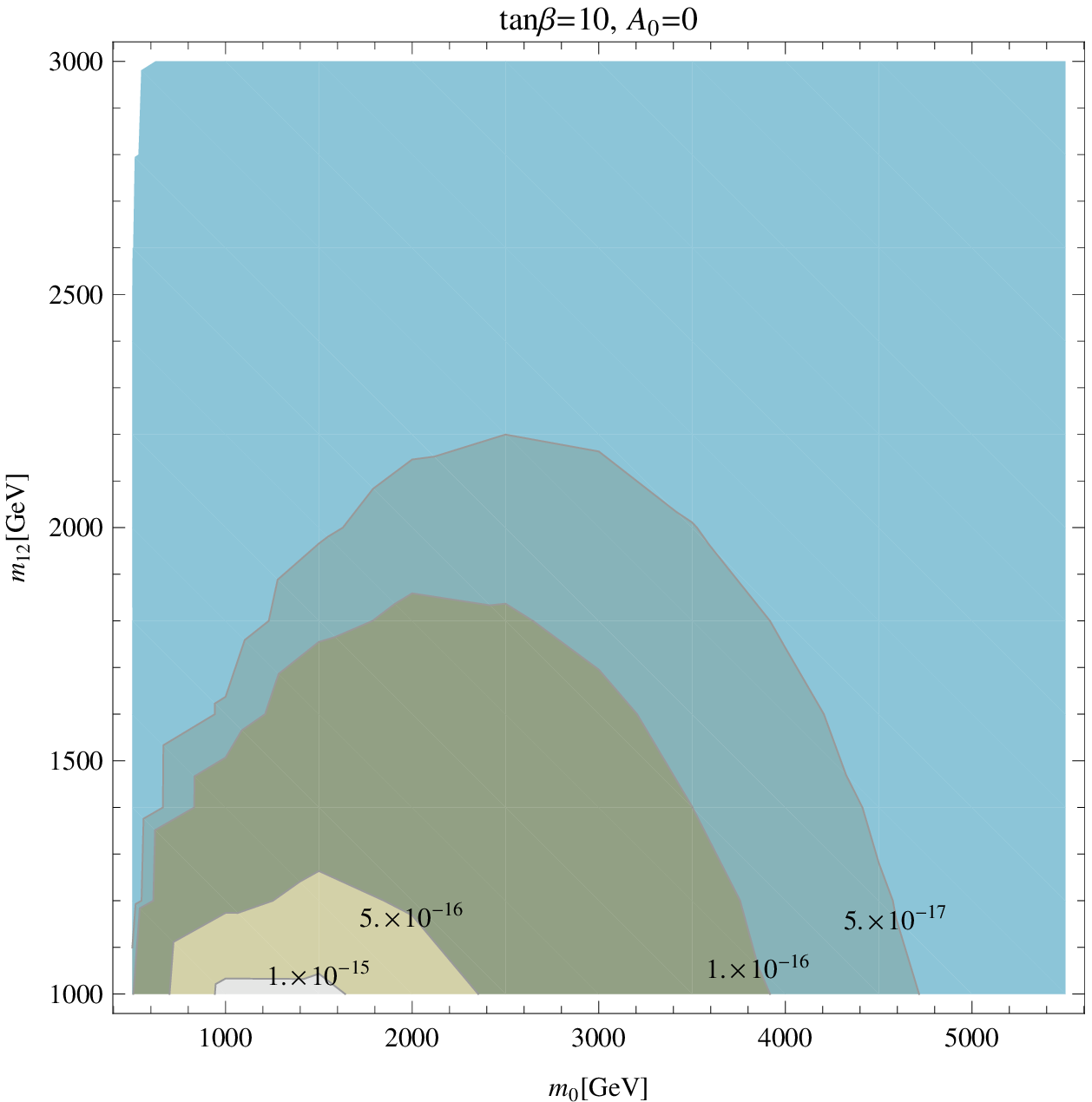  ,scale=0.51,angle=0,clip=}
\psfig{file=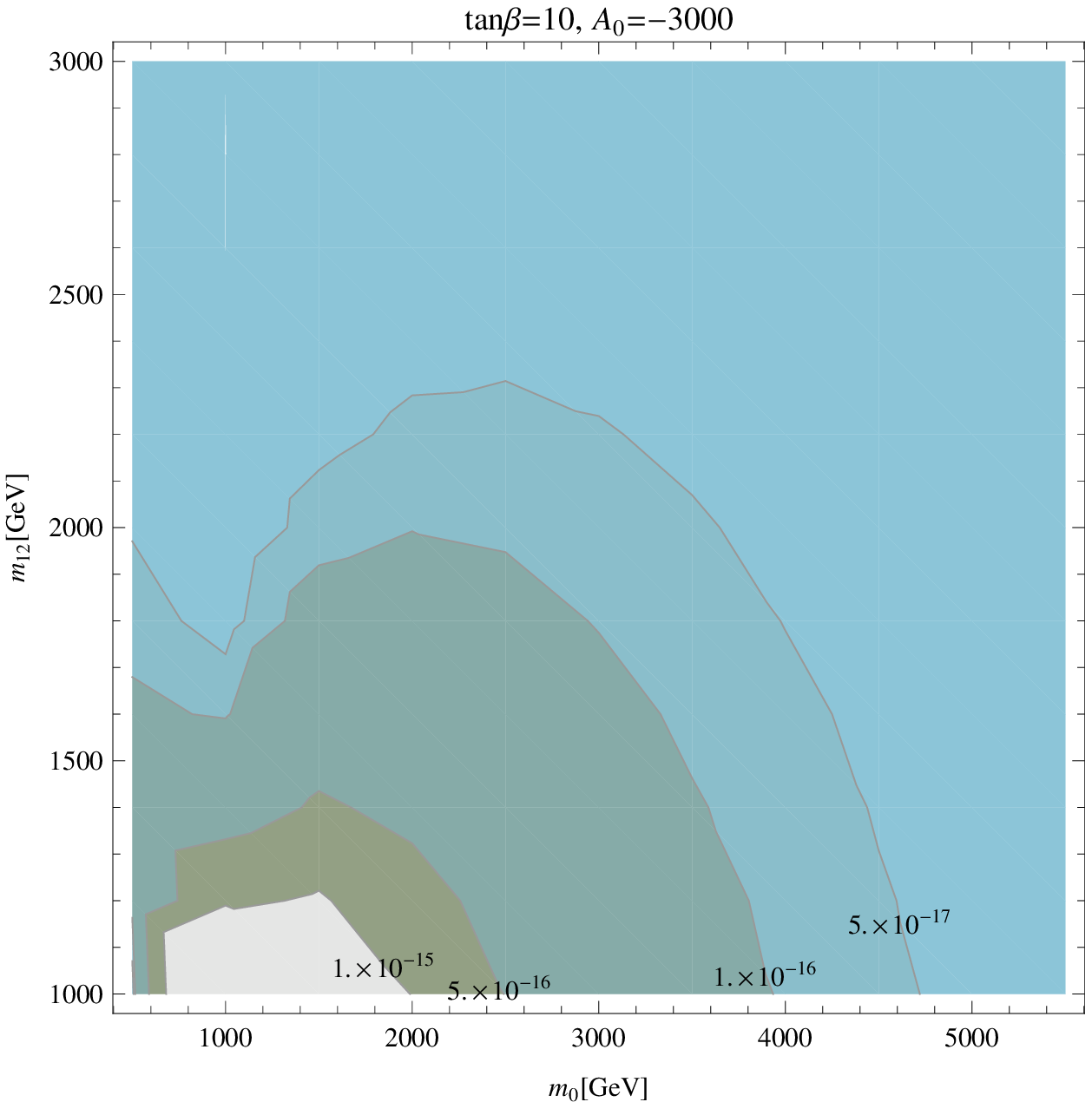  ,scale=0.51,angle=0,clip=}\\
\vspace{0.2cm}
\psfig{file=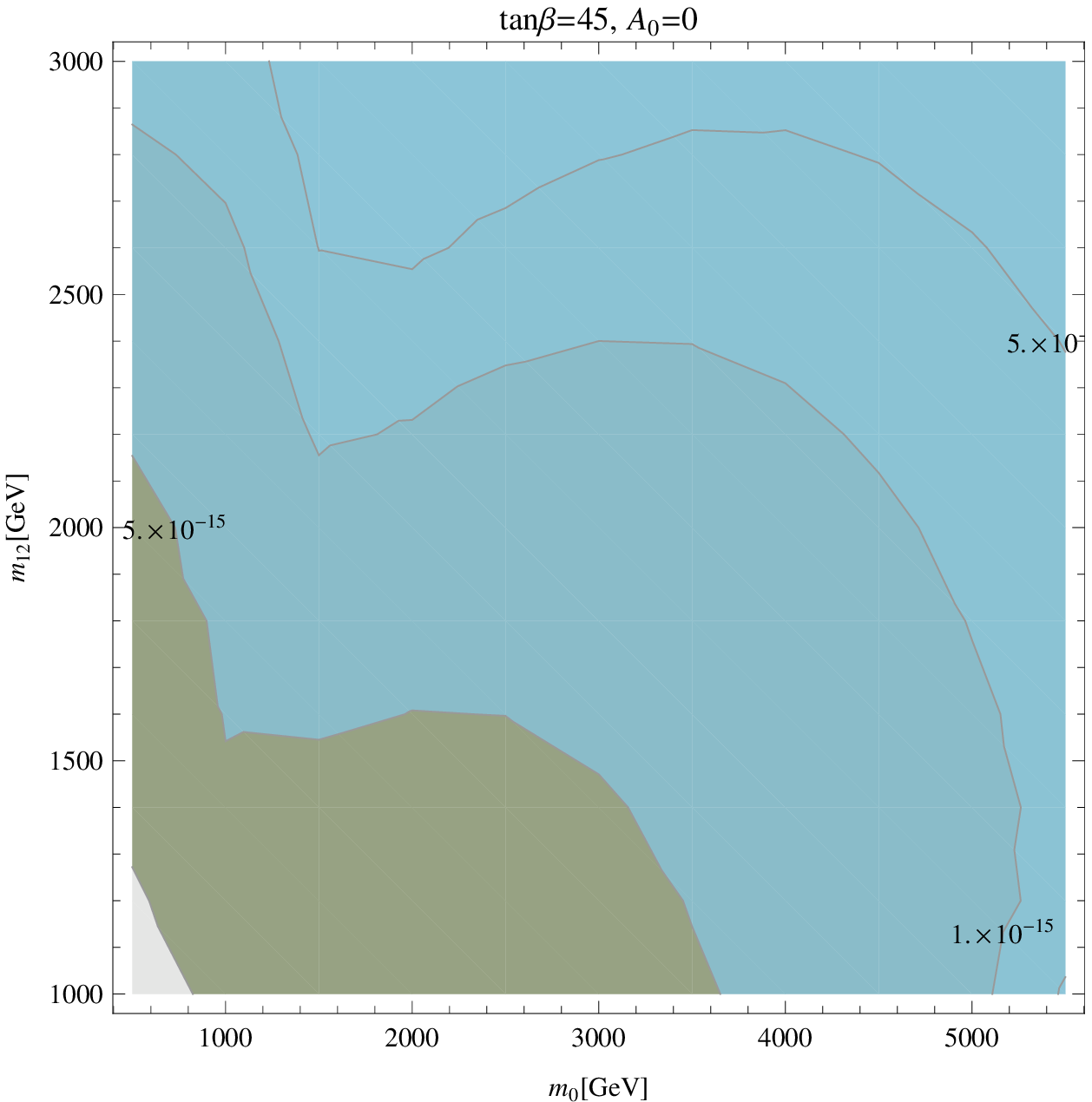 ,scale=0.51,angle=0,clip=}
\psfig{file=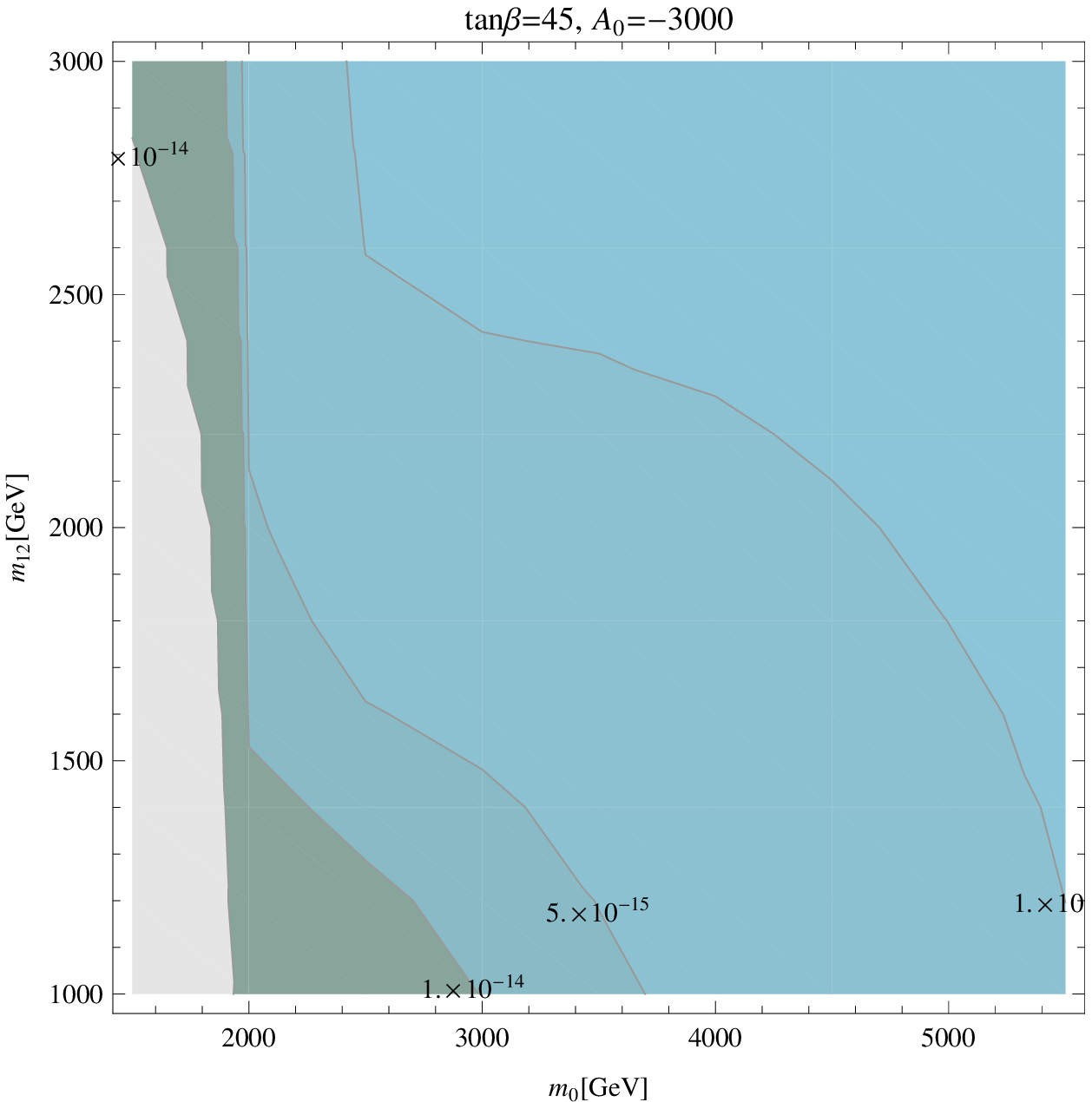   ,scale=0.51,angle=0,clip=}\\
\vspace{-0.2cm}
\end{center}
\caption[Contours of BR($h \to \tau \mu$)  in the
  $m_0$--$m_{1/2}$ plane]{Contours of BR($h \to \tau \mu$)  in the
  $m_0$--$m_{1/2}$ plane for different values of $\tb$ and $A_0$ in
  the \CMSSMI.}   
\label{fig:HTauMueSSI}
\end{figure}

\reffi{fig:HMueESSI} shows the results for BR($h \to e \mu$). The largest value is of the \order{10^{-16}} for low $m_0$ and $m_{1/2}$ values but is excluded from BR($\mu \to e \gamma$). In the allowed range they are typically \order{10^{-18}}. Similarly \reffi{fig:HTauESSI} and \reffi{fig:HTauMueSSI} shows the predictions for BR($h \to e \tau$) and BR($h \to \tau \mu$) respectively. Predictions of the \order{10^{-14}} and \order{10^{-12}} are possible for BR($h \to e \tau$) and BR($h \to \tau \mu$) in the lower left region of the $m_0$--$m_{1/2}$ plane respectively but are excluded from BR($\mu \to e \gamma$) bound. In the allowed region they are of the \order{10^{-16}} or less. These results are in a clear contradiction to the recently reported CMS excess\cite{CMSLFVHD}. If this excess seen in the CMS is confirmed in the future analysis, we will need models other than the \CMSSMI\ to explain this excess. However our findings are in agreement with the ATLAS reports\cite{ATLAS-LFVHD}, where they do not see any significant excess over background. It remains to be seen how these results will develop with the LHC Run II.  
\chapter{Summary \& Conclusions}
\label{sec:conclusions}

SUSY proves to be a very powerful and technically well equiped theory as it successfully explains some of the major deficiencies
of the SM, but still lacks experimental endorsement. Direct searches for
sparticles at LHC did not succeed so far.
The other way around is to probe SUSY via virtual effects of additional
particles to the precision observables. 
For example, in the MSSM, the fermion-sfermion misallignment can generate flavor changing effects that can dominate the SM effects by several orders of magnitude. Any possible experimental deviation from the SM results for the precision observales could be a hint of SUSY. Also, as this misallignment arises from the soft SUSY-breaking terms, this may provide guidlines for the SUSY model building. In this thesis, keeping the above mentioned points in mind, we studied the possible phenomenological consequences of flavor mixing to various observables. 

The flavor mixing was parameterized in terms of a set of dimensionless parameters $\deFABij$ ($F=Q,U,D,L,E; A,B = 
L,R$; $i,j = 1,2,3$). In chapter 1, we reviewed some aspacts of the SM, similarly in chapter 2, a general introduction to MSSM and its seesaw extension was discussed. Calculational details for the considered observables were given in chapter 3 where we presented the higher order corrections to the electroweak precision observables (EWPO), higher order corrections to Higgs boson masses, calculational details of the $B$-physics observables (BPO), quark flavor violating Higgs decays (QFVHD) and lepton flavor violating Higgs decays (LFVHD). In order to calculate slepton mixing effects (squark mixing was already present), we prepared an add-on model file for \fa\ to include lepton flavor violation in the already existing MSSM model file of the \fa. \fc\ drivers were also modified accordingly. The inclusion of lepton flavor violation (LFV) into \fa/\fc\ allowed us to 
calculate the one-loop LFV effects on EWPO 
(via the calculation of gauge-boson self-energies) as well on the 
Higgs-boson masses of the MSSM (via the calculation of the Higgs-boson 
self-energies).  The corresponding results have been included in the 
code \fh\ and are publicly available from version 2.10.2 on. 
We have (re-)caculated the decay \hbs\ in the
\fa\ and \fc\ setup. The BPO and EWPO constraints have been evalated
with the help of (a private version of) \fh, taking into account the
full flavor violating one-loop corrections to $\MW$ and to the relevant
BPO (supplemented with further MSSM higher-order
corrections). 

The effects of squarks mixing to EWPO, BPO and QFVHD such as $\hbs$ in the Model Independent (MI) approach were presented in chapter 4. This evaluation improved on existing analyses in various ways. We took into account the full set of SUSY QCD and SUSY EW corrections,
allowing for LL, RL, LR and RR mixing simultaneously. The parameter
space was restricted not only by BPO, but also by
EWPO, in particular the mass of the
$W$~boson. We have shown that $\MW$ can
yield non-trivial, additional restrictions on the parameter space of the
squark flavor violating $\deFABij$. 

In six representative scenarios, which are allowed by current
searches for SUSY particles and heavy Higgs bosons, we have evaluated
the allowed parameter space for the various $\deFABij$ by applying BPO
and EWPO constraints. Within these allowed ranges we have then evaluated
\brhbs. In the case of only one $\deFABij \neq 0$ we have found that
only relatively large values of $\del{DLR}{23}$ could lead to rates of 
$\brhbs \sim 10^{-4}$, which could be in the detectable range of future
$e^+e^-$ colliders. Allowing two $\deFABij \neq 0$
simultaneously lead to larger values up to $\brhbs \sim 10^{-3}$,
which would make the observation at the ILC relatively easy. Allowing
for a third $\deFABij \neq 0$, on the other hand, did not lead to
larger values of the flavor violating branching ratio.

The effects of slepton mixing to EWPO, Higgs boson masses and LFVHD in the MI approach were presented in chapter 5. 
The numerical analysis was performed on the basis of same six benchmark 
points as in the previous chapter. These benchmark 
points represent different combinations of parameters in the sfermion 
sector.  The restrictions on the various slepton $\deFABij$ in these six 
scenarios, provided by experimental limits on LFV processes (such as 
$\mu\to e\ga$) were taken from \citere{Arana-Catania:2013nha}, and 
the effects on EWPO and Higgs-boson masses were evaluated in the 
experimentally allowed ranges.  In this way we were able to provide a general 
overview about the possible size of LFV effects and potential new 
restrictions on the slepton $\deFABij$ from EWPO and Higgs-boson masses.

The LFV effects in the EWPO turned out to be sizable for $\del{LLL}{23}$
but (at least in the scenarios under investigation) negligible for the 
other $\deFABij$.  The effects of varying $\del{LLL}{23}$ in the 
experimentally allowed ranges turned out to exceed the current 
experimental uncertainties of $\MW$ and $\sweff$ in the case of heavy 
sleptons.  No new general bounds could be set on $\del{LLL}{23}$, 
however, since the absolute values of $\MW$ and $\sweff$ strongly depend 
on the choices in the stop/sbottom sector, which is disconnected from 
the slepton sector presently under investigation.  Such bounds could be 
set on a point-by-point basis in the LFV MSSM parameter space, however.  
Looking at the future anticipated accuracies, also lighter sleptons 
yielded contributions exceeding that precision.  It may therefore be 
possible in the future to set bounds on $\del{LLL}{23}$ from EWPO that 
are stronger than from direct LFV processes.

In the Higgs sector, based on evaluations for flavor violation in the 
squark sector, non-negligible corrections to the light $\cp$-even Higgs 
mass as well as to the charged Higgs-boson mass could be expected.  The 
associated theoretical uncertainties exceeded the anticipated future 
precision for $\Mh$ and $\MHp$.  Taking the existing limits on the 
$\deFABij$ from LFV processes into account, however, the corrections 
mostly turned out to be small.  For the light $\cp$-even Higgs mass they 
stay at the few-MeV level.  For the charged Higgs boson mass they can reach 
\order{2 \gev}, which, depending on the choice of the heavy Higgs-boson 
mass scale, could be at the level of the future experimental precision.  
More importantly, the theoretical uncertainty from LFV effects that 
previously existed for the evaluation of the MSSM Higgs-boson masses, 
has been reduced below the level of future experimental accuracy.

The predictions for the LFVHD in the MI approach were also presented in chapter 5. However due to very tight constraints on the slepton $\deFABij$'s from cLFV decays, the BR's for these processes turned out to be very small.     

Effects of squark mixing in the the CMSSM and slepton mixing in CMSSM extended by type I Seesaw under the
Minimal Flavor Violation (MFV) hypothesis were presented in chaptor 6. This work was motivated by the fact that in many
phenomenological analyses of the CMSSM the effects of intergenerational
mixing in the squark and/or slepton sector are neglected. However, such
mixings are naturally induced, assuming no flavor violation at the GUT
scale, by the RGE running from the GUT to the EW scale exactly due to the
presence of the CKM and/or the PMNS matrix. 
In this sense the CMSSM and the \CMSSMI\ represent two simple
``realistic'' GUT based models, in which flavor violation in induced solely by
RGE running. 
The spectra of the CMSSM and \CMSSMI\ have been numerically evaluated
with the help of the program {\tt SPheno} 
by taking the GUT scale input run down via the appropriate
RGEs to the EW scale.

We have evaluated the predictions for BPO, MSSM 
Higgs boson masses, EWPO in the CMSSM and
\CMSSMI. 
In order to numerically analyze the effects of neglecting
intergenerational mixing 
these observables have been evaluated with the full spectrum at the EW
scale, as well as with the spectrum, but with all intergenerational
mixing set artificially to zero (as it has been done in
many phenomenological analyses). 

The difference in the various observables
indicates the possible size of the effects neglected in those
analyses. In this way it can be checked whether neglecting those mixing
effects is a justified approximation. 

Within the CMSSM we have taken a fixed grid of $A_0$ and $\tb$,
while scanning the $m_0$--$m_{1/2}$ plane. We found that the value of 
$\deFABij$ increases with the increase of the $A_0$ or $\tb$ values.
The Higgs boson masses receive corrections below current and future
experimental uncertainties, 
where the shifts in $\MHp$ were found largest at the level of 
\order{100 \mev}. Similarly for the BPO the induced
effects are at least one order of magnitude smaller than the current
experimental uncertainty. For those two groups of observables the
approximation of neglecting intergenerational mixing explicitly is a
viable option. 

The picture changes for the EWPO. 
We find that the masses of the squarks grow with $m_0$, and thus do the mixing terms,
inducing a splitting between masses in an $SU(2)$ doublet, leading 
to a non-decoupling effect. 
For $m_0 \gsim 3 \tev$ the effects induced in $\MW$ and $\sweff$ are
found to be several times larger than the current experimental
uncertainties and could shift the CMSSM prediction outside the allowed
experimental range. In this way, taking the intergenerational mixing
into account could in principle set bounds
on $m_0$ that are not present in recent phenomenological analyses. 
By investigating numerically squark mass differences, we
have shown that this behavior can be traced back to the non-decoupling
effects in the scalar quark mass matrices, provided by {\tt Spheno} when
taking into account the CKM matrix in the RGE running.
However, we would like to point out that this bound only holds
because of the particularly simple structure of the CMSSM and cannot 
be extended easily to other, more complicated model frameworks.

In the final step of the numerical analysis within the CMSSM we have evaluated \brhbs. Here we
have found that for most of parameter space values
of \order{10^{-7}} are found for \brhbs, i.e.\ outside the reach of
current or future collider experiments.

Going to the \CMSSMI\ the numerical results depend on the
concrete model definition. We have chosen a set of parameter that 
reproduces correctly the observed neutrino data and simultaneously 
induces large LFV effects and induces {\em relatively} large corrections
to the calculated observables. Consequently, parts of the parameter
space are excluded by the experimental bounds on $\br(\mu \to e \ga)$. However $\br(\tau \to e \ga)$ and $\br(\tau \to \mu \ga)$ do not reach to their respective experimental limits. Again predictions for the BR of LFVHD turned out very small in \CMSSMI. We can conclude that we will need models other than the \CMSSMI\ to explain the CMS excess (if it persists) for the channel $\br(h \to \mu \tau)$.  
Concerning the precision observables we find that
BPO are not affected, we also find that the
additional effects induced by slepton flavor violation on Higgs boson
masses are negligible. Again the EWPO are found to show the largest
impact, where for $\MW$ effects at the same level as the
current experimental accuracy have been observed for very large values
of $m_0$. As above, we would like to point out that these effects are due to
the relatively simple structure of the \CMSSMI.

To summarize our MFV analysis: we have analyzed two ``realistic'' GUT based models in which
flavor violation is solely induced by the CKM matrix via RGE running (as
evaluated using the {\tt Spheno code}). We find that
artificially setting all flavor violating terms to zero in
the CMSSM and \CMSSMI\ is an acceptable approximation for BPO, Higgs boson masses (evaluated using a private version of 
\fh). However, in the EWPO (also evaluated with \fh) 
in our numerical analysis we find larger effects in the CMSSM and \CMSSMI. 
The numerical contributions are larger than the current 
experimental accuracy in $\MW$ and $\sweff$. Taking those effects
correctly into account could in principle place new bounds on $m_0$ that
are not present in recent phenomenological analyses.

%% file: sum_es.tex
\chapter*{Resumen y Conclusiones}

\addcontentsline{toc}{chapter}{Resumen y Conclusiones}

La teor\'{\i}a supersim\'etrica ha demostrado un  enorme potencial para explicar algunas de los mayores problemas del Modelo Est\'ander (ME), aunque hasta la fecha no se haya encontrado ninguna evidencia experimental de sus predicciones. Por ejemplo,  la b\'usqueda directa de part\'{\i}culas supersim\'etricas  no ha tenido \'exito por el momento. Sin embargo, es posible detectar la presencia de las nuevas part\'{\i}culas en los cambios que \'estas producen en algunos par\'ametros medidos con gran precisi\'on.  En particular, el modelo supersim\'etrico m\'{\i}nimo (MSSM) predice nuevas contribuciones al cambio de sabor (FC)  de los fermiones debido a mezclas entre las masas de sus correspondientes parejas supersim\'etricas. Esta mezcla est\'a originada por los par\'ametros responsables de la rotura de la Supersimetr\'{\i}a, lo cual tiene un gran inter\'es desde el punto de vista del dise\~no de modelos supersim\'etricos concretos. El cambio de sabor derivado  de la no alineaci\'on entre fermiones y sus parejas escalares no se manifiesta en la aproximaci\'on a nivel m\'as bajo (“tree level”) de los c\'alculos, pero s\'{\i} en el primer orden (“one-loop” level) cuya contribuci\'on puede ser importante para ciertos valores de los par\'ametros del MSSM.  

En esta tesis, se ha estudiado la posible contribuci\'on  de la mezcla de sabor fermi\'onico a varias observaciones. La mezcla de sabor se ha introducido por medio de un conjunto de par\'ametros adimensionales denominados  $\deFABij$ ($F=Q,U,D,L,E; A,B = L,R$; $i,j = 1,2,3$).  En el cap\'{\i}tulo 1, se revisaron  algunos aspectos del ME;  en el cap\'{\i}tulo 2, se introdujo el MSSM y la extensi\'on de \'este que incluye un mecanismo del tipo “see-saw” para explicar las oscilaciones de sabor de los neutrinos. Los detalles de la contribuci\'on SUSY a algunos observables de inter\'es se presenta en el cap\'{\i}tulo 3, en concreto se consideran: observables  de la teor\'{\i}a electro-d\'ebil medidos con gran precisi\'on (EWPO), correcciones a la masa del bos\'on de Higgs, detalles en el c\'omputo de la f\'{\i}sica relacionada con el quark b ($B$-physics observables (BPO)), desintegraciones del bos\'on de Higgs con violaci\'on de sabor de quark (QFVHD) y finalmente, desintegraciones del bos\'on de Higgs con violaci\'on de sabor lept\'onico (LFVHD). Para calcular los efectos de la mezcla del sector lept\'onico se elabor\'o  un algoritmo adicional para  \fa, con \'el se incluye LFV en el modelo del MSSM que el paquete ya tiene definido. Con ello ampliamos la capacidad de los programas incluidos en \fa/\fc\ para computar el efecto del LFV en observables como  EWPO (a partir del c\'omputo de las auto-energ\'{\i}as de los bosones gauge) y tambi\'en sobre la masa de los bosones de Higgs del MSSM. Los resultados correspondientes han sido incluidos en el el c\'odigo  \fh\ y est\'an disponibles para su libre distribuci\'on a partir de la versi\'on   2.10.2.  Se revis\'o  el c\'alculo de la desintegraci\'on  $\hbs$ utilizando  los c\'odigos actualizados de
\fa\ y \fc. Los c\'alculos para evaluar los observables  BPO y EWPO se realizaron con la ayuda de \fh\ (utilizando una versi\'on no p\'ublica), teniendo en cuenta la contribuci\'on a la  de todos los t\'erminos que violan sabor, en el caso de $\MW$  y de los m\'as relevantes en el caso de BPO.  

En el cap\'{\i}tulo 4 se han estudiado los efectos de la mezcla de sabor de los squarks en la observaci\'on de  EWPO, BPO y QFVHD (por ejemplo en $\hbs$) de una manera independiente del modelo (MI) que produce la mezcla de sabor. Nuestro c\'alculo mejora otros previos en varios aspectos: se ha tenido  en cuenta el total de las correcciones supersim\'etricas del tipo fuerte y electro-d\'ebil y, adem\'as, se permiti\'o la mezcla simult\'anea de contribuciones del tipo LL, RL, LR y RR. Tambi\'en se   consider\'o la limitaci\'on del valor de los par\'ametros impuesta no solo por los BPO, sino tambi\'en por los EWPO, en particular la masa del bos\'on $\MW$. Se mostr\'o que la contribuci\'on a $\MW$ produce restricciones adicionales al espacio de los par\'ametros $\deFABij$ que mezclan el sabor de los squarks. 

En la  evaluaci\'on de los posibles valores de los par\'ametros $\deFABij$  se han teniendo en cuenta las limitaciones procedentes de los valores de los BPO y EWPO. Para ello,  se consideraron seis escenarios representativos no excluidos ni por la b\'usqueda de part\'{\i}culas SUSY ni  por el valor experimental de la masa del bos\'on de Higgs.  Los valores de $\deFABij$ obtenidos se usaron para calcular $\brhbs$. En el caso de tomar s\'olo uno de los  $\deFABij \neq 0$  se encontr\'o que \'unicamente  valores relativamente grandes de  $\delta^{DLR}_{23}$ predicen valores de $\brhbs \sim 10^{-4}$,  detectables en futuros  colisionadores $e^+e^-$ .  Permitiendo dos $\deFABij \neq 0$ simult\'aneamente se obtiene un valor  $\brhbs \sim 10^{-3}$, que podr\'{\i}a observarse en el ILC.  En cambio, si se permite un tercer $\deFABij \neq 0$, no incrementa m\'as el valor de esas predicciones.

En el cap\'{\i}tulo 5 se estudiaron los efectos de la mezcla de sabor de los sleptones en los EWPO, las masas de los bosones de Higgs de una manera independiente del modelo que origina la mezcla del sabor.  Nuestro an\'alisis num\'erico toma como referencia seis modelos supersim\'etricos  a los que se atribuyen determinados valores de los par\'ametros de modo que las propiedades de los fermiones sean diferentes en cada uno de los casos.  Los valores de los $\deFABij$ en los seis escenarios est\'an limitados por la no observaci\'on de la violaci\'on del sabor lept\'onico (LFV) en procesos como $\mu\to e \ga$ \cite{Arana-Catania:2013nha}, lo que se ha tenido en cuenta en la evaluaci\'on del los EWPO y las masas de los bosones de Higgs.  De este modo hemos podido computar, de una manera general, el posible impacto de las restricciones en el valor de los  $\deFABij$  debido a \'estos observables  en la predicci\'on de procesos con LFV.  Encontramos que éstas son considerables para  $\del{LLL}{23}$ e insignificantes para el resto de los $\deFABij$, al menos en los escenarios considerados en nuestra investigaci\'on.  El efecto de variar los valores de $\del{LLL}{23}$ dentro de los intervalos permitidos experimentalmente implica contribuciones para $\MW$  y  $\sweff$ que pueden exceder  el margen de error con el que est\'an medidos. Sin embargo, esto no implica nuevas restricciones en los valores de $\del{LLL}{23}$, ya que los valores absolutos de $\MW$  y  $\sweff$  dependen en gran medida del  s-top y el  s-bottom. Este sector est\'a desconectado del  de los sleptones, objeto de nuestra investigaci\'on. Sin embargo, en algunos casos,  los limites que resultan de los  EWPO pueden ser m\'as restrictivos que los procedentes de la medida directa de procesos con LFV.  En el sector de los bosones de Higgs, la introducci\'on de violaci\'on de sabor de los s-quarks implica contribuciones no triviales a las masas del  bos\'on de Higgs neutro m\'as ligero $M_h$ y a la  del cargado $M_{H^{\pm}}$. En ambos casos,  la incertidumbre te\'orica en su determinaci\'on es  superior a la experimental. Si consideramos los valores para $\deFABij$ permitidos por los l\'{\i}mites de procesos con LFV , la contribuci\'on a ambas masas es peque\~na. Para $M_h$ es del orden de unos pocos MeVs mientras para $M_{H^{\pm}}$ puede llegar hasta 2 GeV. Esta \'ultima puede alcanzar el valor  de la futura precisi\'on experimental, dependiendo  de la masa del bos\'on de Higgs neutro m\'as pesado. Pero   lo m\'as relevante, es el hecho de que la incertidumbre derivada de los efectos de LFV en la evaluaci\'on de las masas de los bosones de Higgs neutros se ha reducido hasta hacerse del mismo orden  que la que se prev\'e alacanzar en experimentos futuros.
En el cap\'{\i}tulo 5 tambien se presentaron las predicciones para LFVHD siguiendo la t\'ecnica  MI. Sin embargo, las severas restricciones impuestas por los  procesos con  cLFV hacen que las predicciones para esos procesos sean muy peque\~nas.

En el cap\'{\i}tulo 6 se estudiaron los efectos de la mezcla de squarks en el CMSSM y de sleptones en la extensi\'on de \'este con un mecanismo ``see-saw'' de tipo I. En ambos casos se  utiliz\'o  la hip\'otesis de violaci\'on de sabor m\'{\i}nima (MFV). Este trabajo fue motivado por el hecho de que muchos an\'alisis fenomenol\'ogicos del CMSSM no incluyen estos efectos. Sin embargo, aparecen de manera natural  en la evoluci\'on de los par\'ametros del modelo entre las escalas de energ\'{\i}a de gran unficiación (GUT) y 
electro-d\'ebil (EW) debido a la presencia de las matrices CKM y PMNS en las RGE´s. En este sentido, los modelos CMSSM y \CMSSMI\ constituyen dos ejemplos sencillos de modelos con gran unificaci\'on en  los que la violaci\'on de sabor procede \'unicamente de las RGE. El espectro de masas de las part\'{\i}culas supersim\'etricas en ambos casos se ha evaluado num\'ericamente mediante el programa {\tt Spheno}, a partir de los valores a la escala GUT. 
Se calcularon las predicciones para BPO, y la masa de los bosones de Higgs en el  CMSSM y el \CMSSMI.  Se evalu\'o el impacto de incluir la mezcla de sabor comparando el c\'omputo con el caso simple en el que \'esta se desprecia, como ocurre en otros an\'alisis previos al nuestro. Los resultados indican en qu\'e casos pueden ignorarse las mezclas de sabor. 

En el caso del CMSSM se ha hecho un recorrido a trav\'es de una red de valores en el plano $m_0$--$m_{1/2}$ para valores fijos de  $A_0$ y  $\tb$. Se encontr\'o que el valor de $\deFABij$ aumenta al incrementar los valores de \'estos \'ultimos valores. Los valores de las correcciones a las masas de los bosones de Higgs son inferiores a la precisi\'on en su valor experimental (presente y futuro) . De manera an\'aloga el impacto sobre los BPO est\'a por debajo de la incertidumbre en su medida experimental. Por tanto, encontramos que para  estos dos grupos de observables est\'a justificado el ignorar los efectos de la mezcla.  La conclusi\'on es diferente en el caso de los EWPO, aqu\'{\i} encontramos que las masas de los squarks aumentan con $m_0$, y con ello los par\'ametros de mezcla, esto genera una diferencia entre las masas del doblete de $SU(2)$ que debe ser tenida en cuenta en el c\'omputo de estos observables. Para $m_0 \gsim 3 \tev$ encontramos que induce a valores  de $\MW$ y $\sweff$ que superan varias veces la incertidumbre experimental, hasta el punto llevar a las predicciones del CMSSM fuera de los l\'{\i}mites experimentales. De esta manera, nuestro an\'alisis permite establecer l\'{\i}mites para la masa del $m_0$ que no aparec\'{\i}an en trabajos anteriores. El origen de la diferencia en las masas de los squarks responsable de \'esta contribuci\'on fue corroborado num\'ericamente utilizando {\tt Spheno}, con el que se  compararon los efectos de incluir o ignorar la matriz CKM en la integraci\'on de las RGE. Sin embargo, debemos se\~nalar que nuestras conclusiones son dif\'{\i}ciles de extrapolar a modelos m\'as complejos que los aqu\'{\i} utilizados.

El \'ultimo eslab\'on de nuestro an\'alisis con el CMSSM ha sido la evaluaci\'on del  $\brhbs$. En este caso, encontramos valores del orden de ${10^{-7}}$. Esto es, fuera del alcance de su detecci\'on en los experimentos proyectados para un futuro pr\'oximo. En el caso del \CMSSMI\, los resultados num\'ericos dependen de c\'omo est\'e definido el modelo. Se eligi\'o un conjunto de par\'ametros que reproduce correctamente las observaciones referentes a los neutrinos y a su vez induce a contribuciones apreciables de LFV en los observables que estudiamos.  En consecuencia, algunas regiones del espacio de par\'ametros est\'an exlu\'{\i}das por su predicci\'on a  $\br(\mu \to e \ga)$.  En cambio,  las de los $\br(\tau \to e \ga)$ y  $\br(\tau \to \mu \ga)$ no alcanzan sus respectivos l\'{\i}mites experimentales.  Las predicciones para los BR de  LFVHD  son muy peque\~nas tambi\'en en el \CMSSMI. Con ello concluimos que precisamos de modelos diferentes del \CMSSMI\ para explicar la observaci\'on de $\br(h \to \mu \tau)$  en el detector CMS del CERN. En lo tocante a los observables medidos con gran precisi\'on, encontramos que los BPO no est\'an afectados. Tampoco las predicciones de las masas de los bosones de Higgs. El mayor impacto aparece una vez m\'as en los  EWPO, en el caso de la  $\MW$ pueden ser del orden de la incertidumbre experimental. 

En resumen, en nuestro estudio de MFV hemos utilizado dos modelos con gran unificaci\'on ``realistas'' en los que la violaci\'on de sabor es introducida al tener en cuenta la presencia de las  matrices CKM y PMNS en las RGE's. Se encontr\'o que el desestimar los efectos de violaci\'on de sabor es adecuado para los BPO y  las masas de los bosones de Higgs. Sin embargo, para los EWPO se encontraron efectos grandes. El valor de la contribuci\'on a $\MW$ y $\sweff$ es superior a la incertidumbre de su valor experimental. Por tanto,  los efectos de violaci\'on de sabor, objeto de nuestro estudio  imponen un nuevo l\'{\i}mite superior a $m_0$ que no se ha tenido en cuenta en otros an\'alisis fenomenol\'ogicos recientes.